\documentclass[a4paper,12pt,twoside]{book}

\usepackage[english]{babel}
\usepackage[utf8]{inputenc}
\usepackage[super]{nth}
\usepackage{amsmath}
\usepackage{amssymb}
\usepackage{amsthm}
\usepackage{etoolbox}
\usepackage{enumitem}
\usepackage{bbm}
\usepackage{mathtools}
\usepackage{tikz-cd}
\usepackage{xcolor}
\usepackage{colortbl}
\usepackage{authblk}
\usepackage{url}
\usepackage{hyperref}
\usepackage{verbatim}
\usepackage{tabularx}
\usepackage{booktabs}
\usepackage{longtable}
\usepackage[most]{tcolorbox}
\tcbuselibrary{theorems}

\theoremstyle{remark}
\newtheorem{notation}{Notation}

\newtcbtheorem[number within=chapter]{theorem}{Theorem}
{	
	colback= white,
	colframe=red!70!black,
	fonttitle=\bfseries,
	breakable
}
{theorem}

\newtcbtheorem[
use counter from=theorem,
number within=chapter
]{corollary}{Corollary}
{
	colback=white,
	colframe=blue!70!black,
	fonttitle=\bfseries,
	breakable
}
{corollary}

\newtcbtheorem[
use counter from=theorem,
number within=chapter
]{proposition}{Proposition}
{
	colback=white,
	colframe=green!50!black,
	fonttitle=\bfseries,
	breakable
}
{proposition}

\newtcbtheorem[
use counter from=theorem,
number within=chapter
]{lemma}{Lemma}
{
	colback=white,
	colframe=brown!70!black,
	fonttitle=\bfseries,
	breakable
}
{lemma}

\newtcbtheorem[
use counter from=theorem,
number within=chapter
]{definition}{Definition}
{
	colback=white,
	colframe=black,
	fonttitle=\bfseries,
	breakable
}
{definition}

\patchcmd{\endproof}
{\endtrivlist}
{\endtrivlist\par\nobreak\vspace*{\dimexpr-\baselineskip-\parskip}\nobreak\noindent\hrulefill}
{}{}

\newcommand{\naturalNumbers}{\mathbb{N}}
\newcommand{\integers}{\mathbb{Z}}
\newcommand{\realNumbers}{\mathbb{R}}
\newcommand{\complexNumbers}{\mathbb{C}}

\newcommand{\id}{\mathrm{id}}
\newcommand{\dom}{\mathrm{dom}}
\newcommand{\cod}{\mathrm{cod}}
\newcommand{\Lin}{\mathrm{Lin}}
\newcommand{\cat}{\mathcal}
\newcommand{\Set}{\mathrm{Set}}

\newcommand{\lev}{\overleftarrow{ev}}
\newcommand{\rev}{\overrightarrow{ev}}
\newcommand{\lcoev}{\overleftarrow{coev}}
\newcommand{\rcoev}{\overrightarrow{coev}}
\newcommand{\tr}{\mathrm{tr}}
\newcommand{\image}{\mathrm{Im} \,}

\newcommand{\bra}[1]{{\langle#1|}}
\newcommand{\ket}[1]{{|#1\rangle}}
\newcommand{\norm}[1]{||#1||}

\newcommand{\rank}{\mathrm{rank}}

\newcommand{\density}{\mathcal{D}}
\newcommand{\CJ}{Choi-Jamio\l{}kowski }
\newcommand{\cp}{\mathrm{CP}}
\newcommand{\cptp}{\mathrm{CPTP}}
\newcommand{\ucp}{\mathrm{UCP}}
\newcommand{\ucptp}{\mathrm{UCPTP}}
\newcommand{\psd}{\mathrm{PSD}}
\newcommand{\psdi}{\mathrm{PSD}_1 }
\newcommand{\psdii}{\mathrm{PSD}_2 }
\newcommand{\psdiii}{\mathrm{PSD}_{1,2} }
\newcommand{\hermitian}{\mathrm{H}}
\newcommand{\CR}{\mathrm{CR}}
\newcommand{\radius}{\mathrm{radius}}

\newcommand{\mySmallerRestriction}[1]{|_{#1}}

\newcommand{\GL}{\mathrm{GL}}
\newcommand{\U}{\mathrm{U}}
\newcommand{\SU}{\mathrm{SU}}
\newcommand{\finvec}{\mathrm{FinVec}}
\newcommand{\finhilb}{\mathrm{FinHilb}}
\newcommand{\unitObj}{\mathbbm{1}}

\newcommand{\interior}[1]{\mathrm{int} \, #1}

\newcommand{\supp}{\mathrm{supp}}

\allowdisplaybreaks

\title{On quantum channels: extreme points, topology, and categorical properties}

\author{J. M. S. T. da Silva}

\begin{document}

\maketitle

\tableofcontents

\cleardoublepage

\setcounter{page}{1}
\pagenumbering{arabic}

\chapter{Notation}

\section{Symbols}

\begin{flushleft}
	\begin{tabular}{l l}
		$\id_X$ & the identity function over $X$, given by $\id_X(x)=x$\\
		$\ket{x}$ & a ket, which is a vector $x \in X$ in the Dirac notation\\
		$\bra{x}$ & a bra, which is the $\complexNumbers$-linear map $\bra{x} \colon X \to \complexNumbers$ given \\
		& by $\bra{x} (x') = \langle x \mid x' \rangle$ \\
		$\langle x | x' \rangle$ & the inner product $\langle x , x' \rangle$ in the Dirac notation\\
		$\dim_F$ & the dimension of a vector space or manifold with\\
		& respect to the field $F$\\
		$d_X$ & the same as $\dim_\complexNumbers X$\\
		$[A]$ & matrix of a linear transformation $A$ with respect to\\
		& some basis\\
		$A^\dag$ & the adjoint $A^\dag \colon Y \to X$ of a linear transformation \\
		& $A \colon X \to Y$, or the conjugate transpose of a matrix $A$\\
		$\tr(A)$ & (total) trace of an operator or matrix\\
		$\tr_i(A)$ & partial trace of $A$ over the $i$-th factor of $X_1 \otimes \dots \otimes X_n$\\
		$\Lin_F(X,Y)$ & $F$-linear maps $A \colon X \to Y$ \\
		$\Lin_F(X)$ & the same as $\Lin_F(X,X)$ \\
		$\hermitian(X)$ & hermitian operators in $\Lin_\complexNumbers(X)$\\
		$\U(X)$ & unitary operators in $\Lin_\complexNumbers(X)$\\
		$\U(n)$ & unitary matrices in $M_n(\complexNumbers)$\\
		$\SU(n)$ & unitary matrices $U \in \U(n)$ with $\det(U)$ = 1 \\
		$\complexNumbers P^n$ & the complex projective space\\
		$V_{n,r}(\complexNumbers)$ & the (compact) Stiefel manifold of matrices\\
		& $V \in M_{n,r}(\complexNumbers)$ such that $V^\dag V = I_r$\\
	\end{tabular}
	\begin{tabular}{l l}
		$\cp(X,Y)$ & set of completely positive maps\\
		& $\varepsilon \colon \Lin_\complexNumbers(X) \to \Lin_\complexNumbers(Y)$\\
		$\cptp(X,Y)$ & completely positive and trace preserving maps\\
		$\ucp(X,Y)$ & unital completely positive maps\\
		$\ucptp(X,Y)$ & unital completely positive and trace preserving maps\\
		$\CR(\varepsilon)$ & Choi-rank of a CP map $\varepsilon$\\
		$J$ & the \CJ isomorphism\\
		$\psd(X)$ & positive semidefinite operators in $\Lin_\complexNumbers (X)$\\
		$\psdi(X,Y)$ & operators $A \in \psd(X^* \otimes Y)$ with $\tr_1 \, A = \id_Y$\\
		$\psdii(X,Y)$ & operators $A \in \psd(X^* \otimes Y)$ with $\tr_2 \, A = \id_{X^*}$\\
		$\psdiii(X,Y)$ & the same as $\psdi(X,Y)\cap \psdii(X,Y)$\\
		$\psd(X)_r$ & operators in $\psd(X)$ with rank $r$\\
		$\psdi(X,Y)_r$ & operators $A \in \psdi(X,Y)$ with rank $r$ \\
		$\psdii(X,Y)_r$ & operators $A \in \psdii(X,Y)$ with rank $r$ \\
		$\psdiii(X,Y)_r$ & operators $A \in \psdiii(X,Y)$ with rank $r$ \\
		$\density(X)$ & density operators, which are operators $\rho \in \psd(X)$ \\
		& with $\tr(\rho) = 1$ \\
		$\lfloor x \rfloor$ & the floor of $x \in \realNumbers$, which is the unique $n \in \integers$ such that \\
		& $n \leq x < n+1$ \\
		$\lceil x \rceil$ & the ceil of $x \in \realNumbers$, which is the unique $n \in \integers$ such that \\
		& $n -1 < x \leq n$ \\
	\end{tabular}
\end{flushleft}

\section{Dirac notation}

This is a thesis on Mathematics, so knowledge of Dirac notation can't be assumed. Dirac notation is a notation for vectors in Hilbert spaces and linear transformations between Hilbert spaces. There is a review of the Dirac notation in section \ref{section: dirac notation appendix}, but we briefly discuss it here too. We denote the inner product by $\langle \ , \ \rangle$. The convention in Quantum Physics is that the inner product is antilinear in the first argument and linear in the second.

Let $X$ and $Y$ be finite dimensional complex Hilbert spaces. A vector $x \in X$ is represented in Dirac notation as follows:
\begin{equation}
	x = \ket{x}.
\end{equation}
$\ket{x}$ is called a \textbf{ket}. The $\complexNumbers$-linear transformation
\begin{align*}
	x^\dag \colon X &\longrightarrow \complexNumbers \\
	x' &\longmapsto \langle x, x' \rangle
\end{align*}
is represented in Dirac notation as follows:
\begin{equation}
	x^\dag = \bra{x}.
\end{equation}
$\bra{x}$ is called a \textbf{bra}. In this way, we have that
\begin{equation}
	\langle x \mid x' \rangle = \langle x , x' \rangle.
\end{equation}
The names were chosen so that $\langle x \mid x' \rangle$ is a bracket.

We can combine bras and kets in other ways. If $x \in X$ and $y \in Y$, then $\ket{y}\bra{x}$ represents the $\complexNumbers$-linear transformation
\begin{align*}
	\ket{y}\bra{x} \colon X &\longrightarrow Y \\
	x' &\longmapsto \langle x, x' \rangle y.
\end{align*}
In Dirac notation, this corresponds to
\begin{equation}
	\ket{y}\langle x \mid x' \rangle = \langle x \mid x' \rangle \ket{y}.
\end{equation}

We can also interpret a vector $x \in X$ as the $\complexNumbers$-linear transformation
\begin{align*}
	x \colon \complexNumbers &\longrightarrow X \\
	\lambda &\longmapsto \lambda x.
\end{align*}
The dual of this linear transformation is $x^\dag$. In Dirac notation, this says that kets and bras are dual to each other:
\begin{equation}
	\ket{x}^\dag = \bra{x}.
\end{equation}

In the book Picturing Quantum Processes \cite{Coecke_Kissinger_2017} is explored the relation between the Dirac notation and string diagrams. It may be helpful to understand Dirac notation from the perspective of string diagrams, but we won't use it in this thesis.

\cleardoublepage
\chapter{Introduction}

This is a simplified version of the thesis. The thesis can be obtained at \url{https://teses.usp.br/teses/disponiveis/45/45131/tde-09072026-172437/en.html}. Starting from the Contents page, there is no difference from the thesis, except for the addition of this paragraph.

We'll discuss the historical origins of CPTP maps, also known as quantum channels, their relation with Quantum Computation and Quantum Information, and then explain the research objectives. Lastly, we'll show how this thesis is organized in chapters.

Some mathematical, physical and computational concepts will appear in this chapter. The definitions necessary for this thesis will be given in the next chapters. An exception are concepts from Algebraic Topology, such as CW structures and homology groups. These are harder to explain, so it is recommended to learn about them by taking postgraduate courses on Algebraic Topology or studying books on Singular Homology Theory. The book Singular Homology Theory from Massey \cite{masseySHT} was used as reference. Knowledge of Differential Geometry will be assumed throughout the text, but it isn't strictly necessary, since the new results don't require it.

\section{Historical origins of CPTP maps}

CPTP stands for completely positive trace preserving. The completely positive maps originated from the study of abstract algebras. In the theory of C*-algebras there is an important result called the GNS construction, which is about states of C*-algebras. If $\mathcal{A}$ is a C*-algebra, a state of $\mathcal{A}$ is a positive linear functional $\rho \colon \mathcal{A} \to \complexNumbers$ with norm 1. Given a Hilbert space $H$, denote by $B(H)$ the set of bounded operators over $H$. Instead of states of $\mathcal{A}$, we can consider maps from $\mathcal{A}$ to $B(H)$, for some Hilbert space $H$. A generalization of the GNS construction couldn't be found for positive linear maps. In 1955, Stinespring solved this problem by introducing a stronger property of complete positivity (CP). He proved his dilation theorem in \cite{c8814692-9976-379a-9ad9-042fab94d853} (theorem 1 of the reference), which successfully generalized the GNS construction for CP maps, instead of positive linear maps.

The CPTP maps also originated from works about the foundations of Quantum Physics. At the time of writing this text, Quantum Physics has multiple interpretations, the standard one being the Copenhagen interpretation. According to it, the state of a quantum system collapses to another one when a measurement is made. There is still discussion about what it means to measure something and what happens when something is measured. This is known as the measurement problem. In the 1970s, Kraus published a work related to the measurement problem, in which he studied what operations are allowed on quantum systems \cite{KRAUS1971311}. He studied a quantum system, called the object, coupled to another, called the apparatus. He considered the operation in which both systems evolve in time by a unitary operator, followed by a measurement, described mathematically by a projection operator, and then the apparatus is discarded from the mathematical description by taking a partial trace. He showed that such operation is a CP map and that it transforms the initial density operator $\rho$ of the object to
\begin{equation*}
	\varepsilon(\rho) = \sum_{k} A_k \rho A_k^\dag,
\end{equation*}
for some operators $A_k$. This last result was proved using Stinespring's dilation theorem and is called the Kraus representation of the operation. He was interested not only on the Hilbert space formalism, but also considered generalizations to abstract algebras and worked with infinite dimensional spaces. Later Kraus organized the results about operations in his book \textit{States, Effects, and Operations: Fundamental Notions of Quantum Theory} \cite{stateseffects}. 

Still in the 1970s, the works of Jamio\l{}kowski and Choi \cite{JAMIOLKOWSKI1972275, CHOI1975285} established what is now known as the \CJ isomorphism, which makes a correspondence between CP maps and positive semi-definite (PSD) operators. Choi also showed that a linear map between complex matrices is a CP map iff it is of the form
\begin{equation*}
	\varepsilon(A) = \sum_{k} E_k A E_k^\dag, 
\end{equation*}
for a finite number of matrices $(E_k)_k$ (theorem 1 of \cite{CHOI1975285}). Kraus worked with infinite dimensional spaces, but Choi proved these results for finite dimensions. The results of Stinespring, Jamio\l{}kowski and Choi showed that CP maps have a treatable mathematical structure.

All these results formed the basis for the theory of quantum operations, as in \cite{nielsen_chuang_2010, Wilde_2017}. A quantum operation may not be trace preserving, but the ones that do preserve it are CPTP maps. The preservation of the trace is related to the preservation of probability. For this reason, a quantum operation that doesn't preserve the trace is obtained from some CPTP map, by discarding terms from some of its Kraus representations (see exercise 8.8 of \cite{nielsen_chuang_2010}, about non-trace-preserving quantum operations). Then, it is natural to study CPTP maps, which is the main theme of this thesis.

\section[CPTP maps in QC and QIT]{CPTP maps in Quantum Computation and Quantum Information Theory}

CPTP maps gained prominence in the last years for serving as discrete time models for open quantum systems. They are particularly important for Quantum Computation and Quantum Information Theory. Quantum Computation originated in the early 1980s. In 1980, Benioff published an article where a computer, in the sense of a Turing machine, was implemented by a quantum system. The physical system was quantum, but he didn't argue that quantum computers could be much more efficient than Turing machines. In the same year, Yuri I. Manin published the book \textit{Computable and Uncomputable}, in which he suggested the development of a theory of quantum automata. The suggestion is present in the introduction of the book. The book is in Russian, an English translation of the introduction can be found in another book called \textit{Mathematics as Metaphor: Selected Essays of Yuri I. Manin} \cite{manin2007mathematics}. He argued that there is an exponential growth on the cost to simulate a quantum system. This is due to the following. John von Neumann formalized Quantum Mechanics in a series of papers, which were then summarized in his book \textit{Mathematical Foundations of Quantum Mechanics} \cite{vonneumann}. The formalization used Hilbert spaces, but later he pursued other ideas \cite{REDEI1996493}. At the time of writing this text, the Hilbert space formalism is still largely used. According to it, a quantum system is described by a complex Hilbert space. For simplicity, let's consider that the space is finite dimensional. In a bipartite system, if the subsystems are described by Hilbert spaces $H_1$ and $H_2$, then the composite system is described by the tensor product $H_1 \otimes H_2$. Since the dimension is given by
\begin{equation*}
	\dim_\complexNumbers (H_1 \otimes H_2)  = \dim_\complexNumbers H_1 \dim_\complexNumbers H_2,
\end{equation*}
it gives rise to an exponential growth on the dimension with respect to the number of parts. That is, for $n$ copies of a subsystem described by a Hilbert space $H$, we have that
\begin{equation*}
	\dim_\complexNumbers H^{\otimes^n} = (\dim_\complexNumbers H)^n .
\end{equation*}
In 1981, Richard Feynman gave a keynote speech \cite{Feynman1982} in which he also noted this exponential growth. He then suggested using quantum systems to simulate other quantum systems, which led to the theory of Quantum Computation.

Computability Theory already has multiple equivalent models of computation, such as the (deterministic) Turing Machine, the nondeterministic Turing machine and the lambda calculus. The Church-Turing thesis says that anything that we can compute, can be computed by a Turing machine. In particular, all these computational models can be simulated by a Turing machine. The quantum computer doesn't challenge the Church-Turing thesis, because it can be simulated by a Turing machine, at least approximately. But the Church-Turing thesis doesn't account for efficiency. For this reason was formulated the Strong Church-Turing thesis, which says that any physically realizable model of computation can be simulated by a Turing machine in polynomial time. We don't know of any way to simulate a nondeterministic Turing machine by a deterministic Turing machine in polynomial time, or of any physical realization. Quantum computers challenge the Strong Church-Turing thesis, but it remains unproven whether they can be efficiently simulated or not.

In the 1990s, Peter Shor developed a quantum algorithm to factor numbers into prime factors in polynomial time \cite{doi:10.1137/S0097539795293172}. To this date, this algorithm offers an exponential speed up over the best known algorithms for standard computers. It is also still one of the most important quantum algorithms. Because of the importance of prime factorization for practical applications, this result largely increased the interest on Quantum Computing. Much of the challenge to develop quantum computers is controlling its errors. The theory of error correction couldn't be directly applied to quantum computers, because of the no-cloning theorem. In 1995, Shor publish an article with the first quantum error correction code  \cite{PhysRevA.52.R2493}. These codes allow a quantum computer to be more reliable by encoding the information in multiple qubits. CPTP maps are important for quantum error correction, because they model the noise on quantum computers. They are also used in Quantum Information Theory as models for noisy quantum communication. For this reason, CPTP maps are also called quantum channels.

\section{Research objectives}

The purpose of this thesis is to study the CPTP maps from a mathematical point of view. Since the work was made in a Category Theory research group, the initial proposal was to study the category of CPTP maps. This idea guided the research, but other ones were incorporated as seen appropriate. Even though the main objective was to understand categorical structures, mathematical structures from another areas other than Category Theory also appear and are important even for categorical problems. These other mathematical structures are also important by themselves.

There are multiple mathematical structures which appear in the theory of CPTP maps. We'll study them in more detail in the next chapters. These are some known results about CPTP maps:
\begin{itemize}
	\item The category of CPTP maps is a \textbf{semicartesian symmetric monoidal category}, whose projections are the partial traces \cite{huot2019universal} (in the first paragraph of section 4, \textit{Discussion and outlook}, of this reference);
	
	\item The hom-sets of the category of CPTP maps are \textbf{stratified spaces}, with a smooth manifold for each Choi-rank \cite{iten_colbeck};
	
	\item The hom-sets are also \textbf{compact convex sets}, hence they are the convex closure of their extreme points, by Krein-Milman's theorem \cite{barvinokcourse};
	
	\item The closure of the set of extreme points of the hom-set is the subset of CPTP maps with at most a certain Choi-rank \cite{ruskai2007open}. It is also a \textbf{closed and bounded semi-algebraic set}, hence it has a finite CW structure \cite{algorithms_in_RAG}.
\end{itemize}
Not much more is known about the category of CPTP maps other than that it is a semicartesian category. Understanding this category can be useful to develop functional programming languages for quantum computing. There is a result called the Curry-Howard-Lambek correspondence, which relates the lambda calculus, intuitionistic propositional logic and cartesian closed categories. In other words, it relates programming, logic and categorical models. It had a great influence on the development of functional programming languages, such as Haskell. Some functional programming languages have already been developed for quantum computing. A notable one that explores this relation between programming and categorical logic is Quipper \cite{Fu_2025, 10.1145/3373718.3394765}. For quantum computing, it has been argued that we should use linear logic instead of intuitionistic logic. This is essentially related to the replacement of cartesian closed categories by some monoidal category. In this thesis, the monoidal category would be the semicartesian category of CPTP maps. One of the objectives of this research is to understand in more the detail this category, which could help develop some logic or programming language for quantum computing, following ideas similar to the Curry-Howard-Lambek correspondence. An immediate question to ask about the category of CPTP maps is what are its limits and colimits. We successfully classified its binary products using topological techniques. Some information was obtained about binary coproducts, again using topological techniques, but not a full classification. The classification of limits and colimits seems to be a very hard problem to solve. The study of other mathematical structures can help to solve this problem.

The hom-sets of the category of CPTP maps are convex sets. Some information can be obtained from its set of extreme points. We can also work with the closure of the set of extreme points. The closure has the advantage of being compact, which guarantees many desired mathematical properties. Both the sets of extreme points and its closure are also real semi-algebraic sets. In the case of the closure, this guarantees that it has some finite CW structure \cite{algorithms_in_RAG}. A CW structure is a type of decomposition of the topological space, in which it is expressed as a gluing of balls, also called cells. One of the objectives is to search for such a CW structure. This decomposition can help to compute some topological invariants, such as the singular homology groups, which can then be used to obtain constraints for the existence of solutions to problems related to CPTP maps. For instance, it can help decide whether some limit or colimit exists, or at least reduce the search space of possible solutions. We focused on topological techniques, mainly due to familiarity, but other techniques can also be helpful. Since the set of extreme points and its closure are semi-algebraic sets, results from Real Algebraic Geometry may be helpful. In fact, there already exists algorithms from Real Algebraic Geometry that are capable of computing a CW decomposition, by first computing a Cylindrical Algebraic Decomposition (CAD) \cite{algorithms_in_RAG}. Unfortunately, the best general algorithms for it can take double exponential time to execute, which quickly becomes unfeasible.

There is also a subclass of CPTP maps that are unital, called UCPTP maps. There is some interest for these maps in Quantum Information Theory, so we'll also prove some results for them. We constructed many examples of extreme CPTP and UCPTP maps. We also investigated if the tensor product preserves extremality. That is, if a tensor product of extreme maps is extreme, for the convex sets of CPTP or UCPTP maps. 

In summary, the objectives of this thesis are the following:
\begin{itemize}
	\item To construct examples of extreme CPTP and UCPTP maps;
	
	\item To analyze if the tensor product preserves extremality;
	
	\item To study the limits and colimits of the category of CPTP maps;
	
	\item To study the topology of the hom-sets of the category of CPTP maps, and of its set of extreme points, and of the closure of its set of extreme points. This includes the search for a CW structure.
\end{itemize}
For the first two items, much progress has being made. We were able to construct extreme CPTP maps for every Choi-rank possible. For UCPTP maps, several examples were constructed, but we don't have a full classification by Choi-rank. We also proved that extremality is preserved for CPTP maps, but gave counter-examples for the case UCPTP maps. For the last two items, some progress was made, but there is still much work to be done. The terminal object is already known, since the category of CPTP maps is known to be semicartesian. The initial object is trivial to determine. We successfully classified its binary products. We obtained some information about the binary coproducts, but didn't classify them. The same topological technique applied for the products may be applied to the coproducts if we compute the homology groups of the closure of the set of extreme points of the hom-sets. We obtained some cells for this set, but more work is needed to obtain a full CW decomposition. Such decomposition may help compute the homology groups.

\section{Organization of the text}

We'll be focusing on the theory of CPTP maps from the perspective of Mathematics. The subjects of Linear Algebra, General Topology, Algebraic Topology, Differential Geometry and Category Theory will be the most used in this thesis. We'll be working mostly with finite dimensional spaces. The text was made to be as self contained as possible, hence the number of pages, but not all topics are covered. Knowledge of Differential Geometry will be assumed, but it isn't required for the new results proved in this thesis. For Algebraic Topology, the recommendation is to take postgraduate courses on the subject or study books on Singular Homology Theory, such as the one from Massey \cite{masseySHT}. The chapters with new results were organized such that Algebraic Topology is only used in the last ones.

This thesis starts with a review of CPTP maps in chapter \ref{chapter: cp maps}, by first defining CP maps and then specializing to the CPTP, UCP and UCPTP cases. The review gathers results from multiple books and articles. Some of the results in the review were intended to be used in the thesis, but ended up not being used, because of the difficulty in proving new results. This is the case for the stratifications into smooth manifolds. These known results were kept in the review for those who are interested. We also review some results about the category of CPTP maps in chapter \ref{section: category of cptp maps}. After the review, we show our contribution to the theory of CPTP maps. The new results of this thesis are mainly in chapters \ref{chapter: examples extreme channels}, \ref{section: Extremality of the tensor product}, \ref{section: limits and colimits of CPTP}, \ref{chapter: projection properties}, \ref{chapter: cells extreme cptp} and appendix \ref{appendix: examples extreme ucptp}. All other chapters are support material.

In chapter \ref{chapter: examples extreme channels} we construct examples of extreme quantum channels for the CPTP and UCPTP cases. The hardest part is constructing extreme UCPTP maps with high Choi-rank. We successfully construct hundreds of these examples. For extreme UCPTP maps with Choi-rank 2 we construct examples for each possible dimension. Tables of examples were put in appendix \ref{appendix: examples extreme ucptp}, because of their large number.

In chapter \ref{section: Extremality of the tensor product} we study when the tensor product preserves the extremality of quantum channels. We prove that the tensor product of extreme CPTP maps is an extreme CPTP map. By duality, the same is true for UCP maps. For the UCPTP case we prove that extremality isn't always preserved. To construct counterexamples, we use the examples of extreme UCPTP maps with high Choi-rank from chapter \ref{chapter: examples extreme channels}. We also show that the tensor product of multiple copies of a UCPTP map that has high Choi-rank isn't extreme if the number of copies is large enough.

In chapter \ref{section: limits and colimits of CPTP} we show new results for the category of CPTP maps. It is known that the category is semicartesian. We try to go further by studying its limits and colimits. We successfully classified all its binary products. They only exist for trivial cases, in which one of the Hilbert spaces has dimension 0 or 1. To prove this result was used the topology of the extreme points of the set of density operators. It is known that these extreme points are homeomorphic to a complex projective space. We used the homology groups of the complex projective space in the proof of the classification of products. For the coproducts, we successfully determined the dimension it must have, if it exists. In this case, we used the dimensions of the convex sets of CPTP maps. For both products and coproducts we made progress by applying techniques from General Topology and Algebraic Topology. 

Density operators can be seen as CPTP maps with a 1-dimensional domain. Since the topology of the convex set of density operators and of their extreme points were helpful, we could expect the topology of the convex set of CPTP maps and of its extreme points to help prove theorems. The extreme points of the set of density operators is compact, so it is a closed subspace. For CPTP maps, the extreme points don't need to be closed, so we also have the closure of the set of extreme points to study. Ruskai showed that this closure is the set of all CPTP maps with Choi-rank at most the dimension of the domain \cite{ruskai2007open}. There she also posed the study of this set as an open problem of Quantum Information Theory. Working with the closure has at least the advantage that the set is compact, so we have theorems that guarantee good properties, such as the existence of a finite CW structure.

There is a known parameterization of the set of CPTP maps with at most some Choi-rank \cite{iten_colbeck}. In chapter \ref{chapter: projection properties} we show that this parameterization is an open function. For a fixed Choi-rank it was already known that the parameterization was open, but we extend the result by allowing the Choi-rank to vary.

In chapter \ref{chapter: cells extreme cptp} we use the same parameterization from chapter \ref{chapter: projection properties} to construct cells for the closure of set of the extreme points of the set of CPTP maps. If the domain is 1-dimensional, the problem reduces to the extreme points of the density operators, which is homeomorphic to a complex projective space. The parameterization has a similarity with the projection $\pi \colon S^{2n+1} \to \complexNumbers P^n$. This similarity is used to guess cells for the closure of extreme points. We view the closure of extreme points as a generalization of the complex projective space, in which we replace complex numbers by complex matrices. The non commutativity of matrices makes the problem much harder, but we succeed in mimicking the CW structure of the complex projective space. This turns out to be not enough, the cells obtained in this way don't cover the entire space. This gives a partial decomposition of the spaces into cells. Further work is needed to get a full CW decomposition.

In chapter \ref{chapter: conclusions} we present the conclusions and ideas for further study. After this chapter we have the appendix, with mostly support material. In appendix \ref{appendix: tensor product} we review the tensor product of vector spaces. It uses the same definition from the theory of modules, as an universal property, but specialized to vector spaces. Some properties are proven that are related to monoidal categories with duals, such as spherical categories \cite{MonoidalCatsAndTFT}. In appendix \ref{appendix: hilbert spaces} we review finite dimensional complex Hilbert spaces. We restrict to complex spaces, because these are the spaces used in Quantum Mechanics. The restriction to finite dimensional spaces is mainly because that's what is used in the Quantum Circuit model for Quantum Computation. In the finite dimensional case, the tensor product of Hilbert spaces is the algebraic tensor product, coinciding with the one for vector spaces. This appendix also includes a section about Dirac notation, which is a notation commonly used in Quantum Physics. In appendix \ref{appendix: matrices} we collect some results from Matrix Analysis. In appendix \ref{appendix: category theory} we have a review of Category Theory, focused on what is used in this thesis. There is a section about string diagrams. String diagrams are very useful, but they are hard to explain properly. For this reason, they are not used often in this work. We use almost the same string diagrams from the book Monoidal Categories and Topological Field Theory \cite{MonoidalCatsAndTFT}, with the only difference that we use the opposite orientation for the strings. We don't study them in their full generality, rather we focus on finite dimensional vectors spaces. String diagrams can make some arguments much simpler, for further study we suggest the previously cited book and the book Picturing Quantum Processes \cite{Coecke_Kissinger_2017}. In appendix \ref{appendix: convex sets} we review some results about convex sets in finite dimensions. Convex sets are common in Quantum Mechanics due to the probabilistic nature of the theory. As said earlier, the appendix \ref{appendix: examples extreme ucptp} have tables with examples of extreme UCPTP maps. This is part of the new results, but was put in the appendix because of the size.

\cleardoublepage
\chapter{Completely Positive Maps}
\label{chapter: cp maps}

In this chapter we review the theory of Completely Positive (CP) maps, with emphasis on CPTP maps, which is our main subject of study. This chapter is meant to put in one place results from different books and articles. Many results can be found in the books Quantum Computation and Quantum Information (QCQI) \cite{nielsen_chuang_2010} and The Theory of Quantum Information \cite{watrous_2018}. The book Quantum Computation and Quantum Information is a standard reference on the subject. This book doesn't delve too much on the mathematical aspects of the theory. The book The Theory of Quantum Information has a better mathematical development, but neither of them explain too much the geometric and topological results. The reference QCQI gives a glimpse of the geometry of the set of density operators, when in exercise 2.72 it says that the density matrices over $\complexNumbers^2$ (a single qubit) correspond to vectors in the closed unit ball of $\realNumbers^3$. A similar result holds for every dimension, the set of density operators is compact convex, so it is homeomorphic to a closed ball. Some geometric information can be found in the book Geometry of Quantum States \cite{geometry_quantum}. For example, from section 8.4 of this book we know that the (real) dimension of the set of density operators over $\complexNumbers^n$ is $n^2-1$, and that the set of vector states is homeomorphic to the complex projective space $\complexNumbers P^{n-1}$. Other geometric results can be found in articles, such as \cite{FRIEDLAND2016553,iten_colbeck}. In the first we have results from the point of view of Real Algebraic Geometry, and in the second the point of view is of Differential Geometry.

This chapter starts with CP maps. This is done to gather results that will be used for other types of maps later. After the CP maps, we study the CPTP (Completely Positive and Trace Preserving), UCP (Unital Completely Positive) and UCPTP (Unital Completely Positive and Trace Preserving) maps. We focus primarily on CPTP maps, because they are a model for the time evolution of open quantum systems, or for quantum communication subject to noise. They are also called quantum channels. The UCP maps are studied because they are dual to CPTP maps. The UCPTP maps are studied because there is some interest in Quantum Information Theory on this special type of CPTP maps.

The definition of CP maps is hard to use. Instead, we typically use other equivalent descriptions of CP maps. One of them is the Kraus representation, which presents a CP map by a finite sequence of operators $(E_k)_k$. These operators are not unique, for this reason  we avoid using them when possible. Another equivalent description is obtained by the \CJ isomorphism. This isomorphism gives a correspondence between CP maps and positive semidefinite (PSD) operators. The PSD operators are often simpler to work with. They also have the advantage over the Kraus representation of being unique. Typically, the books define a \CJ isomorphism that is basis dependent. We instead use another version of the isomorphism that is very similar, but basis independent. This other version of the \CJ isomorphism is known, but not commonly used. Both the Kraus representation and the \CJ isomorphism specialize to the CPTP, UCP and UCPTP maps. For this reason, we explain for each of these types of maps the special properties that the Kraus representation and \CJ isomorphism have. We also have a duality given by the adjoint, which gives a correspondence with the CP property with itself, and of the trace preserving property with the unital property. For this reason, CP maps are dual to themselves, CPTP maps are dual to UCP maps, and UCPTP maps are dual to themselves. This duality enables us to transfer results. All the results presented in this paragraph can be found in \cite{watrous_2018}.

For each type of map we show some geometric or topological results about the sets of maps from a finite dimensional complex Hilbert space $X$ to another $Y$, that is, for the hom-sets of the corresponding categories. For CP maps we only show that their hom-sets are stratified into smooth submanifolds of the space of operators. This is done mainly as an auxiliary result, to later show that the hom-sets of the category of CPTP maps are also stratified into smooth submanifolds. This result was proved in \cite{iten_colbeck}. By duality, the same result holds for UCP maps. In each case, we also characterize the tangent spaces. These are the results from the perspective of Differential Geometry. We also show that the hom-sets of the categories of CPTP, UCP and UCPTP maps are compact convex. This isn't true for CP maps, because positive semidefinite operators can have arbitrarily large norm, so the hom-sets are not bounded, and therefore not compact. A compact convex set in $\realNumbers^n$ is homeomorphic to a closed ball. In particular, they have a dimension, as topological manifolds with boundary. For each case, we give an expression for the dimension. Also, being compact convex sets, they are the convex hull of their extreme points. We present the known characterizations of extremality in terms of Kraus representations, due to \cite{CHOI1975285,LANDAU1993107}. For CPTP maps, we also present alternative characterizations of extremality. For $\cptp(X,Y)$, we also show that the closure of the set of extreme points is the sets of CPTP maps with rank at most $d_X$. This is a known result, present in \cite{ruskai2007open}. Because of this result, we show in section \ref{section: a surjective map from the purification} how to parameterize the set of CPTP maps with rank at most $r$ by a Stiefel manifold. This type of parametrization is related to the result that any mixed state can be purified to a vector state on a bigger Hilbert space. The same idea is also present in \cite{iten_colbeck}. The main difference is that in \cite{iten_colbeck} was given emphasis on CPTP maps with a fixed rank, while we allow the rank to vary. We show that this parameterization by a Stiefel manifold gives a continuous surjective function. A new result that the function is also open is postponed to chapter \ref{chapter: projection properties}. The parameterization is important in chapter \ref{chapter: cells extreme cptp}.

Something that can be sometimes very useful are string diagrams for spherical categories. In \cite{Coecke_Kissinger_2017} is shown how to use string diagrams in Quantum Computation and Quantum Information. To keep the pre-requisites for reading this text lower, we opted to avoid using string diagrams, but they are used in some proofs.

In this chapter will be assumed knowledge of Differential Geometry, but it is only used for the stratifications into smooth manifolds. It isn't required for the new results, the stratifications were kept in the text for those who are interested. Also, knowledge of complex projective spaces will be required. They can be studied in courses about Topology, Algebraic Topology or Differential Geometry. The book Singular Homology Theory from Massey \cite{masseySHT} defines the projective spaces, but it is being used primarily as a reference for Algebraic Topology.

\section{CP maps}

Completely Positive (CP) maps form a category CP whose objects are finite dimensional complex Hilbert spaces and arrows are CP maps. To define what complete positivity means, we need first to introduce an alternative view on the tensor product of linear maps. Let $X,X',Y,Y'$ be finite dimensional complex Hilbert spaces. Let also $\varepsilon \colon \Lin_\complexNumbers(X) \to \Lin_\complexNumbers(Y)$ and $\varepsilon' \colon \Lin_\complexNumbers(X') \to \Lin_\complexNumbers(Y')$ be $\complexNumbers$-linear maps. The definition of the tensor product of linear maps then given a $\complexNumbers$-linear map
\begin{equation*}
	\varepsilon \otimes \varepsilon' \colon \Lin_\complexNumbers(X)\otimes \Lin_\complexNumbers(X') \to \Lin_\complexNumbers(Y)\otimes \Lin_\complexNumbers(Y').
\end{equation*}
Instead, we want the domain and codomain to be sets of endomorphisms. We have the $\complexNumbers$-bilinear map
\begin{align*}
	\Lin_\complexNumbers(X) \times \Lin_\complexNumbers(X') &\longrightarrow \Lin_\complexNumbers(X\otimes X') \\
	(f,g) &\longmapsto f \otimes g,
\end{align*}
where by $f\otimes g$ we mean the tensor product of linear maps. By the universal property of the tensor product, this bilinear map induces the $\complexNumbers$-linear map
\begin{align*}
	\Lin_\complexNumbers(X) \otimes \Lin_\complexNumbers(X') &\longrightarrow \Lin_\complexNumbers(X\otimes X') \\
	f \otimes g &\longmapsto f \otimes g.
\end{align*}
Here we only present the definition of this function for elementary tensors $f\otimes g$. The general case is obtained by linearity, since every element of $\Lin_\complexNumbers(X) \times \Lin_\complexNumbers(X')$ is a linear combination of elementary tensors. The definition of the function may seem trivial, due to the notation. But the $f\otimes g$ in the input and the $f \otimes g$ in the output have different meanings. The first is the tensor product between an element of $\Lin_\complexNumbers(X)$ and another of $\Lin_\complexNumbers(X')$, that is, we only see $f$ and $g$ as vectors. The second is the tensor product of the linear maps. We won't prove this here, but the function $\Lin_\complexNumbers(X) \otimes \Lin_\complexNumbers(X') \to \Lin_\complexNumbers(X\otimes X')$ defined above is an isomorphism. Composing $\varepsilon \otimes \varepsilon'$ with this isomorphism, we have a $\complexNumbers$-linear map
\begin{align*}
	\varepsilon \otimes \varepsilon' \colon \Lin_\complexNumbers(X\otimes X') &\longrightarrow \Lin_\complexNumbers(Y\otimes Y') \\
	f \otimes g &\longmapsto \varepsilon(f) \otimes \varepsilon'(g).
\end{align*}
We keep the isomorphisms implicit and use the same notation $\varepsilon \otimes \varepsilon'$. This is what we'll typically mean by a tensor product in the context of CP maps. Now we can define what are CP maps.

\begin{definition}{}{}
	Let $X$ and $Y$ be finite dimensional complex Hilbert spaces. A CP map $\varepsilon \colon X \to Y$ is a $\complexNumbers$-linear map $\varepsilon \colon \Lin_\complexNumbers(X) \to \Lin_\complexNumbers(Y)$ satisfying the following property: for each finite dimensional complex Hilbert space $Z$, the linear map $\id_{\Lin_\complexNumbers(Z)} \otimes \varepsilon \colon \Lin_\complexNumbers(Z \otimes X) \to \Lin_\complexNumbers(Z\otimes Y)$ sends PSD operators to PSD operators.
\end{definition}

\begin{notation}
	We denote the set of CP maps $\varepsilon \colon X \to Y$ by $\cp(X,Y)$.
\end{notation}

A simple consequence of complete positivity is that CP maps send hermitian operators to hermitian operators. This is a property that we'll use sometimes, so we'll briefly prove it here for later reference. The proof only requires positivity, but we'll assume complete positivity to avoid multiple definitions.
\begin{proposition}{}{cp(A dag) = cp(A) dag}
	Let $X$ and $Y$ be finite dimensional complex Hilbert spaces, and let $\varepsilon \in \cp(X,Y)$. Then
	\begin{equation}
		\varepsilon(A^\dag) = \varepsilon(A)^\dag,
	\end{equation}
	for any $A \in \Lin_\complexNumbers(X)$. In particular, $\varepsilon$ sends hermitian operators to hermitian operators.
\end{proposition}
\begin{proof}
	First we show that $\varepsilon$ sends hermitian operators to hermitian operators. Let $H \in \hermitian(X\otimes Y)$. It is hermitian, so it can be diagonalized and its eigenvalues are real. Separating the positive from the negative eigenvalues, we can write $H$ as
	\begin{equation}
		H = P-Q,
	\end{equation}
	where $P,Q \in \Lin_\complexNumbers(X\otimes Y)$ are positive semidefinite operators. Since $\varepsilon$ is completely positive, both $\varepsilon(P)$ and $\varepsilon(Q)$ are positive semidefinite. In particular, both are hermitian, so $\varepsilon(P) - \varepsilon(Q)$ is also hermitian. But
	\begin{equation}
		\varepsilon(H) = \varepsilon(P-Q) = \varepsilon(P) - \varepsilon(Q),
	\end{equation}
	so $\varepsilon(H)$ is hermitian.
	
	Now let $A \in \Lin_\complexNumbers(X\otimes Y)$. It can be written as
	\begin{equation}
		A = H_1 + i H_2,
	\end{equation}
	where
	\begin{equation}
		H_1 = \frac{A+A^\dag}{2},
	\end{equation}
	and
	\begin{equation}
		H_2 = \frac{A-A^\dag}{2i}.
	\end{equation}
	Both operators $H_1$ and $H_2$ are hermitian. In fact,
	\begin{equation}
		H_1^\dag = \frac{A^\dag+A}{2} = H_1,
	\end{equation}
	and
	\begin{equation}
		H_2^\dag = \frac{A^\dag-A}{-2i} = H_2.
	\end{equation}
	Therefore,
	\begin{equation}
		A^\dag = H_1 -i H_2.
	\end{equation}
	Applying $\varepsilon$, we get
	\begin{align*}
		\varepsilon(A^\dag) &= \varepsilon(H_1 - i H_2)\\
		&= \varepsilon(H_1) - i \varepsilon(H_2)\\
		&= (\varepsilon(H_1)^\dag + i \varepsilon(H_2)^\dag)^\dag.
	\end{align*}
	Since $\varepsilon$ send hermitian operators to hermitian operators, we have that $\varepsilon(H_1)^\dag $ $=$ $\varepsilon(H_1)$ and $\varepsilon(H_2)^\dag = \varepsilon(H_2)$. Therefore
	\begin{align*}
		\varepsilon(A^\dag) &= (\varepsilon(H_1) + i \varepsilon(H_2))^\dag\\
		&= \varepsilon(H_1 + i H_2)^\dag\\
		&= \varepsilon(A)^\dag.
	\end{align*}
	This concludes the proof.
\end{proof}

This definition of CP maps is hard to use in practice. There are equivalent descriptions of CP maps which may be easier to use. One is the Kraus representation, another is the correspondence of CP maps with PSD operators given by the \CJ isomorphism. Next we'll study these other descriptions.

\subsection{The \CJ isomorphism, standard definition}

It is common to formulate the \CJ isomorphism as the following trick, which won't be the definition we'll use. Let $X$ and $Y$ be finite dimensional complex Hilbert spaces. Let also $\varepsilon \colon \Lin_\complexNumbers (X) \to \Lin_\complexNumbers (Y)$ be a CPTP map. Let $d_X$ be the $\complexNumbers$-dimension of $X$ and $(\ket{i})_{i=1, \dots, d_X}$ be an orthonormal basis for $X$. Then we have the maximally entangled (vector) state
\begin{equation}
	\label{equation: maximally entangled state}
	\ket{\Omega} = \sum_{i=1}^{d_X} \frac{1}{\sqrt{d_X}}\ket{i} \otimes \ket{i}.
\end{equation}
Taking a tensor product, we have the linear transformation
\begin{equation*}
	\id_{\Lin_\complexNumbers (X)} \otimes \varepsilon \colon \Lin_\complexNumbers (X\otimes X) \to \Lin_\complexNumbers (X \otimes Y).	
\end{equation*}
Applying $\id_{\Lin_\complexNumbers (X)} \otimes \varepsilon$ to the (density operator) state $\ket{\Omega} \bra{\Omega}$, we obtain the state
\begin{equation}
	\label{equation: bad CJ isomorphism definition}
	(\id_{\Lin_\complexNumbers (X)} \otimes \varepsilon)(\ket{\Omega} \bra{\Omega}) \in \density(X\otimes Y).
\end{equation}
The \CJ isomorphism is commonly defined as the correspondence between the CPTP maps $\varepsilon$ and these density operators in $\density(X\otimes Y)$. We won't define the \CJ isomorphism in this way.

\subsection{The \CJ isomorphism, basis independent definition}

From the point of view of Mathematics, the \CJ isomorphism is a natural isomorphism of vector spaces
\begin{equation}
	J \colon \Lin_\complexNumbers (\Lin_\complexNumbers (X), \Lin_\complexNumbers (Y)) \stackrel{\cong}{\to} \Lin_\complexNumbers (X^* \otimes Y). 
\end{equation}
Making a basis independent version of the \CJ isomorphism isn't a new idea. The isomorphism defined by Jamio\l{}kowski \cite{JAMIOLKOWSKI1972275} already didn't depend on a basis. Also, a basis independent version of the \CJ isomorphism was already present in \cite{PhysRevX.7.031021}.

Except for a normalization factor, the difference between $J$ and the isomorphism from equation \ref{equation: bad CJ isomorphism definition} is the change from $X$ to $X^*$. Since $X$ is finite dimensional, $X$ and $X^*$ are isomorphic, but the problem is that we want a natural isomorphism, independent of a choice of basis. Also, the definition of the maximally entangled state $\ket{\Omega}$ depends on the choice of basis, that is, a different basis may give a different state. Working with $X^*$ allows defining an \textbf{isomorphism that is basis independent}.

There is also a natural isomorphism with codomain $\Lin_\complexNumbers (X \otimes Y)$. Much of the usefulness of the \CJ isomorphism is that it is a bijection between CP maps and PSD operators, as we'll see later. The natural isomorphism with codomain $\Lin_\complexNumbers (X \otimes Y)$ doesn't have this property, mainly because the symmetry 
\begin{align*}
	\tau_{X^*, X^*} \colon X^* \otimes X^* &\longrightarrow X^* \otimes X^* \\
	f \otimes g &\longmapsto g \otimes f
\end{align*}
isn't a PSD operator.

Before writting the isomorphism $J$ explicitly, let's understand where it comes from. The \CJ isomorphism can be obtained by combining simpler known isomorphisms. Let $X$, $Y$ and $Z$ be any finite dimensional $\complexNumbers$-vector spaces. We have the following:
\begin{enumerate}
	
	\item An isomorphism
	\begin{align*}
		\Lin_\complexNumbers (\complexNumbers, X) &\longrightarrow X\\
		f &\longmapsto f(1).
	\end{align*}
	
	\item Unitality isomorphisms
	\begin{align*}
		l_X \colon \complexNumbers \otimes X &\longrightarrow X\\
		\lambda \otimes x &\longmapsto \lambda x,
	\end{align*}
	\begin{align*}
		r_X \colon X \otimes \complexNumbers &\longrightarrow X\\
		x \otimes \lambda &\longmapsto \lambda x,
	\end{align*}
	where $\lambda \in \complexNumbers$ and $x \in X$. This is proved in propositions \ref{proposition:left unitality} and \ref{proposition:right unitality}.
	
	\item A symmetry isomorphism
	\begin{align*}
		\tau_{X, Y} \colon X \otimes Y &\longrightarrow Y \otimes X\\
		x \otimes y &\longmapsto y \otimes x,
	\end{align*}
	where $x \in X$ and $y \in Y$. This is proved in proposition \ref{proposition:tensor symmetry}.
	
	\item An isomorphism
	\begin{equation}
		\Lin_\complexNumbers (X \otimes Y, Z) \cong \Lin_\complexNumbers (X, Y^* \otimes Z).
	\end{equation}
	In Category Theory terminology, the category of finite dimensional complex vector spaces is closed monoidal, with $[Y, Z] \coloneqq Y^* \otimes Z$ as internal hom. There is a proof of this isomorphism in proposition \ref{proposition:tensor monoidal closed}.
\end{enumerate}
Combining these isomorphisms, we get the following sequence of isomorphisms:
\begin{align*}
	\Lin_\complexNumbers (\Lin_\complexNumbers (X), \Lin_\complexNumbers (Y)) &\stackrel{(i)}{=} \Lin_\complexNumbers (\Lin_\complexNumbers (X,X), \Lin_\complexNumbers (Y,Y)) \\
	&\stackrel{(ii)}{\cong} \Lin_\complexNumbers (\Lin_\complexNumbers (\complexNumbers \otimes X,X), \Lin_\complexNumbers (\complexNumbers \otimes Y,Y)) \\
	&\stackrel{(iii)}{\cong} \Lin_\complexNumbers (\Lin_\complexNumbers (\complexNumbers, X^* \otimes X), \Lin_\complexNumbers (\complexNumbers, Y^* \otimes Y) )\\
	&\stackrel{(iv)}{\cong} \Lin_\complexNumbers (X^* \otimes X, Y^* \otimes Y)\\
	&\stackrel{(v)}{\cong} \Lin_\complexNumbers (X^*, X^* \otimes Y^* \otimes Y)\\
	&\stackrel{(vi)}{\cong} \Lin_\complexNumbers (X^*, Y^* \otimes X^* \otimes Y)\\
	&\stackrel{(vii)}{\cong} \Lin_\complexNumbers (X^* \otimes Y, X^* \otimes Y)\\
	&\stackrel{(viii)}{=} \Lin_\complexNumbers (X^* \otimes Y).
\end{align*}
In $(i)$ was used that $\Lin_\complexNumbers (Z) = \Lin_\complexNumbers (Z,Z)$, by definition; in $(ii)$ was used the left unitality isomorphisms $l_X$ and $l_Y$; in $(iii)$ was used that the category of finite dimensional complex vector spaces is closed monoidal; in $(iv)$ was used that $\Lin_\complexNumbers (\complexNumbers, Z) \cong Z$; in $(v)$ was used that the category is closed monoidal, to move $X$ from the domain to the codomain as $X^*$; in $(vi)$ was used that $\otimes$ is symmetric; in $(vii)$ was used that the category is closed monoidal, to move $Y^*$ from the codomain to the domain as $Y$; in $(viii)$ was used again that $\Lin_\complexNumbers (Z) = \Lin_\complexNumbers (Z,Z)$.

The \CJ isomorphism is the composition of all these isomorphisms. Next we find an expression for it.

\begin{proposition}{\CJ isomorphism}{CJ expression}
	Let $X$ and $Y$ be finite dimensional complex Hilbert spaces, and let $\varepsilon \in \Lin_\complexNumbers(\Lin_\complexNumbers(X),\Lin_\complexNumbers(Y))$. Let $(x_i)_i$ be a basis for $X$. Let also $(x^i)_i$ be the dual basis for $X^*$, characterized as the one that satisfies the equation $x^i(x_{i'}) = \delta_{i,i'}$. Also, let 
	\begin{equation}
		\ket{\Omega} \coloneqq \sum_i x^i \otimes x_i \in X^* \otimes X,
	\end{equation}
	\begin{equation}
		\ket{\tilde{\Omega}} \coloneqq \sum_i x_i \otimes x^i \in X \otimes X^*,
	\end{equation}
	both of which are independent of the choice of basis. Then the \CJ isomorphism can be expressed as
	\begin{equation}
		\label{equation: J}
		J(\varepsilon) = (\id_{\Lin_\complexNumbers(X^*)}\otimes \varepsilon)(\ket{\Omega} \bra{\Omega}) = \sum_{i,i'} \ket{x^i}\bra{x^{i'}} \otimes \varepsilon(\ket{x_i}\bra{x_{i'}}).
	\end{equation}
	The projection operator $\ket{\Omega} \bra{\Omega}$ is represented in string diagrams as follows:
	\begin{center}
		\tikzset{every picture/.style={line width=0.75pt}} 
		
		\begin{tikzpicture}[x=0.75pt,y=0.75pt,yscale=-1,xscale=1]
			
			\draw  [draw opacity=0] (160,0) .. controls (160,0) and (160,0) .. (160,0) .. controls (160,16.57) and (146.57,30) .. (130,30) .. controls (113.43,30) and (100,16.57) .. (100,0) -- (130,0) -- cycle ; \draw   (160,0) .. controls (160,0) and (160,0) .. (160,0) .. controls (160,16.57) and (146.57,30) .. (130,30) .. controls (113.43,30) and (100,16.57) .. (100,0) ;  
			\draw    (130,30) -- (137,30) ;
			\draw [shift={(140,30)}, rotate = 180] [fill={rgb, 255:red, 0; green, 0; blue, 0 }  ][line width=0.08]  [draw opacity=0] (10.72,-5.15) -- (0,0) -- (10.72,5.15) -- (7.12,0) -- cycle    ;
			\draw  [draw opacity=0] (100,80) .. controls (100,63.43) and (113.43,50) .. (130,50) .. controls (146.57,50) and (160,63.43) .. (160,80) -- (130,80) -- cycle ; \draw   (100,80) .. controls (100,63.43) and (113.43,50) .. (130,50) .. controls (146.57,50) and (160,63.43) .. (160,80) ;  
			\draw    (130,50) -- (123,50) ;
			\draw [shift={(120,50)}, rotate = 360] [fill={rgb, 255:red, 0; green, 0; blue, 0 }  ][line width=0.08]  [draw opacity=0] (10.72,-5.15) -- (0,0) -- (10.72,5.15) -- (7.12,0) -- cycle    ;
			
			\draw (1,32.4) node [anchor=north west][inner sep=0.75pt]    {$| \Omega \rangle \langle \Omega |\ =$};
			\draw (165,2.4) node [anchor=north west][inner sep=0.75pt]    {$X$};
			\draw (165,62.4) node [anchor=north west][inner sep=0.75pt]    {$X$};

		\end{tikzpicture}
	\end{center}
	
	For any $A \in \Lin_\complexNumbers(X^* \otimes Y)$ and $B \in \Lin_\complexNumbers(X)$ we have that
	\begin{align*}
		(J^{-1}(A))(B) &= (\bra{\tilde{\Omega}}\otimes \id_Y)(B\otimes A)(\ket{\tilde{\Omega}} \otimes \id_Y)\\
		&= \sum_{i,i'} \bra{x_i} B \ket{x_{i'}} (\bra{x^i}\otimes \id_Y)A(\ket{x^{i'}} \otimes \id_Y).
	\end{align*}
	It is represented in string diagrams as follows:
	\begin{center}

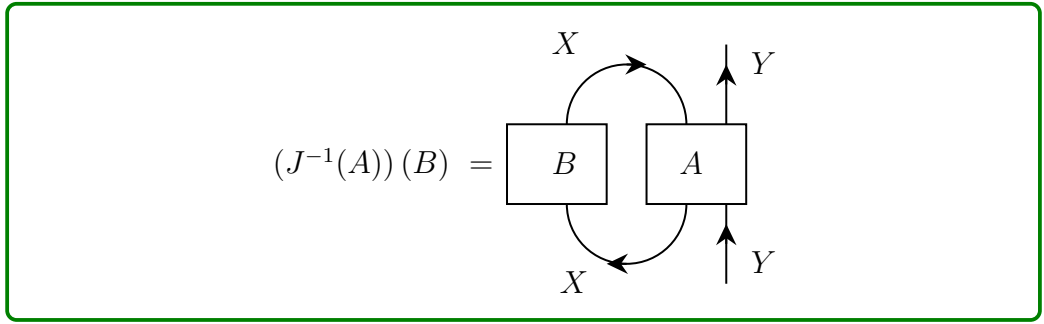
		\tikzset{every picture/.style={line width=0.75pt}} 
		
		\begin{tikzpicture}[x=0.75pt,y=0.75pt,yscale=-1,xscale=1]
			
			\draw   (190,50) -- (240,50) -- (240,90) -- (190,90) -- cycle ;
			\draw    (230,10) -- (230,50) ;
			\draw    (230,90) -- (230,130) ;
			\draw    (230,110) -- (230,103) ;
			\draw [shift={(230,100)}, rotate = 90] [fill={rgb, 255:red, 0; green, 0; blue, 0 }  ][line width=0.08]  [draw opacity=0] (10.72,-5.15) -- (0,0) -- (10.72,5.15) -- (7.12,0) -- cycle    ;
			\draw    (230,30) -- (230,23) ;
			\draw [shift={(230,20)}, rotate = 90] [fill={rgb, 255:red, 0; green, 0; blue, 0 }  ][line width=0.08]  [draw opacity=0] (10.72,-5.15) -- (0,0) -- (10.72,5.15) -- (7.12,0) -- cycle    ;
			\draw   (120,50) -- (170,50) -- (170,90) -- (120,90) -- cycle ;
			\draw  [draw opacity=0] (150,50) .. controls (150,33.43) and (163.43,20) .. (180,20) .. controls (196.57,20) and (210,33.43) .. (210,50) -- (180,50) -- cycle ; \draw   (150,50) .. controls (150,33.43) and (163.43,20) .. (180,20) .. controls (196.57,20) and (210,33.43) .. (210,50) ;  
			\draw    (180,20) -- (187,20) ;
			\draw [shift={(190,20)}, rotate = 180] [fill={rgb, 255:red, 0; green, 0; blue, 0 }  ][line width=0.08]  [draw opacity=0] (10.72,-5.15) -- (0,0) -- (10.72,5.15) -- (7.12,0) -- cycle    ;
			\draw  [draw opacity=0] (210,90) .. controls (210,90) and (210,90) .. (210,90) .. controls (210,90) and (210,90) .. (210,90) .. controls (210,106.57) and (196.57,120) .. (180,120) .. controls (163.43,120) and (150,106.57) .. (150,90) -- (180,90) -- cycle ; \draw   (210,90) .. controls (210,90) and (210,90) .. (210,90) .. controls (210,90) and (210,90) .. (210,90) .. controls (210,106.57) and (196.57,120) .. (180,120) .. controls (163.43,120) and (150,106.57) .. (150,90) ;  
			\draw    (180,120) -- (173,120) ;
			\draw [shift={(170,120)}, rotate = 360] [fill={rgb, 255:red, 0; green, 0; blue, 0 }  ][line width=0.08]  [draw opacity=0] (10.72,-5.15) -- (0,0) -- (10.72,5.15) -- (7.12,0) -- cycle    ;
			
			\draw (205,62.4) node [anchor=north west][inner sep=0.75pt]    {$A$};
			\draw (241,112.4) node [anchor=north west][inner sep=0.75pt]    {$Y$};
			\draw (241,12.4) node [anchor=north west][inner sep=0.75pt]    {$Y$};
			\draw (1,60.4) node [anchor=north west][inner sep=0.75pt]    {$\left( J^{-1}( A)\right)( B) \ =$};
			\draw (141,62.4) node [anchor=north west][inner sep=0.75pt]    {$B$};
			\draw (141,2.4) node [anchor=north west][inner sep=0.75pt]    {$X$};
			\draw (145,122.4) node [anchor=north west][inner sep=0.75pt]    {$X$};

		\end{tikzpicture}
	\end{center}
\end{proposition}
\begin{proof}
	This will be one of the few proofs in this work that will use string diagrams. String diagrams are a powerful tool, they can simplify many computations and make some ideas more clear. They also rely on results about monoidal categories, such as Mac Lane's coherence theorem, which can take a lot of time to explain properly. The difficulty in explaining why string diagrams are correct is the main reason for not using them often in this work. Details about string diagrams can be found in \cite{Coecke_Kissinger_2017, MonoidalCatsAndTFT} and a brief explanation is given in the appendix for Category Theory. Of course, we can also prove the same results without string diagrams, but such proof can be long and tedious. For example, see the proofs for proposition \ref{proposition:tensor monoidal closed}.
	
	The isomorphisms that compose the \CJ isomorphism correspond to the following diagrammatic tricks. The isomorphism
	\begin{equation*}
		\Lin_\complexNumbers (\Lin_\complexNumbers (X, X), \Lin_\complexNumbers (Y, Y)) \to \Lin_\complexNumbers (X^* \otimes X, Y^* \otimes Y)
	\end{equation*}
	is obtained as follows. Given $t \in X^* \otimes X$, we have the linear transformation
	\begin{align*}
		T \colon \complexNumbers &\longrightarrow X^* \otimes X \\
		\lambda &\longmapsto \lambda t.
	\end{align*}
	$T$ is represented diagrammatically as follows:
	\begin{center}

		\tikzset{every picture/.style={line width=0.75pt}} 
		
		\begin{tikzpicture}[x=0.75pt,y=0.75pt,yscale=-1,xscale=1]
			
			\draw   (54,40) -- (124,40) -- (124,80) -- (54,80) -- cycle ;
			\draw    (114,0) -- (114,40) ;
			\draw    (114,20) -- (114,13) ;
			\draw [shift={(114,10)}, rotate = 90] [fill={rgb, 255:red, 0; green, 0; blue, 0 }  ][line width=0.08]  [draw opacity=0] (10.72,-5.15) -- (0,0) -- (10.72,5.15) -- (7.12,0) -- cycle    ;
			\draw    (64,0) -- (64,40) ;
			\draw    (64,20) -- (64,27) ;
			\draw [shift={(64,30)}, rotate = 270] [fill={rgb, 255:red, 0; green, 0; blue, 0 }  ][line width=0.08]  [draw opacity=0] (10.72,-5.15) -- (0,0) -- (10.72,5.15) -- (7.12,0) -- cycle    ;
			
			\draw (85,52.4) node [anchor=north west][inner sep=0.75pt]    {$T$};
			\draw (125,2.4) node [anchor=north west][inner sep=0.75pt]    {$X$};
			\draw (39,2.4) node [anchor=north west][inner sep=0.75pt]    {$X$};
			\draw (1,42.4) node [anchor=north west][inner sep=0.75pt]    {$T\ =$};

		\end{tikzpicture}
		
	\end{center}
	We get a linear transformation in $\Lin_\complexNumbers (X)$ by pulling down a string:
	\begin{center}

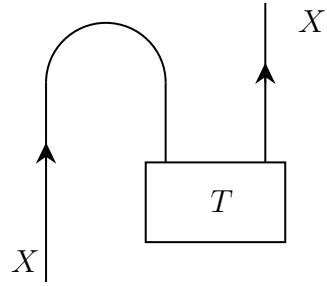
		\tikzset{every picture/.style={line width=0.75pt}} 
		
		\begin{tikzpicture}[x=0.75pt,y=0.75pt,yscale=-1,xscale=1]
			
			\draw   (70,80) -- (140,80) -- (140,120) -- (70,120) -- cycle ;
			\draw    (130,0) -- (130,80) ;
			\draw    (130,40) -- (130,33) ;
			\draw [shift={(130,30)}, rotate = 90] [fill={rgb, 255:red, 0; green, 0; blue, 0 }  ][line width=0.08]  [draw opacity=0] (10.72,-5.15) -- (0,0) -- (10.72,5.15) -- (7.12,0) -- cycle    ;
			\draw    (80,40) -- (80,80) ;
			\draw  [draw opacity=0] (20,40) .. controls (20,23.43) and (33.43,10) .. (50,10) .. controls (66.57,10) and (80,23.43) .. (80,40) -- (50,40) -- cycle ; \draw   (20,40) .. controls (20,23.43) and (33.43,10) .. (50,10) .. controls (66.57,10) and (80,23.43) .. (80,40) ;  
			\draw    (20,40) -- (20,140) ;
			\draw    (20,80) -- (20,73) ;
			\draw [shift={(20,70)}, rotate = 90] [fill={rgb, 255:red, 0; green, 0; blue, 0 }  ][line width=0.08]  [draw opacity=0] (10.72,-5.15) -- (0,0) -- (10.72,5.15) -- (7.12,0) -- cycle    ;
			
			\draw (101,92.4) node [anchor=north west][inner sep=0.75pt]    {$T$};
			\draw (145,2.4) node [anchor=north west][inner sep=0.75pt]    {$X$};
			\draw (1,122.4) node [anchor=north west][inner sep=0.75pt]    {$X$};

		\end{tikzpicture}
		
	\end{center}
	Applying $\varepsilon$ to this diagram we obtain a linear transformation in $\Lin_\complexNumbers (Y)$:
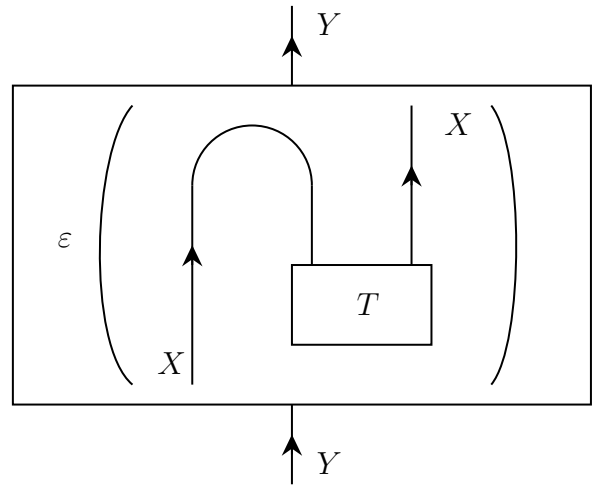
	\begin{center}

		\tikzset{every picture/.style={line width=0.75pt}} 
		
		\begin{tikzpicture}[x=0.75pt,y=0.75pt,yscale=-1,xscale=1]
			
			\draw   (140,130) -- (210,130) -- (210,170) -- (140,170) -- cycle ;
			\draw    (200,50) -- (200,130) ;
			\draw    (200,90) -- (200,83) ;
			\draw [shift={(200,80)}, rotate = 90] [fill={rgb, 255:red, 0; green, 0; blue, 0 }  ][line width=0.08]  [draw opacity=0] (10.72,-5.15) -- (0,0) -- (10.72,5.15) -- (7.12,0) -- cycle    ;
			\draw    (150,90) -- (150,130) ;
			\draw  [draw opacity=0] (90,90) .. controls (90,73.43) and (103.43,60) .. (120,60) .. controls (136.57,60) and (150,73.43) .. (150,90) -- (120,90) -- cycle ; \draw   (90,90) .. controls (90,73.43) and (103.43,60) .. (120,60) .. controls (136.57,60) and (150,73.43) .. (150,90) ;  
			\draw    (90,90) -- (90,190) ;
			\draw    (90,130) -- (90,123) ;
			\draw [shift={(90,120)}, rotate = 90] [fill={rgb, 255:red, 0; green, 0; blue, 0 }  ][line width=0.08]  [draw opacity=0] (10.72,-5.15) -- (0,0) -- (10.72,5.15) -- (7.12,0) -- cycle    ;
			\draw    (60,50) .. controls (39.7,72.6) and (36.7,170.6) .. (60,190) ;
			\draw    (240,50) .. controls (256.7,71.6) and (256.7,169.6) .. (240,190) ;
			\draw   (0,40) -- (290,40) -- (290,200) -- (0,200) -- cycle ;
			\draw    (140,200) -- (140,240) ;
			\draw    (140,225) -- (140,218) ;
			\draw [shift={(140,215)}, rotate = 90] [fill={rgb, 255:red, 0; green, 0; blue, 0 }  ][line width=0.08]  [draw opacity=0] (10.72,-5.15) -- (0,0) -- (10.72,5.15) -- (7.12,0) -- cycle    ;
			\draw    (140,0) -- (140,40) ;
			\draw    (140,25) -- (140,18) ;
			\draw [shift={(140,15)}, rotate = 90] [fill={rgb, 255:red, 0; green, 0; blue, 0 }  ][line width=0.08]  [draw opacity=0] (10.72,-5.15) -- (0,0) -- (10.72,5.15) -- (7.12,0) -- cycle    ;
			
			\draw (171,142.4) node [anchor=north west][inner sep=0.75pt]    {$T$};
			\draw (215,52.4) node [anchor=north west][inner sep=0.75pt]    {$X$};
			\draw (71,172.4) node [anchor=north west][inner sep=0.75pt]    {$X$};
			\draw (21,112.4) node [anchor=north west][inner sep=0.75pt]    {$\varepsilon $};
			\draw (151,222.4) node [anchor=north west][inner sep=0.75pt]    {$Y$};
			\draw (151,2.4) node [anchor=north west][inner sep=0.75pt]    {$Y$};

		\end{tikzpicture}
		
	\end{center}
	Pulling up a string we get a linear transformation in $\Lin_\complexNumbers (\complexNumbers, Y^* \otimes Y)$:
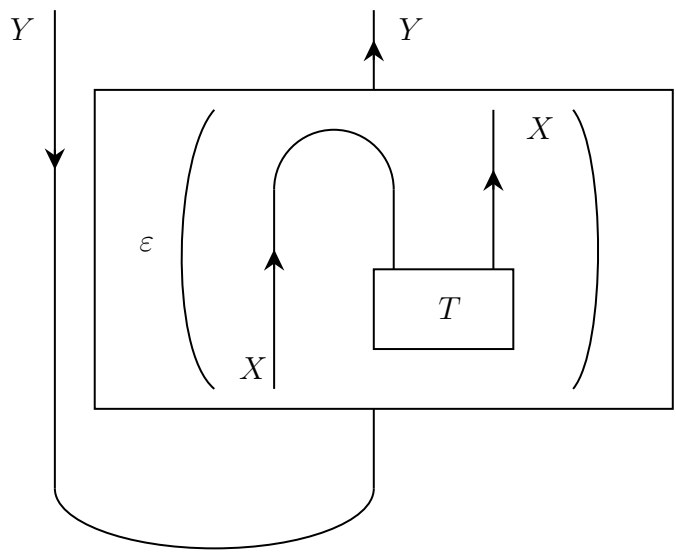
	\begin{center}

		\tikzset{every picture/.style={line width=0.75pt}} 
		
		\begin{tikzpicture}[x=0.75pt,y=0.75pt,yscale=-1,xscale=1]
			
			\draw   (185,130) -- (255,130) -- (255,170) -- (185,170) -- cycle ;
			\draw    (245,50) -- (245,130) ;
			\draw    (245,90) -- (245,83) ;
			\draw [shift={(245,80)}, rotate = 90] [fill={rgb, 255:red, 0; green, 0; blue, 0 }  ][line width=0.08]  [draw opacity=0] (10.72,-5.15) -- (0,0) -- (10.72,5.15) -- (7.12,0) -- cycle    ;
			\draw    (195,90) -- (195,130) ;
			\draw  [draw opacity=0] (135,90) .. controls (135,73.43) and (148.43,60) .. (165,60) .. controls (181.57,60) and (195,73.43) .. (195,90) -- (165,90) -- cycle ; \draw   (135,90) .. controls (135,73.43) and (148.43,60) .. (165,60) .. controls (181.57,60) and (195,73.43) .. (195,90) ;  
			\draw    (135,90) -- (135,190) ;
			\draw    (135,130) -- (135,123) ;
			\draw [shift={(135,120)}, rotate = 90] [fill={rgb, 255:red, 0; green, 0; blue, 0 }  ][line width=0.08]  [draw opacity=0] (10.72,-5.15) -- (0,0) -- (10.72,5.15) -- (7.12,0) -- cycle    ;
			\draw    (105,50) .. controls (84.7,72.6) and (81.7,170.6) .. (105,190) ;
			\draw    (285,50) .. controls (301.7,71.6) and (301.7,169.6) .. (285,190) ;
			\draw   (45,40) -- (335,40) -- (335,200) -- (45,200) -- cycle ;
			\draw    (185,200) -- (185,240) ;
			\draw    (25,70) -- (25,77) ;
			\draw [shift={(25,80)}, rotate = 270] [fill={rgb, 255:red, 0; green, 0; blue, 0 }  ][line width=0.08]  [draw opacity=0] (10.72,-5.15) -- (0,0) -- (10.72,5.15) -- (7.12,0) -- cycle    ;
			\draw    (185,0) -- (185,40) ;
			\draw    (185,25) -- (185,18) ;
			\draw [shift={(185,15)}, rotate = 90] [fill={rgb, 255:red, 0; green, 0; blue, 0 }  ][line width=0.08]  [draw opacity=0] (10.72,-5.15) -- (0,0) -- (10.72,5.15) -- (7.12,0) -- cycle    ;
			\draw  [draw opacity=0] (185,240) .. controls (185,240) and (185,240) .. (185,240) .. controls (185,256.57) and (149.18,270) .. (105,270) .. controls (60.82,270) and (25,256.57) .. (25,240) -- (105,240) -- cycle ; \draw   (185,240) .. controls (185,240) and (185,240) .. (185,240) .. controls (185,256.57) and (149.18,270) .. (105,270) .. controls (60.82,270) and (25,256.57) .. (25,240) ;  
			\draw    (25,0) -- (25,240) ;
			
			\draw (216,142.4) node [anchor=north west][inner sep=0.75pt]    {$T$};
			\draw (260,52.4) node [anchor=north west][inner sep=0.75pt]    {$X$};
			\draw (116,172.4) node [anchor=north west][inner sep=0.75pt]    {$X$};
			\draw (66,112.4) node [anchor=north west][inner sep=0.75pt]    {$\varepsilon $};
			\draw (1,2.4) node [anchor=north west][inner sep=0.75pt]    {$Y$};
			\draw (196,2.4) node [anchor=north west][inner sep=0.75pt]    {$Y$};

		\end{tikzpicture}
		
	\end{center}
	Finally, we get a vector in $Y^* \otimes Y$ by evaluating this linear transformation at 1. This function of $t$ determines a linear transformation
	\begin{equation}
		B \in \Lin_\complexNumbers (X^* \otimes X, Y^* \otimes Y).
	\end{equation}
	The other isomorphisms necessary to obtain $J(\varepsilon)$ correspond to pulling a string up and another string down:
	\begin{center}

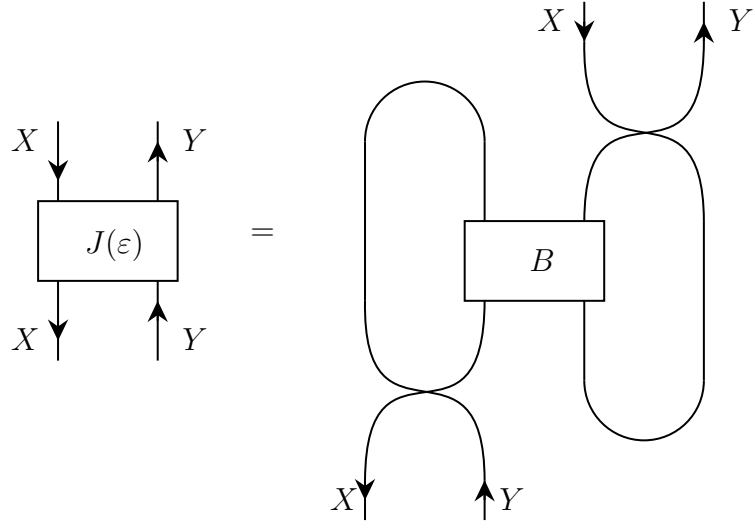
		\tikzset{every picture/.style={line width=0.75pt}} 
		
		\begin{tikzpicture}[x=0.75pt,y=0.75pt,yscale=-1,xscale=1]
			
			\draw  [draw opacity=0] (180,70) .. controls (180,53.43) and (193.43,40) .. (210,40) .. controls (226.57,40) and (240,53.43) .. (240,70) -- (210,70) -- cycle ; \draw   (180,70) .. controls (180,53.43) and (193.43,40) .. (210,40) .. controls (226.57,40) and (240,53.43) .. (240,70) ;  
			\draw    (26,80) -- (26,87) ;
			\draw [shift={(26,90)}, rotate = 270] [fill={rgb, 255:red, 0; green, 0; blue, 0 }  ][line width=0.08]  [draw opacity=0] (10.72,-5.15) -- (0,0) -- (10.72,5.15) -- (7.12,0) -- cycle    ;
			\draw   (16,100) -- (86,100) -- (86,140) -- (16,140) -- cycle ;
			\draw    (76,60) -- (76,100) ;
			\draw    (26,60) -- (26,100) ;
			\draw    (76,140) -- (76,180) ;
			\draw    (26,140) -- (26,180) ;
			\draw    (26,160) -- (26,167) ;
			\draw [shift={(26,170)}, rotate = 270] [fill={rgb, 255:red, 0; green, 0; blue, 0 }  ][line width=0.08]  [draw opacity=0] (10.72,-5.15) -- (0,0) -- (10.72,5.15) -- (7.12,0) -- cycle    ;
			\draw    (76,160) -- (76,153) ;
			\draw [shift={(76,150)}, rotate = 90] [fill={rgb, 255:red, 0; green, 0; blue, 0 }  ][line width=0.08]  [draw opacity=0] (10.72,-5.15) -- (0,0) -- (10.72,5.15) -- (7.12,0) -- cycle    ;
			\draw    (76,80) -- (76,73) ;
			\draw [shift={(76,70)}, rotate = 90] [fill={rgb, 255:red, 0; green, 0; blue, 0 }  ][line width=0.08]  [draw opacity=0] (10.72,-5.15) -- (0,0) -- (10.72,5.15) -- (7.12,0) -- cycle    ;
			\draw    (290,10) -- (290,17) ;
			\draw [shift={(290,20)}, rotate = 270] [fill={rgb, 255:red, 0; green, 0; blue, 0 }  ][line width=0.08]  [draw opacity=0] (10.72,-5.15) -- (0,0) -- (10.72,5.15) -- (7.12,0) -- cycle    ;
			\draw   (230,110) -- (300,110) -- (300,150) -- (230,150) -- cycle ;
			\draw    (240,70) -- (240,110) ;
			\draw    (290,150) -- (290,190) ;
			\draw    (240,240) -- (240,260) ;
			\draw    (180,240) -- (180,247) ;
			\draw [shift={(180,250)}, rotate = 270] [fill={rgb, 255:red, 0; green, 0; blue, 0 }  ][line width=0.08]  [draw opacity=0] (10.72,-5.15) -- (0,0) -- (10.72,5.15) -- (7.12,0) -- cycle    ;
			\draw    (350,20) -- (350,13) ;
			\draw [shift={(350,10)}, rotate = 90] [fill={rgb, 255:red, 0; green, 0; blue, 0 }  ][line width=0.08]  [draw opacity=0] (10.72,-5.15) -- (0,0) -- (10.72,5.15) -- (7.12,0) -- cycle    ;
			\draw    (240,250) -- (240,243) ;
			\draw [shift={(240,240)}, rotate = 90] [fill={rgb, 255:red, 0; green, 0; blue, 0 }  ][line width=0.08]  [draw opacity=0] (10.72,-5.15) -- (0,0) -- (10.72,5.15) -- (7.12,0) -- cycle    ;
			\draw  [draw opacity=0] (350,190) .. controls (350,190) and (350,190) .. (350,190) .. controls (350,190) and (350,190) .. (350,190) .. controls (350,206.57) and (336.57,220) .. (320,220) .. controls (303.43,220) and (290,206.57) .. (290,190) -- (320,190) -- cycle ; \draw   (350,190) .. controls (350,190) and (350,190) .. (350,190) .. controls (350,190) and (350,190) .. (350,190) .. controls (350,206.57) and (336.57,220) .. (320,220) .. controls (303.43,220) and (290,206.57) .. (290,190) ;  
			\draw    (350,110) -- (350,190) ;
			\draw    (180,70) -- (180,150) ;
			\draw    (180,150) .. controls (178.7,230.6) and (239.7,160.6) .. (240,240) ;
			\draw    (180,240) .. controls (180.7,159.6) and (239.7,232.6) .. (240,150) ;
			\draw    (180,240) -- (180,260) ;
			\draw    (290,20) .. controls (288.7,100.6) and (349.7,30.6) .. (350,110) ;
			\draw    (290,110) .. controls (290.7,29.6) and (349.7,102.6) .. (350,20) ;
			\draw    (290,0) -- (290,20) ;
			\draw    (350,0) -- (350,20) ;
			
			\draw (38,112.4) node [anchor=north west][inner sep=0.75pt]    {$J( \varepsilon )$};
			\draw (1,162.4) node [anchor=north west][inner sep=0.75pt]    {$X$};
			\draw (1,62.4) node [anchor=north west][inner sep=0.75pt]    {$X$};
			\draw (87,162.4) node [anchor=north west][inner sep=0.75pt]    {$Y$};
			\draw (87,62.4) node [anchor=north west][inner sep=0.75pt]    {$Y$};
			\draw (121,112.4) node [anchor=north west][inner sep=0.75pt]    {$=$};
			\draw (261,122.4) node [anchor=north west][inner sep=0.75pt]    {$B$};
			\draw (161,242.4) node [anchor=north west][inner sep=0.75pt]    {$X$};
			\draw (246,242.4) node [anchor=north west][inner sep=0.75pt]    {$Y$};
			\draw (265,2.4) node [anchor=north west][inner sep=0.75pt]    {$X$};
			\draw (361,2.4) node [anchor=north west][inner sep=0.75pt]    {$Y$};

		\end{tikzpicture}
		
	\end{center}

To obtain the desired expression for $J(\varepsilon)$, we'll first compute a matrix element of $J(\varepsilon)$. Let $(x_i)_i$ be an orthonormal basis for $X$ and let $(y_j)_j$ be an orthonormal basis for $Y$. We use orthonormal basis because of proposition \ref{proposition:orthonormal dual basis}, to make the proof simpler, but the expression for $J(\varepsilon)$ doesn't require the basis to be orthonormal. We have the dual basis $(x^i)_i$ for $X^*$. Because of the orthonormality, we have that
\begin{equation}
	x^i = x_i^\dag.
\end{equation}
A matrix element is
\begin{equation}
	\langle x^i \otimes y_j \mid J(\varepsilon) \mid x^{i'} \otimes y_{j'} \rangle.
\end{equation}
We can equivalently write it as
\begin{equation}
	((x^i)^\dag \otimes (y_j)^\dag)(J(\varepsilon)(x^{i'} \otimes y_{j'})).
\end{equation}
If $z \in Z$, we also denote by $z$ the linear transformation
\begin{align*}
	z \colon \complexNumbers &\longrightarrow Z \\
	\lambda &\longmapsto \lambda z.
\end{align*}
The meaning of $z$ will be determined from the context. Then the matrix element corresponds to the following diagram:
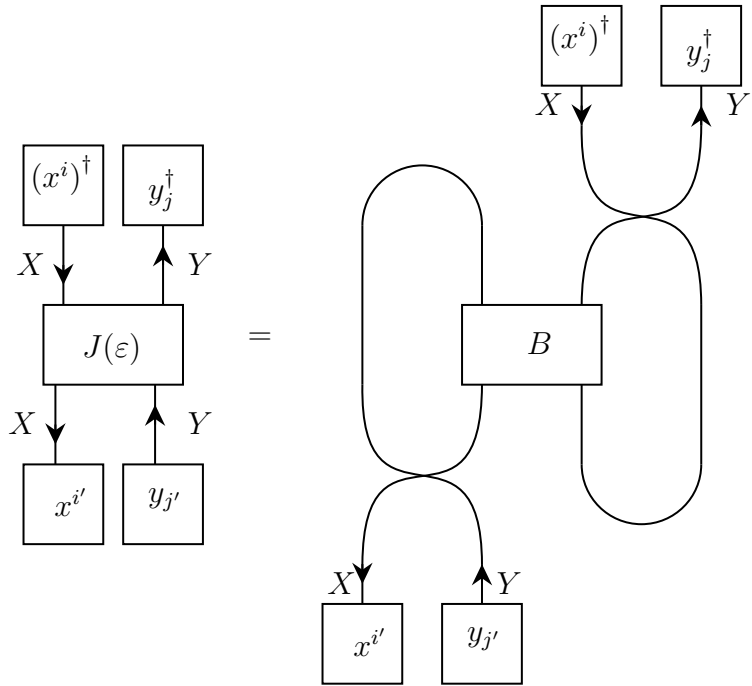
\begin{center}

	\tikzset{every picture/.style={line width=0.75pt}} 
	
	\begin{tikzpicture}[x=0.75pt,y=0.75pt,yscale=-1,xscale=1]
		
		\draw  [draw opacity=0] (180,110) .. controls (180,93.43) and (193.43,80) .. (210,80) .. controls (226.57,80) and (240,93.43) .. (240,110) -- (210,110) -- cycle ; \draw   (180,110) .. controls (180,93.43) and (193.43,80) .. (210,80) .. controls (226.57,80) and (240,93.43) .. (240,110) ;  
		\draw    (30,130) -- (30,137) ;
		\draw [shift={(30,140)}, rotate = 270] [fill={rgb, 255:red, 0; green, 0; blue, 0 }  ][line width=0.08]  [draw opacity=0] (10.72,-5.15) -- (0,0) -- (10.72,5.15) -- (7.12,0) -- cycle    ;
		\draw   (20,150) -- (90,150) -- (90,190) -- (20,190) -- cycle ;
		\draw    (80,110) -- (80,150) ;
		\draw    (30,110) -- (30,150) ;
		\draw    (76,190) -- (76,230) ;
		\draw    (26,190) -- (26,230) ;
		\draw    (26,210) -- (26,217) ;
		\draw [shift={(26,220)}, rotate = 270] [fill={rgb, 255:red, 0; green, 0; blue, 0 }  ][line width=0.08]  [draw opacity=0] (10.72,-5.15) -- (0,0) -- (10.72,5.15) -- (7.12,0) -- cycle    ;
		\draw    (76,210) -- (76,203) ;
		\draw [shift={(76,200)}, rotate = 90] [fill={rgb, 255:red, 0; green, 0; blue, 0 }  ][line width=0.08]  [draw opacity=0] (10.72,-5.15) -- (0,0) -- (10.72,5.15) -- (7.12,0) -- cycle    ;
		\draw    (80,130) -- (80,123) ;
		\draw [shift={(80,120)}, rotate = 90] [fill={rgb, 255:red, 0; green, 0; blue, 0 }  ][line width=0.08]  [draw opacity=0] (10.72,-5.15) -- (0,0) -- (10.72,5.15) -- (7.12,0) -- cycle    ;
		\draw    (290,50) -- (290,57) ;
		\draw [shift={(290,60)}, rotate = 270] [fill={rgb, 255:red, 0; green, 0; blue, 0 }  ][line width=0.08]  [draw opacity=0] (10.72,-5.15) -- (0,0) -- (10.72,5.15) -- (7.12,0) -- cycle    ;
		\draw   (230,150) -- (300,150) -- (300,190) -- (230,190) -- cycle ;
		\draw    (240,110) -- (240,150) ;
		\draw    (290,190) -- (290,230) ;
		\draw    (240,280) -- (240,300) ;
		\draw    (180,280) -- (180,287) ;
		\draw [shift={(180,290)}, rotate = 270] [fill={rgb, 255:red, 0; green, 0; blue, 0 }  ][line width=0.08]  [draw opacity=0] (10.72,-5.15) -- (0,0) -- (10.72,5.15) -- (7.12,0) -- cycle    ;
		\draw    (350,60) -- (350,53) ;
		\draw [shift={(350,50)}, rotate = 90] [fill={rgb, 255:red, 0; green, 0; blue, 0 }  ][line width=0.08]  [draw opacity=0] (10.72,-5.15) -- (0,0) -- (10.72,5.15) -- (7.12,0) -- cycle    ;
		\draw    (240,290) -- (240,283) ;
		\draw [shift={(240,280)}, rotate = 90] [fill={rgb, 255:red, 0; green, 0; blue, 0 }  ][line width=0.08]  [draw opacity=0] (10.72,-5.15) -- (0,0) -- (10.72,5.15) -- (7.12,0) -- cycle    ;
		\draw  [draw opacity=0] (350,230) .. controls (350,230) and (350,230) .. (350,230) .. controls (350,230) and (350,230) .. (350,230) .. controls (350,246.57) and (336.57,260) .. (320,260) .. controls (303.43,260) and (290,246.57) .. (290,230) -- (320,230) -- cycle ; \draw   (350,230) .. controls (350,230) and (350,230) .. (350,230) .. controls (350,230) and (350,230) .. (350,230) .. controls (350,246.57) and (336.57,260) .. (320,260) .. controls (303.43,260) and (290,246.57) .. (290,230) ;  
		\draw    (350,150) -- (350,230) ;
		\draw    (180,110) -- (180,190) ;
		\draw    (180,190) .. controls (178.7,270.6) and (239.7,200.6) .. (240,280) ;
		\draw    (180,280) .. controls (180.7,199.6) and (239.7,272.6) .. (240,190) ;
		\draw    (180,280) -- (180,300) ;
		\draw    (290,60) .. controls (288.7,140.6) and (349.7,70.6) .. (350,150) ;
		\draw    (290,150) .. controls (290.7,69.6) and (349.7,142.6) .. (350,60) ;
		\draw    (290,40) -- (290,60) ;
		\draw    (350,40) -- (350,60) ;
		\draw   (10,70) -- (50,70) -- (50,110) -- (10,110) -- cycle ;
		\draw   (60,70) -- (100,70) -- (100,110) -- (60,110) -- cycle ;
		\draw   (10,230) -- (50,230) -- (50,270) -- (10,270) -- cycle ;
		\draw   (60,230) -- (100,230) -- (100,270) -- (60,270) -- cycle ;
		\draw   (270,0) -- (310,0) -- (310,40) -- (270,40) -- cycle ;
		\draw   (330,0) -- (370,0) -- (370,40) -- (330,40) -- cycle ;
		\draw   (160,300) -- (200,300) -- (200,340) -- (160,340) -- cycle ;
		\draw   (220,300) -- (260,300) -- (260,340) -- (220,340) -- cycle ;
		
		\draw (38,162.4) node [anchor=north west][inner sep=0.75pt]    {$J( \varepsilon )$};
		\draw (1,202.4) node [anchor=north west][inner sep=0.75pt]    {$X$};
		\draw (5,122.4) node [anchor=north west][inner sep=0.75pt]    {$X$};
		\draw (91,202.4) node [anchor=north west][inner sep=0.75pt]    {$Y$};
		\draw (91,122.4) node [anchor=north west][inner sep=0.75pt]    {$Y$};
		\draw (121,162.4) node [anchor=north west][inner sep=0.75pt]    {$=$};
		\draw (261,162.4) node [anchor=north west][inner sep=0.75pt]    {$B$};
		\draw (161,282.4) node [anchor=north west][inner sep=0.75pt]    {$X$};
		\draw (246,282.4) node [anchor=north west][inner sep=0.75pt]    {$Y$};
		\draw (265,42.4) node [anchor=north west][inner sep=0.75pt]    {$X$};
		\draw (361,42.4) node [anchor=north west][inner sep=0.75pt]    {$Y$};
		\draw (12,73.4) node [anchor=north west][inner sep=0.75pt]    {$\left( x^{i}\right)^{\dagger }$};
		\draw (71,80.4) node [anchor=north west][inner sep=0.75pt]    {$y_{j}^{\dagger }$};
		\draw (24,240.4) node [anchor=north west][inner sep=0.75pt]    {$x^{i'}$};
		\draw (71,240.4) node [anchor=north west][inner sep=0.75pt]    {$y_{j'}$};
		\draw (272,3.4) node [anchor=north west][inner sep=0.75pt]    {$\left( x^{i}\right)^{\dagger }$};
		\draw (341,10.4) node [anchor=north west][inner sep=0.75pt]    {$y_{j}^{\dagger }$};
		\draw (174,310.4) node [anchor=north west][inner sep=0.75pt]    {$x^{i'}$};
		\draw (231,310.4) node [anchor=north west][inner sep=0.75pt]    {$y_{j'}$};

	\end{tikzpicture}
	
\end{center}
Using the naturality of the symmetry and the functoriality of the tensor product, we get the following:
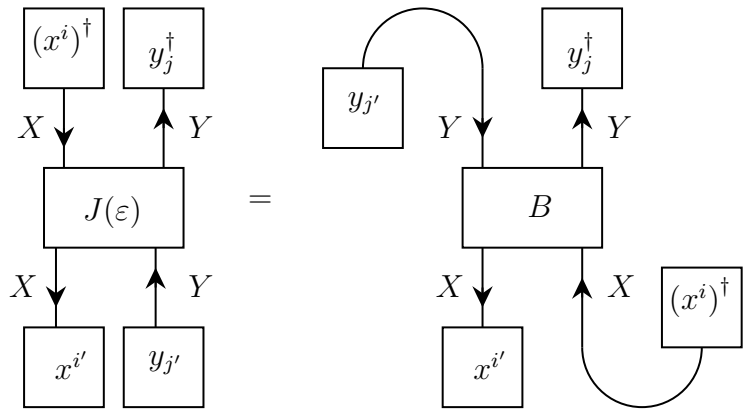
\begin{center}

	\tikzset{every picture/.style={line width=0.75pt}} 
	
	\begin{tikzpicture}[x=0.75pt,y=0.75pt,yscale=-1,xscale=1]
		
		\draw  [draw opacity=0] (180,30) .. controls (180,13.43) and (193.43,0) .. (210,0) .. controls (226.57,0) and (240,13.43) .. (240,30) -- (210,30) -- cycle ; \draw   (180,30) .. controls (180,13.43) and (193.43,0) .. (210,0) .. controls (226.57,0) and (240,13.43) .. (240,30) ;  
		\draw    (30,60) -- (30,67) ;
		\draw [shift={(30,70)}, rotate = 270] [fill={rgb, 255:red, 0; green, 0; blue, 0 }  ][line width=0.08]  [draw opacity=0] (10.72,-5.15) -- (0,0) -- (10.72,5.15) -- (7.12,0) -- cycle    ;
		\draw   (20,80) -- (90,80) -- (90,120) -- (20,120) -- cycle ;
		\draw    (80,40) -- (80,80) ;
		\draw    (30,40) -- (30,80) ;
		\draw    (76,120) -- (76,160) ;
		\draw    (26,120) -- (26,160) ;
		\draw    (26,140) -- (26,147) ;
		\draw [shift={(26,150)}, rotate = 270] [fill={rgb, 255:red, 0; green, 0; blue, 0 }  ][line width=0.08]  [draw opacity=0] (10.72,-5.15) -- (0,0) -- (10.72,5.15) -- (7.12,0) -- cycle    ;
		\draw    (76,140) -- (76,133) ;
		\draw [shift={(76,130)}, rotate = 90] [fill={rgb, 255:red, 0; green, 0; blue, 0 }  ][line width=0.08]  [draw opacity=0] (10.72,-5.15) -- (0,0) -- (10.72,5.15) -- (7.12,0) -- cycle    ;
		\draw    (80,60) -- (80,53) ;
		\draw [shift={(80,50)}, rotate = 90] [fill={rgb, 255:red, 0; green, 0; blue, 0 }  ][line width=0.08]  [draw opacity=0] (10.72,-5.15) -- (0,0) -- (10.72,5.15) -- (7.12,0) -- cycle    ;
		\draw   (230,80) -- (300,80) -- (300,120) -- (230,120) -- cycle ;
		\draw    (240,30) -- (240,80) ;
		\draw    (290,120) -- (290,170) ;
		\draw    (240,140) -- (240,147) ;
		\draw [shift={(240,150)}, rotate = 270] [fill={rgb, 255:red, 0; green, 0; blue, 0 }  ][line width=0.08]  [draw opacity=0] (10.72,-5.15) -- (0,0) -- (10.72,5.15) -- (7.12,0) -- cycle    ;
		\draw    (290,60) -- (290,53) ;
		\draw [shift={(290,50)}, rotate = 90] [fill={rgb, 255:red, 0; green, 0; blue, 0 }  ][line width=0.08]  [draw opacity=0] (10.72,-5.15) -- (0,0) -- (10.72,5.15) -- (7.12,0) -- cycle    ;
		\draw  [draw opacity=0] (350,170) .. controls (350,170) and (350,170) .. (350,170) .. controls (350,170) and (350,170) .. (350,170) .. controls (350,186.57) and (336.57,200) .. (320,200) .. controls (303.43,200) and (290,186.57) .. (290,170) -- (320,170) -- cycle ; \draw   (350,170) .. controls (350,170) and (350,170) .. (350,170) .. controls (350,170) and (350,170) .. (350,170) .. controls (350,186.57) and (336.57,200) .. (320,200) .. controls (303.43,200) and (290,186.57) .. (290,170) ;  
		\draw    (240,120) -- (240,160) ;
		\draw    (290,40) -- (290,80) ;
		\draw   (10,0) -- (50,0) -- (50,40) -- (10,40) -- cycle ;
		\draw   (60,0) -- (100,0) -- (100,40) -- (60,40) -- cycle ;
		\draw   (10,160) -- (50,160) -- (50,200) -- (10,200) -- cycle ;
		\draw   (60,160) -- (100,160) -- (100,200) -- (60,200) -- cycle ;
		\draw   (330,130) -- (370,130) -- (370,170) -- (330,170) -- cycle ;
		\draw   (270,0) -- (310,0) -- (310,40) -- (270,40) -- cycle ;
		\draw   (220,160) -- (260,160) -- (260,200) -- (220,200) -- cycle ;
		\draw   (160,30) -- (200,30) -- (200,70) -- (160,70) -- cycle ;
		\draw    (240,55) -- (240,62) ;
		\draw [shift={(240,65)}, rotate = 270] [fill={rgb, 255:red, 0; green, 0; blue, 0 }  ][line width=0.08]  [draw opacity=0] (10.72,-5.15) -- (0,0) -- (10.72,5.15) -- (7.12,0) -- cycle    ;
		\draw    (290,145) -- (290,138) ;
		\draw [shift={(290,135)}, rotate = 90] [fill={rgb, 255:red, 0; green, 0; blue, 0 }  ][line width=0.08]  [draw opacity=0] (10.72,-5.15) -- (0,0) -- (10.72,5.15) -- (7.12,0) -- cycle    ;
		
		\draw (38,92.4) node [anchor=north west][inner sep=0.75pt]    {$J( \varepsilon )$};
		\draw (1,132.4) node [anchor=north west][inner sep=0.75pt]    {$X$};
		\draw (5,52.4) node [anchor=north west][inner sep=0.75pt]    {$X$};
		\draw (91,132.4) node [anchor=north west][inner sep=0.75pt]    {$Y$};
		\draw (91,52.4) node [anchor=north west][inner sep=0.75pt]    {$Y$};
		\draw (121,92.4) node [anchor=north west][inner sep=0.75pt]    {$=$};
		\draw (261,92.4) node [anchor=north west][inner sep=0.75pt]    {$B$};
		\draw (215,132.4) node [anchor=north west][inner sep=0.75pt]    {$X$};
		\draw (301,52.4) node [anchor=north west][inner sep=0.75pt]    {$Y$};
		\draw (12,3.4) node [anchor=north west][inner sep=0.75pt]    {$\left( x^{i}\right)^{\dagger }$};
		\draw (71,10.4) node [anchor=north west][inner sep=0.75pt]    {$y_{j}^{\dagger }$};
		\draw (24,170.4) node [anchor=north west][inner sep=0.75pt]    {$x^{i'}$};
		\draw (71,170.4) node [anchor=north west][inner sep=0.75pt]    {$y_{j'}$};
		\draw (332,133.4) node [anchor=north west][inner sep=0.75pt]    {$\left( x^{i}\right)^{\dagger }$};
		\draw (281,10.4) node [anchor=north west][inner sep=0.75pt]    {$y_{j}^{\dagger }$};
		\draw (234,170.4) node [anchor=north west][inner sep=0.75pt]    {$x^{i'}$};
		\draw (171,40.4) node [anchor=north west][inner sep=0.75pt]    {$y_{j'}$};
		\draw (301,132.4) node [anchor=north west][inner sep=0.75pt]    {$X$};
		\draw (216,52.4) node [anchor=north west][inner sep=0.75pt]    {$Y$};

	\end{tikzpicture}
	
\end{center}
There are two boxes with a cap or cup attached. We can simplify them to boxes without caps or cups. For the upper diagram
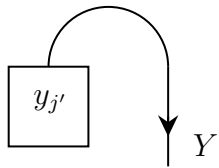
\begin{center}

	\tikzset{every picture/.style={line width=0.75pt}} 
	
	\begin{tikzpicture}[x=0.75pt,y=0.75pt,yscale=-1,xscale=1]
		
		\draw  [draw opacity=0] (20,30) .. controls (20,13.43) and (33.43,0) .. (50,0) .. controls (66.57,0) and (80,13.43) .. (80,30) -- (50,30) -- cycle ; \draw   (20,30) .. controls (20,13.43) and (33.43,0) .. (50,0) .. controls (66.57,0) and (80,13.43) .. (80,30) ;  
		\draw    (80,30) -- (80,80) ;
		\draw   (0,30) -- (40,30) -- (40,70) -- (0,70) -- cycle ;
		\draw    (80,55) -- (80,62) ;
		\draw [shift={(80,65)}, rotate = 270] [fill={rgb, 255:red, 0; green, 0; blue, 0 }  ][line width=0.08]  [draw opacity=0] (10.72,-5.15) -- (0,0) -- (10.72,5.15) -- (7.12,0) -- cycle    ;
		
		\draw (11,40.4) node [anchor=north west][inner sep=0.75pt]    {$y_{j'}$};
		\draw (91,62.4) node [anchor=north west][inner sep=0.75pt]    {$Y$};

	\end{tikzpicture}
	
\end{center}
it can be simplified to the $\complexNumbers$-linear transformation
\begin{equation}
	(y^{j'})^\dag \colon Y^* \to \complexNumbers .
\end{equation}
This simplification won't be necessary, so we won't use it in this proof and won't prove it.

For the lower diagram
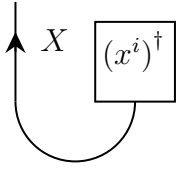
\begin{center}

	\tikzset{every picture/.style={line width=0.75pt}} 
	
	\begin{tikzpicture}[x=0.75pt,y=0.75pt,yscale=-1,xscale=1]
		
		\draw    (0,0) -- (0,50) ;
		\draw  [draw opacity=0] (60,50) .. controls (60,50) and (60,50) .. (60,50) .. controls (60,50) and (60,50) .. (60,50) .. controls (60,66.57) and (46.57,80) .. (30,80) .. controls (13.43,80) and (0,66.57) .. (0,50) -- (30,50) -- cycle ; \draw   (60,50) .. controls (60,50) and (60,50) .. (60,50) .. controls (60,50) and (60,50) .. (60,50) .. controls (60,66.57) and (46.57,80) .. (30,80) .. controls (13.43,80) and (0,66.57) .. (0,50) ;  
		\draw   (40,10) -- (80,10) -- (80,50) -- (40,50) -- cycle ;
		\draw    (0,25) -- (0,18) ;
		\draw [shift={(0,15)}, rotate = 90] [fill={rgb, 255:red, 0; green, 0; blue, 0 }  ][line width=0.08]  [draw opacity=0] (10.72,-5.15) -- (0,0) -- (10.72,5.15) -- (7.12,0) -- cycle    ;
		
		\draw (42,13.4) node [anchor=north west][inner sep=0.75pt]    {$\left( x^{i}\right)^{\dagger }$};
		\draw (11,12.4) node [anchor=north west][inner sep=0.75pt]    {$X$};

	\end{tikzpicture}
	
\end{center}
it corresponds to the linear transformation
\begin{align*}
	h \colon \complexNumbers &\longrightarrow X \\
	\lambda &\longmapsto r_X (\id_X \otimes (x^i)^\dag) \lcoev_X (\lambda).
\end{align*}
By linearity, we have that
\begin{equation}
	h(\lambda) = \lambda h(1).
\end{equation}
Let's compute $h(1)$. We have that
\begin{align*}
	h(1) &\stackrel{(i)}{=} r_X (\id_X \otimes (x^i)^\dag) \lcoev_X (1) \\
	&\stackrel{(ii)}{=} r_X (\id_X \otimes (x^i)^\dag) \left( \sum_{i'} x_{i'} \otimes x^{i'} \right) \\
	&\stackrel{(iii)}{=}  \sum_{i'} r_X (\id_X \otimes (x^i)^\dag) ( x_{i'} \otimes x^{i'} ) \\
	&\stackrel{(iv)}{=}  \sum_{i'} r_X (x_{i'} \otimes (x^i)^\dag (x^{i'}) )  \\
	&\stackrel{(v)}{=}  \sum_{i'} r_X (x_{i'} \otimes \delta_{i,i'}) \\
	&\stackrel{(vi)}{=}  r_X (x_i \otimes 1) \\
	&\stackrel{(vii)}{=}  x_i .
\end{align*}
In $(i)$ was used the definition of $h$; in $(ii)$ was used proposition \ref{definition:coevaluations}; in $(iii)$ was used the linearity to pass the summation to outside; in $(iv)$ the term $\id_X \otimes (x^i)^\dag$  was applied to $x_{i'} \otimes x^{i'} $; in $(v)$ was used that $(x^i)^i$ is an orthonormal basis, as proved in proposition \ref{proposition:orthonormal dual basis}; in $(vi)$ the delta was used to eliminate the summation; in $(vii)$ was applied $r_X$, which multiplies $x_i$ by 1, according to proposition \ref{proposition:right unitality}. This proves that
\begin{equation}
	h(\lambda) = \lambda x_i .
\end{equation}
We also denote this linear transformation by $x_i$, then the meaning of $x_i$ is deduced from the context.

With these results, we can rewrite the diagrammatic expression for matrix element as
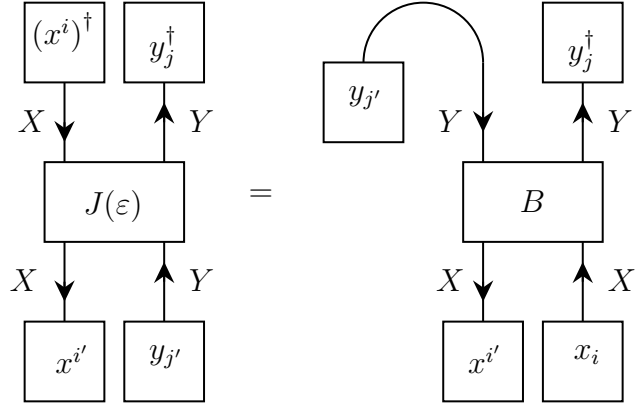
\begin{center}

	\tikzset{every picture/.style={line width=0.75pt}} 
	

	
\end{center}

Next we compute $B(x^{i'} \otimes x_i)$ using the expression found previously for $B$ in terms of $t \in X^* \otimes X$. In this case, we have that
\begin{equation}
	t = x^{i'} \otimes x_i .
\end{equation}
Diagrammatically, we have that
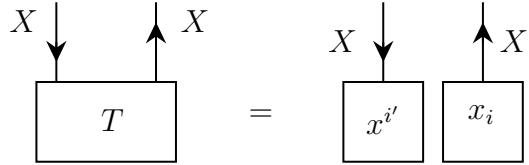
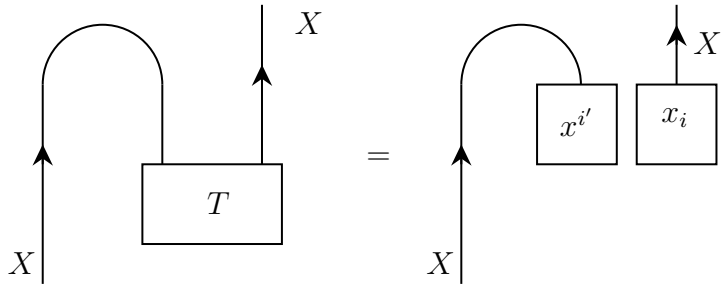
\begin{center}

	\tikzset{every picture/.style={line width=0.75pt}} 
	
	%
	
\end{center}
Part of the diagram is the adjoint of another diagram we already computed:
\begin{center}
	\tikzset{every picture/.style={line width=0.75pt}} 
	
	%
	
\end{center}
it follows that
\begin{center}

	\tikzset{every picture/.style={line width=0.75pt}} 
	
	%
	
\end{center}
Therefore, we have that
\begin{center}

	\tikzset{every picture/.style={line width=0.75pt}} 
	
	%
	
\end{center}
In Dirac notation, the right hand side can be written as
\begin{equation}
	\ket{x_i }\bra{ x_{i'} }.
\end{equation}
	
The next step to compute $B$ is to apply $\varepsilon$, for which we obtain
\begin{equation}
	\varepsilon (\ket{x_i }\bra{ x_{i'} }) \in \Lin_\complexNumbers(Y,Y).
\end{equation}
The final step to compute $B (x^{i'} \otimes x_i)$ is to use the isomorphism
\begin{equation}
	\Lin_\complexNumbers (Y, Y) \cong \Lin_\complexNumbers (\complexNumbers, Y^* \otimes Y),
\end{equation}
which is given by pulling a string upwards. We get that
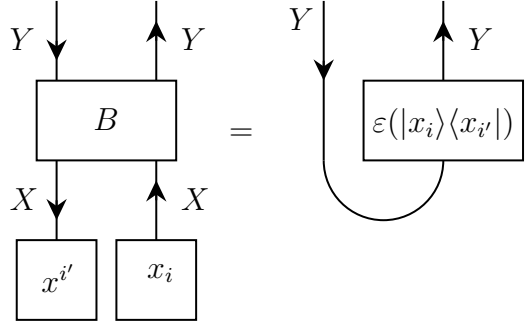
\begin{center}

	\tikzset{every picture/.style={line width=0.75pt}} 
	
	\begin{tikzpicture}[x=0.75pt,y=0.75pt,yscale=-1,xscale=1]
		
		\draw   (180,40) -- (260,40) -- (260,80) -- (180,80) -- cycle ;
		\draw    (160,30) -- (160,37) ;
		\draw [shift={(160,40)}, rotate = 270] [fill={rgb, 255:red, 0; green, 0; blue, 0 }  ][line width=0.08]  [draw opacity=0] (10.72,-5.15) -- (0,0) -- (10.72,5.15) -- (7.12,0) -- cycle    ;
		\draw    (220,0) -- (220,40) ;
		\draw    (220,20) -- (220,13) ;
		\draw [shift={(220,10)}, rotate = 90] [fill={rgb, 255:red, 0; green, 0; blue, 0 }  ][line width=0.08]  [draw opacity=0] (10.72,-5.15) -- (0,0) -- (10.72,5.15) -- (7.12,0) -- cycle    ;
		\draw  [draw opacity=0] (220,80) .. controls (220,80) and (220,80) .. (220,80) .. controls (220,80) and (220,80) .. (220,80) .. controls (220,96.57) and (206.57,110) .. (190,110) .. controls (173.43,110) and (160,96.57) .. (160,80) -- (190,80) -- cycle ; \draw   (220,80) .. controls (220,80) and (220,80) .. (220,80) .. controls (220,80) and (220,80) .. (220,80) .. controls (220,96.57) and (206.57,110) .. (190,110) .. controls (173.43,110) and (160,96.57) .. (160,80) ;  
		\draw    (160,0) -- (160,80) ;
		\draw   (16,40) -- (86,40) -- (86,80) -- (16,80) -- cycle ;
		\draw    (76,80) -- (76,120) ;
		\draw    (26,100) -- (26,107) ;
		\draw [shift={(26,110)}, rotate = 270] [fill={rgb, 255:red, 0; green, 0; blue, 0 }  ][line width=0.08]  [draw opacity=0] (10.72,-5.15) -- (0,0) -- (10.72,5.15) -- (7.12,0) -- cycle    ;
		\draw    (26,80) -- (26,120) ;
		\draw   (56,120) -- (96,120) -- (96,160) -- (56,160) -- cycle ;
		\draw   (6,120) -- (46,120) -- (46,160) -- (6,160) -- cycle ;
		\draw    (76,100) -- (76,93) ;
		\draw [shift={(76,90)}, rotate = 90] [fill={rgb, 255:red, 0; green, 0; blue, 0 }  ][line width=0.08]  [draw opacity=0] (10.72,-5.15) -- (0,0) -- (10.72,5.15) -- (7.12,0) -- cycle    ;
		\draw    (76,0) -- (76,40) ;
		\draw    (26,20) -- (26,27) ;
		\draw [shift={(26,30)}, rotate = 270] [fill={rgb, 255:red, 0; green, 0; blue, 0 }  ][line width=0.08]  [draw opacity=0] (10.72,-5.15) -- (0,0) -- (10.72,5.15) -- (7.12,0) -- cycle    ;
		\draw    (26,0) -- (26,40) ;
		\draw    (76,20) -- (76,13) ;
		\draw [shift={(76,10)}, rotate = 90] [fill={rgb, 255:red, 0; green, 0; blue, 0 }  ][line width=0.08]  [draw opacity=0] (10.72,-5.15) -- (0,0) -- (10.72,5.15) -- (7.12,0) -- cycle    ;
		
		\draw (183,52.4) node [anchor=north west][inner sep=0.75pt]    {$\varepsilon ( |x_{i} \rangle \langle x_{i'}|)$};
		\draw (141,2.4) node [anchor=north west][inner sep=0.75pt]    {$Y$};
		\draw (231,12.4) node [anchor=north west][inner sep=0.75pt]    {$Y$};
		\draw (43,52.4) node [anchor=north west][inner sep=0.75pt]    {$B$};
		\draw (1,92.4) node [anchor=north west][inner sep=0.75pt]    {$X$};
		\draw (70,130.4) node [anchor=north west][inner sep=0.75pt]    {$x_{i}$};
		\draw (17,130.4) node [anchor=north west][inner sep=0.75pt]    {$x^{i'}$};
		\draw (87,92.4) node [anchor=north west][inner sep=0.75pt]    {$X$};
		\draw (1,12.4) node [anchor=north west][inner sep=0.75pt]    {$Y$};
		\draw (87,12.4) node [anchor=north west][inner sep=0.75pt]    {$Y$};
		\draw (111,62.4) node [anchor=north west][inner sep=0.75pt]    {$=$};

	\end{tikzpicture}
	
\end{center}
Then the matrix element have the following diagrammatic expression
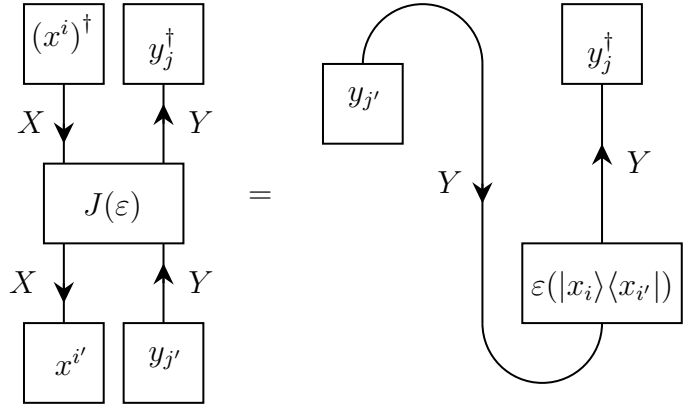
\begin{center}

	\tikzset{every picture/.style={line width=0.75pt}} 
	
	\begin{tikzpicture}[x=0.75pt,y=0.75pt,yscale=-1,xscale=1]
		
		\draw    (30,60) -- (30,67) ;
		\draw [shift={(30,70)}, rotate = 270] [fill={rgb, 255:red, 0; green, 0; blue, 0 }  ][line width=0.08]  [draw opacity=0] (10.72,-5.15) -- (0,0) -- (10.72,5.15) -- (7.12,0) -- cycle    ;
		\draw   (20,80) -- (90,80) -- (90,120) -- (20,120) -- cycle ;
		\draw    (80,40) -- (80,80) ;
		\draw    (30,40) -- (30,80) ;
		\draw    (80,120) -- (80,160) ;
		\draw    (30,120) -- (30,160) ;
		\draw    (30,140) -- (30,147) ;
		\draw [shift={(30,150)}, rotate = 270] [fill={rgb, 255:red, 0; green, 0; blue, 0 }  ][line width=0.08]  [draw opacity=0] (10.72,-5.15) -- (0,0) -- (10.72,5.15) -- (7.12,0) -- cycle    ;
		\draw    (80,140) -- (80,133) ;
		\draw [shift={(80,130)}, rotate = 90] [fill={rgb, 255:red, 0; green, 0; blue, 0 }  ][line width=0.08]  [draw opacity=0] (10.72,-5.15) -- (0,0) -- (10.72,5.15) -- (7.12,0) -- cycle    ;
		\draw    (80,60) -- (80,53) ;
		\draw [shift={(80,50)}, rotate = 90] [fill={rgb, 255:red, 0; green, 0; blue, 0 }  ][line width=0.08]  [draw opacity=0] (10.72,-5.15) -- (0,0) -- (10.72,5.15) -- (7.12,0) -- cycle    ;
		\draw    (300,80) -- (300,73) ;
		\draw [shift={(300,70)}, rotate = 90] [fill={rgb, 255:red, 0; green, 0; blue, 0 }  ][line width=0.08]  [draw opacity=0] (10.72,-5.15) -- (0,0) -- (10.72,5.15) -- (7.12,0) -- cycle    ;
		\draw    (300,40) -- (300,57.6) -- (300,120) ;
		\draw   (10,0) -- (50,0) -- (50,40) -- (10,40) -- cycle ;
		\draw   (60,0) -- (100,0) -- (100,40) -- (60,40) -- cycle ;
		\draw   (10,160) -- (50,160) -- (50,200) -- (10,200) -- cycle ;
		\draw   (60,160) -- (100,160) -- (100,200) -- (60,200) -- cycle ;
		\draw   (280,0) -- (320,0) -- (320,40) -- (280,40) -- cycle ;
		\draw  [draw opacity=0] (180,30) .. controls (180,13.43) and (193.43,0) .. (210,0) .. controls (226.57,0) and (240,13.43) .. (240,30) -- (210,30) -- cycle ; \draw   (180,30) .. controls (180,13.43) and (193.43,0) .. (210,0) .. controls (226.57,0) and (240,13.43) .. (240,30) ;  
		\draw   (160,30) -- (200,30) -- (200,70) -- (160,70) -- cycle ;
		\draw   (260,120) -- (340,120) -- (340,160) -- (260,160) -- cycle ;
		\draw    (240,90) -- (240,97) ;
		\draw [shift={(240,100)}, rotate = 270] [fill={rgb, 255:red, 0; green, 0; blue, 0 }  ][line width=0.08]  [draw opacity=0] (10.72,-5.15) -- (0,0) -- (10.72,5.15) -- (7.12,0) -- cycle    ;
		\draw  [draw opacity=0] (300,160) .. controls (300,160) and (300,160) .. (300,160) .. controls (300,160) and (300,160) .. (300,160) .. controls (300,176.57) and (286.57,190) .. (270,190) .. controls (253.43,190) and (240,176.57) .. (240,160) -- (270,160) -- cycle ; \draw   (300,160) .. controls (300,160) and (300,160) .. (300,160) .. controls (300,160) and (300,160) .. (300,160) .. controls (300,176.57) and (286.57,190) .. (270,190) .. controls (253.43,190) and (240,176.57) .. (240,160) ;  
		\draw    (240,30) -- (240,160) ;
		
		\draw (38,92.4) node [anchor=north west][inner sep=0.75pt]    {$J( \varepsilon )$};
		\draw (1,132.4) node [anchor=north west][inner sep=0.75pt]    {$X$};
		\draw (5,52.4) node [anchor=north west][inner sep=0.75pt]    {$X$};
		\draw (91,132.4) node [anchor=north west][inner sep=0.75pt]    {$Y$};
		\draw (91,52.4) node [anchor=north west][inner sep=0.75pt]    {$Y$};
		\draw (121,92.4) node [anchor=north west][inner sep=0.75pt]    {$=$};
		\draw (311,72.4) node [anchor=north west][inner sep=0.75pt]    {$Y$};
		\draw (12,3.4) node [anchor=north west][inner sep=0.75pt]    {$\left( x^{i}\right)^{\dagger }$};
		\draw (71,10.4) node [anchor=north west][inner sep=0.75pt]    {$y_{j}^{\dagger }$};
		\draw (24,170.4) node [anchor=north west][inner sep=0.75pt]    {$x^{i'}$};
		\draw (71,170.4) node [anchor=north west][inner sep=0.75pt]    {$y_{j'}$};
		\draw (291,12.4) node [anchor=north west][inner sep=0.75pt]    {$y_{j}^{\dagger }$};
		\draw (171,40.4) node [anchor=north west][inner sep=0.75pt]    {$y_{j'}$};
		\draw (263,132.4) node [anchor=north west][inner sep=0.75pt]    {$\varepsilon ( |x_{i} \rangle \langle x_{i'} |)$};
		\draw (216,82.4) node [anchor=north west][inner sep=0.75pt]    {$Y$};

	\end{tikzpicture}
	
\end{center}
The yanking equations say that we can straighten the curved string, obtaining that
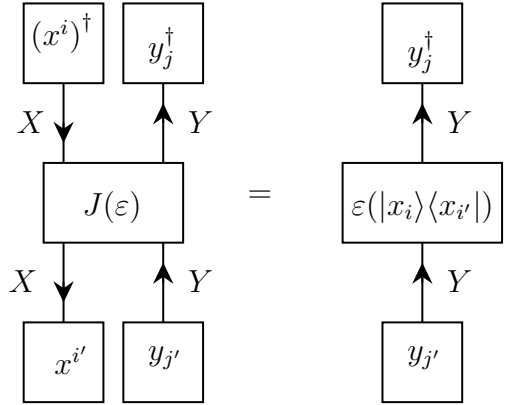
\begin{center}

	\tikzset{every picture/.style={line width=0.75pt}} 
	
	\begin{tikzpicture}[x=0.75pt,y=0.75pt,yscale=-1,xscale=1]
		
		\draw    (30,60) -- (30,67) ;
		\draw [shift={(30,70)}, rotate = 270] [fill={rgb, 255:red, 0; green, 0; blue, 0 }  ][line width=0.08]  [draw opacity=0] (10.72,-5.15) -- (0,0) -- (10.72,5.15) -- (7.12,0) -- cycle    ;
		\draw   (20,80) -- (90,80) -- (90,120) -- (20,120) -- cycle ;
		\draw    (80,40) -- (80,80) ;
		\draw    (30,40) -- (30,80) ;
		\draw    (80,120) -- (80,160) ;
		\draw    (30,120) -- (30,160) ;
		\draw    (30,140) -- (30,147) ;
		\draw [shift={(30,150)}, rotate = 270] [fill={rgb, 255:red, 0; green, 0; blue, 0 }  ][line width=0.08]  [draw opacity=0] (10.72,-5.15) -- (0,0) -- (10.72,5.15) -- (7.12,0) -- cycle    ;
		\draw    (80,140) -- (80,133) ;
		\draw [shift={(80,130)}, rotate = 90] [fill={rgb, 255:red, 0; green, 0; blue, 0 }  ][line width=0.08]  [draw opacity=0] (10.72,-5.15) -- (0,0) -- (10.72,5.15) -- (7.12,0) -- cycle    ;
		\draw    (80,60) -- (80,53) ;
		\draw [shift={(80,50)}, rotate = 90] [fill={rgb, 255:red, 0; green, 0; blue, 0 }  ][line width=0.08]  [draw opacity=0] (10.72,-5.15) -- (0,0) -- (10.72,5.15) -- (7.12,0) -- cycle    ;
		\draw   (10,0) -- (50,0) -- (50,40) -- (10,40) -- cycle ;
		\draw   (60,0) -- (100,0) -- (100,40) -- (60,40) -- cycle ;
		\draw   (10,160) -- (50,160) -- (50,200) -- (10,200) -- cycle ;
		\draw   (60,160) -- (100,160) -- (100,200) -- (60,200) -- cycle ;
		\draw   (190,0) -- (230,0) -- (230,40) -- (190,40) -- cycle ;
		\draw   (190,160) -- (230,160) -- (230,200) -- (190,200) -- cycle ;
		\draw   (170,80) -- (250,80) -- (250,120) -- (170,120) -- cycle ;
		\draw    (210,40) -- (210,80) ;
		\draw    (210,60) -- (210,53) ;
		\draw [shift={(210,50)}, rotate = 90] [fill={rgb, 255:red, 0; green, 0; blue, 0 }  ][line width=0.08]  [draw opacity=0] (10.72,-5.15) -- (0,0) -- (10.72,5.15) -- (7.12,0) -- cycle    ;
		\draw    (210,120) -- (210,160) ;
		\draw    (210,140) -- (210,133) ;
		\draw [shift={(210,130)}, rotate = 90] [fill={rgb, 255:red, 0; green, 0; blue, 0 }  ][line width=0.08]  [draw opacity=0] (10.72,-5.15) -- (0,0) -- (10.72,5.15) -- (7.12,0) -- cycle    ;
		
		\draw (38,92.4) node [anchor=north west][inner sep=0.75pt]    {$J( \varepsilon )$};
		\draw (1,132.4) node [anchor=north west][inner sep=0.75pt]    {$X$};
		\draw (5,52.4) node [anchor=north west][inner sep=0.75pt]    {$X$};
		\draw (91,132.4) node [anchor=north west][inner sep=0.75pt]    {$Y$};
		\draw (91,52.4) node [anchor=north west][inner sep=0.75pt]    {$Y$};
		\draw (121,92.4) node [anchor=north west][inner sep=0.75pt]    {$=$};
		\draw (221,52.4) node [anchor=north west][inner sep=0.75pt]    {$Y$};
		\draw (12,3.4) node [anchor=north west][inner sep=0.75pt]    {$\left( x^{i}\right)^{\dagger }$};
		\draw (71,10.4) node [anchor=north west][inner sep=0.75pt]    {$y_{j}^{\dagger }$};
		\draw (24,170.4) node [anchor=north west][inner sep=0.75pt]    {$x^{i'}$};
		\draw (71,170.4) node [anchor=north west][inner sep=0.75pt]    {$y_{j'}$};
		\draw (201,12.4) node [anchor=north west][inner sep=0.75pt]    {$y_{j}^{\dagger }$};
		\draw (201,170.4) node [anchor=north west][inner sep=0.75pt]    {$y_{j'}$};
		\draw (173,92.4) node [anchor=north west][inner sep=0.75pt]    {$\varepsilon ( |x_{i} \rangle \langle x_{i'} |)$};
		\draw (221,132.4) node [anchor=north west][inner sep=0.75pt]    {$Y$};

	\end{tikzpicture}
	
\end{center}
In Dirac notation, this can be written as
\begin{equation}
	\label{equation: CJ matrix element}
	\bra{ x^i \otimes y_j} J(\varepsilon) \ket{ x^{i'} \otimes y_{j'} } = \bra{y_j} \varepsilon(\ket{x_i} \bra{x_{i'}} ) \ket{y_{j'}}.
\end{equation}
From this equation we can derive the desired expression for $J(\varepsilon)$. We have that
\begin{align*}
	J(\varepsilon) &\stackrel{(i)}{=} \id_{X^* \otimes y} J(\varepsilon) \id_{X^* \otimes y} \\
	&\stackrel{(ii)}{=} \left( \sum_{i,j}  \ket{x^i \otimes y_j} \bra{x^i \otimes y_j} \right) J(\varepsilon) \left( \sum_{i',j'}  \ket{x^{i'} \otimes y_{j'}} \bra{x^{i'} \otimes y_{j'}} \right) \\
	&\stackrel{(iii)}{=} \sum_{i,j, i', j'}   \bra{x^i \otimes y_j} J(\varepsilon)  \ket{x^{i'} \otimes y_{j'}} \ \ket{x^i \otimes y_j} \bra{x^{i'} \otimes y_{j'}} \\
	&\stackrel{(iv)}{=} \sum_{i,j, i', j'} \bra{y_j} \varepsilon(\ket{x_i} \bra{x_{i'}} ) \ket{y_{j'}} \ \ket{x^i \otimes y_j} \bra{x^{i'} \otimes y_{j'}} \\
	&\stackrel{(v)}{=} \sum_{i,j,i',j'} \ket{x^i} \bra{x^{i'}} \otimes \ket{y_j} \bra{y_j} \varepsilon(\ket{x_i} \bra{x_{i'}} ) \ket{y_{j'}} \bra{y_{j'}}  \\
	&\stackrel{(vi)}{=} \sum_{i,i'} \ket{x^i} \bra{x^{i'}} \otimes \left(\sum_j \ket{y_j} \bra{y_j} \right) \varepsilon(\ket{x_i} \bra{x_{i'}} ) \left( \sum_{j'} \ket{y_{j'}} \bra{y_{j'}} \right) \\
	&\stackrel{(vii)}{=} \sum_{i,i'} \ket{x^i} \bra{x^{i'}} \otimes \varepsilon(\ket{x_i} \bra{x_{i'}} ).
\end{align*}
In $(i)$ were inserted identity functions; in $(ii)$ the identity functions were written in the basis; in $(iii)$ $\bra{x^i \otimes y_j}$ and $\ket{x^{i'} \otimes y_{j'}}$ were applied to $J(\varepsilon)$, obtaining the matrix element $\bra{x^i \otimes y_j} J(\varepsilon) \ket{x^{i'} \otimes y_{j'}} \in \complexNumbers$ which was moved to the left; in $(iv)$ was used equation \ref{equation: CJ matrix element}; in $(v)$ the bras and kets of $\ket{x^i \otimes y_j} \bra{x^{i'} \otimes y_{j'}}$ were reorganized; in $(vi)$ the summations over $j$ and $j'$ were moved to make appear identity functions; in $(vii)$ was used that $\sum_j \ket{y_j}\bra{y_j} = \id_Y$. This proves that
\begin{equation}
	\label{equation: CJ first expression}
	J(\varepsilon) = \sum_{i,i'} \ket{x^i} \bra{x^{i'}} \otimes \varepsilon(\ket{x_i} \bra{x_{i'}} ),
\end{equation}
which is one of the desired expressions. We can rewrite this equation to make more explicit its relation with the standard definition of the \CJ isomorphism. We have that
\begin{align*}
	J(\varepsilon) &\stackrel{(i)}{=} \sum_{i,i'} \ket{x^i} \bra{x^{i'}} \otimes \varepsilon(\ket{x_i} \bra{x_{i'}} ) \\
	&\stackrel{(ii)}{=} (\id_{\Lin_\complexNumbers(X^*)} \otimes \varepsilon) \left( \sum_{i,i'} \ket{x^i} \bra{x^{i'}} \otimes \ket{x_i} \bra{x_{i'}} \right) \\
	&\stackrel{(iii)}{=} (\id_{\Lin_\complexNumbers(X^*)} \otimes \varepsilon) \left( \sum_{i,i'} \ket{x^i \otimes x_i} \bra{x^{i'} \otimes x_{i'}} \right) \\
	&\stackrel{(iv)}{=} (\id_{\Lin_\complexNumbers(X^*)} \otimes \varepsilon) \left( \left( \sum_i \ket{x^i \otimes x_i} \right) \left( \sum_{i'} \bra{x^{i'} \otimes x_{i'}} \right) \right) \\
	&\stackrel{(v)}{=} (\id_{\Lin_\complexNumbers(X^*)} \otimes \varepsilon) \left( \ket{\Omega} \bra{\Omega} \right).
\end{align*}
In $(i)$ was used equation \ref{equation: CJ first expression}; in $(ii)$ the application of $\varepsilon$ was factorized together with $\id_{\Lin_\complexNumbers(X^*)}$, so that the summation is next to the basis vectors; in $(iii)$ was used that $\ket{x^i} \otimes \ket{x_i} = \ket{x^i \otimes x_i}$, and the same for the adjoints; in $(iv)$ the summations over $i$ and $i'$ were isolated; in $(v)$ was used that the summation over $i$ is the vector
\begin{equation}
	\ket{\Omega} = \sum_i \ket{x^i \otimes x_i}.
\end{equation}
The summation over $i'$ is the adjoint, which is $\bra{\Omega}$. This proves that
\begin{equation}
	\label{equation: CJ second expression}
	J(\varepsilon) = (\id_{\Lin_\complexNumbers(X^*)} \otimes \varepsilon) \left( \ket{\Omega} \bra{\Omega} \right),
\end{equation}
which is the other desired expression.

It remains to prove the expression for $J^{-1}(A)$, for a given $A \in \Lin_\complexNumbers (X^* \otimes Y)$. For this, we'll use equation \ref{equation: CJ first expression}. This equation has the term $\varepsilon(\ket{x_i} \bra{x_{i'}})$. Isolating this term, we can find an expression for $\varepsilon(B)$ using that $\varepsilon$ is linear. It is easier to understand the arguments using string diagrams. Equation \ref{equation: CJ first expression} have the following diagrammatic expression:
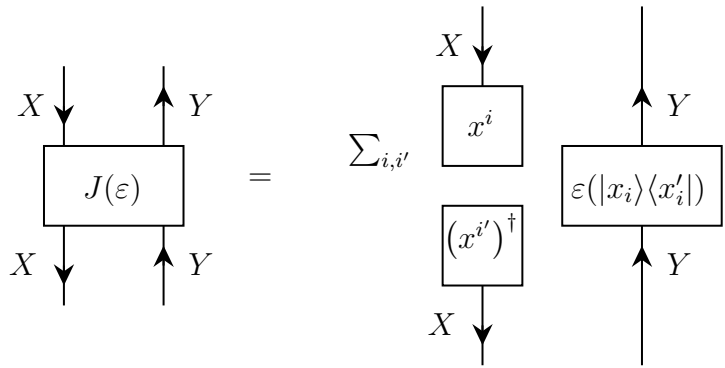
\begin{center}

	\tikzset{every picture/.style={line width=0.75pt}} 
	
	\begin{tikzpicture}[x=0.75pt,y=0.75pt,yscale=-1,xscale=1]
		
		\draw    (30,50) -- (30,57) ;
		\draw [shift={(30,60)}, rotate = 270] [fill={rgb, 255:red, 0; green, 0; blue, 0 }  ][line width=0.08]  [draw opacity=0] (10.72,-5.15) -- (0,0) -- (10.72,5.15) -- (7.12,0) -- cycle    ;
		\draw   (20,70) -- (90,70) -- (90,110) -- (20,110) -- cycle ;
		\draw    (80,30) -- (80,70) ;
		\draw    (30,30) -- (30,70) ;
		\draw    (80,110) -- (80,150) ;
		\draw    (30,110) -- (30,150) ;
		\draw    (30,130) -- (30,137) ;
		\draw [shift={(30,140)}, rotate = 270] [fill={rgb, 255:red, 0; green, 0; blue, 0 }  ][line width=0.08]  [draw opacity=0] (10.72,-5.15) -- (0,0) -- (10.72,5.15) -- (7.12,0) -- cycle    ;
		\draw    (80,130) -- (80,123) ;
		\draw [shift={(80,120)}, rotate = 90] [fill={rgb, 255:red, 0; green, 0; blue, 0 }  ][line width=0.08]  [draw opacity=0] (10.72,-5.15) -- (0,0) -- (10.72,5.15) -- (7.12,0) -- cycle    ;
		\draw    (80,50) -- (80,43) ;
		\draw [shift={(80,40)}, rotate = 90] [fill={rgb, 255:red, 0; green, 0; blue, 0 }  ][line width=0.08]  [draw opacity=0] (10.72,-5.15) -- (0,0) -- (10.72,5.15) -- (7.12,0) -- cycle    ;
		\draw   (220,40) -- (260,40) -- (260,80) -- (220,80) -- cycle ;
		\draw   (220,100) -- (260,100) -- (260,140) -- (220,140) -- cycle ;
		\draw   (280,70) -- (360,70) -- (360,110) -- (280,110) -- cycle ;
		\draw    (320,0) -- (320,70) ;
		\draw    (320,50) -- (320,43) ;
		\draw [shift={(320,40)}, rotate = 90] [fill={rgb, 255:red, 0; green, 0; blue, 0 }  ][line width=0.08]  [draw opacity=0] (10.72,-5.15) -- (0,0) -- (10.72,5.15) -- (7.12,0) -- cycle    ;
		\draw    (320,110) -- (320,180) ;
		\draw    (320,130) -- (320,123) ;
		\draw [shift={(320,120)}, rotate = 90] [fill={rgb, 255:red, 0; green, 0; blue, 0 }  ][line width=0.08]  [draw opacity=0] (10.72,-5.15) -- (0,0) -- (10.72,5.15) -- (7.12,0) -- cycle    ;
		\draw    (240,20) -- (240,27) ;
		\draw [shift={(240,30)}, rotate = 270] [fill={rgb, 255:red, 0; green, 0; blue, 0 }  ][line width=0.08]  [draw opacity=0] (10.72,-5.15) -- (0,0) -- (10.72,5.15) -- (7.12,0) -- cycle    ;
		\draw    (240,0) -- (240,40) ;
		\draw    (240,140) -- (240,180) ;
		\draw    (240,160) -- (240,167) ;
		\draw [shift={(240,170)}, rotate = 270] [fill={rgb, 255:red, 0; green, 0; blue, 0 }  ][line width=0.08]  [draw opacity=0] (10.72,-5.15) -- (0,0) -- (10.72,5.15) -- (7.12,0) -- cycle    ;
		
		\draw (38,82.4) node [anchor=north west][inner sep=0.75pt]    {$J( \varepsilon )$};
		\draw (1,122.4) node [anchor=north west][inner sep=0.75pt]    {$X$};
		\draw (5,42.4) node [anchor=north west][inner sep=0.75pt]    {$X$};
		\draw (91,122.4) node [anchor=north west][inner sep=0.75pt]    {$Y$};
		\draw (91,42.4) node [anchor=north west][inner sep=0.75pt]    {$Y$};
		\draw (121,82.4) node [anchor=north west][inner sep=0.75pt]    {$=$};
		\draw (331,42.4) node [anchor=north west][inner sep=0.75pt]    {$Y$};
		\draw (231,50.4) node [anchor=north west][inner sep=0.75pt]    {$x^{i}$};
		\draw (219,102.4) node [anchor=north west][inner sep=0.75pt]    {$\left( x^{i'}\right)^{\dagger }$};
		\draw (283,82.4) node [anchor=north west][inner sep=0.75pt]    {$\varepsilon ( |x_{i} \rangle \langle x_{i} '|)$};
		\draw (331,122.4) node [anchor=north west][inner sep=0.75pt]    {$Y$};
		\draw (215,12.4) node [anchor=north west][inner sep=0.75pt]    {$X$};
		\draw (211,152.4) node [anchor=north west][inner sep=0.75pt]    {$X$};
		\draw (171,62.4) node [anchor=north west][inner sep=0.75pt]    {$\sum _{i,i'}$};

	\end{tikzpicture}
\end{center}
To isolate $\varepsilon(\ket{x_i} \bra{x_{i'}})$ we just have to apply a $\bra{x_i }$ from above and a $\ket{x_{i'}}$ from below. It follows that
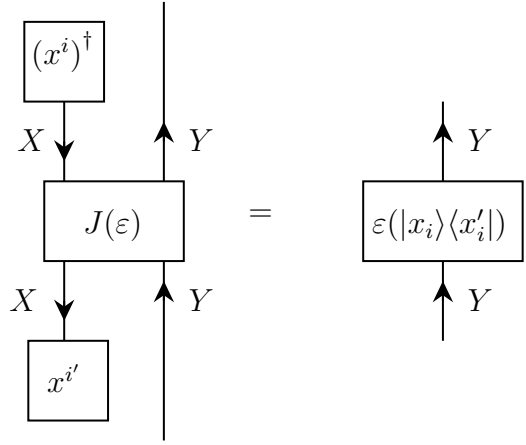
\begin{center}

	\tikzset{every picture/.style={line width=0.75pt}} 
	
	\begin{tikzpicture}[x=0.75pt,y=0.75pt,yscale=-1,xscale=1]
		
		\draw    (30,70) -- (30,77) ;
		\draw [shift={(30,80)}, rotate = 270] [fill={rgb, 255:red, 0; green, 0; blue, 0 }  ][line width=0.08]  [draw opacity=0] (10.72,-5.15) -- (0,0) -- (10.72,5.15) -- (7.12,0) -- cycle    ;
		\draw   (20,90) -- (90,90) -- (90,130) -- (20,130) -- cycle ;
		\draw    (80,0) -- (80,90) ;
		\draw    (30,50) -- (30,90) ;
		\draw    (80,130) -- (80,220) ;
		\draw    (30,130) -- (30,170) ;
		\draw    (30,150) -- (30,157) ;
		\draw [shift={(30,160)}, rotate = 270] [fill={rgb, 255:red, 0; green, 0; blue, 0 }  ][line width=0.08]  [draw opacity=0] (10.72,-5.15) -- (0,0) -- (10.72,5.15) -- (7.12,0) -- cycle    ;
		\draw    (80,150) -- (80,143) ;
		\draw [shift={(80,140)}, rotate = 90] [fill={rgb, 255:red, 0; green, 0; blue, 0 }  ][line width=0.08]  [draw opacity=0] (10.72,-5.15) -- (0,0) -- (10.72,5.15) -- (7.12,0) -- cycle    ;
		\draw    (80,70) -- (80,63) ;
		\draw [shift={(80,60)}, rotate = 90] [fill={rgb, 255:red, 0; green, 0; blue, 0 }  ][line width=0.08]  [draw opacity=0] (10.72,-5.15) -- (0,0) -- (10.72,5.15) -- (7.12,0) -- cycle    ;
		\draw   (10,10) -- (50,10) -- (50,50) -- (10,50) -- cycle ;
		\draw   (12,170) -- (52,170) -- (52,210) -- (12,210) -- cycle ;
		\draw   (180,90) -- (260,90) -- (260,130) -- (180,130) -- cycle ;
		\draw    (220,50) -- (220,90) ;
		\draw    (220,70) -- (220,63) ;
		\draw [shift={(220,60)}, rotate = 90] [fill={rgb, 255:red, 0; green, 0; blue, 0 }  ][line width=0.08]  [draw opacity=0] (10.72,-5.15) -- (0,0) -- (10.72,5.15) -- (7.12,0) -- cycle    ;
		\draw    (220,130) -- (220,170) ;
		\draw    (220,150) -- (220,143) ;
		\draw [shift={(220,140)}, rotate = 90] [fill={rgb, 255:red, 0; green, 0; blue, 0 }  ][line width=0.08]  [draw opacity=0] (10.72,-5.15) -- (0,0) -- (10.72,5.15) -- (7.12,0) -- cycle    ;
		
		\draw (38,102.4) node [anchor=north west][inner sep=0.75pt]    {$J( \varepsilon )$};
		\draw (1,142.4) node [anchor=north west][inner sep=0.75pt]    {$X$};
		\draw (5,62.4) node [anchor=north west][inner sep=0.75pt]    {$X$};
		\draw (91,142.4) node [anchor=north west][inner sep=0.75pt]    {$Y$};
		\draw (91,62.4) node [anchor=north west][inner sep=0.75pt]    {$Y$};
		\draw (121,102.4) node [anchor=north west][inner sep=0.75pt]    {$=$};
		\draw (231,62.4) node [anchor=north west][inner sep=0.75pt]    {$Y$};
		\draw (12,13.4) node [anchor=north west][inner sep=0.75pt]    {$\left( x^{i}\right)^{\dagger }$};
		\draw (20,180.4) node [anchor=north west][inner sep=0.75pt]    {$x^{i'}$};
		\draw (183,102.4) node [anchor=north west][inner sep=0.75pt]    {$\varepsilon ( |x_{i} \rangle \langle x_{i} '|)$};
		\draw (231,142.4) node [anchor=north west][inner sep=0.75pt]    {$Y$};

	\end{tikzpicture}
	
\end{center}
In Dirac notation, this says that
\begin{equation}
	\label{equation: CJ isolated epsilon}
	(\bra{x^i} \otimes \id_Y) J(\varepsilon) ( \ket{x^{i'}} \otimes \id_Y) = \varepsilon(\ket{x_i} \bra{x_{i'}}).
\end{equation}
Let $B \in \Lin_\complexNumbers (X)$, it is expressed in the basis as
\begin{equation}
	B = \sum_{i,i'} \bra{x_i} B \ket{x_{i'}} \ \ket{x_i} \bra{x_{i'}}.
\end{equation}
Applying $\varepsilon$ and using equation \ref{equation: CJ isolated epsilon}, we get that
\begin{equation}
	\label{equation: CJ inverse expression in basis}
	\varepsilon(B) = \sum_{i,i'}  \bra{x_i} B \ket{x_{i'}} \ (\bra{x^i} \otimes \id_Y) J(\varepsilon) ( \ket{x^{i'}} \otimes \id_Y).
\end{equation}
Diagrammatically, this is expressed as follows:
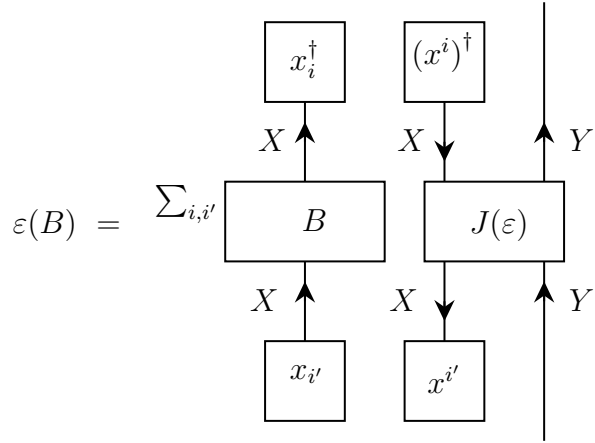
\begin{center}

	\tikzset{every picture/.style={line width=0.75pt}} 
	
	\begin{tikzpicture}[x=0.75pt,y=0.75pt,yscale=-1,xscale=1]
		
		\draw    (220,70) -- (220,77) ;
		\draw [shift={(220,80)}, rotate = 270] [fill={rgb, 255:red, 0; green, 0; blue, 0 }  ][line width=0.08]  [draw opacity=0] (10.72,-5.15) -- (0,0) -- (10.72,5.15) -- (7.12,0) -- cycle    ;
		\draw   (210,90) -- (280,90) -- (280,130) -- (210,130) -- cycle ;
		\draw    (270,0) -- (270,90) ;
		\draw    (220,50) -- (220,90) ;
		\draw    (270,130) -- (270,220) ;
		\draw    (220,130) -- (220,170) ;
		\draw    (220,150) -- (220,157) ;
		\draw [shift={(220,160)}, rotate = 270] [fill={rgb, 255:red, 0; green, 0; blue, 0 }  ][line width=0.08]  [draw opacity=0] (10.72,-5.15) -- (0,0) -- (10.72,5.15) -- (7.12,0) -- cycle    ;
		\draw    (270,150) -- (270,143) ;
		\draw [shift={(270,140)}, rotate = 90] [fill={rgb, 255:red, 0; green, 0; blue, 0 }  ][line width=0.08]  [draw opacity=0] (10.72,-5.15) -- (0,0) -- (10.72,5.15) -- (7.12,0) -- cycle    ;
		\draw    (270,70) -- (270,63) ;
		\draw [shift={(270,60)}, rotate = 90] [fill={rgb, 255:red, 0; green, 0; blue, 0 }  ][line width=0.08]  [draw opacity=0] (10.72,-5.15) -- (0,0) -- (10.72,5.15) -- (7.12,0) -- cycle    ;
		\draw   (200,10) -- (240,10) -- (240,50) -- (200,50) -- cycle ;
		\draw   (200,170) -- (240,170) -- (240,210) -- (200,210) -- cycle ;
		\draw    (150,70) -- (150,63) ;
		\draw [shift={(150,60)}, rotate = 90] [fill={rgb, 255:red, 0; green, 0; blue, 0 }  ][line width=0.08]  [draw opacity=0] (10.72,-5.15) -- (0,0) -- (10.72,5.15) -- (7.12,0) -- cycle    ;
		\draw   (110,90) -- (190,90) -- (190,130) -- (110,130) -- cycle ;
		\draw    (150,50) -- (150,90) ;
		\draw    (150,130) -- (150,170) ;
		\draw    (150,150) -- (150,143) ;
		\draw [shift={(150,140)}, rotate = 90] [fill={rgb, 255:red, 0; green, 0; blue, 0 }  ][line width=0.08]  [draw opacity=0] (10.72,-5.15) -- (0,0) -- (10.72,5.15) -- (7.12,0) -- cycle    ;
		\draw   (130,10) -- (170,10) -- (170,50) -- (130,50) -- cycle ;
		\draw   (130,170) -- (170,170) -- (170,210) -- (130,210) -- cycle ;
		
		\draw (231,102.4) node [anchor=north west][inner sep=0.75pt]    {$J( \varepsilon )$};
		\draw (191,142.4) node [anchor=north west][inner sep=0.75pt]    {$X$};
		\draw (195,62.4) node [anchor=north west][inner sep=0.75pt]    {$X$};
		\draw (281,142.4) node [anchor=north west][inner sep=0.75pt]    {$Y$};
		\draw (281,62.4) node [anchor=north west][inner sep=0.75pt]    {$Y$};
		\draw (202,13.4) node [anchor=north west][inner sep=0.75pt]    {$\left( x^{i}\right)^{\dagger }$};
		\draw (210,180.4) node [anchor=north west][inner sep=0.75pt]    {$x^{i'}$};
		\draw (2,102.4) node [anchor=north west][inner sep=0.75pt]    {$\varepsilon ( B) \ =$};
		\draw (73,90.4) node [anchor=north west][inner sep=0.75pt]    {$\sum _{i,i'}$};
		\draw (147,102.4) node [anchor=north west][inner sep=0.75pt]    {$B$};
		\draw (121,142.4) node [anchor=north west][inner sep=0.75pt]    {$X$};
		\draw (125,62.4) node [anchor=north west][inner sep=0.75pt]    {$X$};
		\draw (141,17.4) node [anchor=north west][inner sep=0.75pt]    {$x_{i}^{\dagger }$};
		\draw (141,180.4) node [anchor=north west][inner sep=0.75pt]    {$x_{i'}$};

	\end{tikzpicture}
\end{center}
The indices $i$ only appear in the top of the diagram, and the indices $i'$ only appear at the bottom, so we can separate the corresponding summations. This leads to the following diagrammatic equation:
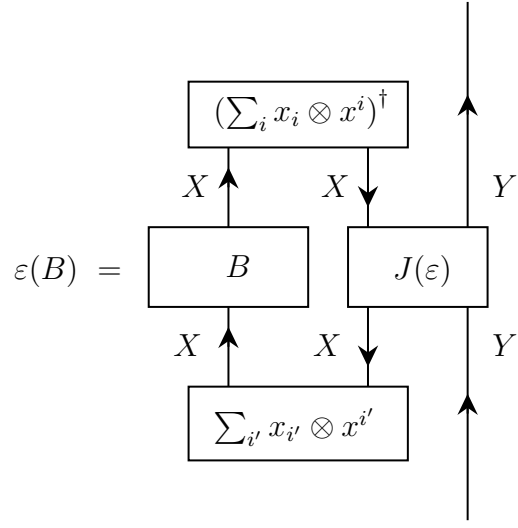
\begin{center}

	\tikzset{every picture/.style={line width=0.75pt}} 
	
	\begin{tikzpicture}[x=0.75pt,y=0.75pt,yscale=-1,xscale=1]
		
		\draw    (180,93) -- (180,100) ;
		\draw [shift={(180,103)}, rotate = 270] [fill={rgb, 255:red, 0; green, 0; blue, 0 }  ][line width=0.08]  [draw opacity=0] (10.72,-5.15) -- (0,0) -- (10.72,5.15) -- (7.12,0) -- cycle    ;
		\draw   (170,113) -- (240,113) -- (240,153) -- (170,153) -- cycle ;
		\draw    (230,0) -- (230,113) ;
		\draw    (180,73) -- (180,113) ;
		\draw    (230,153) -- (230,260) ;
		\draw    (180,153) -- (180,193) ;
		\draw    (180,173) -- (180,180) ;
		\draw [shift={(180,183)}, rotate = 270] [fill={rgb, 255:red, 0; green, 0; blue, 0 }  ][line width=0.08]  [draw opacity=0] (10.72,-5.15) -- (0,0) -- (10.72,5.15) -- (7.12,0) -- cycle    ;
		\draw    (230,206.5) -- (230,199.5) ;
		\draw [shift={(230,196.5)}, rotate = 90] [fill={rgb, 255:red, 0; green, 0; blue, 0 }  ][line width=0.08]  [draw opacity=0] (10.72,-5.15) -- (0,0) -- (10.72,5.15) -- (7.12,0) -- cycle    ;
		\draw    (230,56.5) -- (230,49.5) ;
		\draw [shift={(230,46.5)}, rotate = 90] [fill={rgb, 255:red, 0; green, 0; blue, 0 }  ][line width=0.08]  [draw opacity=0] (10.72,-5.15) -- (0,0) -- (10.72,5.15) -- (7.12,0) -- cycle    ;
		\draw    (110,93) -- (110,86) ;
		\draw [shift={(110,83)}, rotate = 90] [fill={rgb, 255:red, 0; green, 0; blue, 0 }  ][line width=0.08]  [draw opacity=0] (10.72,-5.15) -- (0,0) -- (10.72,5.15) -- (7.12,0) -- cycle    ;
		\draw   (70,113) -- (150,113) -- (150,153) -- (70,153) -- cycle ;
		\draw    (110,73) -- (110,113) ;
		\draw    (110,153) -- (110,193) ;
		\draw    (110,173) -- (110,166) ;
		\draw [shift={(110,163)}, rotate = 90] [fill={rgb, 255:red, 0; green, 0; blue, 0 }  ][line width=0.08]  [draw opacity=0] (10.72,-5.15) -- (0,0) -- (10.72,5.15) -- (7.12,0) -- cycle    ;
		\draw   (90,40) -- (200,40) -- (200,73) -- (90,73) -- cycle ;
		\draw   (90,193) -- (200,193) -- (200,230) -- (90,230) -- cycle ;
		
		\draw (191,125.4) node [anchor=north west][inner sep=0.75pt]    {$J( \varepsilon )$};
		\draw (151,165.4) node [anchor=north west][inner sep=0.75pt]    {$X$};
		\draw (155,85.4) node [anchor=north west][inner sep=0.75pt]    {$X$};
		\draw (241,165.4) node [anchor=north west][inner sep=0.75pt]    {$Y$};
		\draw (241,85.4) node [anchor=north west][inner sep=0.75pt]    {$Y$};
		\draw (1,125.4) node [anchor=north west][inner sep=0.75pt]    {$\varepsilon ( B) \ =$};
		\draw (107,125.4) node [anchor=north west][inner sep=0.75pt]    {$B$};
		\draw (81,165.4) node [anchor=north west][inner sep=0.75pt]    {$X$};
		\draw (85,85.4) node [anchor=north west][inner sep=0.75pt]    {$X$};
		\draw (101,42.4) node [anchor=north west][inner sep=0.75pt]    {$\left(\sum_{i} x_{i} \otimes x^{i}\right)^{\dagger }$};
		\draw (101,202.4) node [anchor=north west][inner sep=0.75pt]    {$\sum_{i'} x_{i'} \otimes x^{i'}$};

	\end{tikzpicture}
	
\end{center}
We can see in this equation the presence of the vector
\begin{equation}
	\tilde{\Omega} = \sum_i x_i \otimes x^i .
\end{equation}
Instead of using $\varepsilon$, let's change variables to get an expression for $J^{-1}$. Let $A \in \Lin_\complexNumbers (X^* \otimes Y)$ and take
\begin{equation}
	\varepsilon = J^{-1} (A).
\end{equation}
Then the diagrammatic equation becomes the following]
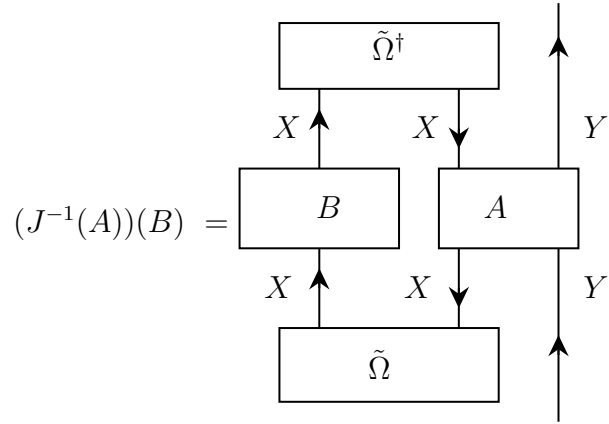
\begin{center}

	\tikzset{every picture/.style={line width=0.75pt}} 
	
	\begin{tikzpicture}[x=0.75pt,y=0.75pt,yscale=-1,xscale=1]
		
		\draw    (226,63) -- (226,70) ;
		\draw [shift={(226,73)}, rotate = 270] [fill={rgb, 255:red, 0; green, 0; blue, 0 }  ][line width=0.08]  [draw opacity=0] (10.72,-5.15) -- (0,0) -- (10.72,5.15) -- (7.12,0) -- cycle    ;
		\draw   (216,83) -- (286,83) -- (286,123) -- (216,123) -- cycle ;
		\draw    (276,0) -- (276,83) ;
		\draw    (226,43) -- (226,83) ;
		\draw    (276,123) -- (276,210) ;
		\draw    (226,123) -- (226,163) ;
		\draw    (226,143) -- (226,150) ;
		\draw [shift={(226,153)}, rotate = 270] [fill={rgb, 255:red, 0; green, 0; blue, 0 }  ][line width=0.08]  [draw opacity=0] (10.72,-5.15) -- (0,0) -- (10.72,5.15) -- (7.12,0) -- cycle    ;
		\draw    (276,176.5) -- (276,169.5) ;
		\draw [shift={(276,166.5)}, rotate = 90] [fill={rgb, 255:red, 0; green, 0; blue, 0 }  ][line width=0.08]  [draw opacity=0] (10.72,-5.15) -- (0,0) -- (10.72,5.15) -- (7.12,0) -- cycle    ;
		\draw    (276,26.5) -- (276,19.5) ;
		\draw [shift={(276,16.5)}, rotate = 90] [fill={rgb, 255:red, 0; green, 0; blue, 0 }  ][line width=0.08]  [draw opacity=0] (10.72,-5.15) -- (0,0) -- (10.72,5.15) -- (7.12,0) -- cycle    ;
		\draw    (156,63) -- (156,56) ;
		\draw [shift={(156,53)}, rotate = 90] [fill={rgb, 255:red, 0; green, 0; blue, 0 }  ][line width=0.08]  [draw opacity=0] (10.72,-5.15) -- (0,0) -- (10.72,5.15) -- (7.12,0) -- cycle    ;
		\draw   (116,83) -- (196,83) -- (196,123) -- (116,123) -- cycle ;
		\draw    (156,43) -- (156,83) ;
		\draw    (156,123) -- (156,163) ;
		\draw    (156,143) -- (156,136) ;
		\draw [shift={(156,133)}, rotate = 90] [fill={rgb, 255:red, 0; green, 0; blue, 0 }  ][line width=0.08]  [draw opacity=0] (10.72,-5.15) -- (0,0) -- (10.72,5.15) -- (7.12,0) -- cycle    ;
		\draw   (136,10) -- (246,10) -- (246,43) -- (136,43) -- cycle ;
		\draw   (136,163) -- (246,163) -- (246,200) -- (136,200) -- cycle ;
		
		\draw (237,95.4) node [anchor=north west][inner sep=0.75pt]    {$A$};
		\draw (197,135.4) node [anchor=north west][inner sep=0.75pt]    {$X$};
		\draw (201,55.4) node [anchor=north west][inner sep=0.75pt]    {$X$};
		\draw (287,135.4) node [anchor=north west][inner sep=0.75pt]    {$Y$};
		\draw (287,55.4) node [anchor=north west][inner sep=0.75pt]    {$Y$};
		\draw (1,100.4) node [anchor=north west][inner sep=0.75pt]    {$(J^{-1}(A))( B) \ =$};
		\draw (153,95.4) node [anchor=north west][inner sep=0.75pt]    {$B$};
		\draw (127,135.4) node [anchor=north west][inner sep=0.75pt]    {$X$};
		\draw (131,55.4) node [anchor=north west][inner sep=0.75pt]    {$X$};
		\draw (181,12.4) node [anchor=north west][inner sep=0.75pt]    {$\tilde{\Omega }^{\dagger }$};
		\draw (179,172.4) node [anchor=north west][inner sep=0.75pt]    {$\tilde{\Omega }$};

	\end{tikzpicture}
\end{center}
In Dirac notation, this is written as
\begin{equation}
	(J^{-1} (A))(B) = (\bra{\tilde{\Omega}} \otimes \id_Y) (B \otimes A) (\ket{\tilde{\Omega}} \otimes \id_Y).
\end{equation}
This is one of the expressions for $(J^{-1} (A))(B)$ that we wanted to prove. The other one was already proved in equation \ref{equation: CJ inverse expression in basis}, but written with other variables. It can be rewritten as
\begin{equation}
	(J^{-1} (A))(B) = \sum_{i,i'}  \bra{x_i} B \ket{x_{i'}} \ (\bra{x^i} \otimes \id_Y) A ( \ket{x^{i'}} \otimes \id_Y).
\end{equation}

The reader may recognize that
\begin{equation}
	\Omega = \rcoev_X (1),
\end{equation}
and
\begin{equation}
	\tilde{\Omega} = \lcoev_X (1),
\end{equation}
as in definition \ref{definition:coevaluations}. Since the coevaluations are basis independent, as proved in proposition \ref{proposition:coevaluations basis independent}, it follows that the expressions for $J$ and $J^{-1}$ are valid for any basis. The hypothesis that the basis are orthonormal isn't necessary, but it simplifies the computations. Also, using the diagrammatic representation of evaluations and coevaluations as cap and cups, we get the following diagrammatic equation:
\begin{center}

	\tikzset{every picture/.style={line width=0.75pt}} 
	
	\begin{tikzpicture}[x=0.75pt,y=0.75pt,yscale=-1,xscale=1]
		
		\draw   (190,50) -- (240,50) -- (240,90) -- (190,90) -- cycle ;
		\draw    (230,10) -- (230,50) ;
		\draw    (230,90) -- (230,130) ;
		\draw    (230,110) -- (230,103) ;
		\draw [shift={(230,100)}, rotate = 90] [fill={rgb, 255:red, 0; green, 0; blue, 0 }  ][line width=0.08]  [draw opacity=0] (10.72,-5.15) -- (0,0) -- (10.72,5.15) -- (7.12,0) -- cycle    ;
		\draw    (230,30) -- (230,23) ;
		\draw [shift={(230,20)}, rotate = 90] [fill={rgb, 255:red, 0; green, 0; blue, 0 }  ][line width=0.08]  [draw opacity=0] (10.72,-5.15) -- (0,0) -- (10.72,5.15) -- (7.12,0) -- cycle    ;
		\draw   (120,50) -- (170,50) -- (170,90) -- (120,90) -- cycle ;
		\draw  [draw opacity=0] (150,50) .. controls (150,33.43) and (163.43,20) .. (180,20) .. controls (196.57,20) and (210,33.43) .. (210,50) -- (180,50) -- cycle ; \draw   (150,50) .. controls (150,33.43) and (163.43,20) .. (180,20) .. controls (196.57,20) and (210,33.43) .. (210,50) ;  
		\draw    (180,20) -- (187,20) ;
		\draw [shift={(190,20)}, rotate = 180] [fill={rgb, 255:red, 0; green, 0; blue, 0 }  ][line width=0.08]  [draw opacity=0] (10.72,-5.15) -- (0,0) -- (10.72,5.15) -- (7.12,0) -- cycle    ;
		\draw  [draw opacity=0] (210,90) .. controls (210,90) and (210,90) .. (210,90) .. controls (210,90) and (210,90) .. (210,90) .. controls (210,106.57) and (196.57,120) .. (180,120) .. controls (163.43,120) and (150,106.57) .. (150,90) -- (180,90) -- cycle ; \draw   (210,90) .. controls (210,90) and (210,90) .. (210,90) .. controls (210,90) and (210,90) .. (210,90) .. controls (210,106.57) and (196.57,120) .. (180,120) .. controls (163.43,120) and (150,106.57) .. (150,90) ;  
		\draw    (180,120) -- (173,120) ;
		\draw [shift={(170,120)}, rotate = 360] [fill={rgb, 255:red, 0; green, 0; blue, 0 }  ][line width=0.08]  [draw opacity=0] (10.72,-5.15) -- (0,0) -- (10.72,5.15) -- (7.12,0) -- cycle    ;
		
		\draw (205,62.4) node [anchor=north west][inner sep=0.75pt]    {$A$};
		\draw (241,112.4) node [anchor=north west][inner sep=0.75pt]    {$Y$};
		\draw (241,12.4) node [anchor=north west][inner sep=0.75pt]    {$Y$};
		\draw (1,60.4) node [anchor=north west][inner sep=0.75pt]    {$\left( J^{-1}( A)\right)( B) \ =$};
		\draw (141,62.4) node [anchor=north west][inner sep=0.75pt]    {$B$};
		\draw (141,2.4) node [anchor=north west][inner sep=0.75pt]    {$X$};
		\draw (145,122.4) node [anchor=north west][inner sep=0.75pt]    {$X$};

	\end{tikzpicture}
	
\end{center}
For $J(\varepsilon)$ we don't have such a nice diagrammatic representation, but $\ket{\Omega} \bra{\Omega}$ can be represented as follows:
\begin{center}

	\tikzset{every picture/.style={line width=0.75pt}} 
	
	\begin{tikzpicture}[x=0.75pt,y=0.75pt,yscale=-1,xscale=1]
		
		\draw  [draw opacity=0] (160,0) .. controls (160,0) and (160,0) .. (160,0) .. controls (160,16.57) and (146.57,30) .. (130,30) .. controls (113.43,30) and (100,16.57) .. (100,0) -- (130,0) -- cycle ; \draw   (160,0) .. controls (160,0) and (160,0) .. (160,0) .. controls (160,16.57) and (146.57,30) .. (130,30) .. controls (113.43,30) and (100,16.57) .. (100,0) ;  
		\draw    (130,30) -- (137,30) ;
		\draw [shift={(140,30)}, rotate = 180] [fill={rgb, 255:red, 0; green, 0; blue, 0 }  ][line width=0.08]  [draw opacity=0] (10.72,-5.15) -- (0,0) -- (10.72,5.15) -- (7.12,0) -- cycle    ;
		\draw  [draw opacity=0] (100,80) .. controls (100,63.43) and (113.43,50) .. (130,50) .. controls (146.57,50) and (160,63.43) .. (160,80) -- (130,80) -- cycle ; \draw   (100,80) .. controls (100,63.43) and (113.43,50) .. (130,50) .. controls (146.57,50) and (160,63.43) .. (160,80) ;  
		\draw    (130,50) -- (123,50) ;
		\draw [shift={(120,50)}, rotate = 360] [fill={rgb, 255:red, 0; green, 0; blue, 0 }  ][line width=0.08]  [draw opacity=0] (10.72,-5.15) -- (0,0) -- (10.72,5.15) -- (7.12,0) -- cycle    ;
		
		\draw (1,32.4) node [anchor=north west][inner sep=0.75pt]    {$| \Omega \rangle \langle \Omega |\ =$};
		\draw (165,2.4) node [anchor=north west][inner sep=0.75pt]    {$X$};
		\draw (165,62.4) node [anchor=north west][inner sep=0.75pt]    {$X$};

	\end{tikzpicture}
	
\end{center}
\end{proof}

Notice the similarity between the $\Omega$ of this proposition and the maximally entangled state defined in equation \ref{equation: maximally entangled state}. They differ by the replacement of $x_i$ by $x^i$ and the normalization factor $\sqrt{d_X}$. Also, equation \ref{equation: CJ second expression} follows the same pattern of equation \ref{equation: bad CJ isomorphism definition}.

The importance of the \CJ isomorphism is that it is a bijection between CP maps and PSD operators, which we prove next.

\begin{proposition}{}{bijection between CP and PSD}
	A $\complexNumbers$-linear map $\varepsilon \in \Lin_\complexNumbers(\Lin_\complexNumbers(X), \Lin_\complexNumbers(Y))$ is CP iff $J(\varepsilon)$ is PSD.
\end{proposition}
\begin{proof}
	$(\implies)$ Let $\varepsilon \in \cp(X,Y)$. It is a $\complexNumbers$-linear map $\varepsilon \in\Lin_\complexNumbers(\Lin_\complexNumbers(X),$ $\Lin_\complexNumbers(Y))$ which is completely positive. By proposition \ref{proposition:CJ expression}, we can write $J(\varepsilon)$ as $J(\varepsilon) = (\id_{\Lin_\complexNumbers(X^*)}\otimes \varepsilon)(\ket{\Omega} \bra{\Omega})$. Since $\varepsilon$ is CP and $\ket{\Omega}\bra{\Omega}$ is a PSD operator, we have that $J(\varepsilon)$ is also a PSD operator.
	
	$(\impliedby)$ Let $A \in \Lin_\complexNumbers(X^* \otimes Y)$ be a PSD operator. This is equivalent to being able to write $A$ as $A = B^\dag B$ for some operator $B \in \Lin_\complexNumbers(X^* \otimes Y)$. The linear map $J^{-1}(A)$ is CP if $(\id_{\Lin_\complexNumbers(Z)} \otimes J^{-1}(A))(C)$ is PSD for any PSD operator $C \in \Lin_\complexNumbers(Z\otimes X)$. By proposition \ref{proposition:CJ expression}, for any $C\in \Lin_\complexNumbers(X)$ we have $(J^{-1}(A))(C) = (\bra{\tilde{\Omega}}\otimes \id_Y)(C\otimes A)(\ket{\tilde{\Omega}} \otimes \id_Y)$. From this, we have that
	\begin{equation}
		(\id_{\Lin_\complexNumbers(Z)}\otimes J^{-1}(A))(C) = (\id_Z \otimes \bra{\tilde{\Omega}}\otimes \id_Y)(C\otimes A)(\id_z \otimes \ket{\tilde{\Omega}} \otimes \id_Y),
	\end{equation}
	for any $C \in \Lin_\complexNumbers(Z \otimes X)$. Taking $C$ as a PSD operator, we can write it as $C = D^\dag D$. Subtituting the equations $A = B^\dag B$ and $C = D^\dag D$ in the expression for $J^{-1}(A)$, we have
	\begin{align*}
		(\id_{\Lin_\complexNumbers(Z)} &\otimes J^{-1}(A))(C) = (\id_Z \otimes \bra{\tilde{\Omega}}\otimes \id_Y)(C\otimes A)(\id_Z \otimes \ket{\tilde{\Omega}} \otimes \id_Y)\\
		&\stackrel{}{=} (\id_Z \otimes \bra{\tilde{\Omega}}\otimes \id_Y)(D^\dag D\otimes B^\dag B)(\id_Z \otimes \ket{\tilde{\Omega}} \otimes \id_Y)\\
		&\stackrel{}{=} (\id_Z \otimes \bra{\tilde{\Omega}}\otimes \id_Y)(D^\dag \otimes B^\dag)(D\otimes B)(\id_Z \otimes \ket{\tilde{\Omega}} \otimes \id_Y).
	\end{align*}
	The last expression is of the form $E^\dag E$, for $E = (D\otimes B)(\id_Z \otimes \ket{\tilde{\Omega}} \otimes \id_Y)$, from which we conclude that $(\id_{\Lin_\complexNumbers(Z)} \otimes J^{-1}(A))(C)$ is PSD. This proves that $J^{-1}(A)$ is a CP map.
\end{proof}

These results shows that we can study PSD operators instead of CP maps. This will simplify many proofs, so we'll define some notation for these PSD operators. 

\begin{notation}
	We'll denote the subset of PSD operators of $\Lin_\complexNumbers(X)$ by $\psd(X)$.
\end{notation}

\subsection{Choi-rank}

The PSD operators have ranks. We can define the Choi-rank of a CP map as the rank of the correspondent PSD operator.

\begin{definition}{}{Choi rank}
	Let $X$ and $Y$ be finite dimensional complex Hilbert spaces. The \CJ isomorphism gives a bijection $J \colon \cp(X,Y) \to \psd(X^* \otimes Y)$. We define the Choi-rank of a CP map $\varepsilon \in \cp(X,Y)$ as
	\begin{equation}
		\CR(\varepsilon) \coloneqq \rank(J(\varepsilon)).
	\end{equation} 
\end{definition}

Every PSD operator is diagonalizable, and can be more generally written as a sum of PSD operators of rank at most 1. From this we get the Kraus representation of CP maps, which we study next.

\subsection{The Kraus representation}

In this section we'll prove that every CP map $\varepsilon \in \cp(X,Y)$ can be expressed as
\begin{equation}
	\label{equation: Kraus representation}
	\varepsilon(B) = \sum_k E_k B E_k^\dag,
\end{equation}
for any $B \in \Lin_\complexNumbers(X)$. In this expression, $(E_k)_k$ is a finite sequence of linear maps $E_k \colon X \to Y$. Equation \ref{equation: Kraus representation} is called a Kraus representation of $\varepsilon$, and each $E_k$ is called a Kraus operator. The converse is also true, the linear map defined by equation \ref{equation: Kraus representation} is always a CP map, for any finite sequence $(E_k)_k$. To find a Kraus representation, it can be done by diagonalizing the PSD operator $J(\varepsilon)$. This also computes the Choi-rank of $\varepsilon$, which we'll define later.

\begin{proposition}{}{cp maps vs kraus representation}
	Every CP map admits a Kraus representation, that is, it can be expressed as in equation \ref{equation: Kraus representation}. Also, the linear map defined by equation \ref{equation: Kraus representation} is always a CP map.
\end{proposition}
\begin{proof}
	$(\implies)$ Let $\varepsilon\in  \cp(X,Y)$. By proposition \ref{proposition:bijection between CP and PSD}, $J(\varepsilon)$ is a PSD operator. Therefore, it can be written in the form
	\begin{equation}
		J(\varepsilon) = \sum_k \ket{\psi_k} \bra{\psi_k},
	\end{equation}
	for some finite sequence $(\ket{\psi_k})_k$ of vectors $\ket{\psi_k} \in X^*\otimes Y$. This sequence of vectors can be obtained by diagonalizing $J(\varepsilon)$, but they don't need to come from a diagonalization. We can then express the CP map $\varepsilon$ in terms of the vectors $\ket{\psi_k}$ by taking the inverse of the \CJ isomorphism. For any $B \in \Lin_\complexNumbers(X)$, we have
	\begin{align*}
		\varepsilon(B) &= J^{-1}(J(\varepsilon))(B)\\
		&= (\bra{\tilde{\Omega}}\otimes \id_Y)(B\otimes J(\varepsilon))(\ket{\tilde{\Omega}} \otimes \id_Y)\\
		&= (\bra{\tilde{\Omega}}\otimes \id_Y)(B\otimes \sum_k \ket{\psi_k} \bra{\psi_k})(\ket{\tilde{\Omega}} \otimes \id_Y)\\
		&= \sum_k (\bra{\tilde{\Omega}}\otimes \id_Y)(B\otimes \ket{\psi_k} \bra{\psi_k})(\ket{\tilde{\Omega}} \otimes \id_Y)\\
		&= \sum_k (\bra{\tilde{\Omega}}\otimes \id_Y)(\id_X \otimes \ket{\psi_k})B(\id_X \otimes  \bra{\psi_k})(\ket{\tilde{\Omega}} \otimes \id_Y).
	\end{align*}
	Defining $E_k \coloneqq (\bra{\tilde{\Omega}}\otimes \id_Y)(\id_X \otimes \ket{\psi_k})$, which is an operator $E_k \colon X \to Y$, we get
	\begin{equation}
		\varepsilon(B) = \sum_k E_k B E_k^\dag.
	\end{equation}
	
	$(\impliedby)$ Let $(E_k)_k$ be a finite sequence of linear maps $E_k \colon X \to Y$, and define a linear map $\varepsilon \colon \Lin_\complexNumbers(X) \to \Lin_\complexNumbers(Y)$ by
	\begin{equation}
		\varepsilon(B) = \sum_k E_k B E_k^\dag.
	\end{equation}
	Let $Z$ be a finite dimensional complex Hilbert space and $C \in \Lin_\complexNumbers(Z\otimes X)$ be a PSD operator. Because $C$ is PSD, we can write it as $C = D^\dag D$, for some $D \in \Lin_\complexNumbers(Z \otimes X)$. Then we have
	\begin{align*}
		(\id_{\Lin_\complexNumbers(Z)} \otimes \varepsilon)(C) &= \sum_k (\id_Z \otimes E_k) C (\id_Z \otimes E_k^\dag)\\
		&= \sum_k (\id_Z \otimes E_k) D^\dag D (\id_Z \otimes E_k^\dag).
	\end{align*}
	The last expression is of the form $\sum_k F_k^\dag F_k$, for $F_k = D (\id_Z \otimes E_k^\dag)$. Each term $F_k^\dag F_k$ is a PSD operator, and a sum of PSD operators is PSD. Therefore, $(\id_{\Lin_\complexNumbers(Z)} \otimes \varepsilon)(C)$ is PSD. This proves that $\varepsilon$ is a CP map.
\end{proof}

\begin{corollary}{}{bijection between kraus operators and psis}
	Let $\varepsilon \colon X \to Y$ be a CP map. The Kraus operators $(E_k)_k$ for $\varepsilon$ are in bijection with the vectors $(\ket{\psi_k})_k$ such that $J(\varepsilon) = \sum_k \ket{\psi_k} \bra{\psi_k}$.
\end{corollary}
\begin{proof}
	We'll use the same notation from proposition \ref{proposition:cp maps vs kraus representation}. Given a finite sequence of vectors $(\ket{\psi_k})_k$ such that $J(\varepsilon) = \sum_k \ket{\psi_k} \bra{\psi_k}$, the corresponding Kraus operators $(E_k)_k$ are given by
	\begin{equation}
		E_k \coloneqq (\bra{\tilde{\Omega}}\otimes \id_Y)(\id_X \otimes \ket{\psi_k}).
	\end{equation}
	This amounts to the natural isomorphism $\Lin_\complexNumbers(X,Y) \cong X^* \otimes Y$. This isomorphism has a simple representation using string diagrams:
	\begin{center}	
		\tikzset{every picture/.style={line width=0.75pt}} 
		
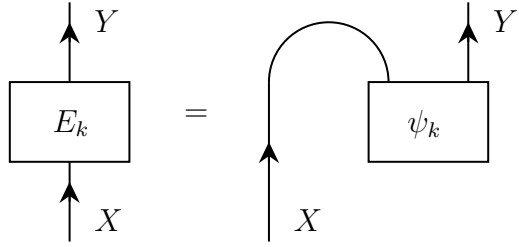
		\begin{tikzpicture}[x=0.75pt,y=0.75pt,yscale=-1,xscale=1]
			
			\draw   (0,40) -- (60,40) -- (60,80) -- (0,80) -- cycle ;
			\draw    (30,80) -- (30,120) ;
			\draw    (30,100) -- (30,93) ;
			\draw [shift={(30,90)}, rotate = 90] [fill={rgb, 255:red, 0; green, 0; blue, 0 }  ][line width=0.08]  [draw opacity=0] (10.72,-5.15) -- (0,0) -- (10.72,5.15) -- (7.12,0) -- cycle    ;
			\draw    (30,0) -- (30,40) ;
			\draw    (30,20) -- (30,13) ;
			\draw [shift={(30,10)}, rotate = 90] [fill={rgb, 255:red, 0; green, 0; blue, 0 }  ][line width=0.08]  [draw opacity=0] (10.72,-5.15) -- (0,0) -- (10.72,5.15) -- (7.12,0) -- cycle    ;
			\draw   (180,40) -- (240,40) -- (240,80) -- (180,80) -- cycle ;
			\draw    (130,40) -- (130,120) ;
			\draw    (130,80) -- (130,73) ;
			\draw [shift={(130,70)}, rotate = 90] [fill={rgb, 255:red, 0; green, 0; blue, 0 }  ][line width=0.08]  [draw opacity=0] (10.72,-5.15) -- (0,0) -- (10.72,5.15) -- (7.12,0) -- cycle    ;
			\draw    (230,0) -- (230,40) ;
			\draw    (230,20) -- (230,13) ;
			\draw [shift={(230,10)}, rotate = 90] [fill={rgb, 255:red, 0; green, 0; blue, 0 }  ][line width=0.08]  [draw opacity=0] (10.72,-5.15) -- (0,0) -- (10.72,5.15) -- (7.12,0) -- cycle    ;
			\draw  [draw opacity=0] (130,40) .. controls (130,23.43) and (143.43,10) .. (160,10) .. controls (176.57,10) and (190,23.43) .. (190,40) -- (160,40) -- cycle ; \draw   (130,40) .. controls (130,23.43) and (143.43,10) .. (160,10) .. controls (176.57,10) and (190,23.43) .. (190,40) ;  
			
			\draw (20,52.4) node [anchor=north west][inner sep=0.75pt]    {$E_{k}$};
			\draw (41,102.4) node [anchor=north west][inner sep=0.75pt]    {$X$};
			\draw (41,2.4) node [anchor=north west][inner sep=0.75pt]    {$Y$};
			\draw (86,52.4) node [anchor=north west][inner sep=0.75pt]    {$=$};
			\draw (198,52.4) node [anchor=north west][inner sep=0.75pt]    {$\psi _{k}$};
			\draw (141,102.4) node [anchor=north west][inner sep=0.75pt]    {$X$};
			\draw (241,2.4) node [anchor=north west][inner sep=0.75pt]    {$Y$};

		\end{tikzpicture}
	\end{center}
\end{proof}

This corollary lets us find a lower bound for the number of Kraus operators needed for a Kraus decomposition. This lower bound is the Choi-rank.
\begin{proposition}{}{}
	The Choi-rank of a CP map is the minimum number of Kraus operators required for its Kraus representations.
\end{proposition}
\begin{proof}
	Let's call $m$ such minimum number of Kraus operators needed for a Kraus representation of some CP map $\varepsilon \in \cp(X,Y)$. By corollary \ref{corollary:bijection between kraus operators and psis}, Kraus representations of $\varepsilon$ correspond to finite sequences of vectors $(\ket{\psi_k})_k$ in $X^* \otimes Y$ such that $J(\varepsilon) = \sum_k \ket{\psi_k} \bra{\psi_k}$. Therefore, $m$ is the same as the minimum length required for such sequence $(\ket{\psi_k})_k$. The image of $\sum_k \ket{\psi_k} \bra{\psi_k}$ is the vector space generated by the vectors $\ket{\psi_k}$. The dimension of such image is $\rank(J(\varepsilon))$, and $m$ vectors $\ket{\psi_k}$ can generate a vector space of dimension at most $m$. This implies that $\rank(J(\varepsilon)) \leq m$. On the other hand, since $J(\varepsilon)$ is a PSD operator, we can diagonalize it, from which we get an expression
	\begin{equation}
		J(\varepsilon) = \sum_{k=1}^r \lambda_k \ket{v_k} \bra{v_k},
	\end{equation}
	where $r = \rank(J(\varepsilon))$, $\lambda_k > 0$ and $\ket{v_k} \in X^* \otimes Y$, with $(\ket{v_k})_k$ orthonormal. Taking $\ket{\psi_k} = \sqrt{\lambda_k} \ket{v_k}$, we get
	\begin{equation}
		J(\varepsilon) = \sum_{k=1}^r \ket{\psi_k} \bra{\psi_k}.
	\end{equation} 
	This shows that $m \leq \rank(J(\varepsilon))$, so
	\begin{equation}
		m = \rank(J(\varepsilon)).
	\end{equation}
\end{proof}

\subsection{Duality}
\label{section: CP duality}

Since $X$ is a finite dimensional complex Hilbert space, we can also turn $\Lin_\complexNumbers(X)$ into a Hilbert space by using the Hilbert-Schmidt inner product. In this way, Functional Analysis says that each linear map $\varepsilon \colon \Lin_\complexNumbers(X) \to \Lin_\complexNumbers(Y)$ has a unique adjoint $\varepsilon^\dag \colon \Lin_\complexNumbers(Y) \to \Lin_\complexNumbers(X)$, which is a $\complexNumbers$-linear map characterized by satisfying the equation
\begin{equation}
	\langle \varepsilon(A), B \rangle = \langle A, \varepsilon^\dag(B) \rangle, 
\end{equation}
for any $A \in \Lin_\complexNumbers(X)$ and $B \in \Lin_\complexNumbers(Y)$.

The operation of taking the adjoint defines an anti-linear map
\begin{align*}
	(-)^\dag \colon \Lin_\complexNumbers(\Lin_\complexNumbers(X), \Lin_\complexNumbers(Y)) &\longrightarrow \Lin_\complexNumbers(\Lin_\complexNumbers(Y), \Lin_\complexNumbers(X))\\
	\varepsilon &\longmapsto \varepsilon^\dag.
\end{align*}
This map is a bijection between CP maps. In this sense, CP maps are dual to themselves. This result will be useful to transfer results from CPTP maps to UCP maps.

\begin{proposition}{}{cp iff adjoint is cp}
	A $\complexNumbers$-linear map $\varepsilon \colon \Lin_\complexNumbers(X) \to \Lin_\complexNumbers(Y)$ is CP iff $\varepsilon^\dag \colon \Lin_\complexNumbers(Y) \allowbreak \to \Lin_\complexNumbers(X)$ is CP. Also, if $\varepsilon$ have $(E_k)_k$ as Kraus operators, then we can take $(E_k^\dag)_k$ as Kraus operators for $\varepsilon^\dag$.
\end{proposition}
\begin{proof}
	By proposition \ref{proposition:cp maps vs kraus representation}, having a Kraus representation is equivalent to being CP, so we'll just find a Kraus representation for $\varepsilon^\dag$, and vice-versa.
	
	Suppose that $\varepsilon$ is a CP map with Kraus operators $(E_k)_k$. For any $A \in \Lin_\complexNumbers(X)$ and $B \in \Lin_\complexNumbers(Y)$, we have
	\begin{align*}
		\langle A, \varepsilon^\dag(B) \rangle &= \langle \varepsilon(A), B \rangle\\
		&=  \left\langle \sum_k E_k A E_k^\dag, B \right\rangle\\
		&= \sum_k \langle E_k A E_k^\dag, B \rangle\\
		&= \sum_k \tr((E_k A E_k^\dag)^\dag B)\\
		&= \sum_k \tr(E_k A^\dag E_k^\dag B)\\
		&= \sum_k \tr(A^\dag E_k^\dag B E_k)\\
		&= \tr\left(A^\dag \sum_k E_k^\dag B E_k\right)\\
		&= \left\langle A, \sum_k E_k^\dag B E_k \right\rangle.
	\end{align*}
	By the uniqueness of $\varepsilon^\dag$. we conclude that $\varepsilon^\dag (B) = \sum_k E_k^\dag B E_k$. This is a Kraus representation of $\varepsilon^\dag$, therefore $\varepsilon^\dag$ is CP and has $(E_k^\dag)_k$ as Kraus operators.
	
	If we assume instead that $\varepsilon^\dag$ is CP, using that $\varepsilon^{\dag\dag} = \varepsilon$ we conclude, by the previous results, that $\varepsilon$ is completely positive. Therefore, $\varepsilon$ is CP if, and only if, $\varepsilon^\dag$ is CP.
\end{proof}

With this result, we can restrict the operation of taking adjoints to CP maps. That is, this defines an anti-linear bijection
\begin{align*}
	(-)^\dag \colon \cp(X,Y) &\longrightarrow \cp(Y,X)\\
	\varepsilon &\longmapsto \varepsilon^\dag.
\end{align*}

\subsection{Stratification into smooth manifolds}

Here we work with PSD operators instead of CP maps. The set $\psd(X)$ is stratified into smooth submanifolds of $\Lin_\complexNumbers(X)$. These manifolds are the sets $\psd(X)_r$ of operators of rank $r$.

\begin{notation}
	We denote by $\psd(X)_r$ the set of operators $A \in \psd(X)$ with $\rank(A) = r$.
\end{notation}

We'll use a little bit of Real Algebraic Geometry (RAG). Since we are not using too much RAG in this work, we'll state the relevant definitions here. The definitions are in general done for real closed fields. We'll take $\realNumbers$ as the real closed field. Both algebraic and semi-algebraic sets are defined as subsets of $\realNumbers^k$, for arbitrary $k \geq 0$.  An algebraic subset of $\realNumbers^k$ is the set of solutions of a system of polynomial equations, with coefficients in $\realNumbers$ and $k$ variables. It is useful to extend this definition admitting also inequations. A semi-algebraic subset of $\realNumbers^k$ is the set of solutions of a system of polynomial inequations, with coefficients in $\realNumbers$ and $k$ variables. In both cases, the number of polynomials in the system of equations or inequations is finite. Let $S \subseteq \realNumbers^k$ and $T \subseteq \realNumbers^l$ be semi-algebraic subsets. A function $f \colon S \to T$ is semi-algebraic if its graph
\begin{equation}
	\mathrm{Graph}(f) = \{(x,f(x)) \in \realNumbers^{k+l} \mid x \in S\}
\end{equation}
is a semi-algebraic subset of $\realNumbers^{k,l}$.

To prove that $\psd(X)_r$ is a smooth submanifold it is used the following result of \cite{gibson1979singular} (result (B4) of appendix B in this reference):

\begin{theorem}{}{orbits of semi-algebraic actions are submanifolds}
	Let $\Phi \colon G \times M \to M$ be a smooth action of a Lie group $G$ on a smooth manifold $M$. Suppose that $G \subseteq \realNumbers^m$ and $M \subseteq \realNumbers^n$, for some $m,n \in \naturalNumbers$. Also suppose that $G$ and $M$ are semi-algebraic sets and that the action $\Phi$ is a semi-algebraic function. Then all the orbits are smooth submanifolds of $M$.
\end{theorem}

The next theorem is an adaptation for complex spaces of proposition 2.1 of \cite{embedded_geometry_of_psdr}. Notice that $\psd(X)_r$ may be empty, depending on the rank $r$. To cover this case, we consider that $\varnothing$ is a smooth manifold with any dimension.

\begin{theorem}{}{psdr is submanifold}
	$\psd(X)_r$ is a smooth submanifold of $\Lin_\complexNumbers(X)$.
\end{theorem}
\begin{proof}
	That it is a submanifold, this is a consequence of PSD operators of a fixed rank being the orbit of a semi-algebraic action. We only have to find such semi-algebraic action and then apply theorem \ref{theorem:orbits of semi-algebraic actions are submanifolds}.
	
	Let $A \in \psd(X)_r$, that is, $A \in \Lin_\complexNumbers(X)$ and is a PSD operator of rank $r$. Let $P$ be the orthogonal projection onto $\image A$. Then we can write
	\begin{align*}
		A &= A^{1/2} P A^{1/2}\\
		&= (A^{1/2}+P^\perp) P (A^{1/2}+P^\perp).
	\end{align*}
	In the first line it was used that $PA^{1/2}=A^{1/2}$, since $A$ and $A^{1/2}$ have the same image. In the second line, the term $P^\perp \coloneqq \id_X-P$ was added to replace $A^{1/2}$ by an invertible operator. It doesn't change the operator, because $P^\perp P = P P^\perp = 0$. In this way, $A$ is of the form
	\begin{equation}
		A = V P V^\dag,
	\end{equation}
	for $V = (A^{1/2}+P^\perp) \in \GL(X)$. Every operator in $\psd(X)_r$ can be written in this form using the same projection $P$, but with different invertible operator $V$. To show this, let $A' \in \psd(X)_r$ be another operator. Let $Q$ be the orthogonal projection onto the image of $A'$. By the same argument made for $A$, we can write $A'$ as
	\begin{equation}
		A' = W Q W^\dag,
	\end{equation}
	for some $W \in \GL(X)$. Since $P$ and $Q$ are orthogonal projections with the same rank $r$, they are related by a unitary operator. In fact, we can diagonalize $P$ and $Q$ as
	\begin{equation}
		P = \sum_{k=1}^r \ket{\psi_k} \bra{\psi_k},
	\end{equation}
	\begin{equation}
		Q = \sum_{k=1}^r \ket{\varphi_k} \bra{\varphi_k},
	\end{equation}
	where both $(\ket{\psi_k})_{k=1,\dots,r}$ and $(\ket{\varphi_k})_{k=1,\dots,r}$ are orthonormal sequences of vectors in $X$. We can complete these orthonormal vectors to orthonormal basis $(\ket{\psi_k})_{k=1,\dots,d_X}$ and $(\ket{\varphi_k})_{k=1,\dots,d_X}$ of $X$. Then define a unitary operator $U \in \U(X)$ by
	\begin{equation}
		U\ket{\psi_k} = \ket{\varphi_k},
	\end{equation}
	for $k=1,\dots,d_X$. It follows that
	\begin{equation}
		Q = UPU^\dag,
	\end{equation}
	from which we get that
	\begin{align*}
		A' &= WQW^\dag\\
		&= W UPU^\dag W^\dag\\
		&= (WU)P(WU)^\dag.
	\end{align*}
	Since $U$ is unitary and $W$ is invertible, the composition $WU$ is invertible. This shows that every $A' \in \psd(X)_r$ can be written as $A' = V P V^\dag$, for a fixed orthogonal projection $P \in \Lin_\complexNumbers(X)$ and some $V \in \GL(X)$. The converse is also true, for every $V \in \GL(X)$ the operator $VPV^\dag$ is in $\psd(X)_r$. That $VPV^\dag$ is PSD, this is because
	\begin{align*}
		VPV^\dag &= VP P^\dag V^\dag\\
		&= (VP)(VP)^\dag,
	\end{align*}
	and any operator of the form $M M^\dag$ is PSD. The rank of $VPV^\dag$ is $r$ because conjugation by an invertible operator ($V$, in this case) doesn't change the rank, and $\rank(P) = r$. This shows that $VPV^\dag \in \psd(X)_r$.
	
	Define the action
	\begin{align*}
		\Phi\colon \GL(X) \times \Lin_\complexNumbers(X) &\longrightarrow \Lin_\complexNumbers(X)\\
		(V,M) &\longmapsto VMV^\dag.
	\end{align*}
	Then $\psd(X)_r$ is the orbit of $P$ by this action, for any orthogonal projection $P \in \Lin_\complexNumbers(X)$ with $\rank(P)=r$. To conclude the proof, we just have to argue that this action is smooth and semi-algebraic. Of course, the set $\GL(X)$ is a Lie group and $\Lin_\complexNumbers(X)$ is a real manifold, if we interpret it as a real vector space. It's straightforward to show that $\Phi$ is an action. That it is smooth, we can show this by expressing the operators as matrices with respect to some basis, and then use that multiplication of matrices and taking the Hermitian conjugate are smooth operations. It remains to show that the action is semi-algebraic. We know that $\Lin_\complexNumbers(X)$ is isomorphic to $\realNumbers^{2 d_X^2}$, so we'll use $\Lin_\complexNumbers(X)$ instead of $\realNumbers^{2 d_X^2}$. The set $\realNumbers^k$ is always a semi-algebraic subset of itself. To see this, just take $p(x_1,\dots,x_k)=0$, then
	\begin{equation}
		\realNumbers^k = \{x \in \realNumbers^k \mid p(x) \geq 0\}.
	\end{equation}
	Also, the set $\GL(X)$ is a semi-algebraic subset of $\Lin_\complexNumbers(X)$. The set $\GL(X)$ is isomorphic to the space of invertible complex matrices $\GL(d_X,\complexNumbers) \subseteq M_{d_X}(\complexNumbers)$. A matrix is invertible iff it has a non zero determinant, so 
	\begin{equation}
		\GL(d_X,\complexNumbers) = \{M \in M_{d_X}(\complexNumbers) \mid \det(M) \neq 0\}.
	\end{equation}
	Notice that $\det(M)$ is a polynomial over the entries of $M$, but with coefficients in $\complexNumbers$. To obtain a real polynomial, it suffices to take the square of the modulus. We have that
	\begin{equation}
		\GL(d_X,\complexNumbers) = \{M \in M_{d_X}(\complexNumbers) \mid |\det(M)|^2 > 0\}.
	\end{equation}
	Then separate the real and imaginary parts of the entries of $M$, which are the real variables in $\realNumbers^{2d_X^2}$. We have that $|\det(M)|^2$ is a real polynomial of these variables. From this it follows that $\GL(X)$ is a semi-algebraic subset of $\Lin_\complexNumbers(X)$. A simple trick shows that $\GL(X) \times \Lin_\complexNumbers(X)$ is also semi-algebraic. First use the isomorphism $\Lin_\complexNumbers(X) \cong M_{d_X}(\complexNumbers)$. Then take the polynomial
	\begin{equation}
		q(M_1,M_2) \coloneqq |\det(M_1)|^2,
	\end{equation}
	for $(M_1, M_2) \in M_{d_X}(\complexNumbers) \times M_{d_X}(\complexNumbers)$. Again, separating the real and imaginary parts of each entry of the matrices $M_1$ and $M_2$, we get the real variables in $\realNumbers^{4 d_X^2}$. Also, $q$ is a real polynomial of these variables. Then
	\begin{equation}
		\GL(d_X,\complexNumbers) \times M_{d_X}(\complexNumbers) = \{(M_1,M_2) \in M_{d_X}(\complexNumbers) \times M_{d_X}(\complexNumbers) \mid q(M_1,M_2) > 0\}.
	\end{equation}
	This is a semi-algebraic subset of $M_{d_X}(\complexNumbers) \times M_{d_X}(\complexNumbers)$. Therefore, $\GL(X) \times \Lin_\complexNumbers(X)$ is a semi-algebraic subset of $\Lin_\complexNumbers(X) \times \Lin_\complexNumbers(X)$. This concludes that the domain and codomain of $\Phi$ are semi-algebraic sets. Then, $\Phi$ is semi-algebraic if its graph is a semi-algebraic subset of $\Lin_\complexNumbers(X) \times \Lin_\complexNumbers(X) \times \Lin_\complexNumbers(X)$. It is known that rational functions are semi-algebraic (proposition (B1) of appendix B of \cite{gibson1979singular}). The action is defined as $\Phi(V,M) = VMV^\dag$. Using a basis, we can see these operators as matrices. Separating the real and imaginary parts of the entries of $V,M$ and $\Phi(V,M)$, $\Phi$ corresponds to a function $\realNumbers^{4d_X^2} \to \realNumbers^{2d_X^2}$. Both multiplication of matrices and taking the Hermitian conjugate are polynomial functions of the real entries of the matrices. From this it follows that the function $\realNumbers^{4d_X^2} \to \realNumbers^{2d_X^2}$ is a real polynomial. In particular, it is rational, so $\Phi$ is a semi-algebraic function. Applying theorem \ref{theorem:orbits of semi-algebraic actions are submanifolds} we conclude that $\psd(X)_r$ is a smooth submanifold of $\Lin_\complexNumbers(X)$.
\end{proof}

This theorem shows that $\psd(X)_r$ is a submanifold of $\Lin_\complexNumbers(X)$, but it doesn't tell its dimension. We'll find its dimension by studying the tangent space, since the dimension of the tangent space equals the dimension of the manifold. The next theorem is based on section 2.1 of \cite{embedded_geometry_of_psdr}, but adapted for complex spaces.

\begin{theorem}{}{TPSDr}
	Let $A \in \psd(X)_r$. The tangent space $T_A \psd(X)_r$ can be expressed as any of the following:
	\begin{align*}
		T_A \psd(X)_r &= \{ WA+(WA)^\dag \mid W \in \Lin_\complexNumbers(X) \},\\
		&= \{ WP_{\image A}+(W P_{\image A})^\dag \mid W \in \Lin_\complexNumbers(X) \}, \\
		&= \{ B \in \hermitian(X) \mid P_{\image A}^\perp B P_{\image A}^\perp = 0 \},
	\end{align*}
	where $P_{\image A}$ is the orthogonal projection onto $\image A$ and $P_{\image A}^\perp = \id - P_{\image A}$, the orthogonal projection onto $(\image A)^\perp$. The dimension of the tangent space is
	\begin{equation}
		\dim_\realNumbers \psd(X)_r = 2 d_X r -r^2.
	\end{equation}
\end{theorem}
\begin{proof}
	In the proof of theorem \ref{theorem:psdr is submanifold} was defined the smooth action\\
	\begin{align*}
		\Phi\colon \GL(X) \times \Lin_\complexNumbers(X) &\longrightarrow \Lin_\complexNumbers(X)\\
		(V,M) &\longmapsto VMV^\dag.
	\end{align*}
	It was also proved that $\psd(X)_r$ is the orbit of any orthogonal projection $P \in \Lin_\complexNumbers(X)$ with rank $r$. Given such $P$, the orbit of $P$ is the image of the function
	\begin{align*}
		\Phi_P \colon \GL(X) &\longrightarrow \psd(X)_r\\
		V &\longmapsto VPV^\dag.
	\end{align*}
	We use this function to find the tangent space of $\psd(X)_r$. Of course, $\Phi_P$ can be used to find some tangent vectors. For it to give all tangent vectors, we need $\Phi_P$ to be a submersion. By theorem 4.14 of \cite{lee2012introduction} (the Global Rank Theorem), if $\Phi_P$ is smooth, surjective and has constant rank, then it is a submersion. We already know that it is smooth and surjective. Let's prove that it has constant rank. Let $V \in \GL(X)$, then the differential of $\Phi_P$ at $V$ is a $\realNumbers$-linear map
	\begin{equation*}
		d_V \Phi_P \colon T_V \GL(X) \to T_{\Phi_P (V)} \psd(X)_r. 
	\end{equation*}
	$\GL(X)$ is an open subset of $\Lin_\complexNumbers(X)$, so $T_V \GL(X) = T_V \Lin_\complexNumbers(X)$. Since $\Lin_\complexNumbers(X)$ is a vector space, we have a natural isomorphism $T_V \Lin_\complexNumbers(X) \cong \Lin_\complexNumbers(X)$. From the embedding $\psd(X)_r \to \Lin_\complexNumbers(X)$ we have a $\realNumbers$-linear map $T_{\Phi_P (V)} \psd(X)_r \to T_{\Phi_P (V)} \Lin_\complexNumbers(X)$, which is injective. Using the isomorphism $T_{\Phi_P (V)} \Lin_\complexNumbers(X) \allowbreak \cong \Lin_\complexNumbers(X)$, we can interpret $T_{\Phi_P (V)} \psd(X)_r$ as a subspace of $\Lin_\complexNumbers(X)$. Putting everything together, we can interpret the differential as a $\realNumbers$-linear map
	\begin{equation*}
		d_V \Phi_P \colon \Lin_\complexNumbers(X) \to T_{\Phi_P (V)} \psd(X)_r,
	\end{equation*}
	where we treat $T_{\Phi_P (V)} \psd(X)_r$ as a subspace of $\Lin_\complexNumbers(X)$. This function can be computed using curves. Let $W \in \Lin_\complexNumbers(X)$, then define the smooth curve
	\begin{equation}
		\gamma(t) \coloneqq V+tW,
	\end{equation}
	defined over an open interval $(-\varepsilon, \varepsilon)$, with $\varepsilon>0$ small enough such that $\gamma(t) \in \GL(X)$. Recall that $\GL(X)$ is open, so such $\varepsilon$ exists. Then the differential can be computed as
	\begin{equation}
		(d_V \Phi_P)(W) = \frac{d}{dt}\bigg|_0 \Phi_P(\gamma(t)).
	\end{equation}
	We have that
	\begin{align*}
		\Phi_P (\gamma(t)) &= (V+tW) P (V+tW)^\dag \\
		&= (V+tW) P (V^\dag+tW^\dag) \\
		&= VPV^\dag +t(VPW^\dag+WPV^\dag) + t^2 W P W^\dag.
	\end{align*}
	The linear part is $t(VPW^\dag+WPV^\dag)$, so
	\begin{equation}
		\label{equation: dVPhiP}
		(d_V \Phi_P)(W) = VPW^\dag+WPV^\dag .
	\end{equation}
	To prove that $\Phi_P$ has constant rank, let's relate $d_V \Phi_P$ to $d_{\id_X} \Phi_P$. Taking $V= \id_X$ in the last equation, we get
	\begin{equation}
		(d_{\id_X} \Phi_P)(W) = PW^\dag+WP .
	\end{equation}
	Conjugating by $V$, we get
	\begin{align*}
		V \, (d_{\id_X} \Phi_P)(W) \, V^\dag  &= VPW^\dag V^\dag+VWP V^\dag \\
		&= VP(VW)^\dag+(VW)P V^\dag \\
		&= (d_V \Phi_P)(VW).
	\end{align*}
	Therefore,
	\begin{equation}
		\label{equation: dVPhiP = V dIPhiP etc}
		(d_V \Phi_P)(W) = V \, (d_{\id_X} \Phi_P)(V^{-1} W) \, V^\dag .
	\end{equation}
	Since $V$ is invertible, both the functions $W \mapsto V^{-1} W$ and $W \mapsto VWV^\dag$ are automorphisms of $\Lin_\complexNumbers(X)$. From equation \ref{equation: dVPhiP = V dIPhiP etc} we get that the images of $d_V \Phi_P$ and $d_{\id_X} \Phi_P$ are isomorphic. Therefore, $\Phi_P$ has constant rank. Then, by theorem 4.14 of \cite{lee2012introduction} we know that $\Phi_P$ is a submersion. This means that
	\begin{equation}
		\image d_V \Phi_P = T_{\Phi_P(V)} \psd(X)_r.
	\end{equation}
	We get an expression for $ T_{\Phi_P(V)} \psd(X)_r$ by equation \ref{equation: dVPhiP}. Let $A \in \psd(X)_r$, since $\Phi_P$ is surjective, there exists $V \in \GL(X)$ such that $A = \Phi_P(V) = VPV^\dag$. In equation \ref{equation: dVPhiP} we have $VP$ and $PV^\dag$, we then make appear $VPV^\dag$ by inserting $V$ and its inverse. We get
	\begin{align*}
		(d_V \Phi_P)(W) &= VPW^\dag+WPV^\dag \\
		&= VPV^\dag (V^{-1})^\dag W^\dag + W V^{-1} VPV^\dag \\
		&= VPV^\dag (W V^{-1})^\dag + (W V^{-1}) VPV^\dag. 
	\end{align*}
	Since $A = VPV^\dag$, we have that
	\begin{equation}
		(d_V \Phi_P)(W) = A (W V^{-1})^\dag + (W V^{-1}) A.
	\end{equation}
	Therefore, the image of $d_V \Phi_P$ is
	\begin{equation}
		\image d_V \Phi_P = \{ (W V^{-1}) A + A (W V^{-1})^\dag \mid W \in \Lin_\complexNumbers(X) \}.
	\end{equation}
	Since $V$ is invertible and $W$ can be any operator in $\Lin_\complexNumbers(X)$, we can absorb $V^{-1}$ into $W$, from which we get that
	\begin{equation}
		\image d_V \Phi_P = \{ W A + A W^\dag \mid W \in \Lin_\complexNumbers(X) \}.
	\end{equation}
	Since $A$ is positive semidefinite, it is hermitian, so $A^\dag = A$. Then $AW^\dag = (WA)^\dag$, so
	\begin{equation}
		\image d_V \Phi_P = \{ W A + (WA)^\dag \mid W \in \Lin_\complexNumbers(X) \}.
	\end{equation}
	This concludes that
	\begin{equation}
		T_A \psd(X)_r = \{ W A + (WA)^\dag \mid W \in \Lin_\complexNumbers(X) \}.
	\end{equation}
	
	The other expressions for the tangent space can be obtained from this last one. Diagonalize $A$ as
	\begin{equation}
		A = \sum_k \lambda_k \ket{v_k}\bra{v_k},
	\end{equation}
	where $\lambda_k > 0$ and $(\ket{v_k})_k$ is a sequence of orthonormal vectors in $X$. Then the Moore-Penrose inverse $A^+$ of $A$ is obtained by inverting the non negative eigenvalues:
	\begin{equation}
		A^+ = \sum_k \lambda_k^{-1} \ket{v_k}\bra{v_k}.
	\end{equation}
	It has the property that $A A^+ = A^+ A = P_{\image A}$. Taking $W = W' A^+$, for any $W' \in \Lin_\complexNumbers(X)$, we get that
	\begin{equation}
		WA + (WA)^\dag = W' A^+ A + (W' A^+ A)^\dag = W' P_{\image A} + (W' P_{\image A})^\dag.
	\end{equation}
	Therefore,
	\begin{equation}
		\{ W' P_{\image A} + (W' P_{\image A})^\dag \mid W' \in \Lin_\complexNumbers(X) \} \subseteq T_A \psd(X)_r.
	\end{equation}
	Conversely, if we take $W' = WA$, for any $W \in \Lin_\complexNumbers(X)$, then
	\begin{equation}
		W' P_{\image A} + (W' P_{\image A})^\dag = W A P_{\image A} + (W A P_{\image A})^\dag = WA + (WA)^\dag.
	\end{equation}
	Therefore,
	\begin{equation}
		T_A \psd(X)_r \subseteq \{ W' P_{\image A} + (W' P_{\image A})^\dag \mid W' \in \Lin_\complexNumbers(X) \} .
	\end{equation}
	This proves that
	\begin{equation}
		T_A \psd(X)_r = \{ W P_{\image A} + (W P_{\image A})^\dag \mid W \in \Lin_\complexNumbers(X) \} .
	\end{equation}
	The operator $B = W P_{\image A} + (W P_{\image A})^\dag$ is, of course, hermitian, so $B \in \hermitian(X)$. Also,
	\begin{equation}
		P_{\image A}^\perp B P_{\image A}^\perp = P_{\image A}^\perp W P_{\image A} P_{\image A}^\perp + P_{\image A}^\perp P_{\image A} W^\dag P_{\image A}^\perp .
	\end{equation}
	Since $P_{\image A} P_{\image A}^\perp = P_{\image A}^\perp P_{\image A} = 0$, we get that
	\begin{equation}
		P_{\image A}^\perp B P_{\image A}^\perp = 0.
	\end{equation}
	Therefore,
	\begin{equation}
		T_A \psd(X)_r \subseteq \{B \in \hermitian(X) \mid P_{\image A}^\perp B P_{\image A}^\perp = 0 \}.
	\end{equation}
	To prove the other inclusion, let $B \in \hermitian(X)$ such that $P_{\image A}^\perp B P_{\image A}^\perp = 0$. Then take
	\begin{equation}
		W = P_{\image A}^\perp B P_{\image A} + \frac{1}{2}P_{\image A} B P_{\image A}.
	\end{equation}
	Of course, we have that $W P_{\image A} = W$. Then
	\begin{align*}
		W P_{\image A} + (W P_{\image A})^\dag &= W+W^\dag\\
		&= P_{\image A}^\perp B P_{\image A} + \frac{1}{2}P_{\image A} B P_{\image A}+P_{\image A} B P_{\image A}^\perp \\
		& \quad + \frac{1}{2}P_{\image A} B P_{\image A}\\
		&=  P_{\image A} B P_{\image A} + P_{\image A}^\perp B P_{\image A} + P_{\image A} B P_{\image A}^\perp\\
		&= B - P_{\image A}^\perp B P_{\image A}^\perp.
	\end{align*}
	Since $P_{\image A}^\perp B P_{\image A}^\perp = 0$, we have that
	\begin{equation}
		B = W P_{\image A} + (W P_{\image A})^\dag.
	\end{equation}
	This prove the other inclusion, so
	\begin{equation}
		T_A \psd(X)_r = \{B \in \hermitian(X) \mid P_{\image A}^\perp B P_{\image A}^\perp = 0 \}.
	\end{equation}
	
	We can use this last characterization to compute the dimension of the tangent space. For any $B \in \Lin_\complexNumbers(X)$, it can be written uniquely as
	\begin{equation}
		B = B_{11} + B_{12} + B_{21} + B_{22},
	\end{equation}
	where
	\begin{equation}
		B_{11} = P_{\image A} B P_{\image A},
	\end{equation}
	\begin{equation}
		B_{12} = P_{\image A} B P_{\image A}^\perp,
	\end{equation}
	\begin{equation}
		B_{21} = P_{\image A}^\perp B P_{\image A},
	\end{equation}
	\begin{equation}
		B_{22} = P_{\image A}^\perp B P_{\image A}^\perp.
	\end{equation}
	In other words, we can break the matrix of $B$ into four blocks. Then $B$ is in the tangent space iff the following holds:
	\begin{itemize}
		\item $B_{11}$ is hermitian;
		
		\item $B_{12}^\dag = B_{21}$;
		
		\item $B_{22} = 0$.
	\end{itemize}
	Therefore, only $B_{11}$ and $B_{12}$ contribute to the dimension. The hermitian operators $B_{12}$ contribute with $r^2$, which is the real dimension of $\hermitian(r)$. The operators $B_{12}$ correspond to matrices in $M_{r, d_X-r}(\complexNumbers)$. Its real dimension is
	\begin{equation}
		\dim_\realNumbers M_{r, d_X-r}(\complexNumbers) = 2 r (d_X-r) = 2 d_X r-2r^2.
	\end{equation}
	Summing both contributions, we get that
	\begin{equation}
		\dim_\realNumbers T_A \psd(X)_r = r^2 + 2 d_X r-2r^2 = 2 d_X r-r^2. 
	\end{equation}
\end{proof}

\subsection{Tensor product of CP maps}

We prove in this section simple results about the tensor product of CP maps. We prove that the tensor product of CP maps is a CP map. Then we show that the Choi-rank of the tensor product is the product of the Choi-ranks. Lastly, we show that dual distributes over the tensor product.

\begin{proposition}{}{tensor product operation elements}
	If $\varepsilon \in \cp(X,Y)$ and $\varepsilon' \in \cp(X',Y')$ have Kraus operators $(E_k)_k$ and $(F_l)_l$, respectively, then $\varepsilon \otimes \varepsilon'$ have $(E_k \otimes F_l)_{k,l}$ as Kraus operators. In particular, $\varepsilon \in \cp(X \otimes X', Y \otimes Y')$.
\end{proposition}
\begin{proof}
	$\varepsilon\otimes \varepsilon'$ is characterized as the map that satisfies $(\varepsilon \otimes \varepsilon')(A\otimes B) = \varepsilon(A) \otimes \varepsilon'(B)$, for any $A \in \Lin_\complexNumbers(X)$ and $B \in \Lin_\complexNumbers(X')$. We have that
	\begin{align*}
		(\varepsilon \otimes \varepsilon')(A\otimes B) &= \varepsilon(A) \otimes \varepsilon'(B) \\
		&= \left(\sum_k E_k A E_k^\dag\right) \otimes \left(\sum_l F_l B F_l^\dag \right) \\
		&= \sum_{k,l} E_k A E_k^\dag \otimes F_l B F_l^\dag \\
		&\stackrel{(*)}{=} \sum_{k,l} (E_k \otimes F_l)(A\otimes B) (E_k^\dag \otimes F_l^\dag) \\
		&= \sum_{k,l} (E_k \otimes F_l)(A\otimes B) (E_k \otimes F_l)^\dag .
	\end{align*}
	In the equality ($*$) we used the identity $(f' \otimes g')(f \otimes g) = f'f \otimes g'g$. The identity $(\varepsilon \otimes \varepsilon')(A\otimes B) = \sum_{k,l} (E_k \otimes F_l)(A\otimes B)(E_k \otimes F_l)^\dag$ together with the linearity of $\varepsilon \otimes \varepsilon'$ implies that $(\varepsilon \otimes \varepsilon')(C) = \sum_{k,l} (E_k \otimes F_l)C(E_k \otimes F_l)^\dag$, for any $C \in \Lin_\complexNumbers(X \otimes X')$. Therefore $(E_k \otimes F_l)_{k,l}$ are Kraus operators for $\varepsilon \otimes \varepsilon'$. Since $\varepsilon \otimes \varepsilon'$ is of the form $(\varepsilon \otimes \varepsilon')(C) = \sum_{k,l} (E_k \otimes F_l)C(E_k \otimes F_l)^\dag$, it is completely positive.
\end{proof}

\begin{corollary}{}{CR of tensor product}
	For any CP maps $\varepsilon$ and $\varepsilon'$ we have $\CR(\varepsilon \otimes \varepsilon') = \CR(\varepsilon) \CR(\varepsilon')$.
\end{corollary}
\begin{proof}
	Let $(E_k)_k$ be linearly independent Kraus operators for $\varepsilon$ and let $(F_l)_l$ be linearly independent Kraus operators for $\varepsilon'$. By propositions \ref{proposition:tensor of l.i. vectors} and \ref{proposition:tensor product operation elements} we know that $(E_k \otimes F_l)_{k,l}$ are linearly independent Kraus for $\varepsilon \otimes \varepsilon'$. Since the Kraus operators are linearly independent, the sequences $(E_k)_k$ and $(F_l)_l$ have $\CR(\varepsilon)$ and $\CR(\varepsilon')$ operators, respectively. Therefore $(E_k \otimes F_l)_{k,l}$ is a sequence with $\CR(\varepsilon) \CR(\varepsilon')$ operators. But $(E_k \otimes F_l)_{k,l}$ is linearly independent, so it must have $\CR(\varepsilon \otimes \varepsilon')$ operators. Therefore $\CR(\varepsilon \otimes \varepsilon') = \CR(\varepsilon) \CR(\varepsilon')$.
\end{proof}

In section \ref{section: CP duality} we showed that there is a duality between CP maps and themselves. This duality is given by the operation $(-)^\dag$ of taking the adjoint, with respect to the Hilbert-Schmidt inner product. For later use, we'll show that $(\varepsilon \otimes \varepsilon')^\dag = \varepsilon^\dag \otimes \varepsilon'^\dag$.

\begin{proposition}{}{dual of tensor of cp maps}
	Let $\varepsilon$ and $\varepsilon'$ be completely positive maps. We have that $(\varepsilon \otimes \varepsilon')^\dag = \varepsilon^\dag \otimes \varepsilon'^\dag$.
\end{proposition}
\begin{proof}
	Lets write that $\varepsilon \in \cp(X,Y)$ and $\varepsilon' \in \cp(X',Y')$. Both $(\varepsilon \otimes \varepsilon')^\dag$ and $\varepsilon^\dag \otimes \varepsilon'^\dag$ are elements of $\cp(Y\otimes Y', X \otimes X')$. Their domain is $\Lin_\complexNumbers(Y\otimes Y')$, which is spanned by the maps $A \otimes A'$, for $A \in \Lin_\complexNumbers(Y)$ and $A' \in \Lin_\complexNumbers(Y')$. Similarly for the codomain $\Lin_\complexNumbers(X \otimes X')$, it is spanned by the maps $B \otimes B'$ for $B \in \Lin_\complexNumbers(X)$ and $B' \in \Lin_\complexNumbers(X')$. By linearity, we'll only need to consider these operators $A\otimes A'$ and $B\otimes B'$. By definition of the dual, $(\varepsilon \otimes \varepsilon')^\dag$ is the unique linear map that satisfies $\langle (\varepsilon \otimes \varepsilon')(A \otimes A') , B \otimes B' \rangle = \langle A \otimes A', (\varepsilon \otimes \varepsilon')^\dag(B \otimes B') \rangle$. We have that
	\begin{align*}
		\langle (\varepsilon \otimes \varepsilon')(A \otimes A') , B \otimes B' \rangle &= \langle \varepsilon(A) \otimes \varepsilon'(A') , B \otimes B' \rangle\\
		&= \langle \varepsilon(A), B \rangle \langle \varepsilon'(A'), B' \rangle\\
		&= \langle A, \varepsilon^\dag(B) \rangle \langle A', \varepsilon^\dag(B') \rangle\\
		&= \langle A \otimes A', \varepsilon(B) \otimes \varepsilon'(B') \rangle\\
		&= \langle A \otimes A', (\varepsilon \otimes \varepsilon')(B \otimes B') \rangle.
	\end{align*}
	By uniqueness it follows that $(\varepsilon \otimes \varepsilon')^\dag = \varepsilon^\dag \otimes \varepsilon'^\dag$.
\end{proof}

\section{CPTP maps}

CPTP maps are our main subject of study. We define them next.

\begin{definition}{}{}
	A CPTP map is a CP map $\varepsilon \in \cp(X,Y)$ which preserves the trace. Recall that $\varepsilon$ is a $\complexNumbers$-linear map $\varepsilon \colon \Lin_\complexNumbers(X) \to \Lin_\complexNumbers(Y)$ which is completely positive. That it preserves the trace, this means that
	\begin{equation}
		\tr(\varepsilon(A)) = \tr(A),
	\end{equation}
	for any $A \in \Lin_\complexNumbers(X)$. We call this the TP property, for being trace preserving.
\end{definition}

CPTP maps are a model for quantum communication, for this reason they are also called quantum channels. We'll not delve into how they are useful in Quantum Computation and Quantum Information, instead we'll focus onto its mathematical properties. Next we'll specialize the results obtained earlier for CP maps to the case of CPTP maps.

\subsection{The \CJ isomorphism}

For CP maps, the \CJ isomorphism is a bijection between CP maps and PSD operators. For CPTP maps, the TP property corresponds to operators $A$ satisfying $\tr_2 A = \id_{X^*}$. We defined the Choi-rank of a CP map in definition \ref{definition:Choi rank}. The TP property, or equivalently the equation $\tr_2 A = \id_{X^*}$, imposes a restriction over the CPTP maps that can't be satisfied for any rank. From this we get a lower bound on the rank of a CPTP map. We introduce some notation, and then prove these results.

\begin{notation}
	The equation $\tr_2 A = \id_{X^*}$ being satisfied will be denoted by a 2 subindex:
	\begin{itemize}
		\item $\psdii(X,Y) \coloneqq \{A \in \psd(X^* \otimes Y) \mid \tr_2 A = \id_{X^*}\}$.
		\item $\psdii(X,Y)_r \coloneqq \{A \in \psd(X^* \otimes Y)_r \mid \tr_2 A = \id_{X^*}\}$.
	\end{itemize}
	This time we use a comma in $\psdii(X,Y)$, because of its interpretation as a hom-set of a category isomorphic to the category of CPTP maps. Also, we'll use the abbreviations
	\begin{itemize}
		\item $\psdii(X) \coloneqq \psdii(X,X)$.
		\item $\psdii(X)_r \coloneqq \psdii(X,X)_r$.
	\end{itemize}
\end{notation}

\begin{proposition}{}{CJ isomorphism for cptp}
	The \CJ isomorphism
	\begin{equation*}
		J \colon \Lin_\complexNumbers(\Lin_\complexNumbers(X), \Lin_\complexNumbers(Y)) \to \Lin_\complexNumbers(X^* \otimes Y)
	\end{equation*}
	restricts to a bijection between $\cptp(X,Y)$ and $\psdii(X,Y)$.
\end{proposition}
\begin{proof}
	By proposition \ref{proposition:bijection between CP and PSD}, we already know that CP maps correspond to PSD operators. It remains to show that the TP property corresponds to the partial trace equation. We use the notation and results of proposition \ref{proposition:CJ expression}.
	
	$(\implies)$ Let $\varepsilon \in \cptp(X,Y)$. Then
	\begin{align*}
		\tr_2 J(\varepsilon) &= \tr_2 \left( \sum_{i,i'} \ket{x^i} \bra{x^{i'}} \otimes \varepsilon(\ket{x_i} \bra{x_{i'}}) \right)\\
		&= \sum_{i,i'} \ket{x^i} \bra{x^{i'}} \ \tr ( \varepsilon(\ket{x_i} \bra{x_{i'}}))\\
		&= \sum_{i,i'} \ket{x^i} \bra{x^{i'}} \ \tr ( \ket{x_i} \bra{x_{i'}})\\
		&= \sum_{i,i'} \ket{x^i} \bra{x^{i'}} \ \delta_{i,i'}\\
		&= \id_{X^*}.
	\end{align*}
	
	$(\impliedby)$ Let $A = J(\varepsilon)$ and $B \in \Lin_\complexNumbers(X)$. We have that
	\begin{align*}
		\tr(\varepsilon(B)) &= \tr((J^{-1}(A))(B))\\
		&= \tr \left(\sum_{i,i'} \bra{x_i} B \ket{x_{i'}} \ (\bra{x^i}\otimes \id_Y)A(\ket{x^{i'}} \otimes \id_Y) \right)\\
		&= \sum_j \bra{y_j} \left(\sum_{i,i'} \bra{x_i} B \ket{x_{i'}} \ (\bra{x^i}\otimes \id_Y)A(\ket{x^{i'}} \otimes \id_Y) \right) \ket{y_j}\\
		&=   \sum_{i,i'} \bra{x_i} B \ket{x_{i'}} \sum_j (\bra{x^i}\otimes \bra{y_j})A(\ket{x^{i'}} \otimes \ket{y_j}) \\
		&=   \sum_{i,i'} \bra{x_i} B \ket{x_{i'}} \ \bra{x^i} \left( \sum_j (\id_{X^*} \otimes \bra{y_j}) A (\id_{X^*} \otimes \ket{y_j}) \right)\ket{x^{i'}} \\
		&=   \sum_{i,i'} \bra{x_i} B \ket{x_{i'}} \ \bra{x^i} \tr_2 A \ket{x^{i'}} \\
		&=   \sum_{i,i'} \bra{x_i} B \ket{x_{i'}} \ \bra{x^i} \id_{X^*} \ket{x^{i'}} \\
		&=   \sum_{i,i'} \bra{x_i} B \ket{x_{i'}} \ \langle x^i \mid x^{i'} \rangle \\
		&=   \sum_{i,i'} \bra{x_i} B \ket{x_{i'}} \ \delta_{i,i'} \\
		&=   \sum_i \bra{x_i} B \ket{x_i} \\
		&= \tr B.
	\end{align*}
	
\end{proof}

\subsection{Choi-rank bounds}

The \CJ isomorphism gives a correspondence between the hom-set $\cptp(X,Y)$ and $\psdii(X,Y)$. The Choi-rank of a CPTP map is the rank of the corresponding PSD operator, as seen in definition \ref{definition:Choi rank}. The equation $\tr_2 A = \id_{X^*}$ that the PSD operator $A$ must satisfy implies a lower bound on its rank, which we prove next.
\begin{proposition}{}{general CPTP rank bounds}
	Let $X$ and $Y$ be finite dimensional complex Hilbert spaces with $d_X \geq 1$ and $d_Y \geq 1$. Then, for any $A \in \psdii(X,Y)$, we have that
	\begin{equation}
		\frac{d_X}{d_Y} \leq \rank (A) \leq d_X d_Y.
	\end{equation}
	Equivalently, for any $\varepsilon \in \cptp(X,Y)$, we have that
	\begin{equation}
		\frac{d_X}{d_Y} \leq \CR(\varepsilon) \leq d_X d_Y.
	\end{equation}
\end{proposition}
\begin{proof}
	Let $A \in \psdii(X,Y)$. Since $\psdii(X,Y) \subseteq \Lin_\complexNumbers (X^* \otimes Y)$, of course
	\begin{equation}
		\rank (A) \leq \dim_\complexNumbers(X^* \otimes Y) = d_X d_Y.
	\end{equation}
	The other bound comes from the equation
	\begin{equation}
		\tr_2 A = \id_{X^*}.
	\end{equation}
	Diagonalize $A$ as
	\begin{equation}
		A = \sum_{k=1}^{\rank A} \lambda_k \ket{v_k}\bra{v_k},
	\end{equation}
	where $\lambda_k > 0$ and $(\ket{v_k})_k$ is an orthonormal sequence in $X^* \otimes Y$. Then define
	\begin{equation}
		\ket{\psi_k} \coloneqq \sqrt{\lambda_k} \ket{v_k},
	\end{equation}
	so
	\begin{equation}
		A = \sum_k \ket{\psi_k}\bra{\psi_k}.
	\end{equation}
	Let $(x^i)_i$ be an orthonormal basis for $X^*$, and let $(y_j)_j$ be an orthonormal basis for $Y$. Then $\ket{\psi_k}$ is written in the basis $(\ket{x^i} \otimes \ket{y_j})_{i,j}$ as
	\begin{equation}
		\ket{\psi_k} = \sum_{i,j} \psi_{i,j,k} \ket{x^i}\otimes \ket{y_j}.
	\end{equation}
	Substituting this in the expression for $A$, we get
	\begin{equation}
		A = \sum_{i,i',j,j',k} \psi_{i,j,k} \overline{\psi_{i',j',k}} \ket{x^i}\bra{x^{i'}} \otimes \ket{y_j}\bra{y_{j'}}.
	\end{equation}
	Its partial trace is then
	\begin{equation}
		\tr_2 A = \sum_{i,i',j,k} \psi_{i,j,k} \overline{\psi_{i',j,k}} \ket{x^i}\bra{x^{i'}} = \sum_{i,i'} \left(\sum_{j,k} \psi_{i,j,k} \overline{\psi_{i',j,k}} \right) \ket{x^i}\bra{x^{i'}}.
	\end{equation}
	Since
	\begin{equation}
		\id_{X^*} = \sum_{i,i'} \delta_{i,i'} \ket{x^i}\bra{x^{i'}},
	\end{equation}
	the equation $\tr_2 A = \id_{X^*}$ is equivalent to
	\begin{equation}
		\sum_{j,k} \psi_{i,j,k} \overline{\psi_{i',j,k}} = \delta_{i,i'},
	\end{equation}
	for every $i,i'$. The left hand side can be interpreted as an inner product. Define
	\begin{equation}
		\psi_i \coloneqq (\psi_{i,j,k})_{j,k} \in \complexNumbers^{d_Y \rank (A)}.
	\end{equation}
	Then we have that
	\begin{equation}
		\langle \psi_i , \psi_{i'} \rangle = \delta_{i,i'}.
	\end{equation}
	In other words, $(\psi_i)_i$ must be an orthonormal sequence in $\complexNumbers^{d_Y \rank (A)}$. If the size of the sequence is greater than the dimension of the space, it can't be linearly independent. But it is orthonormal, so it is linearly independent. The sequence has length $d_X$, so we must have
	\begin{equation}
		d_X \leq d_Y \rank (A).
	\end{equation}
	Dividing by $d_Y$, we get that
	\begin{equation}
		\frac{d_X}{d_Y} \leq \rank (A).
	\end{equation}
\end{proof}

\subsection{The Kraus representation}

For the Kraus representation, the TP property corresponds to the Kraus operators satisfying an equation, which we prove next.

\begin{proposition}{}{kraus TP condition}
	Let $\varepsilon \in \cp(X,Y)$ be a CP map with Kraus operators $(E_k)_k$. Then $\varepsilon$ is TP (trace preserving) iff
	\begin{equation}
		\sum_k E_k^\dag E_k = \id_X.
	\end{equation}
\end{proposition}
\begin{proof}
	Recall that an element $\varepsilon \in \cp(X,Y)$ is a $\complexNumbers$-linear map $\varepsilon \colon \Lin_\complexNumbers(X) \to \Lin_\complexNumbers(Y)$ which is completely positive. Next we prove each implication.
	
	$(\impliedby)$ Let $A \in \Lin_\complexNumbers(X)$. From the Kraus representation, we know that
	\begin{equation}
		\varepsilon (A) = \sum_k E_k A E_k^\dag.
	\end{equation} 
	Using the linearity and the cyclic property of the trace, we get that
	\begin{align*}
		\tr \, \varepsilon (A) &= \tr\left(\sum_k E_k A E_k^\dag \right)\\
		&= \tr\left(\left(\sum_k E_k^\dag E_k \right)A\right).
	\end{align*}
	Since $\sum_k E_k^\dag E_k = \id_X$, it follows that
	\begin{equation}
		\varepsilon(A) = \tr (A),
	\end{equation}
	for any $A$, that is, $\varepsilon$ is TP.
	
	$(\implies)$ Now suppose $\varepsilon$ is TP and let $(\ket{x_i})_i$ be an orthonormal basis for $X$. The previous calculations show that
	\begin{equation}
		\tr(\varepsilon(A)) =  \tr\left(\left(\sum_k E_k^\dag E_k \right)A\right).
	\end{equation}
	Taking $A = \ket{x_i}\bra{x_j}$, we get that
	\begin{equation}
		\tr(\varepsilon(A)) = \tr\left(\left(\sum_k E_k^\dag E_k \right)\ket{x_i} \bra{x_j}\right) = \bra{x_j} \left(\sum_k E_k^\dag E_k \right)\ket{x_i}.
	\end{equation}
	On the other hand,
	\begin{equation}
		\tr(A) = \tr(\ket{x_i}\bra{x_j}) = \langle x_j | x_i \rangle = \delta_{i,j}.
	\end{equation}
	Since $\varepsilon$ is TP, we have that $\tr(\varepsilon(A)) = \tr(A)$, so
	\begin{equation}
		\bra{x_j} \left(\sum_k E_k^\dag E_k \right)\ket{x_i} = \delta_{i,j}.
	\end{equation}
	This implies that
	\begin{equation}
		\sum_k E_k^\dag E_k = \id_X.
	\end{equation}
\end{proof}

\subsection{The Stinespring representation}

For completeness, we'll present the Stinespring representation. It will not be used in this thesis. It says that any CPTP map can be obtained from some isometry. The article \textit{Universal Properties in Quantum Theory} \cite{Huot_2019} also shows that the category of CPTP maps satisfies a universal property related to the category of isometries. The Stinespring representation also exists for CP maps, but instead of an isometry we have some linear transformation. More information about the Stinespring representation can be found in \cite{watrous_2018}.

\begin{proposition}{Stinespring representation}{}
	Let $X$ and $Y$ be finite dimensional complex Hilbert spaces, and let $\varepsilon \in \cptp(X,Y)$. Then there exists an isometry $V \colon X \to Y \otimes \complexNumbers^{\CR(\varepsilon)}$ such that
	\begin{equation}
		\label{equation: stinespring representation}
		\varepsilon(A) = \tr_{\complexNumbers^{\CR(\varepsilon)}} (V A V^\dag),
	\end{equation}
	for all $A \in \Lin_\complexNumbers(X)$.
\end{proposition}
\begin{proof}
	By proposition \ref{proposition:kraus TP condition}, the CPTP map has some Kraus representation with Kraus operators $(E_{k})_{k=1,\dots, \CR(\varepsilon)}$, where $E_k$ has signature $E_k \colon X \to Y$. It satisfies that
	\begin{equation}
		\label{equation: stinespring kraus}
		\varepsilon(A) = \sum_k E_k A E_k^\dag,
	\end{equation}
	for all $A \in \Lin_\complexNumbers(X)$, and
	\begin{equation}
		\label{equation: stinespring tp}
		\sum_k E_k^\dag E_k = \id_X.
	\end{equation}
	We'll construct the isometry $V$ from the Kraus operators. Let $(\ket{k})_{k=1, \dots, \CR(\varepsilon)}$ be an orthonormal basis for $\complexNumbers^{\CR(\varepsilon)}$. Define $V$ as
	\begin{equation}
		\label{equation: V stinespring}
		V \coloneqq \sum_k E_k \otimes \ket{k},
	\end{equation}
	which is to be interpreted as the function given by
	\begin{equation}
		V(x) \coloneqq \sum_k E_k(x) \otimes \ket{k},
	\end{equation}
	for any $x \in X$. Its dual is given by
	\begin{equation}
		\label{equation: Vdag stinespring}
		V^\dag = \sum_k E_k^\dag \otimes \bra{k}.
	\end{equation}
	Its codomain is to be interpreted as $X$ instead of $X \otimes \complexNumbers$. We are using repeatedly the natural isomorphism $X \otimes \complexNumbers \cong X$ from proposition \ref{proposition:right unitality}.
	
	Before proving equation \ref{equation: stinespring representation}, let's show an intermediary result. For any $m \in \{1, \dots, \CR(\varepsilon)\}$, we have that
	\begin{align*}
		(\id_Y \otimes \bra{m}) V &\stackrel{(i)}{=} (\id_Y \otimes \bra{m}) \left(\sum_k E_k \otimes \ket{k} \right) \\
		&\stackrel{(ii)}{=} \sum_k \id_Y E_k \otimes \langle m \mid k \rangle \\
		&\stackrel{(iii)}{=} \sum_k \delta_{k,m} E_k \\
		&\stackrel{(iv)}{=} E_m.
	\end{align*}
	In $(i)$ was used equation \ref{equation: V stinespring}; in $(ii)$ the summation over $k$ was moved to the left; in $(iii)$ was used that $(\ket{k})_k$ is an orthonormal basis for $\complexNumbers^{\CR(\varepsilon)}$; in $(iv)$ the delta was used to eliminate the summation. This proves that
	\begin{equation}
		\label{equation: intermediary stinespring}
		(\id_Y \otimes \bra{m}) V = E_m .
	\end{equation}
	
	Next we prove equation \ref{equation: stinespring representation}. For any $A \in \Lin_\complexNumbers (X)$, we have that
	\begin{align*}
		\tr_{\complexNumbers^{\CR(\varepsilon)}} (V A V^\dag) &\stackrel{(i)}{=} \sum_k (\id_Y \otimes \bra{k}) VAV^\dag (\id_Y \otimes \ket{k}) \\
		&\stackrel{(ii)}{=} \sum_k E_k A E_k^\dag \\
		&\stackrel{(iii)}{=} \varepsilon(A).
	\end{align*}
	In $(i)$ the partial trace was expressed with respect to the basis $(\ket{k})_k$; in $(ii)$ was used equation \ref{equation: intermediary stinespring}; in $(iii)$ was used equation \ref{equation: stinespring kraus}. This proves equation \ref{equation: stinespring representation}.
	
	Lastly, we prove that $V$ is an isometry, that is, $V^\dag V = \id_X$. We have that
	\begin{align*}
		V^\dag V &\stackrel{(i)}{=} \left( \sum_k E_k^\dag \otimes \bra{k} \right) \left( \sum_l E_l \otimes \ket{l} \right) \\
		&\stackrel{(ii)}{=} \sum_{k,l} E_k^\dag E_l \ \langle k \mid l \rangle \\
		&\stackrel{(iii)}{=} \sum_{k,l} E_k^\dag E_l \ \delta_{k,l} \\
		&\stackrel{(iv)}{=} \sum_k E_k^\dag E_k \\
		&\stackrel{(v)}{=} \id_X.
	\end{align*}
	In $(i)$ were used equations \ref{equation: V stinespring} and \ref{equation: Vdag stinespring}; in $(ii)$ the summations were moved to the left; in $(iii)$ was used that $(\ket{k})_k$ is an orthonormal basis for $\complexNumbers^{\CR(\varepsilon)}$; in $(iv)$ the delta was used to eliminate the summation over $l$; in $(v)$ was used equation \ref{equation: stinespring tp}.
\end{proof}

\subsection{Duality}

For CP maps we saw that the operation of taking the adjoint is a duality between CP maps. In the case of CPTP maps, this duality restricts to a correspondence betwenn CPTP maps and UCP maps. We'll postpone this proof for the section about UCP maps.

\subsection{Some results about partial traces}

There are two important results about partial traces that we'll show in this section. One is the already known fact that partial traces are CPTP maps (section 8.3.1 of \cite{nielsen_chuang_2010}). The other one, maybe already known, is that we can compose a partial trace with another linear map to obtain an orthogonal projection. This orthogonal projection is sometimes easier to work with.

Next we prove that the partial traces are CPTP maps, and compute the operators they correspond to by the \CJ isomorphism. Such operators have a nice diagrammatic expression using string diagrams, more details can be found in \cite{Coecke_Kissinger_2017, MonoidalCatsAndTFT}.
\begin{proposition}{}{partial trace as cptp map}
	Let $X$ and $Y$ be finite dimensional complex Hilbert spaces. Then the partial traces
	\begin{align*}
		\tr_1 \colon \Lin_\complexNumbers(X \otimes Y) &\to \Lin_\complexNumbers(Y),\\
		\tr_2 \colon \Lin_\complexNumbers(X \otimes Y) &\to \Lin_\complexNumbers(X),
	\end{align*}
	are CPTP maps $\tr_1 \in \cptp(X\otimes Y, Y)$ and $\tr_2 \in \cptp(X\otimes Y, X)$. By the \CJ isomorphism, these correspond to operators $J(\tr_1) \in \psdii((X\otimes Y)^*, Y)$ and $J(\tr_2) \in \psdii((X\otimes Y)^*, X)$. Using the natural isomorphism $(X\otimes Y)^* \cong X^* \otimes Y^*$, we can interpret these operators as a $J(\tr_1) \in \psdii(X^*\otimes Y^*, Y)$ and a $J(\tr_2) \in \psdii(X^*\otimes Y^*, X)$. Let $(x_i)_i$ and $(y_j)_j$ be orthonormal basis for $X$ and $Y$, and let $(x^i)_i$ and $(y^j)_j$ be the corresponding dual basis for $X^*$ and $Y^*$. Similarly to proposition \ref{proposition:CJ expression}, define the vectors
	\begin{equation}
		\ket{\Omega_X} \coloneqq \sum_i \ket{x^i \otimes x_i} \in X^* \otimes X,
	\end{equation} 
	\begin{equation}
		\ket{\Omega_Y} \coloneqq \sum_j \ket{y^j \otimes y_j} \in Y^* \otimes Y.
	\end{equation}
	Then we have that
	\begin{equation}
		J(\tr_1) = \id_{X^*} \otimes \ket{\Omega_Y}\bra{\Omega_Y},
	\end{equation}
	\begin{equation}
		J(\tr_2) = (\id_{X^*} \otimes \tau_{ X, Y^*})(\ket{\Omega_X}\bra{\Omega_X} \otimes \id_{Y^*})(\id_{X^*} \otimes \tau_{Y^*, X}),
	\end{equation}
	where $\tau_{V,W}$ is the natural isomorphism $\tau_{V,W} \colon V \otimes W \to W \otimes V$.
\end{proposition}
\begin{proof}
	The proof for both cases are very similar, so we'll just prove for $\tr_1$ and explain what must be done for $\tr_2$. Notice that $(x^i \otimes y^j)_{i,j}$ is the basis dual to $(x_i \otimes y_j)_{i,j}$. Notice also that the tensor product in $x^i \otimes y^j$ is in the sense of linear maps, which would give a linear map $x^i \otimes y^j \colon X \otimes Y \to \complexNumbers \otimes \complexNumbers$. By the natural isomorphism
	\begin{align*}
		\mu \colon \complexNumbers \otimes \complexNumbers &\longrightarrow \complexNumbers\\
		\lambda_1 \otimes \lambda_2 &\longmapsto \lambda_1 \lambda_2, 
	\end{align*}
	we can interpret $x^i \otimes y^j$ as an element of $(X \otimes Y)^*$. We'll implicitly be using the isomorphism $(X \otimes Y)^* \cong X^* \otimes Y^*$, which amounts to reinterpreting the tensor product $x^i \otimes y^j$ as a tensor product of elements or of linear maps. Then, by proposition \ref{proposition:CJ expression}, we have that
	\begin{align*}
		J(\tr_1) &= \sum_{i,i',j,j'} \ket{x^i \otimes y^j} \bra{x^{i'} \otimes y^{j'}} \otimes \tr_1 (\ket{x_i \otimes y_j} \otimes \bra{x_{i'} \otimes y_{j'}})\\
		&= \sum_{i,i',j,j'} \delta_{i,i'} \ket{x^i \otimes y^j} \bra{x^{i'} \otimes y^{j'}} \otimes \ket{y_j} \bra{y_{j'}}\\
		&= \sum_i \ket{x^i}\bra{x^i} \otimes \left( \sum_j \ket{y^j \otimes y_j} \right) \left( \sum_{j'} \bra{y^{j'} \otimes y_{j'}} \right) \\
		&= \id_{X^*} \otimes \ket{\Omega_Y} \bra{\Omega_Y}.
	\end{align*}
	
	By proposition \ref{proposition:CJ isomorphism for cptp}, $\tr_1$ is a CPTP map iff $J(\tr_1)$ is positive semidefinite and
	\begin{equation}
		\label{equation: trY J(tr1)}
		\tr_Y J(\tr_1) = \id_{X^* \otimes Y^*}.
	\end{equation}
	Since $\id_{X^*}$ and $\ket{\Omega_Y}\bra{\Omega_Y}$ are positive semidefinite, and since a tensor product of positive semidefinite operators is also positive semidefinite, it follows that $J(\tr_1)$ is positive semidefinite. If we were using string diagrams, equation \ref{equation: trY J(tr1)} would amount to a simple application of the yanking equations. Since we're not using string diagrams, we prove it as follows:
	\begin{align*}
		\tr_Y (J(\tr_1)) &= \tr_Y (\id_{X^*} \otimes \ket{\Omega_Y}\bra{\Omega_Y})\\
		&= \id_{X^*} \otimes \tr_Y( \ket{\Omega_Y}\bra{\Omega_Y})\\
		&= \id_{X^*} \otimes \sum_{j,j'} \tr_Y (\ket{y^j \otimes y_j} \bra{y^{j'} \otimes y_{j'}})\\
		&= \id_{X^*} \otimes \sum_{j,j'} \tr_Y ( \ket{y^j}\bra{y^{j'}} \otimes \ket{y_j}\bra{y_{j'}} )\\
		&= \id_{X^*} \otimes \sum_{j,j'}  \ket{y^j}\bra{y^{j'}} \ \tr(\ket{y_j}\bra{y_{j'}})\\
		&= \id_{X^*} \otimes \sum_{j,j'}  \ket{y^j}\bra{y^{j'}} \ \delta_{j,j'} \\
		&= \id_{X^*} \otimes \sum_j \ket{y^j}\bra{y^j} \\
		&= \id_{X^*} \otimes \id_{Y^*} \\
		&= \id_{X^* \otimes Y^*}.
	\end{align*}
	This concludes that $\tr_1$ is a CPTP map.
	
	Similarly, $\tr_2$ is also a CPTP map. The proof for $\tr_2$ is very similar, but this time we only make the vector $\Omega_X$ appear after we prove that $\tr_2$ is CPTP. We can write $J(\tr_2)$ with respect to the basis, obtaining
	\begin{equation}
		J(\tr_2) = \sum_{i,i',j} \ket{x^i \otimes y^j \otimes x_i} \bra{x^{i'} \otimes y^{j} \otimes x_{i'}}.
	\end{equation}
	Each term $\ket{x^i \otimes y^j \otimes x_i} \bra{x^{i'} \otimes y^{j} \otimes x_{i'}}$ is a positive semidefinite operator, and a sum of positive semidefinite operators is also a positive semidefinite operator. Therefore, $J(\tr_2)$ is a positive semidefinite operator. The rest of the proof is very similar to the one done above for $\tr_1$. A simple calculation shows that $\tr_X (J(\tr_2)) = \id_{X^* \otimes Y^*}$. From this, it follows that $\tr_2$ is a CPTP map. Then we use the braiding $\tau_{X,Y^*}$ just to make the vector $\Omega_X$ appear in the expression for $J(\tr_2)$.
\end{proof}

Now we show how the partial traces correspond to orthogonal projections. It may be the case that these projections are already known, but we didn't find a reference for that yet. They are defined as follows.

\begin{definition}{}{partial traces as projections}
	Let $X$ and $Y$ be finite dimensional complex Hilbert spaces. We define linear maps $p_1,p_2 \colon \Lin_\complexNumbers(X \otimes Y) \to \Lin_\complexNumbers(X\otimes Y)$ as follows:
	\begin{equation}
		p_1 (A) = \tr_2 (A) \otimes \frac{\id_Y}{d_Y},
	\end{equation}
	\begin{equation}
		p_2 (A) =  \frac{\id_X}{d_X} \otimes \tr_1 (A),
	\end{equation}
	for $A \in \Lin_\complexNumbers(X \otimes Y)$.
\end{definition}
Notice that $p_1$ is the composition of $\tr_2$ with the $\complexNumbers$-linear map
\begin{align*}
	\Lin_\complexNumbers(Y) &\longrightarrow \Lin_\complexNumbers(X\otimes Y)\\
	A &\longmapsto A \otimes \frac{\id_Y}{d_Y}.
\end{align*}
If we replaced $d_Y$ by $\sqrt{d_Y}$ in this linear map, it would become an isometry, but then $p_1$ wouldn't be an orthogonal projection. Similarly, $p_2$ is the composition of $\tr_1$ with the $\complexNumbers$-linear map
\begin{align*}
	\Lin_\complexNumbers(X) &\longrightarrow \Lin_\complexNumbers(X\otimes Y)\\
	A &\longmapsto \frac{\id_X}{d_X} \otimes A.
\end{align*}

\begin{proposition}{}{partial traces as projections}
	The linear maps $p_1, p_2 \colon \Lin_\complexNumbers(X \otimes Y) \to \Lin_\complexNumbers(X\otimes Y)$ defined above are orthogonal projections, with respect to the Hilbert-Schmidt inner product. Their kernel and image are
	\begin{equation}
		\ker p_1 = \ker \tr_2,
	\end{equation}
	\begin{equation}
		\image p_1 = \Lin_\complexNumbers(X) \otimes \id_Y,
	\end{equation}
	\begin{equation}
		\ker p_2 = \ker \tr_1,
	\end{equation}
	\begin{equation}
		\image p_2 = \id_X \otimes \Lin_\complexNumbers(Y).
	\end{equation}
	By $\Lin_\complexNumbers(X) \otimes \id_Y$ we mean the set
	\begin{equation}
		\Lin_\complexNumbers(X) \otimes \id_Y = \{A \otimes \id_Y \mid A \in \Lin_\complexNumbers(X) \},
	\end{equation}
	and similarly for $\id_X \otimes \Lin_\complexNumbers(Y)$.
\end{proposition}
\begin{proof}
	We'll do the proof only for $p_1$, since the proof for $p_2$ is similar. The linear map $p_1$ is an orthogonal projection iff $p_1^\dag = p_1$ and $p_1^2 = p_1$. We first show the second equation. For any $A \in \Lin_\complexNumbers(X\otimes Y)$, we have that
	\begin{align*}
		p_1^2(A) &= \tr_2 (p_1(A)) \otimes \frac{\id_Y}{d_Y}\\
		&= \tr_2 \left( \tr_2 A \otimes \frac{\id_Y}{d_Y} \right) \otimes \frac{\id_Y}{d_Y}\\
		&= \tr_2 A \ \tr\left(\frac{\id_Y}{d_Y} \right) \otimes \frac{\id_Y}{d_Y}\\
		&= \tr_2 A \otimes \frac{\id_Y}{d_Y}\\
		&= p_1(A).
	\end{align*}
	Now let's show that $p_1^\dag = p_1$. By sesquilinearity of the inner product and uniqueness of the adjoint, it is enough to show that, for any $A, C \in \Lin_\complexNumbers(X)$ and $B,D \in \Lin_\complexNumbers(Y)$, the following equation holds
	\begin{equation}
		\langle p_1^\dag( A\otimes B), C\otimes D\rangle = \langle p_1( A\otimes B), C\otimes D\rangle.
	\end{equation}
	Applying the definition of the adjoint, we have
	\begin{align*}
		\langle p_1^\dag( A\otimes B), C\otimes D\rangle &= \langle A\otimes B, p_1(C\otimes D) \rangle\\ &= \langle A\otimes B, \tr_2(C\otimes D) \otimes \id_Y/d_Y \rangle\\
		&= \langle A\otimes B, C \ \tr D \otimes \id_Y / d_Y \rangle \\
		&= \langle A\otimes B, C \otimes \id_Y \rangle \ \tr D / d_Y\\
		&= \tr( (A^\dag \otimes B^\dag) (C \otimes \id_Y)) \ \tr D / d_Y\\
		&= \tr( (A^\dag C) \otimes (B^\dag \id_Y)) \ \tr D / d_Y\\
		&= \tr(A^\dag C) \ \tr (B^\dag) \ \tr D / d_Y.
	\end{align*}
	Since the trace is a CPTP map, as proved in \ref{proposition:partial trace as cptp map}, by proposition \ref{proposition:cp(A dag) = cp(A) dag} we know that $\tr(B^\dag) = \overline{\tr(B)}$. So we get
	\begin{equation}
		\langle p_1^\dag( A\otimes B), C\otimes D\rangle =  \tr(A^\dag C) \ \overline{\tr (B)} \ \tr D / d_Y.
	\end{equation}
	
	On the other hand, we have
	\begin{align*}
		\langle p_1( A\otimes B), C\otimes D\rangle &= \langle \tr_2(A\otimes B) \otimes \id_Y/d_Y , C\otimes D \rangle\\
		&= \langle A \ \tr(B) \otimes \id_Y/d_Y , C\otimes D \rangle\\
		&= \langle A \otimes \id_Y , C\otimes D \rangle\ \overline{\tr(B)} / d_Y\\
		&= \tr ( (A^\dag \otimes \id_Y^\dag) (C\otimes D)) \ \overline{\tr(B)} / d_Y\\
		&= \tr ( (A^\dag C) \otimes (\id_Y^\dag D)) \ \overline{\tr(B)} / d_Y\\
		&= \tr (A^\dag C) \ \tr D \ \overline{\tr(B)} / d_Y.
	\end{align*}
	This shows that
	\begin{equation}
		\langle p_1^\dag( A\otimes B), C\otimes D\rangle = \langle p_1( A\otimes B), C\otimes D\rangle,
	\end{equation}
	from which we conclude that $p_1^\dag = p_1$. This also concludes that $p_1$ is an orthogonal projection.
	
	Now let's compute $\ker p_1$ and $\image p_1$. $p_1$ is the composition of $\tr_2$ with the linear map
	\begin{align*}
		\eta \colon \Lin_\complexNumbers(X) &\longrightarrow \Lin_\complexNumbers(X\otimes Y)\\
		A &\longmapsto A \otimes \frac{\id_Y}{d_Y}.
	\end{align*}
	We'll show that $\eta$ is a multiple of an isometry. Let $A,B \in \Lin_\complexNumbers(X)$, then
	\begin{align*}
		\langle \eta(A), \eta(B) \rangle &= \left\langle A \otimes \frac{\id_Y}{d_Y}, B \otimes \frac{\id_Y}{d_Y} \right\rangle \\
		&= \tr \left( \left( A^\dag \otimes \frac{\id_Y^\dag}{d_Y} \right) \left( B \otimes \frac{\id_Y}{d_Y} \right) \right) \\
		&= \tr \left( A^\dag B \otimes \frac{\id_Y}{d_Y^2} \right) \\
		&= \frac{\tr(A^\dag B) \ \tr(\id_Y) }{d_Y^2}\\
		&= \frac{\tr(A^\dag B) \ d_Y }{d_Y^2}\\
		&= \frac{\langle A , B \rangle}{d_Y}.
	\end{align*}
	Therefore, $\sqrt{d_Y} \, \eta$ is an isometry. Every isometry is injective. Since $p_1$ is the composition of $\tr_2$ with an injective linear map, this implies that
	\begin{equation}
		\ker p_1 = \ker \tr_2.
	\end{equation}
	Now let's compute $\image p_1$. By definition of $p_1$, it immediately follows that $\image p_1 \subseteq \Lin_\complexNumbers(X) \otimes \id_Y$. Let's prove the other inclusion. Any element $B \in \Lin_\complexNumbers(X) \otimes \id_Y$ can be written as
	\begin{equation}
		B = A \otimes \frac{\id_Y}{d_Y},
	\end{equation}
	for some $A \in \Lin_\complexNumbers(X)$. Of course, we have
	\begin{equation}
		B = p_1(A),
	\end{equation}
	so
	\begin{equation}
		\image p_1 = \Lin_\complexNumbers(X) \otimes \id_Y.
	\end{equation}
\end{proof}

\begin{corollary}{}{}
	Let $X$ and $Y$ be finite dimensional complex Hilbert spaces. Then the orthogonal complements of $\ker \tr_1$ and $\ker \tr_2$ in $\Lin_\complexNumbers(X\otimes Y)$ are
	\begin{equation}
		(\ker \tr_1)^\perp = \id_X \otimes \Lin_\complexNumbers(Y),
	\end{equation}
	\begin{equation}
		(\ker \tr_2)^\perp = \Lin_\complexNumbers(X) \otimes \id_Y.
	\end{equation}
\end{corollary}
\begin{proof}
	For any orthogonal projection $p$, we have that $\image p = (\ker p)^\perp$.
\end{proof}

We see that the orthogonal complement of $\ker \tr_i$ has a simpler description than the kernel. For this reason, we sometimes pass to the orthogonal complement in some proofs.

The projections $p_1$ and $p_2$ are not orthogonal to each other, because $\image p_1 \cap \image p_2 = \complexNumbers \ \id_{X\otimes Y}$. But we can break both projections into smaller ones, making explicit the smaller projection they have in common. This will be useful in some proofs later.

\begin{proposition}{}{p'1, p'2, p1p2}
	The orthogonal projections $p_1$ and $p_2$ as defined in \ref{definition:partial traces as projections} commute with each other, that is,
	\begin{equation}
		[p_1,p_2] = p_1 p_2 - p_2 p_1 = 0.
	\end{equation}
	Moreover, we have
	\begin{equation}
		p_1 p_2 (A) = \tr A \ \frac{\id_{X \otimes Y}}{d_X d_Y},
	\end{equation}
	for any $A \in \Lin_\complexNumbers(X \otimes Y)$. The composition $p_1 p_2$ is the orthogonal projection onto
	\begin{equation}
		\image(p_1 p_2) = \complexNumbers \ \id_{X \otimes Y} = \{\lambda \, \id_{X\otimes Y} \mid \lambda \in \complexNumbers\}.
	\end{equation}
	We also have that the linear maps
	\begin{equation}
		p'_1 \coloneqq p_1-p_1 p_2,
	\end{equation}
	\begin{equation}
		p'_2 \coloneqq p_2-p_1 p_2,
	\end{equation}
	are orthogonal projections, and all projections $p_1 p_2$, $p'_1$ and $p'_2$ commute with each other and are orthogonal to each other.
\end{proposition}
\begin{proof}
	For any $A \in \Lin_\complexNumbers(X \otimes Y)$, we have
	\begin{align*}
		p_1 p_2 (A) &= p_1 \left( \frac{\id_X}{d_X} \otimes \tr_1 A \right)\\
		&= \tr_2 \left( \frac{\id_X}{d_X} \otimes \tr_1 A \right) \otimes \frac{\id_Y}{d_Y}\\
		&= \left( \frac{\id_X}{d_X} \tr A \right) \otimes \frac{\id_Y}{d_Y}\\
		&= \tr A \ \frac{\id_{X \otimes Y}}{d_X d_Y}.
	\end{align*}
	Similarly, we have
	\begin{align*}
		p_2 p_1 (A) &= p_2 \left(\tr_2 A \otimes \frac{\id_Y}{d_Y} \right)\\
		&= \frac{\id_X}{d_X} \otimes \tr_1 \left(\tr_2 A \otimes \frac{\id_Y}{d_Y} \right)\\
		&= \frac{\id_X}{d_X} \otimes \left(\tr A \ \frac{\id_Y}{d_Y} \right)\\
		&= \tr A \ \frac{\id_{X \otimes Y}}{d_X d_Y}.
	\end{align*}
	In particular, we have that $[p_1, p_2] = 0$. Since they commute, we can easily show that $p_1 p_2$ is an orthogonal projection. We have that
	\begin{equation}
		(p_1 p_2)^2 = p_1 p_2 p_1 p_2 = p_1^2 p_2^2 = p_1 p_2.
	\end{equation}
	Also
	\begin{equation}
		(p_1 p_2)^\dag = p_2^\dag p_1^\dag = p_2 p_1 = p_1 p_2.
	\end{equation}
	This proves that $p_1 p_2$ is an orthogonal projection.
	
	Now let's compute the image of $p_1 p_2$. Of course, we have that $\image (p_1 p_2) \subseteq \complexNumbers \ \id_{X \otimes Y}$. Taking
	\begin{equation}
		A = \frac{\lambda \, \id_{X \otimes Y}}{d_X d_Y},
	\end{equation}
	for an arbitrary $\lambda \in \complexNumbers$, we have that
	\begin{equation}
		p_1 p_2 (A) = \lambda \, \id_{X \otimes Y}.
	\end{equation}
	Therefore,
	\begin{equation}
		\image(p_1 p_2) = \complexNumbers \, \id_{X \otimes Y}.
	\end{equation}
	
	Now let's prove that $p'_1$ and $p'_2$ are orthogonal projections. For $p'_1$ we have
	\begin{align*}
		{p'_1}^2 &= (p_1-p_1 p_2)^2 \\
		&= p_1^2 -2 p_1^2 p_2 +p_1^2 p_2^2 \\
		&= p_1 -2 p_1 p_2 +p_1 p_2 \\
		&= p_1 - p_1 p_2 \\
		&= p'_1.
	\end{align*}
	Also,
	\begin{align*}
		{p'_1}^\dag &= (p_1-p_1 p_2)^\dag \\
		&= p_1^\dag -p_1^\dag p_2^\dag \\
		&= p_1 -p_1 p_2 \\
		&= p'_1.
	\end{align*}
	This shows that $p'_1$ is an orthogonal projection. Similarly, $p'_2$ is also an orthogonal projection. Since $[p_1, p_2]=0$, it follows that the projections $p'_1$, $p'_2$ and $p_1 p_2$ commute with each other. Finally, let's show that they are also pairwise orthogonal. We can write
	\begin{equation}
		p'_1 = p_1(\id_{X\otimes Y}-p_2),
	\end{equation}
	\begin{equation}
		p'_2 = p_2(\id_{X\otimes Y}-p_1).
	\end{equation}
	Then, we have
	\begin{equation}
		\label{equation: p'1p'2}
		p'_1 p'_2 = p_1(\id_{X\otimes Y}-p_2) p_2(\id_{X\otimes Y}-p_1) = 0,
	\end{equation}
	\begin{equation}
		\label{equation: p'1p1p2}
		p'_1 (p_1 p_2) = p_1(\id_{X\otimes Y}-p_2) p_1 p_2 = p_1(\id_{X\otimes Y}-p_2) p_2 p_1 =0,
	\end{equation}
	\begin{equation}
		\label{equation: p'2p1p2}
		p'_2 (p_1 p_2) = p_2(\id_{X\otimes Y}-p_1) p_1 p_2 = 0.
	\end{equation}
	In equations \ref{equation: p'1p'2} and \ref{equation: p'1p1p2} was used that $(\id_{X\otimes Y}-p_2) p_2 = 0$, and in equation \ref{equation: p'2p1p2} was used that $(\id_{X\otimes Y}-p_1) p_1 = 0$. The equations \ref{equation: p'1p'2}, \ref{equation: p'1p1p2} and \ref{equation: p'2p1p2} prove that $p'_1$, $p'_2$ and $p_1 p_2$ are pairwise orthogonal.
\end{proof}

\subsection{Hom-sets as compact convex sets}

The hom-set $\cptp(X,Y)$ is compact convex. By corollary \ref{corollary:compact convex set is homeomorphic to a closed ball}, it is homeomorphic to a closed ball of some dimension at most $\dim_{\realNumbers} \Lin_\complexNumbers(X^* \otimes Y) = 2 d_X^2 d_Y^2$. In this section we show this fact and compute the dimension. By the \CJ isomorphism, we can prove these results for $\psdii(X,Y)$ and then transfer them to $\cptp(X,Y)$.

\begin{proposition}{}{cptp is compact convex}
	For any finite dimensional complex Hilbert spaces $X$ and $Y$, the sets $\cptp(X,Y)$ and $\psdii(X,Y)$ are compact convex.
\end{proposition}
\begin{proof}
	Because of the \CJ isomorphism between $\cptp(X,Y)$ and $\psdii(X,Y)$, it is enough to prove the result for one of them. We prove it for $\psdii(X,Y)$.
	
	First we prove that it is convex. Let $A_1,A_2 \in \psdii(X,Y)$, and let $t_1,t_2 \in [0,1]$ such that $t_1+t_2=1$. Then we have the convex combination $A = t_1 A_1 + t_2 A_2$. Since $A_i$ is PSD and $t_i \geq 0$, then $t_i A_i$ is also PSD. A sum of PSD operators is PSD, which implies that $A$ is PSD. Also, its partial trace over $Y$ is
	\begin{align*}
		\tr_2 A &=  \tr_2 \left(\sum_i t_i A_i \right)\\
		&= \sum_i t_i \, \tr_2 A_i \\
		&= \sum_i t_i \, \id_{X^*}\\
		&= \id_{X^*}.
	\end{align*}
	We proved that $A$ is PSD and $\tr_2 A = \id_{X^*}$, which shows that $A \in \psdii(X,Y)$. Therefore, $\psdii(X,Y)$ is convex.
	
	Now we prove that $\psdii(X,Y)$ is closed. Let $(A_n)_{n \geq 1}$ be a sequence in $\psdii(X,Y)$ which converges to some $A \in \Lin_\complexNumbers(X^* \otimes Y)$. Let's show that $A$ is PSD. Let $v \in X^* \otimes Y$, then
	\begin{align*}
		\bra{v} A \ket{v} &= \bra{v} \lim_n A_n \ket{v} \\
		&= \lim_n \bra{v} A_n \ket{v}.
	\end{align*}
	Since $A_n$ is PSD, then $\bra{v} A_n \ket{v} \geq 0$. Therefore, its limit is also non negative, so
	\begin{equation}
		\bra{v} A \ket{v} \geq 0.
	\end{equation}
	This shows that $A$ is PSD. Now let's show that $\tr_2 A = \id_{X^*}$. We have
	\begin{align*}
		\tr_2 A &= \tr_2 (\lim_n A_n)\\
		&= \lim_n \tr_2 A_n \\
		&= \lim_n \id_{X^*}\\
		&= \id_{X^*}.
	\end{align*}
	We showed that $A$ is PSD and $\tr_2 A = \id_{X^*}$, so $A \in \psdii(X,Y)$. This proves that $\psdii(X,Y)$ is closed.
	
	Next we prove that $\psdii(X,Y)$ is bounded. Let $A \in \psdii(X,Y)$, then we know that $\tr_2 A = \id_{X^*}$. This implies that the total trace is
	\begin{equation}
		\tr A = \tr(\tr_2 A) = \tr(\id_{X^*}) = d_X.
	\end{equation}
	Since $A$ is PSD, it can be diagonalized as
	\begin{equation}
		A = \sum_k \lambda_k \ket{v_k}\bra{v_k},
	\end{equation}
	where $\lambda_k \geq 0$ and $(v_k)_k$ are orthonormal vectors in $X^* \otimes Y$. The trace is the sum of the eigenvalues, so
	\begin{equation}
		\sum_k \lambda_k = d_X.
	\end{equation}
	On the other hand, the square of the Frobenius norm of $A$ is
	\begin{equation}
		||A||^2 = \tr(A^\dag A) = \tr(A^2).
	\end{equation}
	Using the diagonalization of $A$ we get a diagonalization for $A^2$, which is
	\begin{equation}
		A^2 = \sum_k \lambda_k^2 \ket{v_k}\bra{v_k}.
	\end{equation}
	Therefore,
	\begin{equation}
		||A||^2 = \tr(A^2) = \sum_k \lambda_k^2.
	\end{equation}
	We can estimate this sum as follows. We have
	\begin{equation}
		\left( \sum_k \lambda_k \right)^2 = \sum_{k,l} \lambda_k \lambda_l \geq \sum_k \lambda_k^2.
	\end{equation}
	This implies that
	\begin{equation}
		||A||^2 = \sum_k \lambda_k^2 \leq \left( \sum_k \lambda_k \right)^2 =d_X^2,
	\end{equation}
	so
	\begin{equation}
		||A|| \leq d_X.
	\end{equation}
	This shows that $\psdii(X,Y) \subseteq \overline{B(0,d_X)}$, so $\psdii(X,Y)$ is bounded.
	
	In $\realNumbers^n$, a compact set is one that is closed and bounded. Since $\Lin_\complexNumbers(X^* \otimes Y)$ is isomorphic to $\realNumbers^{2 d_X d_Y}$, as $\realNumbers$-vector spaces, the compact sets of $\Lin_\complexNumbers(X^* \otimes Y)$ are also the ones which are closed and bounded. We showed that $\psdii(X,$ $Y)$ is closed and bounded, so it is compact.
	
	This concludes that $\psdii(X,Y)$ is a compact convex set.
	
\end{proof}

If $X=\{0\}$ or $Y = \{0\}$ the convex set is trivial, it is $\varnothing$ or $\{0\}$. We show this in the next proposition.

\begin{proposition}{}{psdii(X,Y) for X or Y = 0}
	Let $X$ and $Y$ be finite dimensional complex Hilbert spaces. Then
	\begin{equation}
		\psdii(X,\{0\}) = \begin{cases}
			\{0\} & \text{ if } X = \{0\},\\
			\ \varnothing &\text{ if } X \neq \{0\}.
		\end{cases}
	\end{equation}
	Also,
	\begin{equation}
		\psdii(\{0\},Y) = \{0\}.
	\end{equation}
\end{proposition}
\begin{proof}
	An element $A \in \psdii(X,\{0\})$ should satisfy the equation
	\begin{equation}
		\tr_2 A = \id_{X^*}.
	\end{equation}
	Also $X^* \otimes \{0\} = \{0\}$, which implies that $\Lin_\complexNumbers(X^* \otimes \{0\}) = \{0\}$, so $A$ must be 0. This also implies that $\tr_2 A = 0$. We only have $\id_{X^*} = 0$ if $X = \{0\}$. Therefore, $\psdii(X,\{0\}) = \varnothing$ for $X \neq \{0\}$. 
	
	It remains to prove that $\psdii(\{0\},Y) = \{0\}$. Of course, $0$ is a positive semidefinite operator. Also,
	\begin{equation}
		\tr_2 0 = 0 = \id_{X^*}.
	\end{equation}
	Therefore, $\psdii(\{0\},Y) = \{0\}$.
\end{proof}

Now we compute the dimension of the hom-sets for $Y \neq \{0\}$. To do it, we use corollary \ref{corollary:compact convex set is homeomorphic to a closed ball}. To use this corollary, we need at least one element of the compact convex set. We can take as element the operator $\id_{X^* \otimes Y}/d_Y$. We can easily show that it is an element of $\psdii(X,Y)$. Of course, the identity function is a PSD operator. Also
\begin{align*}
	\tr_2 \left(\frac{\id_{X^* \otimes Y}}{d_Y}\right) &= \frac{\tr_2 (\id_{X^*} \otimes \id_Y)}{d_Y}\\
	&= \frac{\id_{X^*} \tr(\id_Y)}{d_Y} \\
	&= \frac{\id_{X^*} d_Y}{d_Y} \\
	&= \id_{X^*}.
\end{align*}
This shows that
\begin{equation}
	\frac{\id_{X^* \otimes Y}}{d_Y} \in \psdii(X,Y).
\end{equation}

\begin{proposition}{}{span of psdii-id/dy}
	For any finite dimensional complex Hilbert spaces $X$ and $Y$, with $d_Y \geq 1$, we have that
	\begin{equation}
		\left\langle \psdii(X,Y)-\frac{\id_{X^* \otimes Y}}{d_Y} \right\rangle_\realNumbers = \hermitian(X^* \otimes Y) \cap \ker \tr_2.
	\end{equation}
\end{proposition}
\begin{proof}
	$(\subseteq)$ Let $A \in \psdii(X,Y)$, then it is a PSD operator. In particular, it is hermitian. Also $-\id_{X^* \otimes Y}/d_Y$ is hermitian, and a sum of hermitian operators is a hermitian operator. This implies that $A-\id_{X^* \otimes Y}/d_Y$ is hermitian. Also, its partial trace over $Y$ is
	\begin{equation}
		\tr_2 (A - \id_{X^* \otimes Y}/d_Y) = \tr_2 A - \tr_2( \id_{X^* \otimes Y} / d_Y ) = \id_{X^*} - \id_{X^*} = 0.
	\end{equation}
	This shows that $A - \id_{X^* \otimes Y}/d_Y \in \hermitian(X^*\otimes Y) \cap \ker \tr_2$.
	
	$(\supseteq)$ We'll show that
	\begin{equation}
		\overline{B(0,1/d_Y)} \cap \hermitian(X^* \otimes Y) \cap \ker \tr_2 \subseteq \psdii(X,Y)-\frac{\id_{X^* \otimes Y}}{d_Y} .
	\end{equation}
	Let $B \in \overline{B(0,1/d_Y)} \cap \hermitian(X^* \otimes Y) \cap \ker \tr_2$, and define
	\begin{equation}
		A \coloneqq B + \frac{\id_{X^* \otimes Y}}{d_Y}.
	\end{equation} 
	We just need to show that $A \in \psdii(X,Y)$. For the partial trace, we have
	\begin{align*}
		\tr_2 A &= \tr_2 B + \tr_2(\id_{X^* \otimes Y}/d_Y)\\
		&= 0+\id_{X^*}\\
		&= \id_{X^*}.
	\end{align*}
	Now let's show that $A$ is PSD. Since $B$ is hermitian, we can diagonalize it as
	\begin{equation}
		B = \sum_k \lambda_k \ket{v_k} \bra{v_k},
	\end{equation}
	where $\lambda_k \in \realNumbers$ and $(v_k)_k$ are orthonormal vectors in $X^* \otimes Y$. Any orthonormal basis diagonalizes the identity function, so we also have
	\begin{equation}
		\frac{\id_{X^* \otimes Y}}{d_Y} = \sum_k \frac{1}{d_Y} \ket{v_k} \bra{v_k}.
	\end{equation}
	Combining these results we get a diagonalization of $A$:
	\begin{equation}
		A = \sum_k \left(\lambda_k+\frac{1}{d_Y} \right) \ket{v_k} \bra{v_k}.
	\end{equation}
	$A$ is PSD iff its eigenvalues are non negative, that is, iff
	\begin{equation}
		\lambda_k + \frac{1}{d_Y} \geq 0.
	\end{equation}
	A sufficient condition is to have
	\begin{equation}
		|\lambda_k| \leq \frac{1}{d_Y}.
	\end{equation}
	Since $B \in \overline{B(0,1/d_Y)}$, the square of its Frobenius norm satisfies
	\begin{equation}
		||B||^2 \leq \frac{1}{d_Y^2}.
	\end{equation}
	Using the definition of the norm and that the trace is the sum of the eigenvalues, we get
	\begin{equation}
		||B||^2 = \tr(B^\dag B) = \tr(B^2) = \sum_k \lambda_k^2.
	\end{equation}
	From this we have that
	\begin{equation}
		\lambda_k^2 \leq \sum_l \lambda_l^2 = ||B||^2 \leq \frac{1}{d_Y^2},
	\end{equation}
	so
	\begin{equation}
		|\lambda_k| \leq \frac{1}{d_Y}.
	\end{equation}
	This shows that $A$ is PSD, which concludes that $A \in \psdii(X,Y)$. This also shows that
	\begin{equation}
		\label{equation: ball inside psdii-id/dY}
		\overline{B(0,1/d_Y)} \cap \hermitian(X^* \otimes Y) \cap \ker \tr_2 \subseteq \psdii(X,Y)-\frac{\id_{X^* \otimes Y}}{d_Y} .
	\end{equation}
	The left side generates the whole vector space $\hermitian(X^* \otimes Y) \cap \ker \tr_2$, so we have the inclusion
	\begin{equation}
		\hermitian(X^* \otimes Y) \cap \ker \tr_2 \subseteq \left\langle \psdii(X,Y)-\frac{\id_{X^* \otimes Y}}{d_Y} \right\rangle.
	\end{equation}
	
\end{proof}

\begin{corollary}{}{balls inside and outside psdii}
	For any finite dimensional complex Hilbert spaces $X$ and $Y$, with $d_Y \geq 1$, we have that
	\begin{multline}
		\frac{\id_{X^* \otimes Y}}{d_Y} + \overline{B\left( 0,\frac{1}{d_Y} \right)} \cap \hermitian(X^* \otimes Y) \cap \ker \tr_2 \subseteq \\ \psdii(X,Y) \subseteq \overline{B(0,d_X)},
	\end{multline}
	where the closed balls are in $\Lin_\complexNumbers(X^* \otimes Y)$.
\end{corollary}
\begin{proof}
	The inclusion $\psdii(X,Y) \subseteq \overline{B(0,d_X)}$ was proved in proposition \ref{proposition:cptp is compact convex}, and the inclusion
	\begin{equation}
		\frac{\id_{X^* \otimes Y}}{d_Y} + \overline{B\left(0,\frac{1}{d_Y} \right)}\cap \hermitian(X^* \otimes Y) \cap \ker \tr_2  \subseteq \psdii(X,Y)
	\end{equation}
	is equivalent to equation \ref{equation: ball inside psdii-id/dY}, which was proved in \ref{proposition:span of psdii-id/dy}.
\end{proof}

\begin{corollary}{}{dimension of psdii}
	For any finite dimensional complex Hilbert spaces $X$ and $Y$, with $d_Y \geq 1$, the compact convex set $\psdii(X,Y)$ is homeomorphic to a closed ball of dimension
	\begin{equation}
		\dim_{\realNumbers} \psdii(X,Y) = d_X^2(d_Y^2-1).
	\end{equation}
	Of course, this is the dimension of $\psdii(X,Y)$ as a topological manifold with boundary.
\end{corollary}
\begin{proof}
	From proposition \ref{proposition:span of psdii-id/dy} and corollary \ref{corollary:compact convex set is homeomorphic to a closed ball}, we know that
	\begin{equation}
		\dim_\realNumbers \psdii(X,Y) = \dim_\realNumbers \hermitian(X^* \otimes Y) \cap \ker \tr_2.
	\end{equation}
	Define the $\realNumbers$-linear map
	\begin{align*}
		T \colon \hermitian(X^* \otimes Y) &\longrightarrow \Lin_\complexNumbers(X^*)\\
		A &\longmapsto \tr_2 A,
	\end{align*}
	that is, $T$ is the restriction of $\tr_2$ to $\hermitian(X^* \otimes Y)$. Then $\ker T = \hermitian(X^* \otimes Y) \cap \ker \tr_2$. From Linear Algebra, we know that
	\begin{equation}
		\dim_\realNumbers \ker T + \dim_{\realNumbers} \image T = \dim_\realNumbers \hermitian(X^* \otimes Y).
	\end{equation}
	It is known that the space of hermitian operators has dimension
	\begin{equation}
		\dim_\realNumbers \hermitian(X^* \otimes Y) = (\dim_\complexNumbers(X^* \otimes Y) )^2= (d_X d_Y)^2 = d_X^2 d_Y^2.
	\end{equation}
	Now let's compute the image of $T$. We proved in proposition \ref{proposition:partial trace as cptp map} that the partial trace is a CPTP map. By proposition \ref{proposition:cp(A dag) = cp(A) dag}, the partial trace sends hermitian operators to hermitian operators. Then, we have that
	\begin{equation}
		\image T \subseteq \hermitian(X^*).
	\end{equation}
	On the other hand, if $H \in \hermitian(X^*)$, take
	\begin{equation}
		A \coloneqq H \otimes \frac{\id_Y}{d_Y}.
	\end{equation}
	Then
	\begin{equation}
		T(A) = \tr_2\left( H \otimes \frac{\id_Y}{d_Y} \right) = H \ \tr\left(\frac{\id_Y}{d_Y} \right) = H \frac{d_Y}{d_Y} = H,
	\end{equation}
	so $H \in \image T$. Therefore,
	\begin{equation}
		\image T = \hermitian(X^*),
	\end{equation}
	and
	\begin{equation}
		\dim_\realNumbers \image T = \dim_\realNumbers \hermitian(X^*) = d_X^2.
	\end{equation}
	Putting everything together, we get
	\begin{equation}
		\dim_\realNumbers \ker T = d_X^2 d_Y^2-d_X^2 = d_X^2(d_Y^2-1).
	\end{equation}
	
\end{proof}

\subsection{Stratification into smooth manifolds}

By the \CJ isomorphism, we can work with PSD operators instead of CP maps. This is more convenient to study the stratification into manifolds. In the CP case, $\psd(X)$ is stratified into the smooth submanifolds $\psd(X)_r$. Similarly, $\psdii(X,Y)$ is stratified into the submanifolds $\psd(X, Y)_r$. Recall that $\psdii(X,Y)_r$ is the set of operators $A \in \psdii(X,Y)$ with $\rank A = r$. The next theorem is proved in \cite{iten_colbeck}. Notice that $\psdii(X,Y)_r$ can be empty, depending on the rank $r$. To cover this case, we consider that $\varnothing$ is a smooth manifold with any dimension.

\begin{theorem}{}{psdiir is smooth manifold}
	$\psdii(X,Y)_r$ is a smooth submanifold of $\Lin_\complexNumbers(X^* \otimes Y)$. Its dimension is
	\begin{equation}
		\label{equation: dim psdiir}
		\dim_\realNumbers \psdii(X,Y)_r	= 2 d_X d_Y -r^2 -d_X^2.
	\end{equation}
\end{theorem}
\begin{proof}
	We'll prove that $\psdii(X,Y)_r$ is a submanifold of $\psd(X^* \otimes Y)_r$, and use that a submanifold of a submanifold of $\Lin_\complexNumbers (X^* \otimes Y)$ is also a submanifold of $\Lin_\complexNumbers (X^* \otimes Y)$. The proof is an application of the Regular Level Set Theorem (theorem 9.9 of \cite{tu2011Introduction}).
	
	The Regular Level Set Theorem requires a smooth function, which in this case will be the partial trace. The partial trace is a linear map
	\begin{equation}
		\tr_2 \colon \Lin_\complexNumbers (X^* \otimes Y) \to \Lin_\complexNumbers (X^*).
	\end{equation}
	The partial trace is itself an example of CPTP map, as proved in \ref{proposition:partial trace as cptp map}. Then, by proposition \ref{proposition:cp(A dag) = cp(A) dag} we know that it sends hermitian operator to hermitian operators. Therefore, the partial trace restricts to a smooth function
	\begin{equation}
		\tr_2 \colon \psd(X^* \otimes Y)_r \to \hermitian(X^*).
	\end{equation}
	With respect to this restricted partial trace, we can write
	\begin{equation}
		\psdii(X, Y)_r = \tr_2^{-1}(\id_{X^*}).
	\end{equation}
	We just need to show that $\id_{X^*}$ is a regular value of the restricted partial trace. Since the non restricted partial trace is a linear map between vector spaces, its differential is given by itself, due to the natural isomorphism between the vector space and its tangent space. This gives the following commutative diagram:
	\begin{center}
		\begin{tikzcd}
			T_A \psd(X^* \otimes Y)_r \arrow[r, "d_A \tr_2"] \arrow[d, "d_A i"] & T_{\tr_2(A)} \hermitian(X^*) \arrow[d, "d_{\tr_2(A)} i"]\\
			T_A \Lin_\complexNumbers (X^* \otimes Y) \arrow[r, "d_A \tr_2"] \arrow[d, "\cong"] & T_{\tr_2(A)} \Lin_\complexNumbers (X^*) \arrow[d, "\cong"]\\
			\Lin_\complexNumbers (X^* \otimes Y) \arrow[r, "\tr_2"] & \Lin_\complexNumbers (X^*)
		\end{tikzcd}
	\end{center}
	In the first line we have the differential of the restricted partial trace, while in the second line the partial trace is over the full vector space $\Lin_\complexNumbers (X^* \otimes Y)$. The inclusions $\psd(X^*\otimes Y)_r \subseteq \Lin_\complexNumbers (X^*\otimes Y)$ and $\hermitian(X^*) \subseteq \Lin_\complexNumbers (X^*)$ are both denoted as $i$. The two vertical isomorphisms (both denoted as $\cong$ in the diagram) are the natural isomorphisms between the vector space and its tangent space. We want to show that $\id_{X^*}$ is a regular value of the restricted partial trace. That is, we have to show that
	\begin{equation}
		\label{equation: partial trace submersion}
		d_A \tr_2 (T_A \psd(X^* \otimes Y)_r) = T_{\id_{X^*}} \hermitian(X^*),
	\end{equation}
	for every $A \in \psdii(X,Y)_r$. Since all vertical arrows in this diagram are at least injective linear maps, this lets us identify both $d_A \tr_2 (T_A \psd(X^* \otimes Y)_r)$ and $T_{\id_{X^*}} \hermitian(X^*)$ with subspaces of $\Lin_\complexNumbers (X^*)$. The tangent space $T_{\id_{X^*}} \hermitian(X^*)$ is identified with $\hermitian(X^*)$. As seen in theorem \ref{theorem:TPSDr}, we have the identification $T_A \psd(X^* \otimes Y)_r = \{WA+(WA)^\dag \mid W \in \Lin_\complexNumbers (X^* \otimes Y)\}$. Using the commutative diagram, the equation \ref{equation: partial trace submersion} can be rewritten as
	\begin{equation}
		\tr_2(\{WA+(WA)^\dag \mid W \in \Lin_\complexNumbers (X^* \otimes Y)\}) = \hermitian(X^*).
	\end{equation}
	That the left hand side is contained in $\hermitian(X^*)$, this is because the partial trace sends hermitian operators to hermitian operators. The reason for this is that $\tr_2$ is a CPTP map, as proved in proposition \ref{proposition:partial trace as cptp map}, and because of proposition \ref{proposition:cp(A dag) = cp(A) dag}. For the other inclusion, the difficulty is having to compute a partial trace. The idea is to take some $W$ which let us use that $\tr_2 A = \id_{X^*}$. Let $H \in \hermitian(X^*)$ and take
	\begin{equation}
		W = \frac{1}{2} H \otimes \id_Y.
	\end{equation}
	Then
	\begin{align*}
		\tr_2(WA) &= \tr_2 \left( \left( \frac{1}{2} H \otimes \id_Y \right) A \right)\\
		&= \frac{1}{2} H \tr_2(A)\\
		&= \frac{1}{2} H \id_{X^*}\\
		&= \frac{1}{2} H.
	\end{align*}
	This implies that
	\begin{equation}
		\tr_2(WA+(WA)^\dag) = H/2+H^\dag/2 = H,
	\end{equation}
	which proves the inclusion. This concludes the proof that $\id_{X^*}$ is a regular value of the restricted partial trace. By the Regular Level Set Theorem, $\psdii(X,Y)_r$ is a smooth submanifold of $\psd(X^* \otimes Y)_r$. Since $\psd(X^* \otimes Y)_r$ is a smooth submanifold of $\Lin_\complexNumbers (X^* \otimes Y)$, it follows that $\psdii(X,Y)_r$ is also a smooth submanifold of $\Lin_\complexNumbers (X^* \otimes Y)$. Also, by the same theorem, the dimension of $\psdii(X,Y)_r$ is
	\begin{align*}
		\dim_\realNumbers \psdii(X,Y)_r &= \dim_\realNumbers \psd(X^* \otimes Y)_r - \dim_\realNumbers \hermitian(X^*) \\
		&= 2 d_X d_Y -r^2 -d_X^2.
	\end{align*}
	Notice that the dimension of $\psd(X^* \otimes Y)_r$ was computed in theorem \ref{theorem:TPSDr}.
\end{proof}

Next, we give an expression for the tangent space of $\psdii(X,Y)_r$. The same technique will be used again later, so we'll prove a more general result first as a lemma.

\begin{lemma}{}{tangent space of preimage by linear map}
	Let $p,q \geq 0$, let $M \subseteq \realNumbers^p$ be a smooth submanifold, let $L \colon \realNumbers^p \to \realNumbers^q$ be a $\realNumbers$-linear map and $c \in \realNumbers^q$. Suppose that $c$ is a regular value of $L\mySmallerRestriction{M} \colon M \to \realNumbers^q$, so that the subset $S \coloneqq M \cap L^{-1}(c)$ is a smooth submanifold of $M$. Then, after identifying the tangent spaces as subspaces of $\realNumbers^p$, we have that
	\begin{equation}
		T_x S = (T_x M) \cap \ker L \subseteq \realNumbers^p,
	\end{equation}
	for any $x \in S$.
\end{lemma}
\begin{proof}
	We have the commutative diagram below:
	\begin{center}
		\begin{tikzcd}
			T_x S \arrow[r,"d_x L\mySmallerRestriction{S}"] \arrow[d,swap,"d_x i"] & T_{L(x)} \realNumbers^q \arrow[ddd,"\cong"]\\
			T_x M \arrow[d,swap,"d_x j"] \arrow[ru,swap,"d_x L\mySmallerRestriction{M}"] & {}\\
			T_x \realNumbers^p \arrow[d,swap,"\cong"] & {}\\
			\realNumbers^p \arrow[r,"L"] & \realNumbers^q
		\end{tikzcd}
	\end{center}
	The embeddings are denoted as $i \colon S \to M$ and $j \colon M \to \realNumbers^p$. The differential $d_x L \colon T_x \realNumbers^p \to T_{L(x)} \realNumbers^q$ is omitted from the diagram to be become easier to read. The natural isomorphism of a vector space and its tangent space is denoted by $\cong$ is this diagram.
	
	Since $S = (L\mySmallerRestriction{M})^{-1}(c)$, of course $L$ is constant over $S$, so $d_x L\mySmallerRestriction{S} = 0$. By the commutative of the triangle, this implies that
	\begin{equation}
		d_x L\mySmallerRestriction{M} \ d_x i = 0,
	\end{equation}
	so
	\begin{equation}
		d_x i(T_x S) \subseteq \ker d_x L\mySmallerRestriction{M}.
	\end{equation}
	On the other hand, $c$ is a regular value of $L\mySmallerRestriction{M}$ and $x \in S = L\mySmallerRestriction{M}^{-1}(c)$, so  $d_x L\mySmallerRestriction{M}$ is surjective. Then, the image-kernel theorem says that
	\begin{align*}
		\dim_\realNumbers T_x M &= \dim_\realNumbers \ker d_x L\mySmallerRestriction{M} + \dim_\realNumbers \image d_x L\mySmallerRestriction{M}\\
		&=  \dim_\realNumbers \ker d_x L\mySmallerRestriction{M} + \dim_\realNumbers T_{L(x)} \realNumbers^q.
	\end{align*}
	Also, the Regular Level Set Theorem says how to compute the dimension of $S$. From this, we have
	\begin{equation}
		\dim_\realNumbers T_x S = \dim_\realNumbers T_x M - \dim_\realNumbers T_{L(X)} \realNumbers^q = \dim_\realNumbers \ker d_x L\mySmallerRestriction{M}.
	\end{equation}
	Therefore, both $T_x S$ and $\ker d_x L\mySmallerRestriction{M}$ have the same dimension. This, together with the fact that $d_x i$ is injective, proves that we have an equality
	\begin{equation}
		d_x i(T_x S) = \ker d_x L\mySmallerRestriction{M}.
	\end{equation}
	Finally, the 
	commutative diagram, specially the natural isomorphisms, let us reinterpret this equation as
	\begin{equation}
		T_x S = T_x M \cap \ker L,
	\end{equation}
	interpreting everything as subspaces of $\realNumbers^p$.
\end{proof}

\begin{corollary}{}{tangent space of psdiir}
	The tangent space of $\psdii(X,Y)_r$ at $A$ is
	\begin{equation}
		T_A \psdii(X,Y)_r = T_A \psd(X^* \otimes Y)_r \cap \ker \tr_2,
	\end{equation}
	where every space in this equation is interpreted as a subspace of $\Lin_\complexNumbers(X^* \otimes Y)$.
\end{corollary}
\begin{proof}
	Just apply lemma \ref{lemma:tangent space of preimage by linear map}.
\end{proof}

\subsection{Extreme points}

In proposition \ref{proposition:cptp is compact convex} we saw that the hom-set $\cptp(X,Y)$ is compact convex. By the Krein-Milman theorem (theorem 3.3 of \cite{barvinokcourse}), we know that it is the convex hull of its extreme points. We are interested in studying these extreme points to gain information about the convex set. In theorem 5 of \cite{CHOI1975285}, Choi obtained a characterization of the extreme points of the set of CP maps that send the identity matrix $I$ to some other matrix $K$. Taking $K=I$, this gives the characterization for UCP maps, that is, the completely positive and unital maps. The duality between CPTP and UCP maps then gives a characterization of the extreme points of $\cptp(X,Y)$. We present this characterization in the next theorem. It can also be found in theorem 2.31 of \cite{watrous_2018}. In proposition \ref{proposition:psdii(X,Y) for X or Y = 0}, we proved that if $X=\{0\}$ or $Y = \{0\}$, then $\cptp(X,Y)$ is $\varnothing$ or $\{0\}$. Both cases are trivial, so we restrict attention to spaces of higher dimension.

\begin{theorem}{}{extreme cptp maps in Kraus representation}
	Let $X$ and $Y$ be finite dimensional complex Hilbert spaces, with $d_X \geq 1$ and $d_Y \geq 1$. Let $\varepsilon \in \cptp(X,Y)$, and let $(E_k)_k$ be linearly independent Kraus operators for $\varepsilon$. Then $\varepsilon$ is an extreme point of the convex set $\cptp(X,Y)$ iff $(E_k^\dag E_l)_{k,l}$ is linearly independent. 
\end{theorem}

Using the \CJ isomorphism, we can transfer this characterization from $\cptp(X,Y)$ to $\psdii(X,Y)$.
\begin{corollary}{}{extreme psdii in terms of psik}
	Let $X$ and $Y$ be finite dimensional complex Hilbert spaces, with $d_X \geq 1$ and $d_Y \geq 1$. Let $A \in \psdii(X,Y)$, and write it as
	\begin{equation}
		A = \sum_k \ket{\psi_k} \bra{\psi_k},
	\end{equation}
	where $(\psi_k)_k$ is a sequence of linearly independent vectors in $X^* \otimes Y$. Then $A$ is an extreme point of $\psdii(X,Y)$ iff $(\tr_2 \ket{\psi_k} \bra{\psi_l})_{k,l}$ is linearly independent.
\end{corollary}
\begin{proof}
	This proof is simpler using string diagrams. String diagrams are a powerful tool, but they take too much time to explain properly. Details about string diagrams can be found in \cite{Coecke_Kissinger_2017, MonoidalCatsAndTFT}, also a brief explanation is given in the appendix about Category Theory.
	
	In corollary \ref{corollary:bijection between kraus operators and psis} we saw that the Kraus operators $(E_k)_k$ for $J^{-1}(A)$ are in bijection with the sequences of vectors $(\ket{\psi_k})_k$ in $X^* \otimes Y$ such that the equation
	\begin{equation*}
		A = \sum_k \ket{\psi_k} \bra{\psi_k}
	\end{equation*}
	holds. Such $E_k$ is represented diagrammatically as follows:
	\begin{center}
		\tikzset{every picture/.style={line width=0.75pt}} 
		
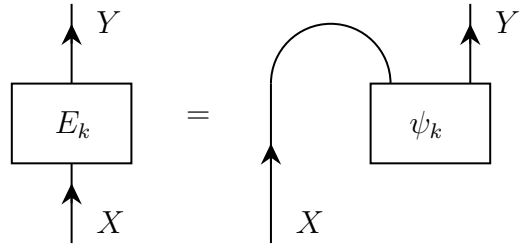
		\begin{tikzpicture}[x=0.75pt,y=0.75pt,yscale=-1,xscale=1]
			
			\draw   (0,40) -- (60,40) -- (60,80) -- (0,80) -- cycle ;
			\draw    (30,80) -- (30,120) ;
			\draw    (30,100) -- (30,93) ;
			\draw [shift={(30,90)}, rotate = 90] [fill={rgb, 255:red, 0; green, 0; blue, 0 }  ][line width=0.08]  [draw opacity=0] (10.72,-5.15) -- (0,0) -- (10.72,5.15) -- (7.12,0) -- cycle    ;
			\draw    (30,0) -- (30,40) ;
			\draw    (30,20) -- (30,13) ;
			\draw [shift={(30,10)}, rotate = 90] [fill={rgb, 255:red, 0; green, 0; blue, 0 }  ][line width=0.08]  [draw opacity=0] (10.72,-5.15) -- (0,0) -- (10.72,5.15) -- (7.12,0) -- cycle    ;
			\draw   (180,40) -- (240,40) -- (240,80) -- (180,80) -- cycle ;
			\draw    (130,40) -- (130,120) ;
			\draw    (130,80) -- (130,73) ;
			\draw [shift={(130,70)}, rotate = 90] [fill={rgb, 255:red, 0; green, 0; blue, 0 }  ][line width=0.08]  [draw opacity=0] (10.72,-5.15) -- (0,0) -- (10.72,5.15) -- (7.12,0) -- cycle    ;
			\draw    (230,0) -- (230,40) ;
			\draw    (230,20) -- (230,13) ;
			\draw [shift={(230,10)}, rotate = 90] [fill={rgb, 255:red, 0; green, 0; blue, 0 }  ][line width=0.08]  [draw opacity=0] (10.72,-5.15) -- (0,0) -- (10.72,5.15) -- (7.12,0) -- cycle    ;
			\draw  [draw opacity=0] (130,40) .. controls (130,23.43) and (143.43,10) .. (160,10) .. controls (176.57,10) and (190,23.43) .. (190,40) -- (160,40) -- cycle ; \draw   (130,40) .. controls (130,23.43) and (143.43,10) .. (160,10) .. controls (176.57,10) and (190,23.43) .. (190,40) ;  
			
			\draw (20,52.4) node [anchor=north west][inner sep=0.75pt]    {$E_{k}$};
			\draw (41,102.4) node [anchor=north west][inner sep=0.75pt]    {$X$};
			\draw (41,2.4) node [anchor=north west][inner sep=0.75pt]    {$Y$};
			\draw (86,52.4) node [anchor=north west][inner sep=0.75pt]    {$=$};
			\draw (198,52.4) node [anchor=north west][inner sep=0.75pt]    {$\psi _{k}$};
			\draw (141,102.4) node [anchor=north west][inner sep=0.75pt]    {$X$};
			\draw (241,2.4) node [anchor=north west][inner sep=0.75pt]    {$Y$};

		\end{tikzpicture}
	\end{center}
	Since $(\psi_k )_k$ is linearly independent, the same is true for $(E_k )_k$, which is a direct consequence of the isomorphism $\Lin_\complexNumbers (X,Y) \cong X^* \otimes Y$. Then we just have to translate the extremality condition for $(E_k)_k $ to a condition for $(\psi_k)_k $.
	
	$(E_k)_k$ represents an extreme point of $\cptp(X,Y)$ iff $(E_k^\dag E_l)_{k,l}$ is linearly independent. We have the following diagrammatic equation:
	\begin{center}

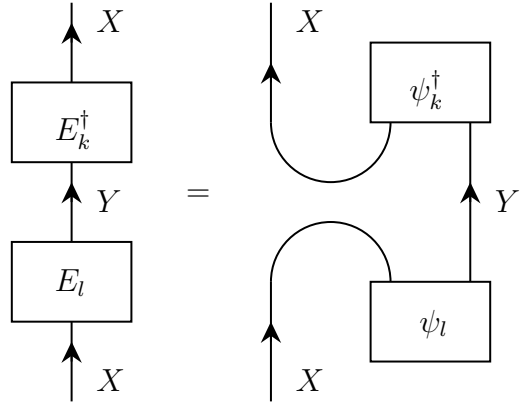
		\tikzset{every picture/.style={line width=0.75pt}} 
		
		\begin{tikzpicture}[x=0.75pt,y=0.75pt,yscale=-1,xscale=1]
			
			\draw   (0,40) -- (60,40) -- (60,80) -- (0,80) -- cycle ;
			\draw    (30,80) -- (30,120) ;
			\draw    (30,100) -- (30,93) ;
			\draw [shift={(30,90)}, rotate = 90] [fill={rgb, 255:red, 0; green, 0; blue, 0 }  ][line width=0.08]  [draw opacity=0] (10.72,-5.15) -- (0,0) -- (10.72,5.15) -- (7.12,0) -- cycle    ;
			\draw    (30,0) -- (30,40) ;
			\draw    (30,20) -- (30,13) ;
			\draw [shift={(30,10)}, rotate = 90] [fill={rgb, 255:red, 0; green, 0; blue, 0 }  ][line width=0.08]  [draw opacity=0] (10.72,-5.15) -- (0,0) -- (10.72,5.15) -- (7.12,0) -- cycle    ;
			\draw   (180,140) -- (240,140) -- (240,180) -- (180,180) -- cycle ;
			\draw    (130,140) -- (130,200) ;
			\draw    (130,170) -- (130,163) ;
			\draw [shift={(130,160)}, rotate = 90] [fill={rgb, 255:red, 0; green, 0; blue, 0 }  ][line width=0.08]  [draw opacity=0] (10.72,-5.15) -- (0,0) -- (10.72,5.15) -- (7.12,0) -- cycle    ;
			\draw    (230,60) -- (230,140) ;
			\draw    (230,100) -- (230,93) ;
			\draw [shift={(230,90)}, rotate = 90] [fill={rgb, 255:red, 0; green, 0; blue, 0 }  ][line width=0.08]  [draw opacity=0] (10.72,-5.15) -- (0,0) -- (10.72,5.15) -- (7.12,0) -- cycle    ;
			\draw  [draw opacity=0] (130,140) .. controls (130,123.43) and (143.43,110) .. (160,110) .. controls (176.57,110) and (190,123.43) .. (190,140) -- (160,140) -- cycle ; \draw   (130,140) .. controls (130,123.43) and (143.43,110) .. (160,110) .. controls (176.57,110) and (190,123.43) .. (190,140) ;  
			\draw   (0,120) -- (60,120) -- (60,160) -- (0,160) -- cycle ;
			\draw    (30,160) -- (30,200) ;
			\draw    (30,180) -- (30,173) ;
			\draw [shift={(30,170)}, rotate = 90] [fill={rgb, 255:red, 0; green, 0; blue, 0 }  ][line width=0.08]  [draw opacity=0] (10.72,-5.15) -- (0,0) -- (10.72,5.15) -- (7.12,0) -- cycle    ;
			\draw   (180,20) -- (240,20) -- (240,60) -- (180,60) -- cycle ;
			\draw    (130,0) -- (130,60) ;
			\draw    (130,40) -- (130,33) ;
			\draw [shift={(130,30)}, rotate = 90] [fill={rgb, 255:red, 0; green, 0; blue, 0 }  ][line width=0.08]  [draw opacity=0] (10.72,-5.15) -- (0,0) -- (10.72,5.15) -- (7.12,0) -- cycle    ;
			\draw  [draw opacity=0] (190,60) .. controls (190,60) and (190,60) .. (190,60) .. controls (190,60) and (190,60) .. (190,60) .. controls (190,76.57) and (176.57,90) .. (160,90) .. controls (143.43,90) and (130,76.57) .. (130,60) -- (160,60) -- cycle ; \draw   (190,60) .. controls (190,60) and (190,60) .. (190,60) .. controls (190,60) and (190,60) .. (190,60) .. controls (190,76.57) and (176.57,90) .. (160,90) .. controls (143.43,90) and (130,76.57) .. (130,60) ;  
			
			\draw (20,52.4) node [anchor=north west][inner sep=0.75pt]    {$E_{k}^{\dagger }$};
			\draw (41,92.4) node [anchor=north west][inner sep=0.75pt]    {$Y$};
			\draw (41,2.4) node [anchor=north west][inner sep=0.75pt]    {$X$};
			\draw (86,92.4) node [anchor=north west][inner sep=0.75pt]    {$=$};
			\draw (203,152.4) node [anchor=north west][inner sep=0.75pt]    {$\psi _{l}$};
			\draw (141,182.4) node [anchor=north west][inner sep=0.75pt]    {$X$};
			\draw (241,92.4) node [anchor=north west][inner sep=0.75pt]    {$Y$};
			\draw (20,132.4) node [anchor=north west][inner sep=0.75pt]    {$E_{l}$};
			\draw (41,182.4) node [anchor=north west][inner sep=0.75pt]    {$X$};
			\draw (198,32.4) node [anchor=north west][inner sep=0.75pt]    {$\psi _{k}^{\dagger }$};
			\draw (141,2.4) node [anchor=north west][inner sep=0.75pt]    {$X$};

		\end{tikzpicture}
	\end{center}
	The properties of string diagrams lets us move the box with $\psi_k^\dag$ to below the box with $\psi_l$. This move introduces a crossing of strings, which represents a symmetry. We get the following:
	\begin{center}

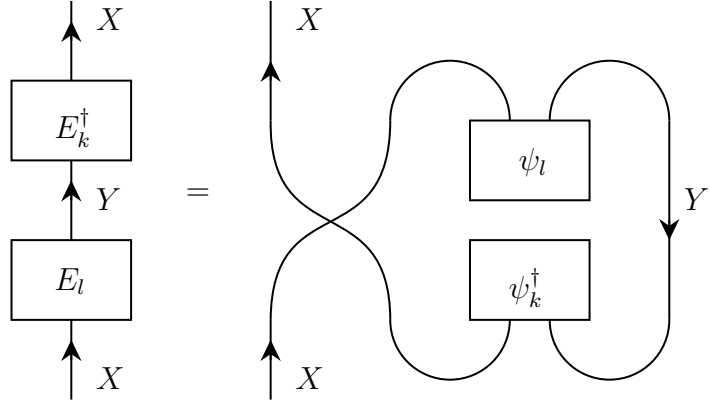
		\tikzset{every picture/.style={line width=0.75pt}} 
		
		\begin{tikzpicture}[x=0.75pt,y=0.75pt,yscale=-1,xscale=1]
			
			\draw   (0,40) -- (60,40) -- (60,80) -- (0,80) -- cycle ;
			\draw    (30,80) -- (30,120) ;
			\draw    (30,100) -- (30,93) ;
			\draw [shift={(30,90)}, rotate = 90] [fill={rgb, 255:red, 0; green, 0; blue, 0 }  ][line width=0.08]  [draw opacity=0] (10.72,-5.15) -- (0,0) -- (10.72,5.15) -- (7.12,0) -- cycle    ;
			\draw    (30,0) -- (30,40) ;
			\draw    (30,20) -- (30,13) ;
			\draw [shift={(30,10)}, rotate = 90] [fill={rgb, 255:red, 0; green, 0; blue, 0 }  ][line width=0.08]  [draw opacity=0] (10.72,-5.15) -- (0,0) -- (10.72,5.15) -- (7.12,0) -- cycle    ;
			\draw   (230,60) -- (290,60) -- (290,100) -- (230,100) -- cycle ;
			\draw    (130,160) -- (130,200) ;
			\draw    (130,180) -- (130,173) ;
			\draw [shift={(130,170)}, rotate = 90] [fill={rgb, 255:red, 0; green, 0; blue, 0 }  ][line width=0.08]  [draw opacity=0] (10.72,-5.15) -- (0,0) -- (10.72,5.15) -- (7.12,0) -- cycle    ;
			\draw    (330,60) -- (330,160) ;
			\draw    (330,110) -- (330,117) ;
			\draw [shift={(330,120)}, rotate = 270] [fill={rgb, 255:red, 0; green, 0; blue, 0 }  ][line width=0.08]  [draw opacity=0] (10.72,-5.15) -- (0,0) -- (10.72,5.15) -- (7.12,0) -- cycle    ;
			\draw  [draw opacity=0] (190,60) .. controls (190,43.43) and (203.43,30) .. (220,30) .. controls (236.57,30) and (250,43.43) .. (250,60) -- (220,60) -- cycle ; \draw   (190,60) .. controls (190,43.43) and (203.43,30) .. (220,30) .. controls (236.57,30) and (250,43.43) .. (250,60) ;  
			\draw   (0,120) -- (60,120) -- (60,160) -- (0,160) -- cycle ;
			\draw    (30,160) -- (30,200) ;
			\draw    (30,180) -- (30,173) ;
			\draw [shift={(30,170)}, rotate = 90] [fill={rgb, 255:red, 0; green, 0; blue, 0 }  ][line width=0.08]  [draw opacity=0] (10.72,-5.15) -- (0,0) -- (10.72,5.15) -- (7.12,0) -- cycle    ;
			\draw   (230,120) -- (290,120) -- (290,160) -- (230,160) -- cycle ;
			\draw    (130,0) -- (130,60) ;
			\draw    (130,40) -- (130,33) ;
			\draw [shift={(130,30)}, rotate = 90] [fill={rgb, 255:red, 0; green, 0; blue, 0 }  ][line width=0.08]  [draw opacity=0] (10.72,-5.15) -- (0,0) -- (10.72,5.15) -- (7.12,0) -- cycle    ;
			\draw  [draw opacity=0] (250,160) .. controls (250,160) and (250,160) .. (250,160) .. controls (250,160) and (250,160) .. (250,160) .. controls (250,176.57) and (236.57,190) .. (220,190) .. controls (203.43,190) and (190,176.57) .. (190,160) -- (220,160) -- cycle ; \draw   (250,160) .. controls (250,160) and (250,160) .. (250,160) .. controls (250,160) and (250,160) .. (250,160) .. controls (250,176.57) and (236.57,190) .. (220,190) .. controls (203.43,190) and (190,176.57) .. (190,160) ;  
			\draw  [draw opacity=0] (270,60) .. controls (270,43.43) and (283.43,30) .. (300,30) .. controls (316.57,30) and (330,43.43) .. (330,60) -- (300,60) -- cycle ; \draw   (270,60) .. controls (270,43.43) and (283.43,30) .. (300,30) .. controls (316.57,30) and (330,43.43) .. (330,60) ;  
			\draw  [draw opacity=0] (330,160) .. controls (330,160) and (330,160) .. (330,160) .. controls (330,160) and (330,160) .. (330,160) .. controls (330,176.57) and (316.57,190) .. (300,190) .. controls (283.43,190) and (270,176.57) .. (270,160) -- (300,160) -- cycle ; \draw   (330,160) .. controls (330,160) and (330,160) .. (330,160) .. controls (330,160) and (330,160) .. (330,160) .. controls (330,176.57) and (316.57,190) .. (300,190) .. controls (283.43,190) and (270,176.57) .. (270,160) ;  
			\draw    (130.01,60) .. controls (129.04,127.17) and (189.79,93.83) .. (190.02,160) ;
			\draw    (130.01,160) .. controls (130.54,93) and (189.79,128.83) .. (190.02,60) ;
			
			\draw (20,52.4) node [anchor=north west][inner sep=0.75pt]    {$E_{k}^{\dagger }$};
			\draw (41,92.4) node [anchor=north west][inner sep=0.75pt]    {$Y$};
			\draw (41,2.4) node [anchor=north west][inner sep=0.75pt]    {$X$};
			\draw (86,92.4) node [anchor=north west][inner sep=0.75pt]    {$=$};
			\draw (253,72.4) node [anchor=north west][inner sep=0.75pt]    {$\psi _{l}$};
			\draw (141,182.4) node [anchor=north west][inner sep=0.75pt]    {$X$};
			\draw (336,92.4) node [anchor=north west][inner sep=0.75pt]    {$Y$};
			\draw (20,132.4) node [anchor=north west][inner sep=0.75pt]    {$E_{l}$};
			\draw (41,182.4) node [anchor=north west][inner sep=0.75pt]    {$X$};
			\draw (248,132.4) node [anchor=north west][inner sep=0.75pt]    {$\psi _{k}^{\dagger }$};
			\draw (141,2.4) node [anchor=north west][inner sep=0.75pt]    {$X$};

		\end{tikzpicture}
	\end{center}
	To remove the crossing of strings, we pull down a string decorated with $X$ and pull up the other string. This corresponds to an isomorphism $\Lin_\complexNumbers (X,X) \cong \Lin_\complexNumbers (X^*,X^*)$. We obtain the following:
	\begin{center}

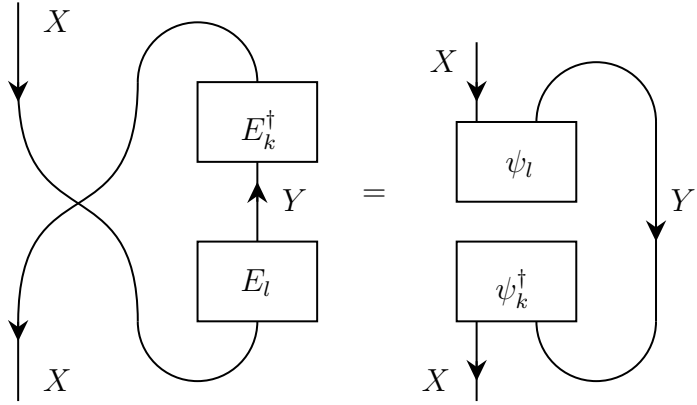
		\tikzset{every picture/.style={line width=0.75pt}} 
		
		\begin{tikzpicture}[x=0.75pt,y=0.75pt,yscale=-1,xscale=1]
			
			\draw   (90,40) -- (150,40) -- (150,80) -- (90,80) -- cycle ;
			\draw    (120,80) -- (120,120) ;
			\draw    (120,100) -- (120,93) ;
			\draw [shift={(120,90)}, rotate = 90] [fill={rgb, 255:red, 0; green, 0; blue, 0 }  ][line width=0.08]  [draw opacity=0] (10.72,-5.15) -- (0,0) -- (10.72,5.15) -- (7.12,0) -- cycle    ;
			\draw    (0,0) -- (0,40) ;
			\draw    (0,40) -- (0,47) ;
			\draw [shift={(0,50)}, rotate = 270] [fill={rgb, 255:red, 0; green, 0; blue, 0 }  ][line width=0.08]  [draw opacity=0] (10.72,-5.15) -- (0,0) -- (10.72,5.15) -- (7.12,0) -- cycle    ;
			\draw   (220,60) -- (280,60) -- (280,100) -- (220,100) -- cycle ;
			\draw    (320,60) -- (320,160) ;
			\draw    (320,110) -- (320,117) ;
			\draw [shift={(320,120)}, rotate = 270] [fill={rgb, 255:red, 0; green, 0; blue, 0 }  ][line width=0.08]  [draw opacity=0] (10.72,-5.15) -- (0,0) -- (10.72,5.15) -- (7.12,0) -- cycle    ;
			\draw  [draw opacity=0] (60,40) .. controls (60,23.43) and (73.43,10) .. (90,10) .. controls (106.57,10) and (120,23.43) .. (120,40) -- (90,40) -- cycle ; \draw   (60,40) .. controls (60,23.43) and (73.43,10) .. (90,10) .. controls (106.57,10) and (120,23.43) .. (120,40) ;  
			\draw   (90,120) -- (150,120) -- (150,160) -- (90,160) -- cycle ;
			\draw    (0,160) -- (0,200) ;
			\draw    (0,160) -- (0,167) ;
			\draw [shift={(0,170)}, rotate = 270] [fill={rgb, 255:red, 0; green, 0; blue, 0 }  ][line width=0.08]  [draw opacity=0] (10.72,-5.15) -- (0,0) -- (10.72,5.15) -- (7.12,0) -- cycle    ;
			\draw   (220,120) -- (280,120) -- (280,160) -- (220,160) -- cycle ;
			\draw    (230,20) -- (230,60) ;
			\draw    (230,40) -- (230,47) ;
			\draw [shift={(230,50)}, rotate = 270] [fill={rgb, 255:red, 0; green, 0; blue, 0 }  ][line width=0.08]  [draw opacity=0] (10.72,-5.15) -- (0,0) -- (10.72,5.15) -- (7.12,0) -- cycle    ;
			\draw  [draw opacity=0] (120,160) .. controls (120,160) and (120,160) .. (120,160) .. controls (120,160) and (120,160) .. (120,160) .. controls (120,176.57) and (106.57,190) .. (90,190) .. controls (73.43,190) and (60,176.57) .. (60,160) -- (90,160) -- cycle ; \draw   (120,160) .. controls (120,160) and (120,160) .. (120,160) .. controls (120,160) and (120,160) .. (120,160) .. controls (120,176.57) and (106.57,190) .. (90,190) .. controls (73.43,190) and (60,176.57) .. (60,160) ;  
			\draw  [draw opacity=0] (260,60) .. controls (260,43.43) and (273.43,30) .. (290,30) .. controls (306.57,30) and (320,43.43) .. (320,60) -- (290,60) -- cycle ; \draw   (260,60) .. controls (260,43.43) and (273.43,30) .. (290,30) .. controls (306.57,30) and (320,43.43) .. (320,60) ;  
			\draw  [draw opacity=0] (320,160) .. controls (320,160) and (320,160) .. (320,160) .. controls (320,160) and (320,160) .. (320,160) .. controls (320,176.57) and (306.57,190) .. (290,190) .. controls (273.43,190) and (260,176.57) .. (260,160) -- (290,160) -- cycle ; \draw   (320,160) .. controls (320,160) and (320,160) .. (320,160) .. controls (320,160) and (320,160) .. (320,160) .. controls (320,176.57) and (306.57,190) .. (290,190) .. controls (273.43,190) and (260,176.57) .. (260,160) ;  
			\draw    (0.01,40) .. controls (-0.96,120.6) and (59.79,80.6) .. (60.02,160) ;
			\draw    (0.01,160) .. controls (0.54,79.6) and (59.79,122.6) .. (60.02,40) ;
			\draw    (230,160) -- (230,200) ;
			\draw    (230,180) -- (230,187) ;
			\draw [shift={(230,190)}, rotate = 270] [fill={rgb, 255:red, 0; green, 0; blue, 0 }  ][line width=0.08]  [draw opacity=0] (10.72,-5.15) -- (0,0) -- (10.72,5.15) -- (7.12,0) -- cycle    ;
			
			\draw (110,52.4) node [anchor=north west][inner sep=0.75pt]    {$E_{k}^{\dagger }$};
			\draw (131,92.4) node [anchor=north west][inner sep=0.75pt]    {$Y$};
			\draw (11,2.4) node [anchor=north west][inner sep=0.75pt]    {$X$};
			\draw (171,92.4) node [anchor=north west][inner sep=0.75pt]    {$=$};
			\draw (243,72.4) node [anchor=north west][inner sep=0.75pt]    {$\psi _{l}$};
			\draw (201,182.4) node [anchor=north west][inner sep=0.75pt]    {$X$};
			\draw (326,92.4) node [anchor=north west][inner sep=0.75pt]    {$Y$};
			\draw (110,132.4) node [anchor=north west][inner sep=0.75pt]    {$E_{l}$};
			\draw (11,182.4) node [anchor=north west][inner sep=0.75pt]    {$X$};
			\draw (238,132.4) node [anchor=north west][inner sep=0.75pt]    {$\psi _{k}^{\dagger }$};
			\draw (205,22.4) node [anchor=north west][inner sep=0.75pt]    {$X$};

		\end{tikzpicture}
	\end{center}
	Since we applied an isomorphism, the linear independence is maintained. The right hand side of the equation is a diagrammatic representation of $\tr_2 \ket{\psi_l} \bra{ \psi_k }$. Therefore, $A$ is an extreme point of $\psdii (X,Y)$ iff $( \tr_2 \ket{\psi_l} \bra{ \psi_k } )_{k,l}$ is linearly independent. Renaming the variables by swapping $k$ and $l$, we get that $A$ is an extreme point of $\psdii(X,Y)$ iff $( \tr_2 \ket{\psi_k} \bra{ \psi_l } )_{k,l}$ is linearly independent.
\end{proof}

Both these characterizations use representatives $(E_k)_k$ or $(\ket{\psi_k})_k$, which are not unique. But there is still another characterization which doesn't use representatives. We'll give this characterization in the next corollary, but first we define the conjugation map. The corollary \ref{corollary:extreme psdii basis independent} isn't a completely new result. A similar result can already be found in theorem 4 of \cite{FRIEDLAND2016553}, but we present it in another form.

\begin{definition}{}{}
	Let $X$ be a finite dimensional complex Hilbert space and  $A \in \Lin_\complexNumbers (X)$. We define the $\complexNumbers$-linear map that conjugates by $A$ as follows:
	\begin{align*}
		\varphi_A \colon \Lin_\complexNumbers (X) &\longrightarrow \Lin_\complexNumbers (X)\\
		B &\longmapsto ABA^\dag.
	\end{align*}
\end{definition}

\begin{corollary}{}{extreme psdii basis independent}
	Let $X$ and $Y$ be finite dimensional complex Hilbert spaces with $d_X \geq 1$ and $d_Y \geq 1$, and let $A \in \psdii(X,Y)$. Then $A$ is an extreme point of $\psdii(X,Y)$ iff
	\begin{equation}
		\ker \tr_2 \cap \image \varphi_A = \{0\}.
	\end{equation}
	This is also equivalent to
	\begin{equation}
		\ker \tr_2 \cap \image \varphi_{P_{\image A}} = \{0\},
	\end{equation}
	where $P_{\image A}$ is the orthogonal projection over $\image A$.
\end{corollary}
\begin{proof}
	This result is obtained by reinterpreting corollary \ref{corollary:extreme psdii in terms of psik}. Let $A \in \psdii(X,\allowbreak Y)$. It is a positive semidefinite operator, so it can be diagonalized as
	\begin{equation}
		A = \sum_k \lambda_k \ket{v_k}\bra{v_k},
	\end{equation}
	where $\lambda_k > 0$ is a positive eigenvalue and $(\ket{v_k})_k$ is an orthonormal sequence in $X^* \otimes Y$. Then define
	\begin{equation}
		\label{equation: psik = sqrt lambdak vk}
		\ket{\psi_k} = \sqrt{\lambda_k} \ket{v_k},
	\end{equation}
	and we have that
	\begin{equation}
		A = \sum_k \ket{\psi_k}\bra{\psi_k}.
	\end{equation}
	The vectors $(\ket{v_k})_k$ are orthonormal and $\lambda_k > 0$, so $(\ket{\psi_k})_k$ is linearly independent. Then, by corollary \ref{corollary:extreme psdii in terms of psik}, $A$ is an extreme point of $\psdii(X,Y)$ iff $(\tr_2 \ket{\psi_k}\bra{\psi_l})_{k,l}$ is linearly independent. Using equation \ref{equation: psik = sqrt lambdak vk} and that $\lambda_k > 0$, it follows that $A$ is extreme iff $(\tr_2 \ket{v_k}\bra{v_l})_{k,l}$ is linearly independent. The vectors $(\ket{v_k})_k$ are an orthonormal basis for $\image A$. Then the orthogonal projection over $\image A$ is
	\begin{equation}
		P_{\image A} = \sum_k \ket{v_k}\bra{v_k}.
	\end{equation}
	Here we see appear the operators $\ket{v_k}\bra{v_k}$. We want to make appear the operators $\ket{v_k}\bra{v_l}$, with possibly different indices $k$ and $l$. This is obtained from $\varphi_{P_{\image A}}$. Let's compute its image. If $B \in \Lin_\complexNumbers (X^* \otimes Y)$, then
	\begin{align*}
		\varphi_{P_{\image A}}(B) &= P_{\image A} B P_{\image A} \\
		&= \left(\sum_k \ket{v_k}\bra{v_k}\right) B \left(\sum_l \ket{v_l}\bra{v_l}\right)\\
		&= \sum_{k,l} \bra{v_k} B \ket{v_l} \ \ket{v_k}\bra{v_l}.
	\end{align*}
	The complex number $\bra{v_k} B \ket{v_l}$ can be any number that we want. In fact, given a sequence of complex numbers $(c_{k,l})_{k,l}$, if we take
	\begin{equation}
		B = \sum_{k,l} c_{k,l} \ket{v_k} \bra{v_l},
	\end{equation}
	then $\bra{v_k} B \ket{v_l} = c_{k,l}$. This implies that $\image \varphi_{P_{\image A}}$ is generated by $(\bra{v_k} B \ket{v_l})_{k,l}$. But $(\ket{v_k} \bra{v_l})_{k,l}$ is orthonormal. In fact, the Hilbert-Schmidt inner product of these operators is
	\begin{align*}
		\langle \ket{v_k} \bra{v_l}, \ket{v_{k'}} \bra{v_{l'}} \rangle &= \tr((\ket{v_k} \bra{v_l})^\dag \ket{v_{k'}} \bra{v_{l'}})\\
		&= \tr(\ket{v_l} \langle v_k \mid v_{k'} \rangle \bra{v_{l'}})\\
		&= \langle v_k \mid v_{k'} \rangle \langle v_{l'} \mid v_l \rangle\\
		&= \delta_{k,k'} \delta_{l,l'}\\
		&= \delta_{(k,l),(k',l')}.
	\end{align*}
	Therefore, $(\ket{v_k} \bra{v_l})_{k,l}$ is an orthonormal basis for $\image \varphi_{P_{\image A}}$. The statement that $(\tr_2 \ket{v_k} \bra{v_l})_{k,l}$ is linearly independent is then equivalent to say that $\tr_2$ sends the linearly independent vectors $(\ket{v_k} \bra{v_l})_{k,l}$ to linearly independent vectors. This is equivalent to say that the function
	\begin{align*}
		T \colon \image \varphi_{P_{\image A}} &\longrightarrow \Lin_\complexNumbers (X^*)\\
		B &\longmapsto \tr_2 B
	\end{align*}
	is injective. At this point we have found another characterization of extremality that doesn't uses a basis explicitly. That is, we proved that $A$ is an extreme point of $\psdii(X,Y)$ iff $\ker T = \{0\}$. To finish the proof, we only have to give another expression for $\ker T$. $T$ is the restriction of $\tr_2 \colon \Lin_\complexNumbers (X^* \otimes Y) \to \Lin_\complexNumbers (X^*)$ to $\image \varphi_{P_{\image A}}$. From this, it follows that
	\begin{equation}
		\ker T = \ker \tr_2 \cap \image \varphi_{P_{\image A}}.
	\end{equation}
	If we replace $P_{\image A}$ with $A$, this only rescales the vectors $\ket{v_k}$ to $\ket{\psi_k}$. This rescaling doesn't change the image, that is,
	\begin{equation}
		\image \varphi_{A} = \image \varphi_{P_{\image A}}.
	\end{equation}
	This finishes the proof.
\end{proof}

\subsection{Choi-rank bounds for extreme points}

The characterizations of the extreme points of $\cptp(X,Y)$ or $\psdii(X,Y)$ lets us find an upper bound for the rank. We show this in the next corollary. 

\begin{corollary}{}{bounds on rank of extreme cptp}
	Let $X$ and $Y$ be finite dimensional complex Hilbert spaces with $d_X \geq 1$ and $d_Y \geq 1$. If $A \in \psdii(X,Y)$ is an extreme point, then
	\begin{equation}
		\frac{d_X}{d_Y} \leq \rank(A) \leq d_X.
	\end{equation}
\end{corollary}
\begin{proof}
	Write $A$ as
	\begin{equation}
		A = \sum_k \ket{\psi_k}\bra{\psi_k},
	\end{equation}
	where $(\ket{\psi_k})_k$ is a linearly independent sequence in $X^* \otimes Y$. This sequence may be obtained from a diagonalization of $A$, using only the eigenvectors with non zero eigenvalue. By corollary \ref{corollary:extreme psdii in terms of psik}, $A$ is an extreme point iff $(\tr_2 \ket{\psi_k}\bra{\psi_l})_{k,l}$ is linearly independent. Since $(\tr_2 \ket{\psi_k}\bra{\psi_l})_{k,l}$ is a sequence of operators in $\Lin_\complexNumbers (X^*)$, it can't be linearly independent if the number of pairs $(k,l)$ is bigger than $\dim_\complexNumbers \Lin_\complexNumbers (X^*) = d_X^2$. Since we are supposing that $(\ket{\psi_k})_k$ is linearly independent, the number of indices $k$ is $\rank (A)$. Then the number of indices $(k,l)$ is $(\rank (A))^2$. Therefore, we must have
	\begin{equation}
		(\rank (A))^2  \leq d_X^2,
	\end{equation}
	so
	\begin{equation}
		\rank A \leq d_X.
	\end{equation}
	This proves the upper bound. The lower bound doesn't depend on the operator being an extreme point. It was already proved in proposition \ref{proposition:general CPTP rank bounds}.
\end{proof}

This corollary raises the question if there are extreme points for each of the ranks in the interval $[d_X/d_Y,d_X]$. Later in section \ref{section: extreme cptp} we'll construct an extreme CPTP map for each Choi-rank allowed by the bound. For the moment, we give a simpler example that has Choi-rank $d_X$.

\subsubsection{Example of extreme CPTP map with maximum Choi-rank}
\label{example: extreme cptp with rank dX}

Let $X$ and $Y$ be finite dimensional complex Hilbert spaces with $d_X \geq 1$ and $d_Y \geq 1$. Let $\rho \in \density(Y)$ be any density operator over $Y$. That is, $\rho$ is a $\complexNumbers$-linear map $\rho \colon Y \to Y$ which is positive semidefinite and $\tr \, \rho = 1$. Then consider the operator
\begin{equation}
	A \coloneqq \id_{X^*} \otimes \rho \in \Lin_\complexNumbers (X^* \otimes Y).
\end{equation}
It is a tensor product of two positive semidefinite operators, so $A$ is also positive semidefinite, that is, $A \in \psd(X^* \otimes Y)$. Its partial trace over $Y$ is
\begin{align*}
	\tr_2 A &= \tr_2 ( \id_{X^*} \otimes \rho) \\
	&= \id_{X^*} \tr \, \rho\\
	&= \id_{X^*}.
\end{align*}
Therefore $A \in \psdii(X,Y)$. The rank of a tensor product is the product of the ranks, so
\begin{equation}
	\rank A = \rank \, \id_{X^*} \ \rank \, \rho = d_X \, \rank \, \rho.
\end{equation}
By corollary \ref{corollary:bounds on rank of extreme cptp}, if $\rank \, \rho \geq 2$ then $A$ isn't an extreme point of $\psdii(X,Y)$. If $\rank \, \rho = 1$, then it is a vector state
\begin{equation}
	\rho = \ket{\psi}\bra{\psi},
\end{equation}
where $\ket{\psi} \in Y$ is a unit vector. In this case, we have
\begin{equation}
	\rank \, A = d_X,
\end{equation}
so it can be an extreme point. We show next that $A$ is indeed an extreme point for any unit vector $\ket{\psi} \in Y$. We prove this using corollary \ref{corollary:extreme psdii in terms of psik}. Let $(x^i)_i$ be an orthonormal basis for $X^*$. Then the identity function can be expressed as
\begin{equation}
	\id_{X^*} = \sum_i \ket{x^i} \bra{x^i}.
\end{equation}
Then we can express $A$ as
\begin{align*}
	A &= \id_{X^*} \otimes \ket{\psi}\bra{\psi} \\
	&= \sum_i \ket{x^i}\bra{x^i} \otimes \ket{\psi}\bra{\psi}\\
	&= \sum_i \ket{x^i \otimes \psi}\bra{x^i \otimes \psi}.
\end{align*}
Defining $\ket{\psi_i} \coloneqq \ket{x^i \otimes \psi}$, we have
\begin{equation}
	A = \sum_i \ket{\psi_i}\bra{\psi_i}.
\end{equation}
The vectors $\ket{\psi_i}$ are orthonormal. In fact, their inner products are
\begin{align*}
	\langle \psi_i \mid \psi_{i'} \rangle &= \langle x^i \otimes \psi \mid x^{i'} \otimes \psi \rangle \\
	&= \langle x^i \mid x^{i'} \rangle \langle \psi \mid \psi \rangle \\
	&= \langle x^i \mid x^{i'} \rangle \\
	&= \delta_{i,i'}.
\end{align*}
In particular, $(\ket{\psi_i})_i$ is linearly independent. By corollary \ref{corollary:extreme psdii in terms of psik}, $A$ is extreme iff $(\tr_2 \ket{\psi_i}\bra{\psi_{i'}})_{i,i'}$ is linearly independent. Let's compute this partial trace. We have that
\begin{align*}
	\tr_2 \ket{\psi_i}\bra{\psi_{i'}} &= \tr_2 (\ket{x^i \otimes \psi}\bra{x^{i'} \otimes \psi})\\
	&= \tr_2 (\ket{x^i}\bra{x^{i'}} \otimes \ket{\psi}\bra{\psi})\\
	&= \ket{x^i}\bra{x^{i'}} \ \tr (\ket{\psi}\bra{\psi})\\
	&= \ket{x^i}\bra{x^{i'}} \ \langle \psi \mid \psi \rangle\\
	&= \ket{x^i}\bra{x^{i'}}.
\end{align*}
The operators $\ket{x^i}\bra{x^{i'}}$ are an orthonormal basis for $\Lin_\complexNumbers(X^*)$. In particular, they are linearly independent. Then, by corollary \ref{corollary:extreme psdii in terms of psik}, $A = \id_{X^*} \otimes \ket{\psi}\bra{\psi}$ is an extreme point of $\psdii(X,Y)$, for any unit vector $\ket{\psi} \in Y$.

\subsection{Extremality of the channels with lowest Choi-rank}

In theorem 9 of \cite{FRIEDLAND2016553} was proved that any non extreme CPTP map is a convex combination of other two CPTP maps with strictly lower Choi-ranks. For the lowest possible Choi-rank there are no CPTP maps with lower Choi-ranks to make the convex combination, from which it was concluded that all CPTP maps with the lowest possible Choi-rank are extreme points.

\begin{theorem}{}{}
	Let $X$ and $Y$ be finite dimensional complex Hilbert spaces, with $d_Y \geq 1$. Then all $\varepsilon \in \cptp(X,Y)$ with Choi-rank
	\begin{equation}
		\CR(\varepsilon) = \left\lceil \frac{d_X}{d_Y} \right\rceil,
	\end{equation}
	the lowest possible Choi-rank, are extreme points of $\cptp(X,Y)$.
\end{theorem}
\begin{proof}
	See theorem 9 of \cite{FRIEDLAND2016553}. This theorem supposes that $d_X , d_Y \geq 2$, but the remaining cases are trivial.
\end{proof}

\subsection{CPTP maps with 1-dimensional domain}

One of our objectives is to understand the geometry and topology of the set of extreme points of $\psdii(X,Y)$ (or, equivalently, $\cptp(X,Y)$), or of its closure. The case $d_X= 1$ is already understood, but for $d_X \geq 2$ it is still an open problem. Next we prove known results about the case $d_X =1$. For simplicity of notation, we consider that $X = \complexNumbers$. In this case, the set $\psdii(\complexNumbers,Y)$ is isomorphic to the set $\density(Y)$ of density operators over $Y$. It is known that the extreme points of $\density(Y)$ are the vector states $\ket{\psi}\bra{\psi}$, where $\ket{\psi} \in Y$ is any unit vector. It is also known that the set of vector states is homeomorphic to the complex projective space $\complexNumbers P^{d_Y-1}$. In particular, the set of extreme points is already closed for $d_X = 1$. Some partial results for $d_X \geq 2$ will be discussed in section \ref{section: partial decomposition into cells}.

\begin{proposition}{}{psdii(C,Y) is isomorphic to D(Y)}
	For any finite dimensional complex Hilbert space $Y$, with $d_Y \geq 1$, we have an isomorphism
	\begin{equation}
		\psdii(\complexNumbers,Y) \cong \density(Y).
	\end{equation}
\end{proposition}
\begin{proof}
	The isomorphism with $\density(Y)$ consists of discarding $\complexNumbers^*$, that is, taking the partial trace over $\complexNumbers^*$. We have two linear maps,
	\begin{equation*}
		\tr_1 \colon \Lin_\complexNumbers(\complexNumbers^* \otimes Y) \longrightarrow \Lin_\complexNumbers(Y)
	\end{equation*}
	and
	\begin{align*}
		f \colon \Lin_\complexNumbers(Y) &\longrightarrow \Lin_\complexNumbers(\complexNumbers^* \otimes Y)\\
		A &\longmapsto \id_{\complexNumbers^*} \otimes A.
	\end{align*}
	Of course, both are $\complexNumbers$-linear maps. Also, they are inverse to each other. In fact, if $A \in \Lin_\complexNumbers(\complexNumbers^* \otimes Y)$ then it can be written as $A = \id_{\complexNumbers^*} \otimes B$, for some $B \in \Lin_\complexNumbers(Y)$. Then
	\begin{align*}
		f(\tr_1(A)) &= \id_{\complexNumbers^*} \otimes \tr_1 (\id_{\complexNumbers^*} \otimes B) \\
		&= \id_{\complexNumbers^*} \otimes \tr(\id_{\complexNumbers^*}) \, B  \\
		&= \id_{\complexNumbers^*} \otimes B\\
		&= A.
	\end{align*}
	On the other hand, for any $B \in \Lin_\complexNumbers(Y)$, we have
	\begin{align*}
		\tr_1(f(B)) &= \tr_1(\id_{\complexNumbers^*} \otimes B)\\
		&= \tr(\id_{\complexNumbers^*}) B\\
		&= B.
	\end{align*}
	This proves that $f = (\tr_1)^{-1}$. Next we just have to show that $\tr_1(\psdii(\complexNumbers,Y))$ $=$ $\density(Y)$. The partial trace is a CPTP map, as seen in \ref{proposition:partial trace as cptp map}, so $\tr_1(\psdii(\complexNumbers,$ $X))$ only contains positive semidefinite operators. For any $A \in \psdii(\complexNumbers,Y)$ we have that $\tr_2 A = \id_{\complexNumbers^*}$. The total trace can be computed from any partial trace, which implies that
	\begin{equation}
		\tr(\tr_1 A) = \tr(\tr_2 A) = \tr(\id_{\complexNumbers^*}) = 1.
	\end{equation}
	Therefore
	\begin{equation}
		\tr_1(\psdii(\complexNumbers,Y)) \subseteq \density(Y).
	\end{equation}
	On the other hand, let $B \in \density(Y)$, then $f(B) = \id_{\complexNumbers^*} \otimes B$. It is positive semidefinite, because it is a tensor product of positive semidefinite operators. Also,
	\begin{equation}
		\tr_2(f(B)) = \tr_2 (\id_{\complexNumbers^*} \otimes B) = \id_{\complexNumbers^*} \tr B = \id_{\complexNumbers^*}.
	\end{equation}
	Notice that $\tr B =1$, since $B \in \density(Y)$. We proved that
	\begin{equation}
		f(\density(Y)) \subseteq  \psdii(\complexNumbers,Y).
	\end{equation}
	Since $f = (\tr_1)^{-1}$, it follows that
	\begin{equation}
		\density(Y) \subseteq \tr_1 (\psdii(\complexNumbers,Y)).
	\end{equation}
	We proved both inclusions, so
	\begin{equation}
		\tr_1 (\psdii(\complexNumbers,Y)) = \density(Y).
	\end{equation}
\end{proof}

Next we show that the set of density operators is compact convex and compute its dimension. This follows easily from previous results.

\begin{corollary}{}{density ops dimension}
	Let $Y$ be a finite dimensional complex Hilbert space, with $d_Y \geq 1$. The set $\density(Y)$ of density operators over $Y$ is compact convex. In particular, it is homeomorphic to a closed ball, whose dimension is
	\begin{equation}
		\dim_\realNumbers \density(Y) = d_Y^2-1.
	\end{equation}
\end{corollary}
\begin{proof}
	In proposition \ref{proposition:psdii(C,Y) is isomorphic to D(Y)} we proved that $\density(Y)$ is isomorphic to $\psdii(\complexNumbers,Y)$. In proposition \ref{proposition:cptp is compact convex} we proved that $\psdii(\complexNumbers,Y)$ is compact convex. By the isomorphism, the same is true for $\density(Y)$. Then the dimension of $\density(Y)$ is obtained as a special case of the dimension of $\psdii(\complexNumbers,Y)$. This dimension was computed in corollary \ref{corollary:dimension of psdii}, from which it follows that $\dim_\realNumbers \density(Y) = d_Y^2-1$. 
\end{proof}

Being a compact convex set, the Krein-Milman theorem says that $\density(Y)$ is the convex hull of its extreme points. Next we show that its extreme points are the vector states. Recall that a vector state over $Y$ is an operator of the form $\ket{\psi} \bra{\psi}$, where $\psi \in Y$ is a unit vector.

\begin{proposition}{}{extreme points of density operators are the pure states}
	Let $Y$ be a finite dimensional complex Hilbert space, with $d_Y \geq 1$. The extreme points of $\density(Y)$ are the vector states. 
\end{proposition}
\begin{proof}
	Let $\rho \in \density(Y)$ be a density operator. By definition, it is a positive semidefinite operator with $\tr \, \rho = 1$. Then $\rho$ can be diagonalized as
	\begin{equation}
		\rho = \sum_k \lambda_k \ket{v_k} \bra{v_k},
	\end{equation}
	where $\lambda_k \geq 0$ and $(v_k)_k$ is an orthonormal basis for $Y$. Also, since $\tr \, \rho = 1$, we have
	\begin{equation}
		\sum_k \lambda_k = 1.
	\end{equation}
	In other words, $\rho$ is a convex combination of the vector states $\ket{v_k}\bra{v_k}$. Of course, each vector state is also an element of $\density(Y)$. Also, since the vectors $v_k$ are linearly independent, the operators $\ket{v_k}\bra{v_k}$ are different. If $\rho$ is an extreme point of $\density(Y)$, then this convex combination must be trivial, which implies that $\lambda_l = 1$ for some $l$, and $\lambda_k = 0$ for $k\neq l$. This implies that every extreme point is a vector state.
	
	On the other hand, every vector state is an extreme point. Let $\ket{\psi} \in Y$ be a unit vector, from which we have the vector state $\ket{\psi} \bra{\psi}$. Write it as a convex combination
	\begin{equation}
		\ket{\psi}\bra{\psi} = t_1 \rho_1 + t_2 \rho_2,
	\end{equation}
	where $\rho_1, \rho_2 \in \density(Y)$, $t_1, t_2 \in (0,1)$ and $t_1+t_2 = 1$. This convex combination can only be trivial if $\rho_1 = \rho_2$. Diagonalizing both $\rho_1$ and $\rho_2$, and absorbing the eigenvalues into the eigenvectors, we can write them as
	\begin{equation}
		\rho_i = \sum_k \ket{\psi_k^i} \bra{\psi_k^i},
	\end{equation}
	for some vectors $\ket{\psi_k^i} \in Y$. Notice that we can have $\ket{\psi_k^i} = 0$. Then the convex combination is written as
	\begin{equation}
		\sum_i t_i \rho_i = \sum_i \sum_k t_i \ket{\psi_k^i} \bra{\psi_k^i}.
	\end{equation}
	Then, by theorem 2.6 of \cite{nielsen_chuang_2010}, the vectors $\sqrt{t_i} \ket{\psi_k^i}$ are multiples of $\ket{\psi}$. Therefore, $\rho_i$ is a multiple of $\ket{\psi} \bra{\psi}$. If
	\begin{equation}
		\rho_i = \lambda_i \ket{\psi} \bra{\psi},
	\end{equation} 
	then
	\begin{equation}
		\tr \rho_i = \lambda_i \tr(\ket{\psi} \bra{\psi}) = \lambda_i.
	\end{equation}
	Since $\tr \rho_i = 1$, we have $\lambda_i = 1$. This shows that
	\begin{equation}
		\rho_i = \ket{\psi}\bra{\psi},
	\end{equation}
	so the convex combination is trivial. Therefore, every vector state is an extreme point of $\density(Y)$.
\end{proof}

Next we show that the set of vector states is homeomorphic to a complex projective space. To simplify the notation, we'll prove this for $Y = \complexNumbers^n$, where $n\geq 1$. Before proving this result, let's remind the definition of the complex projective spaces. Projective spaces are studied in example 3.6 of chapter IV of \cite{masseySHT}.
\begin{definition}{}{CPn}
	For any natural number $n\geq 0$, define the complex projective space $\complexNumbers P^n$ as the quotient
	\begin{equation}
		\complexNumbers P^n = S^{2n+1}/\U(1),
	\end{equation}
	where
	\begin{equation}
		S^{2n+1} = \{(z_1,\dots, z_{n+1}) \in \complexNumbers^{n+1} \mid ||z|| = 1 \},
	\end{equation}
	\begin{equation}
		\U(1) = \{ \lambda \in \complexNumbers \mid |\lambda| = 1 \}.
	\end{equation}
	The equivalence relation over $S^{2n+1}$ is
	\begin{equation}
		z \sim z' \iff \exists \lambda \in \U(1) \text{ such that } z' = \lambda z.
	\end{equation}
	$\complexNumbers P^n$ is a compact Hausdorff topological space. Also, the projection $\pi \colon S^{2n+1} \to \complexNumbers P^n$ is a surjective, continuous and open function.
\end{definition}

\begin{proposition}{}{pure states is homeomorphic to CPn}
	Let $n\geq 1$ be a natural number. The function
	\begin{align*}
		f\colon \complexNumbers P^{\, n-1} &\longrightarrow \{ \ket{\psi}\bra{\psi} \in \density(\complexNumbers^n) \mid \psi \in S^{2n-1} \}\\
		[\psi] &\longmapsto \ket{\psi}\bra{\psi},
	\end{align*}
	is a homeomorphism.
\end{proposition}
\begin{proof}
	By theorem 2.6 of \cite{nielsen_chuang_2010}, two vector states $\ket{\psi} \bra{\psi}$ and $\ket{\varphi} \bra{\varphi}$ are equal iff they only differ by a phase. That is,
	\begin{equation}
		\ket{\psi} \bra{\psi} = \ket{\varphi} \bra{\varphi} \iff \exists \lambda \in \U(1) \text{ such that } \varphi = \lambda \psi.
	\end{equation}
	This is the same equivalence relation over $S^{2n-1}$ from the definition of $\complexNumbers P^{n-1}$. The set of vector states is the image of the function
	\begin{align*}
		p \colon S^{2n-1} &\longrightarrow  \{ \ket{\psi}\bra{\psi} \in \density(\complexNumbers^n) \mid \psi \in S^{2n-1} \}\\
		\psi &\longmapsto \ket{\psi}\bra{\psi}.
	\end{align*}
	Of course, $p$ is surjective. It is also continuous, because it amounts to taking products of the components of $\psi$. The previous results shows that the relation of $p$ equals the relation of $\pi \colon S^{2n-1} \to \complexNumbers P^{n-1}$.
	From this, it follows that we can define the function $f$ above and that it is a bijection. These functions form the commutative diagram below.
	\begin{center}
		\begin{tikzcd}
			S^{2n-1} \arrow[r,"p"] \arrow[d,swap,"\pi"]& \{\ket{\psi}\bra{\psi} \in \density(\complexNumbers^n) \mid \psi \in S^{2n-1} \}\\
			\complexNumbers P^{n-1} \arrow[ru,swap,"f"] & {}
		\end{tikzcd}
	\end{center}
	Since $\pi$ is an open function and $p$ is continuous, it follows that $f$ is continuous. In fact, let $U$ be an open subset of the set of vector states. Since $\pi$ is surjective, we have that
	\begin{equation}
		f^{-1}(U) = \pi \pi^{-1} f^{-1}(U).
	\end{equation}
	The inverse image $\pi^{-1} f^{-1}(U)$ is the same as $(f \pi )^{-1}(U)$. Since $p = f \pi$, we have that
	\begin{equation}
		f^{-1}(U) = \pi p^{-1} (U).
	\end{equation}
	Since $p$ is continuous, the set $p^{-1}(U)$ is open. Then, since $\pi$ is an open function, $\pi(p^{-1}(U))$ is open. This shows that $f^{-1}(U)$ is open, so $f$ is continuous.
	
	We have that $f\colon \complexNumbers P^{n-1} \to  \{ \ket{\psi}\bra{\psi} \in \density(\complexNumbers^n) \mid \psi \in S^{2n-1} \}$ is a continuous bijection from a compact topological space to a Hausdorff space. From General Topology we know that $f$ must be a homeomorphism.
\end{proof}

\begin{corollary}{}{extreme points of psdii(C,Y) = CPn}
	Let $Y$ be a finite dimensional complex Hilbert space of dimension $d_Y \geq 1$. Then the set of extreme points of $\psdii(\complexNumbers,Y)$ is homeomorphic to $\complexNumbers P^{d_Y-1}$. In particular, it is a closed set.
\end{corollary}
\begin{proof}
	Proposition \ref{proposition:psdii(C,Y) is isomorphic to D(Y)} says that $\psdii(\complexNumbers,Y)$ is isomorphic to $\density(Y)$. By proposition \ref{proposition:linear bijection between extreme points}, such isomorphism restricts to a homeomorphism between the sets of extreme points. Proposition \ref{proposition:extreme points of density operators are the pure states} says that the extreme points of $\density(Y)$ are the vector states, and proposition \ref{proposition:pure states is homeomorphic to CPn} says that the set of vector states is homeomorphic to $\complexNumbers P^{d_Y-1}$ (using that $Y \cong \complexNumbers^{d_Y}$). Therefore, the set of extreme points of $\psdii(\complexNumbers,Y)$ is homeomorphic to $\complexNumbers P^{d_Y-1}$.
	
	Notice that $\complexNumbers P^{d_Y-1}$ is a compact topological space. Compactness is preserved by homeomorphisms, so the set of extreme points of $\psdii(\complexNumbers,Y)$ is compact. As any compact set, it is closed.
\end{proof}

\subsection{The closure of the set of extreme points}

Ruskai proved in \cite{ruskai2007open} that the closure of the set of extreme points of $\cptp(X,Y)$ comprises of the CPTP maps with Choi-rank at most $d_X$. Some authors call this closure the set of generalized extreme points. We review this result in this section. We'll prove a slight generalization of the result. Instead of working with the sets of all extreme points, we'll work with the extreme points with at most some Choi-rank. The proof is essentially the same from theorem 1 of \cite{ruskai2007open}, but using $\psdii(X,Y)$ instead of $\cptp(X, Y)$. The main difference is that the more general proof requires examples of extreme CPTP maps for each Choi-rank, while the original result only requires an example for Choi-rank $d_X$. Such examples will be constructed in section \ref{section: extreme cptp}. The case where $X = \{0\}$ or $Y= \{0\}$ is trivial, so we restrict attention to spaces of higher dimension.

\begin{proposition}{}{closure of extreme points of cptp(X,Y)}
	Let $X$ and $Y$ be finite dimensional complex Hilbert spaces, with $d_X \geq 1$ and $d_Y \geq 1$. Let also $r \in \naturalNumbers$ satisfying the bound for the possible Choi-ranks of extreme CPTP maps:
	\begin{equation}
		\frac{d_X}{d_Y} \leq r \leq d_X ,
	\end{equation}
	as in corollary \ref{corollary:bounds on rank of extreme cptp}. Denote by $E_r$ the set of extreme points of $\psdii(X,Y)$ with rank at most $r$. Then its closure is the set of all points with rank at most $r$:
	\begin{equation}
		\overline{E_r} = \bigcup_{d_X /d_Y \leq s \leq r} \psdii(X,Y)_s .
	\end{equation}
\end{proposition}
\begin{proof}
	It is immediate that we have the inclusion
	\begin{equation}
		E_r \subseteq \bigcup_{d_X / d_Y \leq s \leq r} \psdii(X,Y)_s .
	\end{equation}
	The $s$ is at least $d_X / d_Y$, because there is no CPTP map with smaller Choi-rank, as proved in proposition \ref{proposition:general CPTP rank bounds}. Taking the closure, we get that
	\begin{equation}
		\overline{E_r } \subseteq \overline {\bigcup_{d_X / d_Y \leq s \leq r} \psdii(X,Y)_s }.
	\end{equation}
	
	Next we show that the union $\bigcup_{d_X / d_Y \leq s \leq r} \psdii(X,Y)_s$ is a closed subset of $\psdii(X,Y)$. If $A \in \psdii(X,Y)$ and $\rank (A) \geq r +1$, the same is true for any other operator close enough to $A$. One way to prove this is the following. Let $[A]$ be the matrix of $A$ with respect to some fixed basis. If $\rank (A) \geq r +1 $, then there is some square submatrix of $[A]$, with $r+1$ lines and columns, whose determinant is not 0. Since the determinant is a continuous function, there exists some $\varepsilon > 0$ such that if $||B-A|| < \varepsilon$, then the correspondent submatrix of $[B]$ has also a non zero determinant. Therefore $\rank (B) \geq r+1$. This shows that $\psdii(X,Y) \setminus \bigcup_{d_X / d_Y \leq s \leq r} \psdii(X,Y)_s$ is an open subset of $\psdii(X,Y)$. Therefore, $ \bigcup_{d_X / d_Y \leq s \leq r} \psdii(X,Y)_s$ is a closed subset of $\psdii(X,Y)$. The closure of a closed set is itself, so
	\begin{equation}
		\overline{\bigcup_{d_X / d_Y \leq s \leq r} \psdii(X,Y)_s} = \bigcup_{d_X / d_Y \leq s \leq r} \psdii(X,Y)_s.
	\end{equation}
	Combining this result with the inclusion above, we get that
	\begin{equation}
		\overline{E_r} \subseteq \bigcup_{d_X / d_Y \leq s \leq r} \psdii(X,Y)_s.
	\end{equation}
	
	Now we prove the converse statement, that is,
	\begin{equation}
		\bigcup_{d_X / d_Y \leq s \leq r} \psdii(X,Y)_s \subseteq \overline{E_r} .
	\end{equation}
	Let $A \in \bigcup_{d_X / d_Y \leq s \leq r} \psdii(X,Y)_s$. The rank of $A$ is at most $r$, so it can be written as
	\begin{equation}
		A = \sum_{k=1}^{r} \ket{\psi_k} \bra{\psi_k},
	\end{equation}
	where $(\ket{\psi_k})_{k=1,\dots,r}$ is a finite sequence of vectors in $X^* \otimes Y$, which don't need to be linearly independent. Let also $B \in \psdii(X,Y)$ be any extreme point with rank $r$, which exists by the examples to be shown in section \ref{section: extreme cptp}. Write it as
	\begin{equation}
		B = \sum_{k=1}^{r} \ket{\varphi_k} \bra{\varphi_k},
	\end{equation}
	where $(\ket{\varphi_k})_{k=1,\dots,r}$ is a finite sequence of linearly independent vectors in $X^* \otimes Y$. The idea is to use $B$ to construct a curve that starts at $A$ and immediately enters the set of extreme points. To do this, first define the vectors
	\begin{equation}
		\ket{\gamma_k (t)} \coloneqq (1-t)\ket{\psi_k}+t\ket{\varphi_k},
	\end{equation}
	for $t \in [0,1]$. Then define the operator
	\begin{equation}
		\label{equation: closure of extremes C(t)}
		C(t) \coloneqq \sum_k \ket{\gamma_k (t)} \bra{\gamma_k (t)}.
	\end{equation}
	It is a continuous curve that starts at $A$ and ends at $B$, that is,
	\begin{equation}
		C(0) = A,
	\end{equation}
	\begin{equation}
		C(1) = B.
	\end{equation}
	Each term $\ket{\gamma_k (t)} \bra{\gamma_k (t)}$ is a positive semidefinite operator, which implies that $C(t)$ is positive semidefinite. But $\tr_2 C(t)$ doesn't need to be $\id_{X^*}$. Fortunately, we can correct the operator to get the partial trace we want. Since
	\begin{equation}
		\tr_2 C(0) = \tr_2 A = \id_{X^*},
	\end{equation}
	in particular the operator $\tr_2 C(0)$ is invertible. The set $\mathrm{GL}(X^* \otimes Y)$ of invertible operators is an open subset of $\Lin_\complexNumbers (X^* \otimes Y)$. This implies that, for some $\varepsilon_1 \in (0,1]$, $\tr_2 C(t)$ is invertible for $t \in [0, \varepsilon_1)$. Also, since $C(t)$ is positive semidefinite and $\tr_2$ is a CPTP map, as seen in proposition \ref{proposition:partial trace as cptp map}, it follows that $\tr_2 C(t)$ is positive semidefinite. Therefore, we have the operator $(\tr_2 C(t))^{-1/2}$, for $t \in [0,\varepsilon_1)$. We use this operator to correct the partial trace. Define
	\begin{equation}
		P(t) \coloneqq (\tr_2 C(t))^{-1/2} \otimes \id_Y,
	\end{equation}
	\begin{equation}
		\label{equation: closure of extremes D(t)}
		D(t) \coloneqq P(t) C(t) P(t).
	\end{equation}
	It is important to check that these are continuous functions of $t$. The operation of inverting matrices is a continuous function. Taking the square root of PSD operators is also a continuous function, as proved in \ref{proposition:square root of psd operators is continuous}. Then it follows that $P(t)$ is a continuous function. $D(t)$ is constructed by multiplying continuous functions, so it is also continuous.
	
	$D(t)$ is the conjugation of the positive semidefinite operator $C(t)$ by the operator $P(t)$, which implies that $D(t)$ is positive semidefinite. Also,
	\begin{align*}
		\tr_2 D(t) &= \tr_2\left\{[(\tr_2 C(t))^{-1/2} \otimes \id_Y] \ C(t) \ [(\tr_2 C(t))^{-1/2} \otimes \id_Y]\right\}\\
		&= (\tr_2 C(t))^{-1/2} \ \tr_2 C(t) \ (\tr_2 C(t))^{-1/2}\\
		&= \id_{X^*},
	\end{align*}
	so $D(t)$ has the correct partial trace. This shows that $D(t) \in \psdii(X,Y)$, for $t \in [0,\varepsilon_1)$.
	
	Next we want to show that $D(t)$ is an extreme point for small enough $t>0$. $D(t)$ can be written as
	\begin{equation}
		D(t) = \sum_k \ket{\eta_k (t)} \bra{\eta_k (t)},
	\end{equation}
	where
	\begin{equation}
		\ket{\eta_k (t)} = P(t) \ket{\gamma_k (t)}.
	\end{equation}
	Since $P(t)$ is invertible, the vectors $\ket{\eta_k (t)}$ are linearly independent iff the vectors $\ket{\gamma_k (t)}$ are linearly independent. By proposition \ref{proposition:gram matrix}, $(\ket{\gamma_k (t)})_k$ is linearly independent iff its Gram matrix $G(t)$ is invertible, which is also equivalent to $\det G(t)$ being non zero. The function $\det G(t)$ is a polynomial over the variable $t$. As such, the polynomial is identically zero or it has a finite number of roots. It isn't identically zero, because $\ket{\gamma_k (1)} = \ket{\varphi_k}$, and $(\ket{\varphi_k})_k$ is linearly independent. Therefore, $\det G(1) \neq 0$. Then $\det G(t)$ is a polynomial with a finite number of roots. This implies that there exists some $\varepsilon_2$ with
	\begin{equation}
		0 < \varepsilon_2 \leq \varepsilon_1,
	\end{equation}
	such that
	\begin{equation}
		\det G(t) \neq 0,
	\end{equation}
	for any $t \in (0,\varepsilon_2)$. Therefore, $(\ket{\eta_k (t)})_k$ is linearly independent for any $t \in (0, \varepsilon_2)$. We can apply a similar reasoning for $\tr_2 \ket{\eta_k (t)} \bra{\eta_l (t)}$. We have that
	\begin{align*}
		\tr_2 &\ket{\eta_k (t)} \bra{\eta_l (t)} = \tr_2\left(\ P(t) \ \ket{\gamma_k (t)} \bra{\gamma_l (t)} \ P(t) \ \right)\\
		&= \tr_2 \left\{\left((\tr_2 C(t))^{-1/2} \otimes \id_Y \right) \ \ket{\gamma_k (t)} \bra{\gamma_l (t)} \  \left((\tr_2 C(t))^{-1/2} \otimes \id_Y \right) \right\} \\
		&= (\tr_2 C(t))^{-1/2} \ \tr_2 \ket{\gamma_k (t)} \bra{\gamma_l (t)} \ (\tr_2 C(t))^{-1/2}.
	\end{align*}
	We have that $\tr_2 \ket{\eta_k (t)} \bra{\eta_l (t)}$ is the conjugation of $\tr_2 \ket{\gamma_k (t)} \bra{\gamma_l (t)}$ by the invertible operator $(\tr_2 C(t))^{-1/2}$. This implies that $(\tr_2 \ket{\eta_k (t)} \bra{\eta_l (t)})_{k,l}$ is linearly independent iff $(\tr_2 \ket{\gamma_k (t)} \bra{\gamma_l (t)})_{k,l}$ is linearly independent. Let $G'(t)$ be the Gram matrix of $(\tr_2 \ket{\gamma_k (t)} \bra{\gamma_l (t)})_{k,l}$. For $t=1$ we have that
	\begin{equation*}
		\tr_2 \ket{\gamma_k (1)} \bra{\gamma_l (1)} = \tr_2\ket{\varphi_k} \bra{\varphi_l} .
	\end{equation*}
	Since $B$ is an extreme point, by corollary \ref{corollary:extreme psdii in terms of psik} we know that $(\tr_2 \ket{\varphi_k}\bra{\varphi_l})_{k,l}$ is linearly independent. This implies that $\det G'(1) \neq 0$. Therefore, $\det G'(t)$ is a polynomial over $t$ with a finite number of roots. This implies that there exists some $\varepsilon_3$ with
	\begin{equation}
		0 < \varepsilon_3 \leq \varepsilon_2,
	\end{equation}
	such that
	\begin{equation}
		\det G'(t) \neq 0,
	\end{equation}
	for any $t \in (0,\varepsilon_3)$. This proves that both $(\ket{\eta_k (t)})_k$ and $(\ket{\eta_k (t)} \bra{\eta_l (t)})_{k,l}$ are linearly independent, for any $t \in (0, \varepsilon_3)$. By corollary \ref{corollary:extreme psdii in terms of psik}, it follows that $D(t)$ is an extreme point of $\psdii(X,Y)$, for any $t \in (0, \varepsilon_3)$. Lastly, notice that
	\begin{equation}
		D(0) = P(0) C(0) P(0) = \id_{X^* \otimes Y} \, A \, \id_{X^* \otimes Y} = A,
	\end{equation}
	so $D(t)$ is a continuous curve that starts at $A$ and is an extreme point for $t \in (0,\varepsilon_3)$. This implies that $A$ is in the closure of the set of extreme points of $\psdii(X,Y)$.
	
	Lastly, we want to prove that
	\begin{equation}
		\rank(D(t)) \leq r,
	\end{equation}
	for any $t \in [0, \varepsilon_3)$. From equation \ref{equation: closure of extremes D(t)} and since $P(t)$ is invertible, it follows that
	\begin{equation}
		\rank(D(t)) = \rank(C(t)).
	\end{equation}
	From equation \ref{equation: closure of extremes C(t)}, $C(t)$ is a sum of $r$ operators $\ket{\gamma_k(t)} \bra{\gamma_k(t)}$, each of these operators having rank at most 1. The rank of the sum is at most the sum of the ranks, so
	\begin{equation}
		\rank(C(t)) \leq r.
	\end{equation}
	This proves that
	\begin{equation}
		\rank(D(t)) \leq r,
	\end{equation}
	for any $t \in [0, \varepsilon_3 )$.
	
	We showed that $D(t)$ is a continuous curve such that
	\begin{equation}
		D(0) = A,
	\end{equation}
	and that $D(t)$ is an extreme point with rank at most $r$ for any $t \in (0, \varepsilon_3)$. Therefore
	\begin{equation}
		A \in \overline{E_r}.
	\end{equation}
	This proves the inclusion
	\begin{equation}
		\bigcup_{d_X / d_Y \leq s \leq r} \psdii(X,Y)_s \subseteq \overline{E_r}.
	\end{equation}
	
\end{proof}

\subsection{A surjective map from the purification of quantum states}
\label{section: a surjective map from the purification}

This section has the objective of constructing a continuous and surjective function $\pi \colon S^{d_Y}_{d_X,r} \to \bigcup_{d_X / d_Y \leq k \leq r} \psdii(X,Y)_k$, for $d_X / d_Y \leq r \leq d_X d_Y$. The set $S^{d_Y}_{d_X, r}$ will be defined later in definition \ref{definition:generalized sphere}, but it is isomorphic to a (compact) Stiefel manifold. The function $\pi$ is also open, but such proof is new, so it will be postponed to chapter \ref{chapter: projection properties}. It is constructed in such a way that it resembles the projection $\pi \colon S^{2n+1} \to \complexNumbers P^n$ of definition \ref{definition:CPn}. This will be done by first constructing a surjective function that comes from the idea of purification of mixed states. Then we modify the presentation to obtain the desired function. This will be useful in section \ref{section: partial decomposition into cells}. The idea isn't new, it already exists in \cite{iten_colbeck}, but a new perspective is given to it. Also, we don't work with a fixed Choi-rank.

In section 2.5 of \cite{nielsen_chuang_2010} is shown that density operators can be purified. This means that any density operator $\rho \in \density(X)$ can be expressed as a partial trace of a vector state over a bigger Hilbert space. It can be shown that there exists a unit vector $\ket{\psi} \in X\otimes X$ such that
\begin{equation}
	\rho = \tr_2 \ket{\psi}\bra{\psi}.
\end{equation}
In fact, $\rho$ can be diagonalized as
\begin{equation}
	\rho = \sum_k \lambda_k \ket{\psi_k}\bra{\psi_k},
\end{equation}
where $\lambda_k \geq 0$, $\sum_k \lambda_k = 1$ and $(\psi_k)_k$ is an orthonormal basis of $X$. Then take
\begin{equation}
	\ket{\psi} = \sum_k \sqrt{\lambda_k} \, \ket{\psi_k}\otimes \ket{\psi_k}.
\end{equation}
The squared norm of this vector is
\begin{equation}
	||\psi||^2 = \sum_k |\sqrt{\lambda_k}|^2 = \sum_k \lambda_k = 1.
\end{equation}
We get that
\begin{align*}
	\tr_2 \ket{\psi}\bra{\psi} &= \tr_2 \left( \sum_k \sqrt{\lambda_k} \, \ket{\psi_k}\otimes \ket{\psi_k} \right) \left( \sum_l \sqrt{\lambda_l} \, \bra{\psi_l}\otimes \bra{\psi_l} \right)\\
	&= \sum_{k,l} \sqrt{\lambda_k \lambda_l} \, \tr_2 (\ket{\psi_k}\bra{\psi_l} \otimes \ket{\psi_k}\bra{\psi_l})\\
	&= \sum_{k,l} \sqrt{\lambda_k \lambda_l} \, \ket{\psi_k}\bra{\psi_l} \, \tr( \ket{\psi_k}\bra{\psi_l})\\
	&= \sum_{k,l} \sqrt{\lambda_k \lambda_l} \, \ket{\psi_k}\bra{\psi_l} \, \langle \psi_l \mid \psi_k \rangle\\
	&= \sum_{k,l} \sqrt{\lambda_k \lambda_l} \, \ket{\psi_k}\bra{\psi_l} \, \delta_{k,l}\\
	&= \sum_k \lambda_k \ket{\psi_k}\bra{\psi_k}\\
	&= \rho.
\end{align*}

From this idea, we have the continuous function
\begin{align*}
	S(X\otimes X) &\longrightarrow \density(X)\\
	\ket{\psi} &\longmapsto \tr_2 \ket{\psi} \bra{\psi},
\end{align*}
where $S(X\otimes X) = \{\psi \in X\otimes X \mid ||\psi||=1\}$ is the sphere in $X\otimes X$. That every density operator $\rho \in \density(X)$ can be purified to a vector state $\ket{\psi}\bra{\psi}$, this is equivalent to say that the function $S(X\otimes X) \to \density(X)$ is surjective.

Every element $A \in \psdii(X,Y)$ is a multiple of a density operator. This is because $A$ must satisfy that $\tr_2 A = \id_{X^*}$, which implies that
\begin{equation}
	\tr \, A = \tr(\tr_2 A) = \tr(\id_{X^*}) = d_X.
\end{equation}
Therefore $A/d_X$ is a density operator, and we have an inclusion
\begin{equation}
	\frac{1}{d_X} \psdii(X,Y) \subseteq \density(X^* \otimes Y).
\end{equation}
We can then apply the same idea of purifying density operators to construct a surjective function from a space of vectors $\ket{\psi}$ to $\psdii(X,Y)$. Instead of doing this only for $\psdii(X,Y)$, we'll construct a surjective function onto $\bigcup_{d_X / d_Y \leq k\leq r} \psdii(X,Y)_k$, the set of operators with $\rank \, A \leq r$, for $d_X / d_Y \leq r \leq d_X d_Y$. A reason for doing this is that, by proposition \ref{proposition:closure of extreme points of cptp(X,Y)}, if we take $r=d_X$, this union is the closure of the set of extreme points of $\psdii(X,Y)$.

We'll first define a function similar to the function $S(X\otimes X) \to \density(X)$ defined above. Then we'll modify slightly the definition by applying an isomorphism, obtaining an equivalent function, but which resembles the projection $S^{2n+1} \to \complexNumbers P^n$ of the complex projective space of definition \ref{definition:CPn}. This other function will be important in section \ref{section: partial decomposition into cells}.

\begin{definition}{}{pi: X Y Cr -> End(X Y)}
	Let $X$ and $Y$ be finite dimensional complex Hilbert spaces, with $d_X \geq 1$ and $d_Y \geq 1$, and let $r \geq 1$. We define the function
	\begin{align*}
		\pi \colon X^* \otimes Y \otimes \complexNumbers^r &\longrightarrow \Lin_\complexNumbers (X^* \otimes Y)\\
		\ket{\psi} &\longmapsto \tr_3 \ket{\psi}\bra{\psi}.
	\end{align*} 
\end{definition}
Of course, this function $\pi$ is continuous, since $\tr_3$ is $\complexNumbers$-linear and $\ket{\psi}\bra{\psi}$ comprises of taking products of the components of $\psi$.

Even though we are relating this function to the purification of quantum states, it also appears in Mathematics in another form. We have an isomorphism $X^* \otimes Y \otimes \complexNumbers^r \cong \Lin_\complexNumbers ((\complexNumbers^r)^* , X^* \otimes Y)$. Every positive semidefinite operator $A \in \psd(X^* \otimes Y)$ with $\rank(A) \leq r$ can be written as $A = B B^\dag$, for some $B \in \Lin_\complexNumbers ((\complexNumbers^r)^* , X^* \otimes Y)$. The function $\pi$ defined above is another form for the function
\begin{align*}
	\Lin_\complexNumbers ((\complexNumbers^r)^* , X^* \otimes Y) &\longrightarrow \psd(X^* \otimes Y) \\
	B &\longmapsto B B^\dag .
\end{align*}
This relation between the functions will become more clear when we start working with matrices. We can also notice that the vector $\ket{\psi}$ encodes a Kraus representation. This is because of corollary \ref{corollary:bijection between kraus operators and psis} and the vectors $\ket{\psi_k} \in X^* \otimes Y$ that will be defined in the proof of the next proposition.

\begin{proposition}{}{image of non restricted projection}
	The image of the function $\pi$ of definition \ref{definition:pi: X Y Cr -> End(X Y)} is
	\begin{equation}
		\image \pi = \bigcup_{1 \leq k \leq r}\psd(X^* \otimes Y)_k ,
	\end{equation}
	that is, the positive semidefinite operators over $X^* \otimes Y$ with rank at most $r$.
\end{proposition}
\begin{proof}
	Let $(\ket{x^i})_i$, $(\ket{y_j})_j$ and  $(\ket{k})_k$ be orthonormal basis for $X^*$, $Y$ and $\complexNumbers^r$, respectively. Then any vector $\ket{\psi} \in X^* \otimes Y \otimes \complexNumbers^r$ can be uniquely written as
	\begin{equation}
		\ket{\psi} = \sum_{i,j,k} \psi_{i,j,k} \, \ket{x^i} \otimes \ket{y_j} \otimes \ket{k},
	\end{equation}
	where $\psi_{i,j,k} \in \complexNumbers$. Then
	\begin{align*}
		\pi(\psi) &= \tr_3 (\ket{\psi}\bra{\psi})\\
		&= \tr_3 \left[ \left(\sum_{i,j,k} \psi_{i,j,k} \ket{x^i} \otimes \ket{y_j} \otimes \ket{k} \right)\left(\sum_{i',j',k'} \overline{\psi_{i',j',k'}} \bra{x^{i'}} \otimes \bra{y_{j'}} \otimes \bra{k'} \right) \right]\\
		&= \sum_{i,j,k,i',j',k'} \psi_{i,j,k} \overline{\psi_{i',j',k'}} \, \tr_3 [ \, (\ket{x^i} \otimes \ket{y_j} \otimes \ket{k})(\bra{x^{i'}} \otimes \bra{y_{j'}} \otimes \bra{k'}) \, ]\\
		&= \sum_{i,j,k,i',j',k'} \psi_{i,j,k} \overline{\psi_{i',j',k'}} \, \tr_3 [ \, \ket{x^i}\bra{x^{i'}} \otimes \ket{y_j}\bra{y_{j'}} \otimes \ket{k}\bra{k'} \, ]\\
		&= \sum_{i,j,k,i',j',k'} \psi_{i,j,k} \overline{\psi_{i',j',k'}} \, \ket{x^i}\bra{x^{i'}} \otimes \ket{y_j}\bra{y_{j'}} \, \tr(\ket{k}\bra{k'})\\
		&= \sum_{i,j,k,i',j',k'} \psi_{i,j,k} \overline{\psi_{i',j',k'}} \, \ket{x^i}\bra{x^{i'}} \otimes \ket{y_j}\bra{y_{j'}} \, \langle k' \mid k \rangle \\
		&= \sum_{i,j,k,i',j',k'} \psi_{i,j,k} \overline{\psi_{i',j',k'}} \, \ket{x^i}\bra{x^{i'}} \otimes \ket{y_j}\bra{y_{j'}} \, \delta_{k,k'} \\
		&= \sum_{i,j,i',j',k} \psi_{i,j,k} \overline{\psi_{i',j',k}} \, \ket{x^i}\bra{x^{i'}} \otimes \ket{y_j}\bra{y_{j'}} \\
		&= \sum_k \left( \sum_{i,j} \psi_{i,j,k} \, \ket{x^i} \otimes \ket{y_j}  \right) \left( \sum_{i',j'} \overline{\psi_{i',j',k}} \, \bra{x^{i'}} \otimes \bra{y_{j'}}  \right).
	\end{align*}
	Define the vector
	\begin{equation}
		\ket{\psi_k} \coloneqq  \sum_{i,j} \psi_{i,j,k} \, \ket{x^i} \otimes \ket{y_j} \in X^* \otimes Y.
	\end{equation}
	The previous results shows that
	\begin{equation}
		\pi(\psi) = \sum_{k=1}^r \ket{\psi_k}\bra{\psi_k}.
	\end{equation}
	This is a sum of $r$ positive semidefinite operators of rank at most 1. This implies that $\pi(\psi)$ is positive semidefinite and $\rank(\pi(\psi))\leq r$. This proves that
	\begin{equation}
		\image \, \pi \subseteq \bigcup_{0\leq k \leq r}\psd(X^* \otimes Y)_k .
	\end{equation}
	
	Now let's prove the other inclusion. Let $A \in \bigcup_{1 \leq k \leq r} \psd(X^* \otimes Y)_r$, then we can write it as
	\begin{equation}
		A = \sum_{k=1}^r \ket{\psi_k}\bra{\psi_k},
	\end{equation}
	where $\psi_k \in X^* \otimes Y$. The vectors $\psi_k$ may be obtained from a diagonalization of $A$. More precisely, since $\rank(A) \leq r$, at most $r$ of its eigenvalues are non zero. Because of this, it admits a diagonalization
	\begin{equation}
		A = \sum_{k=1}^r \lambda_k \ket{v_k}\bra{v_k},
	\end{equation}
	where $\lambda_k \geq 0$ and $(v_k)_{k=1,\dots,r}$ is a sequence of orthonormal vectors in $X^* \otimes Y$. Then take
	\begin{equation}
		\ket{\psi_k} = \sqrt{\lambda_k} \ket{v_k}.
	\end{equation}
	Now define
	\begin{equation}
		\ket{\psi} \coloneqq \sum_k \ket{\psi_k} \otimes \ket{k} \in X^* \otimes Y \otimes \complexNumbers^r .
	\end{equation}
	Then
	\begin{align*}
		\pi(\psi) &= \tr_{\complexNumbers^*} \left[ \left( \sum_k \ket{\psi_k} \otimes \ket{k} \right) \left( \sum_{k'} \bra{\psi_{k'}} \otimes \bra{k'} \right) \right]\\
		&= \sum_{k,k'} \ket{\psi_k}\bra{\psi_{k'}} \, \tr(\ket{k}\bra{k'})\\
		&= \sum_{k,k'} \ket{\psi_k}\bra{\psi_{k'}} \, \delta_{k,k'}\\
		&= \sum_k \ket{\psi_k}\bra{\psi_k}\\
		&= A.
	\end{align*}
	This proves that
	\begin{equation}
		\bigcup_{1\leq k \leq r}\psd(X^* \otimes Y)_k \subseteq \image \, \pi.
	\end{equation}
\end{proof}

We want the image to be $\bigcup_{d_X / d_Y \leq k \leq r} \psdii(X,Y)_k$ (with the subscript 2). Notice that CPTP maps have Choi-rank of at least $d_X / d_Y$, hence taking $k \geq d_X / d_Y$. Such image is obtained by restricting $\pi$ to
\begin{equation*}
	\pi^{-1}\left( \bigcup_{d_X / d_Y \leq k \leq r} \psdii(X,Y)_k \right) .
\end{equation*}
In \cite{iten_colbeck} was already proved that this pre-image corresponds to a Stiefel manifold. Here we give yet another presentation for the pre-image which resembles the definition of a sphere. This other presentation will help us see how the projection $\pi$ of definition \ref{definition:pi: X Y Cr -> End(X Y)} is a generalization of the projection $\pi \colon S^{2n+1} \to \complexNumbers P^n$ of definition \ref{definition:CPn}. This gives an insight which will be important in section \ref{section: partial decomposition into cells}.

\begin{proposition}{}{pre-image of pi is stiefel manifold}
	Let $X$ and $Y$ be finite dimensional complex Hilbert spaces, with $d_X \geq 1$ and $d_Y \geq 1$. Let $\pi$ be the function of definition \ref{definition:pi: X Y Cr -> End(X Y)}. Let $(\ket{x^i})_i$, $(\ket{y_j})_j$ and $(\ket{k})_k$ be orthonormal basis for $X^*$, $Y$ and $\complexNumbers^r$, respectively. Then any vector $\ket{\psi} \in X^* \otimes Y \otimes \complexNumbers^r$ can be uniquely written as
	\begin{equation}
		\ket{\psi} = \sum_{i,j,k} \psi_{i,j,k} \ket{x^i} \otimes \ket{y_j} \otimes \ket{k}.
	\end{equation}
	Denote by $\psi_i$ the vector obtained by fixing the component $i$, that is,
	\begin{equation}
		\psi_i \coloneqq (\psi_{i,j,k})_{j,k} \in \complexNumbers^{d_Y r}.
	\end{equation}
	The pair of indices $(j,k)$ can be interpreted as a single index, ordered by the lexicographic order. Then we have that
	\begin{multline}
		\label{equation: pre-image is stiefel manifold}
		\pi^{-1}\left( \bigcup_{d_X / d_Y \leq k \leq r} \psdii(X,Y)_k \right) = \\
		\{\psi \in X^* \otimes Y \otimes \complexNumbers^r \mid (\psi_i )_i \text{ is orthonormal }\}.
	\end{multline}
	Organizing the vectors $\psi_i$ as columns of a matrix in $M_{d_Y r,d_X}(\complexNumbers)$, this identifies the pre-image with a Stiefel manifold:
	\begin{equation}
		\pi^{-1}\left( \bigcup_{d_X / d_Y \leq k \leq r} \psdii(X,Y)_k \right) \cong V_{d_Y r, d_X}(\complexNumbers).
	\end{equation}
	Also denote by $\psi_j$ the matrix
	\begin{equation}
		\psi_j \coloneqq (\psi_{i,j,k})_{i,k} \in M_{d_X,r}(\complexNumbers).
	\end{equation}
	Then we have that
	\begin{multline}
		\label{equation: pre-image is generalized sphere}
		\pi^{-1}\left( \bigcup_{d_X / d_Y \leq k \leq r} \psdii(X,Y)_k \right) =\\ \left\{\psi \in X^* \otimes Y \otimes \complexNumbers^r \, \bigg| \, \sum_j \psi_j \psi_j^\dag = I_{d_X} \right\}.
	\end{multline}
\end{proposition}
\begin{proof}
	By proposition \ref{proposition:image of non restricted projection}, we already know that
	\begin{equation}
		\image \pi = \bigcup_{1 \leq k \leq r} \psd(X,Y)_k.
	\end{equation}
	Therefore,
	\begin{equation}
		\pi(\psi) \in \bigcup_{d_X / d_Y \leq k \leq r} \psdii(X,Y)_k \iff \tr_2 \pi(\psi) = \id_{X^*}.
	\end{equation}
	We just have to find a condition over $\psi$ that is equivalent to the equation $\tr_2 \pi(\psi) = \id_{X^*}$.
	
	Given $\psi \in X^* \otimes Y \otimes \complexNumbers^r$, we have
	\begin{align*}
		\tr_2 &\pi(\psi) = \tr_2 \tr_3 \ket{\psi}\bra{\psi} \\ 
		&= \tr_{2,3} \left[ \left(\sum_{i,j,k} \psi_{i,j,k} \ket{x^i} \otimes \ket{y_j} \otimes \ket{k} \right)\left(\sum_{i',j',k'} \overline{\psi_{i',j',k'}} \bra{x^{i'}} \otimes \bra{y_{j'}} \otimes \bra{k'} \right) \right]\\
		&= \sum_{i,j,k,i',j',k'} \psi_{i,j,k} \overline{\psi_{i',j',k'}} \, \tr_{2,3} [ \, \ket{x^i}\bra{x^{i'}} \otimes \ket{y_j}\bra{y_{j'}} \otimes \ket{k}\bra{k'} \, ]\\
		&= \sum_{i,j,k,i',j',k'} \psi_{i,j,k} \overline{\psi_{i',j',k'}} \, \ket{x^i}\bra{x^{i'}} \, \tr(\ket{y_j}\bra{y_{j'}}) \, \tr(\ket{k}\bra{k'})\\
		&= \sum_{i,j,k,i',j',k'} \psi_{i,j,k} \overline{\psi_{i',j',k'}} \, \ket{x^i}\bra{x^{i'}} \, \delta_{j,j'} \delta_{k,k'} \\
		&= \sum_{i,i',j,k} \psi_{i,j,k} \overline{\psi_{i',j,k}} \, \ket{x^i}\bra{x^{i'}}  \\
		&= \sum_{i,i'} \left( \sum_{j,k} \psi_{i,j,k} \overline{\psi_{i',j,k}} \right) \, \ket{x^i}\bra{x^{i'}}  \\
		&= \sum_{i,i'} \overline{\langle \psi_i \mid \psi_{i'} \rangle} \,  \ket{x^i}\bra{x^{i'}}.
	\end{align*}
	The identity function is written on the basis as
	\begin{equation}
		\id_{X^*} = \sum_{i,i'} \delta_{i,i'} \ket{x^i}\bra{x^{i'}}.
	\end{equation}
	Therefore, the equation
	\begin{equation}
		\tr_2 \pi(\psi) = \id_{X^*}
	\end{equation}
	is equivalent to
	\begin{equation}
		\sum_{i,i'} \overline{\langle \psi_i \mid \psi_{i'} \rangle} \,  \ket{x^i}\bra{x^{i'}} = \sum_{i,i'} \delta_{i,i'} \ket{x^i}\bra{x^{i'}}.
	\end{equation}
	The last equation is equivalent to
	\begin{equation}
		\langle \psi_i \mid \psi_{i'} \rangle = \delta_{i,i'},
	\end{equation}
	for $i,i' \in \{1,\dots,d_X\}$. This proves that
	\begin{equation}
		\pi(\psi) \in \bigcup_{d_X / d_Y \leq k \leq r} \psdii(X,Y)_k \iff (\ket{\psi_i})_i \text{ is orthonormal}.
	\end{equation}
	This proves equation \ref{equation: pre-image is stiefel manifold}.
	
	Lastly, we give another characterization in terms of the matrices $\psi_j$. We do this by expressing the inner product $\langle \psi_i, \psi_{i'} \rangle$ is terms of these matrices. We have that
	\begin{align*}
		\langle \psi_i, \psi_{i'} \rangle &= \sum_{j,k} \overline{\psi_{i,j,k}} \psi_{i',j,k}\\
		&= \sum_{j,k} \overline{(\psi_j)_{i,k}} (\psi_j)_{i',k}\\
		&= \sum_{j,k} (\psi_j^\dag)_{k,i} (\psi_j)_{i',k}\\
		&= \sum_j \sum_k  (\psi_j)_{i',k} (\psi_j^\dag)_{k,i}\\
		&= \sum_j (\psi_j \psi_j^\dag)_{i',i}\\
		&= \left(\sum_j \psi_j \psi_j^\dag \right)_{i',i} .
	\end{align*}
	Therefore, $(\ket{\psi_i})_i$ is orthonormal iff $\sum_j \psi_j \psi_j^\dag = I_{d_X}$. This proves equation \ref{equation: pre-image is generalized sphere}.
\end{proof}

The pre-image as a Stiefel manifold is useful for constructing examples, as in chapter \ref{chapter: examples extreme channels}. The last characterization given in this proposition is convenient for proving topological results, because of the similarity with the projection for the complex projective space.

In this section we'll focus on the last characterization. We have that
\begin{equation*}
	\pi \colon X^* \otimes Y \otimes \complexNumbers^r \to \Lin_\complexNumbers (X^* \otimes Y)
\end{equation*}
restricts to a surjection
\begin{equation*}
	\pi \colon \left\{\psi \in X^* \otimes Y \otimes \complexNumbers^r \, \bigg| \, \sum_j \psi_j \psi_j^\dag = I_{d_X} \right\} \to \bigcup_{d_X / d_Y \leq k \leq r} \psdii(X,Y)_k.
\end{equation*}
The advantage of this presentation is how it is related to a sphere. Of course, the Stiefel manifold is also a generalization of the sphere, where a single normalized vector is generalized to a matrix whose columns are orthonormal. But we'll think of generalizing in another way. The equation
\begin{equation}
	\sum_j \psi_j \psi_j^\dag = I_{d_X}
\end{equation}
is a generalization of the equation of a sphere
\begin{equation}
	\sum_j z_j \overline{z_j} = 1,
\end{equation}
where $z_j \in \complexNumbers$ is replaced with $\psi_j \in M_{d_X,r}(\complexNumbers)$ and $1$ is replaced with $I_{d_X}$. Let's introduce a notation for the generalization of the sphere.
\begin{definition}{}{generalized sphere}
	Let $m,n,r$ be non zero natural numbers, we define
	\begin{equation}
		S_{m,r}^n \coloneqq \left\{(\psi_j)_{j=1,\dots,n} \in M_{m,r}(\complexNumbers)^n \ \bigg| \ \sum_{j=1}^n \psi_j \psi_j^\dag = I_m \right\}.
	\end{equation}
\end{definition}

We have the isomorphism 
\begin{align*}
	X^* \otimes Y \otimes \complexNumbers^r &\longrightarrow M_{d_X,r}(\complexNumbers)^{d_Y}\\
	\psi &\longmapsto (\psi_j)_j,
\end{align*}
where $\psi_j \in M_{d_X,r}(\complexNumbers)$ is the matrix defined in proposition \ref{proposition:pre-image of pi is stiefel manifold}. Of course, this isomorphism depends on the basis for $X^*$, $Y$ and $\complexNumbers^r$. Composing $\pi$ with the inverse of this isomorphism, we get another projection, which we also call $\pi$:
\begin{equation*}
	\pi \colon S_{d_X,r}^{d_Y} \to \bigcup_{d_X / d_Y \leq k \leq r} \psdii(X,Y)_k.
\end{equation*}
It is convenient to expresses $\psi$ in terms of these matrices $\psi_j$. In the proof of proposition \ref{proposition:image of non restricted projection} we obtained the expression
\begin{equation}
	\pi(\psi) = \sum_{i,j,i',j',k} \psi_{i,j,k} \overline{\psi_{i',j',k}} \, \ket{x^i}\bra{x^{i'}} \otimes \ket{y_j}\bra{y_{j'}}.
\end{equation}
We just have to express this in terms of the matrices $\psi_j$. We have that
\begin{align*}
	\pi(\psi) &= \sum_{i,j,i',j'} \left(\sum_k (\psi_j)_{i,k} \overline{(\psi_{j'})_{i',k}} \right) \, \ket{x^i}\bra{x^{i'}} \otimes \ket{y_j}\bra{y_{j'}}\\
	&= \sum_{i,j,i',j'} \left(\sum_k (\psi_j)_{i,k} (\psi_{j'}^\dag)_{k,i'} \right) \, \ket{x^i}\bra{x^{i'}} \otimes \ket{y_j}\bra{y_{j'}}\\ 
	&= \sum_{i,j,i',j'} (\psi_j \psi_{j'}^\dag)_{i,i'} \, \ket{x^i}\bra{x^{i'}} \otimes \ket{y_j}\bra{y_{j'}}\\
\end{align*}
Composing with the isomorphism $M_{d_X,r}(\complexNumbers)^{d_Y} \cong X^* \otimes Y \otimes \complexNumbers^r$ we then get
\begin{align*}
	\pi \colon S_{d_X,r}^{d_Y} &\longrightarrow \bigcup_{d_X / d_Y \leq k \leq r} \psdii(X,Y)_k\\
	(\psi_j)_j &\longmapsto \sum_{i,j,i',j'} (\psi_j \psi_{j'}^\dag)_{i,i'} \, \ket{x^i}\bra{x^{i'}} \otimes \ket{y_j}\bra{y_{j'}}.
\end{align*}
This is the projection we'll be using, because of its similarity with the projection $\pi \colon S^{2n+1} \to \complexNumbers P^n$ of definition \ref{definition:CPn}.

We can order the pairs $(i,j)$ in the colexicographic order, that is,
\begin{equation}
	(i,j) < (i',j') \iff j<j' \text{ or } (j=j' \text{ and } i<i').
\end{equation}
Then, for any operator $A \in \Lin_\complexNumbers (X^*\otimes Y)$, its matrix $[A]$ with respect to the basis $\ket{x^i}\bra{x^{i'}} \otimes \ket{y_j}\bra{y_{j'}}$ takes the form
\begin{equation}
	[A] = \begin{pmatrix}
		A_{1,1} & \dots & A_{1,d_y}\\
		\vdots & \ddots & \vdots\\
		A_{d_Y,1} & \dots & A_{d_Y,d_Y}
	\end{pmatrix} \in M_{d_X d_Y}(\complexNumbers),
\end{equation}
where
\begin{equation}
	A_{j,j'} \coloneqq (A_{(i,j),(i',j')}) \in M_{d_X}(\complexNumbers).
\end{equation}
Take $A = \pi((\psi_j)_j)$. The expression for $\pi((\psi_j)_j)$ then says that
\begin{equation}
	A_{j,j'} = \psi_j \psi_{j'}^\dag,
\end{equation}
so
\begin{equation}
	\label{equation: [A]}
	[\pi((\psi_j)_j)] = \begin{pmatrix}
		\psi_1 \psi_{1}^\dag & \dots & \psi_1 \psi_{d_Y}^\dag\\
		\vdots & \ddots & \vdots\\
		\psi_{d_Y} \psi_{1}^\dag & \dots & \psi_{d_Y} \psi_{d_Y}^\dag
	\end{pmatrix}.
\end{equation}
This last matrix can also be expressed as
\begin{equation}
	\begin{pmatrix}
		\psi_1 \psi_{1}^\dag & \dots & \psi_1 \psi_{d_Y}^\dag\\
		\vdots & \ddots & \vdots\\
		\psi_{d_Y} \psi_{1}^\dag & \dots & \psi_{d_Y} \psi_{d_Y}^\dag
	\end{pmatrix} =
	\begin{pmatrix}
		\psi_1 \\
		\vdots \\
		\psi_{d_Y}
	\end{pmatrix}
	\begin{pmatrix}
		\psi_1^\dag & \dots & \psi_{d_Y}^\dag
	\end{pmatrix}.
\end{equation}
This is an expression of the form $[A] = B B^\dag$, which we anticipated in the beginning of this section. This expression could lead us to just concatenate the matrices $\psi_j$ vertically to form a single matrix. This is something useful to do, but the left hand side of the equation $\sum_j \psi_j \psi_j^\dag = I_{d_X}$ is not expressed as a product of such matrix with its adjoint. Rather, the equation  $\sum_j \psi_j \psi_j^\dag = I_{d_X}$ is equivalent to
\begin{equation}
	\begin{pmatrix}
		\psi_1 & \dots & \psi_{d_Y}
	\end{pmatrix}
	\begin{pmatrix}
		\psi_1^\dag \\
		\vdots \\
		\psi_{d_Y}^\dag
	\end{pmatrix} = I_{d_X}.
\end{equation}
For this reason, it is convenient to keep the matrices $\psi_j$ as separate objects, and then concatenate them into a single matrix when necessary.

Next we want to determine the relation of $\pi$ and prove that the projection $\pi$ is an open function, that is, it sends open sets to open sets. We said earlier that $\psi$ encodes a Kraus representation. In theorem 8.2 of \cite{nielsen_chuang_2010} is proved that two Kraus representations $(E_k)_{k=1,\dots,r}$ and $(F_k)_{k=1,\dots,r}$ represent the same CPTP map iff there exists a unitary matrix $V \in \U(r)$ such that
\begin{equation}
	E_k = \sum_{l} V_{k,l} F_l .
\end{equation} 
This matrix may not be unique, but it is unique when the Kraus operators $(E_k)_k$ and $(F_l)_l$ are linearly independent. The relation of $\pi$ is the same of the Kraus representations, but presented in a different way. It is also a simple consequence of Theorem 2.6 of \cite{nielsen_chuang_2010}.
\begin{proposition}{}{relation of pi}
	The relation of the function 
	\begin{equation*}
		\pi \colon S_{d_X,r}^{d_Y} \to \bigcup_{d_X / d_Y \leq k \leq r} \psdii(X,Y)_k,
	\end{equation*}
	defined above in this section, is the following:
	\begin{align*}
		\pi((\psi_j)_j) = \pi((\psi'_j)_j) &\iff \exists V \in \U(r) \text{ such that } \forall j \ \psi'_j = \psi_j V^\dag \\
		&\iff \exists V \in \U(r) \text{ such that } \begin{pmatrix}
			\psi'_1\\
			\vdots\\
			\psi'_{d_Y}
		\end{pmatrix}
		= \begin{pmatrix}
			\psi_1\\
			\vdots\\
			\psi_{d_Y}
		\end{pmatrix} V^\dag.
	\end{align*}
\end{proposition}
\begin{proof}
	$(\implies)$ We start with the first definition of $\pi$, given in definition \ref{definition:pi: X Y Cr -> End(X Y)}. In the proof of proposition \ref{proposition:image of non restricted projection} we obtained the following expression for $\pi(\psi)$:
	\begin{equation}
		\pi(\psi) = \sum_{k=1}^r \ket{\psi_k} \bra{\psi_k},
	\end{equation}
	where
	\begin{equation}
		\ket{\psi_k} = \sum_{i,j} \psi_{i,j,k} \ket{x^i \otimes y_j} \in X^* \otimes Y.
	\end{equation}
	Let $\psi, \psi' \in \pi^{-1}(\psdii(X,Y))$. Notice that $\pi(\psi)/d_X$ and $\pi(\psi')/d_X$ are density operators. Then theorem 2.6 of \cite{nielsen_chuang_2010} says that the equation
	\begin{equation}
		\sum_k \ket{\psi_k} \bra{\psi_k} = \sum_k \ket{\psi'_k} \bra{\psi'_k}
	\end{equation}
	holds iff there exists some $U \in \U(r)$ such that
	\begin{equation}
		\ket{\psi'_k} = \sum_l U_{k,l} \ket{\psi_l},
	\end{equation}
	for all $k$. We just need to rephrase this equation in terms of the matrices $\psi_j$. In components, this equation is written as
	\begin{equation}
		\psi'_{i,j,k} = \sum_l U_{k,l} \psi_{i,j,l},
	\end{equation}
	for all $i,j,k$. Making explicit the matrices $\psi_j$, we get
	\begin{equation}
		(\psi'_j)_{i,k} = \sum_l U_{k,l} (\psi_j)_{i,l}.
	\end{equation}
	The right hand side is the $(i,k)$ component of $\psi_j U^T$, so we get
	\begin{equation}
		\psi'_j = \psi_j U^T .
	\end{equation}
	Take $V = \overline{U}$, then $V \in \U(r)$ and
	\begin{equation}
		\psi'_j = \psi_j V^\dag .
	\end{equation}
	
	$(\impliedby)$ Now let's prove that if $\psi'_j = \psi_j V^\dag$ for all $j$, then $\pi((\psi'_j)_j) = \pi((\psi_j)_j)$. We showed in this section that the matrix $[\pi((\psi_j)_j)]$ is built from the matrix products $\psi_j \psi_{j'}^\dag$. We have that
	\begin{equation}
		{\psi'}_j {\psi'}_{j'}^\dag = \psi_j V^\dag V \psi_{j'}^\dag = \psi_j \psi_{j'}^\dag,
	\end{equation}
	because $V^\dag V = I$. Therefore, we have
	\begin{equation}
		\pi((\psi'_j)_j) = \pi((\psi_j)_j).
	\end{equation}
\end{proof}

We proved known results about $\pi$, but some new results will be postponed to chapter \ref{chapter: projection properties}.

\subsection{Closure under tensor product}

We showed that the tensor product of CP maps is a CP map. Here we specialize this result for CPTP maps.

\begin{proposition}{}{tensor product of cptp}
	If $\varepsilon \in \cptp(X,Y)$ and $\varepsilon' \in \cptp(X',Y')$ then $\varepsilon \otimes \varepsilon' \in \cptp(X \otimes X',Y \otimes Y')$.
\end{proposition}
\begin{proof}
	By proposition \ref{proposition:tensor product operation elements} we already know that $\varepsilon \otimes \varepsilon' \in CP(X \otimes X',Y \otimes Y')$. It remains to check that $\varepsilon \otimes \varepsilon'$ is trace preserving. Lets use the same notation of proposition \ref{proposition:tensor product operation elements}. Being trace preserving means that $\sum_{k,l} (E_k \otimes F_l)^\dag (E_k \otimes F_l) = \id_{X \otimes X'}$. We prove it as follows:
	\begin{align*}
		\sum_{k,l} (E_k \otimes F_l)^\dag (E_k \otimes F_l) &= \sum_{k,l} (E_k^\dag \otimes F_l^\dag) (E_k \otimes F_l) \\
		&= \sum_{k,l} E_k^\dag E_k \otimes F_l^\dag F_l \\
		&=  \left(\sum_k E_k^\dag E_k \right) \otimes \left(\sum_l F_l^\dag F_l \right) \\
		&= \id_X \otimes \id_{X'} \\
		&= \id_{X \otimes X'}.
	\end{align*}
\end{proof}

\section{UCP maps}

UCP maps aren't used directly in Quantum Information Theory, but they are dual do CPTP maps, so they are of mathematical interest. We begin by defining what is a unital map.

\begin{definition}{}{}
	A linear map $\varepsilon \colon \Lin_\complexNumbers (X) \to \Lin_\complexNumbers (Y)$ is unital if
	\begin{equation}
		\varepsilon(\id_X) = \id_Y.
	\end{equation}
\end{definition}

\begin{definition}{}{}
	A UCP map is a linear map $\varepsilon \colon \Lin_\complexNumbers (X) \to \Lin_\complexNumbers (Y)$ which is both unital and CP.
\end{definition}

Next we study the duality between CPTP and UCP maps, from which we can transfer results from CPTP maps to UCP maps.

\subsection{Duality}

Here we prove that UCP maps are dual do CPTP maps. Recall that the duality of CP maps was studied in section \ref{section: CP duality}.

\begin{proposition}{}{cptp iff adjoint is ucp}
	The operation of taking the adjoint
	\begin{align*}
		(-)^\dag \colon \Lin_\complexNumbers ( \Lin_\complexNumbers (X), \Lin_\complexNumbers (Y)) &\longrightarrow \Lin_\complexNumbers (\Lin_\complexNumbers (Y), \Lin_\complexNumbers (X))\\
		\varepsilon &\longmapsto \varepsilon^\dag
	\end{align*}
	restricts to a bijection between CPTP and UCP maps:
	\begin{equation*}
		(-)^\dag \colon \cptp(X,Y) \longrightarrow \ucp(Y,X).
	\end{equation*}
\end{proposition}
\begin{proof}
	We already know that the adjoint gives a bijection between CP maps. It remains to show that a CP map is TP (trace preserving) iff its adjoint is unital.
	
	Let $\varepsilon \in \cp(X,Y)$ and let $A \in \Lin_\complexNumbers (X)$. Then
	\begin{align*}
		\langle A, \varepsilon^\dag(\id_y) \rangle &\stackrel{(i)}{=} \langle \varepsilon(A), \id_Y \rangle\\
		&\stackrel{(ii)}{=} \tr(\varepsilon(A)^\dag \id_Y)\\
		&\stackrel{(iii)}{=} \tr(\varepsilon(A))^\dag.
	\end{align*}
	In $(i)$ was used the definition of the dual, in $(ii)$ was used the definition of the Hilbert-Schmidt inner product, and in $(iii)$ was used that $\id_Y$ is an identity function. On the hand, we have
	\begin{align*}
		\langle A, \id_X \rangle	&\stackrel{(i)}{=} \tr(A^\dag \id_X) \\
		&\stackrel{(ii)}{=} (\tr A)^\dag.
	\end{align*}
	In $(i)$ was used the definition of the Hilbert-Schmidt inner product, and in $(ii)$ was used that $\id_X$ is an identity function. Therefore
	\begin{align*}
		\varepsilon\text{ is TP} &\stackrel{(i)}{\iff} \forall A \in \Lin_\complexNumbers (X) \quad \tr(\varepsilon(A)) = \tr(A)\\
		&\stackrel{(ii)}{\iff} \forall A \in \Lin_\complexNumbers (X) \quad \tr(\varepsilon(A))^\dag = \tr(A)^\dag\\
		&\stackrel{(iii)}{\iff} \forall A \in \Lin_\complexNumbers (X) \quad \langle A, \varepsilon^\dag(\id_Y) \rangle = \langle A , \id_X \rangle\\
		&\stackrel{(iv)}{\iff} \varepsilon^{\dag}(\id_Y) = \id_X\\
		&\stackrel{(v)}{\iff} \varepsilon^\dag \text{ is unital}.
	\end{align*}
	In $(i)$ was used the definition of the TP property. In $(ii)$ was taken the adjoint, to get the same expression of previous equations. In $(iii)$ was used the previous results to replace $\varepsilon$ with $\varepsilon^\dag$ and to write an inner product with $A$. In $(iv)$  the quantifier and inner product with $A$ were eliminated. In $(v)$ was used the definition of being unital. This concludes that
	\begin{equation}
		\varepsilon\text{ is TP} \iff \varepsilon^\dag \text{ is unital}.
	\end{equation}
	
\end{proof}

With this result we can then do proofs for CPTP maps and transfer these results to UCP maps.

\subsection{The Kraus representation}

In the case of CP maps we saw that the dual map has dual Kraus operators. We already know that the TP property corresponds to the equation $\sum_k E_k^\dag E_k = \id$. Using the duality, we get a similar equation that characterizes the unital property.

\begin{proposition}{}{kraus unital condition}
	Let $\varepsilon \colon X \to Y$ be a CP map with Kraus operators $(E_k)_k$. Then $\varepsilon$ is unital iff
	\begin{equation}
		\sum_k E_k E_k^\dag = \id_Y.
	\end{equation}
\end{proposition}
\begin{proof}
	By proposition \ref{proposition:cptp iff adjoint is ucp}, $\varepsilon$ is unital iff $\varepsilon^\dag$ is TP. Since $(E_k)_k$ are Kraus operators for $\varepsilon$, by proposition \ref{proposition:cp iff adjoint is cp} we know that $(E_k^\dag)_k$ are Kraus operators for $\varepsilon\dag$. By proposition \ref{proposition:kraus TP condition}, $\varepsilon^\dag$ is TP iff
	\begin{equation}
		\sum_k E_k E_k^\dag = \id_Y.
	\end{equation}
	This concludes the proof.
\end{proof}

\subsection{The \CJ isomorphism}

For CPTP maps we saw that the \CJ isomorphism gives a bijection between $\cptp(X,Y)$ and $\psdii(X,Y)$. For UCP maps we have a similar result. First we introduce new notation.

\begin{notation}
	We denote the validity of the equation $\tr_1 A = \id_Y$ by a subscript 1:
	\begin{itemize}
		\item $\psdi(X,Y) \coloneqq \{A \in \psd(X^* \otimes Y) \mid \tr_1 A = \id_Y\}$;
		\item $\psdi(X,Y)_r \coloneqq \{A \in \psd(X^* \otimes Y)_r \mid \tr_1 A = \id_Y\}$.
	\end{itemize}
\end{notation}
Then we have the following:

\begin{proposition}{}{}
	The \CJ isomorphism restricts to a bijection
	\begin{equation*}
		J \colon \ucp(X,Y) \to \psdi(X,Y).
	\end{equation*}
\end{proposition}
\begin{proof}
	From the CP case, we know that the \CJ isomorphism gives a bijection $J \colon \cp(X,Y) \to \psd(X^* \otimes Y)$. For this reason, we only have to show that the unital property corresponds to the equation $\tr_1 J(\varepsilon) = \id_Y$. Let $\varepsilon \in \cp(X,Y)$. From proposition \ref{proposition:CJ expression} we have the expression
	\begin{equation}
		J(\varepsilon) = \sum_{i,i'} \ket{x^i} \bra{x^{i'}} \otimes \varepsilon(\ket{x_i} \bra{x_{i'}}).
	\end{equation}
	Taking the partial trace over $X^*$, we get
	\begin{align*}
		\tr_1 J(\varepsilon) &= \sum_{i,i'} \tr(\ket{x^i} \bra{x^{i'}}) \ \varepsilon(\ket{x_i} \bra{x_{i'}})\\
		&=  \sum_{i,i'} \delta_{i,i'} \ \varepsilon(\ket{x_i} \bra{x_{i'}})\\
		&= \varepsilon \left(\sum_i \ket{x_i} \bra{x_i}\right)\\
		&= \varepsilon(\id_X).
	\end{align*}
	This shows that
	\begin{equation}
		\tr_1 J(\varepsilon) = \id_Y \iff \varepsilon(\id_X) = \id_Y.
	\end{equation}
	Therefore, $\varepsilon \in \ucp(X,Y) \iff J(\varepsilon) \in \psdi(X,Y)$. 
\end{proof}

\subsection{Hom-sets as compact convex sets}

By the duality between CPTP and UCP maps, we can transfer the results from CPTP maps to UCP maps. We have the following results.

\begin{proposition}{}{}
	For any finite dimensional complex Hilbert spaces $X$ and $Y$, with $d_X \geq 1$, the sets $\ucp(X,Y)$ and $\psdi(X,Y)$ are compact convex. Therefore, they are homeomorphic to a closed ball, whose dimension is
	\begin{equation}
		\dim_\realNumbers \ucp(X,Y) = \dim_\realNumbers \psdi(X,Y) = d_Y^2 (d_X^2-1).
	\end{equation}
	The dimensions are for topological manifolds with boundary.
\end{proposition}
\begin{proof}
	By duality between CPTP and UCP maps, we know that $\ucp(X,Y)$ is homeomorphic to $\cptp(Y,X)$, which was proved in proposition \ref{proposition:cptp iff adjoint is ucp}. By corollary \ref{corollary:dimension of psdii}, $\cptp(Y,X)$ is homeomorphic to a closed ball of dimension $d_Y^2 (d_X^2-1)$.
\end{proof}

\subsection{Stratification into smooth manifolds}

The set $\psdi(X,Y)$ has a definition very similar to $\psdii(X,Y)$. They differ only by taking a partial trace over $X^*$ instead of $Y$. We can easily transfer the stratification of $\psdii(X,Y)$ to one for $\psdi(X,Y)$.

\begin{theorem}{}{psdir is smooth manifold}
	$\psdi(X,Y)_r$ is a smooth submanifold of $\Lin_\complexNumbers (X^* \otimes Y)$. Its dimension is
	\begin{equation}
		\label{equation: dim psdir}
		\dim_\realNumbers \psdi(X,Y)_r	= 2 d_X d_Y -r^2 -d_Y^2.
	\end{equation}
\end{theorem}
\begin{proof}
	The proof is almost the same as the one for theorem \ref{theorem:psdiir is smooth manifold}. The main difference is replacing $\tr_2$ with $\tr_1$.
\end{proof}

\begin{proposition}{}{tangent space of psdir}
	Let $X$ and $Y$ be finite dimensional complex Hilbert spaces, with $d_X \geq 1$ and $d_Y \geq 1$. Let also $A \in \psdi(X,Y)_r$. The tangent space of $\psdi(X,Y)_r$ at $A$ is
	\begin{equation}
		T_A \psdi(X,Y)_r = T_A \psd(X^* \otimes Y)_r \cap \ker \tr_1,
	\end{equation}
	where every space in this equation is interpreted as a subspace of $\Lin_\complexNumbers (X^* \otimes Y)$.
\end{proposition}
\begin{proof}
	It is the same proof of corollary \ref{corollary:tangent space of psdiir}, but replacing $\tr_2$ with $\tr_1$.
\end{proof}

\subsection{Extreme points}

By the duality between CPTP maps and UCP maps, we obtain a characterization of the extreme points for UCP maps from the one for CPTP maps. Historically, Choi proved in theorem 5 of \cite{CHOI1975285} the characterization of the extreme points of a convex set that has $\ucp(X,Y)$ as a special case.

\begin{theorem}{}{}
	Let $\varepsilon \in \ucp(X,Y)$ be a UCP map with linearly independent Kraus operators $(E_k)_k$. Then it is an extreme point of $\ucp(X,Y)$ if, and only if, $(E_k E_l^\dag)_{k,l}$ is linearly independent.
\end{theorem}
\begin{proof}
	By proposition \ref{proposition:cptp iff adjoint is ucp}, we have an isomorphism between $\ucp(X,Y)$ and $\cptp(Y,X)$. Then, by proposition \ref{proposition:linear bijection between extreme points}, this restricts to a bijection between their extreme points. Then $\varepsilon$ is an extreme point of $\ucp(X,Y)$ if, and only if, $\varepsilon^\dag$ is an extreme point of $\cptp(Y,X)$. By proposition \ref{proposition:cp iff adjoint is cp}, $(E_k^\dag)_k$ are Kraus operators for $\varepsilon^\dag$. Since $(E_k)_k$ is linearly independent, it follows that $(E_k^\dag)_k$ is also linearly independent. Then, by theorem \ref{theorem:extreme cptp maps in Kraus representation}, $\varepsilon^\dag$ is an extreme point of $\cptp(Y,X)$ if, and only if, $(E_k^{\dag\dag} E_l^\dag)_{k,l}$ is linearly independent. But $E_k^{\dag\dag} = E_k$, so $\varepsilon$ is an extreme point of $\ucp(X,Y)$ if, and only if, $(E_k E_l^\dag)_{k,l}$ is linearly independent.
\end{proof}

\subsection{Closure under tensor product}

For both CP and CPTP maps we showed that they are closed by tensor product. We show the same result for UCP maps.

\begin{proposition}{}{tensor product of ucp}
	If $\varepsilon \in \ucp(X,Y)$ and $\varepsilon' \in \ucp(X',Y')$ then $\varepsilon \otimes \varepsilon' \in \ucp(X \otimes X',Y \otimes Y')$.
\end{proposition}
\begin{proof}
	We can prove this result by using the duality between UCP and CPTP maps and proposition \ref{proposition:tensor product of cptp}. Using proposition \ref{proposition:dual of tensor of cp maps}, we have that $\varepsilon \otimes \varepsilon' = (\varepsilon^\dag \otimes \varepsilon'^\dag)^\dag$. We know that $\varepsilon \in \ucp(X,Y)$ and $\varepsilon' \in \ucp(X',Y')$, then proposition \ref{proposition:cptp iff adjoint is ucp} implies that $\varepsilon^\dag \in \cptp(Y,X)$ and $\varepsilon'^\dag \in \cptp(Y',X')$. By proposition \ref{proposition:tensor product of cptp} we know that $\varepsilon^\dag \otimes \varepsilon'^\dag \in \cptp(Y \otimes Y', X \otimes X')$. Using proposition \ref{proposition:cptp iff adjoint is ucp} again we conclude that $(\varepsilon^\dag \otimes \varepsilon'^\dag)^\dag \in  \ucp(X \otimes X',Y \otimes Y')$.
\end{proof}

\section{UCPTP maps}

An UCPTP map is one that is both unital and trace preserving, that is,
\begin{equation}
	\ucptp(X,Y) \coloneqq \ucp(X,Y) \cap \cptp(X,Y).
\end{equation}
In this case, we restrict attention to endomorphisms ($X=Y$), since $X$ and $Y$ have necessarily the same dimension. In fact, let $\varepsilon \in \ucptp(X,Y)$, then $\varepsilon (\id_X) = \id_Y$, because it is unital. But it is also trace preserving, so $\tr(\varepsilon(\id_X)) = \tr(\id_X)$, so $\tr(\id_Y) = \tr(\id_X)$. But $\tr(\id_X) = \dim X$, so $\dim X = \dim Y$. We'll denote $\ucptp(X,X)$ as $\ucptp(X)$.

\subsection{The \CJ isomorphism}

From the CP case, we know that the \CJ isomorphism gives a bijection between $\cp(X,Y)$ and $\psd(X^* \otimes Y)$. For the CPTP case we proved that it stablishes a correspondence between the TP (trace preserving) property, that is,
\begin{equation}
	\tr(\varepsilon(B)) = \tr B,
\end{equation}
and the equation
\begin{equation}
	\tr_2 J(\varepsilon) = \id_{X^*}.
\end{equation}
In the UCP case we proved a similar result. In this case, the isomorphism stablishes a correspondence between unitality, that is,
\begin{equation}
	\varepsilon(\id_X) = \id_Y,
\end{equation}
and the equation
\begin{equation}
	\tr_1 J(\varepsilon) = \id_Y.
\end{equation}
Combining these results, we get that the isomorphism restricts to a bijection between $\ucptp(X)$ and $\psdi(X) \cap \psdii(X)$.

\begin{notation}
	$\psdiii(X,Y) \coloneqq \psdi(X,Y) \cap \psdii(X,Y)$.
\end{notation}

\begin{proposition}{}{}
	For any finite dimensional complex Hilbert space $X$, the \CJ isomorphism restricts to a bijection
	\begin{equation}
		J \colon \ucptp(X) \to \psdiii(X).
	\end{equation}
\end{proposition}
\begin{proof}
	Done above.
\end{proof}
\subsection{The Kraus representation}

As any CP map, a UCPTP map admits a Kraus representation. Combining the results for CPTP and UCP maps, we get the following result.

\begin{proposition}{}{kraus representation of ucptp}
	Let $X$ be a finite dimensional complex Hilbert space. Then any $\varepsilon \in \ucptp(X)$ admits a Kraus representation where the Kraus operators $(E_k)_k$, with $E_k \in \Lin_\complexNumbers (X)$, satisfy the following equations:
	\begin{equation}
		\sum_k E_k^\dag E_k = \id_X,
	\end{equation}
	\begin{equation}
		\sum_k E_k E_k^\dag = \id_X.
	\end{equation}
	Conversely, any finite sequence $(E_k)_k$ of operators $E_k \in \Lin_\complexNumbers (X)$ that satisfy both equations determine an $\varepsilon \in \ucptp(X)$. It is given by
	\begin{equation}
		\varepsilon(A) = \sum_k E_k A E_k^\dag.
	\end{equation}
\end{proposition}
\begin{proof}
	By proposition \ref{proposition:cp maps vs kraus representation}, we know that every CP map admits a Kraus representation. Also, it says that every finite sequence $(E_k)_k$ of operators $E_k \colon X \to X$ determines the CP map $\varepsilon\in \cp(X)$ given by
	\begin{equation}
		\varepsilon(A) = \sum_k E_k A E_k^\dag.
	\end{equation}
	In proposition \ref{proposition:kraus TP condition} we proved that $\varepsilon$ is trace preserving iff
	\begin{equation}
		\sum_k E_k^\dag E_k = \id_X.
	\end{equation}
	Also, in proposition \ref{proposition:kraus unital condition} we proved that $\varepsilon$ is unital iff
	\begin{equation}
		\sum_k E_k E_k^\dag = \id_X.
	\end{equation}
	This concludes the proof.
\end{proof}

\subsection{Duality}

For UCPTP maps, the duality given by taking the adjoint restricts to a bijection from $\ucptp(X)$ to itself.

\begin{proposition}{}{}
	Let $X$ be a finite dimensional complex Hilbert space. The operation of taking the adjoint
	\begin{align*}
		(-)^\dag \colon \Lin_\complexNumbers (\Lin_\complexNumbers(X)) &\longrightarrow \Lin_\complexNumbers (\Lin_\complexNumbers (X))\\
		\varepsilon &\longmapsto \varepsilon^\dag
	\end{align*}
	restricts to a bijection
	\begin{equation*}
		(-)^\dag \colon \ucptp(X) \to \ucptp(X).
	\end{equation*}
\end{proposition}
\begin{proof}
	We have seen in proposition \ref{proposition:cp iff adjoint is cp} that the duality given by taking the adjoint of a CP map given a bijection $(-)^\dag \colon \cp(X) \to \cp(X)$. Also, in proposition \ref{proposition:cptp iff adjoint is ucp} we saw that $\varepsilon$ is trace preserving iff $\varepsilon^\dag$ is unital. Since $\varepsilon^{\dag\dag} = \varepsilon$, it implies that $\varepsilon$ is unital iff $\varepsilon^\dag$ is trace preserving. Therefore, $\varepsilon \in \ucptp(X)$ iff $\varepsilon^\dag \in \ucptp(X)$. 
	
\end{proof}

\subsection{Hom-sets as compact convex sets}

We proved that the hom-sets of CPTP and UCP maps are compact convex. From this, it follows easily that the same is true for UCPTP maps. As any compact convex set in $\realNumbers^n$, it is homeomorphic to a closed ball and is the convex hull of its extreme points. In this section we compute the dimension of this closed ball. We study the extreme points in the next section.

\begin{proposition}{}{ucptp is compact convex}
	For any finite dimensional complex Hilbert space $X$,  the set $\ucptp(X)$ is compact convex.
\end{proposition}
\begin{proof}
	Since both $\ucp(X)$ and $\cptp(X)$ are convex sets, its intersection \linebreak $\ucptp(X)$ is also convex. Since, both $\ucp(X)$ and $\cptp(X)$ are bounded sets, $\ucptp(X)$ is also bounded. And since both $\ucp(X)$ and $\cptp(X)$ are closed sets, $\ucptp(X)$ is also closed. Therefore $\ucptp(X)$ is closed and bounded, so it is compact. 
\end{proof}

The convex set is trivial if $X = \{0\}$, as we prove next.

\begin{proposition}{}{ucptp(0)}
	$\ucptp(\{0\}) = \{0\}$.
\end{proposition}
\begin{proof}
	From proposition \ref{proposition:psdii(X,Y) for X or Y = 0}, we already know that $\cptp(\{0\}) = \{0\}$, so $\ucptp(\{0\})$ can only be $\varnothing$ or $\{0\}$. We just have to check that $0 \in \ucptp(\{0\})$. This is the case if $0 \colon \Lin_\complexNumbers (\{0\}) \to \Lin_\complexNumbers (\{0\})$ is unital. This is trivially true, because it sends 0 to 0, and $\id_{\{0\}} = 0$.
\end{proof}

Now we focus on the case where $X \neq \{0\}$. Being a compact convex set, it is homeomorphic to a closed ball. Next we compute the dimension of this ball. To do this, we use again corollary \ref{corollary:compact convex set is homeomorphic to a closed ball}. To apply this corollary, a point of the compact convex set is required. In the CPTP case we showed that $\id_{X^* \otimes Y}/d_Y \in \psdii(X,Y)$. In the UCPTP case we have $d_X = d_Y$, which implies that $\id_{X^* \otimes Y}/d_Y \in \psdiii(X,Y)$. In fact, we have
\begin{equation}
	\tr_1 \left( \frac{\id_{X^* \otimes Y}}{d_Y} \right) = \frac{d_X \id_Y}{d_Y} = \id_Y.
\end{equation}
Even though we wrote it for a $Y$ possibly different from $X$, they are necessarily isomorphic in the UCPTP case, so we'll take $Y=X$ to keep notation simpler.

From corollary \ref{corollary:balls inside and outside psdii} it follows almost immediately the next corollary.
\begin{corollary}{}{}
	Let $X$ be a finite dimensional complex Hilbert space, with $d_X \geq 1$. We have that
	\begin{multline}
		\frac{\id_{X^* \otimes X}}{d_X} + \overline{B(0,1/d_X) } \cap \hermitian(X^* \otimes X) \cap \ker \tr_1 \cap \ker \tr_2 \subseteq \\ \psdiii(X) \subseteq \overline{B(0,d_X)},
	\end{multline}
	where the closed balls are in $\Lin_\complexNumbers (X^* \otimes X)$.
\end{corollary}
\begin{proof}
	Just take intersections with $\id_{X^* \otimes X}/d_X + \ker \tr_1$ in corollary \ref{corollary:balls inside and outside psdii}, and use that 
	\begin{equation}
		\psdiii(X) =  \psdii(X) \cap \left( \frac{\id_{X^* \otimes X}}{d_X} + \ker \tr_1 \right).
	\end{equation}
	Notice that
	\begin{equation}
		\frac{\id_{X^* \otimes X}}{d_X} + \ker \tr_1 = \{A \in \Lin_\complexNumbers (X^* \otimes X) \mid \tr_1 A = \id_{X} \}.
	\end{equation}
	In fact, we have
	\begin{align*}
		A \in \frac{\id_{X^* \otimes X}}{d_X} + \ker \tr_1 &\iff A-\frac{\id_{X^* \otimes X}}{d_X} \in \ker \tr_1 \\
		&\iff \tr_1\left( A-\frac{\id_{X^* \otimes X}}{d_X} \right) = 0 \\
		&\iff \tr_1 A -\id_X = 0\\
		&\iff \tr_1 A = \id_X.
	\end{align*}
\end{proof}

From this result we can compute the $\realNumbers$-vector space generated by the set $\psdiii(X)$ $-\id_{X^*\otimes X}/d_X$. Computing its dimension, we get the dimension of the ball that $\psdiii(X)$ is homeomorphic to.
\begin{proposition}{}{span of psdiii-id/dX}
	Let $X$ be a finite dimensional complex Hilbert space, with $d_X \geq 1$. We have that
	\begin{equation}
		\left\langle \psdiii(X) - \frac{\id_{X^* \otimes X}}{d_X} \right\rangle_\realNumbers = \hermitian(X^* \otimes X) \cap \ker \tr_1 \cap \ker \tr_2.
	\end{equation}
	The dimension of this space is
	\begin{equation}
		\dim_\realNumbers \hermitian(X^* \otimes X) \cap \ker \tr_1 \cap \ker \tr_2 = d_X^4-2 d_X^2 +1 = (d_X^2-1)^2.
	\end{equation}
\end{proposition}
\begin{proof}
	Since
	\begin{equation}
		\overline{B(0,1/d_X) } \cap \hermitian(X^* \otimes X) \cap \ker \tr_1 \cap \ker \tr_2 \subseteq \psdiii(X)- \frac{\id_{X^* \otimes X}}{d_X},
	\end{equation}
	it follows immediately that
	\begin{equation}
		\hermitian(X^* \otimes X) \cap \ker \tr_1 \cap \ker \tr_2 \subseteq \left\langle \psdiii(X)- \frac{\id_{X^* \otimes X}}{d_X} \right\rangle_\realNumbers.
	\end{equation}
	On the other hand, let $A \in \psdiii(X)$ and let
	\begin{equation}
		B = A- \frac{\id_{X^* \otimes X}}{d_X}.
	\end{equation}
	We have that:
	\begin{itemize}
		\item $B$ is hermitian: this follows immediately from $A$ and $-\id_{X^* \otimes X}/d_X$ being hermitian, and because a sum of hermitian operators is also hermitian.
		
		\item $\tr_1 B = 0$: this follows from both $A$ and $\id_{X^* \otimes X}/d_X$ having $\id_X$ as their partial trace over $X^*$.
		
		\item $\tr_2 B = 0$: this follows from both $A$ and $\id_{X^* \otimes X}/d_X$ having $\id_{X^*}$ as their partial trace over $X$.
	\end{itemize} 
	Therefore
	\begin{equation}
		\left\langle \psdiii(X)- \frac{\id_{X^* \otimes X}}{d_X} \right\rangle_\realNumbers = \hermitian(X^* \otimes X) \cap \ker \tr_1 \cap \ker \tr_2 .
	\end{equation}
	
	It is easy, but tedious, to find a basis for this space and then compute the dimension. Instead, we'll compute it by constructing a linear map that has such space as its kernel. Then we can find the dimension of the kernel by first finding the dimension of the image. The linear map is
	\begin{align*}
		L \colon \hermitian(X^* \otimes X) &\longrightarrow \Lin_\complexNumbers (X^* \otimes X) \\
		A &\longmapsto p'_1(A) + p'_2(A) + p_1 p_2 (A),
	\end{align*}
	where $p'_1$, $p'_2$ and $p_1 p_2$ are the orthogonal projections defined in proposition \ref{proposition:p'1, p'2, p1p2}. $L$ is the orthogonal projection $p'_1+p'_2+p_1 p_2$ restricted to $\hermitian(X^* \otimes X)$. We have that $\ker L = \hermitian(X^* \otimes X) \cap \ker \tr_1 \cap \ker \tr_2$. In fact, if $A \in \ker L$, then
	\begin{equation}
		p'_1(A) + p'_2(A) + p_1 p_2 (A) = 0.
	\end{equation}
	By proposition \ref{proposition:p'1, p'2, p1p2}, we know that $p'_1$, $p'_2$ and $p_1 p_2$ are orthogonal to each other. Therefore, we have
	\begin{equation}
		p'_1(A) = p'_2(A) = p_1 p_2 (A) = 0.
	\end{equation}
	Since $p_1(A) = p'_1 (A)+p_1 p_2 (A)$ and $p_2(A) = p'_2 (A)+p_1 p_2 (A)$, we have that
	\begin{equation}
		A \in \ker p_1 \cap \ker p_2.
	\end{equation}
	By proposition \ref{proposition:partial traces as projections}, $\ker p_1 = \ker \tr_2$ and $\ker p_2 = \ker \tr_1$, so
	\begin{equation}
		A \in \ker \tr_1 \cap \ker \tr_2,
	\end{equation}
	which shows that
	\begin{equation}
		\ker L \subseteq \hermitian(X^* \otimes X) \cap \ker \tr_1 \cap \ker \tr_2.
	\end{equation}
	On the other hand, if $A \in \hermitian(X^* \otimes X) \cap \ker \tr_1 \cap \ker \tr_2$, then $p_1(A) = 0$ and $p_2(A) = 0$. Since $p_1 = p'_1+p_1 p_2$, and since $p'_1$ and $p_1 p_2$ are orthogonal to each other, it follows that
	\begin{equation}
		p'_1(A) = p_1 p_2 (A) = 0.
	\end{equation}
	Similarly, $p_2(A) = 0$ implies that
	\begin{equation}
		p'_2(A) = p_1 p_2 (A) = 0.
	\end{equation}
	Therefore,
	\begin{equation}
		L(A) = 0.
	\end{equation}
	This concludes that
	\begin{equation}
		\ker L = \hermitian(X^* \otimes X) \cap \ker \tr_1 \cap \ker \tr_2 .
	\end{equation}
	
	Now let's compute the image of $L$. Let $H \in \hermitian(X^*)$, then we have the element
	\begin{equation}
		A = H \otimes \id_X \in \hermitian(X^* \otimes X).
	\end{equation}
	By proposition \ref{proposition:partial traces as projections}, we know that $A \in \image p_1$. Since $p_1$ is an orthogonal projection, it follows that
	\begin{equation}
		p_1(A) = A.
	\end{equation}
	Also, $p'_2$ is orthogonal to both $p'_1$ and $p_1 p_2$. This implies that $p'_2$ is orthogonal to $p_1$. Since $A \in  \image p_1$ and $p'_2$ is orthogonal to $p_1$, we have that
	\begin{equation}
		p'_2 (A) = 0.
	\end{equation}
	Therefore,
	\begin{equation}
		L(A) = p_1(A) + p'_2(A) = A = H \otimes \id_X.
	\end{equation}
	This shows that
	\begin{equation}
		\hermitian(X^*) \otimes \id_X \subseteq \image L.
	\end{equation}
	Similarly, taking $H \in \hermitian(X)$ and $A = \id_{X^*} \otimes H$, it follows that $L(A) = A$, so
	\begin{equation}
		\id_{X^*} \otimes \hermitian(X) \subseteq \image L.
	\end{equation}
	Putting everything together, we have that
	\begin{equation}
		(\hermitian(X^*) \otimes \id_X)+(\id_{X^*} \otimes \hermitian(X)) \subseteq \image L.
	\end{equation}
	Next we show that this inclusion is an equality. Let $A \in \hermitian(X^* \otimes X)$. In proposition \ref{proposition:partial trace as cptp map} we showed that $\tr_1$ and $\tr_2$ are CPTP maps. By proposition \ref{proposition:cp(A dag) = cp(A) dag}, we know that $\tr_1 A$ and $\tr_2 A$ are hermitian. Since
	\begin{equation}
		p_1(A) = \tr_2 A \otimes \frac{\id_X}{d_X},
	\end{equation}
	\begin{equation}
		p_2(A) = \frac{\id_{X^*}}{d_X} \otimes \tr_1 A,
	\end{equation}
	we have that
	\begin{equation}
		p_1 (A) \in \hermitian(X^*)\otimes \id_X,
	\end{equation}
	\begin{equation}
		p_2 (A) \in \id_{X^*} \otimes \hermitian(X),
	\end{equation}
	so
	\begin{equation}
		p_1 (A) + p_2 (A) \in (\hermitian(X^*) \otimes \id_X)+(\id_{X^*} \otimes \hermitian(X)).
	\end{equation}
	Also, by proposition \ref{proposition:p'1, p'2, p1p2}, we have
	\begin{equation}
		p_1 p_2 (A) = \tr A \ \frac{\id_{X^* \otimes X}}{d_X}.
	\end{equation}
	Since $A$ is hermitian, then $\tr A \in \realNumbers$. It follows that
	\begin{equation}
		p_1 p_2 (A) = (\hermitian(X^*) \otimes \id_X)+(\id_{X^*} \otimes \hermitian(X)).
	\end{equation}
	We can write $L(A) = p_1(A)+p_2(A)-p_1p_2(A)$, from which we get that
	\begin{equation}
		L(A) \in (\hermitian(X^*) \otimes \id_X)+(\id_{X^*} \otimes \hermitian(X)).
	\end{equation}
	This concludes that
	\begin{equation}
		\image L = (\hermitian(X^*) \otimes \id_X)+(\id_{X^*} \otimes \hermitian(X)).
	\end{equation}
	Next we compute the dimension of this image. A simple exercise in Linear Algebra shows that, for any finite dimensional $\realNumbers$-vector spaces $V_1$ and $V_2$, we have
	\begin{equation}
		\dim_\realNumbers (V_1 + V_2) = \dim_{\realNumbers} V_1 + \dim_{\realNumbers} V_2 - \dim_\realNumbers (V_1 \cap V_2).
	\end{equation}
	This can be proven by applying the kernel-image theorem to the linear map
	\begin{align*}
		V_1 \oplus V_2 &\longrightarrow V_1+V_2\\
		v_1 \oplus v_2 &\longmapsto v_1+v_2.
	\end{align*}
	We have that
	\begin{equation}
		(\hermitian(X^*) \otimes \id_X)\cap(\id_{X^*} \otimes \hermitian(X)) = \realNumbers \, \id_{X^* \otimes X}.
	\end{equation}
	Therefore,
	\begin{align*}
		\dim_\realNumbers \image L &= \dim_{\realNumbers} (\hermitian(X^*) \otimes \id_X) + \dim_{\realNumbers} (\id_{X^*} \otimes \hermitian(X)) - \dim_{\realNumbers} (\realNumbers \, \id_{X^* \otimes X})\\
		&= d_X^2+d_X^2-1\\
		&= 2 d_X^2-1.
	\end{align*}
	Then
	\begin{align*}
		\dim_{\realNumbers} \ker L &= \dim_{\realNumbers} \hermitian(X^* \otimes X) - \dim_{\realNumbers} \image L\\
		&= d_X^4-( 2 d_X^2-1)\\
		&= d_X^4-2 d_X^2+1\\
		&= (d_X^2-1)^2.
	\end{align*}
	This concludes the proof.
\end{proof}

\begin{corollary}{}{dimension of psdiii}
	For any finite dimensional complex Hilbert space $X$, with $d_X \geq 1$, the compact convex set $\psdiii(X)$ is homeomorphic to a closed ball of dimension
	\begin{equation}
		\dim_{\realNumbers} \psdiii(X) = (d_X^2-1)^2.
	\end{equation}
	This is the dimension of $\psdiii(X)$ as a topological manifold with boundary.
\end{corollary}
\begin{proof}
	By proposition \ref{proposition:ucptp is compact convex} we know that $\psdiii(X)$ is compact convex. By corollary \ref{corollary:compact convex set is homeomorphic to a closed ball} and proposition \ref{proposition:span of psdiii-id/dX}, it follows that
	\begin{equation}
		\dim_{\realNumbers} \psdiii(X) = (d_X^2-1)^2.
	\end{equation}
\end{proof}

\subsection{Extreme points}

As $\ucptp(X)$ is compact convex, it is the convex hull of its extreme points. The following characterization of the extreme points of $\ucptp(X)$ was proven is \cite{LANDAU1993107}. It can also be found in theorem 4.21 of \cite{watrous_2018}.

\begin{theorem}{}{characterization of extreme points for ucptp}
	Let $X$ be a finite dimensional complex Hilbert space, with $d_X \geq 1$. Let $\varepsilon \in \ucptp(X)$ be a UCPT map with linearly independent Kraus operators $(E_k)_k$. Then it is an extreme point of $\ucptp(X)$ iff $(E_k^\dag E_l \oplus E_l E_k^\dag)_{k,l}$ is linearly independent as vectors of $\Lin_\complexNumbers (X) \oplus \Lin_\complexNumbers (X)$.
\end{theorem}

A trivial example of extreme UCPTP map is a unitary channel, which is a channel whose Kraus representation is $(U)$, for some unitary operator $U \in \U(X)$. It is a channel with Choi-rank 1. Examples with higher Choi-rank will be given in chapter \ref{chapter: examples extreme channels}.

\subsection{Choi-rank bounds for extreme points}

For CPTP maps we saw that an extreme point can't have a rank higher than $d_X$. Repeating the same reasoning, we find an upper bound for the highest rank an extreme point can have for UCPTP maps.

\begin{corollary}{}{bound on rank of extreme ucptp}
	Let $X$ be a finite dimensional complex Hilbert space, with $d_X \geq 1$. If $\varepsilon \in \ucptp(X)$ is an extreme point of $\ucptp(X)$, then
	\begin{equation}
		1 \leq \CR (\varepsilon) \leq \sqrt{2} \, d_X.
	\end{equation}
\end{corollary}
\begin{proof}
	Let $(E_k)_k$ be linearly independent Kraus operators for $\varepsilon$. The length of this sequence is $\CR (\varepsilon)$, the Choi-rank of $\varepsilon$. If $\varepsilon$ is extreme, then the sequence $(E_k^\dag E_l \oplus E_l E_k^\dag)_{k,l}$ is linearly independent. For this to be possible, the length of the sequence, which is $\CR(\varepsilon)^2$, can't be bigger than $\dim_\complexNumbers (\Lin_\complexNumbers (X)\oplus \Lin_\complexNumbers (X)) = 2d_X^2$. Therefore, we must have
	\begin{equation}
		\CR(\varepsilon)^2 \leq 2 d_X^2,
	\end{equation}
	so
	\begin{equation}
		\CR(\varepsilon) \leq \sqrt{2} \, d_X.
	\end{equation}
\end{proof}

For CPTP maps, we showed in example \ref{example: extreme cptp with rank dX} that the upper bound $d_X$ is optimal. For UCPTP maps, we have the upper bound $\sqrt{2} \, d_X$. A slightly better upper bound is given in \cite{ohno}, which says that $\CR(\varepsilon) \leq \sqrt{2 d_X^2-1}$. Since $\CR(\varepsilon) \in \naturalNumbers$, only the integer part of the upper bound is important. We prove next that both upper bounds have the same integer part.

\begin{proposition}{}{}
	For any natural number $d \geq 1$, we have that
	\begin{equation}
		\lfloor \sqrt{2d^2-1} \rfloor = \lfloor \sqrt{2}d \rfloor.
	\end{equation}
\end{proposition}
\begin{proof}
	The integer part $\lfloor x \rfloor$ of $x$ is the unique integer such that
	\begin{equation}
		\lfloor x \rfloor \leq x < \lfloor x \rfloor+1.
	\end{equation}
	Let's denote the integer parts as
	\begin{equation}
		m \coloneqq \lfloor \sqrt{2d^2-1} \rfloor,
	\end{equation}
	\begin{equation}
		n \coloneqq \lfloor \sqrt{2}d \rfloor.
	\end{equation}
	By the definition of integer part, we have
	\begin{equation}
		\label{equation: integer part upper bound smaller}
		m \leq \sqrt{2d^2-1} < m+1.
	\end{equation}
	\begin{equation}
		\label{equation: integer part upper bound bigger}
		n \leq \sqrt{2}d < n+1.
	\end{equation}
	Both square roots are non negative numbers, so $m,n\geq 0$. Since $\sqrt{2d^2-1}<\sqrt{2}d$, we must have $m\leq n$. In fact, if that wasn't the case, then $n<m$ implies that $n+1\leq m$, so $\sqrt{2}d < n+1 \leq m \leq \sqrt{2d^2-1}$, which is false. Since $m\leq n$, we can write
	\begin{equation}
		n = m+p,
	\end{equation}
	for some integer $p\geq 0$. Squaring the inequations (\ref{equation: integer part upper bound smaller}) and (\ref{equation: integer part upper bound bigger}), we get
	\begin{equation}
		\label{equation: integer part upper bound smaller squared}
		m^2 \leq 2d^2-1 < m^2+2m+1.
	\end{equation}
	\begin{equation}
		\label{equation: integer part upper bound bigger squared}
		n^2 \leq 2d^2 < n^2+2n+1.
	\end{equation}
	Subtracting 1 from the last inequation, we get
	\begin{equation}
		\label{equation: integer part upper bound bigger squared minus 1}
		n^2-1 \leq 2d^2-1 < n^2+2n.
	\end{equation}
	The inequations (\ref{equation: integer part upper bound smaller squared}) and (\ref{equation: integer part upper bound bigger squared minus 1}) imply that
	\begin{equation}
		n^2-1 < m^2+2m+1.
	\end{equation}
	Using that $n=m+p$, we get
	\begin{align*}
		m^2+2pm+p^2-1 &< m^2+2m+1\\
		\implies 2(p-1)m &< 2-p^2\\
	\end{align*}
	If $p\geq 2$, then $2(p-1)m \geq 0$, but $2-p^2 \leq 2-4=-2$, contradicting the inequation above. Therefore, $p=0$ or $p=1$.
	
	Suppose that $p=1$, that is, $n=m+1$. Combining the inequations (\ref{equation: integer part upper bound smaller squared}) and (\ref{equation: integer part upper bound bigger squared}) we get
	\begin{equation}
		m^2 \leq 2d^2-1 < (m+1)^2 \leq 2d^2 < (m+2)^2.
	\end{equation}
	The inequality $2d^2-1 < (m+1)^2$ only has integers, so it is equivalent to  $2d^2 \leq (m+1)^2$. From this we get that
	\begin{equation}
		2d^2 \leq (m+1)^2 \leq 2d^2,
	\end{equation}
	which implies that
	\begin{equation}
		(m+1)^2 = 2d^2.
	\end{equation}
	Solving for $m$, we get
	\begin{equation}
		m = \sqrt{2}d-1.
	\end{equation}
	The expression $\sqrt{2}d-1$ is an irrational number, but $m$ must be a natural number. Therefore, $p$ can't be 1. This proves that $p=0$, that is, $m=n$.
\end{proof}

For any unitary operator $U \in \U(X)$ we have the UCPTP map whose Kraus representation is $(U)$. Its Choi-rank is 1, so it is an extreme point and the lower bound is always achieved. We'll see in section \ref{section: inexistence of extreme points} that for $X$ with low dimension there are no extreme points with some Choi-ranks in the bounds. For higher dimensions, it is still an open problem to construct examples with the Choi-ranks allowed by the bounds or to prove their non-existence. We'll construct many examples in chapter \ref{chapter: examples extreme channels}.

\subsection{Non-existence of extreme points for some Choi-ranks}

\label{section: inexistence of extreme points}

Corollary \ref{corollary:bound on rank of extreme ucptp} shows that an extreme UCPTP map $\varepsilon\in \ucptp(X)$ must have a Choi-rank smaller than $\sqrt{2}d_X$. A question we might ask is if for any $r \in [1,\lfloor\sqrt{2}d_X\rfloor]$ there is an extreme UCPTP map $\varepsilon$ with Choi-rank $\CR(\varepsilon) = r$. The answer for this equation is no. We'll prove next that there are no extreme UCPTP maps with $\CR(\varepsilon)=2$ for $d_X=2$ and $d_X = 3$. We follow the proof from \cite{watrous_2018}. This reference only shows the proof for $d_X=2$, that is, qubit channels, but the same proof can easily be adapted for $d_X=3$.

First we'll prove lemma 4.22 from \cite{watrous_2018}.
\begin{lemma}{}{ucptp channels with rank at most 2}
	Let $\varepsilon \in \ucptp(X)$ be a UCPTP map with Choi-rank $\CR(\varepsilon) \leq 2$. Let $(E_0, E_1)$ be a Kraus decomposition of $\varepsilon$. Then there exists unitary matrices $U,V \in \U(d_X)$ such that $V[E_0] U^\dag$ and $V[E_1] U^\dag$ are both diagonal matrices.
\end{lemma}
\begin{proof}
	To simplify notation, let's define the matrices
	\begin{equation}
		A_0 \coloneqq [E_0],
	\end{equation}
	\begin{equation}
		A_1 \coloneqq [E_1].
	\end{equation}
	The strategy is to first find a unitary matrix $W\in \U(d_X)$ such that $WA_0$ and $WA_1$ are both normal and commute with each other. In such conditions, it follows that $WA_0$ and $WA_1$ can be diagonalized simultaneously. Therefore, there exists $U\in \U(d_X)$ such that $UWA_0 U^\dag$ and $UWA_1 U^\dag$ are both diagonal. Then take $V\coloneqq UW$ and the proof is finished.
	
	Take polar decompositions of $A_0$ and $A_1$:
	\begin{equation}
		A_0 = U_0 P_0,
	\end{equation}
	\begin{equation}
		A_1 = U_1 P_1,
	\end{equation}
	where $U_0, U_1 \in \U(d_X)$ and $P_0, P_1$ are positive semidefinite matrices. Since $\varepsilon$ is a UCPTP map, we have the following equations:
	\begin{equation}
		A_0^\dag A_0 + A_1^\dag A_1 = I_{d_X},
	\end{equation}
	\begin{equation}
		A_0 A_0^\dag + A_1 A_1^\dag = I_{d_X}.
		\label{equation: unitality for rank at most 2}
	\end{equation}
	
	The first equation says that the channel is trace preserving and the second says that it is unital. Using the polar decomposition in the first equation we get
	\begin{align*}
		I_{d_X} &= \sum_k A_k^\dag A_k\\
		&= \sum_k P_k^\dag U_k^\dag U_k P_k\\
		&= \sum _k P_k^\dag P_k\\
		&= \sum_k P_k^2.
	\end{align*}
	In the last equation we used that $P_k$ is positive semidefinite, and therefore hermitian. Isolating $P_1^2$, we get
	\begin{equation}
		P_1^2 = I_{d_X}-P_0^2.
	\end{equation}
	From this equation it follows that $P_0^2$ and $P_1^2$ commute with each other. Since both are positive semidefinite and commute, they are simultaneously diagonalizable. If $P_k^2 = \sum_i \lambda_i \ket{v_i}\bra{v_i}$ is any diagonalization of $P_k^2$, where $\lambda_i \geq 0$ and $(v_i)_i$ is orthonormal, then $P_k = \sum_i \sqrt{\lambda_i}\ket{v_i}\bra{v_i}$ is a diagonalization of $P_k$. Therefore, $P_0$ and $P_1$ are also simultaneously diagonalizable, so they commute with each other.
	
	Now let's analyze the equation (\ref{equation: unitality for rank at most 2}). Using the polar decomposition, we get
	\begin{align*}
		I_{d_X} &= \sum_k A_k A_k^\dag \\
		&= \sum_k U_k P_k P_k^\dag U_k^\dag \\
		&= \sum_k U_k P_k^2 U_k^\dag \\
		&= U_0 P_0^2 U_0^\dag + U_1 P_1^2 U_1^\dag.
	\end{align*}
	Using that $P_1^2 = I_{d_X}-P_0^2$, we get
	\begin{align*}
		I_{d_X} &= U_0 P_0^2 U_0^\dag + U_1 P_1^2 U_1^\dag \\
		&=  U_0 P_0^2 U_0^\dag + U_1 (I_{d_X}-P_0^2) U_1^\dag \\
		&= U_0 P_0^2 U_0^\dag + U_1 I_{d_X} U_1^\dag - U_1 P_0^2 U_1^\dag\\
		&= U_0 P_0^2 U_0^\dag + I_{d_X} - U_1 P_0^2 U_1^\dag .
	\end{align*}
	From this we get the equation
	\begin{equation}
		U_0 P_0^2 U_0^\dag = U_1 P_0^2 U_1^\dag .
	\end{equation}
	It can be written as a commutation equation. Multiply both sides of the equation by $U_0^\dag$ from the left, and by $U_1$ from the right. Then we get
	\begin{align*}
		U_0^\dag U_0 P_0^2 U_0^\dag U_1 &= U_0^\dag U_1 P_0^2 U_1^\dag U_1 \\
		\implies P_0^2 (U_0^\dag U_1) &= (U_0^\dag U_1) P_0^2.
	\end{align*}
	That is, $P_0^2$ commutes with $U_0^\dag U_1$. Since $P_0$ is positive semidefinite and $U_0^\dag U_1$ is unitary, both are diagonalizable. They commute with each other, so they can be diagonalized simultaneously. As shown previously in this proof, any basis of eigenvectors for $P_0^2$ is a basis of eigenvectors for $P_0$. From this we have that $P_0$ and $U_0^\dag U_1$ are also simultaneously diagonalizable, so they commute with each other.
	
	With these results we can now prove that $WA_0$ and $WA_1$ are normal and commute with each other. First let's prove that they are normal.
	\begin{itemize}
		\item $WA_0$ is normal: using that $W=U_0^\dag$ and the polar decomposition of $A_0$, we have that
		\begin{equation}
			WA_0 = U_0^\dag U_0 P_0 = P_0.
		\end{equation}
		$P_0$ is positive semidefinite, so it is diagonalizable, and every diagonalizable matrix is normal.

		\item $WA_1$ is normal: using that $W=U_0^\dag$ and the polar decomposition of $A_0$, we have that
		\begin{equation}
			WA_1 = U_0^\dag U_1 P_1.
		\end{equation}
		Let's show that $(WA_1)^\dag$ commutes with $WA_1$. We have
		\begin{align*}
			(WA_1)^\dag WA_1 &= (U_0^\dag U_1 P_1)^\dag U_0^\dag U_1 P_1\\
			&= P_1^\dag U_1^\dag U_0 U_0^\dag U_1 P_1\\
			&= P_1^2.
		\end{align*}
		If we multiply $(WA_1)^\dag$ and $WA_1$ in the other order, we get
		\begin{align*}
			WA_1 (WA_1)^\dag &= U_0^\dag U_1 P_1 (U_0^\dag U_1 P_1)^\dag \\
			&= U_0^\dag U_1 P_1 P_1^\dag U_1^\dag U_0\\
			&= U_0^\dag U_1 P_1^2 U_1^\dag U_0\\
			&= U_0^\dag U_1 (I_{d_X}-P_0^2) U_1^\dag U_0\\
			&= U_0^\dag U_1 I_{d_X} U_1^\dag U_0-U_0^\dag U_1 P_0^2 U_1^\dag U_0\\
			&= I_{d_X} - U_0^\dag U_1 P_0^2 U_1^\dag U_0\\
			&= I_{d_X} - P_0^2 U_0^\dag U_1 U_1^\dag U_0\\
			&= I_{d_X} - P_0^2\\
			&= P_1^2.
		\end{align*}
		It was used in these equations that $P_1^2 = I_{d_X}-P_0^2$ and that $P_0^2$ commutes with $U_0^\dag U_1$.
		
		This shows that $WA_1$ commutes with $(WA_1)^\dag$, so $WA_1$ is normal.
	\end{itemize}
	
	Lastly we show that $WA_0$ and $WA_1$ commute with each other. We have
	\begin{align*}
		WA_0 WA_1 &= P_0 (U_0^\dag U_1 P_1) \\
		&= (U_0^\dag U_1) P_0 P_1\\
		&= U_0^\dag U_1 P_1 P_0 \\
		&= WA_1 P_0 \\
		&= WA_1 U_0^\dag U_0 P_0 \\
		&= WA_1 WA_0.
	\end{align*}
	We have used in these equations that $P_0$ and $U_0^\dag U_1$ commute with each other, and that $P_0$ commutes with $P_1$.
\end{proof}

Next we use this lemma to prove that there are no extreme UCPTP maps with Choi-rank 2 for $d_X=2$ and $d_X=3$. The next proposition is essentially the same as theorem 4.23 from \cite{watrous_2018}, but with a simple adaptation to include the case $d_X=3$.
\begin{theorem}{}{inexistence of ucptp extreme for rank 2}
	Let $\varepsilon\in \ucptp(X)$ be a UCPTP map with Choi-rank $\CR(\varepsilon)=2$, and $d_X=2$ or $d_X=3$, then $\varepsilon$ isn't an extreme point of the convex set of UCPTP maps.
\end{theorem}
\begin{proof}
	Let $(E_0,E_1)$ be a Kraus representation of $\varepsilon$. It is equivalent to work with the matrix representations of the operators $E_0$ and $E_1$. To simplify the notation, let's define the matrices
	\begin{equation}
		A_k \coloneqq [E_k].
	\end{equation}
	Since $\CR(\varepsilon)=2$, then $(A_0,A_1)$ is linearly independent. By theorem \ref{theorem:characterization of extreme points for ucptp}, $\varepsilon$ is an extreme point of the convex of UCPTP maps iff $(A_k^\dag A_l \oplus A_l A_k^\dag)_{k,l \in \{0,1\}}$ is linearly independent. We'll show that it can't be linearly independent for $d_X=2$ or $d_X=3$. By lemma \ref{lemma:ucptp channels with rank at most 2}, there exists unitary matrices $U,V \in \U(d_X)$ such that both $VA_0 U^\dag$ and $VA_1 U^\dag$ are diagonal matrices. Denote the standard basis of $d_X \times d_X$ matrices by $e_{i,j}$, so that
	\begin{equation}
		(e_{i,j})_{i',j'} = \delta_{i,i'}\delta_{j,j'}.
	\end{equation}
	Since $VA_k U^\dag$ is diagonal, it can be written as
	\begin{equation}
		VA_k U^\dag = \sum_{i=1}^{d_X} \lambda_{k,i} e_{i,i},
	\end{equation}
	so
	\begin{equation}
		A_k = \sum_{i=1}^{d_X} \lambda_{k,i} V^\dag e_{i,i}U.
	\end{equation}
	Then we can write $A_k^\dag A_l$ as
	\begin{align*}
		A_k^\dag A_l &= \left( \sum_{i=1}^{d_X} \lambda_{k,i} V^\dag e_{i,i}U\right)^\dag \left( \sum_{j=1}^{d_X} \lambda_{l,j} V^\dag e_{j,j}U\right)\\
		&= \sum_{i,j=1}^{d_X} \overline{\lambda_{k,i}}  \lambda_{l,j} U^\dag e_{i,i}VV^\dag e_{j,j}U\\
		&=  \sum_{i,j=1}^{d_X} \overline{\lambda_{k,i}}  \lambda_{l,j} U^\dag e_{i,i} e_{j,j}U\\
		&= \sum_{i,j=1}^{d_X} \overline{\lambda_{k,i}}  \lambda_{l,j} U^\dag \delta_{i,j} e_{i,i} U\\
		&= \sum_{i=1}^{d_X} \overline{\lambda_{k,i}}  \lambda_{l,i} U^\dag e_{i,i} U.
	\end{align*}
	The result for $A_l A_k^\dag$ is similar, but $U$ is replaced by $V$. We get
	\begin{equation}
		A_l A_k^\dag = \sum_{i=1}^{d_X} \overline{\lambda_{k,i}}  \lambda_{l,i} V^\dag e_{i,i} V.
	\end{equation}
	With these results, we get an expression for $A_k^\dag A_l \oplus A_l A_k^\dag$:
	\begin{align*}
		A_k^\dag A_l \oplus A_l A_k^\dag &= \left( \sum_{i=1}^{d_X} \overline{\lambda_{k,i}}  \lambda_{l,i} U^\dag e_{i,i} U \right)\oplus \left( \sum_{i=1}^{d_X} \overline{\lambda_{k,i}}  \lambda_{l,i} V^\dag e_{i,i} V\right)\\
		&=  \sum_{i=1}^{d_X} \overline{\lambda_{k,i}}  \lambda_{l,i} \left(U^\dag e_{i,i} U \oplus V^\dag e_{i,i} V \right).
	\end{align*}
	We've written $A_k^\dag A_l \oplus A_l A_k^\dag$ as a linear combination of the matrices $U^\dag e_{i,i} U \oplus V^\dag e_{i,i} V$. The matrices $U^\dag e_{i,i} U \oplus V^\dag e_{i,i} V$ span a complex vector space of dimension at most $d_X$, since the index $i$ ranges from 1 to $d_X$. If the matrices $A_k^\dag A_l \oplus A_l A_k^\dag$ are linearly independent, they span a subspace of dimension $\CR(\varepsilon)^2 = 2^2 = 4$. Therefore, we must have
	\begin{equation}
		d_X \geq 4.
	\end{equation}
	This proves that $\varepsilon$ isn't extreme for $d_X=2$ or $d_X=3$.
\end{proof}

\subsection{Closure under tensor product}

As done for the other types of CP, CPTP and UCP maps, we prove that UCPTP maps are closed under the tensor product.

\begin{proposition}{}{tensor product of ucptp}
	If $\varepsilon \in \ucptp(X,Y)$ and $\varepsilon' \in \ucptp(X',Y')$ then $\varepsilon \otimes \varepsilon' \in \ucptp(X \otimes X',Y \otimes Y')$.
\end{proposition}
\begin{proof}
	By propositions \ref{proposition:tensor product of cptp} and \ref{proposition:tensor product of ucp} we know that $\varepsilon \otimes \varepsilon' \in \cptp(X \otimes X',Y \otimes Y')$ and $\varepsilon \otimes \varepsilon' \in \ucp(X \otimes X',Y \otimes Y')$. Since $\ucptp(X \otimes X',Y \otimes Y') =  \cptp(X \otimes X',Y \otimes Y') \cap \ucp(X \otimes X',Y \otimes Y')$, this concludes the proof.
\end{proof}

\cleardoublepage
\chapter{The Category of CPTP maps}
\label{section: category of cptp maps}

In this chapter we review some known results about the category of CPTP maps. Not much more is known other than that it is a semicartesian category.

CPTP maps form a monoidal category, where the objects are finite dimensional complex Hilbert spaces and whose arrows are the CPTP maps. For any CPTP maps $\varepsilon \in \cptp(X,Y)$ and $\varepsilon' \in \cptp(X',Y')$ we have their tensor product, which can be seen as the linear map
\begin{align*}
	\varepsilon \otimes \varepsilon' \colon \Lin_\complexNumbers (X \otimes X') &\longrightarrow \Lin_\complexNumbers (Y \otimes Y')\\
	A \otimes A' &\longmapsto \varepsilon(A) \otimes \varepsilon' (A'),
\end{align*}
where $A \in \Lin_\complexNumbers (X)$, $A'\in \Lin_\complexNumbers (X')$, and $A \otimes A'$ and $\varepsilon(A) \otimes \varepsilon'(A')$ denote tensor products of linear maps. It is a lengthy calculation to show that this turns the category of CPTP maps into a monoidal category. Of course, the monoidal product is given by the tensor product and the monoidal unit is $\complexNumbers$, or any 1-dimensional complex Hilbert space. There isn't too much known about the limits an colimits in this category. Next we'll review what are its initial and terminal objects. That it is a semicartesian category, this is already known, as we can see in section 4 of \cite{Huot_2019}. We'll also review a known characterization of sections and isomorphisms in this category.

\begin{proposition}{}{category cptp, initial and terminal objects}
	The category of CPTP maps has $\{0\}$ as initial object and $\complexNumbers$ as terminal object. In particular, it is a semicartesian category, with the terminal object coinciding with the monoidal unit.
\end{proposition}
\begin{proof}
	By the \CJ isomorphism, the Hom-set $\cptp(X,Y)$ is isomorphic to $\psdii(X,Y)$. We'll work with $\psdii(X,Y)$, which is simpler. First we show that $\{0\}$ is an initial object. Let $X$ be a finite dimensional complex Hilbert space. By proposition \ref{proposition:psdii(X,Y) for X or Y = 0} we already know that $\psdii(\{0\},X) = \{0\}$. By the \CJ isomorphism, it follows that $\cptp(\{0\},X) = \{0\}$. This proves that $\{0\}$ is an initial object of the category of CPTP maps.
	
	Now let's prove that $\complexNumbers$ is a terminal object of the category. We have that
	\begin{equation}
		\psdii(X,\complexNumbers) = \{A \in \psd(X^* \otimes \complexNumbers) \mid \tr_2 A = \id_{X^*}\}.
	\end{equation}
	Any such $A$ is uniquely written as $A = B \otimes \id_\complexNumbers$, where $B \in \Lin_\complexNumbers (X^*)$. Then $\tr_2 A = B$, which implies that the equation $\tr_2 A = \id_{X^*}$ is equivalent to $B = \id_{X^*}$. The only possible $A$ is then $\id_{X^* \otimes \complexNumbers}$, which is positive semidefinite as has $\tr_2 \id_{X^* \otimes \complexNumbers} = \id_{X^*}$. Therefore,
	\begin{equation}
		\psdii(X,\complexNumbers) = \{\id_{X^* \otimes \complexNumbers }\}.
	\end{equation}
	In proposition \ref{proposition:partial trace as cptp map}, taking $Y = \complexNumbers$ we can identify $\tr_1 \in \cptp(X\otimes \complexNumbers, \complexNumbers)$ with the total trace $\tr \in \cptp(X, \complexNumbers)$, through an isomorphism. From the expression for $J(\tr_1)$ given in that proposition, it follows that
	\begin{equation}
		J(\tr) = \id_{X^* \otimes \complexNumbers}.
	\end{equation}
	Therefore,
	\begin{equation}
		\cptp(X, \complexNumbers) = \{\tr\}.
	\end{equation}
	This shows that $\complexNumbers$ is a terminal object, and that $\tr \colon X \to \complexNumbers$ is the unique CPTP map. Since $\complexNumbers$ is also the monoidal unit, the category is semicartesian.
	
\end{proof}

In any semicartesian category we have a notion of projection, as follows. Let $X$ and $Y$ be objects in such semicartesian category. Denote by 1 the terminal object of the category, which coincides with the monoidal unit. Then we define projections $\pi_X \colon X \otimes Y \to X$ and $\pi_Y \colon X \otimes Y \to Y$ as the compositions
\begin{center}
	\begin{tikzcd}
		\pi_X \colon X \otimes Y  \arrow[r, "\id_X \otimes !"] & X \otimes 1 \arrow[r, "\cong"] & X,
	\end{tikzcd}
\end{center}
\begin{center}
	\begin{tikzcd}
		\pi_Y \colon X \otimes Y  \arrow[r, "! \otimes \id_Y"] & 1 \otimes Y \arrow[r, "\cong"] & Y.
	\end{tikzcd}
\end{center}
We are denoting by $!$ the unique arrow to the terminal object. In the case of the category of CPTP maps, this arrow is the total trace. It follows that these projections are the partial traces
\begin{equation}
	\pi_X = \tr_2,
\end{equation}
\begin{equation}
	\pi_Y = \tr_1.
\end{equation}
Notice that the projection on the first factor $X$ corresponds to the partial trace over the second factor. But the partial traces determine orthogonal projections $p_1$ and $p_2$, as seen in proposition \ref{proposition:partial traces as projections}. The partial trace $\tr_2$ corresponds to $p_1$, and $\tr_1$ corresponds to $p_2$. Therefore, $\pi_X$ corresponds to $p_1$ and $\pi_Y$ corresponds to $p_2$, which reestablishes the order.

Next we'll review a known characterization of sections in the category of CPTP maps, proven in \cite{nayak2006invertible}. As a corollary, we obtain that the isomorphisms are the conjugations by isometric isomorphisms.

\begin{proposition}{}{section}
	Let $X$ and $Y$ be finite dimensional complex Hilbert spaces, with $1 \leq d_X \leq d_Y$. Let also $\varepsilon \in \cptp(X,Y)$ be a section in the category of CPTP maps. That is, there exists $\varphi \in \cptp(Y,X)$ such that
	\begin{equation}
		\varphi \varepsilon = \id_{\Lin_\complexNumbers (X)}.
	\end{equation}
	Then there exists an isometry $U \in \Lin_\complexNumbers (X \otimes \complexNumbers^{\CR(\varepsilon)}, Y)$ and a density operator $\rho \in \density(\complexNumbers^{\CR(\varepsilon)})$ such that
	\begin{equation}
		\label{equation: cptp right inverse}
		\varepsilon(A) = U(A \otimes \rho)U^\dag,
	\end{equation}
	for all $A \in \Lin_\complexNumbers (X)$. Also, the Choi-rank of $\varepsilon$ necessarily satisfies that
	\begin{equation}
		\CR(\varepsilon) \leq \left\lfloor \frac{d_Y}{d_X} \right\rfloor.
	\end{equation}
\end{proposition}
\begin{proof}
	Let $(E_k)_{k=1,\dots,\CR(\varepsilon)}$ be linearly independent Kraus operators for $\varepsilon$ (they can be obtained by diagonalizing $J(\varepsilon)$ and using the eigenvectors with non zero eigenvalue). The number of such Kraus operators is necessarily the Choi-rank $\CR(\varepsilon)$. Let also $(F_l)_l$ be Kraus operators for $\varphi$. Then the composition $\varphi \varepsilon$ have $(F_l E_k)_{k,l}$ as Kraus operators. Since $\varphi \varepsilon = \id_{\Lin_\complexNumbers (X)}$, $(F_l E_k)_{k,l}$ are Kraus operators for $\id_{\Lin_\complexNumbers (X)}$. But $(\id_{\Lin_\complexNumbers (X)})$ is a Kraus operator for $\id_{\Lin_\complexNumbers (X)}$. By theorem 8.2 of \cite{nielsen_chuang_2010}, there exists a unit vector $(v_{l,k})_{k,l}$, that is,
	\begin{equation}
		\sum_{k,l} |v_{l,k}|^2 = 1,
	\end{equation}
	such that
	\begin{equation}
		F_l E_k = v_{l,k} \id_X.
	\end{equation}
	Taking the adjoint, we get that
	\begin{equation}
		E_k^\dag F_l^\dag  = \overline{v_{l,k}} \id_X.
	\end{equation}
	Composing these equations, we get that
	\begin{equation}
		E_k^\dag F_l^\dag F_{l'} E_{k'} = \overline{v_{l,k}} v_{l',k'} \id_X.
	\end{equation}
	Taking $l=l'$ and summing over $l$ we get that
	\begin{equation}
		E_k^\dag \left(\sum_l F_l^\dag F_l\right) E_{k'} = \left(\sum_l \overline{v_{l,k}} v_{l,k'}\right) \id_X.
	\end{equation}
	But $\varphi$ is a CPTP map, so, by proposition \ref{proposition:kraus TP condition}, we have that
	\begin{equation}
		\sum_l F_l^\dag F_l = \id_Y.
	\end{equation}
	Using this equation, we get that
	\begin{equation}
		E_k^\dag  E_{k'} = \left(\sum_l \overline{v_{l,k}} v_{l,k'}\right) \id_X.
	\end{equation}
	The summation on the right can be interpreted as a multiplication of matrices. Let $V=(v_{l,k})_{l,k}$, then
	\begin{equation}
		\sum_l \overline{v_{l,k}} v_{l,k'} = \sum_l V^\dag_{k,l} V_{l,k'} = (V^\dag V)_{k,k'}.
	\end{equation}
	Let
	\begin{equation}
		M \coloneqq V^\dag V \in M_{\CR(\varepsilon)}(\complexNumbers).
	\end{equation}
	Then
	\begin{equation}
		E_k^\dag  E_{k'} = M_{k,k'} \id_X.
	\end{equation}
	
	The trace of $M$ is
	\begin{equation}
		\tr(M) = \sum_k (V^\dag V)_{k,k} = \sum_{k,l} \overline{v_{l,k}} v_{l,k} = \sum_{k,l} |v_{l,k}|^2 =1.
	\end{equation}
	The product $V^\dag V$ is a positive semidefinite matrix. Therefore, it is diagonalizable and has only non negative eigenvalues. Then there exists a unitary matrix $W \in U(\CR(\varepsilon))$ and a diagonal matrix $\Lambda \in M_{\CR(\varepsilon)}(\complexNumbers)$ such that
	\begin{equation}
		WMW^\dag = \Lambda,
	\end{equation}
	and $\Lambda$ only has non negative entries. Let's write
	\begin{equation}
		\Lambda = 	\begin{pmatrix}
			\lambda_1 & \dots & 0\\
			\vdots & \ddots & \vdots\\
			0 & \dots & \lambda_{\CR(\varepsilon)}
		\end{pmatrix},
	\end{equation}
	so
	\begin{equation}
		\Lambda_{k,k'} = \delta_{k,k'} \lambda_k.
	\end{equation}
	Since $W$ is unitary, both $M$ and $\Lambda$ have the same trace. This implies that
	\begin{equation}
		\sum_k \lambda_k = 1.
	\end{equation}
	
	We can use $W$ to make $\Lambda$ appear in the expression for $E_k^\dag E_{k'}$. We have that
	\begin{equation}
		\label{equation: sumkk' Wsk Ekdag etc}
		\sum_{k,k'} W_{s,k} E_k^\dag  E_{k'} W^\dag_{k',s'} = \sum_{k,k'} W_{s,k} M_{k,k'} W^\dag_{k',s'} \id_X.
	\end{equation}
	On the right hand side we have the matrix product $WMW^\dag$:
	\begin{equation}
		\sum_{k,k'} W_{s,k} M_{k,k'} W^\dag_{k',s'} = (WMW^\dag)_{s,s'} = \Lambda_{s,s'} = \delta_{s,s'} \lambda_s.
	\end{equation}
	On the left hand side of equation \ref{equation: sumkk' Wsk Ekdag etc} we can separate the indices $k$ and $k'$. We define the operators
	\begin{equation}
		C_s \coloneq \sum_k E_k W^\dag_{k,s} \in \Lin_\complexNumbers (X,Y).
	\end{equation}
	Then the equation \ref{equation: sumkk' Wsk Ekdag etc} is written as
	\begin{equation}
		C_s^\dag C_{s'} = \delta_{s,s'} \lambda_s \id_X.
	\end{equation}
	These new operators $(C_s)_{s=1,\dots,\CR(\varepsilon)}$ are also Kraus operators for $\varepsilon$. This is because $W^\dag$ is a unitary matrix and because of theorem 8.2 of \cite{nielsen_chuang_2010}.
	
	We know that $\lambda_s \geq 0$ for $s=1,\dots,\CR(\varepsilon)$. If we had $\lambda_s = 0$ for some $s$, then we would have $C_s^\dag C_s = 0$. This would imply that $C_s = 0$, and then we could discard this operator from the Kraus representation. Then we would obtain a Kraus representation with less operators than $\CR(\varepsilon)$. This can't happen, because $\CR(\varepsilon)$ is the size of the smallest Kraus representation of $\varepsilon$. Therefore
	\begin{equation}
		\lambda_s > 0,
	\end{equation}
	for $s=1,\dots,\CR(\varepsilon)$.
	
	We have the equation
	\begin{equation}
		C^\dag_s C_s = \lambda_s \id_X.
	\end{equation}
	This is almost the equation of an isometry, except for the factor $\lambda_s$. Dividing by $\lambda_s$, we get that
	\begin{equation}
		\left(\frac{C_s}{\sqrt{\lambda_s}}\right)^\dag \frac{C_s}{\sqrt{\lambda_s}} = \id_X,
	\end{equation}
	so $C_s / \sqrt{\lambda_s}$ is an isometry. We can combine these isometries into a single isometry $U$. Let $(\ket{s})_{s=1,\dots,\CR(\varepsilon)}$ be any orthonormal basis for $\complexNumbers^{\CR(\varepsilon)}$. We define
	\begin{equation}
		U \coloneqq \sum_{s=1, \dots, \CR(\varepsilon)} \frac{C_s}{\sqrt{\lambda_s}} \otimes \bra{s} \ \in \Lin_\complexNumbers (X \otimes \complexNumbers^{\CR(\varepsilon)},Y).
	\end{equation}
	Notice that we are implicitly using the isomorphism $Y \otimes \complexNumbers \cong Y$. Let's check that $U$ is indeed an isometry. We have that
	\begin{align*}
		U^\dag U &= \left(\sum_s \frac{C_s^\dag}{\sqrt{\lambda_s}} \otimes \ket{s}\right) \left(\sum_{s'} \frac{C_{s'}}{\sqrt{\lambda_{s'}}} \otimes \bra{s'}\right)\\
		&= \sum_{s,s'} \frac{C_s^\dag C_{s'}}{\sqrt{\lambda_s \lambda_{s'}}} \otimes \ket{s}\bra{s'}\\
		&= \sum_{s,s'} \frac{\delta_{s,s'} \lambda_s \id_X}{\sqrt{\lambda_s \lambda_{s'}}} \otimes \ket{s}\bra{s'}\\
		&= \sum_s \frac{\lambda_s \id_X}{\sqrt{\lambda_s \lambda_s}} \otimes \ket{s}\bra{s}\\
		&= \sum_s \id_X \otimes \ket{s}\bra{s}\\
		&= \id_X \otimes \sum_s \ket{s}\bra{s}\\
		&= \id_X \otimes \id_{\complexNumbers^{\CR(\varepsilon)}}\\
		&= \id_{X \otimes \complexNumbers^{\CR(\varepsilon)}}.
	\end{align*}
	This proves that $U^\dag U = \id_{X \otimes \complexNumbers^{\CR(\varepsilon)}}$, so $U$ is an isometry. An isometry is, in particular, an injective linear map. This implies that
	\begin{equation}
		\dim_\complexNumbers(X \otimes \complexNumbers^{\CR(\varepsilon)}) \leq \dim_\complexNumbers Y.
	\end{equation}
	The left hand side is the same as $d_X \CR(\varepsilon)$. The right hand side can also be written as $d_Y$. Therefore, we must have that
	\begin{equation}
		\CR(\varepsilon) \leq \frac{d_Y}{d_X}.
	\end{equation}
	Since the Choi-rank is a natural number, we can take the floor of the right hand side. We get that
	\begin{equation}
		\CR(\varepsilon) \leq \left\lfloor\frac{d_Y}{d_X}\right\rfloor.
	\end{equation}
	
	Now define the density operator
	\begin{equation}
		\rho \coloneqq \sum_s \lambda_s \ket{s}\bra{s} \in \density(\complexNumbers^{\CR(\varepsilon)}).
	\end{equation}
	This is a density operator, because this equation is already a diagonalization of $\rho$ with orthonormal eigenvectors, and because the eigenvalues $\lambda_s$ are positive and $\sum_s \lambda_s = 1$. We can express $\varepsilon(A)$ is terms of both $U$ and $\rho$. For any $A \in \Lin_\complexNumbers (X)$, we have that
	\begin{align*}
		U(A\otimes \rho)U^\dag &= \left(\sum_s \frac{C_s}{\sqrt{\lambda_s}} \otimes \bra{s}\right) (A \otimes \rho) \left(\sum_{s'} \frac{C_{s'}^\dag}{\sqrt{\lambda_{s'}}} \otimes \ket{s'}\right) \\
		&= \sum_{s,s'} \frac{C_s A C_{s'}^\dag}{\sqrt{\lambda_s \lambda_{s'}}} \ \bra{s} \rho \ket{s'}.
	\end{align*}
	The inner product $\bra{s} \rho \ket{s'}$ is
	\begin{align*}
		\bra{s} \rho \ket{s'} &= \bra{s}\left(\sum_{s''} \lambda_{s''} \ket{s''}\bra{s''}\right)\ket{s'}\\
		&= \sum_{s''} \lambda_{s''} \langle s \mid s'' \rangle \langle s'' \mid s' \rangle\\
		&= \sum_{s''} \lambda_{s''} \delta_{s,s''} \delta_{s'',s'} \\
		&= \delta_{s,s' }\lambda_s.
	\end{align*}
	Therefore,
	\begin{align*}
		U(A\otimes \rho)U^\dag &= \sum_{s,s'} \frac{C_s A C_{s'}^\dag}{\sqrt{\lambda_s \lambda_{s'}}} \ \delta_{s,s' }\lambda_s \\
		&= \sum_s  \frac{C_s A C_s^\dag}{\sqrt{\lambda_s \lambda_s}} \ \lambda_s \\
		&= \sum_s C_s A C_s^\dag.
	\end{align*}
	Since $(C_s)_s$ are Kraus operators for $\varepsilon$, this last expression is $\varepsilon(A)$. This proves that
	\begin{equation}
		\varepsilon(A) = U(A\otimes \rho)U^\dag.
	\end{equation}
\end{proof}

\begin{corollary}{}{}
	Let $X$ and $Y$ be finite dimensional complex Hilbert spaces with $d_X = d_Y \geq 1$. If $\varepsilon \in \cptp(X,Y)$ is an isomorphism in the category of CPTP maps, then there exists an isometric isomorphism $U \in \Lin_\complexNumbers (X,Y)$ such that
	\begin{equation}
		\varepsilon(A) = UAU^\dag,
	\end{equation}
	for all $A \in \Lin_\complexNumbers (X)$. Conversely, any linear map of this form is an isomorphism in the category of CPTP maps.
\end{corollary}
\begin{proof}
	$(\implies)$ Suppose that $\varepsilon$ is an isomorphism. Then $\varepsilon$ is a section, whose left inverse is $\varepsilon^{-1} \in \cptp(Y,X)$. By proposition \ref{proposition:section}, there exists an isometry $U \in \Lin_\complexNumbers (X \otimes \complexNumbers^{\CR(\varepsilon)}, Y)$ and a density operator $\rho \in \density(\complexNumbers^{\CR(\varepsilon)})$ such that
	\begin{equation}
		\varepsilon(A) = U(A \otimes \rho) U^\dag,
	\end{equation}
	for all $A \in \Lin_\complexNumbers (X)$. Also, the Choi-rank satisfies that
	\begin{equation}
		\CR(\varepsilon) \leq \frac{d_Y}{d_X} = 1.
	\end{equation}
	Therefore,
	\begin{equation}
		\CR(\varepsilon) = 1.
	\end{equation}
	By the isomorphism $X \otimes \complexNumbers \cong X$, the isometry $U$ corresponds to an isometry in $\Lin_\complexNumbers(X,Y)$. By the isomorphism $\Lin_\complexNumbers (\complexNumbers) \cong \complexNumbers$, the density operator $\rho$ corresponds to 1. Then the results can be restated in the following way: there exists an isometry $U \in \Lin_\complexNumbers (X,Y)$ such that
	\begin{equation}
		\varepsilon(A) = U(A\cdot 1)U^\dag = UAU^\dag.
	\end{equation}
	Since $U$ is an injective linear map between spaces of the same dimension, it is an isomorphism. Therefore, it is a isometric isomorphism.
	
	$(\impliedby)$ Let $U \in \Lin_\complexNumbers (X,Y)$ be an isometric isomorphism and define the $\complexNumbers$-linear map
	\begin{align*}
		\varepsilon_U \colon \Lin_\complexNumbers (X) &\longrightarrow \Lin_\complexNumbers (Y)\\
		A &\longmapsto UAU^\dag .
	\end{align*}
	Let's show that $\varepsilon_U$ is a CPTP map. First we show that it is a completely positive map. Let $Z$ be a finite dimensional complex Hilbert space. Let also $P \in \Lin_\complexNumbers (Z \otimes X)$ be a positive semidefinite operator. Then
	\begin{equation}
		(\id_{\Lin_\complexNumbers (Z)} \otimes \varepsilon_U)(P) = (\id_Z \otimes U)P(\id_Z \otimes U^\dag).
	\end{equation}
	This is the conjugation of the positive semidefinite $P$ by the operator $\id_Z \otimes U$, which implies that $(\id_{\Lin_\complexNumbers (Z)} \otimes \varepsilon_U)(P)$ is positive semidefinite. This proves that $\varepsilon_U$ is completely positive. Next let's prove that $\varepsilon_U$ is trace preserving. For any $A \in \Lin_\complexNumbers (X)$, we have that
	\begin{equation}
		\tr(\varepsilon_U (A)) = \tr(UAU^\dag) = \tr(U^\dag U A) = \tr(A).
	\end{equation}
	This concludes that $\varepsilon_U \in \cptp(X,Y)$.
	
	Since $U$ is an isometric isomorphism, the same is true for $U^\dag$. It follows that $\varepsilon_U$ is an isomorphism in the category of CPTP maps, whose inverse is $\varepsilon_{U^\dag}$.
\end{proof}

\cleardoublepage
\chapter{Extreme Channels by Choi-Rank}
\label{chapter: examples extreme channels}

We'll construct examples of extreme quantum channels for the CPTP and UCPTP cases. They are constructed by Choi-rank, in the attempt to show the existence of channels for each Choi-rank allowed by the known bounds. In the CPTP case we succeed in constructing channels for each Choi-rank allowed by the bounds. For the UCPTP case, we are successful for Choi-rank 2. For Choi-rank 3 and above we find a formula that works for all cases that we could test and that are allowed by the known bounds. Unfortunately, we lack a proof that this formula always works. Nevertheless, we succeed in constructing hundreds of examples of extreme UCPTP maps with Choi-rank at least 3.

\section{Extreme CPTP maps for each Choi-rank}
\label{section: extreme cptp}

As seen in proposition \ref{proposition:pre-image of pi is stiefel manifold}, we can parameterize CPTP maps by (compact) Stiefel manifolds. An element of a Stiefel manifold is an isometry, which is a matrix with orthonormal columns. Constructing orthonormal vectors is a simple task, as least until some additional constraints are imposed. 

By the \CJ isomorphism, a CPTP map corresponds to an operator $A \in \psdii (X,Y)$. The Kraus representation corresponds to expressing $A$ as a sum
\begin{equation}
	A = \sum_{k=0}^{r-1} \ket{\psi_k } \bra{\psi_k },
\end{equation}
where $\psi_k \in X^* \otimes Y$. Notice that we are using indices that start at 0, not 1, which is better for using modular arithmetic later. Following the notation of proposition \ref{proposition:pre-image of pi is stiefel manifold}, we can write $\psi_k$ as
\begin{equation}
	\label{equation: psik in basis}
	\ket{\psi_k} = \sum_{i,j,k} \psi_{i,j,k} \ket{x^i \otimes y_j},
\end{equation}
where $(x^i)_{i=0,\dots, d_X -1}$ is an orthonormal basis for $X^*$, $(y_j)_{j=0,\dots, d_Y -1}$ is an orthonormal basis for $Y$ and $\psi_{i,j,k} \in \complexNumbers$. The condition for having a CPTP map corresponds to the matrices
\begin{equation}
	\psi_i \coloneqq (\psi_{i,j,k})_{j,k} \in M_{d_Y, r} (\complexNumbers)
\end{equation}
being orthonormal with respect to the Euclidean inner product. The CPTP map has Choi-rank $r$ if the vectors $(\psi_k)_k$ are linearly independent. Linear independence is much simpler to obtain than orthonormality. For this reason, the general strategy for constructing examples is to guess orthonormal vectors and hope that $(\psi_k)_k$ will be linearly independent.

Before constructing the examples, recall from corollary \ref{corollary:bounds on rank of extreme cptp} that the Choi-rank of an extreme CPTP map satisfies the following bound
\begin{equation}
	\frac{d_X}{d_Y} \leq \rank(A) \leq d_X .
\end{equation}
We want $r$ to be the Choi-rank, so we'll take $r$ to also satisfy this bound.

We'll construct the examples as follows. The simplest way to obtain orthonormal vectors is to use the vectors from the standard basis. That is, each matrix $\psi_i$ has a single non zero entry, and such entry has value 1. We can start with a 1 at the entry with indices $(0,0)$, and then move the $1$ to a neighbor entry to obtain the next matrix. There are at least two obvious ways of doing this, one is moving the $1$ a row at a time, the other is moving the $1$ a column at a time. When all entries of a row were used, we use the next row, and similarly for columns. Later we'll want all $\psi_k$'s to be non zero, so it is important to use all indices $k$. This leads to the following choice of $\psi_i$. Its entries are given by
\begin{equation}
	\label{equation: extreme cptp example}
	\psi_{i,j,k} \coloneqq \delta_{j, i // r} \delta_{k, i \% r},
\end{equation}
where $i//r$ is the integer division of $i$ by $r$ and $i \% r$ is the remainder of the division of $i$ by $r$. This notation is the same used in the Python programming language. As such, we have that
\begin{equation}
	\label{equation: i integer division}
	i = (i//r) \ r + i \% r .
\end{equation}
If there is a pair $(j,k)$ such that
\begin{equation}
	i = jr+k,
\end{equation}
it is unique. This is because the quotient and remainder of integer division are unique. We have to prove that there exists such a pair $(j,k)$ such that the last equation is satisfied. For $j$ to exist, we need that
\begin{equation}
	0\leq i // r < d_Y ,
\end{equation}
and for $k$ to exist, we need that
\begin{equation}
	0 \leq i \% r < r .
\end{equation}
This last equation is immediate, since the remainder is always less than $r$. For the other equation, of course
\begin{equation}
	0 \leq i // r ,
\end{equation}
because $i \geq 0$. Since $r$ satisfies that
\begin{equation}
	\frac{d_X}{d_Y} \leq r ,
\end{equation}
then 
\begin{equation}
	\frac{1}{r} \leq \frac{d_Y}{d_X} .
\end{equation}
Multiplying by $i$, we get that
\begin{equation}
	\frac{i}{r} \leq \frac{i}{d_X} d_Y .
\end{equation}
Since $i < d_X$, then
\begin{equation}
	\frac{i}{r} < d_Y.
\end{equation}
By equation \ref{equation: i integer division}, and since $i \% r \geq 0$,  then
\begin{equation}
	i // r \leq i/r,
\end{equation}
which implies that
\begin{equation}
	i // r < d_Y .
\end{equation}
This concludes that there exists a unique pair $(j,k)$ such that
\begin{equation}
	i = jr + k,
\end{equation}
for each $i$. From this result it follows that each $\psi_i$ is a matrix with a single non zero entry, whose value is 1, and that there is no repetition of matrices. Therefore, $(\psi_i)_i$ is a sequence of orthonormal matrices and $A \in \psdii(X,Y)$.

Next let's check that $(\psi_k)$ doesn't have zero vectors. For a given $k$, for $\psi_k$ to be non zero it is required that the term
\begin{equation}
	\delta_{k, i \% r} \neq 0
\end{equation}
for some $i$. Since $i$ can be any number in $\{0, \dots, d_X-1\}$, $r \leq d_X$ and $0\leq k < r$, we can take
\begin{equation}
	i = k.
\end{equation}
Then 
\begin{align*}
	\psi_{i,j,k} &\stackrel{(i)}{=} \delta_{j, i//r} \\
	&\stackrel{(ii)}{=} \delta_{j, k // r} \\
	&\stackrel{(iii)}{=} \delta_{j, 0}.
\end{align*}
In $(i)$ was used equation \ref{equation: extreme cptp example}; in $(ii)$ was used that $i=k$; in $(iii)$ was used that $0 \leq k < r$, so the integer division $k // r$ is 0. This shows that
\begin{equation}
	\psi_{k,0,k} = 1,
\end{equation}
for every $k \in \{0, \dots, r-1\}$. In particular, this proves that
\begin{equation}
	\psi_k \neq 0,
\end{equation}
for all $k \in \{0, \dots, r-1\}$.

Instead of first checking that $(\psi_k)_k$ is linearly independent, it is more convenient to check the linear independence of $(\tr_2 \ket{\psi_k } \bra{\psi_l })_{k,l}$, because it already implies that $(\psi_k)_k$ is linearly independent and that the CPTP map is extreme. Let's compute $\tr_2 \ket{\psi_k } \bra{\psi_l }$. We have that
\begin{align*}
	\tr_2 \ket{\psi_k } \bra{\psi_l } &\stackrel{(i)}{=} \tr_2 \left( \left( \sum_{i,j} \psi_{i,j,k} \ket{x^i \otimes y_j} \right) \left( \sum_{i',j'} \overline{\psi_{i',j',l}} \bra{x^{i'} \otimes y_{j'}} \right)  \right) \\
	&\stackrel{(ii)}{=} \sum_{i,i',j,j'} \psi_{i,j,k} \overline{\psi_{i',j',l}} \ \tr_2(\ket{x^i \otimes y_j} \bra{x^{i'} \otimes y_{j'}}) \\
	&\stackrel{(iii)}{=} \sum_{i,i',j,j'} \psi_{i,j,k} \overline{\psi_{i',j',l}} \ \ket{x^i} \bra{x^{i'}} \tr(\ket{y_j} \bra{y_{j'}})\\
	&\stackrel{(iv)}{=} \sum_{i,i',j,j'} \psi_{i,j,k} \overline{\psi_{i',j',l}} \ \delta_{j,j'} \ket{x^i} \bra{x^{i'}} \\
	&\stackrel{(v)}{=} \sum_{i,i',j} \psi_{i,j,k} \overline{\psi_{i',j,l}} \  \ket{x^i} \bra{x^{i'}} \\
	&\stackrel{(vi)}{=} \sum_{i,i'} \left( \sum_j \psi_{i,j,k} \overline{\psi_{i',j,l}} \right) \  \ket{x^i} \bra{x^{i'}}.
\end{align*}
In $(i)$ was used equation \ref{equation: psik in basis}; in $(ii)$ was used that the partial trace is linear; in $(iii)$ was applied the partial trace, which acts as the total trace over a factor; in $(iv)$ was used that $(y_j)_j$ is orthonormal; in $(v)$ the delta was used to eliminate the summation over $j'$; in $(vi)$ the summation over $j$ was separated to make evident the matrix elements of $\tr_2 \ket{\psi_k } \bra{\psi_l }$.

Next we find an expression for the matrix elements. We have that
\begin{align*}
	\sum_j \psi_{i,j,k} \overline{\psi_{i',j,l}} &\stackrel{(i)}{=} \sum_j \delta_{j, i // r} \delta_{k, i \% r} \delta_{j, i' // r} \delta_{l, i' \% r} \\
	&\stackrel{(ii)}{=} \delta_{k, i \% r} \delta_{l, i' \% r} \sum_j \delta_{j, i // r} \delta_{j, i' // r} \\
	&\stackrel{(iii)}{=} \delta_{k, i \% r} \delta_{l, i' \% r} \delta_{i // r, i' // r}.
\end{align*}
In $(i)$ was used equation \ref{equation: extreme cptp example}; in $(ii)$ the terms that don't depend on $j$ were moved to outside the sum; in $(iii)$ the $\delta_{j, i// r}$ was used to eliminate the summation over $j$.

It is important that $\tr_2 \ket{\psi_k } \bra{\psi_l } \neq 0$. Since $k,l \in \{0, \dots, r-1\}$, $r \leq d_X$ and $i,i' \in \{0, \dots, d_X -1\}$, we can take
\begin{equation}
	i = k
\end{equation}
and
\begin{equation}
	i' = l.
\end{equation}
This implies that
\begin{equation}
	i \% r =  k \% r = k
\end{equation}
and
\begin{equation}
	i' \% r = l \% r = l,
\end{equation}
because $k, l < r$. Then the matrix element is
\begin{equation}
	\sum_j \psi_{k,j,k} \overline{\psi_{l,j,l}} = \delta_{k // r, l // r}.
\end{equation}
Since $k,l < r$, then
\begin{equation}
	k // r = 0
\end{equation}
and
\begin{equation}
	l // r = 0.
\end{equation}
Therefore
\begin{equation}
	\sum_j \psi_{k,j,k} \overline{\psi_{l,j,l}} = \delta_{0,0} = 1.
\end{equation}
This proves that
\begin{equation}
	\tr_2 \ket{\psi_k } \bra{\psi_l } \neq 0,
\end{equation}
for every $k,l \in \{0, \dots r-1\}$.

A sufficient condition for non zero vectors to be linearly independent is orthogonality. Orthogonality is trivial when different vectors don't have non zero entries in common. This is the case for our example. The term $\delta_{k, i \% r} \delta_{l, i' \% r}$ selects the non zeros entries as having indices $(i,i')$ with remainders
\begin{equation}
	i \% r = k
\end{equation}
and
\begin{equation}
	i' \% r = l.
\end{equation}
These equations may become more familiar if we use the equivalence relation from the cyclic group $\integers_r$:
\begin{gather}
	i \equiv k \text{ mod } r, \\
	i' \equiv l \text{ mod } r.
\end{gather}
Since $k, l \in \{0, \dots, r-1\}$, both give unique representatives for each equivalence class of $\integers_r$. Therefore, for each $(k,l)$ the vector $\tr_2 \ket{\psi_k } \bra{\psi_l }$ have non zero entries for different pairs $(i,i')$. This implies that $(\tr_2 \ket{\psi_k } \bra{\psi_l } )_{k,l}$ is orthogonal. Since none of the vectors are zero, it follows that $(\tr_2 \ket{\psi_k } \bra{\psi_l } )_{k,l}$ is linearly independent  and that the CPTP map is extreme. This also implies that $(\psi_k )_k$ is linearly independent, so the Choi-rank is $r$.

Notice that this construction doesn't work to produce CPTP maps with Choi-rank higher than $d_X$. If $r > d_X$, then
\begin{equation}
	i \% r = i.
\end{equation}
This implies that
\begin{equation}
	\delta_{k, i \% r} = \delta_{k, i}.
\end{equation}
If $k \geq d_X$, then $k > i$, so
\begin{equation}
	\delta_{k,i} = 0,
\end{equation}
which implies that
\begin{equation}
	\psi_{i,j,k} = 0,
\end{equation}
for every $i$ and $j$. Therefore
\begin{equation}
	\psi_k = 0,
\end{equation}
for every $k \in \{d_X, \dots, r-1\}$. This implies that $(\psi_k)_k$ is linearly dependent and that the Choi-rank is at most $d_X$.

\section{Extreme UCPTP maps with Choi-rank 2}

The convex set of UCPTP maps is contained in the convex set of CPTP maps. Therefore, each UCPTP map that is an extreme point of the set of CPTP maps is also an extreme point of the subset of UCPTP maps. This gives a method for obtaining some examples of extreme UCPTP maps.

Theorem \ref{theorem:inexistence of ucptp extreme for rank 2} shows that there are no extreme UCPTP maps with Choi-rank 2 for $d_X=2$ or $d_X=3$. But we have examples for higher dimensions, as we prove next.

\begin{theorem}{}{extreme ucptp rank 2}
	For $d_X \geq 4$, we can construct as follows a UCPTP map $\varepsilon \in \ucptp(X)$ of Choi-rank 2 that is an extreme point of the set of UCPTP maps. The UCPTP map that will be defined is also an extreme point of the set of CPTP maps. Let $U\in \U(d_X-3)$ and $V \in \U(3)$ be unitary matrices such that $(I_3,V,V^\dag)$ is a linearly independent sequence in $M_3(\complexNumbers)$. Then define the Kraus matrices $(E_0,E_1)$ that represent the UCPTP map as the block matrices
	\begin{equation}
		E_0 \coloneqq
		\begin{pmatrix}
			U & 0\\
			0 & I_3 / \sqrt{2}
		\end{pmatrix},
	\end{equation}
	\begin{equation}
		E_1 \coloneqq
		\begin{pmatrix}
			0 & 0\\
			0 & V/\sqrt{2}
		\end{pmatrix}.
	\end{equation}
	An example of $V$ such that $(I_3,V,V^\dag)$ is linearly independent is
	\begin{equation}
		V = 
		\begin{pmatrix}
			0 & 0 & 1\\
			1 & 0 & 0 \\
			0 & 1 & 0
		\end{pmatrix}.
	\end{equation}
\end{theorem}
\begin{proof}
	First let's compute the products $E_k^\dag E_l$.
	\begin{itemize}
		\item $E_0^\dag E_0$:
		\begin{align*}
			E_0^\dag E_0 &=
			\begin{pmatrix}
				U^\dag & 0\\
				0 & I_3 / \sqrt{2}
			\end{pmatrix}
			\begin{pmatrix}
				U & 0\\
				0 & I_3 / \sqrt{2}
			\end{pmatrix}\\
			&= 
			\begin{pmatrix}
				U^\dag U & 0\\
				0 & I_3 /2
			\end{pmatrix}\\
			&= \begin{pmatrix}
				I_{d_X-3} & 0\\
				0 & I_3 /2
			\end{pmatrix}.
		\end{align*}
		
		\item $E_1^\dag E_1$:
		\begin{align*}
			E_1^\dag E_1 &=
			\begin{pmatrix}
				0 & 0\\
				0 & V^\dag / \sqrt{2}
			\end{pmatrix}
			\begin{pmatrix}
				0 & 0\\
				0 & V/ \sqrt{2}
			\end{pmatrix}\\
			&= 
			\begin{pmatrix}
				0 & 0\\
				0 & V^\dag V /2
			\end{pmatrix}\\
			&= \begin{pmatrix}
				0 & 0\\
				0 & I_3 /2
			\end{pmatrix}.
		\end{align*}
		
		\item $E_0^\dag E_1$:
		\begin{align*}
			E_0^\dag E_1 &=
			\begin{pmatrix}
				U^\dag & 0\\
				0 & I_3 / \sqrt{2}
			\end{pmatrix}
			\begin{pmatrix}
				0 & 0\\
				0 & V/ \sqrt{2}
			\end{pmatrix}\\
			&= 
			\begin{pmatrix}
				0 & 0\\
				0 &  V /2
			\end{pmatrix}.
		\end{align*}
		
		\item $E_1^\dag E_0$:
		\begin{align*}
			E_1^\dag E_0 &=
			\begin{pmatrix}
				0 & 0\\
				0 & V^\dag/ \sqrt{2}
			\end{pmatrix}
			\begin{pmatrix}
				U & 0\\
				0 & I_3 / \sqrt{2}
			\end{pmatrix}\\
			&= 
			\begin{pmatrix}
				0 & 0\\
				0 &  V^\dag /2
			\end{pmatrix}.
		\end{align*}
	\end{itemize}
	
	To compute $E_k E_k^\dag$, we just have to replace $U^\dag U$ by $UU^\dag$ and $V^\dag V$ by $V V^\dag$ in the computation of $E_k^\dag E_k$, but this doesn't change the result, because $U$ and $V$ are unitary matrices. Therefore
	\begin{equation}
		E_k E_k^\dag = E_k^\dag E_k,
	\end{equation}
	for $k=0,1$. In particular, to show that $(E_0,E_1)$ is a Kraus representation of a UCPTP map, we only need to show that $\sum_k E_k^\dag E_k = I_{d_X}$, since the other equation $\sum_k E_k E_k^\dag  = I_{d_X}$ is equivalent. We have that
	\begin{align*}
		\sum_k E_k^\dag E_k &= E_0^\dag E_0 + E_1^\dag E_1\\
		&=  \begin{pmatrix}
			I_{d_X-3} & 0\\
			0 & I_3 /2
		\end{pmatrix}
		+
		\begin{pmatrix}
			0 & 0\\
			0 & I_3 /2
		\end{pmatrix}\\
		&= \begin{pmatrix}
			I_{d_X-3} & 0\\
			0 & I_3
		\end{pmatrix}\\
		&= I_{d_X}.
	\end{align*}
	This proves that $(E_0,E_1)$ represents a UCPTP map.
	
	To show that the UCPTP map is an extreme point of the set of UCPTP maps, it is enough to show that it is an extreme point of the set of CPTP maps. By theorem \ref{theorem:extreme cptp maps in Kraus representation}, if $(E_k)_k$ and $(E_k^\dag E_l)_{k,l}$ are both linearly independent sequences, then $(E_k)_k$ represents an extreme point of the set of CPTP maps. But if $(E_k^\dag E_l)_{k,l}$ is linearly independent, it follows that $(E_k)_k$ is also linearly independent, so it is enough to show that $(E_k^\dag E_l)_{k,l}$ is linearly independent. Let $(\alpha_{k,l})_{k,l \in \{0,1\}}$ be complex numbers such that
	\begin{equation}
		\label{equation: linear independence for Choi rank 2}
		\sum_{k,l} \alpha_{kl} E_k^\dag E_l = 0.
	\end{equation}
	Using the previous computation of $E_k^\dag E_l$, we have that:
	\begin{align*}
		\sum_{k,l} &\alpha_{kl} E_k^\dag E_l =\\ &=\alpha_{00}\begin{pmatrix}
			I_{d_X-3} & 0\\
			0 & I_3 /2
		\end{pmatrix} + \alpha_{11}  \begin{pmatrix}
			0 & 0\\
			0 & I_3 /2
		\end{pmatrix} + \alpha_{01} \begin{pmatrix}
			0 & 0\\
			0 &  V /2
		\end{pmatrix} + \alpha_{10} \begin{pmatrix}
			0 & 0\\
			0 &  V^\dag /2
		\end{pmatrix}\\
		&= \begin{pmatrix}
			\alpha_{00} I_{d_X-3} & 0\\
			0 & \alpha_{00} I_3 /2 + \alpha_{11} I_3 /2 + \alpha_{01} V/2 + \alpha_{10} V^\dag/2
		\end{pmatrix}.
	\end{align*}
	By equation (\ref{equation: linear independence for Choi rank 2}), this matrix is zero. Each block must be zero, so we have two equations,
	\begin{equation}
		\label{equation: alpha00 I equals 0}
		\alpha_{00} I_{d_X-3} = 0,
	\end{equation}
	and
	\begin{equation}
		\label{equation: alpha00I+alpha11I...}
		\alpha_{00} I_3 /2 + \alpha_{11} I_3 /2 + \alpha_{01} V/2 + \alpha_{10} V^\dag/2 = 0.
	\end{equation}
	Equation (\ref{equation: alpha00 I equals 0}) implies that
	\begin{equation}
		\alpha_{00} = 0.
	\end{equation}
	Then equation (\ref{equation: alpha00I+alpha11I...}) simplifies to
	\begin{equation}
		\alpha_{11} I_3 /2 + \alpha_{01} V/2 + \alpha_{10} V^\dag/2 = 0.
	\end{equation}
	Multiplying by 2, it becomes
	\begin{equation}
		\alpha_{11} I_3 + \alpha_{01} V + \alpha_{10} V^\dag = 0.
	\end{equation}
	Therefore, we have an extreme point of the set of CPTP maps iff $(I_3,V,V^\dag)$ is linearly independent.
	
	Finally, let's show that
	\begin{equation}
		V = \begin{pmatrix}
			0 & 0 & 1\\
			1 & 0 & 0 \\
			0 & 1 & 0
		\end{pmatrix}
	\end{equation}
	satisfies that $(I_3,V,V^\dag)$ is linearly independent. Keeping the previous notation, we have that
	\begin{align*}
		\alpha_{11} I_3 + &\alpha_{01} V + \alpha_{10} V^\dag =\\
		&= \begin{pmatrix}
			\alpha_{11} & 0 & 0\\
			0 & \alpha_{11} & 0 \\
			0 & 0 & \alpha_{11}\\
		\end{pmatrix}
		+\begin{pmatrix}
			0 & 0 & \alpha_{01}\\
			\alpha_{01} & 0 & 0 \\
			0 & \alpha_{01} & 0
		\end{pmatrix}
		+\begin{pmatrix}
			0 & \overline{\alpha_{10}} & 0\\
			0 & 0 & \overline{\alpha_{10}} \\
			\overline{\alpha_{10}} & 0 & 0
		\end{pmatrix}\\
		&= \begin{pmatrix}
			\alpha_{11} & \overline{\alpha_{10}} & \alpha_{01}\\
			\alpha_{01} & \alpha_{11} & \overline{\alpha_{10}} \\
			\overline{\alpha_{10}} & \alpha_{01} & \alpha_{11}\\
		\end{pmatrix}.
	\end{align*}
	Since $\alpha_{11} I_3 + \alpha_{01} V + \alpha_{10} V^\dag = 0$, it follows that $\alpha_{11}=\alpha_{01}=\alpha_{10}=0$. This proves that $(I_3,V,V^\dag)$ is linearly independent. Therefore, we have a UCPTP map that is an extreme point of both convex sets of UCPTP maps and CPTP maps. 
\end{proof}

\section{Extreme UCPTP maps with Choi-rank at least 3}

\label{section: extreme ucptp with choi rank at least 3}

There are some examples of extreme UCPTP maps in \cite{ohno,Haagerup2021}. There we have general formulas that work for an infinite number of dimensions $d_X$, but with Choi-rank $r$ of at most $d_X$. For Choi-rank $r \geq d_X$, the examples are for $d_X = 3$ and $d_X = 4$. We contribute with more examples, for $3 \leq d_X \leq 22$ and almost every Choi-rank $r$ with
\begin{equation}
	3 \leq r \leq \sqrt{2} d_X.
\end{equation}
The case $r=1$ is trivial, it consists of unitary channels, and the case $r=2$ was already solved previously. With these results, we have examples of extreme UCPTP maps for almost every Choi-rank allowed by the bounds of corollary \ref{corollary:bound on rank of extreme ucptp}, for $3 \leq d_X \leq 22$. We could produce more examples by the same methods, but it would take more computing time, so it was chosen to stop computing at $d_X = 22$.

Constructing examples of CPTP maps is simpler, because it consists of constructing orthonormal vectors. To construct a UCPTP map, we have to construct orthonormal vectors subject to some constraints. We are interested in extreme points, so we have to further guarantee the linear independence from theorem \ref{theorem:characterization of extreme points for ucptp}. It is hard to guess a solution and check its correctness by hand. Instead, the strategy to find solutions was the following. First was made a C program to find solutions using the brute force method. Having obtained some examples, then try to find a pattern in the examples and rewrite the program to use the guessed pattern. Then generate more examples with the new program and repeat the process. The idea was to first obtain a program that generates as much examples as possible, as fast as possible. Much of the speed was obtained from programming in C and using the Fast Library for Number Theory (FLINT), which is a C library. Having a program that successfully generates examples, a more careful analysis could then lead to conjecture some general formula for constructing extreme UCPTP maps. The last step would be to prove that the formula works. We obtained a formula that works well, failing only for the maximal Choi-rank allowed by the bounds, that is, for $r = \lfloor \sqrt{2} d_X \rfloor$, for some $d_X$, at least up to the values that could be tested computationally. Unfortunately, we still don't have a proof that any of the techniques employed generate examples of extreme UCPTP maps for every $d_X \geq 3$.

There were many ideas that were tested, we'll show only two of them. One is a guess that was useful to generate many examples, from $d_X = 3$ to $d_X = 22$. This guess isn't convenient for mathematical proofs. Having analyzed it, another guess was made that may be more convenient for proofs, but we still don't have a proof that neither of them work for an infinite number of cases. They will be presented in the reversed order, first the one that was obtained last and that may be more convenient for proofs. Then the one that was obtained first, because the corresponding C program ran faster and was able to produce results for higher values of $d_X$.

\subsection{A method to construct UCPTP maps}

\label{section: method 1}

A UCPTP map $\varepsilon \in \ucptp(X)$ with Choi-rank $r$ can be represented by Kraus matrices $(E_k)_{k=0,1,\dots,r-1}$, where $E_k \in M_{d_X}(\complexNumbers)$. As shown in proposition \ref{proposition:kraus representation of ucptp}, these matrices must satisfy the equations
\begin{equation}
	\label{equation: method 1 tp}
	\sum_k E_k^\dag E_k = I_{d_X},
\end{equation}
and
\begin{equation}
	\label{equation: method 1 unital}
	\sum_k E_k E_k^\dag = I_{d_X}.
\end{equation}
The first equation is the condition for the CP map represented by $(E_k)_k$ to be trace preserving. The second equation is the condition for the map to be unital. Furthermore, by theorem \ref{theorem:characterization of extreme points for ucptp}, for the UCPTP map to be an extreme point it is enough to show that the sequence $(E_k^\dag E_l \oplus E_l E_k^\dag)_{k,l}$ is linearly independent. More precisely, if this sequence is linearly independent, then $(E_k)_k$ must also be linearly independent, and then \ref{theorem:characterization of extreme points for ucptp} guarantees that $(E_k)_k$ represents an extreme point of the set of UCPTP maps. In summary, to obtain an extreme UCPTP map, we have to find matrices $(E_k)_k$ that satisfy two equations and such that a sequence of matrices is linearly independent. It is simpler to satisfy an inequality than an equality. Linear independence corresponds to a Gram matrix to have a non zero determinant, which is an inequality. For this reason, it is easier to obtain linear independence than to satisfy equations \ref{equation: method 1 tp} and \ref{equation: method 1 unital}. The strategy is to first construct a parameterized family of UCPTP maps. Then take the simplest choice of parameters and check if it gives an extreme point. If it isn't an extreme point, we may be able to obtain an extreme point by a small perturbation of the parameters. This is because a family of $n$ vectors in $\complexNumbers^n$ can always be perturbed to another linearly independent family by an arbitrarily small perturbation. This works for families of vectors without constraints, but UCPTP maps are constrained by equations \ref{equation: method 1 tp} and \ref{equation: method 1 unital}. Nevertheless, with this idea we are able to construct many examples.

Let's see how we can simplify the two equations. Also, to simplify the notation, let's write just $d$ instead of $d_X$. In \cite{iten_colbeck} was shown that equation \ref{equation: method 1 tp} corresponds to having an isometry matrix, which was also proved in proposition \ref{proposition:pre-image of pi is stiefel manifold}. We repeat the argument here. We'll be using indices that start at 0, because it is more convenient for modular arithmetic and for programming languages such as C. Let's concatenate vertically the matrices $E_k$ into a single matrix $V$:
\begin{equation}
	\label{equation: method 1 concatenating Ek vertically}
	V \coloneqq 
	\begin{pmatrix}
		E_0\\
		E_1\\
		\vdots\\
		E_{r-1}
	\end{pmatrix}.
\end{equation}
Then
\begin{align*}
	V^\dag V &=
	\begin{pmatrix}
		E_0^\dag & E_1^\dag & \dots & E_{r-1}^\dag 
	\end{pmatrix}
	\begin{pmatrix}
		E_0\\
		E_1\\
		\vdots\\
		E_{r-1}
	\end{pmatrix}\\
	&= \sum_k E_k^\dag E_k.
\end{align*}
Therefore,
\begin{equation}
	\sum_k E_k^\dag E_k = I_d \iff V^\dag V = I_d.
\end{equation}
The equation $V^\dag V = I_d$ is the definition of an isometry matrix. It is also equivalent to the statement that the columns of $V$ are orthonormal.

Similarly, we can concatenate vertically the matrices $E_k^\dag$, obtaining
\begin{equation}
	W \coloneqq 
	\begin{pmatrix}
		E_0^\dag\\
		E_1^\dag\\
		\vdots\\
		E_{r-1}^\dag
	\end{pmatrix}.
\end{equation}
Then equation \ref{equation: method 1 unital} is equivalent to the statement that the columns of $W$ are orthonormal.

A simple way to obtain orthogonal vectors is for each coordinate to have at most one vector for which that coordinate is non zero. For the matrices $V$ and $W$, this corresponds to each line of the matrix to have at most one non zero entry. The lines of $V$ are the lines of the matrices $E_k$, and the lines of $W$ are the lines of the matrices $E_k^\dag$. We'll then require that each line of $E_k$ and each line of $E_k^\dag$ have at most one non zero entry.

The matrices $E_k$ that satisfy such requirement can be written in a special form. The requirement says that each line of $E_k$ has at most one non zero entry. We'll be counting lines and columns from 0, so the possible values for the line and column indices will be $0$, $1$, \dots, $d-1$. This will be more convenient for modular arithmetic. Let $L_k$ be the set of lines of $E_k$ for which there is a non zero entry. That is,
\begin{equation}
	L_k \coloneqq \{i \in \{0,\dots,d-1\} \mid \exists j \in \{0,\dots,d-1 \} \text{ such that } (E_k)_{i,j} \neq 0\}.
\end{equation}
Since there is at most one non zero entry for each line, the column $j$ is unique for each $i\in L_k$. In other words, there is a function $f_k \colon L_k \to \{0,\dots,d-1\}$ such that $(E_k)_{i,f_k(i)} \neq 0$, and $(E_k)_{i,j}=0$ for $j \neq f_k(i)$. Then we can write $E_k$ uniquely as
\begin{equation}
	\label{equation: special expression for Ek, first condition}
	E_k = \sum_{i \in L_k} a_{k,i} e_{i,f_k(i)},
\end{equation}
for some $a_{k,i} \in \complexNumbers$. Here $e_{i,j}$ denotes the standard basis vector of the space of matrices, given by $(e_{i,j})_{i',j'} = \delta_{i,i'} \delta_{j,j'}$.

We used one of the conditions of the requirement, now let's use the other. It says that each line of $E_k^\dag$ has at most one non zero entry. From equation (\ref{equation: special expression for Ek, first condition}) we get that
\begin{equation}
	E_k^\dag = \sum_{i \in L_k} \overline{a_{k,i}} e_{f_k(i),i}.
\end{equation}
A line $j$ of $E_k^\dag$ is non zero iff there is some $i \in L_k$ for which $j=f_k(i)$. That is, the non zero lines of $E_k^\dag$ are given by the image of $f_k$. Then, the statement that each line of $E_k^\dag$ has at most one non zero entry is equivalent to say that for any $j \in f_k(L_k)$ there is a unique $i \in L_k$ such that $j = f_k(i)$. In other words, it states that $f_k$ is injective.

We can replace the sets $L_k$ by $\{0,\dots,d-1\}$ if we introduce zero coefficients, that is, $a_{k,i} = 0$ for $i \in \{0,\dots,d-1\}\setminus L_k$. But then we need to extend the function $f_k$ from $L_k$ to $\{0,\dots,d-1\}$. Since $f_k$ is injective, $L_k$ is in bijection with $f_k(L_k)$, so they have the same number of elements. Then the complements $\{0,\dots,d-1\} \setminus L_k$ and $\{0,\dots,d-1\} \setminus f_k(L_k)$ also have the same size. We can extend $f_k$ by using any bijection between these complements. For example, we can order both complements according to the order from $\naturalNumbers$, and then map the elements in order. That is, map the smallest element of one set to the smallest element of the other, and then do the same for the next elements. This shows that $f_k$ can be extended to a bijection $\sigma_k \colon \{0,\dots,d-1\} \to \{0,\dots,d-1\}$. We identify the set of such bijections with $S_d$, the symmetric group on $d$ elements. We showed that matrices $(E_k)_k$ that satisfy the requirement can be written as
\begin{equation}
	\label{equation: method 1 general form}
	E_k =  \sum_{i=0}^{d-1} a_{k,i} e_{i,\sigma_k(i)},
\end{equation}
where $a_{k,i} \in \complexNumbers$ and $\sigma_k \in S_k$ is a permutation.

The converse statement is also true, if $E_k$ is given by equation \ref{equation: method 1 general form}, then $(E_k)_k$ satisfy the requirement that the lines of $E_k$ and $E_k^\dag$ have at most a non zero entry for each line. Since
\begin{equation}
	(E_k)_{i,j} = a_{k,i} \delta_{j, \sigma_k (i)},
\end{equation}
then the only possible non zero entry for the line with index $i$ is at column $j = \sigma_k(i)$. Also, we have that
\begin{equation}
	E_k^\dag = \sum_{i=0}^{d-1} \overline{a_{k,i}} e_{\sigma_k(i),i},
\end{equation}
so
\begin{equation}
	(E_k^\dag)_{j,i} = \overline{a_{k,i}} \delta_{j, \sigma_k (i)}.
\end{equation}
Since $\sigma_k$ is a bijection, we have that
\begin{equation}
	j = \sigma_k (i) \iff i = \sigma_k^{-1} (j),
\end{equation}
so
\begin{equation}
	(E_k^\dag)_{j,i} = \overline{a_{k,i}} \delta_{i, \sigma_k^{-1} (j)}.
\end{equation}
Then the only possible non zero entry for the line with index $j$ is at column $i = \sigma_k^{-1}(j)$. Therefore, it is equivalent to use $f_k$ with non zero coefficients $a_{k,i}$, or to use permutations $\sigma_k$ with coefficients that may be zero. We'll proceed using the permutations instead of $f_k$.

Taking matrices $(E_k)_k$ of this form guarantees the orthogonality part of the two orthonormality conditions, but the normalization conditions still have to be satisfied. That is, the matrices $V$ and $W$ defined previously must have columns with norm 1. The $j$-th column of $V$ is the concatenation of the $j$-th columns of $E_k$. We can rewrite the expression for $E_k$ as a sum over the column $j$ instead of the line $i$. Since $\sigma_k$ is a bijection, we can make a change of variables in the summation, replacing $i$ with $\sigma_k^{-1}(j)$. We get that
\begin{equation}
	E_k =  \sum_{j=0}^{d-1} a_{k,\sigma_k^{-1}(j)} e_{\sigma_k^{-1}(j),j}.
\end{equation}
We can see that the $j$-th column of $E_k$ has at most one non zero entry, with line $i = \sigma_k^{-1}(j)$ and coefficient $a_{k,\sigma_k^{-1}(j)}$. Therefore, the possibly non zero coefficients in the $j$-th column of $V$ are $(a_{k,\sigma_k^{-1}(j)})_{k=0,\dots,r-1}$. Then the $j$-th column of $V$ has norm 1 iff
\begin{equation}
	\sum_{k=0}^{r-1} |a_{k,\sigma_k^{-1}(j)}|^2 = 1.
\end{equation}

Next we repeat the same idea for the matrix $W$. The columns of $W$ must have norm 1. The $i$-th column of $W$ is the concatenation of the $i$-th column of each $E_k^\dag$. The expression obtained for $E_k^\dag$ already has a summation over its columns, so we don't need to make a change of variables this time. We have that
\begin{equation}
	E_k^\dag = \sum_{i=0}^{d-1} \overline{a_{k,i}} e_{\sigma_k(i),i},
\end{equation}
so the $i$-th column of $E_k^\dag$ has at most one non zero entry, whose line is $j = \sigma_k(i)$, and whose coefficient is $\overline{a_{k,i}}$. Then the possibly non zero entries of the $i$-th column of $W$ are $(\overline{a_{k,i}})_{k=0,\dots,r-1}$. Therefore, the $i$-th column of $W$ has norm 1 iff
\begin{equation}
	\sum_{k=0}^{r-1} |a_{k,i}|^2 = 1.
\end{equation}

In summary, to find a UCPTP map, it is enough to find permutations $\sigma_k \in S_d$ and coefficients $a_{k,i} \in \complexNumbers$ such that
\begin{equation}
	\forall j \in \{0,\dots,d-1\} \ \sum_{k=0}^{r-1} |a_{k,\sigma_k^{-1}(j)}|^2 = 1,
\end{equation}
\begin{equation}
	\forall i \in \{0,\dots,d-1\} \ \sum_{k=0}^{r-1} |a_{k,i}|^2 = 1.
\end{equation}
This is a system of linear equations over $|a_{k,i}|^2$.

Let's make the change of variables
\begin{equation}
	\label{equation: method 1 x = |a|^2}
	x_{k,i} \coloneqq |a_{k,i}|^2.
\end{equation}
Also, let's just use the same symbol $i$ in both equations, instead of $i$ and $j$. Then we have the linear system
\begin{equation}
	\forall i \in \{0,\dots,d-1\} \ \sum_{k=0}^{r-1} x_{k,\sigma_k^{-1}(i)} = 1,
\end{equation}
\begin{equation}
	\forall i \in \{0,\dots,d-1\} \ \sum_{k=0}^{r-1} x_{k,i} = 1,
\end{equation}
with the restriction that $x_{k,i} \geq 0$. We'll prove next that the space of solutions is compact convex and find a trivial solution.

\begin{proposition}{}{method 1 convex polytopes}
	Let $d,r\geq 1$, let also $(\sigma_k)_{k=0,\dots,r-1}$ be any sequence of permutations $\sigma_k \in S_d$. The set of points $(x_{k,i})_{\substack{k=0,\dots,r-1\\ i=0,\dots,d-1}}$ such that
	\begin{equation}
		\label{equation: ls1}
		\forall i \ \sum_k x_{i,\sigma_k^{-1}(i)} = 1,
	\end{equation}
	\begin{equation}
		\label{equation: ls2}
		\forall i \ \sum_k x_{k,i} = 1,
	\end{equation}
	\begin{equation}
		\label{equation: xki non negative}
		\forall k \forall i \ x_{k,i} \geq 0,
	\end{equation}
	is a non empty compact convex set. A trivial solution is given by
	\begin{equation}
		x_{k,i}=1/r,
	\end{equation}
	for $k = 0, \dots, r-1$ and $i = 0, \dots, d-1$. All solutions are bounded so that $x_{k,i} \in [0,1]$.
\end{proposition}
\begin{proof}
	Since every summation in equations \ref{equation: ls1} and \ref{equation: ls2} have the same number of terms and the right hand side is the same for of each equation, we can find a solution by taking $x_{k,i}$ to be a constant value $x$ that doesn't depend on $k$ or $i$. Then
	\begin{equation}
		\sum_{k=0}^{r-1} x = 1 \iff rx = 1 \iff x= \frac{1}{r}. 
	\end{equation}
	That is,
	\begin{equation}
		x_{k,i} = \frac{1}{r}
	\end{equation}
	is a solution for the system of linear equations, and it also satisfies the restriction that $x_{k,i} \geq 0$.
	
	Next let's show that the set of solutions is a convex set. Let $(x_{k,i}^0)_{k,i}$ and $(x_{k,i}^1)_{k,i}$ be any two solutions. Let also $t_0,t_1 \in [0,1]$ such that $t_0+t_1=1$. It is straightforward to show that $x_{k_i} \coloneqq t_0 x_{k,i}^0 + t_1 x_{k,i}^1$ is another solution. Indeed, we have the following:
	\begin{itemize}
		\item For equation (\ref{equation: ls1}):
		\begin{align*}
			\sum_k x_{i,\sigma_k^{-1}(i)} &= \sum_k \sum_{j=0}^1 t_j x_{i,\sigma_k^{-1}(i)}^j \\
			&= \sum_{j=0}^1 t_j \sum_k x_{i,\sigma_k^{-1}(i)}^j \\
			&= \sum_{j=0}^1 t_j 1 \\
			&= 1.
		\end{align*}
		
		\item For equation (\ref{equation: ls2}):
		\begin{align*}
			\sum_k x_{i,k} &= \sum_k \sum_{j=0}^1 t_j x_{i,k}^j \\
			&= \sum_{j=0}^1 t_j \sum_k x_{i,k}^j \\
			&= \sum_{j=0}^1 t_j 1 \\
			&= 1.
		\end{align*}
		
		\item For inequation (\ref{equation: xki non negative}):
		\begin{align*}
			x_{k,i} &= \sum_{j=0}^1 t_j x_{k,i}^j \\
			&\geq \sum_{j=0}^1 t_j 0 \\
			&\geq 0.
		\end{align*}
	\end{itemize}
	This shows that the set of solutions is convex.
	
	Next let's show that the set of solutions is closed as a topological subspace of $M_{r,d}(\realNumbers)$. Let $(x^n)_{n\geq 1}$ be a sequence of solutions that converges to a point $x$. Each solution $x^n$ is of the form $x^n = (x^n_{k,i})_{k,i}$. Also the point $x$ is of the form $x = (x_{k,i})$. Since $\lim_{n \to \infty} x^n = x$, the same is true for its components, that is, $\lim_{n \to \infty} x^n_{k,i} = x_{k,i}$. By continuity of the addition, we have that
	\begin{align*}
		\sum_k x_{i,\sigma_k^{-1}(i)} &= \sum_k \lim_{n\to \infty} x^n_{i,\sigma_k^{-1}(i)} \\
		&= \lim_{n\to \infty} \sum_k x^n_{i,\sigma_k^{-1}(i)} \\
		&= \lim_{n\to \infty} 1\\
		&= 1.
	\end{align*}
	Similarly, we have that
	\begin{equation*}
		\sum_k x_{i,k} = 1.
	\end{equation*}
	Also, since $x^n_{k,i} \geq 0$ for every $n$, the same is true for the limit, so
	\begin{align*}
		x_{k,i} \geq 0.
	\end{align*}
	This proves that the set of solutions is a closed topological subspace.
	
	Finally, let's prove that the set of solutions is bounded. This follows immediately from equation (\ref{equation: ls2}) and inequation (\ref{equation: xki non negative}). Since $x_{k,i} \geq 0$, then
	\begin{equation}
		\sum_{k,i} x_{k,i} \geq x_{l,i},
	\end{equation}
	for any $l$. From equation (\ref{equation: ls2}) it follows that
	\begin{equation}
		0 \leq x_{l,i} \leq 1.
	\end{equation}
	Since $l,i$ are arbitrary, this proves that the set of solutions is bounded. Being closed and bounded, it is compact.
\end{proof}

By Krein-Milman's theorem we know that the space of solutions is the convex hull of its extreme points. In this case it is actually a convex polytope, being the convex hull of a finite number of vertices. There are algorithms to compute such vertices, such as the Double Description Method \cite{10.1007/3-540-61576-8_77}. There is a C library called cddlib which implements this algorithm. We can parameterize the space of solutions by first computing all the vertices and then taking convex combinations of them. Unfortunately, the number of vertices can be very high and we still didn't find a simple way for working with them. We'll proceed with another approach, by solving the system of linear equations.

\subsection{The trivial solution}

We found a trivial solution to the linear system, given by
\begin{equation}
	x_{k,i} = \frac{1}{r}.
\end{equation}
We'll show that it doesn't give an extreme point of the set of UCPTP maps with Choi-rank $r$, for $r \geq 2$.

From equation \ref{equation: method 1 general form}, the Kraus representation has the form
\begin{equation}
	E_k = \sum_{i=0}^{d-1} a_{k,i} e_{i, \sigma_k (i)} ,
\end{equation}
for $k = 0, \dots, r-1$. First we find an expression for $E_k^\dag E_k$ that is valid for any choice of parameters. We have that
\begin{align*}
	E_k^\dag E_k &\stackrel{(i)}{=} \left( \sum_{i=0}^{d-1} a_{k,i} e_{i, \sigma_k (i)} \right)^\dag \left( \sum_{j=0}^{d-1} a_{k,j} e_{j, \sigma_k (j)} \right) \\
	&\stackrel{(ii)}{=} \sum_{i,j} \overline{a_{k,i}} a_{k,j} e_{\sigma_k (i), i} e_{j, \sigma_k (j)} \\
	&\stackrel{(iii)}{=} \sum_{i,j} |a_{k,i}|^2 \delta_{i,j} e_{\sigma_k (i), \sigma_k (j)} \\
	&\stackrel{(iv)}{=} \sum_{i = 0}^{d-1} |a_{k,i}|^2 e_{\sigma_k (i), \sigma_k (i)} \\
	&\stackrel{(v)}{=} \sum_{i = 0}^{d-1} |a_{k,\sigma_k^{-1}(i)}|^2 e_{i,i} \\
	&\stackrel{(vi)}{=} \sum_{i = 0}^{d-1} x_{k,\sigma_k^{-1}(i)} e_{i,i}.
\end{align*}
In $(i)$ was used equation \ref{equation: method 1 general form}; in $(ii)$ was applied the hermitian conjugate and was used linearity to group both summations; in $(iii)$ was used that $\overline{z}z = |z|^2$ and that
\begin{equation}
	e_{i,j} e_{k,l} = \delta_{j,k} e_{i,l};
\end{equation}
in $(iv)$ the delta was used to eliminate the summation over $j$; in $(v)$ was made the change of variables $i \mapsto \sigma_{k}^{-1}(i)$, which is possible because $\sigma_k$ is a permutation; in $(vi)$ was used the definition of $x_{k,i}$, as in equation \ref{equation: method 1 x = |a|^2}.

Next we use the trivial choice of parameters, that is,
\begin{equation}
	x_{k,i} = \frac{1}{r}.
\end{equation}
Then we get that
\begin{align*}
	E_k^\dag E_k &\stackrel{(i)}{=} \sum_{i=0}^{d-1} \frac{1}{r} e_{i,i} \\
	&\stackrel{(ii)}{=} \frac{1}{r} \sum_{i=0}^{d-1} e_{i,i} \\
	&\stackrel{(iii)}{=} \frac{1}{r} I_d .
\end{align*}
In $(i)$ was used the trivial solution $x_{k,i} = 1/r$; in $(ii)$ the fraction $1/r$ was moved to the left; in $(iii)$ was used that $\sum_i e_{i,i} = I_d$. This implies that
\begin{equation}
	(\sqrt{r} E_k)^\dag (\sqrt{r} E_k ) = I_d ,
\end{equation}
for $k=0, \dots, r-1$. Therefore $\sqrt{r} E_k$ is a unitary matrix:
\begin{equation}
	\sqrt{r} E_k \in \U(d).
\end{equation}
Denote the unitary matrix by $U_k$:
\begin{equation}
	\label{equation: method 1 unitary matrix}
	U_k \coloneqq \sqrt{r} E_k.
\end{equation}
As a consequence, the UCPTP map represented by $(E_k)_k$ is a mixed unitary channel. Denote UCPTP map by $\varepsilon$, so that
\begin{equation}
	\varepsilon(A) = \sum_{k=0}^{r-1} E_k A E_k^\dag,
\end{equation}
for every $A \in M_d(\complexNumbers)$. We are using a matrix space instead of $\Lin_\complexNumbers (X)$, because $E_k$ was taken to be a matrix. Then we get that
\begin{align*}
	\varepsilon(A) &\stackrel{(i)}{=} \sum_{k=0}^{r-1} \frac{U_k}{\sqrt{r}} A \left( \frac{U_k}{\sqrt{r}} \right)^\dag \\
	&\stackrel{(ii)}{=} \frac{1}{r} \sum_{k=0}^{r-1} U_k A U_k^\dag .
\end{align*}
In $(i)$ was used equation \ref{equation: method 1 unitary matrix}; in $(ii)$ both $\sqrt{r}$ were moved to the left, obtaining the fraction $1/r$. The function $A \mapsto U_k A U_k^\dag $ is a unitary channel, in particular it is a UCPTP map. Therefore, $\varepsilon$ is a convex combination of unitary channels. There are only two possibilities, or all these unitary channels are equal, so that $\varepsilon$ is a unitary channel, or we have a non trivial convex combination, which implies that $\varepsilon$ isn't an extreme point. In both cases, $\varepsilon$ isn't an extreme point with Choi-rank $r$ for $r \geq 2$.

We are interested in the case $r \geq 3$, because we already have examples or proof of nonexistence for $r\leq 2$. To obtain an extreme point, we have to satisfy a condition of linear independence. If some vectors are linearly dependent, a small perturbation may be enough to make them linearly independent. Following this idea, we'll find a neighborhood of the trivial solution for which every point is a solution, then extreme points will be obtained by taking a small perturbation of the trivial solution.

\subsection{Solving the linear system}

Instead of working with convex combinations of vertices, another possibility is to solve the linear system without the non negativity constraint, and then work with a smaller set of solutions that satisfy such constraint. To solve a linear system we need to find one solution and then solve the homogeneous equations. We already found the solution $x_{k,i} = 1/r$. Next we obtain the homogeneous equations. Make the change of variables
\begin{equation}
	x_{k,i} = \frac{1}{r}+y_{k,i}.
\end{equation}
Then the linear system becomes
\begin{equation}
	\forall i \in \{0,\dots,d-1\} \ \sum_{k=0}^{r-1} y_{k,\sigma_k^{-1}(i)} = 0,
\end{equation}
\begin{equation}
	\forall i \in \{0,\dots,d-1\} \ \sum_{k=0}^{r-1} y_{k,i} = 0,
\end{equation}
and the constraint $x_{k,i}\geq 0$ becomes $y_{k,i} \geq -1/r$. Next we solve this homogeneous system of linear equations, but with an assumption over the permutations $\sigma_k$ so that the solution doesn't become too complicated. Fortunately, we don't need to assume too much from the permutations.

\begin{theorem}{}{solution of homogeneous linear system}
	Let $d,r \in \naturalNumbers$ with $d,r \geq 2$, and let $(y_{k,i})_{ \substack{ k=0,\dots,r-1 \\ i=0,\dots,d-1  }}$ be real numbers. Let also $(\sigma_k)_{k=0,\dots,r-1}$ be permutations $\sigma_k \in S_d$ such that $c \coloneqq \sigma_0^{-1} \sigma_1$ is a $d$-cycle. Then consider the homogeneous linear system
	\begin{equation}
		\label{equation: first eq of homogeneous linear system}
		\forall i \in \{0,\dots,d-1\} \ \sum_{k=0}^{r-1} y_{k,\sigma_k^{-1}(i)} = 0,
	\end{equation}
	\begin{equation}
		\label{equation: second eq of homogeneous linear system}
		\forall i \in \{0,\dots,d-1\} \ \sum_{k=0}^{r-1} y_{k,i} = 0.
	\end{equation}
	The solutions of this linear system can be parameterized as follows. We take every $y_{k,i}$ with $k\geq 2$, together with $y_{1,0}$, as the independent variables. The rest are dependent variables, which can be expressed as linear functions of the independent variables. We can visualize it better with the next table:
	\begin{center}
		\begin{tabular} { |c|c|c|c| }
			\hline
			\cellcolor[rgb]{0.9,0.9,0.9} $y_{0,0}$ & \cellcolor[rgb]{0.9,0.9,0.9} $y_{0,1}$ & \cellcolor[rgb]{0.9,0.9,0.9} \dots & \cellcolor[rgb]{0.9,0.9,0.9} $y_{0,d-1}$ \\
			\hline
			$y_{1,0}$ & \cellcolor[rgb]{0.9,0.9,0.9} $y_{1,1}$ & \cellcolor[rgb]{0.9,0.9,0.9} \dots & \cellcolor[rgb]{0.9,0.9,0.9} $y_{1,d-1}$ \\
			\hline
			$y_{2,0}$ & $y_{2,1}$ & \dots & $y_{2,d-1}$ \\
			\hline
			\vdots & \vdots & $\ddots$ & \vdots \\
			\hline
			$y_{r-1,0}$ & $y_{r-1,1}$ & \dots & $y_{r-1,d-1}$ \\
			\hline
		\end{tabular}
	\end{center}
	In this table, the cells colored with gray correspond to the dependent variables. Then we can express the dependent variables as follows:
	\begin{equation}
		\label{equation: first generic expression for dependent y}
		y_{0,i} = -\sum_{k=1}^{r-1} y_{k,i}, \text{ for } i=0,1,\dots,d-1,
	\end{equation}
	\begin{equation}
		\label{equation: second generic expression for dependent y}
		y_{1,c^n(0)} = y_{1,0} + \sum_{m=0,1,\dots,n-1} T_{c^m(0)}, \text{ for } n=1,\dots,d-1,
	\end{equation}
	where
	\begin{equation}
		T_i \coloneqq \sum_{k=2}^{r-1} (y_{k,\sigma_k^{-1}\sigma_1(i)}-y_{k,\sigma_0^{-1}\sigma_1(i)}), \text{ for } i=0,1,\dots,d-1.
	\end{equation}
	
	In particular, the $\realNumbers$-vector space of solutions has dimension $d(r-2)+1$.
\end{theorem}
\begin{proof}
	The proof consists of using the equations of the linear system to isolate one variable at a time and write it as a linear function of the other variables. Some care must be taken to avoid a variable to be expressed as a function of itself. Using equation (\ref{equation: second eq of homogeneous linear system}) we can isolate the variable $y_{0,i}$, obtaining
	\begin{equation}
		y_{0,i} = -\sum_{k=1}^{r-1} y_{k,i},
	\end{equation}
	for $i=0,\dots,d-1$. Then we have to substitute this expression in equation (\ref{equation: first eq of homogeneous linear system}). For $i=0,\dots,d-1$, we get that
	\begin{align*}
		0 &= \sum_{k=0}^{r-1} y_{k,\sigma_k^{-1}(i)}\\
		&= y_{0,\sigma_0^{-1}(i)} + \sum_{k=1}^{r-1} y_{k,\sigma_k^{-1}(i)}\\
		&=-\sum_{k=1}^{r-1} y_{k,\sigma_0^{-1}(i)} + \sum_{k=1}^{r-1} y_{k,\sigma_k^{-1}(i)}\\
		&=\sum_{k=1}^{r-1} (-y_{k,\sigma_0^{-1}(i)} + y_{k,\sigma_k^{-1}(i)}).
	\end{align*}
	Next we have to use these equations to isolate other variables. Since we started isolating variables with $k=0$, we processed isolating other variables with $k=1$. Separating the variables with $k=1$ from the variables with $k\geq 2$, we get that
	\begin{align*}
		0 &= \sum_{k=1}^{r-1} (-y_{k,\sigma_0^{-1}(i)} + y_{k,\sigma_k^{-1}(i)}) \\
		&= -y_{1,\sigma_0^{-1}(i)} + y_{1,\sigma_1^{-1}(i)}+\sum_{k=2}^{r-1} (-y_{k,\sigma_0^{-1}(i)} + y_{k,\sigma_k^{-1}(i)}).
	\end{align*}
	We have to variables with $k=1$ in this equation, which are $y_{1,\sigma_0^{-1}(i)}$ and $y_{1,\sigma_1^{-1}(i)}$. We can isolate any of them, but we'll choose to isolate $y_{1,\sigma_0^{-1}(i)}$. This is because of the choice of permutations that will be made later, so we get a more convenient expression. Then, for $i=0,\dots,d-1$, we get that
	\begin{equation}
		y_{1,\sigma_0^{-1}(i)} = y_{1,\sigma_1^{-1}(i)}+\sum_{k=2}^{r-1} (y_{k,\sigma_k^{-1}(i)}-y_{k,\sigma_0^{-1}(i)}).
	\end{equation}
	One difficulty in proceeding is that, in this equation, a variable with $k=1$ is written as a function of another variable with $k=1$. In the general case, without any assumption about the permutations, we should decompose the permutation $\sigma_0^{-1} \sigma_1$ into cyclic permutations, and then group the equations according to each cyclic permutation. To simplify the solution, we make the assumption that $\sigma_0^{-1} \sigma_1$ is a $d$-cycle, so we don't have to partition the equations. Then we can change variables in the last equation, replacing $i$ with $\sigma_1(i)$, obtaining
	\begin{equation}
		y_{1,\sigma_0^{-1}\sigma_1(i)} = y_{1,i}+\sum_{k=2}^{r-1} (y_{k,\sigma_k^{-1}\sigma_1(i)}-y_{k,\sigma_0^{-1}\sigma_1(i)}),
	\end{equation}
	for $i=0,\dots,d-1$. To simplify the notation, let's write
	\begin{equation}
		c \coloneqq \sigma_0^{-1}\sigma_1,
	\end{equation}
	\begin{equation}
		T_i \coloneqq \sum_{k=2}^{r-1} (y_{k,\sigma_k^{-1}\sigma_1(i)}-y_{k,\sigma_0^{-1}\sigma_1(i)}).
	\end{equation}
	Then we can rewrite the equation as
	\begin{equation}
		y_{1,c(i)} = y_{1,i}+T_i,
	\end{equation} 
	for $i=0,\dots,d-1$. Since $c$ is a $d$-cycle, we can apply $c$ repeatedly over $i$, and the value $i$ will only be repeated when we apply it $d$ times. That is,
	\begin{equation}
		c^n(i) = i \iff n \in d \integers.
	\end{equation}
	Starting from $i=0$, we have that
	\begin{equation}
		y_{1,c(0)} = y_{1,0}+T_0.
	\end{equation} 
	Then we use the equation for $i = c(0)$, obtaining that
	\begin{align*}
		y_{1,c^2(0)} &= y_{1,c(0)}+T_{c(0)}\\
		&= y_{1,0}+T_0+T_{c(0)}\\
		&= y_{1,0} + \sum_{m=0,1} T_{c^m(0)}.
	\end{align*}
	By induction, we have that
	\begin{equation}
		y_{1,c^n(0)} = y_{1,0} + \sum_{m=0,1,\dots,n-1} T_{c^m(0)},
	\end{equation}
	for $n=1,\dots,d$. For $n=1,\dots,d-1$, this equation expresses each $y_{1,i}$ as a function of the independent variables, for with $i= 1,\dots,d-1$. The same equation is also valid for $n=d$, but in this case it only has independent variables, so it should be trivially satisfied, otherwise these variables couldn't be independent. For $n=d$ we have that
	\begin{equation}
		c^n(0) = c^d(0) = 0 .
	\end{equation}
	In this case, the equation becomes
	\begin{equation}
		y_{1,0} = y_{1,0} + \sum_{m=0,1,\dots,d-1} T_{c^m(0)},
	\end{equation}
	which is equivalent to
	\begin{equation}
		\sum_{m=0,1,\dots,d-1} T_{c^m(0)} = 0.
	\end{equation}
	The numbers $c^m(0)$ are a reordering of $0,1,\dots,d-1$. Since the order doesn't matter in the summation, it is equivalent to
	\begin{equation}
		\sum_{i=0,1,\dots,d-1} T_{i} = 0.
	\end{equation}
	We have that
	\begin{align*}
		\sum_{i=0,1,\dots,d-1} T_{i} &= \sum_{i=0,1,\dots,d-1} \sum_{k=2}^{r-1} (y_{k,\sigma_k^{-1}\sigma_1(i)}-y_{k,\sigma_0^{-1}\sigma_1(i)})\\
		&\stackrel{(i)}{=}  \sum_{k=2}^{r-1} \sum_{i=0,1,\dots,d-1} (y_{k,\sigma_k^{-1}\sigma_1(i)}-y_{k,\sigma_0^{-1}\sigma_1(i)})\\
		&\stackrel{(ii)}{=} \sum_{k=2}^{r-1}\left( \sum_{i=0,1,\dots,d-1} y_{k,\sigma_k^{-1}\sigma_1(i)}-\sum_{i=0,1,\dots,d-1}y_{k,\sigma_0^{-1}\sigma_1(i)}\right)\\
		&\stackrel{(iii)}{=} \sum_{k=2}^{r-1}\left( \sum_{i=0,1,\dots,d-1} y_{k,i}-\sum_{i=0,1,\dots,d-1}y_{k,i}\right)\\
		&= 0.
	\end{align*}
	In $(i)$ the summations were swapped; in $(ii)$ the summation over $i$ was distributed over each term; in $(iii)$ was used that the permutations $\sigma_k^{-1} \sigma_1$ and $\sigma_0^{-1}\sigma_1$ reorder the values in $\{0,1,\dots,d-1\}$, but the order doesn't matter for the summations over $i$, so we can replace both $\sigma_k^{-1} \sigma_1(i)$ and $\sigma_0^{-1}\sigma_1(i)$ with $i$.
	
	With these results we used all equations of the linear system to write some variables as linear functions of other variables. These other variables don't have to satisfy any other equation, so they are independent variables. This solves the linear system.
	
	Finally, let's compute the dimension of the $\realNumbers$-vector space of solutions of the linear system. The dimension $\dim_\realNumbers$ is the number of independent variables, so
	\begin{equation}
		\dim_\realNumbers = d(r-2)+1.
	\end{equation}
\end{proof}

This theorem solves the homogeneous linear system, but we need only the solutions with $y_{k,i} \geq -1/r$. One of these solutions is the trivial solution, $y_{k,i}=0$. The next result shows that there is a hypercube centered at 0, such that any choice of the independent variables inside the hypercube gives a solution for which $y_{k,i} \geq -1/r$ for every $(k,i)$.

\begin{theorem}{}{radius of hypercube}
	Let $d,r \in \naturalNumbers$ with $d,r \geq 2$, and let $(y_{k,i})_{ \substack{ k=0,\dots,r-1 \\ i=0,\dots,d-1  }}$ be real numbers. Let also $(\sigma_k)_{k=0,\dots,r-1}$ be permutations $\sigma_k \in S_d$ such that $c \coloneqq \sigma_0^{-1}\sigma_1$ is a $d$-cycle. As in theorem \ref{theorem:solution of homogeneous linear system}, divide the variables $y_{k,i}$ into dependent and independent variables. If each independent variable $y_{k,i}$ is inside the hypercube defined by the inequation
	\begin{equation}
		\label{equation: hypercube inequation}
		|y_{k,i}| \leq \frac{1}{r[1+(2d-1)(r-2)]},
	\end{equation}
	then $y_{k,i} \geq -1/r$ for $k=0,\dots,r-1$ and $i=0,\dots,d-1$.
\end{theorem}
\begin{proof}
	A sufficient condition to have $y_{k,i}\geq -1/r$ is to require that $|y_{k,i}|\leq 1/r$. First suppose that every independent variable $y_{k,i}$ satisfies that
	\begin{equation}
		|y_{k,i}|\leq \varepsilon,
	\end{equation}
	for some $\varepsilon \in [0 , 1/r]$ that we must determine. Next we estimate the module of the dependent variables, and adjust $\varepsilon$ so that $|y_{k,i}|\leq 1/r$ holds also for the dependent variables. We start with the dependent variables with $k=1$. By theorem \ref{theorem:solution of homogeneous linear system}, we can express such dependent variables as
	\begin{equation}
		y_{1,c^n(0)} = y_{1,0} + \sum_{m=0,1,\dots,n-1} T_{c^m(0)},
	\end{equation}
	for $n=1,\dots,d-1$, where
	\begin{equation}
		T_i \coloneqq \sum_{k=2}^{r-1} (y_{k,\sigma_k^{-1}\sigma_1(i)}-y_{k,\sigma_0^{-1}\sigma_1(i)}),
	\end{equation}
	for $i=0,1,\dots,d-1$. Recall that $y_{1,c^n(0)}$ is a dependent variable for $n=1,\dots,d-1$, but $y_{1,0}$ is an independent variable. Also, $y_{k,i}$ is an independent variable for $k\geq 2$. Taking the module and using the triangle inequality, we get that
	\begin{equation}
		|y_{1,c^n(0)}| \leq |y_{1,0}| + \sum_{m=0,1,\dots,n-1} |T_{c^m(0)}|.
	\end{equation}
	Since $y_{1,0}$ is an independent variable, we can use the hypothesis that $|y_{1,0}| \leq \varepsilon$, so
	\begin{equation}
		|y_{1,c^n(0)}| \leq \varepsilon + \sum_{m=0,1,\dots,n-1} |T_{c^m(0)}|.
	\end{equation}
	Each $T_j$ is a linear combination of independent variables. Using the triangle inequality, we get that
	\begin{align*}
		|T_j| &= \left| \sum_{k=2}^{r-1} y_{k,\sigma_k^{-1}\sigma_1(j)}-y_{k,\sigma_0^{-1}\sigma_1(j)} \right|\\
		&\leq \sum_{k=2}^{r-1} |y_{k,\sigma_k^{-1}\sigma_1(j)}-y_{k,\sigma_0^{-1}\sigma_1(j)}|\\
		&\leq \sum_{k=2}^{r-1} (|y_{k,\sigma_k^{-1}\sigma_1(j)}|+|y_{k,\sigma_0^{-1}\sigma_1(j)}|)\\
		&\leq \sum_{k=2}^{r-1}(\varepsilon+\varepsilon)\\
		&= 2(r-2)\varepsilon.
	\end{align*}
	Combining these estimates, we have that
	\begin{equation}
		|y_{1,c^n(0)}| \leq \varepsilon + \sum_{m=0,1,\dots,n-1} 2(r-2)\varepsilon = [1+2n(r-2)]\varepsilon.
	\end{equation}
	The highest value is obtained when $n=d-1$, so
	\begin{equation}
		|y_{1,c^n(0)}| \leq [1+2(d-1)(r-2)]\varepsilon,
	\end{equation}
	for $n=1,\dots,d-1$. Since $c$ is a $d$-cycle, the term $c^n(0)$ goes through all numbers in $1,\dots,d-1$, without repetition, so
	\begin{equation}
		|y_{1,i}| \leq [1+2(d-1)(r-2)]\varepsilon,
	\end{equation}
	for $i=1,\dots,d-1$. Therefore, we can guarantee that $|y_{1,i}| \leq 1/r$ if $[1+2(d-1)(r-2)]\varepsilon \leq 1/r$, which occurs iff
	\begin{equation}
		\varepsilon \leq \frac{1}{r[1+2(d-1)(r-2)]}.
	\end{equation}
	This inequality already implies that $\varepsilon \leq 1/r$, so we don't need to make $\varepsilon$ be the minimum of two numbers.
	
	Next we have to make an estimate of $|y_{0,i}|$. By theorem \ref{theorem:solution of homogeneous linear system}, we can express $y_{0,i}$ as
	\begin{equation}
		y_{0,i} = - \sum_{k=1}^{r-1} y_{k,i}.
	\end{equation}
	Separating the variable with $k=1$ from the others, we get that
	\begin{equation}
		y_{0,i} = - y_{1,i}- \sum_{k=2}^{r-1} y_{k,i}.
	\end{equation}
	Using the triangle inequality, we get that
	\begin{equation}
		|y_{0,i}| \leq |y_{1,i}| + \sum_{k=2}^{r-1} |y_{k,i}|.
	\end{equation}
	Since $y_{k,i}$ is an independent variable for $k\geq 2$, we can use the assumption that $|y_{k,i}| \leq \varepsilon$, so
	\begin{equation}
		\sum_{k=2}^{r-1} |y_{k,i}| \leq (r-2)\varepsilon.
	\end{equation}
	The $y_{1,i}$ may be a dependent or an independent variable. In both cases, we can use the same estimate obtained for the dependent variables, which says that
	\begin{equation}
		|y_{1,i}| \leq [1+2(d-1)(r-2)]\varepsilon.
	\end{equation}
	Combining these results, we have that
	\begin{align*}
		|y_{0,i}| &\leq [1+2(d-1)(r-2)]\varepsilon + (r-2)\varepsilon\\
		&= [1+(2d-2)(r-2)+(r-2)]\varepsilon\\
		& = [1+(2d-1)(r-2)]\varepsilon.
	\end{align*}
	Therefore, if we take
	\begin{equation}
		\varepsilon \leq \frac{1}{r[1+(2d-1)(r-2)]},
	\end{equation}
	then every variable satisfies that $|y_{k,i}| \leq 1/r$, which concludes the proof.
\end{proof}

The fraction in inequation \ref{equation: hypercube inequation} will be used again, so let's give it a name.
\begin{definition}{}{}
	Let $d, r\geq 3$, we define
	\begin{equation}
		\label{equation: radius of hypercube}
		\radius(d,r) \coloneqq \frac{1}{r[1+(2d-1)(r-2)]}.
	\end{equation}
	It is a radius of a ball with respect to the maximum norm.
\end{definition}

From theorem \ref{theorem:radius of hypercube}, now we have a hypercube of parameters for which every parameter choice gives a UCPTP map. The next step would be to guess a good choice of permutations $\sigma_k$ and parameters in the hypercube so that the UCPTP map has Choi-rank $r$ and it is an extreme point of the set of UCPTP maps.

\subsection{Choice of permutations}

The sequence of permutations $(\sigma_k)_{k=0,\dots,r-1}$, with $\sigma_k \in S_d$, should be chosen such that there is some choice of the parameters $(a_{k,i})_{\substack{k=0,\dots,r-1\\ i=0,\dots,d-1}}$ for which the corresponding UCPTP map is extreme. We'll use a guess for the permutations. We don't have a proof that it always works, but it successfully generated many examples of extreme UCPTP maps. The choice of permutations will be the same for every $r$, except for the length of $(\sigma_k)_{k=0,\dots,r-1}$, so we'll focus on the case where $r \geq d$. What will depend on $r$ is the choice of parameters $y_{k,i}$, which will be discussed later.

The basic idea was to try to distribute the non zero entries between the matrices $E_k$. Some care must be taken, because the matrices will be multiplied, we don't want $E_k^\dag E_l \oplus E_l E_k^\dag$ to be zero. A solution that worked well was to distribute the non zero entries in diagonals. The first matrix $d$ permutations were taken to be given by
\begin{equation}
	\sigma_k (i) \coloneqq (k+i) \% d, 
\end{equation}
for $k=0, \dots, d-1$. Here $n \% d$ is the remainder of the division of $n$ by $d$. In this way, for $k=0 \dots, d-1$, the matrices $E_k$ have non zero values at different entries, which trivially implies that they are orthogonal. If we take $a_{k,i}$ to be non zero for every $k,i \in \{0,\dots, d-1\}$, then we already used all possible $d^2$ entries for matrices in $M_d(\complexNumbers)$. In this way, the matrices $(E_k)_{k=0, \dots, r-1}$ won't be trivially orthogonal for $r > d$ as happened for $(E_k)_{k=0, \dots, d-1}$. Orthogonality isn't necessary, only linear independence, but it makes the proofs easier. Another possibility would be to take some coefficients $a_{k,i}$ to be zero, such that $(E_k)_{k=0, \dots, r-1}$ is trivially orthogonal. This wasn't the route taken, but this idea may be useful in the future. We'll be taking every $a_{k,i}$ to be non zero.

It remains to choose the permutations for $d \leq k \leq r-1$. The guess was to use other diagonals, by replacing the $+$ with $-$:
\begin{equation}
	\sigma_k (i) \coloneqq (k+d-i) \% d, 
\end{equation}
for $d \leq k \leq r-1$. We use $d-i$ instead of $-i$ to avoid getting a negative number, but $d-i \equiv -i \text{ mod } d$. These choices of permutations were successful for obtaining many extreme UCPTP maps, but we still lack a proof that it works for an infinite number of dimensions $d$.

In summary, we choose the permutations as
\begin{equation}
	\label{equation: method 1 permutations}
	\sigma_k (i) \coloneqq 
	\begin{cases}
		(k+i) \% d, \text{ if } 0 \leq k \leq d-1, \\
		(k+d-i) \% d, \text{ if } d \leq k \leq r-1.
	\end{cases}
\end{equation}
Again, $\%$ denotes the operation of remainder of integer division.

\subsection{Choice of parameters}
\label{section: method 1 choice of parameters}

From the previous results, we have examples of UCPTP maps which are parameterized by complex numbers $(a_{k,i})_{\substack{k=0,\dots,r-1\\ i=0,\dots,d-1}}$. From the polar decomposition, we can write them as
\begin{equation}
	a_{k,i} = \lambda_{k,i} |a_{k,i}|,
\end{equation}
where $\lambda_{k,i} \in \U(1)$. We'll just take
\begin{equation}
	\lambda_{k,i} = 1.
\end{equation}
The absolute value $|a_{k,i}|$ is parameterized as
\begin{equation}
	|a_{k,i}| = \sqrt{\frac{1}{r} + y_{k,i}}.
\end{equation}

The trivial solution, for which
\begin{equation}
	y_{k,i} = 0,
\end{equation}
we already know that doesn't give an extreme point with Choi-rank $r$ for $r \geq 2$. As said previously, the strategy is to make a small perturbation so that the vectors become linearly independent and we get an extreme point.

The variables $y_{k,i}$ can be grouped into dependent or independent variables. By theorem \ref{theorem:solution of homogeneous linear system}, we can take as independent variables the $y_{k,i}$ with
\begin{equation}
	(k,i) \in \{(1,0)\} \cup \{2, \dots, r-1 \} \times \{0, \dots, r-1\}.
\end{equation}
The remaining $2d-1$ variables are dependent. See theorem \ref{theorem:solution of homogeneous linear system} for a better visualization of how they are organized into a table. We then have to choose values for the independent variables. It is important to choose them so that the proof of linear independence becomes easier. The following choice was made for the \textbf{independent variables}:
\begin{equation}
	y_{k,i} \coloneqq \delta_{k,i} \radius(d,r),
\end{equation}
for $(k,i)=\{(1,0)\} \cup \{2, \dots, r-1 \} \times \{0, \dots, r-1\}$. By theorem \ref{theorem:radius of hypercube}, we know that this choice of $y_{k,i}$ gives a UCPTP map.

A C program was made to test if this choice of parameters gives extreme UCPTP maps. The program used the Fast Library for Number Theory (FLINT) C library to make fast \textbf{exact computations}. The results are summarized in table \ref{table: results method 1}. We can see that the guess failed a single time, for the case with $(d,r)=(5,7)$. All other tested cases were successful. This failure may be due to a limitation of the method employed to generate examples, or it may be the case that the bound $r \leq \lfloor \sqrt{2} d \rfloor$ from corollary \ref{corollary:bound on rank of extreme ucptp} isn't optimal. Further study has yet to be done to try to prove that this method works for an infinite number of dimensions $d$.

\begin{table}
	\caption{Exact computational results for the UCPTP maps given by the choice of parameters of section \ref{section: method 1 choice of parameters}. It was tested if the given UCPTP map has Choi-rank $r$ and is an extreme point of the set of UCPTP maps. Were tested the cases with $3 \leq d \leq 7$ and $3 \leq r \leq \lfloor \sqrt{2}d \rfloor$. The cells of the table with a dash symbol (-) are the ones where $r > \lfloor \sqrt{2}d \rfloor $, for which the bounds of corollary \ref{corollary:bound on rank of extreme ucptp} guarantee that we don't have an extreme point, so no test was made in this case. The cases where the tests were successful, obtaining an extreme UCPTP map with Choi-rank $r$, are indicated by an S. The cases where the test failed are indicated by an F. The cells with a question mark (?) are the ones for which we couldn't make the test, because of the long time necessary for the computer program to finish.}
	\label{table: results method 1}
	
	\begin{center}
		\begin{tabular}{ |c|c|c|c|c|c| }
			\hline
			9 & - & - & - & - & ? \\
			\hline
			8 & - & - & - & S & ? \\
			\hline
			7 & - & - & {\color{red}F} & S & S \\
			\hline
			6 & - & - & S & S & S \\
			\hline
			5 & - & S & S & S & S \\
			\hline
			4 & S & S & S & S & S \\
			\hline
			3 & S & S & S & S & S \\
			\hline
			$r$ \ / \ $d$ & 3 & 4 & 5 & 6 & 7 \\
			\hline
		\end{tabular}
	\end{center}
\end{table}

\subsection{Another method which was tested for higher dimensions}

Many other guesses were made to construct examples of extreme UCPTP maps. One of them was made before the one presented in section \ref{section: method 1}. It isn't as well structured as the one from section \ref{section: method 1}, but the algorithm implemented for it was able to produce examples much faster and for higher dimensions. We'll briefly present the ideas and show some computational results.

The same idea of equation \ref{equation: method 1 concatenating Ek vertically} was used, to concatenate the Kraus matrices $E_k$ vertically to form a single matrix:
\begin{equation}
	V = 
	\begin{pmatrix}
		E_0\\
		E_1\\
		\vdots\\
		E_{r-1}
	\end{pmatrix} \in M_{dr, d}(\complexNumbers).
\end{equation}

The guess was that $V$ would have non zero entries on inclined lines. Modular arithmetic is used for the lines to remain inside the matrix. The idea is summarized as the following:
\begin{center}

	\tikzset{every picture/.style={line width=0.75pt}} 
	
	\begin{tikzpicture}[x=0.75pt,y=0.75pt,yscale=-1,xscale=1]
		
		\draw    (80,210) .. controls (68.7,183.6) and (72.7,34.6) .. (80,10) ;
		\draw    (130,210) .. controls (137.7,188.6) and (139.7,41.6) .. (130,10) ;
		\draw [line width=2.25]    (80,10) -- (130,100) ;
		\draw [line width=2.25]    (80,70) -- (130,160) ;
		\draw [line width=2.25]    (80,130) -- (125,210) ;
		\draw [line width=2.25]    (125,10) -- (130,20) ;
		
		\draw (26,87.4) node [anchor=north west][inner sep=0.75pt]    {$V\ =$};
		\draw (112,27.4) node [anchor=north west][inner sep=0.75pt]    {$0$};
		\draw (101,82.4) node [anchor=north west][inner sep=0.75pt]    {$0$};
		\draw (101,137.4) node [anchor=north west][inner sep=0.75pt]    {$0$};
		\draw (86,182.4) node [anchor=north west][inner sep=0.75pt]    {$0$};

	\end{tikzpicture}
	
\end{center}
In this case, three lines are depicted. The lines start at the first column and proceed downwards as we go to the columns to the right. There is a constant spacing between consecutive lines. They also have the same slope. The third line doesn't fit in the matrix, so modular arithmetic is used for the line to continue from the first row. For this reason, there is a small piece of line at the right top corner of the matrix, that is actually a piece of the third line that didn't fit at the bottom of the matrix. Even though the matrix was depicted with lines, we actually have a discrete structure, so a single line can have row jumps depending on the slope of the line.

With this idea, we have three parameters to describe the lines: the number of lines ($N$), the spacing between the lines ($A$) and the slope ($B$). All of these parameters are taken to be positive integers. All indices will start from 0, which is better for modular arithmetic. There are $N$ lines, indexed by $n \in \{0, \dots, N-1\}$. To describe a single line, we first compute the index of its row $R(n,i) \in \{0, \dots, dr\}$ for a given column $i \in \{0, \dots, d-1\}$. $R(n,i)$ is given by
\begin{equation}
	R(n,i) \coloneqq (An+Bi)) \% (dr).
\end{equation}
The idea is that we jump downwards $n$ times, with steps according to the spacing $A$, and then move downwards as the column index $i$ increases, according to the slope $B$. We may move downwards too much, to the point that $An+Bi \geq dr$. This problem is solved by modular arithmetic, we take the remainder of the integer division by $dr$. The operation of taking the remainder is denoted by $\%$. Also, each line will have a complex number for each column. All these numbers form together a matrix
\begin{equation}
	(a_{n,i})_{\substack{n = 0, \dots, N-1 \\ i = 0, \dots, d-1}} \in M_{N, d}(\complexNumbers).
\end{equation}
Then the line with index $n$ corresponds to the matrix
\begin{equation}
	\sum_{i=0}^{d-1} a_{n,i} e_{R(n,i),i} .
\end{equation}
$V$ is obtained by summing these matrices:
\begin{equation}
	V \coloneqq \sum_{n = 0}^{N-1} \sum_{i=0}^{d-1} a_{n,i} e_{R(n,i),i}.
\end{equation}

\subsection{Choice of parameters}
\label{section: method 2 choice of parameters}

After some analysis of the problem, it was found that there are many examples with the following structure. We take constant coefficients, with
\begin{equation}
	a_{n, i} = \frac{1}{\sqrt{N}}.
\end{equation}
Also, we take the slope $B$ to be the product of the spacing and the number of lines:
\begin{equation}
	B = AN.
\end{equation}
Then we get that
\begin{equation}
	R(n,i) = (A (n+Ni)) \% (dr),
\end{equation}
and
\begin{equation}
	V = \frac{1}{\sqrt{N}} \sum_{n = 0}^{N-1} \sum_{i=0}^{d-1} e_{R(n,i),i}.
\end{equation}
The spacing $A$ is taken to be the biggest such that
\begin{equation}
	d+1 \leq A \leq dr/N
\end{equation}
and
\begin{equation}
	\gcd(A, dr) = 1 ,
\end{equation}
if it exists. Then only the parameter $N$ remains, which is found by a single for loop. We require $N$ to be bounded by
\begin{equation}
	2 \leq N \leq r-1.
\end{equation}

With these choices of $A$ and $B$, $N$ is the only remaining parameter for $V$. Not every $N$ works, but we have reduced three for loops, for the parameters $A$, $B$ and $N$, to a single for loop, only for $N$. We also eliminated the need to find $a_{n,i}$ computationally. Taking care for the lines not to overlay each other, all entries of $V$ are 0 or $1/\sqrt{N}$, which makes computation easier. These are some of the reasons for the algorithm for this guess to be able to compute solutions faster. The disadvantage of this guess is that it has more fails and it may be harder for making mathematical proofs.

Finally, we present the results obtained computationally. A C program was made to test if $V$ gives an extreme UCPTP map. To make fast \textbf{exact computations}, it was used the Fast Library for Number Theory (FLINT) C library. In tables \ref{table: results method 2 part 1} and \ref{table: results method 2 part 2} are shown for which $(d,r)$ the method succeeded or failed to produce an example. The method was applied to the cases where $3 \leq d \leq 22$ and for Choi-ranks with $3 \leq r \leq \sqrt{2}d$. The upper bound $\sqrt{2}d$ is the maximum Choi-rank allowed by corollary \ref{corollary:bound on rank of extreme ucptp}. The method produced a total of 915 examples. Due to the large number of examples, they are presented separately in appendix \ref{appendix: examples extreme ucptp}. The time needed to finish the program increases with $d$ and $r$. It was decided to stop at $d=22$ so that the program would not continue running for many hours, but it is possible to compute results for higher values.

We can see that this method also failed to produce an example for $(d,r) = (5,7)$. It failed in many cases for small $d$, but for $13 \leq d \leq 22$ it only failed for the maximum Choi-rank allowed by corollary \ref{corollary:bound on rank of extreme ucptp}, that is, for $\lfloor \sqrt{2}d \rfloor$. This may be a limitation of the method or it can be the case that the upper bound $\lfloor \sqrt{2}d \rfloor$ isn't optimal.

\begin{table}
	\caption{Exact computational results for the UCPTP maps given by the choice of parameters of section \ref{section: method 2 choice of parameters}. It was tested if the given UCPTP map has Choi-rank $r$ and is an extreme point of the set of UCPTP maps. Were tested the cases with $3 \leq d \leq 22$ and $3 \leq r \leq \lfloor \sqrt{2}d \rfloor$. Because of the table size, we only show the cases with $3 \leq d \leq 17$ in this table, the remaining cases are shown in table \ref{table: results method 2 part 2}. The cells of the table with a dash symbol (-) are the ones where $r > \lfloor \sqrt{2}d \rfloor $, for which the bounds of corollary \ref{corollary:bound on rank of extreme ucptp} guarantee that we don't have an extreme point, so no test was made in this case. The cases where the tests were successful, obtaining an extreme UCPTP map with Choi-rank $r$, are indicated by an S. The cases where the test failed are indicated by an F.}
	\label{table: results method 2 part 1}
	
	\begin{center}
		\begin{tabular}{ |c|c|c|c|c|c|c|c|c|c|c|c|c|c|c|c| }
			\hline
			24&-&-&-&-&-&-&-&-&-&-&-&-&-&-&{\color{red}F}\\
			\hline
			23&-&-&-&-&-&-&-&-&-&-&-&-&-&-&S\\
			\hline
			22&-&-&-&-&-&-&-&-&-&-&-&-&-&S&S\\
			\hline
			21&-&-&-&-&-&-&-&-&-&-&-&-&S&S&S\\
			\hline
			20&-&-&-&-&-&-&-&-&-&-&-&-&S&S&S\\
			\hline
			19&-&-&-&-&-&-&-&-&-&-&-&S&S&S&S\\
			\hline
			18&-&-&-&-&-&-&-&-&-&-&{\color{red}F}&S&S&S&S\\
			\hline
			17&-&-&-&-&-&-&-&-&-&-&S&S&S&S&S\\
			\hline
			16&-&-&-&-&-&-&-&-&-&S&S&S&S&S&S\\
			\hline
			15&-&-&-&-&-&-&-&-&S&{\color{red}F}&S&S&S&S&S\\
			\hline
			14&-&-&-&-&-&-&-&{\color{red}F}&S&S&S&S&S&S&S\\
			\hline
			13&-&-&-&-&-&-&-&{\color{red}F}&S&S&S&S&S&S&S\\
			\hline
			12&-&-&-&-&-&-&S&S&S&S&S&S&S&S&S\\
			\hline
			11&-&-&-&-&-&{\color{red}F}&S&S&S&S&S&S&S&S&S\\
			\hline
			10&-&-&-&-&-&S&S&S&S&S&S&S&S&S&S\\
			\hline
			9&-&-&-&-&S&S&S&S&S&S&S&S&S&S&S\\
			\hline
			8&-&-&-&{\color{red}F}&S&S&S&S&S&S&S&S&S&S&S\\
			\hline
			7&-&-&{\color{red}F}&{\color{red}F}&S&S&S&S&S&S&S&S&S&S&S\\
			\hline
			6&-&-&S&S&{\color{red}F}&S&S&S&S&S&S&S&S&S&S\\
			\hline
			5&-&{\color{red}F}&S&S&S&{\color{red}F}&S&S&S&S&S&S&S&S&S\\
			\hline
			4&{\color{red}F}&S&{\color{red}F}&{\color{red}F}&S&S&S&S&S&S&S&S&S&S&S\\
			\hline
			3&S&{\color{red}F}&S&S&S&S&S&S&S&S&S&S&S&S&S\\
			\hline
			$r$ \ / \ $d$&3&4&5&6&7&8&9&10&11&12&13&14&15&16&17 \\
			\hline
		\end{tabular}
	\end{center}
\end{table}

\begin{table}
	\caption{Continuation of table \ref{table: results method 2 part 1} for $18 \leq d \leq 22$.}
	\label{table: results method 2 part 2}
	
	\begin{center}
		\begin{tabular}{ |c|c|c|c|c|c| }
			\hline
			31&-&-&-&-&{\color{red}F}\\
			\hline
			30&-&-&-&-&S\\
			\hline
			29&-&-&-&S&S\\
			\hline
			28&-&-&S&S&S\\
			\hline
			27&-&-&S&S&S\\
			\hline
			26&-&S&S&S&S\\
			\hline
			25&S&S&S&S&S\\
			\hline
			24&S&S&S&S&S\\
			\hline
			23&S&S&S&S&S\\
			\hline
			22&S&S&S&S&S\\
			\hline
			21&S&S&S&S&S\\
			\hline
			20&S&S&S&S&S\\
			\hline
			19&S&S&S&S&S\\
			\hline
			18&S&S&S&S&S\\
			\hline
			17&S&S&S&S&S\\
			\hline
			16&S&S&S&S&S\\
			\hline
			15&S&S&S&S&S\\
			\hline
			14&S&S&S&S&S\\
			\hline
			13&S&S&S&S&S\\
			\hline
			12&S&S&S&S&S\\
			\hline
			11&S&S&S&S&S\\
			\hline
			10&S&S&S&S&S\\
			\hline
			9&S&S&S&S&S\\
			\hline
			8&S&S&S&S&S\\
			\hline
			7&S&S&S&S&S\\
			\hline
			6&S&S&S&S&S\\
			\hline
			5&S&S&S&S&S\\
			\hline
			4&S&S&S&S&S\\
			\hline
			3&S&S&S&S&S\\
			\hline
			$r$ \ / \ $d$&18&19&20&21&22 \\
			\hline
		\end{tabular}
	\end{center}
\end{table}

\cleardoublepage
\chapter{Extremality of the tensor product}

\label{section: Extremality of the tensor product}

In this chapter prove that the tensor product of extreme CPTP maps is an extreme CPTP map. By duality, the same is true for UCP maps. Then we investigate the case of UCPTP maps. We show that the tensor product of extreme UCPTP maps may not be an extreme UCPTP map. We use extreme UCPTP maps with high rank to construct counterexamples. High rank extreme channels were constructed in chapter \ref{chapter: examples extreme channels}.

\section{The CPTP case}

We prove that the tensor product preserves extramality for CPTP maps.

\begin{theorem}{}{extremality is preserved for cptp}
	If $\varepsilon$ is an extreme point of $\cptp(X,Y)$ and $\varepsilon'$ is an extreme point of $\cptp(X',Y')$, then $\varepsilon \otimes \varepsilon'$ is an extreme point of $\cptp(X \otimes X', Y\otimes Y')$.
\end{theorem}
\begin{proof}
	We always use the Hilbert-Schmidt inner product. Let $(E_k)_k$ be linearly independent Kraus operators for $\varepsilon$, and let $(F_l)_l$ be linearly independent Kraus operators for $\varepsilon'$. By proposition \ref{proposition:tensor product operation elements}, $(E_k \otimes F_l)_{k,l}$ are Kraus operators for $\varepsilon\otimes \varepsilon'$. Also, proposition \ref{proposition:tensor of l.i. vectors} implies that $(E_k \otimes F_l)_{k,l}$ is linearly independent. Then, by theorem \ref{theorem:extreme cptp maps in Kraus representation}, $\varepsilon \otimes \varepsilon'$ is extreme if, and only if, $((E_k \otimes F_l)^\dag (E_{k'}\otimes F_{l'}))_{k,k',l,l'}$ is linearly independent (as vectors of $\Lin_\complexNumbers(X \otimes Y)$). By proposition \ref{proposition:gram matrix}, linear independence is equivalent to the invertibility of the Gram matrix. To prove the invertibility of the Gram matrix, we'll show that it is the Kronecker product of two invertible matrices. Let $G$, $G'$ and $G''$ be the Gram matrices of $(E_k^\dag E_{k'})_{k,k'}$, $(F_l^\dag F_{l'})_{l,l'}$ and  $((E_k \otimes F_l)^\dag (E_{k'}\otimes F_{l'}))_{k,k',l,l'}$, respectively. Its matrix elements are
	\begin{equation}
		G_{(k,k'),(k'',k''')} = \langle E_k^\dag E_{k'}, E_{k''}^\dag E_{k'''} \rangle,
	\end{equation}
	\begin{equation}
		G'_{(l,l'),(l'',l''')} = \langle F_l^\dag F_{l'}, F_{l''}^\dag F_{l'''} \rangle,
	\end{equation}
	\begin{equation}
		G''_{(k,k',l,l'),(k'',k''',l'',l''')} = \langle (E_k \otimes F_l)^\dag (E_{k'}\otimes F_{l'}), (E_{k''}\otimes F_{l''})^\dag (E_{k'''}\otimes F_{l'''}) \rangle.
	\end{equation}
	Since $\varepsilon$ and $\varepsilon'$ are extreme CPTP maps, and since $(E_k)_k$ and $(F_l)_l$ are linearly independent Kraus operators, by theorem \ref{theorem:extreme cptp maps in Kraus representation} we know that $(E_k^\dag E_{k'})_{k,k'}$ and $(F_l^\dag F_{l'})_{l,l'}$ are both linearly independent. By proposition \ref{proposition:gram matrix}, it implies that $G$ and $G'$ are invertible matrices. We can rewrite $G''$ in terms of $G$ and $G'$ as follows:
	\begin{align*}
		G''_{(k,k',l,l'),(k'',k''',l'',l''')} &= \langle (E_k \otimes F_l)^\dag (E_{k'}\otimes F_{l'}), (E_{k''}\otimes F_{l''})^\dag (E_{k'''}\otimes F_{l'''}) \rangle\\
		&= \langle (E_k^\dag \otimes F_l^\dag)(E_{k'}\otimes F_{l'}), (E_{k''}^\dag\otimes F_{l''}^\dag) (E_{k'''}\otimes F_{l'''}) \rangle\\
		&= \langle E_k^\dag E_{k'} \otimes F_l^\dag F_{l'}, E_{k''}^\dag E_{k'''} \otimes F_{l''}^\dag F_{l'''} \rangle\\
		&\stackrel{(*)}{=} \langle E_k^\dag E_{k'}, E_{k''}^\dag E_{k'''} \rangle \langle F_l^\dag F_{l'}, F_{l''}^\dag F_{l'''} \rangle\\
		&= G_{(k,k'),(k'',k''')} G'_{(l,l'),(l'',l''')}.
	\end{align*}
	In equality (*) we used equation (\ref{equation: HS multiplicativity}). Given any order of $(k,k')$ and $(l,l')$, we can give the lexicographic order to $(k,k',l,l')$. In this way, the identity $G''_{(k,k',l,l'),(k'',k''',l'',l''')} = G_{(k,k'),(k'',k''')} G'_{(l,l'),(l'',l''')}$ says that $G'' = G \otimes G'$, that is, $G''$ is the Kronecker product of $G$ and $G'$. Since $G$ and $G'$ are invertible, then $G''$ is also invertible, with inverse $G''^{-1} = G^{-1} \otimes G'^{-1}$. Then proposition \ref{proposition:gram matrix} implies that $((E_k \otimes F_l)^\dag (E_{k'} \otimes F_{l'}))_{k,k',l,l'}$ is linearly independent. Therefore, by theorem \ref{theorem:extreme cptp maps in Kraus representation}, $\varepsilon \otimes \varepsilon'$ is an extreme CPTP map.
\end{proof}

\section{The UCP case}

The result for UCP maps can be obtained by using the duality between CPTP and UCP maps.

\begin{theorem}{}{}
	If $\varepsilon$ is an extreme point of $\ucp(X,Y)$ and $\varepsilon'$ is an extreme point of $\ucp(X',Y')$, then $\varepsilon \otimes \varepsilon'$ is an extreme point of $\ucp(X \otimes X', Y\otimes Y')$.
\end{theorem}
\begin{proof}
	Since $\varepsilon \in \ucp(X,Y)$ and $\varepsilon' \in \ucp(X',Y')$, by proposition \ref{proposition:cptp iff adjoint is ucp}, we know that $\varepsilon^\dag \in \cptp\allowbreak(Y,X)$ and $\varepsilon'^\dag \in \cptp(Y',X')$. Using the same proposition and proposition \ref{proposition:linear bijection between extreme points}, we know that the duality gives a bijection between the extreme points. Since $\varepsilon$ and $\varepsilon'$ are extreme points of $\ucp(X,Y)$ and $\ucp(X',Y')$, then $\varepsilon^\dag$ and $\varepsilon'^\dag$ are extreme points of $\cptp(Y,X)$ and $\cptp(Y',X')$. By theorem \ref{theorem:extremality is preserved for cptp}, we know that $\varepsilon^\dag \otimes \varepsilon'^\dag$ is an extreme point of $\cptp(Y\otimes Y', X \otimes X')$. By proposition \ref{proposition:dual of tensor of cp maps} we know that $\varepsilon^\dag \otimes \varepsilon'^\dag = (\varepsilon \otimes \varepsilon')^\dag$, so $(\varepsilon \otimes \varepsilon')^\dag$ is extreme. Using again the duality between CPTP and UCP maps we conclude that $\varepsilon \otimes \varepsilon'$ is an extreme point of $\ucp(X \otimes X', Y \otimes Y')$.
\end{proof}

\section{The UCPTP case}

The tensor product of extreme UCPTP maps may not be extreme. But it is always extreme if one of the two maps is over a Hilbert space of dimension 2. This is a consequence of the next proposition and the fact that extreme maps are unitary for dimension 2.

\begin{proposition}{}{tensor with unitary}
	Let $\varepsilon \in \ucptp(X)$ and $\mathcal{U} \in \ucptp(Y)$ be a unitary map, that is, $\mathcal{U}(A) = U A U^\dag$  for some unitary operator $U \in \Lin_\complexNumbers (Y)$. We assume that $d_Y > 0$. Then $\varepsilon$ is an extreme point of $\ucptp(X)$ if, and only if, $\varepsilon\otimes \mathcal{U}$ is an extreme point of $\ucptp(X \otimes Y)$.
\end{proposition}
\begin{proof}
	Let $(E_k)_k$ be linearly independent Kraus operators for $\varepsilon$. $\mathcal{U}$ has already $(U)$ as linearly independent Kraus operators, since there is only one operator. By propositions \ref{proposition:tensor of l.i. vectors} and \ref{proposition:tensor product operation elements}, $\varepsilon \otimes \mathcal{U}$ has linearly independent Kraus operators $(E_k \otimes U)$. By theorem \ref{theorem:characterization of extreme points for ucptp}, $\varepsilon \otimes \mathcal{U}$ is an extreme point of $\ucptp(X\otimes Y)$ if, and only if, $((E_k \otimes U)^\dag (E_l \otimes U) \oplus (E_l \otimes U)(E_k \otimes U)^\dag)_{k,l}$ is linearly independent. We have that
	\begin{align*}
		(E_k \otimes U)^\dag (E_l \otimes U) \oplus (E_l \otimes U)(E_k \otimes U)^\dag &= (E_k^\dag \otimes U^\dag) (E_l \otimes U) \oplus (E_l \otimes U)\\
		&\quad \; (E_k^\dag \otimes U^\dag)\\
		&= (E_k^\dag E_l \otimes U^\dag U) \oplus (U U^\dag \otimes E_l E_k^\dag)\\
		&= (E_k^\dag E_l \otimes \id_Y) \oplus (\id_Y \otimes E_l E_k^\dag).
	\end{align*}
	By proposition \ref{proposition:gram matrix}, we can check if $((E_k^\dag E_l \otimes \id_Y) \oplus (\id_Y \otimes E_l E_k^\dag))_{k,l}$ is linearly independent or not by analysing its Gram matrix. Let $G$ be the Gram matrix of $(E_k^\dag E_l \oplus E_l E_k^\dag)_{k,l}$ and $G'$ be the Gram matrix of $((E_k^\dag E_l \otimes \id_Y) \oplus (\id_Y \otimes E_l E_k^\dag))_{k,l}$. The matrix elements of $G$ are
	\begin{align*}
		G_{(k,l),(k',l')} &= \langle E_k^\dag E_l \oplus E_l E_k^\dag, E_{k'}^\dag E_{l'} \oplus E_{l'} E_{k'}^\dag \rangle\\
		&= \langle E_k^\dag E_l, E_{k'}^\dag E_{l'} \rangle + \langle E_l E_k^\dag, E_{l'} E_{k'}^\dag \rangle.
	\end{align*}
	The matrix elements of $G'$ are
	\begin{align*}
		G'_{(k,l),(k',l')} &= \langle (E_k^\dag E_l \otimes \id_Y) \oplus (\id_Y \otimes E_l E_k^\dag), (E_{k'}^\dag E_{l'} \otimes \id_Y) \oplus (\id_Y \otimes E_{l'} E_{k'}^\dag) \rangle\\
		&= \langle E_k^\dag E_l \otimes \id_Y, E_{k'}^\dag E_{l'} \otimes \id_Y \rangle + \langle \id_Y \otimes E_l E_k^\dag, \id_Y \otimes E_{l'} E_{k'}^\dag \rangle\\
		&= \langle E_k^\dag E_l, E_{k'}^\dag E_{l'} \rangle \langle \id_Y, \id_Y\rangle + \langle \id_Y, \id_Y\rangle \langle E_l E_k^\dag, E_{l'} E_{k'}^\dag \rangle \\
		&= d_Y (\langle E_k^\dag E_l, E_{k'}^\dag E_{l'} \rangle + \langle E_l E_k^\dag, E_{l'} E_{k'}^\dag \rangle)\\
		& = d_Y \, G_{(k,l),(k',l')}.
	\end{align*}
	That is, $G' = d_Y \, G$. Therefore $G$ is invertible if, and only if, $G'$ is invertible. By proposition \ref{proposition:gram matrix}, this says that $(E_k^\dag E_l \oplus E_l E_k^\dag)_{k,l}$ is linearly independent if, and only if, $((E_k^\dag E_l \otimes \id_Y) \oplus (\id_Y \otimes E_l E_k^\dag))_{k,l}$ is linearly independent. Then, by theorem \ref{theorem:characterization of extreme points for ucptp}, $\varepsilon$ is an extreme point of $\ucptp(X)$ if, and only if, $\varepsilon \otimes \mathcal{U}$ is an extreme point of $\ucptp(X \otimes Y)$.
\end{proof}

By theorem 4.23 of \cite{watrous_2018}, for $d_X = 2$, the extreme points of $\ucptp(X)$ are the unitary maps. Combining this result with proposition \ref{proposition:tensor with unitary} we obtain the next theorem.

\begin{theorem}{}{}
	Let $\varepsilon$ be an extreme point of $\ucptp(X)$ and $\varepsilon'$ be an extreme point of $\ucptp(Y)$. If $d_X = 2$ or $d_Y = 2$, then $\varepsilon \otimes \varepsilon'$ is an extreme point of $\ucptp(X \otimes Y)$.
\end{theorem}

For higher dimensions the extremality may not be preserved. We can prove this using the upper bound on the Choi-rank proved in corollary \ref{corollary:bound on rank of extreme ucptp}. If $\varepsilon \in \ucptp(X)$ is an extreme point, then $\CR(\varepsilon) \leq \sqrt{2} \, d_X$. If we take two extreme UCPTP maps of high enough Choi-ranks, their tensor product will have a Choi-rank that exceeds the previous upper bound, so the tensor product won't be extreme. This is proven in the next theorem.

\begin{theorem}{}{tensor product of ucptp need not be extreme}
	Let $\varepsilon$ be an extreme point of $\ucptp(X)$ and $\varepsilon'$ be an extreme point of $\ucptp(Y)$. In particular, their Choi-ranks satisfy that $\CR(\varepsilon) \leq \sqrt{2} \, d_X$ and $\CR(\varepsilon') \leq \sqrt{2} \, d_Y$. If they further satisfy that $\sqrt[4]{2} \, d_X < \CR(\varepsilon)$ and $\sqrt[4]{2} \, d_Y < \CR(\varepsilon')$, then $\sqrt{2} \, \dim_\complexNumbers (X\otimes Y) < \CR(\varepsilon \otimes \varepsilon')$, so $\varepsilon \otimes \varepsilon'$ isn't an extreme point of $\ucptp(X \otimes Y)$.
\end{theorem}
\begin{proof}
	By corollary \ref{corollary:CR of tensor product}, we know that $\CR(\varepsilon \otimes \varepsilon') = \CR(\varepsilon)$ $CR(\varepsilon')$. If $\sqrt[4]{2} \, d_X $ $<$ $ \CR(\varepsilon)$ and $\sqrt[4]{2} \, d_Y < \CR(\varepsilon')$, multiplying both inequalities we get \linebreak $\sqrt[4]{2}\, d_X \sqrt[4]{2} \, d_Y < \CR(\varepsilon)\CR(\varepsilon')$, so $\sqrt{2} \dim_\complexNumbers (X \otimes Y) < \CR(\varepsilon \otimes \varepsilon')$. By corollary \ref{corollary:bound on rank of extreme ucptp}, this implies that $\varepsilon \otimes \varepsilon'$ isn't an extreme point of $\ucptp(X \otimes Y)$.
\end{proof}

By theorem \ref{theorem:tensor product of ucptp need not be extreme}, to prove that the tensor product of extreme UCPTP maps may not be extreme, it is sufficient to find extreme maps of high enough Choi-rank. A large number of high rank extreme UCPTP maps were obtained in section \ref{section: extreme ucptp with choi rank at least 3}, so there are many examples for which extremality isn't preserved.

For the tensor product of multiple copies of the same UCPTP map, if the number of copies is high enough, it can't be extreme for high Choi-rank, as we prove next.
\begin{theorem}{}{}
	Let $X$ be a finite dimension complex Hilbert space with dimension $d$. Let also $\varepsilon \in \ucptp(X)$ with $\CR(\varepsilon) > d$, then $\varepsilon^{\otimes^n}$ isn't an extreme point of $\ucptp(X^{\otimes^n})$ for
	\begin{equation}
		n > \frac{1}{2 \log_2 \left( \frac{\CR(\varepsilon)}{d} \right)}.
	\end{equation}
\end{theorem}
\begin{proof}
	By corollary \ref{corollary:bound on rank of extreme ucptp}, if $\varepsilon^{\otimes^n}$ is an extreme point, then
	\begin{equation}
		\label{equation: CR of iterated tensor}
		\CR(\varepsilon^{\otimes^n}) \leq \sqrt{2} \dim_\complexNumbers (X^{\otimes^n}).
	\end{equation}
	By corollary \ref{corollary:CR of tensor product}, we have that
	\begin{equation}
		\CR(\varepsilon^{\otimes^n}) = \CR(\varepsilon)^n .
	\end{equation}
	Also,
	\begin{equation}
		\dim_\complexNumbers (X^{\otimes^n}) = (\dim_\complexNumbers X)^n = d^n.
	\end{equation}
	Substituting these results in equation \ref{equation: CR of iterated tensor}, we get that
	\begin{equation}
		\CR(\varepsilon)^n \leq \sqrt{2} \ d^n.
	\end{equation}
	Taking the $n$th root, we get that
	\begin{equation}
		\CR(\varepsilon) \leq 2^{\frac{1}{2n}} d.
	\end{equation}
	Dividing by $d$ and taking the $\log_2$, we get that
	\begin{equation}
		\frac{1}{2n} \geq \log_2 \left( \frac{\CR(\varepsilon)}{d} \right).
	\end{equation}
	Isolating $n$, we get that
	\begin{equation}
		n \leq \frac{1}{2 \log_2 \left( \frac{\CR(\varepsilon)}{d} \right)} .
	\end{equation}
	
	By the contrapositive argument, if
	\begin{equation}
		n > \frac{1}{2 \log_2 \left( \frac{\CR(\varepsilon)}{d} \right)},
	\end{equation}
	then $\varepsilon^{\otimes^n}$ isn't an extreme point of $\ucptp(X^{\otimes^n})$.
\end{proof}

\cleardoublepage
\chapter{Results for the Category of CPTP Maps}
\label{section: limits and colimits of CPTP}

We saw in proposition \ref{proposition:category cptp, initial and terminal objects} that the category of CPTP maps is semicartesian, having $\{0\}$ as initial object and $\complexNumbers$ as terminal object. Also, $\complexNumbers$ is the monoidal unit of the category. Not much more seems to be known about the limits and colimits of this category. The basic limits are terminal object, binary product, equalizer and pullback, and the basic colimits are initial object, coproduct, coequalizer and pushout. We begin our investigation by these basic limits and colimits. In this chapter we prove that this category only have trivial binary products and find the dimension a coproduct should have, if it exists. The formula of the dimension of the coproduct doesn't always give a natural number, from which we conclude that coproducts don't exist for an infinite number of cases. For both products and coproducts we use topological methods. The technique consists of using the universal property to construct a homeomorphism between two topological spaces, and then compute topological invariants to show that such homeomorphism can't exist.

The topology of the set of extreme points of $\cptp(X,Y)$ isn't yet understood for $d_X \geq 2$ and $d_Y \geq 2$. It also isn't understood for the closure of the set of extreme points. For $X = \complexNumbers$ the set $\cptp(\complexNumbers,Y)$ is isomorphic to the set $\density(Y)$ of density operators over $Y$. In this case, the set of extreme points is closed and is homeomorphic to the complex projective space $\complexNumbers P^{d_Y-1}$.  This knowledge is enough to completely classify the binary products. For coproducts the corresponding reasoning would be to use the hom-set $\cptp(X,\complexNumbers)$, but this set has only one element, since $\complexNumbers$ is a terminal object. Because of this, we are forced to use the hom-sets $\cptp(X,Z)$ for $d_Z \geq 2$, for which we have less topological information. For this reason, we don't have yet a full classification of the coproducts. We expect that the study of the topology of the set of extreme points or of its closure will help classify the limits and colimits. A partial result in this direction will be shown in chapter \ref{chapter: cells extreme cptp}.

\section{Products}

In this section we classify all binary products of the category of CPTP maps. It is recommended to have knowledge of Algebraic Topology, specially singular homology. The book Singular Homology Theory from Massey \cite{masseySHT} is being used as reference. A surface level knowledge about singular homology may be enough.

Recall that an arrow in the category of CPTP maps, which we denote by $\varepsilon \colon X \to Y$, is a $\complexNumbers$-linear map $\varepsilon \colon \Lin_\complexNumbers (X) \to \Lin_\complexNumbers(Y)$ which is completely positive and trace preserving.

\begin{theorem}{}{products in CPTP}
	The category CPTP only have the binary products $X_1 \times X_2$ for which one of the objects $X_i$ is $\{0\}$ or $\complexNumbers$. In this case, we have
	\begin{equation}
		X \times \{0\} \cong \{0\} \times X \cong \{0\},
	\end{equation}
	\begin{equation}
		Y \times \complexNumbers \cong \complexNumbers \times Y \cong Y.
	\end{equation}
\end{theorem}
\begin{proof}
	Let $X_1$ and $X_2$ be finite dimensional complex Hilbert spaces. First we'll consider the case where $d_{X_1} \geq 1$ and $d_{X_2} \geq 1$, so that $\density(X_1) \neq \varnothing$ and $\density(X_2) \neq \varnothing$. Suppose that there exists the binary product $X_1 \times X_2$, which we'll denote by $P$. Let also $p_i \in \cptp(P,X_i)$ be the projection in the sense of the product diagram in a category. The universal property of the binary product says the following: for any finite dimensional complex Hilbert space $Z$, for any CPTP maps $\varepsilon_1 \in \cptp(Z,X_1)$ and $\varepsilon_2 \in \cptp(Z,X_2)$, there exists a unique CPTP map $\varepsilon \in \cptp(Z, P)$ such that the next diagram commutes.
	\begin{center}
		\begin{tikzcd}
			& Z \arrow[ld, swap, "\varepsilon_1"] \arrow[d,dashed,"\varepsilon"] \arrow[dr, "\varepsilon_2"] & \\
			X_1 & P \arrow[l,"p_1"] \arrow[r, swap, "p_2"] & X_2
		\end{tikzcd}
	\end{center}
	In particular, this must be true for $Z = \complexNumbers$. In proposition \ref{proposition:psdii(C,Y) is isomorphic to D(Y)} we showed that $\psdii(\complexNumbers,Y) \cong \density(Y)$. By the \CJ isomorphism, we have that $\cptp(\complexNumbers,Y) \cong \density(Y)$. In this case, the isomorphism is just evaluation at $\id_\complexNumbers$:
	\begin{align*}
		\cptp(\complexNumbers,Y) &\longrightarrow \density(Y)\\
		\varepsilon &\longmapsto \varepsilon(\id_\complexNumbers).
	\end{align*}
	It follows that the universal property for $Z=\complexNumbers$ says the following: for any $\rho_1 \in \density(X_1)$ and $\rho_2 \in \density(X_2)$ there exists a unique $\rho \in \density(P)$ such that
	\begin{equation}
		p_1 (\rho) = \rho_1,
	\end{equation}
	\begin{equation}
		p_2 (\rho) = \rho_2.
	\end{equation}
	Notice that $\density(X_1) \neq \varnothing$ and $\density(X_2) \neq \varnothing$, so the universal property implies that $\density(P)$ must be non empty. This can only be the case if $d_P \geq 1$.
	
	We can combine both CPTP maps $p_1$ and $p_2$ into the $\complexNumbers$-linear map
	\begin{align*}
		(p_1 , p_2) \colon \Lin_\complexNumbers (P) &\longrightarrow \Lin_\complexNumbers (X_1)\times \Lin_\complexNumbers (X_2)\\
		A &\longmapsto (p_1(A), p_2(A)).
	\end{align*}
	Then the universal property for $Z = \complexNumbers$ says that the restriction
	\begin{equation}
		(p_1 , p_2) \colon \density(P) \longrightarrow \density(X_1)\times \density(X_2)
	\end{equation}
	is a bijection. Let's verify this last statement. The existence part says that, for any $\rho_1 \in \density(X_1)$ and $\rho_2 \in \density(X_2)$, there exists a $\rho \in \density(P)$ such that $p_1(\rho)= \rho_1$ and $p_2 (\rho) = \rho_2$. Therefore $(p_1 , p_2) (\rho) = (\rho_1, \rho_2)$. This implies that the restriction $(p_1 , p_2) \colon \density(P) \to \density(X_1) \times \density(X_2)$ is surjective. For the injective property, let $\rho, \rho' \in \density(P)$ be density operators such that $(p_1 , p_2) (\rho) = (p_1 , p_2) (\rho')$. This implies that
	\begin{equation}
		p_1(\rho) = p_1(\rho'),
	\end{equation}
	\begin{equation}
		p_2(\rho) = p_2(\rho').
	\end{equation}
	Define
	\begin{equation}
		\rho_1 = p_1(\rho),
	\end{equation}
	\begin{equation}
		\rho_2 = p_2(\rho).
	\end{equation}
	The previous equations says that we also have
	\begin{equation}
		\rho_1 = p_1(\rho'),
	\end{equation}
	\begin{equation}
		\rho_2 = p_2(\rho').
	\end{equation}
	The uniqueness part of the universal property then says that $\rho = \rho'$. This shows that the restriction $(p_1 , p_2) \colon \density(P) \to \density(X_1) \times \density(X_2)$ is injective, which concludes that it is a bijection. It is linear, so it is continuous. Since it is a continuous bijection between compact Hausdorff topological spaces, from General Topology we know that it is a homeomorphism. By corollary \ref{corollary:density ops dimension}, we know that this is a homeomorphism between topological manifolds with boundary which are homeomorphic to closed balls. In particular, these manifolds must have the same dimension. Using corollary \ref{corollary:density ops dimension}, we get that
	\begin{equation}
		\dim_\realNumbers \density(P) = d_P^2-1,
	\end{equation}
	and
	\begin{equation}
		\dim_{\realNumbers} (\density(X_1) \times \density(X_2)) = (d_{X_1}^2-1)+(d_{X_2}^2-1) = d_{X_1}^2 + d_{X_2}^2-2.
	\end{equation}
	The dimension of the manifolds must be equal, which implies that
	\begin{equation}
		d_P^2 = d_{X_1}^2 + d_{X_2}^2-1.
	\end{equation}
	This determines $P$ up to isomorphism, but still doesn't determine the projections $p_i$. Next we'll show that the product doesn't exists, except for trivial cases.
	
	Since $(p_1 , p_2) \colon \Lin_\complexNumbers (P) \to \Lin_\complexNumbers (X_1) \times \Lin_\complexNumbers (X_2)$ is a linear map which is also a bijection between the convex sets $\density(P)$ and $\density(X_1)\times \density(X_2)$, by proposition \ref{proposition:linear bijection between extreme points} it is a bijection between their extreme points. Since $(p_1 , p_2) \colon \density(P) \to \density(X_1) \times \density(X_2)$ is a homeomorphism, it restricts to a homeomorphism between the extreme points of $\density(P)$ and the extreme points of $\density(X_1)\times \density(X_2)$. By proposition \ref{proposition:extreme points of product of convex sets}, we have that
	\begin{align*}
		\text{extreme points of } &(\density(X_1)\times \density(X_2)) =\\ 
		&(\text{extreme points of } \density(X_1)) \times (\text{extreme points of } \density(X_2)).
	\end{align*}
	Also, by propositions \ref{proposition:extreme points of density operators are the pure states} and \ref{proposition:pure states is homeomorphic to CPn}, the set of extreme points of $\density(Y)$ is homeomorphic to $\complexNumbers P^{d_Y-1}$. Combining these results, we get a homeomorphism
	\begin{equation}
		\complexNumbers P^{d_P-1} \cong \complexNumbers P^{d_{X_1}-1} \times \complexNumbers P^{d_{X_2}-1}.
	\end{equation}
	As we'll see, this restricts the possible dimensions of $X_1$ and $X_2$. Topological invariants are preserved by homeomorphisms. They can be used to test if two topological spaces can be homeomorphic. One such topological invariant is the singular homology. The singular homology (with coefficients in $\integers$) of a spaces $X$ comprises of an infinite sequence $H_0 (X), H_1 (X), H_2 (X),\dots$ of abelian groups. The singular homology groups  of the complex projective space are (see application 5 of theorem 4.2 of chapter IV of \cite{masseySHT})
	\begin{equation}
		H_i (\complexNumbers P^n) \cong \begin{cases}
			\integers, &\text{ if } i \text{ is even and } 0\leq i \leq 2n,\\
			0 & \text{ if } i \text{ is odd or } i>2n.
		\end{cases}
	\end{equation}
	The homology group of a product of spaces is determined by the K\"unneth theorem (theorem 4 of chapter VI of \cite{masseySHT}). This theorem says that we have an isomorphism of abelian groups
	\begin{equation}
		H_k(X\times Y) \cong \bigoplus_{i+j=k} (H_i(X)\otimes H_j(Y)).
	\end{equation}
	In this expression, the tensor product is of abelian groups. We won't need to compute the whole homology of $ \complexNumbers P^{d_{X_1}-1} \times \complexNumbers P^{d_{X_2}-1}$, only one group will be enough. Suppose $d_{X_1} \geq 2$ and $d_{X_2} \geq 2$. Then we have
	\begin{equation}
		H_0 (\complexNumbers P^{d_{X_1}-1}) \cong \integers,
	\end{equation}
	\begin{equation}
		H_2 (\complexNumbers P^{d_{X_1}-1}) \cong \integers,
	\end{equation}
	\begin{equation}
		H_0 (\complexNumbers P^{d_{X_2}-1}) \cong \integers,
	\end{equation}
	\begin{equation}
		H_2 (\complexNumbers P^{d_{X_2}-1}) \cong \integers.
	\end{equation}
	By the K\"unneth theorem, we have
	\begin{align*}
		H_2 &(\complexNumbers P^{d_{X_1}-1} \times \complexNumbers P^{d_{X_2}-1}) \cong \bigoplus_{i+j=2} \left(H_i(\complexNumbers P^{d_{X_1}-1})\otimes H_j(\complexNumbers P^{d_{X_2}-1})\right)\\
		&\cong \left(H_0(\complexNumbers P^{d_{X_1}-1})\otimes H_2(\complexNumbers P^{d_{X_2}-1})\right)
		\oplus \left(H_2(\complexNumbers P^{d_{X_1}-1})\otimes H_0(\complexNumbers P^{d_{X_2}-1})\right)\\
		&\cong (\integers \otimes \integers) \oplus (\integers \otimes \integers)\\
		&\cong \integers \oplus \integers\\
		&= \integers^2.
	\end{align*}
	But $H_2 (\complexNumbers^{d_P-1})$ can only be isomorphic to $\{0\}$ or $\integers$, not $\integers^2$. Therefore, the binary product doesn't exists for $d_{X_1} \geq 2$ and $d_{X_2} \geq 2$.
	
	The previous results shows that a binary product can only exist if $d_{X_i} = 0$ or $d_{X_i} = 1$, for $i=1$ or $i=2$. In these cases the product indeed exists, as we'll show next. By symmetry, we can assume that $i=2$ and let $X_1$ have any dimension. The case $d_{X_2} = 1$ is already covered by Category Theory. This is because $\complexNumbers$ is a terminal object of the category of CPTP maps, and we always have a product between any object and a terminal object. In this case, we can take $P = X_1$, $p_1 = \id_{\Lin_\complexNumbers (X_1)}$ and $p_2 = \tr$. Recall that, as shown in the proof of proposition \ref{proposition:category cptp, initial and terminal objects}, the total trace is the unique arrow to the terminal object. Since $\complexNumbers$ is a terminal object, the universal property of the product reduces to the next statement: for any finite dimensional complex Hilbert space $Z$, for any $\varepsilon_1 \in \cptp(Z,X_1)$ there exists a unique $\varepsilon \in \cptp(Z,X_1)$ such that the diagram below commutes.
	\begin{center}
		\begin{tikzcd}
			& Z \arrow[ld,swap,"\varepsilon_1"] \arrow[d,dashed,"\varepsilon"] \arrow[rd,"\tr"]& \\
			X_1 & X_1 \arrow[l,"\id"] \arrow[r,swap,"\tr"] & \complexNumbers
		\end{tikzcd}
	\end{center}
	The right triangle commutes for any $\varepsilon$, because $\complexNumbers$ is a terminal object. The left triangle commutes iff $\varepsilon = \varepsilon_1$. This proves that $X_1$ is the binary product $X_1 \times \complexNumbers$ in the category of CPTP maps.
	
	Now consider the case $d_{X_2} = 0$, that is, $X_2 = \{0\}$. In this case we take $P = \{0\}$, $p_1 = 0$ and $p_2 = 0$. By proposition \ref{proposition:psdii(X,Y) for X or Y = 0}, if $Z \neq \{0\}$ then $\cptp(Z,\{0\}) = \varnothing$, so the universal property of the product is trivially valid for such $Z$. The only remaining case is  $Z = \{0\}$. By proposition \ref{proposition:category cptp, initial and terminal objects}, we know that $\{0\}$ is an initial object. Then $\cptp(\{0\},Y) = \{0\}$, so we only have to check the commutativity of the diagram below.
	\begin{center}
		\begin{tikzcd}
			& \{0\} \arrow[ld,swap,"0"] \arrow[d,"0"] \arrow[rd,"0"]& \\
			X_1 & \{0\} \arrow[l,"p_1 = 0"] \arrow[r,swap,"p_2 = 0"] & \{0\}
		\end{tikzcd}
	\end{center}
	This diagram only contains identically zero maps, so it trivially commutes. This proves that $\{0\}$ is a binary product $X_1 \times \{0\}$ in the category of CPTP maps. This concludes the proof.
\end{proof}

\section{Coproducts}

In theorem \ref{theorem:products in CPTP} we saw that the dimension of $\density(Y)$ was enough to determine the vertex of the product diagram, up to isomorphism. We used the topology of the extreme points to show that the product can't exist, unless for trivial cases. We can try to apply the same idea for other limits or colimits. We already know the dimension of the hom-sets $\cptp(X,Y)$, but we don't yet understand the topology of the set of its extreme points, or of the closure of the set of its extreme points. Still, we can use the dimension of $\cptp(X,Y)$ to determine the vertex of the coproduct diagram, if it exists. The category already has the initial object $\{0\}$. From Category Theory, we know that the category has the coproduct $X+\{0\}$, which can be taken as $X$. Next we'll consider the case where the objects have non zero dimension.

\begin{theorem}{}{}
	Let $X_1$ and $X_2$ be finite dimensional complex Hilbert spaces, with $d_{X_1} \geq 1$ and $d_{X_2} \geq 1$. If the coproduct $X_1 + X_2$ exists in the category of CPTP maps, its dimension must be
	\begin{equation}
		d_{X_1+X_2} = \sqrt{d_{X_1}^2+d_{X_2}^2}.
	\end{equation}
\end{theorem}
\begin{proof}
	Suppose the coproduct of $X_1$ and $X_2$ exists, and denote it by $C$. Let also $q_i \colon X_i \to C$, for $i\in \{1,2\}$, be the inclusions of the coproduct. The universal property of the coproduct says that for any finite dimensional complex Hilbert space $Z$, for any $\varepsilon_1 \in \cptp(X_1,Z) $ and $\varepsilon_2 \in \cptp(X_2,Z)$, there exists a unique $\varepsilon \in \cptp(C, Z)$ such that the diagram below commutes.
	\begin{center}
		\begin{tikzcd}
			& Z & \\
			X_1 \arrow[ur,"\varepsilon_1"] \arrow[r, swap, "q_1"] & C \arrow[u,dashed,"\varepsilon"] & X_2 \arrow[l,"q_2"] \arrow[ul,swap,"\varepsilon_2"]
		\end{tikzcd}
	\end{center}
	Define the $\complexNumbers$-linear map
	\begin{align*}
		T_Z \colon \Lin_\complexNumbers (\Lin_\complexNumbers (C), \Lin_\complexNumbers (Z)) &\longrightarrow \Pi_{i=1,2} \Lin_\complexNumbers (\Lin_\complexNumbers (X_i),\Lin_\complexNumbers (Z))\\
		\varepsilon &\longmapsto (\varepsilon q_1, \varepsilon q_2).
	\end{align*}
	The universal property is then equivalent to the statement that the restriction
	\begin{equation*}
		T_Z \colon \cptp(C,Z) \longrightarrow \cptp(X_1,Z) \times \cptp(X_2,Z)
	\end{equation*}
	is a bijection, for any $Z$. Since $T_Z$ is linear, it is continuous. Therefore, the restriction is a continuous bijection between compact Hausdorff spaces. From General Topology it follows that it must be a homeomorphism. Then, for any $Z$ we have a homeomorphism
	\begin{equation}
		\cptp(C,Z) \cong \cptp(X_1,Z) \times \cptp(X_2,Z).
	\end{equation}
	We are supposing that $X_1 \neq \{0\}$ and $X_2 \neq \{0\}$. In this case, by proposition \ref{proposition:psdii(X,Y) for X or Y = 0}, we know that $\cptp(X_i,\{0\}) = \varnothing$, so the homeomorphism becomes trivial for $Z = \{0\}$. For $Z \neq \{0\}$ the dimension of the hom-sets was given in corollary \ref{corollary:dimension of psdii}. Using this result, we get that
	\begin{equation}
		\dim_{\realNumbers} \cptp(C,Z) = d_C^2 (d_Z^2-1),
	\end{equation}
	and
	\begin{align*}
		\dim_{\realNumbers} (\cptp(X_1,Z) \times \cptp(X_2,Z)) &= d_{X_1}^2 (d_Z^2-1) + d_{X_2}^2 (d_Z^2-1) \\
		&= (d_{X_1}^2+d_{X_2}^2) (d_Z^2-1).
	\end{align*}
	Therefore, we must have
	\begin{equation}
		d_C^2 (d_Z^2-1) =  (d_{X_1}^2+d_{X_2}^2) (d_Z^2-1),
	\end{equation}
	for any $Z \neq \{0\}$. This can only be the case if
	\begin{equation}
		d^2_C = d_{X_1}^2+d_{X_2}^2.
	\end{equation}
\end{proof}

This result doesn't yet classify the coproducts, but we can already see that coproducts won't exist for an infinite number of pairs. This is because the dimension $d_{X_1+X_2}$ must be a natural number, which isn't the case for $\sqrt{d_{X_1}^2+d_{X_2}^2}$ for an infinite number of cases.

In the case of the product, we have products with $\{0\}$ and $\complexNumbers$. For coproducts, we have the coproduct with $\{0\}$, but we don't have the coproduct with $\complexNumbers$, except for $\{0\}+\complexNumbers$, which is $\complexNumbers$. We prove this in the next corollary.

\begin{corollary}{}{}
	Let $X$ be a finite dimensional complex Hilbert space, with $d_X \geq 1$. The coproduct $X+\complexNumbers$ doesn't exist.
\end{corollary}
\begin{proof}
	Suppose the coproduct exists and denote it by $C$. Then its dimension $d_C$ must satisfy the equation
	\begin{equation}
		d_C^2 = d_X^2+1.
	\end{equation}
	This implies that $d_C^2 > d_X^2$, and therefore $d_C > d_X$, which implies that $d_C \geq d_X+1$. Taking the square of both sides of $d_C \geq d_X+1$, we get
	\begin{equation}
		d_C^2 \geq (d_X+1)^2 = d_X^2+2d_X+1.
	\end{equation}
	But $d_C^2= d_X^2+1$ and $d_X \geq 1$, so the last inequality implies that
	\begin{equation}
		d_C^2 \geq d_X^2+2d_X+1 = d_C^2+2d_X \geq d_C^2+2.
	\end{equation}
	This can't be the case, therefore the coproduct $X + \complexNumbers$ doesn't exists.
\end{proof}

\cleardoublepage
\chapter{Openness of the Projection}

\label{chapter: projection properties}

In this chapter we'll continue the study of the function
\begin{equation*}
	\pi \colon S^{d_Y}_{d_X,r} \to \bigcup_{d_X / d_Y \leq k \leq r} \psdii(X,Y)_k
\end{equation*}
defined in section \ref{section: a surjective map from the purification}, now focusing on new results. We'll prove that $\pi$ is an open function, and moreover that it restricts to open functions. If we work with CPTP maps with a fixed Choi-rank, then $\pi$ gives a surjective submersion and we have a quotient of a smooth manifold by a compact Lie group. From Differential Geometry it is known that such function is open. The novelty is that we work with a variable Choi-rank.

If $\pi$ is open, then it should satisfy the following property. If $O$ is open, then $\pi(O)$ is open. Then, by continuity of $\pi$, $\pi^{-1} \pi(O)$ is also open. We'll prove in lemma \ref{lemma:pi-1pi(U) is open} that if $O$ is open then $\pi^{-1}\pi(O)$ is also open. Then we'll use this result to prove in theorem \ref{theorem:pi is open} that $\pi$ is an open function. We'll prove these results in a little more generality, replacing the codomain of $\pi$ by a possibly smaller subset $Z$. 

\begin{lemma}{}{pi-1pi(U) is open}
	Let $\pi \colon S_{d_X,r}^{d_Y} \to \bigcup_{ d_X / d_Y \leq k \leq r} \psdii(X,Y)_k$ be the projection defined in section \ref{section: a surjective map from the purification}. Let $Z \subseteq \bigcup_{d_X / d_Y \leq k \leq r} \psdii(X,Y)_k$, then we have the restriction of $\pi$ to $\pi^{-1}(Z)$, which gives a continuous surjective function $\pi \colon \pi^{-1}(Z) \to Z$. For any open subset $O \subseteq \pi^{-1}(Z)$, the set $\pi^{-1} \pi (O)$ is an open subset of $\pi^{-1}(Z)$.
\end{lemma}
\begin{proof}
	$\pi^{-1} \pi (O)$ is the set of points in $\pi^{-1}(Z)$ that are related to some point in $O$. By proposition \ref{proposition:relation of pi}, this relation is given by multiplication by a unitary operator $V^\dag \in \U(r)$. Define the function
	\begin{align*}
		\mu_V \colon M_{d_X,r}^{d_Y}(\complexNumbers) &\longrightarrow M_{d_X,r}^{d_Y}(\complexNumbers)\\
		(\psi_j)_j &\longmapsto (\psi_j V^\dag)_j.
	\end{align*}
	Multiplication of matrices is $\complexNumbers$-bilinear, so $\mu_V$ is $\complexNumbers$-linear. Also, it is an isomorphism with inverse $\mu_V^{-1} = \mu_{V^{-1}}$. Since it is linear, it is continuous, so $\mu_V$ is a homeomorphism. Of course, $\pi^{-1}(Z)$ is invariant by multiplication by $V^\dag$ on the right, since this is the relation of $\pi$. Therefore, $\mu_V$ restricts to a homeomorphism
	\begin{equation*}
		\mu_V \colon \pi^{-1}(Z) \to \pi^{-1}(Z). 
	\end{equation*}
	We can use this function to express $\pi^{-1} \pi (O)$ as an union of open sets. We have that
	\begin{align*}
		\pi^{-1} \pi (O) &= \{ (\psi_j V^\dag)_j \mid (\psi_j )_j \in O, \ V \in \U(r) \}\\
		&= \bigcup_{V \in \U(r)} \{ (\psi_j V^\dag)_j \mid (\psi_j )_j \in O \}\\
		&= \bigcup_{V \in \U(r)} \mu_V (O).
	\end{align*}
	Since $\mu_{V^\dag} \colon \pi^{-1}(Z) \to \pi^{-1}(Z)$ is a homeomorphism and $O$ is an open subset of $\pi^{-1}(Z)$, then $\mu_V (O)$ is also an open subset of $\pi^{-1}(Z)$. An union of open sets is open, so $\pi^{-1} \pi (O)$ is an open subset of $\pi^{-1}(Z)$.
\end{proof}

Next we prove that $\pi \colon S_{d_X,r}^{d_Y} \to \bigcup_{d_X / d_Y \leq k \leq r} \psdii(X,Y)_k$ is an open function. Up to now, we assumed in this section that $r$ can be any natural number $r\geq d_X / d_Y$. In the next proposition we suppose that $r \leq d_X d_Y$, so that we can use the polar decomposition of proposition \ref{proposition:polar decomposition standardized}. By proposition \ref{proposition:general CPTP rank bounds}, we know that any $A \in \psdii(X,Y)$ can't have a rank bigger than $d_X d_Y$, so we are not losing generality by taking $r \leq d_X d_Y$.

\begin{theorem}{}{pi is open}
	Let $\pi \colon S_{d_X,r}^{d_Y} \to \bigcup_{d_X / d_Y\leq k \leq r} \psdii(X,Y)_k$ be the projection defined in section \ref{section: a surjective map from the purification}. We suppose here that $d_X / d_Y \leq r \leq d_X d_Y$. Let also $Z \subseteq \bigcup_{d_X / d_Y \leq k \leq r} \psdii(X,Y)_k$ be any subset. Then $\pi$ restricts to a continuous surjection $\pi \colon \pi^{-1}(Z) \to Z$. This restriction is an open function, that is, it sends open sets to open sets.
\end{theorem}
\begin{proof}
	First we discussion how this proof differs from previous works. The proof itself begins in the second paragraph. If we were working with a fixed rank $k$, and took $Z = \psdii(X,Y)_k$, this proposition would amount to the fact that $\psdii(X,Y)_k$ is homeomorphic to a quotient of manifolds $V_{d_X,d_Y,k}/\U(k)$, as proved in lemma 20 of \cite{iten_colbeck}, where $V_{d_X,d_Y,k}$ is an open subset of a Stiefel manifold. Then our restricted projection $\pi \colon \pi^{-1}(Z) \to Z$ would factor through the projection $V_{d_X,d_Y,k}\to V_{d_X,d_Y,k}/\U(k)$, which is an open function, and this would imply that the restricted $\pi$ is open. Instead, we are allowing any rank $k \leq r$. In this case, $\bigcup_{d_X / d_Y \leq k \leq r} \psdii(X,Y)_k$ still corresponds to the quotient $S^{d_Y}_{d_X,r}/\U(r)$, because of the relation of $\pi$, which was studied in proposition \ref{proposition:relation of pi}. As proved in proposition \ref{proposition:pre-image of pi is stiefel manifold}, the set $S^{d_Y}_{d_X,r}$ is another way to present a Stiefel manifold. Therefore we still have a quotient of a smooth manifold by a compact Lie group, but this quotient isn't well behaved and we can't use the Quotient Manifold Theorem (theorem 21.10 of \cite{lee2012introduction}). More precisely, the action is smooth and proper, but not free. It is smooth because it consists of the multiplication of matrices $\psi_j V^\dag$, where $(\psi_j)_j \in S^{d_Y}_{d_X,r}$ and $V\in \U(r)$. It is also proper because  $\U(r)$ is compact. But it fails to be free due to the excessive redundancy in the representation of an operator $A \in \psdii(X,Y)_k$ by a representative $\psi \in S^{d_Y}_{d_X,r}$ for $k<r$. Nevertheless, it is still possible to prove that $\pi$ is an open function, but it becomes harder due to the operators not having a fixed rank. We also add more generality by proving that $\pi \colon \pi^{-1}(Z) \to Z$ is also open for any $Z$, which will be convenient later.
	
	Now we begin the proof of the proposition. Let $O\subseteq \pi^{-1}(Z)$ be an open subset. We want to prove that $\pi(O)$ is an open subset of $Z$. In lemma \ref{lemma:pi-1pi(U) is open} we proved that $\pi^{-1} \pi(O)$ is also open in $\pi^{-1}(Z)$. The function $\pi$ is surjective, so
	\begin{equation}
		\pi(O) = \pi (\pi^{-1} \pi (O)).
	\end{equation}
	Therefore, we can instead prove that $\pi (\pi^{-1} \pi (O))$ is open. Let $A \in \pi (\pi^{-1} \pi (O))$ and $B \in Z$ such that
	\begin{equation}
		||B-A|| < \delta,
	\end{equation}
	for some $\delta>0$ to be determined later. We want to find a $\delta$ such that necessarily $B \in \pi (\pi^{-1} \pi (O))$. Since $A \in \pi (\pi^{-1} \pi (O))$, there exists $\psi = (\psi_j)_j \in \pi^{-1} \pi (O)$ such that $A = \pi(\psi)$. Since $B \in Z$ and $\pi$ is surjective, there exists $\varphi = (\varphi_j)_j \in \pi^{-1}(Z)$ such that $B = \pi(\varphi)$. Instead of working with a sequence of matrices, it is more convenient to stack them into a single matrix, as done earlier in this section. We define
	\begin{equation}
		\Psi \coloneqq \begin{pmatrix}
			\psi_1 \\
			\vdots\\
			\psi_{d_Y}
		\end{pmatrix} \in M_{d_X d_Y,r}(\complexNumbers),
	\end{equation}
	\begin{equation}
		\Phi \coloneqq \begin{pmatrix}
			\varphi_1 \\
			\vdots\\
			\varphi_{d_Y}
		\end{pmatrix} \in M_{d_X d_Y,r}(\complexNumbers).
	\end{equation}
	As shown earlier in this section, we can order the indices $(i,j)$ of the basis $\ket{x^i} \otimes \ket{y_j}$ in the colexicographic order, and then the matrices $[A]$ and $[B]$ are expressed with respect to the basis $\ket{x^i}\bra{x^{i'}} \otimes \ket{y_j}\bra{y_{j'}}$ as
	\begin{equation}
		\label{equation: [A] = PsiPsidag}
		[A] = \Psi \Psi^\dag,
	\end{equation}
	\begin{equation}
		\label{equation: [B] = PhiPhidag}
		[B] = \Phi \Phi^\dag.
	\end{equation}
	The representatives $\Psi$ and $\Phi$ are not unique. We want to find another representatives that are close to each other. Then we'll use that $\pi^{-1} \pi (O)$ is open to prove that the new representative of $B$ is in $\pi^{-1} \pi (O)$. As proved in proposition \ref{proposition:relation of pi}, another representative is obtained by multiplying the representative on the right by a unitary matrix $V^\dag \in \U(r)$. For this reason, it is convenient to take a polar decomposition of $\Psi$ and $\Phi$. These matrices are in $M_{d_X d_Y, r}(\complexNumbers)$, with $r \leq d_X d_Y$, so the number of columns is at most the number of lines. For this reason, the positive semidefinite factor $P$ is a polar decomposition $\Psi = PU$ isn't unique. But we proved in proposition \ref{proposition:polar decomposition standardized}  that we can always take $P = (\Psi \Psi^\dag)^{1/2}$. From this proposition we get polar decompositions
	\begin{equation}
		\Psi = (\Psi \Psi^\dag)^{1/2} U,
	\end{equation}
	\begin{equation}
		\Phi = (\Phi \Phi^\dag)^{1/2} W,
	\end{equation}
	where $U,W \in V_{d_X d_Y, r} (\complexNumbers)$ are isometries. The proposition also shows that we can take $U$ and $W$ such that
	\begin{equation}
		\label{equation: image Psi subseteq image U}
		\image \Psi \subseteq \image U,
	\end{equation}
	\begin{equation}
		\image \Phi \subseteq \image W.
	\end{equation}
	By equations \ref{equation: [A] = PsiPsidag} and \ref{equation: [B] = PhiPhidag}, the positive semidefinite factors are $[A]^{1/2}$ and $[B]^{1/2}$, so we have
	\begin{equation}
		\Psi = [A]^{1/2} U,
	\end{equation}
	\begin{equation}
		\Phi = [B]^{1/2} W.
	\end{equation}
	We proved in proposition \ref{proposition:square root of psd operators is continuous} that the operation of taking the square root of positive semidefinite operators is continuous. Therefore, if $A$ and $B$ are close enough, the same is true for $[A]^{1/2}$ and $[B]^{1/2}$. We won't try in this proof to replace $U$ and $W$ with matrices which are close to each other. This is due to the fact proven in proposition \ref{proposition:polar decomposition standardized} that the relevant parts of these matrices are the products $P_{\image \Psi} U$ and $P_{\image \Phi} W$, where $P_{S}$ denotes the orthogonal projection onto $S$. For any matrix $T$, both $T$ and $T T^\dag$ have the same rank. Of course, the image of $T T^\dag$ is contained in the image of $T$. Since they have the same rank, the images must be equal, that is,
	\begin{equation}
		\image (T T^\dag) = \image T.
	\end{equation}
	Applying this to the matrices $\Psi$ and $\Phi$, and using that $[A] = \Psi \Psi^\dag$ and $[B] = \Phi \Phi^\dag$, it follows that
	\begin{equation}
		\label{equation: image Psi = image [A]}
		\image \Psi = \image [A],
	\end{equation}
	\begin{equation}
		\image \Phi = \image [B].
	\end{equation}
	As in proposition \ref{proposition:polar decomposition standardized}, we can insert $P_{\image [A]}$ and $P_{\image [B]}$ in the polar decompositions, from which we get
	\begin{equation}
		\Psi = [A]^{1/2} P_{\image [A]} U,
	\end{equation}
	\begin{equation}
		\Phi = [B]^{1/2} P_{\image [B]} W.
	\end{equation}
	
	Before proceeding is important to have some information about the rank of $[B]$ and which columns are linearly independent. Let
	\begin{equation}
		r_A \coloneqq \rank(A),
	\end{equation}
	\begin{equation}
		r_B \coloneqq \rank(B).
	\end{equation}
	Then $[A]$ has $r_A$ columns which are linearly independent. Let $j_1$, \dots, $j_{r_A}$ be the indices of these columns.
	Write $[A]$ and $[B]$ as
	\begin{equation}
		[A] = \begin{pmatrix}
			a_1 & \dots & a_r
		\end{pmatrix},
	\end{equation}
	\begin{equation}
		[B] = \begin{pmatrix}
			b_1 & \dots & b_r
		\end{pmatrix},
	\end{equation}
	where $a_1$, \dots, $a_r$ are the columns of $[A]$ and $b_1$, \dots, $b_r$ are the columns of $[B]$. Therefore $(a_{j_k})_{k=1,\dots,r_A}$ is linearly independent. Linearly independence is stable by small perturbations, so there exists a $\delta_1 > 0$ such that if $||B-A||< \delta_1$ then the columns $(b_{j_k})_{k=1,\dots,r_A}$ are also linearly independent. The existence of such $\delta_1$ can also be proved using proposition \ref{proposition:gram matrix}, and using that the determinant of the Gram matrix of $(b_{j_k})_{k=1,\dots,r_A}$ is a continuous function of $[B]$. The linear independence of  $(b_{j_k})_{k=1,\dots,r_A}$ then implies that
	\begin{equation}
		r_A \leq r_B.
	\end{equation}
	If $r_B > r_A$, take more columns of $[B]$ and extend the sequence $(b_{j_k})_{k=1,\dots,r_A}$ to a bigger sequence $(b_{j_k})_{k=1,\dots,r_B}$ of linearly independent columns. Now we have columns of $[A]$ and $[B]$ which are basis for their images.
	
	Next we use the columns $(a_{j_k})_{k=1,\dots,r_A}$ and $(b_{j_k})_{k=1,\dots,r_B}$ to construct new representatives for $[A]$ and $[B]$. Multiplying the isometry $U$ by a unitary matrix $V_1^\dag \in \U(r)$, obtaining $U V_1^\dag$, gives any other isometry with the same image, as was proved in proposition \ref{proposition:isometries are equivalent if they have the same image}. From equations \ref{equation: image Psi subseteq image U} and \ref{equation: image Psi = image [A]}, we have that $\image [A] \subseteq \image U$, so we'll choose $V_1$ such that the first $r_A$ columns of $UV_1^\dag$ are a basis for $\image [A]$. Since $U V_1^\dag$ is an isometry, its columns must orthonormal. We have the basis $(a_{j_k})_{k=1,\dots,r_A}$ for $\image [A]$, but they may not be orthonormal. But we can obtain orthonormal vectors by the Gram-Schmidt process. Recall that if $(v_1,\dots,v_n)$ is a sequence of linearly independent vectors, then the Gram-Schmidt process constructs orthonormal vectors $(v'_1,\dots, v'_n)$ defined recursively by
	\begin{equation}
		v'_k \coloneqq \frac{N_k}{||N_k||},
	\end{equation}
	where
	\begin{equation}
		N_1 \coloneqq v_1,
	\end{equation}
	\begin{equation}
		N_{k+1} \coloneqq v_{k+1} - \sum_{l=1}^k \langle v'_l, v_{k+1} \rangle v'_l .
	\end{equation}
	Notice that the vectors $v'_k$ depend continuously on $(v_1, \dots, v_n)$. Applying the Gram-Schmidt process to the vectors $(a_{j_k})_{k=1,\dots,r_A}$ we get an orthonormal basis $(u_k)_{k=1,\dots,r_A}$ for $\image [A]$, and applying the process for $(b_{j_k})_{k=1,\dots,r_B}$ we get an orthonormal basis $(w_k)_{k=1,\dots,r_B}$ for $\image [B]$. Then we take $V_1, V_2 \in \U(r)$ such that the first $r_A$ columns of $U V_1^\dag$ are $u_1$, \dots, $u_{r_A}$ and the first $r_B$ columns of $W V_2^\dag$ are $w_1, \dots, w_{r_B}$. That is, they take the form
	\begin{equation}
		U V_1^\dag = \begin{pmatrix}
			u_1 & \dots & u_{r_A} & U'
		\end{pmatrix},
	\end{equation}
	\begin{equation}
		W V_2^\dag = \begin{pmatrix}
			w_1 & \dots & w_{r_B} & W'
		\end{pmatrix},
	\end{equation}
	where $U' \in V_{d_X d_Y, r-r_A}(\complexNumbers)$ and $W' \in V_{d_X d_Y, r-r_B}(\complexNumbers)$ are smaller isometries. We can discard the matrices $U'$ and $W'$ by multiplying them by $P_{\image [A]}$ and $P_{\image [B]}$. In fact, write
	\begin{equation}
		U' = \begin{pmatrix}
			u_{r_{A}+1} & \dots & u_r
		\end{pmatrix},
	\end{equation}
	where $u_{r_{A}+1}$, \dots, $u_r$ are the columns of $U'$. Then we have that
	\begin{equation}
		P_{\image [A]} U' = \begin{pmatrix}
			P_{\image [A]} u_{r_{A}+1} & \dots & P_{\image [A]} u_r
		\end{pmatrix}.
	\end{equation}
	Since $U V_1^\dag$ is an isometry, the columns of $U'$ are orthogonal to the columns $(u_k)_{k=1,\dots,r_A}$, and these are a basis for $\image [A]$. Therefore
	\begin{equation}
		P_{\image [A]} u_{r_{A}+1} = \dots = P_{\image [A]} u_r = 0,
	\end{equation}
	so
	\begin{equation}
		P_{\image [A]} U' = 0.
	\end{equation}
	This proves that
	\begin{equation}
		P_{\image [A]} U V_1^\dag = \begin{pmatrix}
			u_1 & \dots & u_{r_A} & 0
		\end{pmatrix},
	\end{equation}
	where 0 is the zero matrix in $M_{d_X d_Y, r-r_A}(\complexNumbers)$. Similarly, we have that
	\begin{equation}
		P_{\image [B]} W V_2^\dag = \begin{pmatrix}
			w_1 & \dots & w_{r_B} & 0
		\end{pmatrix},
	\end{equation}
	where 0 is the zero matrix in $M_{d_X d_Y, r-r_B}(\complexNumbers)$. With this, we have new representatives
	\begin{equation}
		\Psi V_1^\dag = [A]^{1/2} \begin{pmatrix}
			u_1 & \dots & u_{r_A} & 0
		\end{pmatrix},
	\end{equation}
	\begin{equation}
		\Phi V_2^\dag = [B]^{1/2} \begin{pmatrix}
			w_1 & \dots & w_{r_B} & 0
		\end{pmatrix}.
	\end{equation}
	
	Next we want to show that these representatives are close to each other. We break this problem in two parts, one of comparing the matrices $[A]$ and $[B]$, and another of comparing the columns $u_k$ and $w_k$. This is done by summing and subtracting a matrix and using the triangle inequality. We can estimate the distance of the representatives as follows
	\begin{align*}
		||\Psi V_1^\dag - \Phi V_2^\dag|| &= ||[A]^{1/2} \begin{pmatrix}
			u_1 & \dots & u_{r_A} & 0
		\end{pmatrix} -   [B]^{1/2} \begin{pmatrix}
			w_1 & \dots & w_{r_B} & 0
		\end{pmatrix} || \\
		&= ||[A]^{1/2} \begin{pmatrix}
			u_1 & \dots & u_{r_A} & 0
		\end{pmatrix} - [A]^{1/2} \begin{pmatrix}
			w_1 & \dots & w_{r_B} & 0
		\end{pmatrix} \\
		& \quad +[A]^{1/2} \begin{pmatrix}
			w_1 & \dots & w_{r_B} & 0
		\end{pmatrix} -   [B]^{1/2} \begin{pmatrix}
			w_1 & \dots & w_{r_B} & 0
		\end{pmatrix} || \\
		&\leq ||[A]^{1/2} \left\{\begin{pmatrix}
			u_1 & \dots & u_{r_A} & 0
		\end{pmatrix} -  \begin{pmatrix}
			w_1 & \dots & w_{r_B} & 0
		\end{pmatrix}\right\} || \\
		& \quad +||([A]^{1/2}-[B]^{1/2}) \begin{pmatrix}
			w_1 & \dots & w_{r_B} & 0
		\end{pmatrix}||.
	\end{align*}
	Then we have to estimate two terms:
	\begin{equation}
		T_1 \coloneqq ||[A]^{1/2} \left\{\begin{pmatrix}
			u_1 & \dots & u_{r_A} & 0
		\end{pmatrix} - \begin{pmatrix}
			w_1 & \dots & w_{r_B} & 0
		\end{pmatrix}\right\} ||,
	\end{equation}
	\begin{equation}
		T_2 \coloneqq ||([A]^{1/2}-[B]^{1/2}) \begin{pmatrix}
			w_1 & \dots & w_{r_B} & 0
		\end{pmatrix}||.
	\end{equation}
	
	The first term $T_1$ is the least obvious that should be small, because we may have $r_B > r_A$. We'll show why $T_1$ is small, and then the rest of the proof consists of a series of simple, but tedious, arguments to prove that the representatives are close to each other. By the continuity of the Gram-Schmidt process, for any $\varepsilon>0$ there exists $\delta_2 >0$, with $\delta_2 \leq \delta_1$, such that if $||B-A||< \delta_2$ then
	\begin{equation}
		||u_k-w_k|| < \varepsilon,
	\end{equation}
	for $1 \leq k \leq r_A$.
	The vectors $(u_k)_{1\leq k \leq r_A}$ are a basis for $\image [A]$, so we also have that $w_{r_A+1}$, \dots, $w_{r_B}$ are almost orthogonal to $\image [A]$. Therefore $P_{\image [A]} w_k$ is close to 0. More precisely, let $1\leq k \leq r_A$ and $r_A +1 \leq k' \leq r_B$. Using that $\langle w_k , w_{k'} \rangle =0$, the Cauchy-Schwarz inequality and that $||w_{k'}|| = 1$, we get that
	\begin{align*}
		|\langle u_k, w_{k'} \rangle| &= |\langle u_k - w_k, w_{k'} \rangle|\\
		&\leq ||u_k-w_k|| \, ||w_{k'}||\\
		&= ||u_k-w_k||\\
		&< \varepsilon.
	\end{align*}
	The vector $P_{\image [A]} w_{k'}$ can be expressed as
	\begin{equation}
		P_{\image [A]} w_{k'} =\sum_{k=1}^{r_A} \langle u_k, w_{k'} \rangle u_k.
	\end{equation}
	Using the triangle inequality, we can estimate its norm as
	\begin{align*}
		||P_{\image [A]} w_{k'}|| &= || \sum_{k=1}^{r_A} \langle u_k, w_{k'} \rangle u_k ||\\
		&\leq \sum_{k=1}^{r_A}  |\langle u_k, w_{k'} \rangle| \ ||u_k ||\\
		&= \sum_{k=1}^{r_A}  |\langle u_k, w_{k'} \rangle|\\
		&< r_A  \varepsilon.
	\end{align*}
	For this reason, even though $w_{k'}$ can't be chosen to be small, it will only contribute a small amount to $T_1$. From this we get that
	\begin{align*}
		||&P_{\image [A]} \left\{\begin{pmatrix}
			u_1 & \dots & u_{r_A} & 0
		\end{pmatrix} - \begin{pmatrix}
			w_1 & \dots & w_{r_B} & 0
		\end{pmatrix}\right\} ||^2 = \\
		&=||\begin{pmatrix}
			P_{\image [A]}(u_1-w_1) & \dots & P_{\image [A]}(u_{r_A}-w_{r_B}) & -P_{\image [A]} w_{r_A} & \dots 
		\end{pmatrix} ||^2 \\
		&= \sum_{k=1}^{r_A} ||P_{\image [A]}(u_k-w_k) ||^2 +\sum_{k'=r_A+1}^{r_B} ||P_{\image [A]}w_{k'}||^2\\
		&\leq \sum_{k=1}^{r_A} ||u_k-w_k||^2 +\sum_{k'=r_A+1}^{r_B} ||P_{\image [A]}w_{k'}||^2\\
		&< \sum_{k=1}^{r_A} \varepsilon^2 + \sum_{k'=r_A+1}^{r_B} (r_A \varepsilon)^2\\
		&= r_A \varepsilon^2 + (r_B-r_A) (r_A \varepsilon)^2\\
		&= (r_A+(r_B-r_A)r_A^2) \varepsilon^2,
	\end{align*}
	so
	\begin{align*}
		||P_{\image [A]} \left\{\begin{pmatrix}
			u_1 & \dots & u_{r_A} & 0
		\end{pmatrix} - \begin{pmatrix}
			w_1 & \dots & w_{r_B} & 0
		\end{pmatrix}\right\} || &<\\ &\sqrt{r_A+(r_B-r_A)r_A^2} \ \varepsilon .
	\end{align*}
	The term $T_1$ can be rewritten with a projection in the middle, as
	\begin{equation}
		T_1 \coloneqq ||[A]^{1/2} P_{\image [A]} \left\{\begin{pmatrix}
			u_1 & \dots & u_{r_A} & 0
		\end{pmatrix} - \begin{pmatrix}
			w_1 & \dots & w_{r_B} & 0
		\end{pmatrix}\right\} ||.
	\end{equation}
	We have shown that if $||B-A||$ is small enough then the matrix product to the right of $[A]^{1/2}$ is arbitrarily small. Since matrix multiplication and the norm are continuous functions, it follows that, for any $\varepsilon > 0$, there exists some $\delta_3>0$, with $\delta_3 \leq \delta_2$, such that if $||B-A||<\delta_3 $ then
	\begin{equation}
		T_1 < \varepsilon.
	\end{equation}
	
	Next we want to show that $T_2$ is small. First we analyze the matrix $[A]^{1/2}-[B]^{1/2}$. It was proved in proposition \ref{proposition:square root of psd operators is continuous} that the square root of positive semidefinite operators is a uniformly continuous function. We proved there that
	\begin{equation}
		||[A]^{1/2}-[B]^{1/2}|| \leq \sqrt{d_X d_Y} ||A-B||^{1/2}.
	\end{equation}
	For any $\varepsilon> 0 $, let
	\begin{equation}
		\delta_4 \coloneqq \frac{\varepsilon^2}{d_X d_Y}.
	\end{equation}
	If $||B-A|| < \delta_4$, then
	\begin{equation}
		||[A]^{1/2}-[B]^{1/2}|| \leq \sqrt{d_X d_Y} ||B-A||^{1/2} < \sqrt{d_X d_Y} \frac{\varepsilon}{\sqrt{d_X d_Y}} = \varepsilon.
	\end{equation}
	The other matrix $\begin{pmatrix}
		w_1 & \dots & w_{r_B} & 0
	\end{pmatrix}$ have orthonormal columns $w_1$, \dots, $w_{r_B}$, and the other columns are zero. In particular, this matrix is bounded. Then $T_2$ is the norm of the product of the matrix $[A]^{1/2} - [B]^{1/2}$, which is arbitrarily small, with another that is bounded. Since matrix multiplication and the norm are continuous functions, for any $\varepsilon>0$ there exists some $\delta_5 >0$, with $\delta_5 \leq \delta_4$, such that if $||B-A|| < \delta_5$ then
	\begin{equation}
		T_2 < \varepsilon.
	\end{equation}
	
	Taking $\delta = \min\{\delta_3, \delta_5\}$, we have that if $||B-A||< \delta$ then
	\begin{equation}
		||\Psi V_1^\dag - \Phi V_2^\dag|| < 2 \varepsilon,
	\end{equation}
	for an arbitrary $\varepsilon> 0$. The sequences of matrices $(\psi_j V_1^\dag)_j$ and $(\varphi_j V_2^\dag)_j$ are in $S^{d_Y}_{d_X,r}$, and by proposition \ref{proposition:relation of pi} they are other representatives for $A$ and $B$. That is,
	\begin{equation}
		A = \pi((\psi_j V_1^\dag)_j),
	\end{equation}
	\begin{equation}
		B = \pi((\varphi_j V_2^\dag)_j).
	\end{equation}
	Since $(\psi_j)_j \in \pi^{-1}\pi(O)$, the same is true for other equivalent representatives. Therefore
	\begin{equation}
		(\psi_j V_1^\dag)_j \in \pi^{-1} \pi (O).
	\end{equation}
	By lemma \ref{lemma:pi-1pi(U) is open}, and since $O$ is open in $\pi^{-1}(Z)$, we know that $\pi^{-1} \pi (O)$ is also open in $\pi^{-1}(Z)$. Therefore, there exists some $\eta>0$ such that if $(\gamma_j)_j \in \pi^{-1}(Z)$ and $||(\gamma_j)_j - (\psi_j V_1^\dag)_j|| < \eta$, then $(\gamma_j)_j \in \pi^{-1}\pi(O)$. Take $\varepsilon = \eta/3$, then it follows from the previous results that if $||B-A||< \delta$ then $(\varphi_j V_2^\dag)_j \in \pi^{-1}\pi(O)$. Since $B = \pi((\varphi_j V_2^\dag)_j)$, this implies that
	\begin{equation}
		B \in \pi \pi^{-1} \pi(O) = \pi(O).
	\end{equation}
	We proved that for any $A \in \pi(O)$ there exists $\delta>0$ such that
	\begin{equation}
		Z \cap B(A,\delta) \subseteq \pi(O).
	\end{equation}
	This proves that $\pi(O)$ is open in $Z$, and therefore $\pi \colon \pi^{-1}(Z) \to Z$ is an open function.
\end{proof}

\cleardoublepage
\chapter{Cells for Closure of Extreme Channels}
\label{chapter: cells extreme cptp}

This chapter assumes knowledge of Algebraic Topology, specially CW structures. It is recommended to take postgraduate courses on the subject. The book Singular Homology Theory from \cite{masseySHT} is being used as reference. This book defines CW decomposition in section 3 of chapter 4. A CW structure is a type of decomposition of a topological space, in which it is expressed as a gluing of balls. It is useful to compute topological invariants, such as singular homology. These topological invariants serve as constraints for the existence of solutions to problems. In particular, they can help classify the limits and colimits of the category of CPTP maps, but are not limited to problems from Category Theory.

We are interested in understanding the geometry and topology of the set of extreme points of $\cptp(X,Y)$, or of the closure of the set of extreme points. By the \CJ isomorphism, it is equivalent to study $\psdii(X,Y)$. Understanding the closure of the set of extreme points is also posed as an open problem of Quantum Information Theory in problem 1 of \cite{ruskai2007open}. By theorem 8.1 of \cite{FRIEDLAND2016553}, we know that the closure of the set of extreme points is a real semi-algebraic set (with $\realNumbers$ as the real closed field). We also showed in proposition \ref{proposition:cptp is compact convex} that $\psdii(X,Y) \subseteq \overline{B(0,d_X)}$, from which it follows that the closure of the set of extreme points is bounded. Therefore, the closure of the set of extreme points is a closed, bounded and semi-algebraic set. Then, by theorem 5.43 of \cite{algorithms_in_RAG}, there exists a triangulation of this set. In particular, a CW decomposition exists. We want to find a CW decomposition of the closure of the set of extreme points. In this chapter we show some progress we had on this task. The cases where $X = \{0\}$ or $Y = \{0\}$ are trivial, so we assume that $d_X \geq 1$ and $d_Y \geq 1$. We'll also fix orthonormal basis $(x^i)_i$ and $(y_j)_j$ for $X^*$ and $Y$, respectively.

The idea is to use that $\psdii(\complexNumbers, Y)$ is homeomorphic to the complex projective space $\complexNumbers P^{d_Y - 1}$. Then we think of $\psdii(X,Y)_r$ as a generalization of the complex projective space in which we replace complex numbers by complex matrices. We can obtain some cells for the closure of the set of extreme points of $\psdii(X,Y)$ by adapting the techniques used to obtain a CW decomposition of the complex projective space. We succeeded in obtaining cells in this way, but we only got a partial decomposition of the space. Further work is needed to try to complete it to a full CW decomposition of the closure of the set of extreme points.

\section{Mimicking the complex projective space}
\label{section: partial decomposition into cells}

From corollary \ref{corollary:extreme points of psdii(C,Y) = CPn} we know that the set of extreme points of $\psdii(\complexNumbers,Y)$ is closed and is homeomorphic to $\complexNumbers P^{d_Y-1}$. A CW decomposition of the complex projective spaces is already known (see example 3.6 of chapter IV of \cite{masseySHT}), so we could try to adapt the ideas used for them to find a CW decomposition of the closure of the set of extreme points of $\psdii(X,Y)$. We had some success in this direction, having found some characteristic maps that resembles the ones for the complex projective spaces. Unfortunately, we still didn't find enough characteristic maps to have a CW decomposition. For the definition of characteristic map, see section 2 of chapter 4 of \cite{masseySHT}.

As we saw in proposition \ref{proposition:closure of extreme points of cptp(X,Y)}, the closure of the set of extreme points of $\psdii(X,Y)$ is $\bigcup_{d_X / d_Y \leq k\leq d_X}\psdii(X,Y)_k$, that is, the set of operators $A \in \psdii(X,Y)$ with $\rank A \leq d_X$. We also have the projection
\begin{align*}
	\pi \colon S^{d_Y}_{d_X,r} &\longrightarrow \bigcup_{d_X / d_Y \leq k\leq r} \psdii(X,Y)_k\\
	(\psi_j)_j &\longmapsto \sum_{i,i',j,j'} (\psi_j \psi_{j'}^\dag)_{i,i'} \ket{x^i}\bra{x^{i'}} \otimes \ket{y_j} \bra{y_{j'}},
\end{align*}
defined in section \ref{section: a surjective map from the purification}. Also, recall that $S^{d_Y}_{d_X,r}$ is a generalization of the sphere, defined as
\begin{equation}
	S^{d_Y}_{d_X,r} = \left\{(\psi_j)_j \in M_{d_X,r}(\complexNumbers)^{d_Y} \ \bigg| \ \sum_j \psi_j \psi_j^\dag = I_{d_X}\right\}.
\end{equation}
We'll only be interested in the case where $r=d_X$, which is the one that gives a projection onto the closure of the set of extreme points of $\psdii(X,Y)$. As we'll see, this projection has some similarities with the one for complex projective spaces.

The projection for the complex projective space $\complexNumbers P^{n-1}$ is
\begin{align*}
	\pi \colon S^{2n-1} &\longrightarrow \complexNumbers P^{n-1}\\
	(z_1,\dots, z_{n}) &\longmapsto [z_1,\dots, z_n].
\end{align*}
As seen in proposition \ref{proposition:pure states is homeomorphic to CPn}, we have a homeomorphism $f$ between $\complexNumbers P^{n-1}$ and the vector states over $\complexNumbers^n$. Composing $\pi \colon S^{2n-1} \to \complexNumbers P^{n-1}$ with $f$ we get the projection
\begin{align*}
	p \colon S^{2n-1} &\longrightarrow \{ \ket{\psi}\bra{\psi} \in \density(\complexNumbers^n) \mid \psi \in S^{2n-1} \}\\
	\psi &\longmapsto \ket{\psi}\bra{\psi}.
\end{align*}
Writing $\ket{\psi}$ in the standard basis $(\ket{e_j})_j$ of $\complexNumbers^n$, we have
\begin{equation}
	\ket{\psi} = \sum_j \psi_j \ket{e_j}. 
\end{equation}
Then the operator $\ket{\psi}\bra{\psi}$ is expressed as
\begin{equation}
	\ket{\psi}\bra{\psi} = \left(\sum_j \psi_j \ket{e_j}\right) \left(\sum_{j'} \overline{\psi_{j'}} \bra{e_{j'}}\right) = \sum_{j,j'} \psi_j \overline{\psi_{j'}} \ket{e_j} \bra{e_{j'}}.
\end{equation}
We can see that $p$ corresponds to $\pi \colon S^{d_Y}_{d_X,r} \to \bigcup_{d_X / d_Y \leq k \leq r} \psdii(X,Y)_k$ for the special case where $d_X = r = 1$.

A next step would be to guess a skeleton for $\bigcup_{d_X / d_Y \leq k \leq d_X} \psdii(X,Y)_k$ inspired by the one for $\complexNumbers P^{n-1}$. The skeleton of $\complexNumbers P^{n-1}$ is given by the sequence of inclusions
\begin{equation}
	\complexNumbers P^0 \subseteq \complexNumbers P^1 \subseteq \dots \subseteq \complexNumbers P^{n-1},
\end{equation}
where we interpret $\complexNumbers P^l$ as the subset
\begin{equation}
	\complexNumbers P^{l-1} = \{[z_1,\dots, z_n] \in \complexNumbers P^{n-1} \mid \forall j>l, \ z_j = 0\}.
\end{equation}
Inspired by this, we define
\begin{equation}
	K^l \coloneqq \pi\left(\{(\psi_j)_j \in S^{d_Y}_{d_X,d_X} \mid  \forall j>l, \ \psi_j = 0 \}\right),
\end{equation}
for $1\leq l \leq d_Y$. Also, $\complexNumbers P^l$ is obtained from $\complexNumbers P^{l-1}$ by gluing a ball through the characteristic map
\begin{align*}
	f^l \colon B^{2l} &\longrightarrow \complexNumbers P^l\\
	(z_1,\dots,z_l) &\longmapsto [z_1,\dots,z_l,z_{l+1},0,\dots, 0],
\end{align*}
where
\begin{equation}
	B^{2l} = \left\{(z_1,\dots,z_l) \in \complexNumbers^l \ \bigg| \ \sum_{j=1}^l |z_j|^2 \leq 1 \right\}, 
\end{equation}
and
\begin{equation}
	z_{l+1} \coloneqq \left( 1-\sum_{j=1}^l |z_j|^2 \right)^{1/2}.
\end{equation}
We can apply the same reasoning used to find the characteristic map $f^l$, to find a similar function $F^l$ for $K^l$. These functions $F^l$ will give rise to characteristic maps, but we'll see that we won't get yet a CW decomposition. Finding such CW decomposition is still an open problem.

\subsection{Constructing a candidate for a characteristic map}

In this section we construct candidates for characteristic maps $F^l$ that resembles the $f^l$ that we have for complex projective spaces. We'll mostly be concerned with the case where $r=d_X$, but we'll consider also the case $r \leq d_X$ when possible. The reasoning for building such $F^l$ is to make a choice of a representative $(\psi_j)_j \in S^{d_Y}_{d_X,d_X}$ for each operator $A$ in a subset of $K^l \setminus K^{l-1}$. Such subset of $K^l \setminus K^{l-1}$ is determined by the conditions we need to be able to make such choice. In the case of the complex projective spaces, we could just take the whole space as the subspace, but it won't be the case for $K^l \setminus K^{l-1}$.

Let $A \in K^l \setminus K^{l-1}$, then there exists $(\psi_j)_j \in \pi^{-1}( K^l \setminus K^{l-1})$ such that
\begin{equation}
	A = \pi((\psi_j)_j).
\end{equation}
Such $(\psi_j)_j$ is a sequence of square matrices $(\psi_j)_j \in M_{d_X}(\complexNumbers)^{d_Y}$ such that
\begin{equation}
	\psi_l \neq 0,
\end{equation}
\begin{equation}
	\psi_{l+1} = \dots = \psi_{d_Y} = 0,
\end{equation}
and
\begin{equation}
	\sum_{j=1}^{d_Y} \psi_j \psi_j^\dag = I_{d_X}.
\end{equation}
The equation $\psi_l \neq 0$ comes from $A \not\in K^{l-1}$, and the other two from $A \in K^l$. We want to make a choice of representative for $A$. By proposition \ref{proposition:relation of pi}, a change of representative is made by multiplying the matrices by a unitary matrix $V^\dag \in \U(d_X)$. That is, another representative would be
\begin{equation}
	(\psi'_j)_j = (\psi_j V^\dag)_j.
\end{equation}
The polar decomposition for square matrices express a matrix as a product of a positive semidefinite matrix with a unitary matrix. This is convenient, because we are multiplying by a unitary matrix $V^\dag$. The matrix $\psi_l$ is the only one it is guaranteed to be non zero. Because of this, we take a polar decomposition of $\psi_l$, from which we get that
\begin{equation}
	\psi_l = \left(\psi_l \psi_l^\dag \right)^{1/2} U_l,
\end{equation}
for some $U_l \in \U(d_X)$ unitary matrix. Changing the representative will then change $U_l$ to $U_l V^\dag$. The matrix $U_l V^\dag$ can be any unitary matrix we want. The matrix $U_l$ of the polar decomposition is unique only if $\psi_l$ is invertible. As a consequence, if $\psi_l$ isn't invertible, choosing what $U_l V^\dag$ should be won't be enough to determine a representative completely. For this reason, we'll suppose that $\psi_l$ is invertible. Then, take $V = U_l$, from which we get the representative
\begin{equation}
	(\psi'_j)_j = (\psi_j U_l^\dag)_j.
\end{equation}
It is chosen in such a way that
\begin{equation}
	\psi'_l = \left(\psi_l \psi_l^\dag \right)^{1/2} U_l U_l^\dag = \left(\psi_l \psi_l^\dag \right)^{1/2},
\end{equation}
so $\psi'_l$ is positive semidefinite. Notice that the term $\psi_j \psi_j^\dag$ doesn't depend on the representative. That is,
\begin{equation}
	\psi'_j {\psi'}_j^\dag = \psi_j V^\dag V \psi_j^\dag = \psi_j \psi_j^\dag.
\end{equation}
The sequence $(\psi'_j)_j$ is an element of $S^{d_Y}_{d_X,d_X}$, so it satisfies the equation
\begin{equation}
	\label{equation: sum psi' psi'dag = I}
	\sum_{j=1}^{d_Y} \psi'_j {\psi'}_j^\dag = I_{d_X}.
\end{equation}
Of course, since $\psi_{l+1} = \dots = \psi_{d_Y} = 0$, the same is true for $\psi'_j$, that is,
\begin{equation}
	\psi'_{l+1} = \dots = \psi'_{d_Y} = 0.
\end{equation}
Also, $\psi_l \neq 0$ implies that
\begin{equation}
	\psi'_l \neq 0.
\end{equation}
Using this in equation \ref{equation: sum psi' psi'dag = I} and isolating $\psi'_l$, we get
\begin{equation}
	\psi'_l {\psi'}_l^\dag = I_{d_X}-\sum_{j<l} \psi'_j {\psi'}_j^\dag.
\end{equation}
But $\psi'_l$ was chosen to be a positive operator, so $\psi'_l {\psi'}_l^\dag = {\psi'}_l^2$. Taking the square root, we get
\begin{equation}
	\psi'_l = \left(I_{d_X}-\sum_{j<l} \psi'_j {\psi'}_j^\dag \right)^{1/2}.
\end{equation}
We see that such representative $(\psi'_j)_j$ is determined by the first $l-1$ matrices.

We can construct a function that captures this choice of representative. It consists of taking the first $l-1$ matrices $\psi_j$, constructing $\psi_l$ from the previous ones, taking $\psi_j = 0$ for $j>l$ and projecting with $\pi$. This would give a function
\begin{equation}
	F^l (\psi_1,\dots, \psi_{l-1}) = \pi(\psi_1,\dots,\psi_{l-1}, \psi_l, 0, \dots, 0),
\end{equation}
where
\begin{equation}
	\psi_l \coloneqq \left(I_{d_X}-\sum_{j<l} \psi_j \psi_j^\dag \right)^{1/2}.
\end{equation}
For such function to make sense, we must be able to take the square root, so $I_{d_X}-\sum_{j<l} \psi_j \psi_j^\dag$ should be a positive semidefinite operator. We denote this by saying that
\begin{equation}
	\sum_{j<l} \psi_j \psi_j^\dag \leq I_{d_X}.
\end{equation}
More generally, for any Hermitian matrices $H$ and $H'$, we say that
\begin{equation}
	H \leq H'
\end{equation}
if $H'-H$ is a positive semidefinite operator. This defines a partial order over the hermitian matrices, but we won't need this fact. Also, we should have
\begin{equation}
	\sum_{j=1}^l \psi_j \psi_j^\dag = I_{d_X},
\end{equation}
but this follows immediately from the definition of $\psi_l$. Therefore, the domain of $F^l$ should be the set $B^{l-1}_{d_X,d_X}$ defined next. We define a set $B^n_{m,r}$ in more generality, which perhaps will be useful when we try to find more characteristics functions in the future.

\begin{definition}{}{generalized ball}
	Let $m,n,r$ be natural numbers with $m\geq 1$, $n \geq 1$ and $r \geq 1$. We define the set
	\begin{equation}
		B^n_{m,r} \coloneqq \left\{(\psi_j)_{j=1,\dots,n} \in M_{m,r} (\complexNumbers)^n \ \bigg| \ \sum_j \psi_j \psi_j^\dag \leq I_m \right\}.
	\end{equation}
\end{definition}
Notice that $B^n_{m,r}$ is a generalization of the closed unit ball in $\complexNumbers^n$. In fact, if we take $m=r=1$, then the matrices $\psi_j$ are $1\times1$, which are complex numbers. Then the identity matrix $I_m$ is just 1. The sum $\sum_j \psi_j \psi_j^\dag$ becomes $\sum_j |\psi_j|^2$. Therefore, we get
\begin{equation}
	B^n_{1,1} = \left\{(\psi_j)_j \in \complexNumbers^n \ \bigg| \ \sum_j |\psi_j|^2 \leq 1 \right\},
\end{equation}
which is the closed unit ball.

We argued that $B^{l-1}_{d_X,d_X}$ should be the domain of $F^l$. Of course, its codomain can be taken as $K^l$. We then write a complete definition of the function $F^l$. We do it for $l \geq 2$, because we must have $l-1 \geq 1$.
\begin{definition}{}{F^l}
	For $2\leq l \leq d_Y$ we define the function
	\begin{align*}
		F^l \colon B^{l-1}_{d_X,d_X} &\longrightarrow K^l\\
		(\psi_1, \dots, \psi_{l-1}) &\longmapsto \pi(\psi_1,\dots, \psi_{l-1}, \psi_l, 0, \dots, 0),
	\end{align*}
	where
	\begin{equation}
		\psi_l \coloneqq \left( I_{d_X}-\sum_{j<l} \psi_j \psi_j^\dag \right)^{1/2}.
	\end{equation}
\end{definition}

\begin{proposition}{}{F^l is continuous}
	The function $F^l$ of definition \ref{definition:F^l} is continuous.
\end{proposition}
\begin{proof}
	The projection $\pi$ is continuous, because it comprises of computing the matrix products $\psi_j \psi_{j'}^\dag$. The matrix operations are all continuous, so we just have to show that taking the square root of a positive semidefinite operator is a continuous function. This was already done in proposition \ref{proposition:square root of psd operators is continuous}. This proves that $F^l$ is continuous.
\end{proof}

This is the candidate for a characteristic map inspired by the CW decomposition of the complex projective spaces. It is indeed a characteristic map, but we'll see that $K^l$ isn't obtained from $K^{l-1}$ by gluing a ball through $F^l$. This can already be seen from having to suppose earlier that $\psi_l$ is invertible. We'll understand the problem better after we study in more detail the sets $B^n_{m,r}$.

\subsection{Topological properties of $B^n_{m,r}$}

In definition \ref{definition:generalized ball} we defined a set $B^n_{m,r}$ which generalizes the closed unit ball. Here we show that it is compact convex, which implies that it is homeomorphic to a closed ball. Then we compute its dimension and characterized its interior.

We'll first show that $B^n_{m,r}$ is bounded by two closed balls, which will be useful for the compactness property and for computing the dimension of $B^n_{m,r}$.

\begin{proposition}{}{generalized ball bounds}
	Let $B^n_{m,r}$ be the set defined in definition \ref{definition:generalized ball}. We have that
	\begin{equation}
		\overline{B(0,1)} \subseteq B^n_{m,r} \subseteq \overline{B(0,\sqrt{m})}.
	\end{equation}
\end{proposition}
\begin{proof}
	We start by the inclusion $B^n_{m,r} \subseteq \overline{B(0,\sqrt{m})}$. Let $(\psi_j)_j \in B^n_{m,r}$, then the operator
	\begin{equation}
		P \coloneqq I_m - \sum_j \psi_j \psi_j^\dag
	\end{equation}
	is positive semidefinite. Rearranging the terms, we get
	\begin{equation}
		\label{equation: sum psij psijdag = I-P}
		\sum_j \psi_j \psi_j^\dag = I_m -P.
	\end{equation}
	The squared norm of $(\psi_j)_j$ is
	\begin{equation}
		||(\psi_j)_j||^2 = \sum_j ||\psi_j||^2.
	\end{equation}
	The norm of $||\psi_j||$ is the Frobenius norm, which is the same as the Euclidean norm of its entries. The Frobenius norm squared is
	\begin{equation}
		||\psi_j||^2 = \tr(\psi_j^\dag \psi_j).
	\end{equation}
	The trace is cyclic, so we have
	\begin{equation}
		||\psi_j||^2 = \tr(\psi_j \psi_j^\dag).
	\end{equation}
	Taking the sum over $j$, we get
	\begin{equation}
		||(\psi_j)_j||^2 = \sum_j \tr(\psi_j \psi_j^\dag) = \tr\left( \sum_j \psi_j \psi_j^\dag \right).
	\end{equation}
	Substituting equation \ref{equation: sum psij psijdag = I-P}, we get
	\begin{equation}
		||(\psi_j)_j||^2 = \tr(I_m - P) = m-\tr(P).
	\end{equation}
	The trace of a diagonalizable matrix is the sum of its eigenvalues. The matrix $P$ is positive semidefinite, so it is diagonalizable and has only non negative eigenvalues. Therefore $\tr(P)\geq 0$. This implies that
	\begin{equation}
		||(\psi_j)_j||^2 \leq m.
	\end{equation}
	This proves that $B^n_{m,r} \subseteq \overline{B(0,\sqrt{m})}$.
	
	Now let's prove that $\overline{B(0,1)} \subseteq B^n_{m,r}$. The reason for this inclusion is that a positive operator may be perturbed by a small enough hermitian operator and remain positive. Let $(\psi_j)_j \in M_{m,r}(\complexNumbers)^n$ such that
	\begin{equation}
		||(\psi_j)_j|| \leq \varepsilon,
	\end{equation}
	for some $\varepsilon > 0$ to be determined. From the previous paragraph, we know that the norm squared is the trace of a sum, so
	\begin{equation}
		\tr\left( \sum_j \psi_j \psi_j^\dag \right)  = ||(\psi_j)_j ||^2 \leq \varepsilon^2.
	\end{equation}
	Let's denote the sum by $Q$:
	\begin{equation}
		Q \coloneqq \sum_j \psi_j \psi_j^\dag \in M_m(\complexNumbers).
	\end{equation}
	Each term $\psi_j \psi_j^\dag$ is a positive semidefinite operator, and a sum of positive semidefinite operators is positive semidefinite. Therefore, $Q$ is positive semidefinite, so it can be diagonalized as
	\begin{equation}
		Q = \sum_i \lambda_i \ket{v_i}\bra{v_i},
	\end{equation} 
	where $\lambda_i \geq 0$ and $(v_i)_i$ is an orthonormal basis for $\complexNumbers^m$. Since $\tr(Q) \leq \varepsilon^2$, we have
	\begin{equation}
		\sum_i \lambda_i \leq \varepsilon^2.
	\end{equation}
	In particular, the same bound is true for each term:
	\begin{equation}
		\lambda_i \leq \varepsilon^2,
	\end{equation}
	for every $i$. The identity matrix can be diagonalized with any orthonormal basis, so
	\begin{equation}
		I_m = \sum_i \ket{v_i}\bra{v_i}. 
	\end{equation}
	Therefore, we can diagonalize $I_m - Q$ as
	\begin{equation}
		I_m-Q = \sum_i (1-\lambda_i) \ket{v_i}\bra{v_i}.
	\end{equation}
	This is positive semidefinite iff $1-\lambda_i \geq 0$ for every $i$. Therefore, if we take $\varepsilon = 1$, it follows that $I_m - Q$ is positive semidefinite. This proves that $\overline{B(0,1)} \subseteq B^n_{m,r}$.
\end{proof}

Now we show that $B^n_{m,r}$ is compact convex. This implies that it is homeomorphic to a closed ball, which is important to obtain a characteristic map.
\begin{theorem}{}{generalized ball is compact convex}
	The set $B^n_{m,r}$ from definition \ref{definition:generalized ball} is compact convex.
\end{theorem}
\begin{proof}
	A compact set in $\realNumbers^k$ is a closed bounded set. From proposition \ref{proposition:generalized ball bounds} we already know that $B^n_{m,r}$ is bounded. Let's prove that it is closed. Let $(\psi^t)_{t \geq 1}$ be a sequence in $B^n_{m,r}$ that converges to some $\psi \in M_{m,r}(\complexNumbers)^n$. Each $\psi^t$ and $\psi$ are finite sequences of matrices:
	\begin{equation}
		\psi^t = (\psi^t_j)_{j=1,\dots, n},
	\end{equation}
	\begin{equation}
		\psi = (\psi_j)_{j=1,\dots, n}.
	\end{equation}
	Since
	\begin{equation}
		\lim_{t \to \infty} \psi^t = \psi,
	\end{equation}
	we also have
	\begin{equation}
		\lim_{t \to \infty} \psi^t_j = \psi_j,
	\end{equation}
	for $j=1,\dots, n$.
	Now let's check that $\psi \in B^n_{m,r}$. This is the case if $I_m - \sum_j \psi_j \psi_j^\dag$ is positive semidefinite. This is equivalent to say that
	\begin{equation}
		\bra{v} \left( I_m - \sum_j \psi_j \psi_j^\dag \right)\ket{v} \geq 0,
	\end{equation}
	for any unit vector $v \in \complexNumbers^m$. Using that $\psi_j = \lim_{t\to \infty} \psi_j^t$ and the continuity of the operations, we get that
	\begin{equation}
		\bra{v} \left( I_m - \sum_j \psi_j \psi_j^\dag \right)\ket{v}  = \lim_{t \to \infty} \bra{v} \left( I_m - \sum_j \psi_j^t (\psi_j^t)^\dag \right)\ket{v}.
	\end{equation}
	Since $\psi^t \in B^n_{m,r}$, the right hand side is a limit of non negative real numbers, so the limit is also non negative. Therefore
	\begin{equation}
		\bra{v} \left( I_m - \sum_j \psi_j \psi_j^\dag \right)\ket{v} \geq 0,
	\end{equation}
	which proves that $\psi \in B^n_{m,r}$. This concludes that $B^n_{m,r}$ is closed. It is closed and bounded, so it is compact.
	
	Now let's prove that $B^n_{m,r}$ is convex. Let $\psi^1, \psi^2 \in B^n_{m,r}$, and let $t_1, t_2 \in [0,1]$ such that $t_1+t_2=1$. We have the convex combination
	\begin{equation}
		\psi \coloneqq \sum_i t_i \psi^i \in M_{m,r}(\complexNumbers)^n.
	\end{equation}
	Instead of working with $M_{m,r}(\complexNumbers)^n$, it will be a little more convenient to work with $M_{m,nr}(\complexNumbers)$, concatenating the matrices horizontally. We define
	\begin{equation}
		\Psi^i \coloneq \begin{pmatrix}
			\psi^i_1 & \dots & \psi^i_n
		\end{pmatrix},
	\end{equation}
	\begin{equation}
		\Psi \coloneq \sum_i t_i \Psi^i =  \begin{pmatrix}
			\psi_1 & \dots & \psi_n
		\end{pmatrix}.
	\end{equation}
	Since $\psi^i \in B^n_{m,r}$, we have that
	\begin{equation}
		I_m - \sum_j \psi^i_j (\psi^i_j)^\dag \geq 0.
	\end{equation}
	This is equivalent to
	\begin{equation}
		I_m - \Psi^i (\Psi^i)^\dag \geq 0.
	\end{equation}
	We want to prove that
	\begin{equation}
		I_m - \Psi \Psi^\dag \geq 0.
	\end{equation}
	Using that $\Psi = \sum_i t_i \Psi^i$, we get
	\begin{equation}
		I_m - \Psi \Psi^\dag = I_m - t_1^2 \Psi^1 (\Psi^1)^\dag - t_2^2 \Psi^2 (\Psi^2)^\dag - t_1 t_2 \left(\Psi^1 (\Psi^2)^\dag + \Psi^2 (\Psi^1)^\dag \right).
	\end{equation}
	Let $v \in \complexNumbers^m$ be a unit vector. Then
	\begin{multline}
		\bra{v} \left( I_m - \Psi \Psi^\dag \right)\ket{v} = 1 - t_1^2 \bra{v}\Psi^1 (\Psi^1)^\dag \ket{v} - t_2^2 \bra{v} \Psi^2 (\Psi^2)^\dag \ket{v}\\
		- t_1 t_2 \bra{v}\left(\Psi^1 (\Psi^2)^\dag + \Psi^2 (\Psi^1)^\dag \right)\ket{v}.
	\end{multline}
	The last term is a sum of a complex number and its conjugate, which is the double of the real part, so
	\begin{equation}
		\bra{v}\left(\Psi^1 (\Psi^2)^\dag + \Psi^2 (\Psi^1)^\dag \right)\ket{v} = 2 \Re(\bra{v}(\Psi^1 (\Psi^2)^\dag)\ket{v}). 
	\end{equation}
	We can estimate the size of $\bra{v}\Psi^1 (\Psi^2)^\dag \ket{v}$ from the Cauchy-Schwarz inequality. It says that
	\begin{align*}
		|\bra{v}\Psi^1 (\Psi^2)^\dag \ket{v}| &\leq || (\Psi^1)^\dag \ket{v} || \ || (\Psi^2)^\dag \ket{v} || \\ 
		&= \bra{v}\Psi^1 (\Psi^1)^\dag \ket{v}^{1/2} \ \bra{v}\Psi^2 (\Psi^2)^\dag \ket{v}^{1/2}.
	\end{align*}
	Using this results, we get the lower bound
	\begin{multline}
		\bra{v} \left( I_m - \Psi \Psi^\dag \right)\ket{v} \geq 1 - t_1^2 \bra{v}\Psi^1 (\Psi^1)^\dag \ket{v} - t_2^2 \bra{v} \Psi^2 (\Psi^2)^\dag \ket{v}\\
		- 2 t_1 t_2 \bra{v}\Psi^1 (\Psi^1)^\dag \ket{v}^{1/2} \ \bra{v}\Psi^2 (\Psi^2)^\dag \ket{v}^{1/2}.
	\end{multline}
	In the right hand side we have a perfect square, so we get
	\begin{equation}
		\label{equation: sandwich of Im - Psi Psidag}
		\bra{v} \left( I_m - \Psi \Psi^\dag \right)\ket{v} \geq 1-\left(\sum_i t_i \bra{v}\Psi^i (\Psi^i)^\dag \ket{v}^{1/2} \right)^2 .
	\end{equation}
	We supposed that $\Psi^i (\Psi^i)^\dag \leq I_m$. Let
	\begin{equation}
		P^i \coloneqq I_m -\Psi^i (\Psi^i)^\dag,
	\end{equation}
	which is positive semidefinite.
	Then
	\begin{equation}
		\bra{v}\Psi^i (\Psi^i)^\dag \ket{v} = \bra{v}(I_m-P^i)\ket{v} = 1- \bra{v} P^i \ket{v}.
	\end{equation}
	Since $P^i$ is positive semidefinite, we have that $\bra{v} P^i \ket{v} \geq 0$, so
	\begin{equation}
		\bra{v}\Psi^i (\Psi^i)^\dag \ket{v} \leq 1.
	\end{equation}
	The same is true if we take the convex combination:
	\begin{equation}
		\sum_i t_i \bra{v}\Psi^i (\Psi^i)^\dag \ket{v}^{1/2} \leq \sum_i 1^{1/2} = \sum_i t_i = 1.
	\end{equation}
	Using this inequality in equation \ref{equation: sandwich of Im - Psi Psidag}, we get that
	\begin{equation}
		\bra{v} \left( I_m - \Psi \Psi^\dag \right)\ket{v} \geq 0.
	\end{equation}
	This concludes that $\psi \in B^n_{m,r}$, and that $B^n_{m,r}$ is convex.
\end{proof}

Since $B^n_{m,r}$ is a compact convex set, we know that it is homeomorphic to a closed ball. Next we compute the dimension of such ball.
\begin{corollary}{}{}
	The set $B^n_{m,r}$ of definition \ref{definition:generalized ball} is homeomorphic to a closed ball of dimension
	\begin{equation}
		\dim_\realNumbers B^n_{m,r} = 2mnr.
	\end{equation}
\end{corollary}
\begin{proof}
	We proved in theorem \ref{theorem:generalized ball is compact convex} that $B^n_{m,r}$ is compact convex. In proposition \ref{proposition:generalized ball bounds} was also proved that $\overline{B(0,1)} \subseteq B^n_{m,r}$. From this we get that $0 \in B^n_{m,r}$ and that the real vector space generated by $B^n_{m,r}$ is
	\begin{equation}
		\langle B^n_{m,r} \rangle_\realNumbers = M_{m,r}(\complexNumbers)^n.
	\end{equation}
	From corollary \ref{corollary:compact convex set is homeomorphic to a closed ball} it follows that
	\begin{equation}
		\dim_\realNumbers B^n_{m,r} = \dim_\realNumbers M_{m,r}(\complexNumbers)^n = 2mnr.
	\end{equation}
\end{proof}

We proved that $B^n_{m,r}$ is homeomorphic to a closed ball. Next we characterize its interior. This is important for the construction of a characteristic map, since such function should map homeomorphically the interior of a closed ball to the points it adds to a simpler space.
\begin{theorem}{}{interior and boundary of generalized ball}
	The interior and boundary of the set $B^n_{m,r}$ from definition \ref{definition:generalized ball} are
	\begin{equation}
		\interior{B^n_{m,r}} = \left\{(\psi_j)_j \in B^n_{m,r} \ \bigg| \ \rank\left(I_m - \sum_j \psi_j \psi_j^\dag\right) = m \right\},
	\end{equation}
	\begin{equation}
		\partial B^n_{m,r} = \left\{(\psi_j)_j \in B^n_{m,r} \ \bigg| \ \rank\left(I_m - \sum_j \psi_j \psi_j^\dag\right) < m \right\}.
	\end{equation}
	Of course, the matrix $I_m - \sum_j \psi_j \psi_j^\dag$ has rank $m$ iff it is invertible.
\end{theorem}
\begin{proof}
	We first we do some general calculations, then we specialize to the cases we want to prove. Let $(\psi_j)_j \in B^n_{m,r}$ and define
	\begin{equation}
		\label{equation: P = Im - sum psijpsijdag}
		P \coloneqq I_m - \sum_j \psi_j \psi_j^\dag
	\end{equation}
	Let $(\gamma_j)_j \in M_{m,r}(\complexNumbers)^n$ such that
	\begin{equation}
		\label{equation: distance between psi and gamma}
		||(\psi_j)_j - (\gamma_j)_j|| \leq \varepsilon,
	\end{equation}
	for some $\varepsilon > 0$ to be determined later. Define
	\begin{equation}
		\varphi_j \coloneqq \psi_j + \gamma_j,
	\end{equation}
	so that $(\varphi_j)_j$ is a small perturbation of $(\psi_j)_j$. It will be a little more convenient to work with $M_{m,nr}(\complexNumbers)$ instead of $M_{m,r}(\complexNumbers)^n$. Concatenating the matrices horizontally, we get
	\begin{equation}
		\Psi \coloneqq \begin{pmatrix}
			\psi_1 & \dots & \psi_n
		\end{pmatrix},
	\end{equation}
	\begin{equation}
		\Gamma \coloneqq \begin{pmatrix}
			\gamma_1 & \dots & \gamma_n
		\end{pmatrix},
	\end{equation}
	\begin{equation}
		\Phi \coloneqq \begin{pmatrix}
			\varphi_1 & \dots & \varphi_n
		\end{pmatrix},
	\end{equation}
	which are matrices in $M_{m,nr}(\complexNumbers)$. Then equations \ref{equation: P = Im - sum psijpsijdag} and \ref{equation: distance between psi and gamma} become
	\begin{equation}
		P = I_m - \Psi \Psi^\dag,
	\end{equation}
	\begin{equation}
		||\Psi - \Gamma|| \leq \varepsilon.
	\end{equation}
	For $(\varphi_j)_j$ to be in $B^n_{m,r}$ we need that
	\begin{equation}
		I_m - \Phi \Phi^\dag \geq 0.
	\end{equation}
	This is the case if we have
	\begin{equation}
		\bra{v}\left( I_m - \Phi \Phi^\dag \right)\ket{v} \geq 0,
	\end{equation}
	for any unit vector $v \in \complexNumbers^m$. Using that $\Phi = \Psi + \Gamma$, the left hand side is
	\begin{multline}
		\label{equation: sandwich of Im-PhiPhidag}
		\bra{v}\left( I_m - \Phi \Phi^\dag \right)\ket{v} =\\ 1 - \bra{v}\Psi \Psi^\dag\ket{v} - \bra{v}\Psi \Gamma^\dag \ket{v} - \bra{v}\Gamma \Psi^\dag \ket{v} - \bra{v}\Gamma \Gamma^\dag \ket{v}.
	\end{multline}
	The term $\bra{v}\Psi \Gamma^\dag \ket{v}+ \bra{v}\Gamma \Psi^\dag \ket{v}$ is the sum of a complex number and its conjugate, which is the double of the real part:
	\begin{equation}
		\bra{v}\Psi \Gamma^\dag \ket{v}+ \bra{v}\Gamma \Psi^\dag \ket{v} = 2 \Re(\bra{v}\Psi \Gamma^\dag \ket{v}).
	\end{equation}
	We can estimate the size of $|\bra{v}\Psi \Gamma^\dag \ket{v}|$ by the Cauchy-Schwarz inequality, from which we get that
	\begin{equation}
		|\bra{v}\Psi \Gamma^\dag \ket{v}| \leq ||\Psi^\dag \ket{v}|| \ ||\Gamma^\dag \ket{v}|| = \bra{v}\Psi \Psi^\dag \ket{v}^{1/2} \ \bra{v}\Gamma \Gamma^\dag \ket{v}^{1/2}.
	\end{equation}
	Applying these results in equation \ref{equation: sandwich of Im-PhiPhidag}, we get
	\begin{multline}
		\label{equation: estimate for sandwich of Im-PhiPhidag}
		\bra{v}\left( I_m - \Phi \Phi^\dag \right)\ket{v} \geq 1 - \bra{v}\Psi \Psi^\dag\ket{v} -2  \bra{v}\Psi \Psi^\dag \ket{v}^{1/2} \ \bra{v}\Gamma \Gamma^\dag \ket{v}^{1/2} \\
		- \bra{v}\Gamma \Gamma^\dag \ket{v}. 
	\end{multline}
	Having proved this result, now we specialize to each case we want to prove.
	
	Suppose that the $P$ defined earlier is invertible. In this case, we want to find an $\varepsilon$ that is small enough such that $(\varphi_j)_j \in B^n_{m,r}$. Since $P$ is invertible, its eigenvalues are strictly positive. Let $\lambda_{min} > 0$ be the smallest eigenvalue of $P$. Then we have that
	\begin{equation}
		\bra{v}P\ket{v} \geq \lambda_{min},
	\end{equation}
	for any unit vector $v \in \complexNumbers^m$. Since $P=I_m-\Psi \Psi^\dag$, this implies that
	\begin{equation}
		1-\bra{v} \Psi \Psi^\dag \ket{v} \geq \lambda_{min}.
	\end{equation}
	Substituting this inequality in equation \ref{equation: estimate for sandwich of Im-PhiPhidag}, we get
	\begin{equation}
		\bra{v}\left( I_m - \Phi \Phi^\dag \right)\ket{v} \geq \lambda_{min} -2  \bra{v}\Psi \Psi^\dag \ket{v}^{1/2} \ \bra{v}\Gamma \Gamma^\dag \ket{v}^{1/2} - \bra{v}\Gamma \Gamma^\dag \ket{v}. 
	\end{equation}
	Since $\Psi \Psi^\dag \leq I_m$, it follows that $\bra{v} \Psi \Psi^\dag \ket{v} \leq 1$. Using this, we get
	\begin{equation}
		\label{equation: new estimate of sand Im-PhiPhidag}
		\bra{v}\left( I_m - \Phi \Phi^\dag \right)\ket{v} \geq \lambda_{min} -2 \bra{v}\Gamma \Gamma^\dag \ket{v}^{1/2} - \bra{v}\Gamma \Gamma^\dag \ket{v}. 
	\end{equation}
	We can simplify further the right hand side by using that $\Gamma \Gamma^\dag \leq I_m$. This implies that $\bra{v}\Gamma \Gamma^\dag \ket{v} \leq 1$. Taking the square root, we get $\bra{v}\Gamma \Gamma^\dag \ket{v}^{1/2} \leq 1$. Multiplying by $\bra{v}\Gamma \Gamma^\dag \ket{v}^{1/2}$ we get that
	\begin{equation}
		\bra{v}\Gamma \Gamma^\dag \ket{v} \leq \bra{v}\Gamma \Gamma^\dag \ket{v}^{1/2}.
	\end{equation}
	Applying this inequality in equation \ref{equation: new estimate of sand Im-PhiPhidag}, we get
	\begin{align*}
		\bra{v}\left( I_m - \Phi \Phi^\dag \right)\ket{v} &\geq \lambda_{min} -2 \bra{v}\Gamma \Gamma^\dag \ket{v} - \bra{v}\Gamma \Gamma^\dag \ket{v}\\
		&=\lambda_{min} -3 \bra{v}\Gamma \Gamma^\dag \ket{v} .
	\end{align*}
	Therefore, it is enough to have
	\begin{equation}
		\bra{v}\Gamma \Gamma^\dag \ket{v} \leq \frac{\lambda_{min}}{3}.
	\end{equation}
	The left hand side is bounded by $\tr(\Gamma \Gamma^\dag)$. In fact, we can extend $v$ to an orthonormal basis $(v_i)_i$ for $\complexNumbers^m$, with $v_1 = v$. The trace can be expressed with respect to this basis as
	\begin{equation}
		\tr(\Gamma \Gamma^\dag) = \sum_i \bra{v_i} \Gamma \Gamma^\dag \ket{v_i}.
	\end{equation}
	Since $\Gamma \Gamma^\dag$ is positive semidefinite, each term $\bra{v_i} \Gamma \Gamma^\dag \ket{v_i}$ is non negative. Therefore
	\begin{equation}
		\bra{v}\Gamma \Gamma^\dag \ket{v} \leq \tr(\Gamma \Gamma^\dag).
	\end{equation}
	The right hand side is the same as $||\Gamma||^2$. Therefore, if we take
	\begin{equation}
		\varepsilon = \sqrt{\frac{\lambda_{min}}{3}},
	\end{equation}
	then $(\varphi_j)_j \in B^n_{m,r}$. This proves that
	\begin{equation}
		P \text{ invertible} \implies (\psi_j)_j \in \interior{B^n_{m,r}}.
	\end{equation}
	
	Now let's prove the converse statement. We do this by the contrapositive argument. Suppose $P$ isn't invertible. It is positive semidefinite, so its minimum eigenvalue must be 0. Let $v \in \complexNumbers^m$ be a unit vector that is an eigenvector of $P$ with 0 eigenvalue:
	\begin{equation}
		\bra{v}P\ket{v} = 0.
	\end{equation}
	Since $P = I_m-\Psi \Psi^\dag$, this implies that
	\begin{equation}
		1-\bra{v} \Psi \Psi^\dag \ket{v} = 0,
	\end{equation}
	so
	\begin{equation}
		\bra{v} \Psi \Psi^\dag \ket{v} = 1.
	\end{equation}
	We want to prove that $I_m - \Phi \Phi^\dag$ isn't positive semidefinite. It this case we need an equality, so we use equation \ref{equation: sandwich of Im-PhiPhidag}. Using that $\bra{v} \Psi \Psi^\dag \ket{v} = 1$, equation \ref{equation: sandwich of Im-PhiPhidag} simplifies to
	\begin{equation}
		\bra{v}\left( I_m - \Phi \Phi^\dag \right)\ket{v} = - \bra{v}\Psi \Gamma^\dag \ket{v} - \bra{v}\Gamma \Psi^\dag \ket{v} - \bra{v}\Gamma \Gamma^\dag \ket{v}.
	\end{equation}
	We have to make a choice of $\Gamma$ such that the right hand side is negative. Define $\Gamma$ such that
	\begin{equation}
		\Gamma^\dag \ket{v} = \varepsilon \Psi^\dag \ket{v},
	\end{equation}
	and $\Gamma^\dag \ket{w} = 0$ for any $w \in \complexNumbers^m$ that is orthogonal to $v$. Then we have
	\begin{equation}
		\bra{v}\left( I_m - \Phi \Phi^\dag \right)\ket{v} = - (2\varepsilon+\varepsilon^2) \bra{v}\Psi \Psi^\dag \ket{v} = - (2\varepsilon+\varepsilon^2) < 0.
	\end{equation}
	This proves that $I_m - \Phi \Phi^\dag$ isn't positive semidefinite, so $(\varphi_j)_j \not \in B^n_{m,r}$. The squared norm of $(\gamma_j)_j$ is
	\begin{equation}
		||(\gamma_j)_j||^2 = ||\Gamma||^2 = \tr(\Gamma \Gamma^\dag).
	\end{equation}
	We can extend $v$ to an orthonormal basis $(v_i)_i$ of $\complexNumbers^m$, where $v_1 = v$. The trace is expressed with respect to this basis as
	\begin{equation}
		\tr(\Gamma \Gamma^\dag) = \sum_i \bra{v_i} \Gamma \Gamma^\dag \ket{v_i}.
	\end{equation}
	We defined $\Gamma$ such that $\Gamma^\dag \ket{w} = 0$ if $w$ is orthogonal to $v$. For this reason, only the term with $i=1$ contributes to the sum. Therefore,
	\begin{equation}
		\tr(\Gamma \Gamma^\dag) = \bra{v} \Gamma \Gamma^\dag \ket{v}.
	\end{equation}
	We defined $\Gamma^\dag \ket{v} = \varepsilon \Psi^\dag \ket{v}$, and we know that $\bra{v} \Psi \Psi^\dag \ket{v} = 1$. This shows that
	\begin{equation}
		\tr(\Gamma \Gamma^\dag) = \varepsilon^2.
	\end{equation}
	Therefore,
	\begin{equation}
		||(\gamma_j)_j|| = \varepsilon.
	\end{equation}
	For any $\varepsilon > 0$ we constructed a $(\varphi_j)_j \in M_{m,r}(\complexNumbers)^n$ such that
	\begin{equation}
		||(\varphi_j)_j - (\psi_j)_j|| = ||(\gamma_j)_j|| = \varepsilon,
	\end{equation}
	and $(\varphi_j)_j \not \in B^n_{m,r}$. This proves that
	\begin{equation}
		P \text{ not invertible} \implies (\psi_j)_j \not \in \interior{B^n_{m,r}}.
	\end{equation}
	This concludes the proof that
	\begin{equation}
		\interior{B^n_{m,r}} = \left\{(\psi_j)_j \in B^n_{m,r} \ \bigg| \ \rank\left(I_m - \sum_j \psi_j \psi_j^\dag\right) = m \right\}.
	\end{equation}
	Since $\partial B^n_{m,r} = B^n_{m,r} \setminus \interior{B^n_{m,r}}$, the characterization for $\partial B^n_{m,r}$ follows.
\end{proof}

\subsection{Discussion about a possible skeleton}

We defined sets
\begin{equation}
	K^l = \pi\left(\{(\psi_j)_j \in S^{d_Y}_{d_X,d_X} \mid  \forall j>l, \ \psi_j = 0 \}\right).
\end{equation}
which are organized in a sequence of inclusions
\begin{equation}
	K^1 \subseteq \dots \subseteq K^{d_Y} = \bigcup_{ d_X / d_Y \leq k\leq d_X} \psdii(X,Y)_k. 
\end{equation}
This sequence is a generalization of the sequence of inclusions
\begin{equation}
	\complexNumbers P^0 \subseteq \complexNumbers P^1 \subseteq \dots \complexNumbers P^{d_Y-1}.
\end{equation}
More precisely, if $d_X = 1$, we have a homeomorphism
\begin{equation}
	K^l \cong \complexNumbers P^{l-1}.
\end{equation}
Then the inclusion $\complexNumbers P^{l-1} \subseteq \complexNumbers P^l$ corresponds to the inclusion $K^l \subseteq K^{l+1}$.

For the complex projective spaces we also have characteristic maps
\begin{align*}
	f^l \colon B^{2l} &\longrightarrow \complexNumbers P^l\\
	(z_1,\dots,z_l) &\longmapsto \left[z_1,\dots,z_l,\left(1-\sum_{j\leq l} |z_j|^2 \right)^{1/2},0,\dots, 0\right],
\end{align*}
for $1 \leq l \leq d_Y-1$. Inspired by this idea, we constructed functions
\begin{align*}
	F^l \colon B^{l-1}_{d_X,d_X} &\longrightarrow K^l\\
	(\psi_1, \dots, \psi_{l-1}) &\longmapsto \pi \left(\psi_1,\dots, \psi_{l-1}, \left( I_{d_X}-\sum_{j<l} \psi_j \psi_j^\dag \right)^{1/2}, 0, \dots, 0 \right),
\end{align*}
for $2\leq l \leq d_Y$. We have that $f^{l-1}$ generalizes to $F^l$. For $f^l$, its domain is already a closed ball, but for $F^l$ its domain is a compact convex set, so it is homeomorphic to a closed ball.

In the case of complex projective spaces, the sequence of inclusions $\complexNumbers P^0 \subseteq \dots \subseteq \complexNumbers P^{d_Y-1}$ determines a skeleton for a CW decomposition of $\complexNumbers P^{d_Y}$. Each set $\complexNumbers P^l$ is obtained from $\complexNumbers P^{l-1}$ by gluing a $2l$-dimensional ball through the characteristic map $f^l$. We were able to generalize $\complexNumbers P^{l-1}$ to $K^l$, and $f^{l-1}$ to $F^l$, so a natural question to ask is if the sets $K^l$ determine a skeleton for a CW decomposition of $K^{d_Y}$, where $K^l$ would be obtained from $K^{l-1}$ by gluing a closed ball through $F^l$. Unfortunately, the answer for this question is no, except for trivial cases. We'll explain why this is the case, and what can we do about it.

For $F^l$ to be a characteristic map that obtains $K^l$ from $K^{l-1}$ by gluing a ball, it should satisfy the following properties:
\begin{itemize}
	\item $F^l$ should be continuous;
	\item $F^l(\partial B^{l-1}_{d_X,d_X}) \subseteq K^{l-1}$;
	\item $F^l$ should restrict to a homeomorphism $F^l \colon \interior{B^{l-1}_{d_X,d_X}} \stackrel{\cong}{\to} K^l \setminus K^{l-1}$.
\end{itemize}
From proposition \ref{proposition:F^l is continuous}, we already know that $F^l$ is continuous. But, assuming that $d_X \geq 2$ and $d_Y \geq 2$, the last two properties are not satisfied. Let's show that $F^l(\partial B^{l-1}_{d_X,d_X}) \not\subseteq K^{l-1}$ for $l\geq 2$. Just take
\begin{equation}
	\psi_1 = \begin{pmatrix}
		1 & 0_{d_X,d_X-1}\\
		0_{d_X-1,d_X} & 0_{d_X-1,d_X-1}
	\end{pmatrix} \in M_{d_X}(\complexNumbers),
\end{equation}
where $0_{s,t}$ is the zero matrix in $M_{s,t}(\complexNumbers)$, and take
\begin{equation}
	\psi_j = 0 \in M_{d_X}(\complexNumbers) \text{ for } 2\leq j \leq l-1.
\end{equation}
Then
\begin{align*}
	I_{d_X} - \sum_{j < l} \psi_j \psi_j^\dag &= \begin{pmatrix}
		1 & 0_{d_X,d_X-1}\\
		0_{d_X-1,d_X} & I_{d_X-1}
	\end{pmatrix} - \begin{pmatrix}
		1 & 0_{d_X,d_X-1}\\
		0_{d_X-1,d_X} & 0_{d_X-1,d_X-1}
	\end{pmatrix} - 0 \\
	&= \begin{pmatrix}
		0 & 0_{d_X,d_X-1}\\
		0_{d_X-1,d_X} & I_{d_X-1}
	\end{pmatrix}. 
\end{align*}
This matrix is positive semidefinite and its rank is $d_X-1$, which is smaller than $d_X$. Therefore,
\begin{equation}
	(\psi_1,\dots,\psi_{l-1}) \in \partial B^{l-1}_{d_X,d_X}.
\end{equation}
Define
\begin{equation}
	\psi_l \coloneqq I_{d_X} - \sum_{j<l} \psi_j \psi_j^\dag.
\end{equation}
Then $F(\psi_1,\dots,\psi_{l-1})$ is the operator $A$ represented by $\psi = (\psi_1,\dots,\psi_l,0,\dots,$ $0)\allowbreak \in S^{d_Y}_{d_X,d_X}$, that is, $A = \pi(\psi)$. Of course, $\pi(\psi) \in K^l$. Since $\psi_l\neq 0$, we have that $\pi(\psi) \not \in K^{l-1}$. This proves that $F(\partial B^{l-1}_{d_X,d_X}) \not \subseteq K^{l-1}$.

Even though $F^l$ fails to be a characteristic map that obtains $K^l$ from $K^{l-1}$, we can adapt the sequence of inclusions so that $F^l$ obtains $K^l$ from another subset of $K^l$. $F^l$ fails because there were points in $F^l(\partial B^{l-1}_{d_X,d_X})$ that weren't in $K^{l-1}$. We can then augment $K^{l-1}$ with those missing points, which leads to the definition of a set $\widetilde{K}^{l-1}$.

\begin{definition}{}{Ktilde}
	For $2 \leq l \leq d_Y$, we define the set
	\begin{multline}
		\widetilde{K}^{l-1} \coloneqq\\ \pi\left(\{ (\psi_j)_j \in S^{d_Y}_{d_X,d_X} \mid \rank(\psi_l) < d_X, \ \forall j>l \ \psi_j = 0\}\right).
	\end{multline}
\end{definition}
Of course, we have the inclusion
\begin{equation}
	K^{l-1} \subseteq \widetilde{K}^{l-1} \subseteq K^l.
\end{equation}
The inclusion $\widetilde{K}^{l-1} \subseteq K^l$ comes from the condition $\forall j>l \ \psi_j = 0$. The inclusion $K^{l-1} \subseteq \widetilde{K}^{l-1}$ consists of the special case where $\rank(\psi_l) = 0$, that is, $\psi_j = 0$. Notice that $\widetilde{K}^{l-1} = K^{l-1}$ for $d_X = 1$. This is because for $1\times 1$ matrices the only ranks possible are 0 and 1. Therefore, the inequality $\rank(\psi_l)<1$ reduces to $\psi_l = 0$.

With this new set, our sequence of inclusions becomes
\begin{equation}
	\label{equation: sequence of inclusions}
	K^1 \subseteq \widetilde{K}^1 \subseteq \dots \subseteq K^{d_Y-1} \subseteq \widetilde{K}^{d_Y-1} \subseteq K^{d_Y} = \bigcup_{d_X / d_Y \leq k\leq d_X} \psdii(X,Y)_k .
\end{equation}
Let's analyze the beginning of this sequence.
\begin{proposition}{}{}
	The first space in the sequence of inclusions of equation \ref{equation: sequence of inclusions} is
	\begin{equation}
		K^1= \{\id_{X^*} \otimes \ket{y_1}\bra{y_1}\}.
	\end{equation}
\end{proposition}
\begin{proof}
	An element $A \in K^1$ is an operator represented by some $(\psi_j)_j \in S^{d_Y}_{d_X,d_X}$ such that all $\psi_j$ are 0 except possibly $\psi_1$. The representative must satisfy the identity
	\begin{equation}
		\sum_j \psi_j \psi_j^\dag = I_{d_X}.
	\end{equation} 
	Since $\psi_1$ is the only possible non zero matrix, we have
	\begin{equation}
		\psi_1 \psi_1^\dag = I_{d_X}.
	\end{equation}
	Since $\psi_1 \in M_{d_X}(\complexNumbers)$ is a square matrix, this says that $\psi_1$ is a unitary matrix:
	\begin{equation}
		\psi_1 \in \U(d_X).
	\end{equation}
	Then the matrix of $A$ is (see equation \ref{equation: [A]})
	\begin{equation}
		[A] = \begin{pmatrix}
			\psi_1 \psi_1^\dag & 0 & \dots & 0\\
			0 & 0 & \dots & 0\\
			\vdots & \vdots & \ddots & 0\\
			0 & 0 & 0 & 0
		\end{pmatrix} = \begin{pmatrix}
			I_{d_X} & 0 & \dots & 0\\
			0 & 0 & \dots & 0\\
			\vdots & \vdots & \ddots & 0\\
			0 & 0 & 0 & 0
		\end{pmatrix} , 
	\end{equation}
	where each 0 is the zero matrix of $M_{d_X}(\complexNumbers)$. Writing $A$ with respect to the basis $\ket{x^i}\bra{x^{i'}} \otimes \ket{y_j}\bra{y_{j'}}$, we have
	\begin{align*}
		A &= \sum_{i,i',j,j'} (\psi_j \psi_{j'}^\dag)_{i,i'} \ket{x^i}\bra{x^{i'}} \otimes \ket{y_j}\bra{y_{j'}}\\
		&= \sum_{i,i'} (\psi_1 \psi_1^\dag)_{i,i'} \ket{x^i}\bra{x^{i'}} \otimes \ket{y_1}\bra{y_1}\\ 
		&= \sum_{i,i'} (I_{d_X})_{i,i'} \ket{x^i}\bra{x^{i'}} \otimes \ket{y_1}\bra{y_1}\\ 
		&= \sum_{i,i'} \delta_{i,i'} \ket{x^i}\bra{x^{i'}} \otimes \ket{y_1}\bra{y_1}\\
		&= \sum_i \ket{x^i}\bra{x^i} \otimes \ket{y_1}\bra{y_1}\\ 
		&= \id_{X^*} \otimes \ket{y_1}\bra{y_1}.
	\end{align*}
	Therefore $K^1 = \{\id_{X^*} \otimes \ket{y_1} \bra{y_1}\}$.
\end{proof}

Notice that $K^1$ is a singleton, which is also the case for $\complexNumbers P^0$. The next space to study would be $\widetilde{K}^1$, but we don't yet have one or more characteristic maps that relate $K^1$ to $\widetilde{K}^1$.

We'll prove next that the functions $F^l$ are characteristic maps that obtain $K^l$ from $\widetilde{K}^{l-1}$, for $2\leq l \leq d_Y$. We still need to understand how the topology of $\widetilde{K}^{l-1}$ is related to the topology of $K^{l-1}$, for $2\leq l \leq d_Y$. This is still an open problem that we want to address in the future. It may be the case that we need to extended our sequence of inclusions even further, and construct more characteristic maps. This extended sequence may be a skeleton for a CW structure of the closure of the set of extreme points of $\psdii(X,Y)$.

\subsection{Proof that $F^l$ is a characteristic map with respect to $\widetilde{K}^{l-1}$}

As proved earlier, $F^l$ can't be a characteristic map that obtains $K^l$ from $K^{l-1}$, except for trivial cases. For this reason, we introduced new sets $\widetilde{K}^{l-1}$. In this section we prove that $F^l$ is a characteristic map that obtains $K^l$ from $\widetilde{K}^{l-1}$. From proposition \ref{proposition:F^l is continuous} we already know that $F^l \colon B^{l-1}_{d_X,d_X} \to K^l$ is a continuous function. For it to be a characteristic map that obtains $K^l$ from $\widetilde{K}^{l-1}$, it must satisfy two more properties:
\begin{itemize}
	\item $F^l(\partial B^{l-1}_{d_X,d_X}) \subseteq \widetilde{K}^{l-1}$;
	
	\item $F^l$ restricts to a homeomorphism $F^l \colon \interior{B^{l-1}_{d_X,d_X}} \to K^l \setminus \widetilde{K}^{l-1}$.
\end{itemize}
The set $\widetilde{K}^{l-1}$ was defined so that the first item is true. We prove this item first. We prove a slightly stronger statement. We have not just an inclusion, but an equality.
\begin{proposition}{}{Fl(dB) = Ktilde l-1}
	For $2 \leq l \leq d_Y$, we have that
	\begin{equation}
		F^l(\partial B^{l-1}_{d_X,d_X}) = \widetilde{K}^{l-1}.
	\end{equation}
\end{proposition}
\begin{proof}
	From the definition of $F^l$ is immediate that $F^l(\partial B^{l-1}_{d_X,d_X}) \subseteq K^l$. An element $A \in F^l(\partial B^{l-1}_{d_X,d_X})$ is of the form $A = \pi(\psi)$, where $\psi = (\psi_j)_{j=1,\dots,d_Y} \in S^{d_Y}_{d_X,d_X}$, $(\psi_1,\dots,\psi_{l-1}) \in \partial B^{l-1}_{d_X,d_X}$,
	\begin{equation}
		\psi_l = \left( \sum_{j<l} \psi_j \psi_j^\dag \right)^{1/2},
	\end{equation}
	and $\psi_j = 0$ for $j > l$. From theorem \ref{theorem:interior and boundary of generalized ball} and since $(\psi_1,\dots,\psi_{l-1}) \in \partial B^{l-1}_{d_X,d_X}$, we know that
	\begin{equation}
		\rank\left( \sum_{j<l} \psi_j \psi_j^\dag \right) < d_X.
	\end{equation}
	The square root also has the same rank, so
	\begin{equation}
		\rank(\psi_l)<d_X.
	\end{equation}
	This proves that $F^l(\partial B^{l-1}_{d_X,d_X}) \subseteq \widetilde{K}^{l-1}$.
	
	Now let's prove the other inclusion. Let $A \in \widetilde{K}^{l-1}$, then there exists $\psi = (\psi_j)_j \in S^{d_Y}_{d_X,d_X}$ such that $A = \pi (\psi)$, $\rank(\psi_l) < d_X$ and $\psi_j = 0$ for $j > l$. We use $\psi$ to construct another representative $\psi'$ such that $(\psi'_1,\dots,\psi'_{l-1}) \in \partial B^{l-1}_{d_X,d_X}$ and $A = F^l(\psi'_1,\dots,\psi'_{l-1})$. Such representative is obtained by the condition that $\psi'_l$ should be positive semidefinite. Take a polar decomposition of $\psi_l$:
	\begin{equation}
		\psi_l = (\psi_l \psi_l^\dag)^{1/2} U_l,
	\end{equation}
	where $U_l \in \U(d_X)$ is a unitary matrix. Since $\psi_l$ doesn't have full rank, the unitary matrix $U_l$ isn't unique. Then define $\psi' = (\psi'_j)_j$ as
	\begin{equation}
		\label{equation: psi'j = psij Uldag}
		\psi'_j \coloneqq \psi_j U_l^\dag,
	\end{equation}
	for $1\leq j \leq d_Y$. In this way,
	\begin{equation}
		\psi'_l = (\psi_l \psi_l^\dag)^{1/2},
	\end{equation}
	so $\psi'_l$ is positive semidefinite. It is immediate that $\psi'_j = 0$ for $j>l$, since the same is true for $\psi_j$. Also, we have that
	\begin{equation}
		\psi'_j {\psi'}_{j'}^\dag = \psi_j U_l^\dag U_l \psi_{j'}^\dag = \psi_j \psi_{j'}^\dag,
	\end{equation}
	for any $j,j'$. From this it follows that
	\begin{equation}
		\sum_j \psi'_j {\psi'}_j^\dag = \sum_j \psi_j \psi_j^\dag = I_{d_X},
	\end{equation}
	where the last equality comes from $\psi \in S^{d_Y}_{d_X,d_X}$. Of course, only the terms with $j\leq l$ contribute to the sum, since $\psi'_j = 0$ for $j>l$. We defined $\psi'$ so that $\psi'_j = \psi_j U_l^\dag$. Then, from proposition \ref{proposition:relation of pi}, we know that $\pi(\psi') = \pi(\psi) = A$, that is, $\psi'$ is another representative for $A$. The equation $\sum_j \psi'_j {\psi'}_j^\dag = I_{d_X}$ together with $\psi'_l$ being positive semidefinite implies that
	\begin{equation}
		\psi'_l = \left( \sum_{j < l} \psi'_j {\psi'}_j^\dag \right)^{1/2}.
	\end{equation}
	The equation $\sum_j \psi'_j {\psi'}_j^\dag = I_{d_X}$ also implies that
	\begin{equation}
		I_{d_X} - \sum_{j<l} \psi'_j {\psi'}_j^\dag = (\psi'_l)^2,
	\end{equation}
	which is positive semidefinite, so $(\psi'_1,\dots,\psi'_{l-1}) \in B^{l-1}_{d_X,d_X}$. It follows that
	\begin{equation}
		A = \pi\left(\psi'_1,\dots,\psi'_{l-1},\left( \sum_{j < l} \psi'_j {\psi'}_j^\dag \right)^{1/2}, 0, \dots, 0 \right).
	\end{equation}
	This is the same expression in the definition of $F^l$, so
	\begin{equation}
		A = F^l(\psi'_1,\dots,\psi'_{l-1}).
	\end{equation}
	It remains to prove that $(\psi'_1,\dots,\psi'_{l-1}) \in \partial B^{l-1}_{d_X,d_X}$. We need to prove that $\rank\left(I_{d_X} - \sum_{j<l} \psi'_j {\psi'}_j^\dag \right) < d_X$. The previous results shows that
	\begin{equation}
		I_{d_X} - \sum_{j<l} \psi'_j {\psi'}_j^\dag = {\psi'}_l^2 = \psi_l \psi_l^\dag.
	\end{equation}
	We already know that $\rank(\psi_l) < d_X$. Also, for any matrix $B$, its rank is the same of $B B^\dag$. From this it follows that $\rank(\psi_l \psi_l^\dag) < d_X$, so $(\psi'_1,\dots, \psi'_{l-1}) \in \partial B^{l-1}_{d_X,d_X}$. This proves that $\widetilde{K}^{l-1} \subseteq F^l(\partial B^{l-1}_{d_X,d_X})$.
\end{proof} 

Next we prove the last item. We have to prove that $F^l$ restricts to a homeomorphism $F^l \colon \interior{B^{l-1}_{d_X,d_X}} \to K^l \setminus \widetilde{K}^{l-1}$. We already know that $F^l$ is continuous without taking a restriction. A restriction of a continuous function is always continuous. It remains then to show that we have a bijection with continuous inverse. Being a bijection, the condition that the inverse is continuous is equivalent to $F^l$ being an open function, that is, it sends open sets to open sets.

\begin{proposition}{}{F^l restricts to bijection}
	For $2 \leq l \leq d_Y$, the function $F^l$ defined in \ref{definition:F^l} restricts to a bijection
	\begin{equation}
		F^l \colon \interior{B^{l-1}_{d_X,d_X}} \to K^l \setminus \widetilde{K}^{l-1}.
	\end{equation}
\end{proposition}
\begin{proof}
	We first prove that $F^l (\interior{B^{l-1}_{d_X,d_X}}) \subseteq K^l \setminus \widetilde{K}^{l-1}$. From the definition of $F^l$ it is immediate that $F^l (\interior{B^{l-1}_{d_X,d_X}}) \subseteq K^l$. Then we have to prove that the image doesn't have any point in $\widetilde{K}^{l-1}$. Let $(\psi_1, \dots, \psi_{l-1}) \in \interior{B^{l-1}_{d_X,d_X}}$ and $A = F^l(\psi_1, \dots, \psi_{l-1})$. Since $(\psi_1, \dots, \psi_{l-1}) \in B^{l-1}_{d_X,d_X}$, the matrix $I_{d_X} - \sum_{j<l} \psi_j \psi_j^\dag$ is positive semidefinite. Define
	\begin{equation}
		\psi_l \coloneqq \left(I_{d_X} - \sum_{j<l} \psi_j \psi_j^\dag \right)^{1/2},
	\end{equation}
	then we have the element $\psi = (\psi_j)_j \in S^{d_Y}_{d_X,d_X}$, where $\psi_j \coloneqq 0$ for $j > l$.
	From the definition of $F^l$, we have that 
	\begin{equation}
		A = \pi(\psi_1,\dots,\psi_l,0,\dots,0) = \pi(\psi).
	\end{equation}
	For $A$ not to be an element of $\widetilde{K}^{l-1}$ we just need that $\rank(\psi_l) = d_X$. Since $(\psi_1,\dots,\psi_{l-1}) \in \interior{B^{l-1}_{d_X,d_X}}$, from theorem \ref{theorem:interior and boundary of generalized ball} we know that
	\begin{equation}
		\rank\left(  I_{d_X} - \sum_{j<l} \psi_j \psi_j^\dag \right) = d_X.
	\end{equation}
	Since $\psi_l$ is the square root of this matrix, it has the same rank, so
	\begin{equation}
		\rank(\psi_l) = d_X.
	\end{equation}
	This proves that $A \in K^l \setminus \widetilde{K}^{l-1}$. Therefore
	\begin{equation}
		F^l (\interior{B^{l-1}_{d_X,d_X}}) \subseteq K^l \setminus \widetilde{K}^{l-1} .
	\end{equation}
	
	Now let's prove the other inclusion. Let $A \in K^l \setminus \widetilde{K}^{l-1}$, then there exists $\psi = (\psi_j)_j \in S^{d_Y}_{d_X,d_X}$ such that
	\begin{equation}
		\rank(\psi_l) = d_X,
	\end{equation}
	$\psi_j = 0$ for $j > l$, and
	\begin{equation}
		A = \pi(\psi) = \pi(\psi_1,\dots,\psi_l,0,\dots,0).
	\end{equation}
	We just need to construct another representative $\psi'$ such that $\psi'_l$ is positive semidefinite. We'll repeat the same arguments done in the proof of proposition \ref{proposition:Fl(dB) = Ktilde l-1}. We take a polar decomposition of $\psi_l$:
	\begin{equation}
		\psi_l = (\psi_l \psi_l^\dag)^{1/2} U_l^\dag,
	\end{equation}
	where $U_l \in \U(d_X)$ is a unitary matrix. This time the matrix $U_l$ is unique, because $\psi_l$ has full rank. Also, as in equation \ref{equation: psi'j = psij Uldag}, we define
	\begin{equation}
		\psi'_j \coloneqq \psi_j U_l^\dag,
	\end{equation}
	for every $j$. The same arguments made in proposition \ref{proposition:Fl(dB) = Ktilde l-1} shows that
	\begin{equation}
		(\psi'_1, \dots, \psi'_{l-1}) \in B^{l-1}_{d_X,d_X}
	\end{equation}
	and
	\begin{equation}
		A = F^l(\psi'_1,\dots,\psi'_{l-1}).
	\end{equation}
	They also show that
	\begin{equation}
		\psi'_l = \left( I_{d_X} - \sum_{j<l} \psi'_j {\psi'}_j^\dag \right)^{1/2}.
	\end{equation}
	For $(\psi'_1,\dots, \psi'_{l-1})$ to be in $\interior{B^{l-1}_{d_X,d_X}}$, we just need to have $\rank({\psi'}_l^2) = d_X$. The rank of $\psi'_l$ and ${\psi'}_l^2$ are the same, so we need to have $\rank(\psi'_l) = d_X$. Since $\psi'_l = \psi_l U_l^\dag$ and $U_l^\dag$ is invertible, we have
	\begin{equation}
		\rank(\psi'_l) = \rank(\psi_l) = d_X.
	\end{equation}
	This proves that
	\begin{equation}
		K^l \setminus \widetilde{K}^{l-1} \subseteq F^l (\interior{B^{l-1}_{d_X,d_X}}) .
	\end{equation}
	This also concludes that $F^l$ restricts to a surjection
	\begin{equation}
		F^l \colon \interior{B^{l-1}_{d_X,d_X}} \to K^l \setminus \widetilde{K}^{l-1}.
	\end{equation}
	
	Next we prove that this restriction is injective. Let $(\psi_1,\dots,\psi_{l-1}) , (\psi'_1,\dots,$ $\psi'_{l-1}) \in \interior{B^{l-1}_{d_X,d_X}}$ such that
	\begin{equation}
		F^l (\psi_1,\dots,\psi_{l-1}) = F^l (\psi'_1,\dots,\psi'_{l-1}).
	\end{equation}
	Define
	\begin{equation}
		\psi_l \coloneqq \left( I_{d_X} - \sum_{j < l} \psi_j \psi_j^\dag \right)^{1/2},
	\end{equation}
	\begin{equation}
		\psi'_l \coloneqq \left( I_{d_X} - \sum_{j < l} \psi'_j {\psi'}_j^\dag \right)^{1/2}.
	\end{equation}
	From the definition of $F^l$, we have that
	\begin{equation}
		\pi(\psi_1,\dots,\psi_l,0,\dots,0) = \pi(\psi'_1,\dots,\psi'_l,0,\dots,0).
	\end{equation}
	Then proposition \ref{proposition:relation of pi} says that there exists some unitary matrix $V \in \U(d_X)$ such that
	\begin{equation}
		\psi'_j = \psi_j V^\dag,
	\end{equation}
	for every $j$. Using this equation we get that
	\begin{equation}
		\psi'_j {\psi'}_{j'}^\dag = \psi_j V^\dag V \psi_{j'}^\dag = \psi_j \psi_{j'}^\dag,
	\end{equation}
	for any $j,j'$. Applying this result to the definitions of $\psi_l$ and $\psi'_l$, we get that
	\begin{equation}
		\psi'_l = \left( I_{d_X} - \sum_{j < l} \psi'_j {\psi'}_j^\dag \right)^{1/2} = \left( I_{d_X} - \sum_{j < l} \psi_j \psi_j^\dag \right)^{1/2} = 
		\psi_l .
	\end{equation}
	But we also have that $\psi'_l = \psi_l V^\dag$, so
	\begin{equation}
		\psi_l = \psi_l V^\dag.
	\end{equation}
	Since $(\psi_1,\dots,\psi_{l-1}) \in \interior{B^{l-1}_{d_X,d_X}}$, by theorem \ref{theorem:interior and boundary of generalized ball} we have that
	\begin{equation}
		\rank\left( I_{d_X} - \sum_{j < l} \psi_j \psi_j^\dag \right) = d_X .
	\end{equation}
	The square root has the same rank, so
	\begin{equation}
		\rank(\psi_l) = d_X,
	\end{equation}
	that is, $\psi_l$ is invertible. Multiplying the equation $\psi_l = \psi_l V^\dag$ by $\psi_l^{-1}$ it follows that
	\begin{equation}
		V = I_{d_X}.
	\end{equation}
	Then we have that $\psi'_j = \psi_j V^\dag = \psi_j$, for every $j$. This proves that 
	\begin{equation}
		(\psi_1,\dots,\psi_{l-1}) = (\psi'_1,\dots,\psi'_{l-1}).
	\end{equation}
	Therefore, the restriction
	\begin{equation}
		F^l \colon \interior{B^{l-1}_{d_X,d_X}} \to K^l \setminus \widetilde{K}^{l-1}
	\end{equation}
	is injective.
\end{proof}

We proved that $F^l$ restricts to a continuous bijection
\begin{equation}
	F^l \colon \interior{B^{l-1}_{d_X,d_X}} \to K^l \setminus \widetilde{K}^{l-1}.
\end{equation}
Next we prove that this restriction is an open function. This is the same as proving that its inverse is continuous. From this we get that the restricted $F^l$ is a homeomorphism.

\begin{proposition}{}{F^l is open}
	For $2 \leq l \leq d_Y$, the function $F^l$ from definition \ref{definition:F^l} restricts to a function $F^l \colon \interior{B^{l-1}_{d_X,d_X}} \to K^l \setminus \widetilde{K}^{l-1}$. This restriction is an open function, that is, it sends open sets to open sets.
\end{proposition}
\begin{proof}
	Let $O \subseteq \interior{B^{l-1}_{d_X,d_X}}$ be an open subset. We want to show that $F^l (O)$ is an open subset of $K^l \setminus \widetilde{K}^{l-1}$. By theorem \ref{theorem:pi is open}, the restriction $\pi \colon \pi^{-1}(K^l \setminus \widetilde{K}^{l-1})$ $ \to K^l \setminus \widetilde{K}^{l-1}$ is an open function. Of course, it is also surjective, so we have that
	\begin{equation}
		F^l(O) = \pi \pi^{-1} F^l (O).
	\end{equation}
	Therefore, if $\pi^{-1} F^l (O)$ is open in $\pi^{-1}(K^l \setminus \widetilde{K}^{l-1})$ then $F^l (O)$ is open in $K^l \setminus \widetilde{K}^{l-1}$. Let's find another characterization of $\pi^{-1} F^l(O)$. Let $(\psi_j)_j \in \pi^{-1}(K^l \setminus \widetilde{K}^{l-1})$, that is, $\psi_l$ is invertible and $\psi_j = 0$ for $j>l$. We have that
	\begin{equation}
		\label{equation: psi in pi-1Fl iff}
		(\psi_j)_j \in \pi^{-1} F^l(O) \iff \pi((\psi_j)_j) \in F^l(O).
	\end{equation}
	What characterizes $F^l$ is the usage of a positive semidefinite matrix in the $l$ component. Since $\psi_l$ is invertible, we can uniquely find an equivalent representative whose $l$ component is positive semidefinite. We take a polar decomposition of $\psi_l$:
	\begin{equation}
		\psi_l = (\psi_l \psi_l^\dag)^{1/2} U_l,
	\end{equation}
	where $U_l \in \U(d_X)$ is an unitary matrix. Since $\psi_l$ has full rank, the unitary matrix $U_l$ is unique. We are more interested on its inverse $U_l^\dag$. From the equation above it follows immediately that
	\begin{equation}
		\label{equation: Uldag}
		U_l^\dag = \psi_l^{-1} (\psi_l \psi_l^\dag)^{1/2}.
	\end{equation}
	Then we have the sequence $(\psi_j U_l^\dag)_j \in S^{d_Y}_{d_X,d_X}$. By proposition \ref{proposition:relation of pi}, we have that
	\begin{equation}
		\pi((\psi_j)_j) = \pi((\psi_j U_l^\dag)_j) = \pi(\psi_1 U_l^\dag, \dots, \psi_{l-1} U_l^\dag, (\psi_l \psi_l^\dag)^{1/2}, 0, \dots, 0).
	\end{equation}
	Since $(\psi_j)_j \in S^{d_Y}_{d_X,d_X}$, we have that
	\begin{equation}
		\sum_j \psi_j \psi_j^\dag = I_{d_X}.
	\end{equation}
	From this it follows that
	\begin{equation}
		I_{d_X} - \sum_{j < l} (\psi_j U_l^\dag) (\psi_j U_l^\dag)^\dag = I_{d_X} - \sum_{j < l} \psi_j \psi_j^\dag = \psi_l \psi_l^\dag,
	\end{equation}
	which is positive semidefinite and invertible, so
	\begin{equation}
		(\psi_1 U_l^\dag, \dots, \psi_{l-1} U_l^\dag) \in \interior{B^{l-1}_{d_X,d_X}}.
	\end{equation}
	Taking the square root, we get that
	\begin{equation}
		(\psi_l \psi_l^\dag)^{1/2} = \left(I_{d_X} - \sum_{j<l} (\psi_j U_l^\dag) (\psi_j U_l^\dag)^\dag \right)^{1/2}.
	\end{equation}
	This shows that
	\begin{equation}
		\pi((\psi_j U_l^\dag)_j) = F^l (\psi_1 U_l^\dag, \dots, \psi_{l-1} U_l^\dag).
	\end{equation}
	Using this result in equation \ref{equation: psi in pi-1Fl iff}, we get that
	\begin{equation}
		(\psi_j)_j \in \pi^{-1} F^l(O) \iff F^l (\psi_1 U_l^\dag, \dots, \psi_{l-1} U_l^\dag) \in F^l(O).
	\end{equation}
	We supposed that $O \subseteq \interior{B^{l-1}_{d_X,d_X}}$. By proposition \ref{proposition:F^l restricts to bijection}, the restriction $F^l \colon $ $ \interior{B^{l-1}_{d_X,d_X}} \to K^l \setminus \widetilde{K}^{l-1}$ is a bijection. Therefore, we have that
	\begin{equation}
		\label{equation: psi in pi-1Fl iff (psi1Uldag,...) in O}
		(\psi_j)_j \in \pi^{-1} F^l(O) \iff (\psi_1 U_l^\dag, \dots, \psi_{l-1} U_l^\dag) \in O.
	\end{equation}
	The right hand side can be expressed as the preimage of $O$ by a continuous function. Define the function
	\begin{align*}
		h \colon \pi^{-1}(K^l \setminus \widetilde{K}^{l-1}) &\longrightarrow \interior{B^{l-1}_{d_X,d_X}}\\
		(\psi_j)_j &\longmapsto (\psi_1 U_l^\dag, \dots, \psi_{l-1} U_l^\dag),
	\end{align*}
	where $U^\dag$ is given by equation \ref{equation: Uldag}. It is known that matrix multiplication and taking the inverse are continuous functions. Also, we proved in proposition \ref{proposition:square root of psd operators is continuous} that the square root of positive semidefinite operators is continuous. From this it follows that $U_l^\dag$ is a continuous function of $(\psi_j)_j$. Then $h$ is also a continuous function. Equation \ref{equation: psi in pi-1Fl iff (psi1Uldag,...) in O} then shows that
	\begin{equation}
		\pi^{-1} F^l (O) = h^{-1}(O).
	\end{equation}
	Since $O$ is open in $\interior{B^{l-1}_{d_X,d_X}}$ and $h$ is continuous, it follows that $h^{-1}(O)$ is open in $\pi^{-1}(K^l \setminus \widetilde{K}^{l-1})$. This shows that $\pi^{-1} F^l (O)$ is open in $\pi^{-1}(K^l \setminus \widetilde{K}^{l-1})$. Since $\pi \colon \pi^{-1}(K^l \setminus \widetilde{K}^{l-1}) \to K^l \setminus \widetilde{K}^{l-1}$ is open, it follows that $F^l(O)$ is open. This concludes that $F^l \colon \interior{B^{l-1}_{d_X,d_X}} \to K^l \setminus \widetilde{K}^{l-1}$ is an open function.
\end{proof}

Combining the previous results we conclude that $F^l$ is a characteristic map with respect to $\widetilde{K}^{l-1}$.

\begin{theorem}{}{}
	For $2 \leq l \leq d_Y$, the function $F^l$ of definition \ref{definition:F^l} is a characteristic map with the following properties:
	\begin{itemize}
		\item $F^l(\partial B^{l-1}_{d_X,d_X}) = \widetilde{K}^{l-1}$;
		
		\item $F^l$ restricts to a homeomorphism $F^l \colon \interior{B^{l-1}_{d_X,d_X}} \to K^l \setminus \widetilde{K}^{l-1}$.
	\end{itemize}
\end{theorem}
\begin{proof}
	The first item was proved in proposition \ref{proposition:Fl(dB) = Ktilde l-1}. In proposition \ref{proposition:F^l is continuous} we proved that $F^l$ is continuous. In proposition \ref{proposition:F^l restricts to bijection} we proved that $F^l$ restricts to a bijection $F^l \colon \interior{B^{l-1}_{d_X,d_X}} \to K^l \setminus \widetilde{K}^{l-1}$. Since the non restricted $F^l$ is continuous, its restriction is also continuous. Therefore, the restriction is a continuous bijection. In proposition \ref{proposition:F^l is open} we proved that this restriction is an open function. This is equivalent to say that its inverse is continuous. Therefore the restriction is a homeomorphism, which proves the second item. 
\end{proof}

\cleardoublepage
\chapter{Conclusions and Further Directions}
\label{chapter: conclusions}

We succeeded in constructing many examples of extreme CPTP or UCPTP maps. Were constructed examples of extreme CPTP maps for each Choi-rank allowed by the bounds of corollary \ref{corollary:bounds on rank of extreme cptp}. This shows that these bounds are optimal. For extreme UCPTP maps, the non trivial case is for Choi-rank at least 2. There are extreme UCPTP maps with Choi-rank 2 only for dimensions $d \geq 4$. We constructed examples for each dimension in theorem \ref{theorem:extreme ucptp rank 2}. It only remains to construct examples or prove their nonexistence for Choi-rank at least 3. We constructed hundreds of examples of extreme UCPTP maps with Choi-rank at least 3 in section \ref{section: extreme ucptp with choi rank at least 3}. In section \ref{section: method 1 choice of parameters} was shown a formula that succeeded to obtain an extreme UCPTP map for all tested cases except $(d,r) = (5,7)$. Further work has to be done to prove that the presented methods work for an infinite number of dimensions $d$, or to find a simpler way to produce examples. Understanding better the vertices of the convex polytope from proposition \ref{proposition:method 1 convex polytopes} may help on finding simpler examples.

We wanted to understand which operations preserves the extremality of quantum channels. It is known that composition may not preserve extremality. We then investigated if the tensor product preserves it. We showed in chapter \ref{section: Extremality of the tensor product} that the tensor product of extreme CPTP maps is an extreme CPTP map. The same is true for UCP maps, by duality. For UCPTP maps we showed that extremality may not be preserved by the tensor product. From the hundreds of examples of extreme UCPTP maps that we constructed, many of them have high Choi-rank. The tensor product of a high rank extreme UCPTP map with itself isn't an extreme point of the set of UCPTP maps.

For the category of CPTP maps, not much is known other than that it is a semicartesian category. We tried to investigate its limits and colimits. The only successful case was the classification of binary products. The category only has trivial binary products, in which one of the Hilbert spaces have dimension 0 or 1. To prove this, it was used Algebraic Topology, more precisely the Homology groups of the complex projective space. We could try to apply the same technique for other limits or colimits, which motivated studying the topology of the sets of CPTP maps, of its extreme points and of the closure of the set of extreme points.

The study of the closure of the set of extreme points has its own motivations from Quantum Information Theory. It was cited as an open problem by Ruskai \cite{ruskai2007open}. We tried to find a CW decomposition of the closure of the extreme points. We succeeded in finding some cells which resemble the CW structure of the complex projective space. We hope that these cells can be completed to a full CW decomposition, but further study has to be done.

For further directions, much of Quantum Information Theory admits techniques from Real Algebraic Geometry. The set of extreme CPTP maps and its closure, both are semi-algebraic sets. The closure is a closed and bounded semi-algebraic set. There are algorithms from Real Algebraic Geometry that can compute a CW decomposition of such a set by first computing a Cylindrical Algebraic Decomposition (CAD) \cite{algorithms_in_RAG}. Unfortunately, the best general algorithms for CAD are at least double exponential in time. In practice, we can only run such algorithms for small problems. Sometimes there are faster algorithms which are specialized to a subset of the  problems. For example, the boolean satisfiability problem (SAT) is known to be NP-complete. We still don't have any algorithm for SAT that runs in polynomial time, but we have such algorithms for 2-SAT and HORNSAT, which solve the satisfiability problem for special types of boolean formulas. We can try to find faster algorithms, specialized to problems from Quantum Information Theory. Computing a CW decomposition and the homology groups for some dimensions could then help guess a solution for the general case.

Real Algebraic Geometry also has the algorithm of quantifier elimination. Quantifier elimination may be used to prove that a set is empty or not, without having to construct an example in the case that it is not empty. Unfortunately, we don't have fast general algorithms for it. We could try to use quantifier elimination or another algorithm to try to prove that there are no extreme UCPTP maps for some dimensions and Choi-ranks.

Another possibility for studying the closure of the extreme points of the CPTP maps would be to use equivariant cohomology. The set
\begin{equation}
	K_r \coloneqq \bigcup_{d_X / d_Y \leq s \leq r} \psdii(X,Y)_s ,
\end{equation}
with
\begin{equation}
	\frac{d_X}{d_Y}  \leq r \leq d_X,
\end{equation}
is the closure of the set of extreme points with rank at most $r$. It is homeomorphic to a quotient of manifolds:
\begin{equation}
	K_r \cong V_{r d_Y, d_X}(\complexNumbers) / \U(r),
\end{equation}
where $V_{r d_Y, d_X}(\complexNumbers)$ is a (compact) Stiefel manifold. We have a quotient of a smooth manifold ($V_{r d_Y, d_X}(\complexNumbers)$) by a compact Lie Group ($\U(r)$), but the action isn't free. There is a trick to replace this quotient by another whose action is free. The cohomology of this other quotient is then the equivariant cohomology. Some information about the cohomology of $K_r$ may be obtained from its equivariant cohomology.

For compact smooth manifolds, we can find a CW decomposition using Morse theory. There are generalizations of Morse theory for other spaces, such as stratified spaces. Another possibility would be to guess some Morse function and apply some generalization of Morse theory to the closure of the set of extreme points.

Yet another approach for finding a CW decomposition of the closure of the set of extreme points would be to filter the space by rank. Let
\begin{equation}
	K_r \coloneqq \bigcup_{\frac{d_X}{d_Y} \leq s \leq r} \psdii(X,Y)_s ,
\end{equation}
where
\begin{equation}
	\frac{d_X}{d_Y} \leq r \leq d_X .
\end{equation}
Then we can filter the closure of the set of extreme points as
\begin{equation}
	K_{\left\lceil \frac{d_X}{d_Y} \right\rceil} \subseteq \dots \subseteq K_{d_X} = \bigcup_{\frac{d_X}{d_Y} \leq s \leq d_X} \psdii(X,Y)_s .
\end{equation}
This approach has the convenience that the projection
\begin{equation}
	\pi \colon S^{d_Y}_{d_X, r} \to K_r
\end{equation}
from section \ref{section: a surjective map from the purification} is well behaved for operators in $K_r \setminus K_{r-1}$. This is because these operators are parameterized by matrices $\Psi$ with maximal rank, where
\begin{equation}
	\Psi = \begin{pmatrix}
		\psi_1 \\
		\vdots \\
		\psi_{d_Y}
	\end{pmatrix}
\end{equation}
and $(\psi_j)_{j=1, \dots, d_y} \in S^{d_Y}_{d_X, r}$. Having maximal rank implies that we can take a polar decomposition uniquely. A possibility would then be to decompose $K_r$ into cells recursively, starting from $K_{\left\lceil \frac{d_X}{d_Y} \right\rceil}$, and use the projection with domain $S^{d_Y}_{d_X, r}$ to parameterize $K_r$.

Finally, it is known that the set of CPTP maps is stratified into smooth manifolds by Choi-rank. We can try to find a similar stratification into smooth manifolds for the case of UCPTP maps.

\appendix

\cleardoublepage
\chapter{Tensor product of vector spaces}
\label{appendix: tensor product}

In Quantum Mechanics, a bipartite system is described by the tensor product of two complex Hilbert spaces. For this reason, tensor products are widely used. In this appendix is presented the definition of tensor product as seen in Mathematics. It will be defined as done in the theory of modules, but specialized to vector spaces. We focus on finite dimensional spaces, because those are the ones used in the Quantum Circuit model for Quantum Computation. The general case for any vector spaces is also presented for completeness. The general construction has the advantage of not depending on a choice of basis. We also prove some properties that are related to monoidal categories. In particular, we show the existence of non-degenerate pairings, which is important for monoidal categories with duals, such as spherical categories \cite{MonoidalCatsAndTFT}. We also show the property of existence of an internal hom, which lets us move a vector space from the domain to the codomain by taking the dual. This is important to understand the \CJ isomorphism. Even though some concepts from monoidal categories appear, we won't focus on the general case.

\section{Definition and basic results}

Familiarity with fields will be assumed, but the field $F$ can be taken as $F = \realNumbers$ or $F = \complexNumbers$.

\begin{definition}{Tensor product}{}
	Let $F$ be a field and let $X$ and $Y$ be $F$-vector spaces. The tensor product of $X$ and $Y$ is a pair $(P, \otimes)$, where $P$ is an $F$-vector space and $\otimes$ is any $F$-bilinear map 
	\begin{equation*}
		\otimes \colon X \times Y \to P
	\end{equation*}
	with the following universal property: if $B \colon X \times Y \to Z$ is an $F$-bilinear map, then there exists a unique $F$-linear map $u \colon P \to Z$ such that $B = u \otimes$. That is, the following diagram commutes:
	\begin{center}
		\begin{tikzcd}
			X \times Y \arrow[r, "\otimes"] \arrow[rd, swap, "B"] & P \arrow[d, dashed, "u"] \\
			{} & Z
		\end{tikzcd}
	\end{center}
	The space $P$ is denoted as $X \otimes Y$ if there is no ambiguity. The application of the bilinear map $\otimes$ to a pair $(x,y)$ will also be written as $x \otimes y$:
	\begin{equation}
		x \otimes y \coloneqq \otimes(x,y).
	\end{equation}
	The left side is the \textbf{infix notation} for $\otimes$, and the right is the \textbf{prefix notation}. Also, the elements of $X \otimes Y$ of the form $x \otimes y$ are called \textbf{elementary tensors}.
\end{definition}
Notice that this definition doesn't say either that the tensor product exists or is unique. In fact, the tensor product isn't unique, but the universal property guarantees that it is unique up to unique isomorphism. The existence of the tensor product will be proved later.

\begin{proposition}{Tensor product uniqueness up to unique isomorphism}{tensor product uniqueness}
	Let $F$ be a field and let $X$ and $Y$ be $F$-vector spaces. If $(P_1, \otimes_1)$ and $(P_2, \otimes_2)$ are tensor products of $X$ and $Y$, then there exists a unique isomorphism $u \colon P_1 \to P_2$ such that
	$\otimes_2 = u\otimes_1$:
	\begin{center}
		\begin{tikzcd}
			X \times Y \arrow[r, "\otimes_1"] \arrow[rd, swap, "\otimes_2"] & P_1 \arrow[d, dashed, "u"] \\
			{} & P_2
		\end{tikzcd}
	\end{center}
\end{proposition}
\begin{proof}
	Since $\otimes_2$ is bilinear, the universal property for $\otimes_1$ says that there exists a unique linear transformation $u \colon P_1 \to P_2$ such that the following diagram commutes:
	\begin{center}
		\begin{tikzcd}
			X \times Y \arrow[r, "\otimes_1"] \arrow[rd, swap, "\otimes_2"] & P_1 \arrow[d, dashed, "u"] \\
			{} & P_2
		\end{tikzcd}
	\end{center}
	Since $\otimes_1$ is bilinear, the universal property for $\otimes_2$ says that there exists a unique linear transformation $v \colon P_2 \to P_1$ such that the following diagram commutes:
	\begin{center}
		\begin{tikzcd}
			X \times Y \arrow[r, "\otimes_2"] \arrow[rd, swap, "\otimes_1"] & P_2 \arrow[d, dashed, "v"] \\
			{} & P_1
		\end{tikzcd}
	\end{center}
	Combining both diagrams, we get the following commutative diagrams:
	\begin{center}
		\begin{tikzcd}
			X \times Y \arrow[r, "\otimes_1"] \arrow[rd, swap, "\otimes_1"] & P_1 \arrow[d, "vu"] \\
			{} & P_1
		\end{tikzcd}
	\end{center}
	\begin{center}
		\begin{tikzcd}
			X \times Y \arrow[r, "\otimes_2"] \arrow[rd, swap, "\otimes_2"] & P_2 \arrow[d, "uv"] \\
			{} & P_2
		\end{tikzcd}
	\end{center}
	The last two diagrams also commute if we replace $vu$ and $uv$ by the identity maps. The universal property of the tensor product says that the linear transformation that makes the diagram commute is unique. This implies that $vu = \id_{P_1}$ and $uv = \id_{P_2}$, so $v= u^{-1}$.
\end{proof}

\section{Construction for finite dimensions}

Since we only need finite dimensional vector spaces, we can use a basis to construct faster a candidate for the tensor product. This also has the advantage of already giving a basis for the tensor product. The disadvantage is using a basis to define it. The general case will be proven in proposition \ref{proposition:tensor product existence general}.

\begin{proposition}{Construction of the tensor product of finite dimensional vector spaces}{construction of the tensor product finite case}
	Let $F$ be a field and let $X$ and $Y$ be finite dimensional $F$-vector spaces. Let $(x_i)_{i \in I}$ and $(y_j)_{j \in J}$ be basis for $X$ and $Y$, respectively. The tensor product of $X$ and $Y$ can be constructed as follows. Its $F$-vector space $X \otimes Y$ is the space freely generated by the family of pairs $(x_i, y_j)_{i \in I, j \in J}$. Alternatively, we use the notation
	\begin{equation}
		x_i \otimes y_j \coloneqq (x_i, y_j).
	\end{equation}
	Being freely generated means that $(x_i \otimes y_j)_{i\in I, j \in J}$ is interpreted as an $F$-basis for $X \otimes Y$. The $F$-bilinear map
	\begin{equation*}
		\otimes \colon X \times Y \to X \otimes Y
	\end{equation*}
	is defined using the basis. For any $x \in X$ and $y \in Y$, they can be written uniquely as
	\begin{gather}
		x = \sum_i a_i x_i, \\
		y = \sum_j b_j y_j ,
	\end{gather}
	where $a_i, b_j \in F$. Then
	\begin{equation}
		x \otimes y \coloneqq \sum_{i,j} a_i b_j (x_i \otimes y_j).
	\end{equation}
\end{proposition}
\begin{proof}
	We have to prove that $\otimes$ is bilinear and satisfy the universal property of the tensor product. First we prove the bilinearity. We start by the linearity in the first argument. Let $x, x' \in X$ and $y \in Y$. We have to show that
	\begin{equation}
		(x+x') \otimes y = x \otimes y + x' \otimes y .
	\end{equation}
	We can express the vectors uniquely as
	\begin{gather}
		x = \sum_i a_i x_i , \\
		x' = \sum_i a'_i x_i , \\
		y = \sum_j b_j y_j ,
	\end{gather}
	where $a_i, a'_i, b_j \in F$. Then, we have that
	\begin{align*}
		(x+x') \otimes y &\stackrel{(i)}{=} \left(\sum_i (a_i + a'_i)x_i \right) \otimes \left(\sum_j b_j y_j \right) \\
		&\stackrel{(ii)}{=} \sum_{i,j} (a_i + a'_i)b_j (x_i \otimes y_j) \\
		&\stackrel{(iii)}{=} \sum_{i,j} a_i b_j (x_i \otimes y_j) + \sum_{i,j} a'_i b_j (x_i \otimes y_j) \\
		&\stackrel{(iv)}{=} x \otimes y + x' \otimes y .
	\end{align*}
	In $(i)$ the vectors were written with respect to the basis; in $(ii)$ was used the definition of $\otimes$; in $(iii)$ the sum was split so separate the $a_i$ from $a'_i$; in $(iv)$ the definition of $\otimes$ was used again.
	
	Now let $\lambda \in F$. We have that
	\begin{align*}
		(\lambda x) \otimes y &\stackrel{(i)}{=} \left( \sum_i \lambda a_i x_i \right) \otimes \left( \sum_j b_j y_j \right) \\
		&\stackrel{(ii)}{=} \sum_{i,j} \lambda a_i b_j (x_i \otimes y_j) \\
		&\stackrel{(iii)}{=} \lambda \sum_{i,j} a_i b_j (x_i \otimes y_j) \\
		&\stackrel{(iv)}{=} \lambda (x \otimes y).
	\end{align*}
	In $(i)$ the vector were written with respect to the basis; in $(ii)$ was used the definition of $\otimes$; in $(iii)$ the $\lambda$ was passed to outside the summation; in $(iv)$ was used the definition of $\otimes$. This concludes that $\otimes \colon X \times Y \to X \otimes Y$ is linear in the first argument. The linearity in the second argument is proved similarly, so $\otimes$ is bilinear.
	
	Next we prove that the universal property holds for $\otimes$. Let $B \colon X \times Y \to Z$ be a bilinear map. Suppose there exists some linear transformation $u \colon X \otimes Y \to Z$ such that $B = u\otimes$. Then, applying it to the basis vectors, we get that
	\begin{equation}
		u(x_i \otimes y_j) = B(x_i, y_j),
	\end{equation}
	for every $i \in I$ and $j \in J$. Since $(x_i \otimes y_j)_{i \in I, j \in J}$ is, by definition, a basis for $X \otimes Y$, this determines $u$ completely and it is defined by it. In general, if $x \in X$ and $y \in Y$, they can be written uniquely as
	\begin{gather}
		x = \sum_i a_i x_i , \\
		y = \sum_j b_j y_j , 
	\end{gather}
	where $a_i, b_j \in F$. Then
	\begin{align*}
		u(x \otimes y) &\stackrel{(i)}{=} u( \sum_{i,j} a_i b_jx_i \otimes y_j) \\
		&\stackrel{(ii)}{=} \sum_{i,j} a_i b_j u(x_i \otimes y_j) \\
		&\stackrel{(iii)}{=} \sum_{i,j} a_i b_j B(x_i, y_j).
	\end{align*}
	In $(i)$ was used the definition of $\otimes$; in $(ii)$ was used that $u$ is linear; in $(iii)$ was used the definition of $u$ over the basis. Such $u$ satisfies that $B = u \otimes$ and is the unique linear transformation with this property, so the universal property holds for $\otimes$.
\end{proof}

As defined in this proposition, the tensor product $X \otimes Y$ depends on a choice of basis for $X$ and $Y$. If we change the basis we get another tensor product, but it is isomorphic, by proposition \ref{proposition:tensor product uniqueness}. In finite dimensions it is obvious that they would be isomorphic, because they have the same dimension, but the universal property also says that the isomorphism is unique and one bilinear map factors through the other by such isomorphism.

Notice that $x_i \otimes y_j$ has two interpretations, due to the notation. One is as the pair $(x_i, y_j)$, interpreted as a basis vector of $X \otimes Y$. The other is $\otimes(x_i, y_j)$, the application of the bilinear map $\otimes$ to the pair $(x_i, y_j)$. The bilinear map $\otimes$ is defined so that $\otimes(x_i, y_j) = (x_i, y_j)$, the right side being interpret as the basis vector, so there is no incoherence.

If we use basis $(x_i)_{i \in I}$ and $(y_j)_{j \in J}$ to construct the tensor product, but then work with another basis $(x'_i)_{i \in I}$ and $(y'_j)_{j \in J}$, the only possible interpretation for $x'_i \otimes y'_j$ is $\otimes(x'_i, y'_j)$, which is computed using the original basis. Nevertheless, the vectors $(x'_i \otimes y'_j)_{i \in I, j \in J}$ form another basis for $X \otimes Y$. We prove this next, but for any tensor product.

\begin{proposition}{Basis for tensor product}{}
	Let $F$ be a field and let $X$ and $Y$ be finite dimensional $F$-vector spaces. Let $(x_i)_{i \in I}$ and $(y_j)_{j \in J}$ be $F$-basis for $X$ and $Y$, respectively. For any tensor product of $X$ and $Y$, $(x_i \otimes y_j )_{i \in I, j \in J}$ is a basis for $X \otimes Y$. The tensor product doesn't need to be the one from proposition \ref{proposition:construction of the tensor product finite case}. Also, by $x_i \otimes y_j$ we mean the application of the bilinear map $\otimes \colon X \times Y \to X \otimes Y$ to $(x_i,y_j)$.
\end{proposition}
\begin{proof}
	By proposition \ref{proposition:construction of the tensor product finite case}, we can construct a tensor product $X \otimes' Y$ for which $(x_i \otimes' y_j)_{i \in I, j \in J}$ is a basis. By proposition \ref{proposition:tensor product uniqueness}, the tensor product is unique up to isomorphism. More precisely, there exists a unique isomorphism $u \colon X \otimes' Y \to X \otimes Y$ such that $\otimes = u \otimes'$, that is, the following diagram commutes:
	\begin{center}
		\begin{tikzcd}
			X \times Y \arrow[r, "\otimes'"] \arrow[rd, swap, "\otimes"] & X \otimes' Y \arrow[d, dashed, "u"] \\
			{} & X \otimes Y
		\end{tikzcd}
	\end{center}
	Since $u$ is an isomorphism and $(x_i \otimes' y_j)_{i \in I, j \in J}$ is a basis for $X \otimes' Y$, then $(u(x_i \otimes' y_j))_{i \in I, j \in J}$ is a basis for $X \otimes Y$. The commutativity of the diagram then says that
	\begin{align*}
		u(x_i \otimes' y_j) &\stackrel{(i)}{=} u \otimes' (x_i, y_j) \\
		&\stackrel{(ii)}{=} \otimes(x_i , y_j) \\
		&\stackrel{(iii)}{=} x_i \otimes y_j.
	\end{align*}
	In $(i)$ the $\otimes'$ was changed to prefix notation; in $(ii)$ was used that $u \otimes' = \otimes$; in $(iii)$ the $\otimes$ was passed to infix notation. This proves that $(x_i \otimes y_j)_{i \in I, j \in J}$ is a basis for $X \otimes Y$.
\end{proof}

From now on we don't bother about which tensor product construction is being used and just say that $X \otimes Y$ is the tensor product of $X$ and $Y$.

The following corollary holds more generally for modules, but we'll only prove it for finite dimensional vector spaces. It has the convenience of not working with basis.

\begin{corollary}{}{}
	Let $F$ be a field and let $X$, $Y$ and $Z$ be finite dimensional $F$-vector spaces. Let also $f, g \colon X \otimes Y \to Z$ be any linear transformation. Then
	\begin{equation}
		f=g \iff \forall x \in X, \forall y \in Y, \ f(x \otimes y) = g(x \otimes y).
	\end{equation}
\end{corollary}
\begin{proof}
	If $(x_i)_{i \in I}$ and $(y_j)_{j \in J}$ are basis for $X$ and $Y$, then $(x_i \otimes y_j)_{i \in I, j \in J}$ is a basis for $X \otimes Y$. Since $f=g$ iff $f$ and $g$ coincide over the basis $(x_i \otimes y_j)_{i \in I, j \in J}$, of course the equality can also be determined by the elementary tensors $x\otimes y$, since they include the basis.
\end{proof}

\section{The general case}

For completeness, we show the standard proof of existence of the tensor product for general vector spaces. Here it is assumed familiarity with linear algebra in the general case, not only for finite dimensional spaces.

This proof have the advantage of \textbf{not using a basis}. A similar proof works more generally for modules. A proof that doesn't use a basis is mandatory for modules, since not all modules have a basis.

\begin{proposition}{Existence of tensor product for general spaces}{tensor product existence general}
	Let $F$ be a field and let $X$ and $Y$ be $F$-vector spaces. There exists an $F$-vector space $X \otimes Y$ and a bilinear function $\otimes \colon X \times Y \to X \otimes Y$ such that the pair $(X \otimes Y, \otimes)$ satisfies the universal property of the tensor product. Also, the space $X \otimes Y$ is generated by the elementary tensors $x \otimes y$, where $(x,y) \in X \times Y$. That is, every element of $X \otimes Y$ is a finite linear combination of elementary tensors.
\end{proposition}
\begin{proof}
	We define $X \otimes Y$ as follows. Let $V$ be the $F$-vector space freely generated by $X \times Y$. We can implement the freely generated space as the set of functions $f \colon X \times Y \to F$ with finite support. More precisely, the support is the set of pairs for which the function is not 0:
	\begin{equation}
		\supp(f) \coloneqq \{(x,y) \in X \times Y \mid f(x,y) \neq 0\}.
	\end{equation}
	Then
	\begin{equation}
		V = \{f \colon X\times Y \to F \mid \supp(f) \text{ is finite}\}.
	\end{equation}
	The $F$-vector space structure of $V$ is induced by $F$. We define it as follow.
	\begin{itemize}
		\item Addition in $V$: let $f,g \in V$, then $f+g$ is the function given by
		\begin{equation}
			(f+g)(x,y) \coloneqq f(x,y) + g(x,y),
		\end{equation}
		for every $(x,y) \in X \times Y$.
		
		\item Scalar multiplication in $V$: let $f \in V$ and $\lambda \in F$, then $\lambda f$ is the function given by
		\begin{equation}
			(\lambda f)(x,y) \coloneqq \lambda f(x,y),
		\end{equation}
		for every $(x,y) \in X \times Y$.
	\end{itemize}
	It is straightforward to verify that $f+g$ and $\lambda f$ have finite support, so they are elements of $V$. Also, it is straightforward to verify that these operations turn $V$ into an $F$-vector space.
	
	Every pair $(x,y) \in X \times Y$ corresponds in $V$ to the function $\delta_{x,y}$ defined by
	\begin{equation}
		\delta_{x,y} (x',y') = \begin{cases}
			1 \ ,\text{ if } (x',y') = (x,y) \\
			0 \ ,\text{ otherwise}
		\end{cases}
	\end{equation}
	Notice that this isn't the Dirac delta function, which is a distribution. It is more similar to the Kronecker delta function, but using elements of $X \times Y$ instead of integer indices.
	
	The next step is to take a quotient of $V$ to enforce the equivalence class of $\delta_{x,y}$ to have the properties of an elementary tensor $x \otimes y$. This is done as follows. First we define vectors which encode the bilinearity of the tensor product. They are the vectors
	\begin{equation}
		\label{equation: tensor bilinearity 1}
		\delta_{x+x',y} - \delta_{x,y} -\delta_{x',y} ,
	\end{equation}
	\begin{equation}
		\label{equation: tensor bilinearity 2}
		\delta_{x,y+y'} - \delta_{x,y} -\delta_{x,y'} ,
	\end{equation}
	\begin{equation}
		\label{equation: tensor bilinearity 3}
		\delta_{\lambda x,y} - \lambda \delta_{x,y} ,
	\end{equation}
	\begin{equation}
		\label{equation: tensor bilinearity 4}
		\delta_{x,\lambda y} - \lambda \delta_{x,y} ,
	\end{equation}
	for every $x,x' \in X$, $y,y' \in Y$ and $\lambda \in F$. Let $W \leq V$ be the $F$-space generated by these vectors. Then we define
	\begin{equation}
		X \otimes Y \coloneqq V/W .
	\end{equation}
	Also define the function
	\begin{align*}
		\otimes \colon X \times Y &\longrightarrow X \otimes Y \\
		(x,y) &\longmapsto [\delta_{x,y}],
	\end{align*}
	where $[f]$ denotes the equivalence class of $f$. In the infix notation for $\otimes$, we have that
	\begin{equation}
		x \otimes y = [\delta_{x,y}],
	\end{equation}
	for every $x \in X$ and $y \in Y$.
	
	It is immediate from the definition of $W$ that $\otimes$ is bilinear. In fact, the equations \ref{equation: tensor bilinearity 1}, \ref{equation: tensor bilinearity 2}, \ref{equation: tensor bilinearity 3} and \ref{equation: tensor bilinearity 4} imply, respectively, that
	\begin{gather}
		(x+x') \otimes y = x \otimes y + x' \otimes y , \\
		x \otimes (y+y') = x \otimes y + x \otimes y' , \\
		(\lambda x) \otimes y = \lambda (x\otimes y) , \\
		x \otimes (\lambda y) = \lambda (x\otimes y) ,
	\end{gather}
	for every $x,x' \in X$, $y,y' \in Y$ and $\lambda \in F$.
	
	Since the functions $\delta_{x,y}$ generate $V$, it follows that the elementary tensors $x \otimes y$ generate $X \otimes Y$. That is, every tensor $t \in X \otimes Y$ can be written as a finite linear combination
	\begin{equation}
		t = \sum_{i \in I} \lambda_i (x_i \otimes y_i) ,
	\end{equation}
	for some finite set of indices $I$, scalars $\lambda_i \in F$ and vectors $x_i \in X$ and $y_i \in Y$.
	
	To complete the proof, it remains to show that the pair $(X \otimes Y, \otimes)$ satisfy the universal property of the tensor product. Let $B \colon X \times Y \to Z$ be a bilinear function, where $X$, $Y$ and $Z$ are $F$-vector spaces. First suppose that there exists a linear transformation $u \colon X \otimes Y \to Z$ such that the following diagram commutes:
	\begin{center}
		\begin{tikzcd}
			X \times Y \arrow[r,"\otimes"] \arrow[rd, swap, "B"] & X \otimes Y \arrow[d, "u"] \\
			{} & Z
		\end{tikzcd}
	\end{center}
	Then, for any $(x,y) \in X \times Y$, we have that
	\begin{equation}
		u(x \otimes y) = B(x,y).
	\end{equation}
	Since the elementary tensors $x \otimes y$ generate $X \otimes Y$, by linearity, this equation determines $u$. This proves the uniqueness of $u$.
	
	For the proof of existence, we have to ensure that the definition of $u$ is consistent. Instead of proving it directly, we'll first construct a linear transformation over $V$, then we'll show that it induces a linear transformation over the quotient $V/W$. The function over $V/W$ will coincide with $u$, which will prove the existence part. Denote by $i$ the function that includes $X \times Y$ into $V$ as the basis vectors:
	\begin{align*}
		i \colon X \times Y &\longrightarrow V \\
		(x,y) &\longmapsto \delta_{x,y} .
	\end{align*}
	Together with the bilinear function $B \colon X \times Y \to Z$, we have the next diagram
	\begin{center}
		\begin{tikzcd}
			X \times Y \arrow[r, "i"] \arrow[rd, swap, "B"] & V \\
			{} & Z
		\end{tikzcd}
	\end{center}
	We can complete this diagram with a linear transformation $C \colon V \to Z$ as follows. Every $f \in V$ can be uniquely written as a finite linear combination
	\begin{equation}
		f = \sum_{j \in J} \lambda_j \delta_{x_j, y_j},
	\end{equation}
	for some finite set of indices $J$, some scalars $\lambda_j \in F$ and vectors $x_j \in X$ and $y_j \in Y$. Then we define
	\begin{equation}
		C(f) \coloneqq \sum_{j \in J} \lambda_j B(x_j, y_j).
	\end{equation}
	It is simple to prove that $C$ is a linear transformation:
	\begin{itemize}
		\item Additivity: let $f, g \in V$, they can be uniquely written as
		\begin{gather}
			f = \sum_{(x,y) \in X \times Y} \lambda_{x,y} \delta_{x,y}, \\
			g = \sum_{(x,y) \in X \times Y} \mu_{x,y} \delta_{x,y},
		\end{gather}
		where $\lambda_{x,y}, \mu_{x,y} \in F$ and these scalar are non zero for a finite number of pairs $(x,y)$. Then
		\begin{align*}
			C(f+g) &\stackrel{(i)}{=} C\left( \sum_{(x,y) \in X \times Y} \lambda_{x,y} \delta_{x,y} + \sum_{(x,y) \in X \times Y} \mu_{x,y} \delta_{x,y} \right) \\
			&\stackrel{(ii)}{=} C \left( \sum_{(x,y) \in X \times Y} (\lambda_{x,y} + \mu_{x,y}) \delta_{x,y} \right) \\
			&\stackrel{(iii)}{=}  \sum_{(x,y) \in X \times Y} (\lambda_{x,y} + \mu_{x,y}) B(x,y) \\
			&\stackrel{(iv)}{=} \lambda_{x,y} B(x,y) + \mu_{x,y} B(x,y) \\
			&\stackrel{(v)}{=} C(f) + C(g).
		\end{align*}
		In $(i)$ both $f$ and $g$ were written in the basis; in $(ii)$ both summations were merged into a single summation; in $(iii)$ was used the definition of $C$; in $(iv)$ the single summation was split into two; in $(v)$ was used the definition of $C$.
		
		\item Scalar multiplication: let $f \in V$ and $\eta \in F$. Again, $f$ can be uniquely written as
		\begin{equation}
			f = \sum_{(x,y) \in X \times Y} \lambda_{x,y} \delta_{x,y}.
		\end{equation}
		Then
		\begin{align*}
			C(\eta f) &\stackrel{(i)}{=} C \left(\eta \sum_{(x,y) \in X \times Y} \lambda_{x,y} \delta_{x,y} \right) \\
			&\stackrel{(ii)}{=} C \left(\sum_{(x,y) \in X \times Y} \eta \lambda_{x,y} \delta_{x,y} \right) \\
			&\stackrel{(iii)}{=} \sum_{(x,y) \in X \times Y} \eta \lambda_{x,y} B(x,y) \\
			&\stackrel{(iv)}{=} \eta \sum_{(x,y) \in X \times Y} \lambda_{x,y} B(x,y) \\
			&\stackrel{(v)}{=} \eta C(f).
		\end{align*}
		In $(i)$ $f$ was written in the basis; in $(ii)$ $\eta$ was passed to inside the summation; in $(iii)$ was used the definition of $C$; in $(iv)$ $\eta$ was passed to outside the summation; in $(v)$ was used the definition of $C$.
	\end{itemize}
	
	It is immediate from the definition of $C$ that the following diagram commutes
	\begin{center}
		\begin{tikzcd}
			X \times Y \arrow[r, "i"] \arrow[rd, swap, "B"] & V \arrow[d, "C"]\\
			{} & Z
		\end{tikzcd}
	\end{center}
	Next we want to prove that $C$ induces a linear transformation $\overline{C} \colon V/W \to Z$, so that the next diagram commutes:
	\begin{center}
		\begin{tikzcd}
			V \arrow[r, "C"] \arrow[d, swap, "\pi"] & Z\\
			V/W \arrow[ru, dashed, swap, "\overline{C}"] &
		\end{tikzcd}
	\end{center}
	In this diagram, $\pi \colon V \to V/W$ is the function that sends an element of $V$ to its equivalence class, that is
	\begin{align*}
		\pi \colon V &\longrightarrow V/W \\
		f &\longmapsto [f].
	\end{align*}
	$C$ induces $\overline{C}$ iff $W \subseteq \ker{C}$. $W$ was defined as the subspace of $V$ generated by the vectors of equations \ref{equation: tensor bilinearity 1}, \ref{equation: tensor bilinearity 2}, \ref{equation: tensor bilinearity 3} and \ref{equation: tensor bilinearity 4}. As such, we only need to prove that each of these vectors are in $\ker{C}$. We'll prove it only for equations \ref{equation: tensor bilinearity 1} and \ref{equation: tensor bilinearity 3}, the other cases are similar.
	
	For equation \ref{equation: tensor bilinearity 1}, we have that:
	\begin{align*}
		C(\delta_{x+x',y} - \delta_{x,y} -\delta_{x',y}) &\stackrel{(i)}{=} C(\delta_{x+x',y}) - C(\delta_{x,y}) -C(\delta_{x',y}) \\
		&\stackrel{(ii)}{=} B(x+x',y) - B(x,y) - B(x',y) \\
		&\stackrel{(iii)}{=} 0.
	\end{align*}
	In $(i)$ was used that $C$ is linear; in $(ii)$ was used the definition of $C$ over the basis vectors; in $(iii)$ was used that $B$ is bilinear. This proves that the vector from equation \ref{equation: tensor bilinearity 1} is in $\ker{C}$.
	
	For equation \ref{equation: tensor bilinearity 3}, we have that:
	\begin{align*}
		C(\delta_{\lambda x,y} - \lambda \delta_{x,y}) &\stackrel{(i)}{=} C(\delta_{\lambda x,y}) - \lambda C( \delta_{x,y}) \\
		&\stackrel{(ii)}{=} B(\lambda x, y) - \lambda B(x,y) \\
		&\stackrel{(iii)}{=} 0.
	\end{align*}
	In $(i)$ was used that $C$ is linear; in $(ii)$ was used the definition of $C$ over basis vectors; $(iii)$ was used that $B$ is bilinear. This proves that the vector from equation \ref{equation: tensor bilinearity 3} is in $\ker{C}$.
	
	Similarly, the vectors from equations \ref{equation: tensor bilinearity 2} and \ref{equation: tensor bilinearity 4} are also $\ker{C}$. From this it follows that $W \subseteq \ker{C}$, which implies that $C$ induces a unique linear transformation $\overline{C} \colon V/W \to Z$ such that the following diagram commutes:
	\begin{center}
		\begin{tikzcd}
			V \arrow[r, "C"] \arrow[d, swap, "\pi"] & Z\\
			V/W \arrow[ru, dashed, swap, "\overline{C}"] &
		\end{tikzcd}
	\end{center}
	Finally, we just have to check that $\overline{C} = u$. Their codomains are both $Z$. The domain of $\overline{C}$ is $V/W$, which, by definition, is the same as $X \otimes Y$, which is the domain of $u$. Having the same domain and codomain, we can compare both functions for equality. Evaluating at an elementary tensor $x \otimes y$, we have
	\begin{align*}
		\overline{C} (x \otimes y) &\stackrel{(i)}{=} \overline{C} \pi (\delta_{x,y}) \\
		&\stackrel{(ii)}{=} C(\delta_{x,y}) \\
		&\stackrel{(iii)}{=} B(x,y) \\
		&\stackrel{(iv)}{=} u(x \otimes y).
	\end{align*}
	In $(i)$ was used that the elementary tensor $x\otimes y$ is defined as the equivalence class $[\delta_{x,y}]$. Changing notation, we can write this equivalence class as $\pi(\delta_{x,y})$. In $(ii)$ was used that $\overline{C} \pi = C$; in $(iii)$ was used the definition of $C$ over a basis vector; in $(iv)$ was used that $u(x \otimes y)$ is also $B(x,y)$. This shows that $\overline{C}$ and $u$ coincide over the elementary tensors. Since the elementary tensors generate $X \otimes Y$, it follows that $\overline{C} = u$. The existence of $\overline{C}$ then proves the existence of $u$.
	
\end{proof}

\section{Tensor product of linear transformations}

For any linear transformations $f \colon X \to Y$ and $f' \colon X' \to Y'$ we can define their tensor product, which is a linear transformation $f \otimes f' \colon X \otimes X' \to Y \otimes Y'$. It is defined so that
\begin{equation*}
	(f \otimes f')(x \otimes x') = f(x) \otimes f'(x').
\end{equation*}
This is formalized in the next proposition.

\begin{proposition}{Tensor product of linear transformations}{tensor product of linear transformations}
	Let $F$ be a field and let $X$, $Y$, $X'$ and $Y'$ be $F$-vectors spaces. Let also
	\begin{gather*}
		f \colon X \to Y, \\
		f' \colon X' \to Y'
	\end{gather*}
	be linear transformations. There exists a unique linear transformation
	\begin{equation*}
		f \otimes f' \colon X \otimes X' \to Y \otimes Y'
	\end{equation*}
	such that the following diagram commutes.
	\begin{center}
		\begin{tikzcd}
			X \times X' \arrow[r, "\otimes"] \arrow[d, swap, "f \times f'"] & X \otimes X' \arrow[d, dashed, "f \otimes f'"] \\
			Y \times Y' \arrow[r, "\otimes"] & Y \otimes Y'.
		\end{tikzcd}
	\end{center}
	The linear map $f\otimes f'$ satisfies that
	\begin{equation}
		\label{equation: tensor product of linear transformations}
		(f\otimes f')(x \otimes x') = f(x) \otimes f'(x'),
	\end{equation}
	for any $x \in X$ and $x' \in X'$.
\end{proposition}
\begin{proof}
	Since $f$ and $f'$ are linear and $\otimes \colon Y \times Y' \to Y \otimes Y'$ is bilinear, the composition $\otimes (f \times f')$ is also bilinear. The universal property of the tensor product for $X$ and $X'$ then says that there exists a unique linear transformation $f \otimes f' \colon X \otimes X' \to Y \otimes Y'$ such that the following diagram commutes:
	\begin{center}
		\begin{tikzcd}
			X \times X' \arrow[r, "\otimes"] \arrow[dr, swap, "\otimes(f \times f')"] & X \otimes X' \arrow[d, dashed, "f \otimes f'"] \\
			{} & Y \otimes Y'.
		\end{tikzcd}
	\end{center}
	Let $x \in X$ and $x' \in X'$. The commutativity of the diagram implies that
	\begin{align*}
		(f\otimes f')(x \otimes x') &\stackrel{(i)}{=} (f\otimes f')\otimes(x, x') \\
		&\stackrel{(ii)}{=} \otimes (f\times f') (x, x') \\
		&\stackrel{(iii)}{=} \otimes (f(x), f'(x')) \\
		&\stackrel{(iv)}{=} f(x) \otimes f'(x').
	\end{align*}
	In $(i)$ the second $\otimes$ was written in prefix notation; in $(ii)$ was used the commutativity of the diagram; in $(iii)$ was applied $f \times f'$; in $(iv)$ the $\otimes$ was written in infix notation. This concludes the proof.
\end{proof}

Notice that this proof is general, it doesn't require the vector spaces to be finite dimensional. The same proof is valid for modules.

Next we'll prove some properties of the tensor product that are important for the symmetric monoidal category of finite dimensional vector spaces.

\section{Symmetry}

We'll use the universal property of the tensor product to construct an isomorphism $X \otimes Y \cong Y \otimes X$. In finite dimensions it is obvious that some isomorphism must exist, because the dimensions are equal. But we construct an isomorphism that also works in infinite dimensions and for modules.

\begin{proposition}{Symmetry of the tensor product}{tensor symmetry}
	Let $F$ be a field and let $X$ and $Y$ be $F$-vector spaces. Let $\sigma_{X,Y}$ be the transposition function
	\begin{align*}
		\sigma_{X,Y} \colon X \times Y &\longrightarrow Y \times X \\
		(x,y) &\longmapsto (y,x) .
	\end{align*}
	There exists a unique isomorphism
	\begin{equation*}
		\tau_{X, Y} \colon X \otimes Y \to Y \otimes X
	\end{equation*}
	such that the following diagram commutes:
	\begin{center}
		\begin{tikzcd}
			X \times Y \arrow[r, "\otimes"] \arrow[d, swap, "\sigma_{X,Y}"] & X \otimes Y \arrow[d, dashed, "\tau_{X,Y}"] \\
			Y \times X \arrow[r, "\otimes"] & Y \otimes X .
		\end{tikzcd}
	\end{center}
	For any $x \in X$ and $y \in Y$, it satisfies that
	\begin{equation}
		\tau_{X, Y}(x \otimes y) = y \otimes x .
	\end{equation}
\end{proposition}
\begin{proof}
	First we prove that the composition $\otimes \sigma_{X,Y}$ is bilinear. We prove the linearity in the first argument, the proof for the second argument is similar. Let $x,x' \in X$ and $y \in Y$, we have that
	\begin{align*}
		\otimes \sigma_{X,Y}(x+x',y) &\stackrel{(i)}{=} \otimes(y, x+x') \\
		&\stackrel{(ii)}{=} \otimes(y,x) + \otimes(y, x') \\
		&\stackrel{(iii)}{=} \otimes \sigma_{X,Y} (x,y) + \otimes \sigma_{X,Y} (x',y) .
	\end{align*}
	In $(i)$ was applied $\sigma_{X,Y}$; in $(ii)$ was used that $\otimes$ is bilinear; in $(iii)$ the $\sigma_{X,Y}$ was put back.
	
	Now let $\lambda \in F$. We have that
	\begin{align*}
		\otimes \sigma_{X,Y}(\lambda x,y) &\stackrel{(i)}{=} \otimes (y, \lambda x) \\
		&\stackrel{(ii)}{=} \lambda [\otimes (y, x)] \\
		&\stackrel{(iii)}{=} \lambda [\otimes \sigma_{X,Y} (x,y)] .
	\end{align*}
	In $(i)$ was applied $\sigma_{X,Y}$; in $(ii)$ was used that $\otimes$ is bilinear; in $(iii)$ the $\sigma_{X,Y}$ was put back. This concludes that $\otimes \sigma_{X,Y}$ is bilinear.
	
	The universal property of the tensor product then says that there exists a unique linear transformation $\tau_{X,Y} \colon X \otimes Y \to Y \otimes X$ such that the following diagram commutes:
	\begin{center}
		\begin{tikzcd}
			X \times Y \arrow[r, "\otimes"] \arrow[rd, swap, "\otimes \sigma_{X,Y}"] & X \otimes Y \arrow[d, dashed, "\tau_{X,Y}"] \\
			{} & Y \otimes X
		\end{tikzcd}
	\end{center}
	Let $x \in X$ and $y \in Y$. The commutativity of the diagram implies that
	\begin{align*}
		\tau_{X,Y}(x\otimes y) &\stackrel{(i)}{=} \tau_{X,Y} \otimes (x,y) \\
		&\stackrel{(ii)}{=} \otimes \sigma_{X,Y} (x,y) \\
		&\stackrel{(iii)}{=} \otimes (y,x) \\
		&\stackrel{(iv)}{=} y \otimes x .
	\end{align*}
	In $(i)$ the $\otimes$ was written in prefix notation; in $(ii)$ was used the commutativity of the diagram; in $(iii)$ was applied $\sigma_{X,Y}$; in $(iv)$ the $\otimes$ was written in infix notation.
\end{proof}

\section{Unitality}

We prove that $F \otimes X \cong X \cong X \otimes F$. The proof is general and the same ideas apply for modules.

\begin{proposition}{Left Unitality}{left unitality}
	Let $F$ be a field and $X$ be an $F$-vector space. Denote by $m$ the operation of scalar multiplication
	\begin{equation*}
		m \colon F \times X \to X .
	\end{equation*}
	There exists a unique isomorphism
	\begin{equation*}
		l_X \colon F \otimes X \to X
	\end{equation*}
	such that the following diagram commutes:
	\begin{center}
		\begin{tikzcd}
			F \times X \arrow[r, "\otimes"] \arrow[dr, swap, "m"] & F \otimes X \arrow[d, dashed, "l_X"] \\
			{} & X
		\end{tikzcd}
	\end{center}
	For any $\lambda \in F$ and $x \in X$, it satisfies that
	\begin{equation}
		l_X (\lambda \otimes x) = \lambda x .
	\end{equation}
	The inverse is given by
	\begin{equation}
		l_X^{-1} (x) = 1 \otimes x .
	\end{equation}
\end{proposition}
\begin{proof}
	The properties of scalar multiplication imply that $m$ is bilinear. The universal property of the tensor product then says that there exists a unique linear transformation $l_X \colon F \otimes X \to X$ such that the diagram commutes.
	
	Let $\lambda \in F$ and $x \in X$. The commutativity of the diagram implies that
	\begin{align*}
		l_X (\lambda \otimes x) &\stackrel{(i)}{=} l_X \otimes (\lambda, x) \\
		&\stackrel{(ii)}{=} m(\lambda, x) \\
		& \stackrel{(iii)}{=} \lambda x.
	\end{align*}
	In $(i)$ the $\otimes$ was written in prefix notation; in $(ii)$ was used the commutativity of the diagram; in $(iii)$ was just made a change in notation.
	
	Let's check that $l_X^{-1}(x) = 1 \otimes x$ is the inverse of $l_X$. For any $x \in X$, we have
	\begin{align*}
		l_X l_X^{-1}(x) &\stackrel{(i)}{=} l_X(1 \otimes x) \\
		&\stackrel{(ii)}{=} x.
	\end{align*}
	In $(i)$ was applied $l_X^{-1}$ and in $(ii)$ was applied $l_X$. This shows that $l_X^{-1}$ is a right inverse. Now let $\lambda \in F$, we have that
	\begin{align*}
		l_X^{-1} l_X (\lambda \otimes x) &\stackrel{(i)}{=} l_X^{-1}(\lambda x) \\
		&\stackrel{(ii)}{=} 1 \otimes (\lambda x) \\
		&\stackrel{(iii)}{=} \lambda (1\otimes x) \\
		&\stackrel{(iv)}{=} \lambda \otimes x.
	\end{align*}
	In $(i)$ was applied $l_X$; in $(ii)$ was applied $l_X^{-1}$; in $(iii)$ and $(iv)$ were used the bilinearity of $\otimes$ to pass $\lambda$ from the second factor to outside, and then from outside to the first factor. Since the elementary tensors $\lambda \otimes x$ generate $F \otimes X$, it follows that
	\begin{equation}
		l_X^{-1} l_X = \id_{F \otimes X}.
	\end{equation}
	This concludes that $l_X^{-1}$ is the inverse for $l_X$, so $l_X$ is an isomorphism.
\end{proof}

\begin{proposition}{Right Unitality}{right unitality}
	Let $F$ be a field and $X$ be an $F$-vector space. Denote by $m$ the operation
	\begin{align*}
		m \colon X \times F &\longrightarrow X \\
		(x, \lambda) &\longmapsto \lambda x .
	\end{align*}
	There exists a unique linear transformation
	\begin{equation*}
		r_X \colon X \otimes F \to X
	\end{equation*}
	such that the following diagram commutes:
	\begin{center}
		\begin{tikzcd}
			X \times F \arrow[r, "\otimes"] \arrow[dr, swap, "m"] & X \otimes F \arrow[d, dashed, "r_X"] \\
			{} & X
		\end{tikzcd}
	\end{center}
	For any $\lambda \in F$ and $x \in X$, it satisfies that
	\begin{equation}
		r_X (x \otimes \lambda) = \lambda x .
	\end{equation}
	The inverse is given by
	\begin{equation}
		r_X^{-1} (x) = x \otimes 1 .
	\end{equation}
\end{proposition}
\begin{proof}
	The proof follows closely the one for the left unitality isomorphism $l_X$.
\end{proof}

\section{Associativity}

We prove that $(X \otimes Y) \otimes Z \cong X \otimes (Y \otimes Z)$. The general technique, which also applies for modules, will be used. It consists of first generalizing the definition of tensor product from pairs to families of vector spaces. Then we show that both $(X \otimes Y) \otimes Z$ and $X \otimes (Y \otimes Z)$ satisfy the universal product of the tensor product of a triple, which implies that they are isomorphic.

\begin{definition}{Tensor product of multiple spaces}{}
	Let $F$ be a field. Let also $(X_i)_{i \in I}$ be a finite family of $F$-vectors spaces. A tensor product of this family is a pair $(P, \otimes)$, where $P$ is an $F$-vector space and
	\begin{equation*}
		\otimes \colon \Pi_{i \in I} X_i \to P
	\end{equation*}
	is a multilinear function with the following universal property: if
	\begin{equation*}
		M \colon  \Pi_{i \in I} X_i \to Q
	\end{equation*}
	is another multilinear function, then there exists a unique linear transformation $u \colon P \to Q$ such that the following diagram commutes:
	\begin{center}
		\begin{tikzcd}
			\Pi_{i \in I} X_i \arrow[r, "\otimes"] \arrow[rd, swap, "M"] & P \arrow[d, dashed, "u"] \\
			{} & Q
		\end{tikzcd}
	\end{center}
\end{definition}

\begin{proposition}{}{}
	Let $F$ be a field, let also $X$, $Y$ and $Z$ be $F$-vector spaces. The space $(X \otimes Y) \otimes Z$ together with the trilinear function
	\begin{equation*}
		\otimes(\otimes \times \id_Z) \colon X \times Y \times Z \to (X \otimes Y) \otimes Z
	\end{equation*}
	is a tensor product for $X$, $Y$ and $Z$. Also, the space $X \otimes (Y \otimes Z)$ together with the trilinear function
	\begin{equation*}
		\otimes(\id_X \times \otimes) \colon X \times Y \times Z \to X \otimes (Y \otimes Z)
	\end{equation*}
	is a tensor product for $X$, $Y$ and $Z$.
\end{proposition}
\begin{proof}
	We prove the universal property for $(X \otimes Y) \otimes Z$, the proof for $X \otimes (Y \otimes Z)$ is similar. Let $T \colon X \times Y \times Z \to Q$ be a trilinear function. We start by supposing that there exists a linear map $w \colon (X \otimes Y) \otimes Z \to Q$ such that the following diagram commutes:
	\begin{center}
		\begin{tikzcd}
			X \times Y \times Z \arrow[rr, "\otimes(\otimes \times \id_Z)"] \arrow[rrd, swap, "T"] & {} & (X \otimes Y) \otimes Z \arrow[d, dashed, "w"] \\
			{} & {} & Q
		\end{tikzcd}
	\end{center}
	We use this commutative diagram to determine $w$, which will guarantee the uniqueness of $w$. We need to apply the universal properties of the tensor products of pairs. We start with the innermost tensor product, which is $X \otimes Y$. We can use $w$ to construct a function with a single tensor product as
	\begin{align*}
		v \colon (X \otimes Y) \times Z &\longrightarrow Q \\
		(t, z) &\longmapsto w(t \otimes z).
	\end{align*}
	We need further to eliminate the factor $Z$. This leads to a family of functions
	\begin{align*}
		u_z \colon X \otimes Y &\longrightarrow Q \\
		t &\longmapsto w(t \otimes z),
	\end{align*}
	parameterized by $z \in Z$. It is immediate from the bilinearity of $\otimes$ and the linearity of $w$ that $u_z$ is linear. To apply the universal property of $X \otimes Y$ we need a bilinear function. For any $z \in Z$, we have the bilinear function
	\begin{align*}
		B_z \colon X \times Y &\longrightarrow Q \\
		(x,y) &\longmapsto T(x,y,z),
	\end{align*}
	parameterized by $z \in Z$. Then we have the next diagram
	\begin{center}
		\begin{tikzcd}
			X \times Y \arrow[r, "\otimes"] \arrow[rd, swap, "B_z"] & X \otimes Y \arrow[d, "u_z"] \\
			{} & Q
		\end{tikzcd}
	\end{center}
	Let's show that it commutes. Let $x \in X$ and $y \in Y$, we have that
	\begin{align*}
		u_z \otimes (x,y) &\stackrel{(i)}{=} u_z (x \otimes y) \\
		&\stackrel{(ii)}{=} w((x\otimes y) \otimes z) \\
		&\stackrel{(iii)}{=} w\otimes(\otimes \times \id_Z) (x,y,z) \\
		&\stackrel{(iv)}{=} T(x,y,z) \\
		&\stackrel{(v)}{=} B_z (x,y).
	\end{align*}
	In $(i)$ the $\otimes$ was passed to infix notation; in $(ii)$ was used the definition of $u_z$; in $(iii)$ both tensor products were written in prefix notation; in $(iv)$ was used the commutativity of the first diagram in this proof; in $(v)$ was used the definition of $B_z$. This proves that the diagram commutes. The universal property of $X \otimes Y$ then says $u_z$ is the unique linear transformation that makes the diagram commute. We proved that
	\begin{equation}
		w((x\otimes y)\otimes z) = u_z (x \otimes y),
	\end{equation}
	for any $x \in X$, $y \in Y$ and $z \in Z$. Since the elementary tensors $(x\otimes y)\otimes z$ generate $(X \otimes Y) \otimes Z$, and since each $u_z$ is unique, this determines $w$ completely. Therefore, $w$ is unique, if it exists. 
	
	The uniqueness proof hints the construction of $w$. We constructed bilinear maps $B_z$. The universal property of $X \otimes Y$ says that there exists a unique linear transformation $u_z \colon X \otimes Y \to Q$ such that the next diagram commutes
	\begin{center}
		\begin{tikzcd}
			X \times Y \arrow[r, "\otimes"] \arrow[rd, swap, "B_z"] & X \otimes Y \arrow[d, "u_z"] \\
			{} & Q
		\end{tikzcd}
	\end{center}
	for any $z \in Z$. Then we can define the function
	\begin{align*}
		v \colon (X \otimes Y) \times Z &\longrightarrow Q \\
		(t, z) &\longmapsto u_z(t).
	\end{align*}
	To use the universal property of the tensor product for the pair $(X \otimes Y, Z)$, we need this function to be bilinear. The linearity in the first argument follows from the linearity of $u_z$. The linearity in the second argument will be proved using the uniqueness of $u_z$. Let $z, z' \in Z$ and $\lambda, \lambda' \in F$. We'll prove that $u_{\lambda z + \lambda' z'} = \lambda u_z + \lambda' u_{z'}$. Consider the diagram
	\begin{center}
		\begin{tikzcd}
			X \times Y \arrow[r, "\otimes"] \arrow[rd, swap, "B_{\lambda z + \lambda' z'}"] & X \otimes Y \arrow[d, "\lambda u_z + \lambda' u_{z'}"] \\
			{} & Q
		\end{tikzcd}
	\end{center}
	Let's show that it commutes. Let $x\in X$ and $y \in Y$, we have that
	\begin{align*}
		(\lambda u_z + \lambda' u_{z'})\otimes (x,y) &\stackrel{(i)}{=} (\lambda u_z + \lambda' u_{z'}) (x \otimes y) \\
		&\stackrel{(ii)}{=} \lambda u_z(x \otimes y) + \lambda' u_{z'} (x \otimes y) \\
		&\stackrel{(iii)}{=} \lambda B_z(x,y) + \lambda' B_{z'} (x, y) \\
		&\stackrel{(iv)}{=} \lambda T(x,y,z) + \lambda' T(x, y, z') \\
		&\stackrel{(v)}{=} T(x,y,\lambda z + \lambda' z')\\
		&\stackrel{(vi)}{=} B_{\lambda z + \lambda' z'}(x,y).
	\end{align*}
	In $(i)$ the $\otimes$ was written in infix notation; in $(ii)$ was used the definition of sum of linear transformations; in $(iii)$ was used that $u_z \otimes = B_z$ and $u_{z'} \otimes = B_{z'}$; in $(iv)$ was used the definition of $B_z$; in $(v)$ was used that $T$ is linear in the third argument; in $(vi)$ was used the definition of $B_z$ again. This shows that the diagram commutes. But $u_{\lambda z + \lambda' z'}$ is the unique linear transformation that makes the diagram commute, so
	\begin{equation}
		u_{\lambda z + \lambda' z'}  = \lambda u_{z} + \lambda' u_{z'},
	\end{equation}
	for any $z, z' \in Z$ and $\lambda, \lambda' \in F$.
	
	This also shows that $v$ is linear in the second argument, so $v$ is bilinear. The universal property of the tensor product of the pair $(X \otimes Y, Z)$ then says that there exists a unique linear transformation $w \colon (X \otimes Y) \otimes Z \to Q$ such that the next diagram commutes
	\begin{center}
		\begin{tikzcd}
			(X \otimes Y) \times Z \arrow[r, "\otimes"] \arrow[rd, swap, "v"] & (X \otimes Y) \otimes Z \arrow[d, dashed, "w"] \\
			{} & Q
		\end{tikzcd}
	\end{center}
	Now consider the following diagram
	\begin{center}
		\begin{tikzcd}
			X \times Y \times Z \arrow[r, "\otimes \times \id_Z"] \arrow[rd, swap, "T"] & (X \otimes Y) \times Z \arrow[d, "v"] \\
			{} & Q
		\end{tikzcd}
	\end{center}
	Let's show that it commutes. Let $x \in X$, $y \in Y$ and $z \in Z$. We have that
	\begin{align*}
		v(\otimes \times \id_Z) (x,y,z) &\stackrel{(i)}{=} v(x\otimes y, z) \\
		&\stackrel{(ii)}{=} u_z (x \otimes y) \\
		&\stackrel{(iii)}{=} B_z(x,y) \\
		&\stackrel{(iv)}{=} T(x,y,z).
	\end{align*}
	In $(i)$ was applied $\otimes \times \id_Z$ and the tensor product was written in infix notation; in $(ii)$ was used the definition of $v$; in $(iii)$ was used that $u_z \otimes = B_z$; in $(iv)$ was used the definition of $B_z$. This proves that the diagram commutes.
	
	The commutativity of the last two diagrams implies the commutativity of the diagram
	\begin{center}
		\begin{tikzcd}
			X \times Y \times Z \arrow[rr, "\otimes(\otimes \times \id_Z)"] \arrow[rrd, swap, "T"] & {} & (X \otimes Y) \otimes Z \arrow[d, dashed, "w"] \\
			{} & {} & Q
		\end{tikzcd}
	\end{center}
	This concludes that $((X\otimes Y) \otimes Z, \otimes (\otimes \times \id_Z))$ satisfies the universal property of the tensor product for the triple $(X,Y,Z)$.
\end{proof}

\begin{corollary}{Associativity}{associativity}
	Let $F$ be a field and let $X$, $Y$ and $Z$ be $F$-vector spaces. There exists a unique isomorphism
	\begin{equation*}
		a_{X, Y, Z} \colon (X \otimes Y) \otimes Z \to X \otimes (Y \otimes Z)
	\end{equation*}
	such that the following diagram commutes:
	\begin{center}
		\begin{tikzcd}
			X \times Y \times Z \arrow[rr, "\otimes(\otimes \times \id_Z)"] \arrow[rrd, swap, "\otimes(\id_X \times \otimes)"] & {} & (X \otimes Y) \otimes Z \arrow[d, dashed, "a_{X,Y,Z}"] \\
			{} & {} & X \otimes (Y \otimes Z)
		\end{tikzcd}
	\end{center}
	For any $x \in X$, $y \in Y$ and $z \in Z$ it satisfies that
	\begin{equation}
		a_{X,Y,Z}((x \otimes y) \otimes z) = x \otimes (y \otimes z).
	\end{equation}
\end{corollary}
\begin{proof}
	Since both pairs $((X \otimes Y) \otimes Z, \otimes(\otimes \times \id_Z))$ and $(X \otimes (Y \otimes Z), \otimes (\id_X \times \otimes))$ are tensor products for the triple $(X,Y,Z)$, it follows from the universal property that there exists a unique isomorphism
	\begin{equation*}
		a_{X, Y, Z} \colon (X \otimes Y) \otimes Z \to X \otimes (Y \otimes Z)
	\end{equation*}
	such that the diagram commutes. Let $x \in X$, $y \in Y$ and $z \in Z$. We have that
	\begin{align*}
		a_{X,Y,Z} ((x \otimes y) \otimes z) &\stackrel{(i)}{=}  a_{X,Y,Z} \otimes (\otimes \times \id_Z)(x,y,z)\\
		&\stackrel{(ii)}{=} \otimes(\id_X \times \otimes) (x,y,z)\\
		&\stackrel{(iii)}{=} x \otimes (y \otimes z).
	\end{align*}
	In $(i)$ the tensor products were written in prefix notation; in $(ii)$ was used the commutativity of the diagram; in $(iii)$ the tensor products were applied and written in infix notation.
\end{proof}

\section{Evaluations and coevaluations}

The dual of a vector space $X$ is the space of linear functionals
\begin{equation*}
	X^* = \Lin(X, F).
\end{equation*}
As such, we can construct an evaluation function that applies linear functions to vectors:
\begin{align*}
	X^* \times X &\longrightarrow F \\
	(f, x) &\longmapsto f(x).
\end{align*}
This function is bilinear. The universal property of $X^* \otimes X$ then says that it factors uniquely through $\otimes$ and a linear function $X^* \otimes X \to F$. This linear function is also called an evaluation and is important for the diagrammatical calculus of the category of finite dimensional vector spaces (see section \ref{section: string diagrams}). We have two types of evaluations, one for each order of the factors $X^*$ and $X$.

\begin{proposition}{Evaluations}{evaluations}
	Let $F$ be a field and $X$ be an $F$-vector space. The evaluation of functionals define bilinear maps:
	\begin{align*}
		\overleftarrow{Ev}_X \colon X^* \times X &\longrightarrow F\\
		(f, x) &\longmapsto f(x),\\
		\overrightarrow{Ev}_X \colon X \times X^* &\longrightarrow F\\
		(x, f) &\longmapsto f(x).
	\end{align*}
	There exist unique linear transformations
	\begin{equation*}
		\lev_X \colon X^* \otimes X \to F,
	\end{equation*}
	\begin{equation*}
		\rev_X \colon X \otimes X^* \to F,
	\end{equation*}
	such that the following diagrams commute:
	\begin{center}
		\begin{tikzcd}
			X^* \times X \arrow[r, "\otimes"] \arrow[rd, swap, "\overleftarrow{Ev}_X"] & X^* \otimes X \arrow[d, dashed, "\lev_X"] \\
			{} & F
		\end{tikzcd}
	\end{center}
	\begin{center}
		\begin{tikzcd}
			X \times X^* \arrow[r, "\otimes"] \arrow[rd, swap, "\overrightarrow{Ev}_X"] & X \otimes X^* \arrow[d, dashed, "\rev_X"] \\
			{} & F
		\end{tikzcd}
	\end{center}
	For any $x \in X$ and $f \in X^*$, we have that
	\begin{gather}
		\lev_X (f \otimes x) = f(x), \\
		\rev_X (x \otimes f) = f(x).
	\end{gather}
\end{proposition}
\begin{proof}
	That bilinearity of $\overleftarrow{Ev}_X$ and $\overrightarrow{Ev}_X$ follows from the linearity of functionals. The universal property of the tensor product then says that there exists unique linear transformations $\lev_X$ and $\rev_V$ such that the diagrams commute.
	
	Let $x \in X$ and $f \in X^*$. We have that
	\begin{align*}
		\lev_X (f \otimes x) &\stackrel{(i)}{=} \lev_X \otimes (f,x) \\
		&\stackrel{(ii)}{=} \overleftarrow{Ev}_X (f,x) \\
		&\stackrel{(iii)}{=} f(x).
	\end{align*}
	In $(i)$ the $\otimes$ was written in prefix notation; in $(ii)$ was used the commutativity of the diagram for $\lev_X$; in $(iii)$ was used the definition of $\overleftarrow{Ev}_X$. The expression for $\rev_X$ is proved similarly.
\end{proof}

The reason for putting arrows on the evalutions is that they indicate how to draw their respective string diagrams.

For finite dimensional spaces, the evaluations $\lev_X$ and $\rev_X$ are examples of what we call non-degenerate pairings (\cite{MonoidalCatsAndTFT}, section 1.5 Pairings in monoidal categories). In this case, there are linear transformations, called coevaluations, which act as a kind of inverses to the evaluations, but not in the sense of direct composition. More generally, they exist for projective modules (lemma 1.6 of \cite{MonoidalCatsAndTFT}). The coevaluations are very important for string diagrams.

\begin{definition}{Coevaluations}{coevaluations}
	Let $F$ be a field and let $X$ be a \textbf{finite dimensional} $F$-vector space. We define coevalution linear transformations as follows. Given a basis $(x_i)_{i \in I}$ for $X$, we have the corresponding dual basis $(x^i)_{i \in I}$ for $X^*$. It is characterized as the functionals that satisfy the equation
	\begin{equation}
		x^i (x_j) = \delta_{i,j},
	\end{equation}
	for every $i,j \in J$. The coevaluations are defined as
	\begin{align*}
		\rcoev_X \colon F &\longrightarrow X^* \otimes X \\
		\lambda &\longmapsto \lambda \sum_{i \in I} x^i \otimes x_i, \\
		\lcoev_X \colon F &\longrightarrow X \otimes X^* \\
		\lambda &\longmapsto \lambda \sum_{i \in I} x_i \otimes x^i .
	\end{align*}
\end{definition}
Again, the arrows over the coevaluations are to indicate how to draw their string diagrams.

Since a basis was used in the definition of the coevalutions, it may seem that they depend on a choice of basis. This isn't the case, the way the vectors $x_i$ and their duals $x^i$ are combined makes any effect from a change of basis be canceled by the changes in the dual.

\begin{proposition}{Coevaluations are basis independent}{coevaluations basis independent}
	Let $F$ be a field and $X$ be a \textbf{finite dimensional} $F$-vector space. The linear transformations $\lcoev_X$ and $\rcoev_X$ defined in \ref{definition:coevaluations} don't depend on the choice of basis.
\end{proposition}
\begin{proof}
	Let $(x_i)_{i=1,\dots,n}$ be a basis for $X$ and $(x^i)_{i=1,\dots,n}$ be the dual basis for $X^*$. Any other basis for $X$ is related to $(x_i)_{i=1,\dots,n}$ by some invertible matrix $M \in M_n (F)$. That is, for any other basis $(x'_i)_{i=1,\dots,n}$ there exists a unique invertible matrix $M \in M_n (F)$ such that
	\begin{equation}
		x'_i = \sum_j M_{i,j} x_j ,
	\end{equation}
	for $i = 1,\dots,n$. We also have the dual basis $(x'^i)_{i=1,\dots,n}$ for $X^*$. Since $(x^i)_{i=1,\dots,n}$ is a basis for $X^*$, we can write that
	\begin{equation}
		x'^i = \sum_j a_{j,i} x^j ,
	\end{equation}
	for some coefficients $a_{j,i} \in F$. To determine these coefficients, we apply the functional on the basis vector $x_k$:
	\begin{align*}
		x'^i (x_k) &\stackrel{(i)}{=} \sum_j a_{j,i} x^j (x_k) \\
		&\stackrel{(ii)}{=} \sum_j a_{j,i} \delta_{j,k} \\
		&\stackrel{(iii)}{=} a_{k,i} .
	\end{align*}
	In $(i)$ $x'^i$ was written in the basis $(x^i)_{i=1,\dots,n}$; in $(ii)$ was used that $x^j (x_k) = \delta_{j,k}$; in $(iii)$ the delta was used to eliminate the summation. This shows that
	\begin{equation}
		\label{equation: x'^i(x_k)}
		a_{k,i} = x'^i (x_k).
	\end{equation}
	
	Also, $x'^i$ is a vector of the dual basis, so it satisfies that
	\begin{equation}
		\label{equation: dual basis other}
		x'^i (x'_l) = \delta_{i,l},
	\end{equation}
	for $i,l = 1, \dots, n$. We can write $x'_l$ in the basis $(x_i)_{i=1,\dots,n}$ as
	\begin{equation}
		x'_l = \sum_j M_{l,j} x_j .
	\end{equation}
	Substituting this in equation \ref{equation: dual basis other}, we get
	\begin{align*}
		\delta_{i,l} &\stackrel{(i)}{=} x'^i (x'_l)\\
		&\stackrel{(ii)}{=} x'^i \left( \sum_j M_{l,j} x_j \right) \\
		&\stackrel{(iii)}{=} \sum_j M_{l,j} x'^i ( x_j ) \\
		&\stackrel{(iv)}{=} \sum_j M_{l,j} a_{j,i} .
	\end{align*}
	$(i)$ is equation \ref{equation: dual basis other}; in $(ii)$ $x'_l$ was written in the original basis; in $(iii)$ was used that the functional $x'^i$ is a linear function, so we can pass the summation and $M_{l,j}$ to outside; in $(iv)$ was used equation \ref{equation: x'^i(x_k)}. We obtained that
	\begin{equation*}
		\sum_j M_{l,j} a_{j,i} = \delta_{i,l},
	\end{equation*}
	for $i,l = 1,\dots,n$. We can form the matrix of coefficients
	\begin{equation}
		A \coloneqq (a_{j,i})_{j,i=1,\dots,n}.
	\end{equation}
	Then the equation just obtained can be rewritten as:
	\begin{equation}
		M A = I_n .
	\end{equation}
	$M$ is a square matrix, so it has a right inverse iff it is invertible and the right inverse is the inverse. Therefore,
	\begin{equation}
		A = M^{-1}.
	\end{equation}
	
	We showed that if the basis for $X$ changes by a matrix $M$, then the dual basis changes by the inverse $M^{-1}$. This will lead to the coevaluations to be basis independent, which we show next. Let's compute $\lcoev_X$ in the new basis. We have that
	\begin{align*}
		\lcoev_X(1) &\stackrel{(i)}{=} \sum_i x'_i \otimes x'^i \\
		&\stackrel{(ii)}{=} \sum_i \left( \sum_j M_{i,j} x_j \right) \otimes \left( \sum_k a_{k,i} x^k \right) \\
		&\stackrel{(iii)}{=} \sum_{j,k} \sum_i a_{k,i} M_{i,j} (x_j \otimes x^k ) \\
		&\stackrel{(iv)}{=} \sum_{j,k} \sum_i (M^{-1})_{k,i} M_{i,j} (x_j \otimes x^k ) \\
		&\stackrel{(v)}{=} \sum_{j,k} (M^{-1} M)_{k,j} (x_j \otimes x^k ) \\
		&\stackrel{(vi)}{=} \sum_{j,k} (I_n)_{k,j} (x_j \otimes x^k ) \\
		&\stackrel{(vii)}{=} \sum_{j,k} \delta_{k,j} (x_j \otimes x^k ) \\
		&\stackrel{(viii)}{=} \sum_j x_j \otimes x^j.
	\end{align*}
	In $(i)$ was used the definition of $\lcoev_X$, but using the new basis; in $(ii)$ the new basis written in the original basis; in $(iii)$ was used the bilinearity of the tensor product to pass the summations and coefficients to outside; in $(iv)$ was used that $A = M^{-1}$; in $(v)$ the matrix multiplication $M^{-1} M$ was identified; in $(vi)$ was used that $M^{-1} M = I_n$; in $(vii)$ was used that $(I_n)_{k,j} = \delta_{k,j}$; in $(viii)$ the delta was used to eliminate the summation over $k$.
	
	This shows that
	\begin{equation}
		\sum_i x'_i \otimes x'^i = \sum_i x_i \otimes x^i ,
	\end{equation}
	so $\lcoev_X$ doesn't depend on the choice of basis. Both $\lcoev_X$ and $\rcoev_X$ are related by the symmetry $\tau_{X,X^*} \colon X \otimes X^* \stackrel{\cong}{\to} X^* \otimes X$, so $\rcoev_X$ is also basis independent.
\end{proof}

\begin{proposition}{Non-degeneracy of evaluations}{ev is non degenerate pairing}
	Let $F$ be a field and $X$ be a \textbf{finite dimension} $F$-vector space. The evaluations $\lev_X$ and $\rev_X$ are non-degenerate pairings with inverses $\lcoev_X$ and $\rcoev$, in the sense that
	\begin{gather}
		l_{X^*} (\lev_X \otimes \id_{X^*}) a_{X^*, X, X^*}^{-1} (\id_{X^*} \otimes \lcoev_X) r_{X^*}^{-1} = \id_{X^*}, \\
		r_X (\id_X \otimes \lev_X) a_{X, X^*, X} (\lcoev_X \otimes \id_X ) l_X^{-1} = \id_X, \\
		l_X (\rev_X \otimes \id_X) a_{X,X^*,X}^{-1} (\id_X \otimes \rcoev_X) r_X^{-1} = \id_X , \\
		r_{X^*} (\id_{X^*} \otimes \rev_X) a_{X^* ,X ,X^* } (\rcoev_X \otimes \id_{X^*} ) l_{X^*}^{-1} = \id_{X^*} .
	\end{gather}
\end{proposition}
\begin{proof}
	The first equation will be proved, the other three have similar proofs. Let $f \in X^*$, let $(x_i)_i$ be a basis for $X$ and $(x^i)_i$ be the dual basis for $X^*$. We have that
	\begin{align*}
		l_{X^*} (\lev_X \otimes \id_{X^*}) a_{X^*, X, X^*}^{-1} &(\id_{X^*} \otimes \lcoev_X) r_{X^*}^{-1} (f) \\
		&\stackrel{(i)}{=} l_{X^*} (\lev_X \otimes \id_{X^*}) a_{X^*, X, X^*}^{-1} (\id_{X^*} \otimes \lcoev_X) (f \otimes 1) \\
		&\stackrel{(ii)}{=} l_{X^*} (\lev_X \otimes \id_{X^*}) a_{X^*, X, X^*}^{-1} (f \otimes \lcoev_X(1)) \\
		&\stackrel{(iii)}{=} l_{X^*} (\lev_X \otimes \id_{X^*}) a_{X^*, X, X^*}^{-1} \left(f \otimes \left( \sum_i x_i \otimes x^i \right) \right) \\
		&\stackrel{(iv)}{=} \sum_i l_{X^*} (\lev_X \otimes \id_{X^*}) a_{X^*, X, X^*}^{-1} (f \otimes (  x_i \otimes x^i)) \\
		&\stackrel{(v)}{=} \sum_i l_{X^*} (\lev_X \otimes \id_{X^*}) ((f \otimes x_i) \otimes x^i) \\
		&\stackrel{(vi)}{=} \sum_i l_{X^*} (\lev_X(f \otimes x_i) \otimes x^i) \\
		&\stackrel{(vii)}{=} \sum_i l_{X^*} (f(x_i) \otimes x^i) \\
		&\stackrel{(viii)}{=} \sum_i  f(x_i) x^i .
	\end{align*}
	In $(i)$ was applied $r_{X*}^{-1}$; in $(ii)$ was applied $\id_{X^*} \otimes \lcoev_X$; in $(iii)$ was used the definition of $\lcoev_X$; in $(iv)$ was used the linearity to pass the summation to outside; in $(v)$ was applied $a_{X^*, X, X^*}^{-1}$, which reorders the parenthesis; in $(vi)$ was applied $\lev_X \otimes \id_{X^*}$; in $(vii)$ was used the definition of $\lev_X$, which evaluates the functional $f$ on $x_i$; in $(viii)$ was used the definition of $l_{X^*}$.
	
	$\sum_i  f(x_i) x^i$ is $f$ expressed in the basis. In fact, if we evaluate it at a basis vector $x_j$, we get
	\begin{align*}
		\sum_i f(x_i) x^i (x_j) &\stackrel{(i)}{=} \sum_i f(x_i) \delta_{i,j} \\
		&\stackrel{(ii)}{=} f(x_j).
	\end{align*}
	In $(i)$ was used that $x^i(x_j) = \delta_{i,j}$, which is the characterization of the dual basis; in $(ii)$ the delta was used to eliminate the summation.
	
	Since both linear transformations $\sum_i  f(x_i) x^i$ and $f$ coincide on a basis, they are equal:
	\begin{equation}
		\sum_i  f(x_i) x^i = f.
	\end{equation}
	This proves that
	\begin{equation}
		l_{X^*} (\lev_X \otimes \id_{X^*}) a_{X^*, X, X^*}^{-1} (\id_{X^*} \otimes \lcoev_X) r_{X^*}^{-1} = \id_{X^*} .
	\end{equation} 
\end{proof}

This result is represented in string diagrams as yanking equations. Next we show that evaluations and coevaluations are dual to each other. For duality to make sense, we need an inner product, so we restrict to Hilbert spaces.

\begin{proposition}{}{}
	Let $X$ be a finite dimensional complex Hilbert space. Then
	\begin{equation}
		\label{equation: lev dual}
		\lev_X^\dag = \rcoev_X,
	\end{equation}
	\begin{equation}
		\rev_X^\dag = \lcoev_X.
	\end{equation}
\end{proposition}
\begin{proof}
	We'll prove equation \ref{equation: lev dual}, the other equation has a similar proof. Let $(x_i)_i$ be an orthonormal basis for $X$, and let $(x^i)_i$ be the dual basis for $X^*$. We have that
	\begin{align*}
		\langle \lev_X (x^i \otimes x_j), 1 \rangle &\stackrel{(i)}{=} \langle x^i(x_j), 1 \rangle \\
		&\stackrel{(ii)}{=} \langle \delta_{i,j}, 1 \rangle \\
		&\stackrel{(iii)}{=} \delta_{i,j}.
	\end{align*}
	In $(i)$ was used proposition \ref{proposition:evaluations}; in $(ii)$ was used that $(x^i)_i$ is the dual basis, so $x^i(x_j) = \delta_{i,j}$; in $(iii)$ was used the definition of the inner product for $\complexNumbers$, which is given by $\langle \alpha, \beta \rangle = \overline{\alpha} \beta$.
	
	We also have that
	\begin{align*}
		\langle x^i \otimes x_j, \rcoev(1) \rangle &\stackrel{(i)}{=} \left\langle x^i \otimes x_j, \sum_k x^k \otimes x_k \right\rangle \\
		&\stackrel{(ii)}{=} \sum_k \langle x^i \otimes x_j, x^k \otimes x_k \rangle \\
		&\stackrel{(iii)}{=} \sum_k \langle x^i, x^k \rangle \langle x_j, x_k \rangle \\
		&\stackrel{(iv)}{=} \sum_k \langle x_k, x_i\rangle \langle x_j, x_k \rangle \\
		&\stackrel{(v)}{=} \sum_k \delta_{k,i} \delta_{k,j} \\
		&\stackrel{(vi)}{=} \delta_{i,j}.
	\end{align*}
	In $(i)$ was used definition \ref{definition:coevaluations}; in $(ii)$ was used that an inner product is linear in the second argument, to pass the summation to outside; in $(iii)$ was used the definition of the inner product over a tensor product, as in equation \ref{equation: inner product of tensor product}; in $(iv)$ was used that $(x_i)_i$ is orthonormal and propositions \ref{proposition:dual hilbert space} and \ref{proposition:orthonormal dual basis}; in $(v)$ was used that $(x_i)_i$ is orthonormal; in $(vi)$ $\delta_{k,i}$ was used to eliminate the summation. This proves that
	\begin{equation}
		\langle \lev_X (x^i \otimes x_j), 1 \rangle = \langle x^i \otimes x_j, \rcoev(1) \rangle.
	\end{equation}
	It follows that $\lev_X^\dag = \rcoev_X$.
\end{proof}

\section{Internal hom property}

We denote the set of functions $f \colon X \to Y$ as $\Set(X,Y)$ (see appendix \ref{appendix: category theory}). We have the following property for the category of sets: any function $f \colon X \times Y \to Z$ corresponds to the curried function
\begin{align*}
	\tilde{f} \colon X &\longrightarrow \Set(Y,Z)\\
	x &\longmapsto \tilde{f}(x),
\end{align*}
where $\tilde{f}(x)$ is the function
\begin{align*}
	\tilde{f}(x) \colon Y &\longrightarrow Z \\
	y &\longmapsto f(x,y).
\end{align*}
This establishes a bijection
\begin{equation}
	\Set(X \times Y, Z) \cong \Set(X, \Set(Y,Z)).
\end{equation}

A similar idea may be applicable to monoidal categories, but replacing the cartesian product $\times$ by the monoidal product $\otimes$. This leads to the concept of closed monoidal categories. We won't study closed monoidal categories in its full generality, rather we'll focus on the category of finite dimensional vector spaces. A brief exposition of the general ideas will be given for context.

The left side of the bijection has $\Set(X \times Y, Z)$, which is replaced by the hom-set $\cat{C}(X \otimes Y, Z)$, where $\cat{C}$ is some monoidal category. The right side needs a substitute for the hom-set $\Set(Y,Z)$, which must be an object of the category $\cat{C}$. Such object, if it exists, is called an internal hom and is denoted as $[Y,Z]$. The internal hom must satisfy that
\begin{equation}
	\cat{C}(X \otimes Y, Z) \cong \cat{C}(X, [Y,Z]).
\end{equation}

The category of finite dimensional $F$-vector spaces has as internal hom the $F$-vector space
\begin{equation}
	[Y,Z] \coloneqq Y^* \otimes Z
\end{equation}
As such, we have an isomorphism
\begin{equation}
	\label{equation: lin closed monoidal}
	\Lin_F(X \otimes Y, Z) \cong \Lin_F(X, [Y,Z]).
\end{equation}
This will be important for constructing the \CJ isomorphism, which is used in Quantum Information Theory. It is simple to construct the isomorphism from equation \ref{equation: lin closed monoidal} using string diagrams for spherical categories, of which the category of finite dimensional vector spaces is a special case. In proposition \ref{proposition:pulling up or down} we have a proof of an isomorphism that is almost what we need. There was proved that we have an isomorphism
\begin{equation}
	\varphi \colon \Lin_F (X \otimes Y, Z) \to \Lin_F (X, Z \otimes Y^*).
\end{equation}
We then just have to replace $Z \otimes Y^*$ with $Y^* \otimes Z$, which is done by the symmetry
\begin{equation}
	\tau_{Z, Y^*} \colon Z \otimes Y^* \stackrel{\cong}{\to} Y^* \otimes Z . 
\end{equation}
The isomorphism of equation \ref{equation: lin closed monoidal} is proved in the next proposition. It is a simple corollary of \ref{proposition:pulling up or down}, but a direct proof is made in case the reader wants a non diagrammatic proof.

\begin{proposition}{Internal hom property}{tensor monoidal closed}
	Let $F$ be a field and let $X$, $Y$ and $Z$ be \textbf{finite dimensional} $F$-vector spaces. We have an isomorphism
	\begin{align*}
		\widetilde{\varphi} \colon \Lin_F (X \otimes Y, Z) &\longrightarrow \Lin_F (X, Y^* \otimes Z)\\
		f &\longmapsto \tau_{Z, Y^*} (f \otimes \id_{Y^*}) a_{X,Y,Y^*}^{-1} (\id_X \otimes \lcoev_Y ) r_X^{-1},
	\end{align*}
	with inverse
	\begin{align*}
		\widetilde{\varphi}^{-1} \colon \Lin_F (X, Y^* \otimes Z) &\longrightarrow \Lin_F (X \otimes Y, Z) \\
		g &\longmapsto r_Z (\id_Z \otimes \lev_Y) a_{Z, Y^*, Y} (\tau_{Y^*, Z} g \otimes \id_Y).
	\end{align*}
\end{proposition}
\begin{proof}
	From proposition \ref{proposition:pulling up or down}, we have an isomorphism
	\begin{align*}
		\varphi \colon \Lin_F (X \otimes Y, Z) &\longrightarrow \Lin_F (X, Z \otimes Y^*)\\
		f &\longmapsto (f \otimes \id_{Y^*}) a_{X,Y,Y^*}^{-1} (\id_X \otimes \lcoev_Y ) r_X^{-1},
	\end{align*}
	with inverse
	\begin{align*}
		\varphi^{-1} \colon \Lin_F (X, Z \otimes Y^*) &\longrightarrow \Lin_F (X \otimes Y, Z) \\
		g &\longmapsto r_Z (\id_Z \otimes \lev_Y) a_{Z, Y^*, Y} (g \otimes \id_Y) .
	\end{align*}
	Then $\widetilde{\varphi}$ is the composition of $\varphi$ with the isomorphism
	\begin{align*}
		h \colon \Lin_F (X, Z \otimes Y^*) &\longrightarrow \Lin_F (X, Y^* \otimes Z) \\
		g &\longmapsto \tau_{Z, Y^*} g.
	\end{align*}
	That is, we have that
	\begin{equation}
		\widetilde{\varphi} = h \varphi.
	\end{equation}
	It is immediate that $h$ has as inverse the linear transformation
	\begin{align*}
		h^{-1} \colon \Lin_F (X, Y^* \otimes Z) &\longrightarrow \Lin_F (X, Z \otimes Y^*) \\
		g &\longmapsto \tau_{Z, Y^*}^{-1} g.
	\end{align*}
	The symmetry $\tau_{Z, Y^*}$ just swaps the vectors in an elementary tensor, that is,
	\begin{equation}
		\tau_{Z, Y^*} (z \otimes \alpha) = \alpha \otimes z,
	\end{equation}
	for any $z \in Z$ and $\alpha \in Y^*$. As such, $\tau_{Z, Y^*}^{-1}$ is the linear transformation that does the reverse swap, so
	\begin{equation}
		\tau_{Z, Y^*}^{-1} = \tau_{Y^* , Z}.
	\end{equation}
	
	Since $\widetilde{\varphi}$ is the composition of two isomorphism, it is also an isomorphism, with inverse
	\begin{equation}
		\widetilde{\varphi} = \varphi^{-1} h^{-1}.
	\end{equation}
	This concludes the proof.
\end{proof}
\begin{proof}[\protect{Alternative proof of proposition \ref{proposition:tensor monoidal closed}}]
	The previous proof relied on string diagrams. String diagrams can be difficult to understand, because of the use of Mac Lane's coherence theorem. A direct proof will be given that doesn't rely on it, even though the expressions for the isomorphisms were deduced from the diagrammatic calculus. Mac Lane's coherence theorem and string diagrams are powerful tools for monoidal categories. As we'll see, not using them makes the proof much longer. It is still a simple proof, but tedious.
	
	Let $f \in \Lin_F (X \otimes Y, Z)$ and $g \in \Lin_F (X, Y^* \otimes Z)$. We have to prove that
	\begin{equation}
		\widetilde{\varphi}^{-1} \widetilde{\varphi} (f) = f,
	\end{equation}
	and
	\begin{equation}
		\label{equation: tensor monoidal closed identity g}
		\widetilde{\varphi} \widetilde{\varphi}^{-1} (g) = g.
	\end{equation}
	This can be proved by checking that the equations hold over elementary tensors. We'll then prove that
	\begin{equation}
		\label{equation: tensor monoidal closed identity f}
		[\widetilde{\varphi}^{-1} \widetilde{\varphi} (f)] (x \otimes y) = f (x \otimes y),
	\end{equation}
	for any $x \in X$ and $y \in Y$. For $g$ we can evaluate it at any $x \in X$, but it will be helpful to evaluate at basis vectors. Since we'll need basis for the spaces, let $(x_i)_i$ be a basis for $X$, let $(y_j)_j$ be a basis for $Y$, let $(y^j)_j$ be the respective dual basis for $Y^*$ and let $(z_k)_k$ be a basis for $Z$. By definition, the dual basis satisfies that $y^j (y_{j'}) = \delta_{j,j'}$.
	
	Let's start proving equation \ref{equation: tensor monoidal closed identity f}. First, let's expand and simplify the expression for $\widetilde{\varphi}^{-1} (\widetilde{\varphi} (f))$. We have that
	\begin{align*}
		\widetilde{\varphi}^{-1} \widetilde{\varphi} (f) &\stackrel{(i)}{=} \widetilde{\varphi}^{-1}(\tau_{Z, Y^*} (f \otimes \id_{Y^*}) a_{X,Y,Y^*}^{-1} (\id_X \otimes \lcoev_Y ) r_X^{-1}) \\
		&\stackrel{(ii)}{=} r_Z (\id_Z \otimes \lev_Y) a_{Z, Y^*, Y} (\tau_{Y^*, Z} ( \\ 
		& \qquad \qquad \tau_{Z, Y^*} (f \otimes \id_{Y^*}) a_{X,Y,Y^*}^{-1} (\id_X \otimes \lcoev_Y ) r_X^{-1} \\
		& \qquad ) \otimes \id_Y) \\
		&\stackrel{(iii)}{=} r_Z (\id_Z \otimes \lev_Y) a_{Z, Y^*, Y} [(f \otimes \id_{Y^*}) a_{X,Y,Y^*}^{-1} (\id_X \otimes \lcoev_Y ) r_X^{-1} \otimes \id_Y].
	\end{align*}
	In $(i)$ was used the expression for $\widetilde{\varphi}$; in $(ii)$ was used the expression for $\widetilde{\varphi}^{-1}$; in $(iii)$ was used that $\tau_{Y^*, Z} \tau_{Z, Y^*} = \id_{Z \otimes Y^*}$.
	
	Next we have to apply this expression for $\widetilde{\varphi}^{-1} \widetilde{\varphi} (f)$ over an elementary tensor. Let $x \in X$ and $y \in Y$, we'll apply the expression over $x \otimes y$. We have that
	\begin{align*}
		[&\widetilde{\varphi}^{-1} \widetilde{\varphi} (f)] (x \otimes y) \\
		&\stackrel{(i)}{=} r_Z (\id_Z \otimes \lev_Y) a_{Z, Y^*, Y} [(f \otimes \id_{Y^*}) a_{X,Y,Y^*}^{-1} (\id_X \otimes \lcoev_Y ) r_X^{-1} \otimes \id_Y] (x \otimes y) \\
		&\stackrel{(ii)}{=} r_Z (\id_Z \otimes \lev_Y) a_{Z, Y^*, Y} [(f \otimes \id_{Y^*}) a_{X,Y,Y^*}^{-1} (\id_X \otimes \lcoev_Y ) r_X^{-1}(x) \otimes y]\\
		&\stackrel{(iii)}{=} r_Z (\id_Z \otimes \lev_Y) a_{Z, Y^*, Y} [(f \otimes \id_{Y^*}) a_{X,Y,Y^*}^{-1} (\id_X \otimes \lcoev_Y ) (x \otimes 1) \otimes y]\\
		&\stackrel{(iv)}{=} r_Z (\id_Z \otimes \lev_Y) a_{Z, Y^*, Y} [(f \otimes \id_{Y^*}) a_{X,Y,Y^*}^{-1} (x \otimes \lcoev_Y(1) ) \otimes y]\\
		&\stackrel{(v)}{=} r_Z (\id_Z \otimes \lev_Y) a_{Z, Y^*, Y} \left[(f \otimes \id_{Y^*}) a_{X,Y,Y^*}^{-1} \left(x \otimes \sum_j \left(y_j \otimes y^j\right) \right) \otimes y \right]\\
		&\stackrel{(vi)}{=} \sum_j r_Z (\id_Z \otimes \lev_Y) a_{Z, Y^*, Y} [(f \otimes \id_{Y^*}) a_{X,Y,Y^*}^{-1} (x \otimes (y_j \otimes y^j)) \otimes y]\\
		&\stackrel{(vii)}{=} \sum_j r_Z (\id_Z \otimes \lev_Y) a_{Z, Y^*, Y} [(f \otimes \id_{Y^*}) ((x \otimes y_j) \otimes y^j) \otimes y]\\
		&\stackrel{(viii)}{=} \sum_j r_Z (\id_Z \otimes \lev_Y) a_{Z, Y^*, Y} [(f(x \otimes y_j) \otimes y^j) \otimes y]\\
		&\stackrel{(iv)}{=} \sum_j r_Z (\id_Z \otimes \lev_Y) [f(x \otimes y_j) \otimes (y^j \otimes y)]\\
		&\stackrel{(x)}{=} \sum_j r_Z [f(x \otimes y_j) \otimes \lev_Y(y^j \otimes y)]\\
		&\stackrel{(xi)}{=} \sum_j r_Z [f(x \otimes y_j) \otimes y^j(y)]\\
		&\stackrel{(xii)}{=} \sum_j y^j(y) f(x \otimes y_j)\\
		&\stackrel{(xiii)}{=} f\left(x \otimes \sum_j y^j(y) y_j \right)\\
		&\stackrel{(xiv)}{=} f(x \otimes y).
	\end{align*}
	The steps were the following:
	\begin{enumerate}[label=(\roman*)]
		\item Was applied the expression for $\widetilde{\varphi}^{-1} \widetilde{\varphi} (f)$;
		\item The expression inside the square brackets was applied over $x \otimes y$;
		\item Was used that $r^{-1}(x) = x \otimes 1$, as proved in proposition \ref{proposition:right unitality};
		\item $\id_X \otimes \lcoev_Y$ was applied over $x \otimes 1$;
		\item Was used that $\lcoev_Y(1) = \sum_j y_j \otimes y^j$, as in definition \ref{definition:coevaluations};
		\item The $\sum_j$ was passed to outside, which is possible because all functions are linear and $\otimes$ is bilinear;
		\item Was applied the associativity isomorphism $a^{-1}_{X, Y, Y^*}$, which reorders the parathesis from $x \otimes (y_j \otimes y^j)$ to $(x \otimes y_j) \otimes y^j$;
		\item $f \otimes \id_{Y^*}$ was applied to $(x \otimes y_j) \otimes y^j$;
		\item The associativity isomorphism $a_{Z, Y^*, Y}$ was applied to $(f(x \otimes y_j) \otimes y^j) \otimes y$, which reorders the parenthesis to $f(x \otimes y_j) \otimes (y^j \otimes y)$;
		\item $\id_Z \otimes \lev_Y$ was applied to $f(x \otimes y_j) \otimes (y^j \otimes y)$;
		\item Was used the definition of $\lev_Y$, as in proposition \ref{proposition:evaluations};
		\item Was applied $r_Z$, which replaces a $\otimes$ by a scalar multiplication, as proved in proposition \ref{proposition:right unitality};
		\item The $\sum_j y^j (y)$ was passed to inside of $f$, using that $f$ is linear and the bilinearity of $\otimes$;
		\item Was used that $\sum_j y^j(y) y_j = y$. In fact, $(y_j)_j$ is a basis for $Y$, so $y$ can be uniquely expressed as
		\begin{equation}
			y = \sum_k \lambda_k y_k,
		\end{equation}
		for some coefficients $\lambda_j \in F$. By definition, $y^j (y_k) = \delta_{j,k}$. Therefore,
		\begin{equation}
			y^j (y) = \sum_k \lambda_k y^j (y_k) = \sum_j \lambda_k \delta_{j,k} = \lambda_j .
		\end{equation}
	\end{enumerate}
	This proves equation \ref{equation: tensor monoidal closed identity f}. Since the equation holds for any elementary tensor $x \otimes y$, it follows that
	\begin{equation}
		\widetilde{\varphi}^{-1} \widetilde{\varphi} (f) = f.
	\end{equation}
	Since $f$ is arbitrary, we have that
	\begin{equation}
		\widetilde{\varphi}^{-1} \widetilde{\varphi} = \id_{\Lin_F (X \otimes Y, Z)}.
	\end{equation}
	
	Next we prove equation \ref{equation: tensor monoidal closed identity g}. Again, we first write an expression for $\widetilde{\varphi} \widetilde{\varphi}^{-1}(g)$. Using the expressions for $\widetilde{\varphi}$ and $\widetilde{\varphi}^{-1}(g)$, we get
	\begin{align*}
		\widetilde{\varphi} \widetilde{\varphi}^{-1}(g) &= \tau_{Z, Y^*} (\\
		&\qquad\qquad r_Z (\id_Z \otimes \lev_Y) a_{Z, Y^*, Y} (\tau_{Y^*, Z} g \otimes \id_Y) \\
		&\quad \otimes \id_{Y^*}) a_{X,Y,Y^*}^{-1} (\id_X \otimes \lcoev_Y ) r_X^{-1}.
	\end{align*}
	In this case we can't immediately cancel the symmetries $\tau_{Z, Y^*}$ and $\tau_{Y^*, Z}$. Next, let's evaluate this expression at a basis vector $x_i$. We can express $g(x_i)$ uniquely as
	\begin{equation}
		g(x_i) = \sum_{j,k} \lambda_{i,j,k} y^j \otimes z_k,
	\end{equation}
	for some unique scalars $\lambda_{i,j,k} \in F$. Then we have that
	\begin{align*}
		&[\widetilde{\varphi} \widetilde{\varphi}^{-1}(g)](x_i) \\ &\stackrel{(i)}{=} \tau_{Z, Y^*} [r_Z (\id_Z \otimes \lev_Y) a_{Z, Y^*, Y} (\tau_{Y^*, Z} g \otimes \id_Y) \otimes \id_{Y^*}] a_{X,Y,Y^*}^{-1} (\id_X \otimes \lcoev_Y ) r_X^{-1} (x_i) \\
		&\stackrel{(ii)}{=} \tau_{Z, Y^*} [r_Z (\id_Z \otimes \lev_Y) a_{Z, Y^*, Y} (\tau_{Y^*, Z} g \otimes \id_Y) \otimes \id_{Y^*}] a_{X,Y,Y^*}^{-1} (\id_X \otimes \lcoev_Y ) (x_i \otimes 1) \\
		&\stackrel{(iii)}{=} \tau_{Z, Y^*} [r_Z (\id_Z \otimes \lev_Y) a_{Z, Y^*, Y} (\tau_{Y^*, Z} g \otimes \id_Y) \otimes \id_{Y^*}] a_{X,Y,Y^*}^{-1} (x_i \otimes \lcoev_Y(1) )\\
		&\stackrel{(iv)}{=} \tau_{Z, Y^*} [r_Z (\id_Z \otimes \lev_Y) a_{Z, Y^*, Y} (\tau_{Y^*, Z} g \otimes \id_Y) \otimes \id_{Y^*}] a_{X,Y,Y^*}^{-1} \left(x_i \otimes \left(\sum_j y_j \otimes y^j \right) \right)\\
		&\stackrel{(v)}{=} \sum_j \tau_{Z, Y^*} [r_Z (\id_Z \otimes \lev_Y) a_{Z, Y^*, Y} (\tau_{Y^*, Z} g \otimes \id_Y) \otimes \id_{Y^*}] a_{X,Y,Y^*}^{-1} (x_i \otimes (y_j \otimes y^j))\\
		&\stackrel{(vi)}{=} \sum_j \tau_{Z, Y^*} [r_Z (\id_Z \otimes \lev_Y) a_{Z, Y^*, Y} (\tau_{Y^*, Z} g \otimes \id_Y) \otimes \id_{Y^*}] ((x_i \otimes y_j) \otimes y^j )\\
		&\stackrel{(vii)}{=} \sum_j \tau_{Z, Y^*} [r_Z (\id_Z \otimes \lev_Y) a_{Z, Y^*, Y} (\tau_{Y^*, Z} g(x_i) \otimes y_j) \otimes y^j]\\
		&\stackrel{(viii)}{=} \sum_j y^j \otimes r_Z (\id_Z \otimes \lev_Y) a_{Z, Y^*, Y} (\tau_{Y^*, Z} g(x_i) \otimes y_j)\\
		&\stackrel{(iv)}{=} \sum_j y^j \otimes r_Z (\id_Z \otimes \lev_Y) a_{Z, Y^*, Y} \left(\tau_{Y^*, Z} \left( \left(\sum_{j',k} \lambda_{i,j',k} y^{j'} \right) \otimes z_k \right) \otimes y_j \right)\\
		&\stackrel{(x)}{=} \sum_{j, j', k}\lambda_{i,j',k} \ y^j \otimes r_Z (\id_Z \otimes \lev_Y) a_{Z, Y^*, Y} (\tau_{Y^*, Z} (  y^{j'} \otimes z_k ) \otimes y_j )\\
		&\stackrel{(xi)}{=} \sum_{j, j', k}\lambda_{i,j',k} \ y^j \otimes r_Z (\id_Z \otimes \lev_Y) a_{Z, Y^*, Y} ((z_k \otimes y^{j'}) \otimes y_j )\\
		&\stackrel{(xii)}{=} \sum_{j, j', k}\lambda_{i,j',k} \ y^j \otimes r_Z (\id_Z \otimes \lev_Y) (z_k \otimes (y^{j'} \otimes y_j ))\\
		&\stackrel{(xiii)}{=} \sum_{j, j', k}\lambda_{i,j',k} \ y^j \otimes r_Z (z_k \otimes \lev_Y(y^{j'} \otimes y_j ))\\
		&\stackrel{(xiv)}{=} \sum_{j, j', k}\lambda_{i,j',k} \ y^j \otimes r_Z (z_k \otimes y^{j'}(y_j))\\
		&\stackrel{(xv)}{=} \sum_{j, j', k}\lambda_{i,j',k} \ y^j \otimes r_Z (z_k \otimes \delta_{j,j'})\\
		&\stackrel{(xvi)}{=} \sum_{j, k}\lambda_{i,j,k} \ y^j \otimes r_Z (z_k \otimes 1)\\
		&\stackrel{(xvii)}{=} \sum_{j, k}\lambda_{i,j,k} \ y^j \otimes z_k\\
		&\stackrel{(xviii)}{=} g(x_i).
	\end{align*}
	The steps were the following:
	\begin{enumerate}[label=(\roman*)]
		\item Was used the expression for $\widetilde{\varphi} \widetilde{\varphi}^{-1} (g)$;
		\item Was used that $r_X^{-1}(x_i) = x_i \otimes 1$, as proved in proposition \ref{proposition:right unitality};
		\item $\id_X \otimes \lcoev_Y$ was applied to $x_i \otimes 1$;
		\item Was used that $\lcoev_Y (1) = \sum_j y_j \otimes y^j$, as in definition \ref{definition:coevaluations};
		\item The $\sum_j$ was passed to outside, using that the functions are linear and that $\otimes$ is bilinear;
		\item Was applied the associativity isomorphism $a^{-1}_{X,Y,Y^*}$, which reorders the parenthesis from $x_i \otimes (y_j \otimes y^j)$ to $(x_i \otimes y_j) \otimes y^j$;
		\item $x_i$, $y_j$ and $y^j$ were distributed to the respective linear maps;
		\item $\tau_{Z, Y^*}$ was used to swap $y^j$ with the other factor;
		\item $g(x_i)$ was expressed in the basis for $Y^* \otimes Z$;
		\item The summation $\sum_{j', k} \lambda_{i,j',k}$ was passed to outside, using that all functions are linear and the bilinearity of $\otimes$;
		\item Was used $\tau_{Y^*, Z}$ to swap $y^{j'}$ and $z_k$;
		\item Was applied the associativity isomorphism $a_{Z, Y^*, Y}$, which reorders the parenthesis from $(z_k \otimes y^{j'}) \otimes y_j$ to $z_k \otimes  (y^{j'} \otimes y_j )$;
		\item $\id_Z \otimes \lev_Y$ was applied to $z_k \otimes  (y^{j'} \otimes y_j )$;
		\item Was used the definition of $\lev_Y$ as in proposition \ref{proposition:evaluations};
		\item Was used the definition of the dual basis, which says that $y^{j'}(y_j) = \delta_{j,j'}$;
		\item $\delta_{j,j'}$ was used to eliminate the summation over $j'$;
		\item Was used that $r_Z (z_k \otimes 1) = z_k$, as in proposition \ref{proposition:right unitality};
		\item Was used that the summation is $g(x_i)$ expressed in the basis for $Y^* \otimes Z$.
	\end{enumerate}
	This proves that
	\begin{equation}
		[\widetilde{\varphi} \widetilde{\varphi}^{-1}(g)](x_i) = g(x_i),
	\end{equation}
	for any of the basis vectors $x_i$. Since the equality holds for any basis vector, then
	\begin{equation}
		\widetilde{\varphi} \widetilde{\varphi}^{-1}(g) = g.
	\end{equation}
	Since this equality holds for any $g$, we have that
	\begin{equation}
		\widetilde{\varphi} \widetilde{\varphi}^{-1} = \id_{\Lin_F (X, Y^* \otimes Z)}.
	\end{equation}
	This concludes that $\widetilde{\varphi}$ is an isomorphism.
	
\end{proof}

\cleardoublepage
\chapter{Finite Dimensional Hilbert Spaces}
\label{appendix: hilbert spaces}

John von Neumann axiomatized Quantum Mechanics using complex Hilbert spaces. This work is about Quantum Information Theory using finite dimensional spaces. This appendix gives the necessary mathematical background about finite dimensional complex Hilbert spaces. It only exposes the essential features. Familiarity with Linear Algebra will be assumed.

\section{Hilbert spaces}

In finite dimensions, a Hilbert space is the same as an inner product space. As such, a finite dimensional complex Hilbert space is a complex vector space $X$ with an inner product $\langle \ , \ \rangle \colon X \times X \to \complexNumbers$. An inner product is a positive-definite sesquilinear form. The sesquilinear property treats differently the two arguments of the inner product. The tradition in Quantum Physics is to define it as being \textbf{antilinear in the first argument and linear in the second}. This convention is convenient for Dirac notation. As such, being sesquilinear means that the following properties are valid:
\begin{enumerate}
	\item $\forall x,x',x''\in X$, $\langle x,x'+x''\rangle = \langle x,x'\rangle+ \langle x,x''\rangle$;
	\item $\forall x,x' \in X$, $\forall \lambda \in \complexNumbers$, $\langle x,\lambda x'\rangle = \lambda \langle x,x'\rangle$;
	\item $\forall x,x' \in X$, $\langle x,x'\rangle = \overline{\langle x',x\rangle}$.
\end{enumerate}
To be positive-definite means that
\begin{equation}
	\langle x,x \rangle \geq 0,
\end{equation}
for any $x\in X$, with equality only when $x=0$.

\section{Induced norm}

Any inner product on a Hilbert space $X$ induces a norm. It is given by
\begin{equation}
	\label{equation: norm induced by inner product}
	||x|| = \sqrt{\langle x, x\rangle},
\end{equation}
for any $x \in X$. We'll typically use this norm, unless stated otherwise.

\section{The category of Hilbert spaces}

Finite dimensional complex Hilbert spaces form a category $\finhilb$, whose objects are the finite dimensional complex Hilbert spaces and arrows are the linear transformations between these maps.

\section{The adjoint}

Let $A\colon X \to Y$ be a $\complexNumbers$-linear map, where $X$ and $Y$ are finite dimensional complex Hilbert spaces. The adjoint of $A$ is the unique $\complexNumbers$-linear map $A^\dag \colon Y \to X$ such that
\begin{equation}
	\langle A(x), y\rangle = \langle x , A^\dag(y) \rangle,
\end{equation}
for every $x \in X$ and $y \in Y$. In the theory of Operator Algebras, it is common to denote the adjoint by $A^*$, but we won't use this notation for the adjoint. The reason for this is that there is a functor that comes from taking the adjoint, and another that comes from taking the dual space, and they don't coincide. If $X$ is a Hilbert space, its dual is denoted by $X^*$. So if we use the star to denote the dual, we have to use another notation for the adjoint.

\section{The dual space}

Every finite dimensional complex vector space $X$ has a dual space $X^*$, which is also a complex vector space with the same dimension. The dual space is defined as the set of linear functionals over $X$:
\begin{equation}
	X^* \coloneqq \Lin_\complexNumbers(X, \complexNumbers),
\end{equation}
that is, all $\complexNumbers$-linear transformations from $X$ to $\complexNumbers$.

Any basis for $X$ induces a basis for $X^*$, called the dual basis. It is defined as follows. Let $(x_i)_{i \in I}$ be a basis for $X$, then the dual basis for $X^*$ consists of linear functionals $(x^i )_{i \in I}$ characterized by the equation
\begin{equation}
	x^i (x_j) = \delta_{i,j},
\end{equation}
for any $i,j \in I$. The next proposition proves that it is a basis for $X^*$.

\begin{proposition}{Dual basis}{dual basis}
	Let $X$ be a finite dimensional complex vector space, and let $(x_i)_i$ be a basis for $X$. Then $(x^i)_i$ is a basis for $X^*$.
\end{proposition}
\begin{proof}
	Let $f \in X^*$, we'll prove that it can be written as
	\begin{equation}
		\label{equation: f in dual basis}
		f = \sum_i f(x_i) x^i .
	\end{equation}
	Evaluating the right side on a basis vector $x_j$, we get that
	\begin{align*}
		\sum_i f(x_i) x^i (x_j) &\stackrel{(i)}{=} \sum_i f(x_i) \delta_{i,j} \\
		&\stackrel{(ii)}{=} f(x_j).
	\end{align*}
	In $(i)$ was used the definition of $x^i$; in $(ii)$ the delta was used to eliminate the summation. Since both sides of equation \ref{equation: f in dual basis} coincide over a basis, it follows that the equality holds. This proves that $(x^i)_i $ generates $X^*$.
	
	Lastly, we have to prove that $(x^i)_i$ is linearly independent. Let $(a_i)_i$ be complex numbers such that
	\begin{equation}
		\sum_i a_i x^i = 0.
	\end{equation}
	Evaluating the left side on a basis vector $x_j$, we get that
	\begin{align*}
		\sum_i a_i x^i (x_j ) = a_j.
	\end{align*}
	This is the same computation done previously, but replacing $f(x_i)$ with $a_i$. Since the left side of the equation is $0$, it follows that $a_i = 0$, for all $i$. This concludes that $(x^i)_i$ is a basis for $X^*$.
\end{proof}

If $X$ is furthermore a finite dimensional complex Hilbert space, then $X^*$ is also a finite dimensional complex Hilbert space. For it to be a Hilbert space, we need some inner product. The inner product on $X$ induces an inner product over $X^*$. Before proving this, we'll use the adjoints to obtain an antilinear isomorphism $X \cong X^*$. Then this it will be combined with the inner product over $X$ to obtain an inner product over $X^*$. Notice the importance of the spaces being finite dimensional, since we can't have an isomorphism $X \cong X^*$ in the infinite dimensional case.

Let $x \in X$, it induces the linear transformation
\begin{align*}
	\complexNumbers &\longrightarrow X \\
	\lambda &\longmapsto \lambda x.
\end{align*}
Taking the adjoint we get a linear functional $x^\dag \colon X \to \complexNumbers$. The adjoint is characterized by the equation
\begin{equation}
	\langle x^\dag(x'), \lambda\rangle_\complexNumbers = \langle x', \lambda x \rangle_X,
\end{equation}
which must be valid for any $x' \in X$ and $\lambda \in \complexNumbers$. The inner product over $\complexNumbers$ is just the product of complex numbers, but conjugating the first argument:
\begin{equation}
	\langle x^\dag(x'), \lambda\rangle_\complexNumbers = \overline{x^\dag(x')} \lambda.
\end{equation}
Taking $\lambda = 1$ and using that $\overline{\langle x', x \rangle} = \langle x, x' \rangle$, we get that
\begin{equation}
	x^\dag(x') = \langle x, x' \rangle.
\end{equation}
In summary, we have the linear functional
\begin{align*}
	x^\dag \colon X &\longrightarrow \complexNumbers \\
	x' &\longmapsto \langle x, x' \rangle .
\end{align*}
This determines the function
\begin{align*}
	X &\longrightarrow X^* \\
	x &\longmapsto x^\dag .
\end{align*}
Next we prove that it is an antilinear isomorphism.

\begin{proposition}{}{X antilinear iso X*}
	Let $X$ be a finite dimensional complex Hilbert space. For any $x \in X$ we have the linear functional $x^\dag \in X^*$ defined by
	\begin{equation}
		x^\dag (x') = \langle x, x' \rangle,
	\end{equation}
	for any $x' \in X$. This determines the function
	\begin{align*}
		X &\longrightarrow X^* \\
		x &\longmapsto x^\dag,
	\end{align*}
	which is an antilinear isomorphism of $\complexNumbers$-vector spaces.
\end{proposition}
\begin{proof}
	We just have to prove that we have an antilinear isomorphism. The antilinearity is immediate from the inner product being antilinear in the first argument. That it is an isomorphism, it suffices to prove that it is an injection, since $X$ and $X^*$ have the same dimension. Injective linear transformations are characterized as the ones with a zero kernel. The same is valid for antilinear transformations, since this characterization only requires the function to be additive. Let $x \in X$ such that $x^\dag = 0$. Then
	\begin{equation}
		x^\dag (x) = 0.
	\end{equation}
	But
	\begin{equation}
		x^\dag (x) = \langle x, x \rangle,
	\end{equation}
	so 
	\begin{equation}
		\langle x, x \rangle = 0.
	\end{equation}
	The inner product is positive-definite, which implies that $x = 0$. Therefore, the kernel is zero and we have an injective antilinear transformation.
\end{proof}

Next we use this  antilinear isomorphism to contruct an inner product for the dual.
\begin{proposition}{Dual Hilbert space}{dual hilbert space}
	Let $X$ be a finite dimensional complex Hilbert space with inner product $\langle \ , \ \rangle \colon X \times X \to \complexNumbers$. Then $X^*$ is also a finite dimensional complex Hilbert space, whose inner product is given by
	\begin{equation}
		\langle x^\dag, y^\dag \rangle_{X^*} \coloneqq \langle y, x \rangle_X ,
	\end{equation}
	for any $x,y \in X$.
\end{proposition}
\begin{proof}
	The inner product over $X$ is a function $\langle \ , \ \rangle_X \colon X \times X \to \complexNumbers $. Combining it with the antilinear isomorphism $X^* \cong X$ from proposition \ref{proposition:X antilinear iso X*}, we get the function
	\begin{align*}
		X^* \times X^* &\longrightarrow \complexNumbers \\
		(x^\dag, y^\dag ) &\longmapsto \langle x, y \rangle_X . 
	\end{align*}
	Because of the antilinearity of the isomorphism $X \cong X^*$, this function is linear in the first argument and antilinear in the second. We defined the inner product as being antilinear is the first argument, to fix this we swap $x$ and $y$. It is immediate that we get a sesquilinear form. That it is positive-definite, it is immediate from the fact that
	\begin{equation}
		\langle x^\dag , x^\dag \rangle_{X^*} = \langle x, x \rangle_X,
	\end{equation}
	for any $x \in X$.
\end{proof}

We have defined a dual basis for $X^*$ from a basis for $X$. We also have defined an antilinear isomorphism $X \cong X^*$ in proposition \ref{proposition:X antilinear iso X*}. Any isomorphism maps basis to basis, so this gives yet another basis for $X^*$ from a basis for $X$. They don't need to coincide, but they do coincide when the basis for $X$ is orthonormal.

\begin{proposition}{}{orthonormal dual basis}
	Let $X$ be a finite dimensional complex Hilbert space. Let also $(x_i)_i$ be an orthonormal basis for $X$. Then
	\begin{equation}
		x_i^\dag = x^i
	\end{equation}
	and $(x_i^\dag)_i$ is an orthonormal basis for $X^*$.
\end{proposition}
\begin{proof}
	Since $(x_i)_i$ is orthonormal, then
	\begin{equation}
		\langle x_i, x_j \rangle = \delta_{i,j},
	\end{equation}
	for every $i,j$. Rewriting the left hand side in terms of $x_i^\dag$, we get that
	\begin{equation}
		x_i^\dag (x_j) = \delta_{i,j}.
	\end{equation}
	$x^i$ is the unique linear functional that satisfies this equation, so
	\begin{equation}
		x_i^\dag = x^i .
	\end{equation}
	
	Next we prove that $(x_i^\dag)_i$ is an orthonormal basis. That it is a basis, this follows from $(x^i)_i$ being a basis, which was proved in \ref{proposition:dual basis}. By definition of the inner product over $X^*$, as in proposition \ref{proposition:dual hilbert space}, we have that
	\begin{equation}
		\langle x_i^\dag, x_j^\dag \rangle_{X^*} = \langle x_j, x_i \rangle_X .
	\end{equation}
	Since $(x_i)_i$ is orthonormal, then
	\begin{equation}
		\langle x_i, x_j \rangle_X = \delta_{i,j},
	\end{equation}
	so
	\begin{equation}
		\langle x_i^\dag, x_j^\dag \rangle_{X^*} = \delta_{j, i} = \delta_{i,j}.
	\end{equation}
	This concludes that $(x_i^\dag)_i$ is orthonormal.
\end{proof}

\section{Tensor product of Hilbert spaces}

The tensor product of vector spaces was discussed previously. We also have tensor product of Hilbert spaces. The general case is more complicated, because of the completeness property, but in finite dimensions we can still use the same definition used for vector spaces. For the tensor product of Hilbert spaces to be a Hilbert space, we need an inner product. We define the inner product as follows. Let $X$ and $Y$ be finite dimensional complex Hilbert spaces with inner products $\langle \ , \ \rangle_X$ and $\langle \ , \ \rangle_Y$. Then the inner product over $X \otimes Y$, for elementary tensors, is given by
\begin{equation}
	\langle x \otimes y, x' \otimes y' \rangle_{X \otimes Y} = \langle x, x' \rangle_X \langle y, y' \rangle_Y ,
\end{equation}
for every $x, x' \in X$ and $y, y' \in Y$. For general elements of $X \otimes Y$, we just write them as linear combinations of elementary tensors and use the bilinearity of the inner product.

This definition needs to be justified in some way, which can be done by using a basis or the universal property of the tensor product. We'll use the universal property. We need to construct an inner product over $X \otimes Y$, whose function signature is
\begin{equation*}
	\langle \ , \ \rangle_{X \otimes Y} \colon (X \otimes Y) \times (X \otimes Y) \to \complexNumbers.
\end{equation*}
The universal property of the tensor product lets us obtain a linear transformation from $X \otimes Y$ by first constructing a bilinear function over $X \times Y$. We need to construct a function with signature
\begin{equation*}
	X \times Y \times X \times Y \to \complexNumbers.
\end{equation*}
Taking the cartesian product of both inner products, we get a function
\begin{equation*}
	\langle \ , \ \rangle_X \times \langle \ , \ \rangle_Y \colon X \times X \times Y \times Y \to \complexNumbers \times \complexNumbers .
\end{equation*}
The domain needs to be reordered, which can be done by swapping the middle $X$ and $Y$, which corresponds to the function
\begin{align*}
	X \times Y \times X \times Y &\longrightarrow X \times X \times Y \times Y \\
	(x, y, x', y') &\longmapsto (x, x', y, y').
\end{align*}
Instead of $\complexNumbers \times \complexNumbers$ we need $\complexNumbers$. This is achieved by the multiplication function
\begin{align*}
	m \colon \complexNumbers \times \complexNumbers &\longrightarrow \complexNumbers \\
	(z, w) &\longmapsto zw.
\end{align*}
Composing these three functions, we get the function
\begin{align*}
	X \times Y \times X \times Y &\longrightarrow \complexNumbers \\
	(x, y, x', y') &\longmapsto \langle x, x' \rangle_X \langle y, y' \rangle_Y .
\end{align*}
Next we have to use the universal property of the tensor product two times, to replace both $X \times Y$ by $X \otimes Y$. To do this, we need bilinear functions with $X \times Y$ as domain. This is obtained by fixing some of the arguments. Fixing $x$ and $y$, we get the function
\begin{align*}
	X \times Y &\longrightarrow \complexNumbers \\
	(x', y') &\longmapsto \langle x, x' \rangle_X \langle y, y' \rangle_Y .
\end{align*}
Since every inner product is linear in the second argument, it follows that this function in $\complexNumbers$-bilinear. By the universal property of the tensor product, there exists a unique $\complexNumbers$-linear transformation
\begin{equation}
	u_{x, y} \colon X \otimes Y \to \complexNumbers,
\end{equation}
such that
\begin{equation}
	u_{x, y} (x' \otimes y') = \langle x, x' \rangle_X \langle y, y' \rangle_Y ,
\end{equation}
for every $x' \in X$ and $y' \in Y$. From this function we get the function
\begin{align*}
	X \times Y \times (X \otimes Y) &\longrightarrow \complexNumbers \\
	(x, y, t) &\longmapsto u_{x,y}(t) .
\end{align*}
We need to apply again the universal property of the tensor product to replace the remaining $X \times Y$ by $X \otimes Y$. Since the inner product is antilinear in the first argument, we need to take a complex conjugate to obtain a linear function. For each $t \in X \otimes Y$ we have a function
\begin{align*}
	v_t \colon X \times Y &\longrightarrow \complexNumbers \\
	(x, y) &\longmapsto \overline{u_{x,y} (t)}.
\end{align*}
The complex conjugate is taken so that we get a bilinear function. In this case, the bilinearity isn't immediate. Let's first prove the bilinearity for $t = x' \otimes y'$, an elementary tensor. Let $x_1, x_2 \in X$, $y \in Y$ and $\lambda_1, \lambda_2 \in \complexNumbers$. Then
\begin{align*}
	u_{\lambda_1 x_1 + \lambda_2 x_2 , y} (x' \otimes y') &\stackrel{(i)}{=} \langle \lambda_1 x_1 + \lambda_2 x_2, x' \rangle \langle y, y' \rangle \\
	&\stackrel{(ii)}{=} \overline{\lambda_1 } \langle x_1, x' \rangle \langle y, y' \rangle + \overline{\lambda_2} \langle x_2, x' \rangle \langle y, y' \rangle \\
	&\stackrel{(iii)}{=} \overline{\lambda_1 } u_{x_1, y'} (x' \otimes y') + \overline{\lambda_2 } u_{x_2, y'} (x' \otimes y').
\end{align*}
In $(i)$ was used the definition of $u$; in $(ii)$ was used that the inner product is antilinear in the first argument; in $(iii)$ was again used the definition of $u$. Since $u_{x,y}$ is a linear function and the elementary tensors generate $X \otimes Y$, it follows that
\begin{equation}
	u_{\lambda_1 x_1 + \lambda_2 x_2 , y} (t) = \overline{\lambda_1 } u_{x_1, y'} (t) + \overline{\lambda_2 } u_{x_2, y'} (t),
\end{equation}
for any $t \in X \otimes Y$. Rewriting in terms of $v$, we have that
\begin{equation}
	v_t (\lambda_1 x_1 + \lambda_2 x_2 , y) = \lambda_1 v_t (x_1, y') + \lambda_2 v_t (x_2, y'),
\end{equation}
for any $t \in X \otimes Y$. This proves that $v_t$ is linear in the first argument. The linearity in the second argument is proved similarly. Therefore, $v_t$ is $\complexNumbers$-bilinear. The universal property of the tensor product then implies that, for each $t \in X \otimes Y$, there exists a unique $\complexNumbers$-linear transformation
\begin{equation}
	w_t \colon X \otimes Y \to \complexNumbers
\end{equation}
such that
\begin{equation}
	\label{equation: wt x otimes y}
	w_t (x \otimes y) = v_t (x,y) = \overline{u_{x,y} (t)},
\end{equation}
for every $x \in X$ and $y \in Y$. We can now define the inner product over $X \otimes Y$ as
\begin{equation}
	\langle t_1, t_2 \rangle_{X \otimes Y} \coloneqq \overline{ w_{t_2}(t_1) },
\end{equation}
for every $t_1, t_2 \in X \otimes Y$. It is immediate that
\begin{equation}
	\label{equation: inner product of tensor product}
	\langle x \otimes y, x' \otimes y' \rangle_{X \otimes Y} = \langle x , x' \rangle_X \langle y, y' \rangle_Y ,
\end{equation}
for every $x, x' \in X$ and $y, y' \in Y$. Next we need to prove that the properties of an inner product are satisfied. Since $w_t$ is a linear function, it is immediate that $\langle \ , \ \rangle_{X \otimes Y} $ is antilinear in the first argument. For the second argument, we first consider the case where the first argument is an elementary tensor. Let $x \in X$, $y \in Y$, $t_1, t_2 \in X \otimes Y$ and $\lambda_1, \lambda_2 \in \complexNumbers$. Then
\begin{align*}
	\langle x \otimes y, \lambda_1 t_1 + \lambda_2 t_2 \rangle_{X \otimes Y} &\stackrel{(i)}{=} \overline{w_{\lambda_1 t_1 + \lambda_2 t_2} (x \otimes y)} \\
	&\stackrel{(ii)}{=} u_{x,y} (\lambda_1 t_1 + \lambda_2 t_2) \\
	&\stackrel{(iii)}{=} \lambda_1 u_{x,y} (t_1) +  \lambda_2 u_{x,y} (t_2) \\
	&\stackrel{(iv)}{=} \lambda_1 \langle x \otimes y, t_1 \rangle_{X \otimes Y} + \lambda_2 \langle x \otimes y, t_2 \rangle_{X \otimes Y} .
\end{align*}
In $(i)$ was used the definition of $\langle \ , \ \rangle_{X \otimes Y} $; in $(ii)$ was used equation \ref{equation: wt x otimes y}; in $(iii)$ was used that $u_{x,y}$ is linear; in $(iv)$ the expression was written back in terms of $\langle \ , \ \rangle_{X \otimes Y}$. Since $\langle \ , \ \rangle_{X \otimes Y}$ is antilinear in the first argument, it follows that the same equality holds if we replace the elementary tensor $x \otimes y$ by any other tensor in $X \otimes Y$. This proves that $\langle \ , \ \rangle_{X \otimes Y}$ is $\complexNumbers$-linear in the second argument, so it is a sesquilinear form.

Finally, we have to prove that $\langle \ , \ \rangle_{X \otimes Y}$ is positive-definite. Let $(x_i)_i$ be an orthonormal basis for $X$ and let $(y_j)_j$ be an orthonormal basis for $Y$. Then $(x_i \otimes y_j)_{i,j}$ is a basis for $X \otimes Y$. As such, any $t \in X \otimes Y$ can be uniquely written in the form
\begin{equation}
	t = \sum_{i,j} \lambda_{i,j} \ x_i \otimes y_j ,
\end{equation}
for some $\lambda_{i,j} \in \complexNumbers$. Then
\begin{align*}
	\langle t, t \rangle_{X \otimes Y} &\stackrel{(i)}{=} \left\langle \sum_{i,j} \lambda_{i,j} \ x_i \otimes y_j, \sum_{i',j'} \lambda_{i',j'} \ x_{i'} \otimes y_{j'} \right\rangle_{X \otimes Y} \\
	&\stackrel{(ii)}{=} \sum_{i,j,i',j'} \overline{\lambda_{i,j} } \lambda_{i',j'} \langle x_i \otimes y_j, x_{i'} \otimes y_{j'} \rangle_{X \otimes Y} \\
	&\stackrel{(iii)}{=} \sum_{i,j,i',j'} \overline{\lambda_{i,j} } \lambda_{i',j'} \langle x_i , x_{i'} \rangle_X \langle y_j, y_{j'} \rangle_Y \\
	&\stackrel{(iv)}{=} \sum_{i,j,i',j'} \overline{\lambda_{i,j} } \lambda_{i',j'} \delta_{i,i'} \delta_{j,j'} \\
	&\stackrel{(v)}{=} \sum_{i,j} \overline{\lambda_{i,j} } \lambda_{i,j} \\
	&\stackrel{(vi)}{=} \sum_{i,j} |\lambda_{i,j}|^2 \\
	&\stackrel{(vii)}{\geq} 0 .
\end{align*}
In $(i)$ $t$ was written in the basis; in $(ii)$ was used the bilinearity of $\langle \ , \ \rangle_{X \otimes Y}$; in $(iii)$ was used equation \ref{equation: inner product of tensor product}; in $(iv)$ was used that the basis are orthonormal; in $(v)$ the deltas were used to eliminate the summations over $i'$ and $j'$; in $(vi)$ was used that $|z|^2 = \overline{z}z$; in $(vii)$ was used that $|\lambda_{i,j}|^2 \geq 0$ and that a sum of non negative numbers is non negative. This proves that
\begin{equation}
	\langle t, t \rangle_{X \otimes Y} \geq 0,
\end{equation}
for any $t \in X \otimes Y$. If
\begin{equation}
	\langle t, t \rangle_{X \otimes Y} = 0,
\end{equation}
then
\begin{equation}
	\sum_{i,j} |\lambda_{i,j} |^2 = 0,
\end{equation}
which implies that
\begin{equation}
	\lambda_{i,j} = 0,
\end{equation}
for all $i,j$. Therefore,
\begin{equation}
	t = 0,
\end{equation}
so $\langle \ , \ \rangle_{X \otimes Y}$ is positive-definite. This concludes that it is an inner product over $X \otimes Y$.

\section{Dirac notation}

\label{section: dirac notation appendix}

In Quantum Physics we typically use Dirac notation for vectors in a Hilbert space and linear transformations between Hilbert spaces. It works as follows. Let $X$ be a finite dimensional complex Hilbert space. The inner product is denoted with a vertical bar instead of a comma:
\begin{equation}
	\langle x \mid x' \rangle \coloneqq \langle x, x' \rangle,
\end{equation}
for any $x, x' \in X$. From proposition \ref{proposition:X antilinear iso X*}, we have a linear functional $x^\dag \in X^*$ such that
\begin{equation}
	x^\dag (x') = \langle x \mid x' \rangle.
\end{equation}
Inspired by this result, we introduce the following notation:
\begin{gather}
	\langle x | \coloneqq x^\dag, \\
	| x' \rangle \coloneqq x'.
\end{gather}
$\langle x |$ is called a bra and $| x' \rangle$ is called a ket, so that $\langle x \mid x' \rangle$ is a bracket.

The notation is mixing vectors and linear transformations. If we interpret the ket $|x \rangle$ as the linear transformation
\begin{align*}
	\complexNumbers &\longrightarrow X \\
	\lambda &\longmapsto \lambda x ,
\end{align*}
then both bras and kets represent linear transformations. In this sense, we can say that
\begin{equation}
	\langle x |^\dag = | x \rangle.
\end{equation}
Also, this lets us interpret the inner product $\langle x \mid x' \rangle$ as the composition of the linear transformations $\langle x |$ and $| x' \rangle$. This results in a linear transformation in $\Lin_\complexNumbers (\complexNumbers)$, but we have the isomorphism
\begin{align*}
	\complexNumbers &\longrightarrow \Lin_\complexNumbers (\complexNumbers) \\
	\lambda &\longmapsto \lambda \id_\complexNumbers .
\end{align*}
Thinking in terms of functions instead of points is also convenient for Category Theory and is used in string diagrams. The connection between the Dirac notation and string diagrams is explored in \cite{Coecke_Kissinger_2017}. We'll continue to interpret the ket $\ket{x}$ as a vector, but it can be useful to interpret it as a linear transformation.

We can also use a bra and a ket in the reversed order. Let $X$ and $Y$ be finite dimensional complex Hilbert spaces. Let also $x \in X$ and $y \in Y$. Then we have the linear transformation
\begin{align*}
	\ket{y} \bra{x} \colon X &\longrightarrow Y \\
	x' &\longmapsto \ket{y} \langle x \mid x' \rangle = \langle x, x' \rangle y.
\end{align*}
That is, it applies $x^\dag$ and then multiplies the complex number by the vector $y$. This can be used to construct a basis for $\Lin_\complexNumbers (X,Y)$ from basis for $X$ and $Y$. Before that, a useful special case is the identity operator.

\begin{proposition}{}{identity bra ket}
	Let $X$ be a finite dimensional complex Hilbert space, and let $(x_i)_i$ be an orthonormal basis for $X$. Then
	\begin{equation}
		\id_X = \sum_i \ket{x_i} \bra{x_i} .
	\end{equation}
\end{proposition}
\begin{proof}
	Let $x \in X$, it can be uniquely written as
	\begin{equation}
		\ket{x} = \sum_i \lambda_i \ket{x_i},
	\end{equation}
	for some $\lambda_i \in \complexNumbers$. Then
	\begin{align*}
		\left( \sum_i \ket{x_i} \bra{x_i} \right) \ket{x} &\stackrel{(i)}{=} \left( \sum_i \ket{x_i} \bra{x_i} \right) \left( \sum_j \lambda_j \ket{x_j} \right) \\
		&\stackrel{(ii)}{=} \sum_{i,j} \lambda_j \ket{x_i} \langle x_i | x_j \rangle \\
		&\stackrel{(iii)}{=} \sum_{i,j} \lambda_j \delta_{i,j} \ket{x_i} \\
		&\stackrel{(iv)}{=} \sum_i \lambda_i \ket{x_i} \\
		&\stackrel{(v)}{=} \ket{x}.
	\end{align*}
	In $(i)$ the ket $\ket{x}$ was written in the basis; in $(ii)$ the projection operator $\ket{x_i} \bra{x_i}$ was applied to $\ket{x_j}$; in $(iii)$ was used that the basis is orthonormal; in $(iv)$ the delta was used to eliminate the summation over $j$; in $(v)$ was used that the summation is $x$ written in the basis. This proves that
	\begin{equation}
		\sum_i \ket{x_i} \bra{x_i} = \id_X .
	\end{equation}
\end{proof}

\begin{proposition}{}{}
	Let $X$ and $Y$ be finite dimensional complex Hilbert spaces. Let also $(x_j)_j$ be an orthonormal basis for $X$ and let $(y_i)_i$ be an orthonormal basis for $Y$. Then $(\ket{y_i} \bra{x_j})_{i,j}$ is a basis for $\Lin_\complexNumbers (X, Y)$.
\end{proposition}
\begin{proof}
	Let $A \colon X \to Y$ be a $\complexNumbers$-linear transformation. Multiplying $A$ by $\id_X$ and $\id_Y$ and using proposition \ref{proposition:identity bra ket}, we get that
	\begin{align*}
		A &\stackrel{(i)}{=} \id_Y A \id_X \\
		&\stackrel{(ii)}{=} \left(\sum_i \ket{y_i} \bra{y_i} \right) A  \left(\sum_j \ket{x_j} \bra{x_j} \right) \\
		&\stackrel{(iii)}{=} \sum_{i,j} \bra{y_i} A \ket{x_j} \ket{y_i} \bra{x_j}.
	\end{align*}
	In $(i)$ $A$ was multiplied by the identity transformations; in $(ii)$ was used proposition \ref{proposition:identity bra ket} for the identities; in $(iii)$ $A$ was applied to $\ket{x_j }$ and $\bra{y_i }$ was applied to $A \ket{x_i}$, resulting in the complex number $\bra{y_i } A \ket{x_j }$, which was moved to the left of the linear transformation $\ket{y_i} \bra{x_j}$. This proves that $(\ket{y_i } \bra{x_j })_{i,j}$ generates $\Lin_\complexNumbers (X, Y)$.
	
	Next we prove that $(\ket{y_i } \bra{x_j })_{i,j}$ is linearly independent. Let $(a_{i,j})_{i,j}$ be complex numbers such that
	\begin{equation}
		\sum_{i,j} a_{i,j} \ket{y_i } \bra{x_j } = 0.
	\end{equation}
	Applying $\bra{y_{i'}}$ from the left and $\ket{x_{j'}}$ from the right, we get that
	\begin{align*}
		\sum_{i,j} a_{i,j} \langle y_{i'} \mid y_i \rangle \langle x_j \mid x_{j'} \rangle &\stackrel{(i)}{=} \sum_{i,j} a_{i,j} \delta_{i,i'} \delta_{j,j'} \\
		&\stackrel{(ii)}{=} a_{i', j'}.
	\end{align*}
	In $(i)$ was used that the basis are orthonormal; in $(ii)$ the deltas were used to eliminate the summation.
	
	Since we applied $\bra{y_{i'}}$ and $\ket{x_{j'}}$ to a summation which is zero, the result is zero, so
	\begin{equation}
		a_{i,j} = 0,
	\end{equation}
	for all $i,j$. This concludes that $(\ket{y_i} \bra{x_j})_{i,j}$ is a basis.
\end{proof}

The result is actually stronger, the basis $(\ket{y_i }\bra{x_j })_{i,j}$ is orthonormal, but to prove this we need to define an inner product over $\Lin_\complexNumbers (X, Y)$. We'll use the Hilbert-Schmidt inner product, which will be defined later.

\section{Trace definition}

The operations of (total) trace and partial trace are often used in Quantum Information Theory. Here we review the trace.

The trace of a square matrix $M \in M_n (\complexNumbers)$ is the sum of the entries in its diagonal:
\begin{equation}
	\tr M \coloneqq \sum_{i=1}^n M_{i,i} .
\end{equation}
Familiarity with the trace of matrices and its properties will be assumed.

The trace can also be defined for an operator that acts on a Hilbert space. Let $X$ be a \textbf{non zero} finite dimensional complex Hilbert space, and let $(x_i)_i$ be an orthonormal basis for $X$. Let also $A \in \Lin_\complexNumbers (X)$, then its trace is defined as the trace of its matrix $[A]$ with respect to the basis $(x_i)_i$. That is,
\begin{equation}
	\tr \, A \coloneqq \tr \, [A] = \sum_{i} \bra{x_i} A \ket{x_i}.
\end{equation}
The trace of matrices is cyclic, that is, we have the property
\begin{equation}
	\tr(MN) = \tr(NM),
\end{equation}
where $M$ and $N$ are matrices. This implies that the trace of an operator doesn't depends on a choice of basis. That is, another orthonormal basis would be of the form $x'_i = U x_i$, for some unitary operator $U \in U(X)$. Denote by $[A]'$ the matrix of $A$ with respect to the basis $(x'_i)_i$. Then
\begin{align*}
	\tr \, [A]' &\stackrel{(i)}{=} \sum_i \bra{x'_i} A \ket{x'_i}\\
	&\stackrel{(ii)}{=} \sum_i \bra{x_i} U^\dag A U \ket{x_i}\\
	&\stackrel{(iii)}{=} \tr \, [U^\dag A U]\\
	&\stackrel{(iv)}{=} \tr([U]^\dag [A] [U])\\
	&\stackrel{(v)}{=} \tr([U] [U]^\dag [A])\\
	&\stackrel{(vi)}{=} \tr \, [A].
\end{align*}
In $(i)$ the trace of the matrix $[A]'$ was expressed in the basis $(x'_i)_i$; in $(ii)$ the basis was changed from $(x'_i)_i $ to $(x_i)_i $; in $(iii)$ the expression was rewritten as the trace of the matrix of $[U^\dag A U]$, with respect to the basis $(x_i)_i$; in $(iv)$ was used that the matrix of a composition of operators is the product of the respective matrices; in $(v)$ was used the cyclic property of the trace of matrices; in $(vi)$ was used that $[U]$ is an unitary matrix, so $[U] [U]^\dag = I$. This proves that the trace of an operator doesn't depend on a choice of orthonormal basis.

For the special case in which $X$ has dimension 0, we consider that $\tr = 0$. That is,
\begin{align*}
	\tr \colon \{0\} &\longrightarrow \{0\} \\
	0 &\longmapsto 0.
\end{align*}

Having defined the trace, we can define the Hilbert-Schmidt inner product. We'll first define this inner product, because it is used frequently, and then continue studying the trace. 

\section{The Hilbert-Schmidt inner product}

The hom-sets $\Lin_\complexNumbers (X,Y)$ are finite dimensional complex vector spaces. We can turn them into Hilbert spaces by using the Hilbert-Schmidt inner product. It is defined as
\begin{equation}
	\label{equation: hilbert schmidt}
	\langle A, B \rangle_{HS} \coloneqq \tr(A^\dag B),
\end{equation}
for any $A, B \in \Lin_\complexNumbers (A, B)$. Note that $A^\dag B \in \Lin_\complexNumbers (X)$ and that the trace is defined using the inner product over $X$. Since the trace is cyclic, we can also write the inner product as
\begin{equation}
	\langle A, B \rangle_{HS} \coloneqq \tr(B A^\dag).
\end{equation}
We have that $B A^\dag \in \Lin_\complexNumbers (Y)$ and this trace uses the inner product over $Y$. In this sense, the Hilbert-Schmidt inner product is induced by the inner products over $X$ and $Y$.

Next we give a proof that this is indeed an inner product, and that it can be expressed as the Euclidean inner product with respect to a basis.

\begin{proposition}{Hilbert-Schmidt inner product}{hilbert schmidt is inner product}
	Let $X$ and $Y$ be finite dimensional complex Hilbert spaces. Equation \ref{equation: hilbert schmidt} defines an inner product over $\Lin_\complexNumbers (X,Y)$.
\end{proposition}
\begin{proof}
	Let $(x_j)_j$ be an orthonormal basis for $X$ and let $(y_i)_i$ be an orthonormal basis for $Y$. We'll represent the matrix of a linear map $L\in \Lin_\complexNumbers (X,Y)$ with respect to these basis by $[L]$. We have to prove four properties. The first three are about sesquilinearity, while the last is positive-definiteness:
	\begin{enumerate}
		\item Let $A,A',A''$ be linear maps in $\Lin(X,Y)$. Then
		\begin{align*}
			\langle A, A'+A''\rangle &\stackrel{(i)}{=} \tr(A^\dag (A'+A''))\\
			&\stackrel{(ii)}{=} \tr(A^\dag A' + A^\dag A'')\\
			&\stackrel{(iii)}{=} \tr(A^\dag A')+\tr(A^\dag A'')\\
			&\stackrel{(iv)}{=} \langle A, A'\rangle+\langle A, A''\rangle.
		\end{align*}
		In $(i)$ was used the definition of the Hilbert-Schmidt inner product. In $(ii)$ was used that composition is bilinear. In $(iii)$ was used the linearity of the trace. In $(iv)$ was used again the definition of the Hilbert-Schmidt inner product, but this time to recover inner products.
		
		\item Let $A,A'$ be linear maps in $\Lin(X,Y)$ and $\lambda \in \complexNumbers$. Then
		\begin{align*}
			\langle A, \lambda A' \rangle &\stackrel{(i)}{=} \tr(A^\dag (\lambda A'))\\
			&\stackrel{(ii)}{=} \lambda\tr(A^\dag A')\\
			&\stackrel{(iii)}{=} \lambda \langle A, A' \rangle.
		\end{align*}
		In $(i)$ was used the definition of the Hilbert-Schmidt inner product. In $(ii)$ was used the bilinearity of composition and linearity of the trace to pass the $\lambda$ to outside. In $(iii)$ was used again the definition of the Hilbert-Schmidt inner product to recover an inner product.
		
		\item Let $A,A' \in \Lin(X,Y)$, then
		\begin{align*}
			\langle A, A' \rangle &\stackrel{(i)}{=} \tr(A^\dag A')\\
			&\stackrel{(ii)}{=} \tr([A]^\dag [A'])\\
			&\stackrel{(iii)}{=} \sum_j ([A]^\dag [A'])_{j,j}\\
			&\stackrel{(iv)}{=} \sum_{i,j} [A]^\dag_{j,i} [A']_{i,j}\\
			&\stackrel{(v)}{=} \sum_{i,j} \overline{[A]_{i,j}} [A']_{i,j}\\
			&\stackrel{(vi)}{=} \overline{ \sum_{i,j} \overline{[A']_{i,j}} [A]_{i,j} } \\
			&\stackrel{(vii)}{=} \overline{ \sum_{i,j} [A'^\dag]_{j,i} [A]_{i,j} }\\
			&\stackrel{(viii)}{=} \overline{ \sum_j [A'^\dag A]_{j,j} }\\
			&\stackrel{(ix)}{=} \overline{ \tr(A'^\dag A) }\\
			&\stackrel{(x)}{=} \overline{\langle A',A \rangle}.
		\end{align*}
		In $(i)$ was used the definition of the Hilbert-Schmidt inner product. In $(ii)$ was used that the trace of an operator is the trace of its matrix. Of course, if $[A]$ is the matrix of $A$ and the matrix of $A'$ is $[A']$, then the matrix of $A^\dag A'$ is the matrix product $[A]^\dag [A']$. In $(iii)$ was used that the trace of a matrix is the sum of the entries in its diagonal. In $(iv)$ was used the definition of matrix multiplication in terms of matrix elements. In $(v)$ was used that the matrix of the adjoint $A^\dag$ is the conjugate transpose of $[A]$. In $(vi)$ was taken the conjugate of the whole sum, and another conjugation was taken over $[A']_{j,i}$ to cancel the bigger one. The order of $A$ and $A'$ is also switched. This is done to obtain $\langle A', A \rangle$ later. In $(vii)$ was used that $[A'^\dag]$ is the conjugate transpose of $[A']$. In $(viii)$ was identified a matrix product. In $(ix)$ was used that the trace of a matrix is the sum of the entries in its diagonal. In $(x)$ was used the definition of the Hilbert-Schmidt inner product to recover an inner product.
		
		\item Positive-definite: Let $A \in \Lin(X,Y)$. From step $(v)$ of the previous item, we know that
		\begin{equation}
			\langle A , A \rangle = \sum_{i,j} \overline{[A]_{i,j}} [A]_{i,j}.
		\end{equation}
		Since $|z|^2 = \overline{z} z$, we have that
		\begin{equation}
			\langle A , A \rangle = \sum_{i,j} |[A]_{i,j}|^2 .
		\end{equation}
		It is immediate that $\langle A , A \rangle \geq 0$ and that
		\begin{equation}
			\langle A , A \rangle = 0 \iff A = 0.
		\end{equation}
	\end{enumerate}
\end{proof}

\begin{corollary}{Hilbert-Schmidt inner product as an Euclidean inner product}{}
	With respect to any orthonormal basis, the Hilbert-Schmidt inner product is given by the Euclidean inner product. More precisely, let $X$ and $Y$ be finite dimensional complex Hilbert spaces. Let $(x_j)_j$ be an orthonormal basis for $X$ and let $(y_i)_i$ be an orthonormal basis for $Y$. Let also $A,A' \in \Lin_\complexNumbers (X,Y)$, and let $[A]$ and $[A']$ be their matrices with respect to the basis $(x_j)_j$ and $(y_i)_i$. Then we can write that
	\begin{equation}
		\langle A, A' \rangle_{HS} = \sum_{i,j} \overline{[A]_{i,j}}[A']_{i,j}.
	\end{equation}
	The right hand side is the Euclidean inner product of the entries $[A]_{i,j}$ and $[A']_{i,j}$.
\end{corollary}
\begin{proof}
	This was already proved in proposition \ref{proposition:hilbert schmidt is inner product}, as an expression found in the third item.
\end{proof}

The inner product induces the Euclidean norm. In this case, it is given by
\begin{equation}
	\norm{A} = \sqrt{\langle A, A \rangle_{HS}} = \sqrt{\tr(A^\dag A)},
\end{equation}
for any $A \in \Lin_\complexNumbers (X,Y)$. The Euclidean norm induced by the Hilbert-Schmidt inner product is also called the Frobenius norm.

\section{Some properties of the trace}

It follows immediately from the cyclic property of the trace of matrices that the same property holds for operators. That is, for any $A, B \in \Lin_\complexNumbers (X)$, we have that
\begin{equation}
	\tr(AB) = \tr(BA).
\end{equation}
The trace is multiplicative with respect to the tensor product, which is proved in the next proposition.
\begin{proposition}{}{}
	Let $X$ and $Y$ be finite dimensional complex Hilbert spaces. Let also $A \in \Lin(X)$ and $B \in \Lin(Y)$, then
	\begin{equation}
		\tr(A\otimes B) = \tr(A) \tr(B).
	\end{equation}
\end{proposition}
\begin{proof}
	Let $(x_i)_i$ be an orthonormal basis for $X$ and let $(y_j)_j$ be an orthonormal basis for $Y$. Then $(x_i \otimes y_j)_{i,j}$ is an orthonormal basis for $X \otimes Y$ and
	\begin{equation}
		\tr (A \otimes B) = \sum_{i,j} \langle x_i \otimes y_j \mid (A \otimes B) \mid x_i \otimes y_j \rangle. 
	\end{equation}
	Since $A \otimes B$ acts on an elementary tensor by applying each operator separately, we have that
	\begin{equation}
		\tr (A \otimes B) = \sum_{i,j} \langle x_i \otimes y_j \mid (A | x_i \rangle \otimes B | y_j \rangle).
	\end{equation}
	From equation \ref{equation: inner product of tensor product}, we have that
	\begin{equation}
		\tr (A \otimes B) = \sum_{i,j} \langle x_i | A | x_i \rangle \langle y_j | B | y_j \rangle.
	\end{equation}
	Factoring the sums over $i$ and $j$, we get that
	\begin{equation}
		\tr (A \otimes B) = \left( \sum_i \langle x_i | A | x_i \rangle \right) \left( \sum_j \langle y_j | B | y_j \rangle \right).
	\end{equation} 
	The two factors are the traces, so
	\begin{equation}
		\tr (A \otimes B) = \tr(A) \tr(B).
	\end{equation}
\end{proof}

\section{Multiplicativity of the Hilbert-Schmidt inner product}

It was shown previously that the inner product over a tensor product $X \otimes Y$ is multiplicative, as in equation \ref{equation: inner product of tensor product}. We have a similar property for the Hilbert-Schmidt inner product on $\Lin_\complexNumbers (X\otimes Y)$. Any element of $\Lin_\complexNumbers (X\otimes Y)$ is a linear combination of linear maps of the form $A \otimes B$, for $A \in \Lin_\complexNumbers (X)$ and $B \in \Lin_\complexNumbers (Y)$. Therefore, they are sufficient to characterize the inner product on $\Lin_\complexNumbers (X\otimes Y)$. Given $A,A' \in \Lin_\complexNumbers (X)$ and $B,B' \in \Lin_\complexNumbers (Y)$, we have that
\begin{align*}
	\langle A \otimes B, A' \otimes B' \rangle_{HS} &\stackrel{(i)}{=} \tr((A \otimes B)^\dag (A' \otimes B'))\\
	&\stackrel{(ii))}{=} \tr((A^\dag \otimes B^\dag)(A' \otimes B'))\\
	&\stackrel{(iii)}{=} \tr(A^\dag A' \otimes B^\dag B').
\end{align*}
In $(i)$ was used the definition of the Hilbert-Schmidt inner product; in $(ii)$ was used that the adjoint of the tensor product of operators is the tensor product of the adjoints; in $(iii)$ was used the functoriality of $\otimes$.

The trace is multiplicative, that is, we have the property $\tr(f \otimes g) = \tr(f)\tr(g)$. Using this property, we get 
\begin{align*}
	\langle A \otimes B, A' \otimes B' \rangle &\stackrel{(i)}{=} \tr(A^\dag A') \tr(B^\dag B') \\
	&\stackrel{(ii)}{=} \langle A,A' \rangle \langle B,B' \rangle.
\end{align*}
In $(i)$ was used that the trace is multiplicative; in $(ii)$ were recognized the Hilbert-Schmidt inner products.
Therefore 
\begin{equation}
	\label{equation: HS multiplicativity}
	\langle A \otimes B, A' \otimes B' \rangle_{HS} = \langle A,A' \rangle_{HS} \langle B,B' \rangle_{HS},
\end{equation}
for any $A,A' \in \Lin_\complexNumbers (X)$ and $B,B' \in \Lin_\complexNumbers (Y)$.

\section{Partial trace}

Now we define the partial traces. The partial trace can be defined as the tensor product of the (total) trace with identity maps, composed with an unitality isomorphism to replace the tensor product with a scalar by a scalar multiplication. Let $X_1, \dots, X_n$ be finite dimensional complex Hilbert spaces. The partial trace over $X_i$ is a $\complexNumbers$-linear map
\begin{equation*}
	\tr_i \colon \Lin_\complexNumbers (X_1 \otimes \dots \otimes X_n) \to \Lin_\complexNumbers (X_1 \otimes \dots \otimes X_{i-1} \otimes X_{i+1} \otimes \dots \otimes X_n ).
\end{equation*}
That is, it discards the factor $X_i$. It is defined as follows. We have the tensor product
\begin{equation}
	\id_{\Lin_\complexNumbers(X_1 \otimes \dots \otimes X_{i-1})} \otimes \tr \otimes \id_{\Lin_\complexNumbers(X_{i+1} \otimes \dots \otimes X_n)},
\end{equation}
which is a linear transformation with domain
\begin{equation}
	\Lin_\complexNumbers (X_1 \otimes \dots \otimes X_{i-1} \otimes X_i \otimes X_{i+1} \otimes \dots \otimes X_n )
\end{equation}
and codomain
\begin{equation}
	\Lin_\complexNumbers (X_1 \otimes \dots \otimes X_{i-1} \otimes \complexNumbers \otimes X_{i+1} \otimes \dots \otimes X_n ).
\end{equation}
In other words, the total trace replaces the $X_i$ by $\complexNumbers$ in the tensor product. The partial trace is obtained by eliminating the $\complexNumbers$ factor, which is done by the unitality isomorphisms. For example, we have the left unitality isomorphism
\begin{equation}
	l_{X_{i+1} \otimes \dots \otimes X_n } \colon \complexNumbers \otimes X_{i+1} \otimes \dots \otimes X_n \stackrel{\cong}{\to} X_{i+1} \otimes \dots \otimes X_n.
\end{equation}
We can then eliminate the $\complexNumbers$ factor using the isomorphism given by conjugation:
\begin{equation}
	A \mapsto (\id_{X_1 \otimes \dots \otimes X_{i-1}} \otimes l_{X_{i+1} \otimes \dots \otimes X_n }) A (\id_{X_1 \otimes \dots \otimes X_{i-1}} \otimes l_{X_{i+1} \otimes \dots \otimes X_n })^\dag .
\end{equation}
where $A \in \Lin_\complexNumbers (X_1 \otimes \dots \otimes X_{i-1} \otimes \complexNumbers \otimes X_{i+1} \otimes \dots \otimes X_n )$. Instead of using the left unitality isomorphis, we could've used the right unitality isomorphism. By Mac Lane's coherence theorem, the result is the same.

It follows that the partial trace can be computed by the following expression: 
\begin{equation}
	\tr_i(A_1 \otimes \dots \otimes A_n) = \tr(A_i) \ (A_1 \otimes \dots \otimes A_{i-1} \otimes A_{i+1} \otimes \dots \otimes A_n),
\end{equation}
where $A_j \in \Lin(X_j)$, for $j=1,\dots,n$. The general case is obtained using that $\tr_i$ is $\complexNumbers$-linear.

Depending on the context, we'll use the notation $\tr_{X_i}$ instead of $\tr_i$. The choice of notation is made depending on what is less ambiguous. For example, one of the spaces $X_i$ could itself be a tensor product of other spaces, and then the index $i$ may become confusing. Another way to present partial traces is by string diagrams of spherical categories. They have the advantage of don't having any of these ambiguity problems, and also have a nice diagramatic representation. We'll not use them in this text to keep it simpler to understand. For more details about string diagrams, see \cite{Coecke_Kissinger_2017, MonoidalCatsAndTFT}.

\section{The operator norm}

There is another norm over $\Lin_\complexNumbers (X,Y)$ which is the operator norm. We'll avoid using this norm, because the Frobenius norm is given by the trace, which is convenient for the study of CPTP maps, but sometimes the operator norm will be useful. For this norm, we'll restrict our attention to hermitian operators. For any hermitian operator $A \in H(X)$, it can be diagonalized and has real eigenvalues $\lambda_1$, \dots, $\lambda_{d_X}$. In this case, the operator norm can be computed as
\begin{equation}
	\label{equation: operator norm}
	||A||_{op} = \max_{1\leq i \leq d_X} |\lambda_i|.
\end{equation}
There is a general result every norm over a finite dimensional vector space is equivalent. We give next a proof of equivalence between the Frobenius and operator norms for Hermitian operators.
\begin{proposition}{Equivalence of operator and Frobenius norms}{equivalence between frobenius and operator norms for hermitian}
	Let $X$ be a finite dimensional complex Hilbert space, and let $A \in H(X)$ be a Hermitian operator over $X$. Then
	\begin{equation}
		||A||_{op} \leq ||A|| \leq \sqrt{d_X} \ ||A||_{op}.
	\end{equation}
\end{proposition}
\begin{proof}
	$A$ is hermitian, so it is diagonalizable and has real eigenvalues. Let $\lambda_1$, \dots, $\lambda_{d_X}$ be the eigenvalues of $A$. The square of the Frobenius norm of $A$ is $||A||^2 = \tr(A^\dag A)$. $A^\dag A$ is positive semidefinite, and its eigenvalues are $|\lambda_1|^2$, \dots, $|\lambda_{d_X}|^2$. The trace is the sum of the eigenvalues, so
	\begin{equation}
		||A||^2 = \sum_i |\lambda_i|^2.
	\end{equation}
	On the other hand, the squared operator norm is
	\begin{equation}
		||A||_{op}^2 = \left(\max_i |\lambda_i|\right)^2 = \max_i |\lambda_i|^2.
	\end{equation}
	The maximum is one of the terms in the sum $\sum_i |\lambda_i|^2$, so
	\begin{equation}
		||A||_{op}^2 \leq ||A||^2.
	\end{equation}
	But each term of the sum is bounded by $\max_i |\lambda_i|^2$, so
	\begin{equation}
		||A||^2 \leq \sum_i ||A||_{op}^2 = d_X ||A||_{op}^2.
	\end{equation}
	Taking the square root, it follows that
	\begin{equation}
		||A||_{op} \leq ||A|| \leq \sqrt{d_X} \ ||A||_{op}.
	\end{equation}
\end{proof}

\section{Continuity of square root of operators}

Here we review a known result that the square root of positive semidefinite operators is a continuous function. In particular, it is a homeomorphism between the set of positive semidefinite operators and itself. We'll denote by $\psd(X)$ the set of positive semidefinite operators over a finite dimensional complex Hilbert space $X$. The proof of the next proposition is based on \cite{continuity_square_root_of_psd}.

\begin{proposition}{Square root of operators}{square root of psd operators is continuous}
	Let $X$ be a finite dimensional complex Hilbert space. The function
	\begin{align*}
		\psd(X) &\longrightarrow \psd(X)\\
		A &\longmapsto A^{1/2}
	\end{align*}
	is a homeomorphism, whose inverse is
	\begin{align*}
		\psd(X) &\longrightarrow \psd(X)\\
		A &\longmapsto A^2.
	\end{align*}
	With respect to the operator norm $|| \ ||_{op}$, we also have that
	\begin{equation}
		||\sqrt{A}-\sqrt{B}||_{op} \leq ||A-B||_{op}^{1/2},
	\end{equation}
	for any $A,B \in \psd(X)$. With respect to the Frobenius norm, we have that
	\begin{equation}
		||\sqrt{A}-\sqrt{B}|| \leq \sqrt{d_X} ||A-B||^{1/2},
	\end{equation}
	for any $A,B \in \psd(X)$. In particular, the square root of operators is uniformly continuous.
\end{proposition}
\begin{proof}
	For finite dimensional normed spaces every norm is equivalent. We can then show continuity with respect to any norm. We'll do it for the operator norm $|| \ ||_{op}$, because it is related to eigenvalues. The case $X = \{0\}$ is trivial, so we assume that $X \neq \{0\}$. Let $A,B \in \psd(X)$ be positive semidefinite operators over $X$. If $A$ and $B$ commuted with each other, we could write $A-B = (\sqrt{A}-\sqrt{B})(\sqrt{A}+\sqrt{B})$, and then use this equation to relate $||\sqrt{A}-\sqrt{B}||_{op}$ with $||A-B||_{op}$. They may not commute, but we still can write a similar expression. We have that
	\begin{align*}
		A - B &\stackrel{(i)}{=} \left(\sqrt{A}\right)^2 - \left(\sqrt{B}\right)^2 \\
		&\stackrel{(ii)}{=} \left(\sqrt{A}\right)^2 -\sqrt{A}\sqrt{B} +\sqrt{A}\sqrt{B} -\left(\sqrt{B}\right)^2 \\
		&\stackrel{(iii)}{=} \sqrt{A}(\sqrt{A}-\sqrt{B}) + (\sqrt{A}-\sqrt{B})\sqrt{B}. 
	\end{align*}
	In $(i)$ the operators were replaced by the square of the square root, to make square roots appear; in $(ii)$ was added and subtracted the term $\sqrt{A} \sqrt{B}$; in $(iii)$ was factored the term $\sqrt{A} - \sqrt{B}$.
	
	$\sqrt{A}$ and $\sqrt{B}$ are positive semidefinite operators, so they are hermitian. This implies that $\sqrt{A}-\sqrt{B}$ is a hermitian operator, so it is diagonalizable and has real eigenvalues. Similarly, $A-B$ is also hermitian. Let $\lambda \in \realNumbers$ be an eigenvalue of $\sqrt{A} - \sqrt{B}$, and let $\ket{x} \in X$ be an eigenvector of $\sqrt{A} - \sqrt{B}$ with $||x|| = 1$ and eigenvalue $\lambda$, so
	\begin{equation}
		\label{equation: x eigenvector of sqrt A - sqrt B}
		\left(\sqrt{A} - \sqrt{B}\right)\ket{x} = \lambda \ket{x}.
	\end{equation}
	Using the expression obtained for $A-B$ we get that
	\begin{align*}
		\bra{x} (A-B) \ket{x} &\stackrel{(i)}{=} \bra{x}\left(\sqrt{A}\left(\sqrt{A}-\sqrt{B}\right) + \left(\sqrt{A}-\sqrt{B}\right)\sqrt{B}\right)\ket{x}\\
		&\stackrel{(ii)}{=} \bra{x}\sqrt{A}(\sqrt{A}-\sqrt{B})\ket{x} + \bra{x}(\sqrt{A}-\sqrt{B})\sqrt{B}\ket{x}\\
		&\stackrel{(iii)}{=} \lambda \bra{x} \sqrt{A} \ket{x} + \lambda \bra{x} \sqrt{B} \ket{x}\\
		&\stackrel{(iv)}{=} \lambda \bra{x} (\sqrt{A}+\sqrt{B}) \ket{x}.
	\end{align*}
	In $(i)$ was used the expression for $A-B$; in $(ii)$ $\ket{x}$ was distributed between both terms of the sum; in $(iii)$ was applied equation \ref{equation: x eigenvector of sqrt A - sqrt B}; in $(iv)$ the operators were collected between $\bra{x}$ and $\ket{x}$.
	
	Using that $\lambda = \bra{x}\left(\sqrt{A} - \sqrt{B}\right)\ket{x}$, we get that
	\begin{equation}
		\label{equation: x*(A-B)x = etc}
		\bra{x} (A-B) \ket{x} = \bra{x}\left(\sqrt{A} - \sqrt{B}\right)\ket{x}\bra{x}\left(\sqrt{A} + \sqrt{B}\right)\ket{x}.
	\end{equation}
	We want to express $ \bra{x}\left(\sqrt{A} - \sqrt{B}\right)\ket{x}$ in terms of the rest, but we want to avoid dividing by something that could be 0. Instead, we can relate $\bra{x}\left(\sqrt{A} + \sqrt{B}\right)\ket{x}$ with $|\bra{x}\left(\sqrt{A} - \sqrt{B}\right)\ket{x}|$. Using the triangle inequality, we get that
	\begin{align*}
		\left|\bra{x}(\sqrt{A} - \sqrt{B})\ket{x}\right| &\stackrel{(i)}{=} \left|\bra{x}\sqrt{A}\ket{x} - \bra{x}\sqrt{B}\ket{x}\right|\\
		&\stackrel{(ii)}{\leq} \left|\bra{x}\sqrt{A}\ket{x}\right|+\left|\bra{x}\sqrt{B}\ket{x}\right|.
	\end{align*}
	In $(i)$ the $\ket{x}$ was distributed between the operators; in $(ii)$ was used the triangle inequality.
	
	Since $\sqrt{A}$ and $\sqrt{B}$ are positive semidefinite, the numbers $\bra{x}\sqrt{A}\ket{x}$ and $\bra{x}\sqrt{B}\ket{x}$ are real and non negative. Therefore,
	\begin{align*}
		\left|\bra{x}\left(\sqrt{A} - \sqrt{B} \right)\ket{x} \right| &\stackrel{(i)}{\leq} \bra{x}\sqrt{A}\ket{x}+\bra{x}\sqrt{B}\ket{x} \\
		&\stackrel{(ii)}{=} \bra{x}\left(\sqrt{A}+\sqrt{B}\right)\ket{x}.
	\end{align*}
	In $(i)$ was used the previous inequality and that $\bra{x}\sqrt{A}\ket{x}$ and $\bra{x}\sqrt{B}\ket{x}$ are non negative; in $(ii)$ the operators were collected between $\bra{x}$ and $\ket{x}$.
	
	Taking the square of $\left|\bra{x}\left(\sqrt{A}-\sqrt{B}\right)\ket{x}\right|$ and using the previous inequality, we can make appear the product $\bra{x}\left(\sqrt{A} - \sqrt{B}\right)\ket{x} \bra{x}\left(\sqrt{A}+\sqrt{B}\right)\ket{x}$. We have that
	\begin{align*}
		\left|\bra{x}\left(\sqrt{A}-\sqrt{B}\right)\ket{x}\right|^2 &\stackrel{(i)}{=} \left|\bra{x}\left(\sqrt{A}-\sqrt{B}\right)\ket{x}\right|\ \left|\bra{x}\left(\sqrt{A}-\sqrt{B}\right)\ket{x}\right|\\
		&\stackrel{(ii))}{\leq} \left|\bra{x}\left(\sqrt{A}-\sqrt{B}\right)\ket{x}\right|\ \bra{x}\left(\sqrt{A}+\sqrt{B}\right)\ket{x}\\
		&\stackrel{(iii)}{=} \left|\bra{x}\left(\sqrt{A}-\sqrt{B}\right)\ket{x} \bra{x}\left(\sqrt{A}+\sqrt{B}\right)\ket{x}\right|.
	\end{align*}
	In $(i)$ the square was expanded to a product; in $(ii)$ was used that 
	\begin{equation*}
		\left| \bra{x} \left(\sqrt{A} - \sqrt{B}\right) \ket{x} \right| \leq \bra{x}\left(\sqrt{A} + \sqrt{B}\right)\ket{x};
	\end{equation*}
	in $(iii)$ the non negative factor $\bra{x}\left(\sqrt{A} + \sqrt{B}\right)\ket{x}$ was passed to inside the modulus.
	
	Using equation \ref{equation: x*(A-B)x = etc}, we get that
	\begin{equation}
		\left|\bra{x}\left(\sqrt{A}-\sqrt{B}\right)\ket{x}\right|^2 \leq \left|\bra{x}(A-B)\ket{x}\right|,
	\end{equation}
	for any $x$ that is an eigenvector of $\sqrt{A}-\sqrt{B}$ with $||x|| = 1$. We can make the right hand side independent of $x$ by taking the supremum over all unit vectors. This supremum is a maximum, because it is the supremum of a continuous and bounded function. This maximum is the operator norm of $A-B$. We get that
	\begin{equation}
		\left|\bra{x}\left(\sqrt{A}-\sqrt{B}\right)\ket{x}\right|^2 \leq \max_{||y||=1} |\bra{y}(A-B)\ket{y}| = ||A-B||_{op},
	\end{equation}
	for any $x$ that is an eigenvalue of $\sqrt{A} - \sqrt{B}$, with $||x||=1$. For hermitian operators, the operator norm is the highest absolute value of the eigenvalues. $\bra{x}(\sqrt{A}-\sqrt{B})\ket{x}$ can be any eigenvalue of $\sqrt{A}-\sqrt{B}$, so
	\begin{equation}
		||\sqrt{A} - \sqrt{B}||_{op}^2 \leq ||A-B||_{op}.
	\end{equation}
	Taking the square root, we get that
	\begin{equation}
		\label{equation: op norm of square root difference}
		||\sqrt{A} - \sqrt{B}||_{op} \leq ||A-B||_{op}^{1/2}.
	\end{equation}
	Using proposition \ref{proposition:equivalence between frobenius and operator norms for hermitian} we can rewrite this inequality in terms of the Frobenius norm. We have that
	\begin{align*}
		\left|\left|\sqrt{A} - \sqrt{B}\right|\right| &\stackrel{(i)}{\leq} \sqrt{d_X} \ \left|\left|\sqrt{A} - \sqrt{B}\right|\right|_{op}\\
		&\stackrel{(ii)}{\leq} \sqrt{d_X} \ ||A-B||_{op}^{1/2}\\
		&\stackrel{(iii)}{\leq} \sqrt{d_X} \ ||A-B||^{1/2}.
	\end{align*}
	In $(i)$ was used proposition \ref{proposition:equivalence between frobenius and operator norms for hermitian} to replace the Frobenius norm by the operator norm; in $(ii)$ was used equation \ref{equation: op norm of square root difference}; in $(iii)$ was used proposition \ref{proposition:equivalence between frobenius and operator norms for hermitian} again to replace the operator norm by the Frobenius norm.
	
	This implies that the square root of positive semidefinite operators is a continuous function. Of course, the function $A \mapsto A^2$ is continuous and is the inverse of the function $A \mapsto \sqrt{A}$. This shows that the square root is a homeomorphism between $\psd(X)$ and itself.
	
\end{proof}

\cleardoublepage
\chapter{Matrices}

\label{appendix: matrices}

This appendix collects some results about matrices to be used in the main text. We typically work with linear transformations, but they correspond to matrices when using some basis.

\section{Kronecker product}

The Kronecker product arises naturally from the tensor product of linear transformations. An important difference is that the indices for a matrix are ordered, while a basis for a vector space doesn't need to be ordered. This means that some order of the indices must be chosen to obtain a matrix.

Let $X$, $X'$, $Y$, and $Y'$ be finite dimensional complex Hilbert spaces. Let also $A \in \Lin_\complexNumbers (X,X')$ and $B \in \Lin_\complexNumbers (Y,Y')$. Then we have the linear transformation $A \otimes B \in \Lin_\complexNumbers (X \otimes Y, X' \otimes Y')$. The Kronecker product appears when we replace these linear transformations by matrices. First we need some basis. Let:
\begin{itemize}
	\item $(x_1,\dots,x_m)$ be a basis for $X$;
	\item $(x'_1,\dots,x'_{m'})$ be a basis for $X'$;
	\item $(y_1,\dots,y_n)$ be a basis for $Y$;
	\item $(y'_1,\dots,y'_{n'})$ be a basis for $Y'$.
\end{itemize}
Then we have the basis $(x_i \otimes y_j)_{\substack{i=1,\dots,m \\ j=1,\dots,n}}$ for $X \otimes Y$ and $(x'_{i'} \otimes y'_{j'})_{\substack{i'=1,\dots,m' \\ j'=1,\dots,n'}}$ for $X' \otimes Y'$. We can express the matrix $[A\otimes B]$ of $A\otimes B$ with respect to these basis. Computing $A\otimes B$ on $x_i \otimes y_j$, we have
\begin{align*}
	(A\otimes B)(x_i \otimes y_j) &\stackrel{(i)}{=} A(x_i) \otimes B(y_j)\\
	&\stackrel{(ii)}{=} \left(\sum_{i'} [A]_{i',i} x'_{i'} \right) \otimes \left(\sum_{j'} [B]_{j',j} y'_{j'}\right).
\end{align*}
In $(i)$ was used equation \ref{equation: tensor product of linear transformations}; in $(ii)$ the vectors were written in the basis. By bilinearity of $\otimes$, we get that
\begin{equation}
	(A\otimes B)(x_i \otimes y_j) = \sum_{i',j'} [A]_{i',i} [B]_{j',j} \ x'_{i'} \otimes y'_{j'}.
\end{equation}
Instead of considering the indices of the matrix $[A \otimes B]$ to be natural numbers, it is simpler to consider the indices to be pairs $(i,j)$. In this sense, $[A\otimes B]$ has elements 
\begin{equation}
	[A\otimes B]_{(i',j'),(i,j)} = [A]_{i',i} [B]_{j',j}.
\end{equation}
The pair $(i,j)$ can be identified with a natural number by ordering the pairs. This is done by taking the lexicographic order, that is,
\begin{equation*}
	(i,j) < (i',j') \iff i<i' \text{ or } i=i' \text{ and } j<j'.
\end{equation*}
To be obtain a matrix in the usual sense, we just have to replace the pairs by natural numbers, but it won't be done in this work.

The matrix $[A\otimes B]$ is called the Kronecker product of $[A]$ and $[B]$, which we denote as $[A] \otimes  [B]$.

\section{Gram matrices}

Gram matrices give an alternative way to study linear independence of vectors. In this section we review some results regarding them.

\begin{definition}{Gram matrix}{}
	Let $X$ be a finite dimensional complex Hilbert space, and let $(x_1,\dots, x_n)$ be a sequence of vectors in $X$. Define the matrix
	\begin{equation}
		G = \begin{pmatrix}
			\langle x_1, x_1 \rangle & \langle x_1, x_2 \rangle & \cdots & \langle x_1, x_n \rangle \\
			\langle x_2, x_1 \rangle & \langle x_2, x_2 \rangle & \cdots & \langle x_2, x_n \rangle \\
			\vdots & \vdots & \ddots & \vdots\\
			\langle x_n, x_1 \rangle & \langle x_n, x_2 \rangle & \cdots & \langle x_n, x_n \rangle
		\end{pmatrix}.
	\end{equation}
	This matrix is called the Gram matrix of $(x_1,\dots,x_n)$.
\end{definition}

A proof of proposition \ref{proposition:gram matrix} may be found in theorem 7.2.10 of \cite{Horn_Johnson}.

\begin{proposition}{Linear independence from Gram matrix}{gram matrix}
	Let $X$ be a finite dimensional complex Hilbert space, and let $(x_1,\dots, x_n)$ be a sequence of vectors in $X$, with $n \geq 1$. Then $(x_1,\dots, x_n)$ is linearly independent iff its Gram matrix is invertible.
\end{proposition}
\begin{proof}
	First, given vectors $a = (a_1,\dots,a_n)$ and $b = (b_1,\dots,b_n)$, which we'll represent as column vectors, lets compute $a^\dag G b$. We have that
	\begin{align*}
		Gb &\stackrel{(i)}{=} \begin{pmatrix}
			\langle x_1, x_1 \rangle & \langle x_1, x_2 \rangle & \cdots & \langle x_1, x_n \rangle \\
			\langle x_2, x_1 \rangle & \langle x_2, x_2 \rangle & \cdots & \langle x_2, x_n \rangle \\
			\vdots & \vdots & \ddots & \vdots\\
			\langle x_n, x_1 \rangle & \langle x_n, x_2 \rangle & \cdots & \langle x_n, x_n \rangle
		\end{pmatrix}	\begin{pmatrix}
			b_1\\
			b_2\\
			\vdots\\
			b_n
		\end{pmatrix}\\
		&\stackrel{(ii)}{=} 	\begin{pmatrix}
			\sum_{j=1}^n \langle x_1, x_j \rangle b_j\\
			\sum_{j=1}^n \langle x_2, x_j \rangle b_j\\
			\vdots\\
			\sum_{j=1}^n \langle x_n, x_j \rangle b_j\\
		\end{pmatrix}\\
		&\stackrel{(iii)}{=} 	\begin{pmatrix}
			\left\langle x_1,\sum_{j=1}^n b_j x_j \right\rangle\\
			\left\langle x_2,\sum_{j=1}^n b_j x_j \right\rangle\\
			\vdots\\
			\left\langle x_n,\sum_{j=1}^n b_j x_j \right\rangle\\
		\end{pmatrix}.
	\end{align*}
	In $(i)$ were written the elements of the matrices $G$ and $b$; in $(ii)$ the matrices were multiplied; in $(iii)$ was used that the inner product is $\complexNumbers$-linear in the second argument. Then
	\begin{align*}
		a^\dag G b &\stackrel{(i)}{=} \sum_{i=1}^n \overline{a_i} \left\langle x_i,\sum_{j=1}^n b_j x_j \right\rangle \\
		&\stackrel{(ii)}{=} \left\langle \sum_{i=1}^n a_i x_i,\sum_{j=1}^n b_j x_j \right\rangle.
	\end{align*}
	In $(i)$ the row vector $a^\dag$ and the column vector $Gb$ were multiplied; in $(ii)$ was used that the inner product is antilinear in the first argument.
	
	Taking $b=a$, we get that
	\begin{equation}
		\label{equation: a dag G a}
		a^\dag G a = \left|\left|\sum_{i=1}^n a_i x_i \right|\right|^2,
	\end{equation}
	which is non negative. Since $a$ is arbitrary, this proves that $G$ is a positive semidefinite matrix. Therefore, $G$ is diagonalizable and its eigenvalues are real and non-negative.
	
	Next we prove the equivalence of $(x_1,\dots,x_n)$ being linearly independent and $G$ being invertible. Suppose that $(x_1, \dots, x_n)$ is linearly dependent. Then there exists some $a \neq 0$ such that
	\begin{equation}
		\sum_{i=1}^n a_i x_i = 0.
	\end{equation}
	Equation \ref{equation: a dag G a} implies that
	\begin{equation}
		a^\dag G a = 0.
	\end{equation}
	Therefore, $a$ is an eigenvector of $G$ with eigenvalue 0. This implies that $G$ isn't invertible. By the contrapositive argument, if $G$ is invertible, then $(x_1, \dots, x_n)$ is linearly independent.
	
	Now suppose that $(x_1, \dots, x_n)$ is linearly independent and let's prove that $G$ is invertible. $G$ is a positive semidefinite matrix, so it can be diagonalized and have non negative eigenvalues. Let $v_1,\dots,v_n$ be orthonormal eigenvectors of $G$ and $\lambda_i \geq 0$ be the eigenvalue of $v_i$. $G$ is invertible iff each $\lambda_i > 0$, for all $i$. Take the column vector $a$ to be
	\begin{equation}
		\label{equation: a in terms of vi}
		a = \sum_{i=1}^n \alpha_i v_i,
	\end{equation}
	for some $\alpha_i \in \complexNumbers$. Then we have that
	\begin{align*}
		a^\dag G a &\stackrel{(i)}{=} a^\dag G \sum_{i=1}^n \alpha_i v_i \\
		&\stackrel{(ii)}{=} \sum_{i=1}^n \alpha_i a^\dag G v_i \\
		&\stackrel{(iii)}{=} \sum_{i=1}^n \alpha_i a^\dag \lambda_i v_i \\
		&\stackrel{(iv)}{=} \sum_{i=1}^n \lambda_i \alpha_i a^\dag v_i \\
		&\stackrel{(v)}{=} \sum_{i=1}^n \lambda_i \alpha_i \overline{\alpha_i} \\
		&\stackrel{(vi)}{=} \sum_{i=1}^n \lambda_i |\alpha_i|^2.
	\end{align*}
	In $(i)$ was used the definition of $a$; in $(ii)$ was used that the matrix multiplication is bilinear to pass the summation and $\alpha_i$ to the left; in $(iii)$ was used that $v_i$ is an eigenvector of $G$ with eigenvalue $\lambda_i$; in $(iv)$ was used that the matrix multiplication is bilinear to move the $\lambda_i$ to the left; in $(v)$ was used that $(v_i)_i$ is orthonormal and equation \ref{equation: a in terms of vi}; in $(vi)$ was used that $|z|^2 = \overline{z}z$.
	
	By equation \ref{equation: a dag G a}, and since $(x_i)_i$ is linearly independent, we have that
	\begin{equation}
		\label{equation: a neq 0 implis adag G a neq 0}
		a \neq 0 \implies a^\dag G a \neq 0.
	\end{equation}
	Take
	\begin{equation}
		\label{equation: alphai = delta}
		\alpha_i = \delta_{i,j},
	\end{equation}
	for some $j \in \{1, \dots, n\}$. Then $a \neq 0$ and
	\begin{align*}
		a^\dag G a &= \sum_{i=1}^n \lambda_i |\alpha_i|^2 \\
		&\stackrel{(i)}{=} \sum_{i=1}^n \lambda_i \delta_{i,j}^2 \\
		&\stackrel{(ii)}{=} \lambda_j.
	\end{align*}
	In $(i)$ was used equation \ref{equation: alphai = delta}; in $(ii)$ the delta was used to eliminate the summation. Since $a \neq 0$, equation \ref{equation: a neq 0 implis adag G a neq 0} implies that $a^\dag G a \neq 0$, so
	\begin{equation}
		\lambda_j \neq 0,
	\end{equation}
	for every $j$. This proves that $G$ is invertible.
\end{proof}

Gram matrices of tensor products of vectors will be important. For this reason, we prove the following propositions:

\begin{proposition}{Gram matrix of tensor product}{Gram tensor}
	Let $X$ and $Y$ be finite dimensional complex Hilbert spaces. Let also $(x_i)_{i=1,...,m}$ be vectors of $X$ and $(y_j)_{j=1,...,n}$ be vectors of $Y$. Let:
	\begin{itemize}
		\item $G$ be the Gram matrix of $(x_i)_{i=1,...,m}$;
		\item $G'$ be the Gram matrix of $(y_j)_{j=1,...,n}$;
		\item $G''$ be the Gram matrix of $(x_i \otimes y_j)_{\substack{i=1,...,m \\ j=1,...,n}}$.
	\end{itemize}
	We order the pairs $(i,j)$ in lexicographic order, that is,
	\begin{equation*}
		(i,j) < (i',j') \iff i<i' \text{ or } i=i' \text{ and } j<j'.
	\end{equation*}
	This lets us interpret $(x_i \otimes y_j)_{\substack{i=1,...,m \\ j=1,...,n}}$ as a sequence of vectors. Then $G''$ is the Kronecker product of $G$ and $G'$, that is, $G''= G \otimes G'$. Its matrix elements are, therefore,
	\begin{equation}
		G''_{(i,j),(i',j')} = G_{i,i'} G'_{j,j'},
	\end{equation}
	for $i,i' \in \{1, \dots, m\}$ and $j,j' \in \{1, \dots, n\}$.
\end{proposition}
\begin{proof}
	The matrix elements of $G$, $G'$ and $G''$ are, by definition,
	\begin{equation}
		\label{equation: G elemento de matriz}
		G_{i,i'} = \langle x_i , x_{i'} \rangle,
	\end{equation}
	\begin{equation}
		\label{equation: G' elemento de matriz}
		G'_{j,j'} = \langle y_j , y_{j'} \rangle,
	\end{equation}
	\begin{equation}
		G''_{(i,j),(i',j')} = \langle x_i \otimes y_j , x_{i'} \otimes y_{j'} \rangle.
	\end{equation}
	By definition of the inner product over $X \otimes Y$, we have that 
	\begin{equation}
		G''_{(i,j),(i',j')}  =  \langle x_i \otimes y_j , x_{i'} \otimes y_{j'} \rangle  =  \langle x_i, x_{i'}\rangle \langle y_j , y_{j'} \rangle.
	\end{equation}
	By (\ref{equation: G elemento de matriz}) and (\ref{equation: G' elemento de matriz}), we have that
	\begin{equation}
		G''_{(i,j),(i',j')} = G_{i,i'} G'_{j,j'} .
	\end{equation}
	This equation says that $G''$ is the Kronecker product $G'' = G \otimes G'$.
\end{proof}

\begin{proposition}{Linear independence of tensor product}{tensor of l.i. vectors}
	Let $X$ and $Y$ be finite dimensional complex Hilbert spaces. Let also $(x_1, \dots, x_m)$ be a sequence of vectors in $X$, and let $(y_1 , \dots, y_n)$ be a sequence in $Y$, with $m,n \geq 1$. Then $(x_i \otimes y_j)_{i,j}$ is linearly independent iff both $(x_i)_i$ and $(y_j)_j$ are linearly independent.
\end{proposition}
\begin{proof}
	We use the same notation of proposition \ref{proposition:Gram tensor}.  We regard $(x_i \otimes y_j)_{i,j}$ as a sequence by ordering the pairs $(i,j)$ in the lexicographic order. The sequences $(x_i)_i$, $(y_j)_j$ and $(x_i \otimes y_j)_{i,j}$ have Gram matrices which we denote as $G$, $G'$ and $G''$, respectively. By proposition \ref{proposition:Gram tensor}, we have that
	\begin{equation}
		G'' = G \otimes G' .
	\end{equation}
	Also,
	\begin{equation}
		\det(G'') = \det(G \otimes G') = \det(G)^n \det(G')^m ,
	\end{equation}
	so $G''$ is invertible iff both $G$ and $G'$ are invertible. In this case, $G''^{-1} = G^{-1} \otimes G'^{-1}$. Using proposition \ref{proposition:gram matrix} we conclude that $(x_i \otimes y_j)_{i,j}$ is linearly independent iff both $(x_i)_i$ and $(y_j)_j$ are linearly independent.
\end{proof}

\section{Isometries}

An isometry can be an operator or a matrix. For operators, an isometry is a $\complexNumbers$-linear map $U \colon X \to Y$, between finite dimensional complex Hilbert spaces $X$ and $Y$, that satisfies the property
\begin{equation}
	U^\dag U = \id_X.
\end{equation}

For matrices we have a similar idea. A matrix $U \in M_{m,n}(\complexNumbers)$ is an isometry if
\begin{equation}
	U^\dag U = I_n.
\end{equation}
In both cases $U$ have a left inverse, so it is injective and we must have $d_X \leq d_Y$ and $n \leq m$. The equation $U^\dag U = I_n$ is also equivalent to the statement that the columns of $U$ are orthonormal. In this sense, we can think of an isometry matrix as a choice of an orthonormal basis for a subspace of $\complexNumbers^m$ of dimension $n$. We won't show this, but the subset $V_{m,n} (\complexNumbers) \subseteq M_{m,n}(\complexNumbers)$ of isometries is a smooth submanifold called the (compact) Stiefel manifold.

There is an equivalence relation that consists of multiplying an isometry by a unitary operator:
\begin{equation}
	U \sim U' \iff \exists V \in U(n) \text{ such that } U'=UV.
\end{equation}
It is a simple exercise to prove that this defines an equivalence relation over the Stiefel manifold $V_{m,n}(\complexNumbers)$. This is also the same relation that is used to define the Grassmannian manifold. 
\begin{proposition}{Equivalence relation for isometries}{}
	The relation defined above is an equivalence relation over the Stiefel manifold $V_{m,n}(\complexNumbers)$.
\end{proposition}
\begin{proof}
	For the reflexive property, take $V = I_n$, then we get that $U \sim U$, for any $U \in V_{m,n} (\complexNumbers)$.
	
	For the symmetric property, if $U \sim U'$ then there exists $V \in U(n)$ such that $U' = U V$. This implies that $U = U' V^\dag$. Of course, $V^\dag \in U(n)$, so $U' \sim U$.
	
	For the transitivity, suppose that $U \sim U'$ and $U' \sim U''$. Then there exists $V,V' \in U(n)$ such that $U' = U V$ and $U'' = U' {V'}$. Substituting the expression for $U'$ in the expression for $U''$, we get that $U'' = U V V'$. Since $V V' \in U(n)$, we have that $U \sim U''$. This proves that $\sim$ is an equivalence relation over $V_{m,n} (\complexNumbers)$. 
\end{proof}

Still related to the Grassmannian manifold, two isometries are equivalent iff their columns generate the same vector space, which we prove next. Recall that the image of a matrix is the space generated by its columns.
\begin{proposition}{}{isometries are equivalent if they have the same image}
	For any $U,U' \in V_{m,n}(\complexNumbers)$, we have that
	\begin{equation}
		U \sim U' \iff \image U = \image U'.
	\end{equation}
\end{proposition}
\begin{proof}
	$(\implies)$ If $U \sim U'$, then there exists $V \in U(n)$ such that $U' = U V$. Since $V$ is an isomorphism, it follows that $\image U = \image U'$.
	
	$(\impliedby)$ Suppose that $\image U = \image U'$. Then the columns of $U$ and $U'$ are both orthonormal basis to the same space $\image U$. Write these matrices as
	\begin{equation}
		U = \begin{pmatrix}
			u_1 & \dots & u_n
		\end{pmatrix},
	\end{equation}
	\begin{equation}
		U' = \begin{pmatrix}
			u'_1 & \dots & u'_n
		\end{pmatrix},
	\end{equation}
	where $u_1,\dots,u_n$ are the columns of $U$, and $u'_1, \dots,u'_n$ are the columns of $U'$. Then $u'_j$ is a linear combination of $(u_j)_j$, so there exists complex numbers $v_{k,k'} \in \complexNumbers$ such that
	\begin{equation}
		\label{equation: u'j = sum k vkj uk}
		u'_j = \sum_k v_{k,j} u_k.
	\end{equation}
	We can organize these complex numbers in a matrix $V = (v_{k,k'})_{k,k'} \in M_n (\complexNumbers)$. The orthonormality of both basis implies that $V$ is unitary. In fact, for any $j,j'$ we have that
	\begin{align*}
		\delta_{j,j'} &\stackrel{(i)}{=} \langle u'_j, u'_{j'} \rangle \\
		&\stackrel{(ii)}{=} \left\langle \sum_{k} v_{k,j} u_k, \sum_{k'} v_{k',j'} u_{k'} \right\rangle \\
		&\stackrel{(iii)}{=} \sum_{k,k'} \overline{v_{k,j}}v_{k',j'} \langle u_k, u_{k'}\rangle\\
		&\stackrel{(iv)}{=} \sum_{k,k'} \overline{v_{k,j}}v_{k',j'} \delta_{k,k'}\\
		&\stackrel{(v)}{=} \sum_k \overline{v_{k,j}}v_{k,j'}\\
		&\stackrel{(vi)}{=} \sum_{k} (V^\dag)_{j,k} V_{k,j'}\\
		&\stackrel{(vii)}{=} (V^\dag V)_{j,j'}.
	\end{align*}
	In $(i)$ was used that $U$ has orthonormal columns; in $(ii)$ was used equation \ref{equation: u'j = sum k vkj uk}; in $(iii)$ was used that the inner product in sesquilinear; in $(iv)$ was used that $U$ has orthonormal columns; in $(v)$ the delta was used to eliminate the summation over $k'$; in $(vi)$ was used the definition of the matrix $V$; in $(vii)$ the sum was identified as a matrix multiplication.
	
	Therefore, $V^\dag V = I_n$, so $V \in U(n)$. Rewriting equation \ref{equation: u'j = sum k vkj uk} in terms of the components, we get that
	\begin{equation}
		U'_{i,j} = \sum_k v_{k,j} U_{i,k}.
	\end{equation}
	The right hand side is $(UV)_{i,j}$. Therefore $U' = UV$, so $U \sim U'$.
\end{proof}

\section{Polar decomposition}

Any complex number $z$ admits a polar decomposition $z = \lambda|z|$, where $\lambda \in U(1)$. There is a similar result for matrices. In Theorem 7.3.1 of \cite{Horn_Johnson} is proved that matrices admit a polar decomposition. In the next result we extract from that theorem the part which will be relevant for us.

\begin{theorem}{}{polar decomposition}
	Let $m,n \geq 1$ and $M \in M_{m,n} (\complexNumbers)$. Then there exists $P\in M_m (\complexNumbers)$, $Q \in M_n ( \complexNumbers)$, both $P$ and $Q$ positive semidefinite, and $U \in M_{m,n} ( \complexNumbers)$ such that
	\begin{equation}
		M = PU = UQ.
	\end{equation}
	Other properties are valid depending on $m$, $n$ and if $M$ has maximum rank:
	\begin{itemize}
		\item If $m \leq n$, then $U U^\dag = I_m$ and $P = (M M^\dag)^{1/2}$.
	\end{itemize}
	\begin{itemize}
		\item If $m \geq n$, then $U^\dag U = I_n$ and $Q = (M^\dag M)^{1/2}$.
	\end{itemize}
	In both cases, if $M$ has maximum rank, then $U$ is unique.
\end{theorem}

The polar decomposition isn't unique if $M$ doesn't have maximum rank or if $M$ isn't a square matrix. Nevertheless, we can adapt the polar decomposition to write $M$ as a product of two matrices determined uniquely by $M$. We prove this in the next result. We only do it for $m\geq n$, which is the case  we''be interested in. The case where $m \leq n$ can be inferred by applying the result to $M^\dag$.

\begin{proposition}{}{polar decomposition standardized}
	Let $m,n \geq 1$, with $m\geq n$, and $M \in M_{m,n} (\complexNumbers)$. There exists a $U \in M_{m,n}(\complexNumbers)$ such that $U^\dag U = I_n$ and
	\begin{equation}
		M = (M M^\dag)^{1/2} U  = U (M^\dag M)^{1/2}.
	\end{equation}
	Such matrix $U$ necessarily satisfies that
	\begin{equation}
		\image M \subseteq \image U.
	\end{equation}
	Also, we can write $M$ as
	\begin{equation}
		M = (M M^\dag)^{1/2} (P_{\image M} U),
	\end{equation}
	where $P_{\image M}$ is the orthogonal projection over $\image M$. The factor $P_{\image M} U$ is unique, and it is
	\begin{equation}
		P_{\image M} U = \left[(M M^\dag)^{1/2}\right]^+ M,
	\end{equation}
	where $A^+$ denotes the Moore-Penrose inverse of $A$.
\end{proposition}
\begin{proof}
	From theorem \ref{theorem:polar decomposition} we know that $M$ can be written as
	\begin{equation}
		\label{equation: M=UQ}
		M = UQ,
	\end{equation}
	where $Q = (M^\dag M)^{1/2}$ and $U \in M_{m,n}(\complexNumbers)$, with $U^\dag U = I_n$. It is immediate from equation \ref{equation: M=UQ} that $\image M \subseteq \image U$, which proves one of the statements of this proposition. Next we'll prove that, for the same $U$, we have
	\begin{equation}
		M = (M M^\dag)^{1/2} U.
	\end{equation}
	Using that $M = UQ$ we get that
	\begin{align*}
		M M^\dag &\stackrel{(i)}{=} (UQ) (UQ)^\dag\\
		&\stackrel{(ii)}{=} UQQ^\dag U^\dag\\
		&\stackrel{(iii)}{=} U Q^2 U^\dag.
	\end{align*}
	In $(i)$ was used that $M = UQ$; in $(ii)$ the $\dag$ was distributed to the matrices; in $(iii)$ was used that $Q$ is positive semidefinite, so it is hermitian and $Q^\dag = Q$.
	
	$Q$ is positive semidefinite, so it is diagonalizable and has non negative eigenvalues. Therefore, we can write it as
	\begin{equation}
		\label{equation: Q diagonalization}
		Q = \sum_k \lambda_k \ket{v_k}\bra{v_k},
	\end{equation}
	where $\lambda_k > 0$ and $(v_k)_k$ is an orthonormal sequence is $\complexNumbers^n$. The length of this sequence may be zero, in which case $Q = 0$. Taking the square, we get that
	\begin{equation}
		Q^2 = \sum_k \lambda_k^2 \ket{v_k}\bra{v_k}.
	\end{equation}
	Conjugating by $U$, we get that
	\begin{equation}
		M M^\dag = U Q^2 U^\dag = \sum_k \lambda_k^2 U\ket{v_k}\bra{v_k}U^\dag.
	\end{equation}
	Now define the vectors
	\begin{equation}
		\label{equation: wk = Uvk}
		\ket{w_k} \coloneqq U \ket{v_k} \in \complexNumbers^m.
	\end{equation}
	They are also orthonormal. Indeed, for any $k,l$ we have that
	\begin{equation}
		\langle w_k | w_l \rangle = \bra{v_k}U^\dag U \ket{v_l} = \langle v_k | v_l \rangle = \delta_{k,l}.
	\end{equation}
	We used that $U^\dag U = I_n$ and that $(v_k)_k$ is orthonormal. This proves that the equation
	\begin{equation}
		M M^\dag = \sum_k \lambda_k^2 \ket{w_k}\bra{w_k}
	\end{equation}
	is a diagonalization of $M M^\dag$. The square root of $M M^\dag$ is obtained by taking the square root of its eigenvalues:
	\begin{equation}
		\label{equation: diagonalization of (MMdag)1/2}
		(M M^\dag)^{1/2} = \sum_k \lambda_k \ket{w_k}\bra{w_k}.
	\end{equation}
	Using equation \ref{equation: wk = Uvk} we can express $(M M^\dag)^{1/2}$ in terms of $Q$:
	\begin{align*}
		(M M^\dag)^{1/2} &\stackrel{(i)}{=} \sum_k \lambda_k \ket{w_k}\bra{w_k}\\
		&\stackrel{(ii)}{=} \sum_k \lambda_k U\ket{v_k}\bra{v_k}U^\dag\\
		&\stackrel{(iii)}{=} U \left(\sum_k \lambda_k \ket{v_k}\bra{v_k}\right)U^\dag\\
		&\stackrel{(iv)}{=} U Q U^\dag.
	\end{align*}
	In $(i)$ was used equation \ref{equation: diagonalization of (MMdag)1/2}; in $(ii)$ was used equation \ref{equation: wk = Uvk}; in $(iii)$ was used linearity to pass the summation and $\lambda_k$ to the left of $\ket{v_k}$; in $(iv)$ was used equation \ref{equation: Q diagonalization}.
	
	From equation \ref{equation: M=UQ}, this implies that
	\begin{equation}
		(M M^\dag)^{1/2} = M U^\dag.
	\end{equation}
	Multiplying by $U$ and using that $U^\dag U = I_n$, we get
	\begin{equation}
		M = (M M^\dag)^{1/2} U.
	\end{equation}
	This proves that we can always take $P = (M M^\dag)^{1/2}$ in the polar decomposition, even when $P$ is not unique.
	
	The $U$ has a unique part and another part that can be replaced by any matrix. By replacing the non unique part of $U$ with 0, we get a unique matrix. First, we take $P = (M M^\dag)^{1/2}$ and write
	\begin{equation}
		\label{equation: M = PU}
		M = PU.
	\end{equation}
	Then we'll identify the unique part of $U$. A matrix of the form $A^\dag A$ can be interpreted as the Gram matrix of the columns of $A$. The rank of such Gram matrix is the same of $A$, so
	\begin{equation}
		\rank(A^\dag A) = \rank (A) .
	\end{equation}
	From this we know that
	\begin{equation}
		\rank(M M^\dag) = \rank (M) .
	\end{equation}
	Of course, we also have that
	\begin{equation}
		\image(M M^\dag ) \subseteq \image M .
	\end{equation}
	The rank is the dimension of the image, so this inclusion is an equality, that is, 
	\begin{equation}
		\image(M M^\dag ) = \image M .
	\end{equation}
	Taking the square root doesn't change the image, therefore
	\begin{equation}
		\image P = \image M.
	\end{equation}
	Let $P_{\image M}$ be the orthogonal projection over $\image M$. Recall that equation \ref{equation: diagonalization of (MMdag)1/2} gives a diagonalization of $P$. Inverting the non zero eigenvalues of $P$, we get another positive semidefinite operator $P^+$:
	\begin{equation}
		P^+ = \sum_k \lambda_k^{-1} \ket{w_k}\bra{w_k}. 
	\end{equation}
	This operator $P^+$ is the Moore-Pensore inverse of $P$. It has the property that
	\begin{equation}
		P P^+ = P^+ P = P_{\image M}.
	\end{equation}
	Multiplying equation \ref{equation: M = PU} by $P^+$, we get that
	\begin{equation}
		P^+ M = P^+ P U = P_{\image M} U.
	\end{equation}
	We took $P = (M M^\dag)^{1/2}$, so $P^+ M$ is completely determined by $M$. This shows that the product $P_{\image M} U$ is unique. The product $(I_m-P_{\image M}) U$ is the non unique part of $U$. We can easily show that it can be replaced by any matrix without affecting the equation $M = PU$. In fact, let $W\in M_{m,n}(\complexNumbers)$ be any matrix such that
	\begin{equation}
		\label{equation: P U = P W}
		P_{\image M} U = P_{\image M} W.
	\end{equation}
	Then
	\begin{align*}
		P W &\stackrel{(i)}{=} P P_{\image M} W \\
		&\stackrel{(ii)}{=} P P_{\image M} U \\
		&\stackrel{(iii)}{=} P U \\
		&\stackrel{(iv)}{=} M.
	\end{align*}
	In $(i)$ was used that $\image P = \image M$, so $P P_{\image M} = P$; in $(ii)$ was used equation \ref{equation: P U = P W}; in $(iii)$ was used again that $P P_{\image M} = P$; in $(iv)$ was used that $M = PU$. 
	
	This concludes that $M$ can be written as
	\begin{equation}
		M = P \, (P_{\image M} U),
	\end{equation}
	where $P = (M M^\dag)^{1/2}$, $U \in M_{m,n}(\complexNumbers)$ is an isometry and the product $P_{\image M}U$ is unique. This finishes the proof.
\end{proof}

\cleardoublepage
\chapter{Category Theory}
\label{appendix: category theory}

A standard reference for Category Theory is Mac Lane's book \cite{catswork}. This appendix has the purpose of introducing only the aspects of Category Theory necessary for understanding this work. We'll cover the basic definitions of categories, monoidal categories and present the string diagrams.

String diagrams are a notation used arrows in monoidal categories. There are different types of string diagrams depending on the properties and structures of a monoidal category. We'll focus on string diagrams for the category of finite dimensional complex vector spaces. This category is an example of a symmetric spherical category. Our diagrams are almost the same from the book Monoidal Categories and Topological Field Theory \cite{MonoidalCatsAndTFT}, but with the opposite orientation for the strings. Other string diagrams can also be found in the book Picturing Quantum Processes \cite{Coecke_Kissinger_2017}. We recommend the study of string diagrams, because they can make some arguments much simpler. We'll avoid using string diagrams in this thesis, because of difficulty to explain them properly.

\section{Categories}

In Mathematics there are multiple notions of spaces and functions between them. For example, we have
\begin{itemize}
	\item Sets and functions between sets;
	
	\item Groups and homomorphisms of groups;
	
	\item Rings and homomorphisms of rings;
	
	\item Topological spaces and continuous functions;
	
	\item Vector spaces and linear transformations;
	
	\item Modules and homomorphisms of modules;
	
	\item Banach spaces and bounded operators.
\end{itemize}
Category Theory abstracts these notions to categories, where the spaces become objects and the functions become arrows. Arrows are also called morphisms, depending on the author. The definition of category strips away most of the structure, essentially using that functions can be composed, composition is associative and there are identity functions.

\begin{definition}{Category}{category}
	A \textbf{category} $\cat{C}$ consists of the following data:
	\begin{itemize}
		\item A set $\cat{C}_0$ of \textbf{objects} of the category.
		
		\item For each pair of objects $X, Y \in \cat{C}_0$, a set $\cat{C}(X,Y)$ of \textbf{arrows} from $X$ to $Y$. $\cat{C}(X,Y)$ is called a \textbf{hom-set}. For each arrow $f \in \cat{C}(X,Y)$, we call $X$ the \textbf{domain} of $f$ and $Y$ its \textbf{codomain}. We denote the domain as $\dom(f)$ and the codomain as $\cod(f)$. We also denote that $f \in \cat{C}(X,Y)$ by writing $f\colon X \to Y$. The set of all arrows, with any domain or codomain, is denoted as $\cat{C}_1$.
		
		\item An operation of \textbf{composition}, defined for triples of objects $X, Y, Z \in \cat{C}_0$:
		\begin{align*}
			\circ \colon \cat{C}(Y, Z) \times \cat{C}(X, Y) &\longrightarrow \cat{C}(X, Z)\\
			(g, f) &\longmapsto g \circ f.
		\end{align*}
		That is, the composition $g \circ f$ is defined for arrows $f, g \in \cat{C}_1$ with $\cod(f) = \dom(g)$. This operation must be \textbf{associative}, which means that
		\begin{equation}
			h \circ (g \circ f) = (h \circ g) \circ f,
		\end{equation}
		for any $X, Y, Z, W \in \cat{C}_0$, $f \in \cat{C}(X, Y)$, $g \in \cat{C}(Y, Z)$ and $h \in \cat{C}(Z, W)$.
		
		\item For each object $X \in \cat{C}_0$, a special arrow $\id_X \in \cat{C}(X,X)$, called the \textbf{identity arrow}, which is a \textbf{unit for the composition operation}. That is, for any $X,Y \in \cat{C}_0$, $f\in \cat{C}(X,Y)$ and $g \in \cat{C}(Y,X)$, the following holds:
		\begin{equation}
			f \circ \id_X = f,
		\end{equation}
		\begin{equation}
			\id_X \circ g = g.
		\end{equation}
	\end{itemize}
\end{definition}

The composition operation will typically be suppressed, that is, we'll write $gf$ instead of $g \circ f$. Since it is associative, we'll also avoid writing parenthesis. Notice that the definition of category doesn't require the objects to be sets or the arrows to be functions.

The numbers in $\cat{C}_0$ and $\cat{C}_1$ reflect that there are also definitions of higher categories, which have arrows between arrows and etc. Higher categories won't be used in this work. Later we'll define what is a monoidal category, which can be viewed as a special case of a bicategory. Bicategories are higher categories, but we won't use this fact, and also won't define any type of higher category.

In this work we'll be mainly interested in the category of finite dimensional vectors spaces and other categories related to it. Fields won't be defined here, but the cases of most interest are $F = \realNumbers$ and $F = \complexNumbers$.

\begin{definition}{}{finvec category}
	The \textbf{category of finite dimensional vectors spaces} over a field $F$, denoted $\finvec_F$, is the category with the following structure:
	\begin{itemize}
		\item Its objects $(\finvec_F)_0$ are the $F$-vector spaces $X$;
		
		\item The hom-set $\finvec_F (X,Y)$ is the set of linear transformations $f \colon X \to Y$.
		
		\item The operation of composition is the usual composition of functions.
		
		\item The identity arrow $\id_X$ is the identity function
		\begin{align*}
			\id_X \colon X &\longrightarrow X \\
			x &\longmapsto x.
		\end{align*}
	\end{itemize}
\end{definition}

\begin{proposition}{}{finvec is a category}
	Let $F$ be a field. $\finvec_F$ from definition \ref{definition:finvec category} is a category.
\end{proposition}
\begin{proof}
	For it to be a category, the composition of linear transformations must be a linear transformation, which is known from Linear Algebra. In fact, if $X, Y, Z \in (\finvec_F)_0$, and $f \colon X \to Y$ and $g \colon Y \to Z$ are linear transformations, then we have the composite function $g f$. This function is linear:
	\begin{itemize}
		\item For any $x, x' \in X$, we have
		\begin{align*}
			g(f(x+x')) &\stackrel{(i)}{=} g(f(x)+f(x'))\\
			&\stackrel{(ii)}{=} g(f(x))+g(f(x')).
		\end{align*}
		In $(i)$ was used that $f$ is linear, in $(ii)$ was used that $g$ is linear.
		
		\item For any $x \in X$ and $\lambda \in F$, we have
		\begin{align*}
			g(f(\lambda x)) &\stackrel{(i)}{=} g(\lambda f(x)) \\
			&\stackrel{(ii)}{=} \lambda g(f(x)).
		\end{align*}
		In $(i)$ was used that $f$ is linear, in $(ii)$ was used that $g$ is linear.
	\end{itemize}
	In this way, the composition inherited from functions between sets restricts to an operation of composition of linear transformations. Since the composition of functions is associative, of course the same is true for linear transformations. This can be easily proved. Two functions are equal if they have the same values on the same points. Let $X,Y,Z,W \in (\finvec_F)_0$, and let $f \colon X \to Y$, $g \colon Y \to Z$ and $h \colon Z \to W$ be linear transformations. By definition,
	\begin{equation}
		h \circ (g \circ f) = (h \circ g) \circ f \iff \forall x \in X, \ [h \circ (g \circ f)](x) = [(h \circ g) \circ f](x).
	\end{equation}
	We have that
	\begin{align*}
		[h \circ (g \circ f)](x) &= h((g\circ f(x))) \\
		&= h(g(f(x))), 
	\end{align*}
	and
	\begin{align*}
		[(h \circ g) \circ f](x) &= (h \circ g)(f(x))\\
		& h(g(f(x))).
	\end{align*}
	In both cases, the proof amounts to use the definition of composition repeatedly. This proves that $h \circ (g \circ f) = (h \circ g) \circ f$, so composition is associative.
	
	It is trivial that the identity function $\id_X$ is a linear transformation. In fact,
	\begin{equation}
		\id_X(x+x') = x+x' = \id_X(x)+\id_X(x'),
	\end{equation}
	for any $x,x' \in X$, and
	\begin{equation}
		\id_X(\lambda x) = \lambda x = \lambda \id_X(x),
	\end{equation}
	for any $x \in X$ and $\lambda \in F$. This proves that $\id_X$ is linear. Finally, let's prove that it is a unit for the composition. We have that
	\begin{equation}
		(f \circ \id_X)(x) = f(\id_X(x)) = f(x),
	\end{equation}
	for any $x \in X$, so
	\begin{equation}
		f \circ \id_X = f.
	\end{equation}
	Similarly, let $j \colon Y \to X$ be a linear transformation. Then,
	\begin{equation}
		(\id_X \circ j)(y) = \id_X(j(y)) = j(y),
	\end{equation}
	for any $y \in Y$, so
	\begin{equation}
		\id_X \circ j = j.
	\end{equation}
	This concludes the proof that $\finvec_F$ is a category.
\end{proof}

\subsection{Diagrams}

Many times it is more convenient to use some diagrammatic notation to reason about arrows in a category instead of writing regular equations. The most common type of diagram have the following structure. It consists of two sequences of composable arrows
\begin{equation*}
	f_1 \colon Z_1 \to Z_2, \, \dots , f_m \colon Z_m \to Z_{m+1},
\end{equation*}
and
\begin{equation*}
	g_1 \colon W_1 \to W_2, \, \dots , g_n \colon W_n \to W_{n+1},
\end{equation*}
with $m,n \geq 1$, $Z_1 = W_1$ and $Z_{m+1} = W_{n+1}$. In this way, it makes sense to compare the compositions of the arrows in both sequences. That is, it makes sense to compare $f_m \dots f_1$ with $g_n \dots g_1$. We can make a diagram to represent these arrows, as follows:

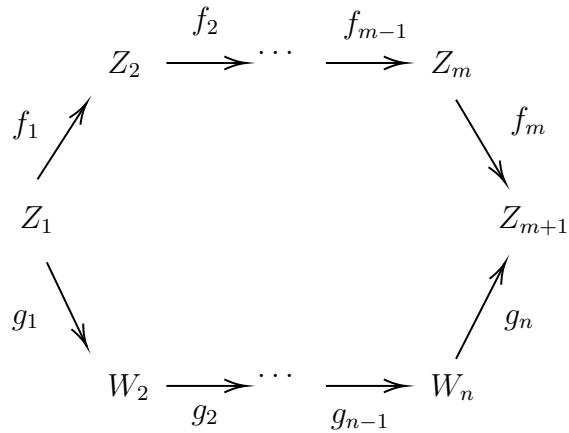
\begin{center}
	\tikzset{every picture/.style={line width=0.75pt}} 
	
	\begin{tikzpicture}[x=0.75pt,y=0.75pt,yscale=-1,xscale=1]
		
		\draw    (15.3,90.4) -- (38.92,53.68) ;
		\draw [shift={(40,52)}, rotate = 122.75] [color={rgb, 255:red, 0; green, 0; blue, 0 }  ][line width=0.75]    (10.93,-3.29) .. controls (6.95,-1.4) and (3.31,-0.3) .. (0,0) .. controls (3.31,0.3) and (6.95,1.4) .. (10.93,3.29)   ;
		\draw    (80,32) -- (118,32) ;
		\draw [shift={(120,32)}, rotate = 180] [color={rgb, 255:red, 0; green, 0; blue, 0 }  ][line width=0.75]    (10.93,-3.29) .. controls (6.95,-1.4) and (3.31,-0.3) .. (0,0) .. controls (3.31,0.3) and (6.95,1.4) .. (10.93,3.29)   ;
		\draw    (160,32) -- (198,32) ;
		\draw [shift={(200,32)}, rotate = 180] [color={rgb, 255:red, 0; green, 0; blue, 0 }  ][line width=0.75]    (10.93,-3.29) .. controls (6.95,-1.4) and (3.31,-0.3) .. (0,0) .. controls (3.31,0.3) and (6.95,1.4) .. (10.93,3.29)   ;
		\draw    (225.3,50.4) -- (248.98,90.28) ;
		\draw [shift={(250,92)}, rotate = 239.3] [color={rgb, 255:red, 0; green, 0; blue, 0 }  ][line width=0.75]    (10.93,-3.29) .. controls (6.95,-1.4) and (3.31,-0.3) .. (0,0) .. controls (3.31,0.3) and (6.95,1.4) .. (10.93,3.29)   ;
		\draw    (20,132) -- (39.14,172.19) ;
		\draw [shift={(40,174)}, rotate = 244.54] [color={rgb, 255:red, 0; green, 0; blue, 0 }  ][line width=0.75]    (10.93,-3.29) .. controls (6.95,-1.4) and (3.31,-0.3) .. (0,0) .. controls (3.31,0.3) and (6.95,1.4) .. (10.93,3.29)   ;
		\draw    (80,194) -- (118,194) ;
		\draw [shift={(120,194)}, rotate = 180] [color={rgb, 255:red, 0; green, 0; blue, 0 }  ][line width=0.75]    (10.93,-3.29) .. controls (6.95,-1.4) and (3.31,-0.3) .. (0,0) .. controls (3.31,0.3) and (6.95,1.4) .. (10.93,3.29)   ;
		\draw    (160,194) -- (198,194) ;
		\draw [shift={(200,194)}, rotate = 180] [color={rgb, 255:red, 0; green, 0; blue, 0 }  ][line width=0.75]    (10.93,-3.29) .. controls (6.95,-1.4) and (3.31,-0.3) .. (0,0) .. controls (3.31,0.3) and (6.95,1.4) .. (10.93,3.29)   ;
		\draw    (225.3,180.4) -- (249.09,133.78) ;
		\draw [shift={(250,132)}, rotate = 117.04] [color={rgb, 255:red, 0; green, 0; blue, 0 }  ][line width=0.75]    (10.93,-3.29) .. controls (6.95,-1.4) and (3.31,-0.3) .. (0,0) .. controls (3.31,0.3) and (6.95,1.4) .. (10.93,3.29)   ;
		
		\draw (5,102.4) node [anchor=north west][inner sep=0.75pt]    {$Z_{1}$};
		\draw (49,24.4) node [anchor=north west][inner sep=0.75pt]    {$Z_{2}$};
		\draw (211,24.4) node [anchor=north west][inner sep=0.75pt]    {$Z_{m}$};
		\draw (245,102.4) node [anchor=north west][inner sep=0.75pt]    {$Z_{m+1}$};
		\draw (124,24.4) node [anchor=north west][inner sep=0.75pt]    {$\dotsc $};
		\draw (1,54.4) node [anchor=north west][inner sep=0.75pt]    {$f_{1}$};
		\draw (91,2.4) node [anchor=north west][inner sep=0.75pt]    {$f_{2}$};
		\draw (251,52.4) node [anchor=north west][inner sep=0.75pt]    {$f_{m}$};
		\draw (167,4.4) node [anchor=north west][inner sep=0.75pt]    {$f_{m-1}$};
		\draw (49,186.4) node [anchor=north west][inner sep=0.75pt]    {$W_{2}$};
		\draw (211,186.4) node [anchor=north west][inner sep=0.75pt]    {$W_{n}$};
		\draw (124,186.4) node [anchor=north west][inner sep=0.75pt]    {$\dotsc $};
		\draw (1,154.4) node [anchor=north west][inner sep=0.75pt]    {$g_{1}$};
		\draw (91,202.4) node [anchor=north west][inner sep=0.75pt]    {$g_{2}$};
		\draw (161,204.4) node [anchor=north west][inner sep=0.75pt]    {$g_{n-1}$};
		\draw (248,154.4) node [anchor=north west][inner sep=0.75pt]    {$g_{n}$};

	\end{tikzpicture}
\end{center}
Such diagram is said to be commutative, or that it commutes, if
\begin{equation}
	f_m \dots f_1 = g_n \dots g_1 .
\end{equation}
This implies that both paths from $Z_1$ to $Z_{m+1}$ are equivalent.

When some arrow of a diagram have some uniqueness property, we may represent such arrow by a dashed arrow:
\begin{center}
	\begin{tikzcd}
		A \arrow[r, dashed, "f"] & B
	\end{tikzcd}
\end{center}
A common scenario is when there exists a unique arrow that completes a diagram to a commutative diagram:
\begin{center}
	\begin{tikzcd}
		A \arrow[r, "f"] \arrow[dr, swap, "g"]& B \arrow[d, dashed, "u"] \\
		{} & C
	\end{tikzcd}
\end{center}
This diagram is to be interpreted as saying that there exists a unique arrow $u \colon B \to C$ such that $g = uf$.

There are other types of diagrams that we can use in Category Theory. An important type of diagram for monoidal categories are the string diagrams, which will be exposed later.

\subsection{Isomorphisms}

An isomorphism in a category $\cat{C}$ is an arrow $f\in \cat{C}(X,Y)$ for which there exists another arrow $f^{-1} \in \cat{C}(Y,X)$ such that
\begin{gather}
	f^{-1} f = \id_X,\\
	f f^{-1} = \id_Y .
\end{gather}

This definition unifies several definitions of isomorphism across Mathematics:
\begin{itemize}
	\item For the category of sets, the isomorphisms are the bijections;
	
	\item For the category of vector spaces over a field $F$, the isomorphisms are the linear isomorphisms of vector spaces;
	
	\item For the category of rings, the isomorphisms are the isomorphisms of rings.
	
	\item For the category of topological spaces, the isomorphisms are the homeomorphisms.
\end{itemize}

\subsection{The product category}

Monoidal categories, to be defined later, will be used frequently. To define monoidal categories, we first need to define product categories.

\begin{definition}{Product category}{product category}
	Let $\cat{C}$ and $\cat{C'}$ be categories. The \textbf{product category} $\cat{C} \times \cat{C'}$ is the category defined by the following data:
	\begin{itemize}
		\item Its objects are the pairs of objects:
		\begin{equation}
			(\cat{C} \times \cat{C'})_0 \coloneqq \cat{C}_0 \times \cat{C'}_0 .
		\end{equation}
		So an object of $\cat{C} \times \cat{C'}$ is a pair $(C, C')$, where $C$ is an object of $\cat{C}$ and $C'$ is an object of $\cat{C'}$. We'll also adopt the product notation:
		\begin{equation}
			C \times C' \coloneqq (C, C').
		\end{equation}
		
		\item Its arrows are pairs of arrows:
		\begin{equation}
			(\cat{C} \times \cat{C'})_1 \coloneqq \cat{C}_1 \times \cat{C'}_1 .
		\end{equation}
		We define the domain and codomain of arrows in $\cat{C} \times \cat{C'}$ as the pairs of domains or codomains of the arrows in $\cat{C}$ and $\cat{C'}$. That is, for any $f \in \cat{C}(X, Y)$ and $f' \in \cat{C'}(X', Y')$, we have the arrow $(f,f') \in (\cat{C}\times \cat{C'})(X\times X', Y \times Y')$. Its domain and codomain are
		\begin{equation}
			\dom(f,f') = (\dom(f), \dom(f')) = (X, X'),
		\end{equation}
		\begin{equation}
			\cod(f,f') = (\cod(f), \cod(f')) = (Y, Y').
		\end{equation}
		We'll also adopt the product notation:
		\begin{equation}
			f\times f' \coloneqq (f, f').
		\end{equation}
		
		\item Composition of arrows in $\cat{C} \times \cat{C'}$ is given by composing the arrows in $\cat{C}$ and $\cat{C'}$ separately. If $f \in \cat{C}(X, Y)$, $f' \in \cat{C'}(X', Y')$, $g \in \cat{C}(Y, Z)$ and $g' \in \cat{C'}(Y', Z')$, we have the arrows $(f, f') \in (\cat{C} \times \cat{C'})(X \times X', Y \times Y')$ and $(g, g') \in (\cat{C} \times \cat{C'})(Y \times Y', Z \times Z')$. The composition is defined as
		\begin{equation}
			(g, g') \circ (f, f') \coloneqq (g\circ f, g' \circ f').
		\end{equation}
		
		\item For any objects $X \in \cat{C}_0$ and $X' \in \cat{C'}_0$, the identity $\id_{(X,X')}$ is defined as the pair of identities:
		\begin{equation}
			\id_{(X,X')} \coloneqq (\id_X, \id_{X'}).
		\end{equation}
	\end{itemize}
\end{definition}

\begin{proposition}{}{}
	The product category defined in \ref{definition:product category} is a category.
\end{proposition}
\begin{proof}
	The associativity of the composition in the product category follows immediately from the associativity for $\cat{C}$ and $\cat{C'}$. Let $f \in \cat{C}(X, Y)$, $g \in \cat{C}(Y, Z)$, $h \in \cat{C}(Z, W)$, $f' \in \cat{C'}(X', Y')$, $g' \in \cat{C'}(Y', Z')$, $h' \in \cat{C'}(Z', W')$. Then,
	\begin{align*}
		(h,h') \circ ((g,g') \circ (f,f')) &\stackrel{(i)}{=} (h,h') \circ (g\circ f, g'\circ f') \\
		&\stackrel{(ii)}{=} (h\circ (g\circ f), h'\circ (g'\circ f')) \\
		&\stackrel{(iii)}{=} ((h \circ g) \circ f, (h' \circ g') \circ f') \\
		&\stackrel{(iv)}{=} (h \circ g, h' \circ g') \circ (f, f') \\
		&\stackrel{(v)}{=} ((h,h') \circ (g,g'))\circ (f,f').
	\end{align*}
	In $(i)$, $(ii)$, $(iv)$ and $(v)$ was used the definition of the composition in $\cat{C} \times \cat{C'}$. In $(iii)$ was used that the composition is associative in both $\cat{C}$ and $\cat{C'}$. This proves that the composition for $\cat{C} \times \cat{C'}$ is associative.
	
	Next, we prove that the pairs of identities are the identities in $\cat{C} \times \cat{C'}$. Let $X,Y \in \cat{C}_0$ and $X', Y' \in \cat{C'}_0$. Let also $f \in \cat{C}(X, Y)$, $f' \in \cat{C}(X', Y')$, $g \in \cat{C}(Y, X)$ and $g' \in \cat{C}(Y', X')$. Then we have the arrows $(f,f') \in (\cat{C} \times \cat{C'})(X \times X', Y \times Y')$ and $(g,g') \in (\cat{C} \times \cat{C'})(Y \times Y', X \times X')$. Composing $(f, f')$ with $(\id_{X}, \id_{X'})$, we get
	\begin{align*}
		(f, f')\circ (\id_X, \id_{X'}) &\stackrel{(i)}{=} (f \circ \id_X, f' \circ \id_{X'}) \\
		&\stackrel{(ii)}{=} (f, f').
	\end{align*}
	In $(i)$ was used the definition of the composition in $\cat{C} \times \cat{C'}$. In $(ii)$ was used that $\id_X$ is an identity arrow in $\cat{C}$ and $\id_{X'}$ is an identity arrow in $\cat{C'}$. Similarly, composing $(g, g')$ with $(\id_X, \id_{X'})$, we get
	\begin{align*}
		(\id_X, \id_{X'}) \circ (g, g') &\stackrel{(i)}{=} (\id_X \circ g, \id_{X'} \circ g')
		\\
		&\stackrel{(ii)}{=} (g, g').
	\end{align*}
	Again, in $(i)$ was used the definition of the composition in $\cat{C} \times \cat{C'}$. In $(ii)$ was used that $\id_X$ is an identity arrow in $\cat{C}$ and $\id_{X'}$ is an identity arrow in $\cat{C'}$. This proves that $(\id_{X}, \id_{X'})$ is an identity arrow for $\cat{C} \times \cat{C'}$, so
	\begin{equation}
		\id_{X \times X'} = (\id_X, \id_{X'}).
	\end{equation}
	This concludes the proof that $\cat{C} \times \cat{C'}$ is a category.
\end{proof}

\section{Some limits}

In this section we'll study some basic limits that were used in this thesis. Later we'll study some basic colimits. Limits and colimits are important for Categorical Logic. We are interested in studying limits and colimits, because understanding them and other categorical properties can help develop an internal logic for the category. Such internal logic may help develop a type theory and a programming language. In particular, it may help develop a functional programming language for quantum computing.

\subsection{Terminal object}

In a category $\cat{C}$, a terminal object $T \in \cat{C}_0$ is an object with the property that, for any object $X \in \cat{C}_0$, there exists a unique arrow in $\cat{C}$ from $X$ to $T$. In other words,
\begin{equation}
	|\cat{C}(X,T)| = 1,
\end{equation}
for every $X \in \cat{C}_0$.

As an example, in the category $\mathrm{Set}$, of sets and functions between them, a terminal object is a set with a single element.

\subsection{Binary product}

In a category $\cat{C}$, a binary product of objects $X_1, X_2 \in \cat{C}_0$ is another object $P \in \cat{C}_0$ together with arrows $p_i \in \cat{C}(P, X_i)$ with the following universal property: for any object $Z \in \cat{C}_0$ and arrows $a_i \in \cat{C}(Z, X_i)$, there exists a unique arrow $u \in \cat{C}(Z, P)$ such that $p_i u = a_i$. In other words, the following diagram commutes:
\begin{center}
	\begin{tikzcd}
		& Z \arrow[ld, swap, "a_1"] \arrow[rd, "a_2"] \arrow[d, dashed, "u"] & \\
		X_1 & P \arrow[l, "p_1"] \arrow[r, swap, "p_2"] & X_2
	\end{tikzcd}
\end{center}

In the category $\mathrm{Set}$, we can take as binary product the cartesian product of sets:
\begin{equation}
	P = X_1 \times X_2 = \{(x_1, x_2) \mid x_1 \in X_1, x_2 \in X_2 \}.
\end{equation}
The arrows $p_i$ we can take as being the projections into the coordinates:
\begin{equation}
	p_1 (x_1, x_2) = x_1,
\end{equation}
\begin{equation}
	p_2 (x_1, x_2) = x_2.
\end{equation}

\section{Some colimits}

In this section we'll study some basic colimits that were used in this thesis.

\subsection{Initial object}

The initial objects are dual to terminal objects. The duality is related to the dual category, which won't be explained here. An initial object in a category $\cat{C}$ is an object $I \in \cat{C}_0$ such that
\begin{equation}
	|\cat{C}(I,X)| = 1,
\end{equation}
for every $X \in \cat{C}_0$.

\subsection{Binary coproduct}

In a category $\cat{C}$, a binary coproduct of objects $X_1, X_2 \in \cat{C}_0$ is another object $C \in \cat{C}_0$ together with arrows $q_i \in \cat{C}(X_i, P)$ with the following universal property: for any object $Z \in \cat{C}_0$ and arrows $a_i \in \cat{C}(X_i, Z)$, there exists a unique arrow $u \in \cat{C}(C, Z)$ such that $u q_i = a_i$. In other words, the following diagram commutes:
\begin{center}
	\begin{tikzcd}
		& Z & \\
		X_1 \arrow[ru, "a_1"] \arrow[r, swap, "q_1"] & C \arrow[u, dashed, swap, "u"] & X_2 \arrow[lu, swap, "a_2"] \arrow[l, "q_2"]
	\end{tikzcd}
\end{center}

In the category $\mathrm{Set}$, we can take as binary coproduct the disjoint union:
\begin{equation}
	C = X_1 \sqcup X_2 = \bigcup_{i=1, 2} X_i \times \{i\} = \bigcup_{i=1, 2} \{ (x_i, i) \mid x_i \in X_i \}.
\end{equation}
Then we can take $q_i$ to be the injections:
\begin{equation}
	q_i(x_i) = (x_i, i).
\end{equation}

\section{Functors}

As there are arrows between objects in a category, we can treat categories as objects and define arrows between them, which are the functors. Functors are defined such that they respect the structures from the definition of category.

\begin{definition}{Functor}{functor}
	Let $\cat{C}$ and $\cat{D}$ be categories. A \textbf{functor} $F \colon \cat{C} \to \cat{D}$ consists of two functions, $F_0 \colon \cat{C}_0 \to \cat{D}_0$ and $F_1 \colon \cat{C}_1 \to \cat{D}_1$, with the following properties:
	\begin{itemize}
		\item If $f \in \cat{C}(X,Y)$, then $F_1 (f) \in \cat{D}(F_0(X), F_0(Y))$. In other words, we have that
		\begin{equation}
			\label{equation: functor respects dom}
			\dom(F_1 (f)) = F_0 (\dom(f)),
		\end{equation}
		\begin{equation}
			\label{equation: functor respects cod}
			\cod(F_1 (f)) = F_0 (\cod(f)).
		\end{equation}
		From this, it follows that composable arrows in $\cat{C}$ are sent to composable arrows in $\cat{D}$.
		
		\item $F$ respects the composition: if $f\in \cat{C}(X,Y)$ and $g \in \cat{C}(Y,Z)$, then
		\begin{equation}
			\label{equation: functor respects composition}
			F_1 (g \circ f) = F_1 (g) \circ F_1 (f).
		\end{equation}
		
		\item $F$ preserves identities: for any $X \in \cat{C}_0$, we have
		\begin{equation}
			\label{equation: functor preserves identities}
			F_1 (\id_X) = \id_{F_0 (X)}.
		\end{equation}
	\end{itemize}
\end{definition}
Typically, the indices in $F_0$ and $F_1$ will be suppressed and we'll just write $F$.

The tensor product is an important example of functor for this work, which we see next. The tensor product of vector spaces is defined by a universal property. In this way, it isn't unique, but all choices of tensor product are isomorphic. For this reason, to have a function $(X,Y) \mapsto X \otimes Y$ that sends a pair of vector spaces to its tensor product, a choice of tensor product must be made. The standard construction of the tensor product shows how to make such a choice (see proposition \ref{proposition:tensor product existence general}).

\begin{proposition}{$\otimes$ as a functor}{tensor is a functor}
	Let $F$ be a field. The tensor product of vectors spaces can be turned into a functor $\otimes \colon \finvec_F \times \finvec_F \to \finvec_F$ as follows. Given a choice of tensor product of vector spaces $X \otimes Y$, for each $X,Y \in (\finvec_F)_0$, it determines a function
	\begin{align*}
		\otimes_0 \colon (\finvec_F)_0 \times (\finvec_F)_0 &\longrightarrow (\finvec_F)_0 \\
		(X,Y) &\longmapsto X \otimes Y .
	\end{align*}
	Such choice can be made using the standard construction of the tensor product from proposition \ref{proposition:tensor product existence general}. As proved in proposition \ref{proposition:tensor product of linear transformations}, we also have tensor products of linear transformations. The tensor product of linear transformations $f \in \finvec_F (X,Y)$ and $f' \in \finvec_F (X', Y')$ is the unique linear transformation $f\otimes f' \in \finvec_F (X \otimes X', Y \otimes Y')$ such that the following diagram commutes:
	\begin{center}
		\begin{tikzcd}
			X \times X' \arrow[r, "\otimes"] \arrow[d, swap, "f \times f'"] & X \otimes X' \arrow[d, dashed, "f \otimes f'"] \\
			Y \times Y' \arrow[r, "\otimes"] & Y \otimes Y'
		\end{tikzcd}
	\end{center}
	This determines a function
	\begin{align*}
		\otimes_1 \colon (\finvec_F)_1 \times (\finvec_F)_1 &\longrightarrow (\finvec_F)_1 \\
		(f,f') &\longmapsto f \otimes f' .
	\end{align*}
	The indices in $\otimes_0$ and $\otimes_1$ will be suppressed, we'll just write $\otimes$. Both $\otimes_0$ and $\otimes_1$ together define a functor
	\begin{equation*}
		\otimes \colon \finvec_F \times \finvec_F \to \finvec_F .
	\end{equation*}
	The equations \ref{equation: functor respects composition} and \ref{equation: functor preserves identities}, from the functor definition, can be restated for $\otimes$ as follows, in infix notation:
	\begin{itemize}
		\item If $f \in \finvec_F (X, Y)$, $f' \in \finvec_F (X', Y')$, $g \in \finvec_F (Y, Z)$ and $g' \in \finvec_F (Y', Z')$, then
		\begin{equation}
			\label{equation: tensor respects composition}
			(g\circ f) \otimes (g' \circ f') = (g \otimes g') \circ (f \otimes f').
		\end{equation}
		
		\item For any $X, X' \in (\finvec_F)_0$, we have that
		\begin{equation}
			\label{equation: tensor preserves identities}
			\id_X \otimes \id_{X'} = \id_{X \otimes X'}.
		\end{equation}
	\end{itemize}
\end{proposition}
\begin{proof}
	We already have functions $\otimes_0 \colon (\finvec_F)_0 \times (\finvec_F)_0 \to (\finvec_F)_0$ and $\otimes_1 \colon (\finvec_F)_1 \times (\finvec_F)_1 \to (\finvec_F)_1$. It remains to show that the properties of a functor are satisfied.
	
	From the definition of $\otimes_1$, it is immediate that equations \ref{equation: functor respects dom} and $\ref{equation: functor respects cod}$ hold for $\otimes$. Next, let's prove that $\otimes$ respects composition. Let $f \in $ $\finvec_F (X,$ $Y)$, $f' \in \finvec_F (X', Y')$, $g \in \finvec_F (Y, Z)$ and $g' \in \finvec_F (Y', Z')$. By definition of $f \otimes f'$ and $g \otimes g'$, we have the following commutative diagrams:
	
	\begin{center}
		\begin{tikzcd}
			X \times X' \arrow[r, "\otimes"] \arrow[d, swap, "f \times f'"] & X \otimes X' \arrow[d, dashed, "f \otimes f'"] \\
			Y \times Y' \arrow[r, "\otimes"] & Y \otimes Y'
		\end{tikzcd}
	\end{center}
	
	\begin{center}
		\begin{tikzcd}
			Y \times Y' \arrow[r, "\otimes"] \arrow[d, swap, "g \times g'"] & Y \otimes Y' \arrow[d, dashed, "g \otimes g'"] \\
			Z \times Z' \arrow[r, "\otimes"] & Z \otimes Z'
		\end{tikzcd}
	\end{center}
	Combining both diagrams, we get the next commutative diagram:
	\begin{center}
		\begin{tikzcd}
			X \times X' \arrow[r, "\otimes"] \arrow[d, swap, "(gf) \times (g'f')"] & X \otimes X' \arrow[d, "(g\otimes g')(f \otimes f')"] \\
			Z \times Z' \arrow[r, "\otimes"] & Z \otimes Z'
		\end{tikzcd}
	\end{center}
	By definition of $(gf)\otimes (g'f')$, it is the unique linear transformation that makes this diagram commute. Therefore, we have that
	\begin{equation}
		(gf)\otimes (g'f') = (g\otimes g')(f \otimes f').
	\end{equation}
	Using prefix notation, we can rewrite this as
	\begin{equation}
		\otimes((g, g')\circ (f, f')) = \otimes(g, g') \circ \otimes(f, f'),
	\end{equation}
	which makes it clear that we have proved equation \ref{equation: functor respects composition} for $\otimes$.
	
	Finally, we prove that $\otimes$ preserves identities. Let $X, Y \in (\finvec_F )_0$. By definition of $\id_X \otimes \id_Y$, we have the following commutative diagram:
	\begin{center}
		\begin{tikzcd}
			X \times X' \arrow[r, "\otimes"] \arrow[d, swap, "\id_X \times \id_{X'}"] & X \otimes X' \arrow[d, dashed, "\id_X \otimes \id_{X'}"] \\
			X \times X' \arrow[r, "\otimes"] & X \otimes X'
		\end{tikzcd}
	\end{center}
	Since $\id_X \times \id_{X'} = \id_{X \times X'}$, it follows immediately that $\id_{X \otimes X'}$ also makes the diagram commute. That is, we have the commutative diagram
	\begin{center}
		\begin{tikzcd}
			X \times X' \arrow[r, "\otimes"] \arrow[d, swap, "\id_{X \times X'}"] & X \otimes X' \arrow[d, "\id_{X \otimes X'}"] \\
			X \times X' \arrow[r, "\otimes"] & X \otimes X'
		\end{tikzcd}
	\end{center}
	Since $\id_X \otimes \id_{X'}$ is the unique linear transformation that makes the diagram commute, we have that
	\begin{equation}
		\id_X \otimes \id_{X'} = \id_{X \otimes X'}.
	\end{equation}
	Rewriting this in prefix notation, we have
	\begin{equation}
		\otimes(\id_X, \id_{X'}) = \id_{X \otimes X'},
	\end{equation}
	which is equation \ref{equation: functor preserves identities} for $\otimes$. This concludes the proof that $\otimes$ is a functor.
\end{proof}

There are other important examples of functors that appear naturally in Mathematics. We won't use all these examples, but they are listed to motivate the study of functors:
\begin{itemize}
	\item In Differential Geometry, the operation of taking the tangent bundle is a functor from the category of smooth manifolds and smooth functions to the category of smooth vector bundles. It sends a smooth manifold $M$ to its tangent bundle $\pi_M \colon TM \to M$. Notice that $TM$ is also a smooth manifold and $\pi_M$ is a smooth function. Let $f\colon M \to N$ be a smooth function. For any $x \in M$, we have the tangent spaces $T_x M$ and $T_{f(x)} N$, which are finite dimensional $\realNumbers$-vector spaces. The derivative at $x$ is a linear transformation $d_x f \colon T_x M \to T_{f(x)} N$. Putting together the derivatives at each point into a single entity, we get a smooth function $Tf \colon TM \to TN$. The functor sends $f$ to the pair $(f, Tf)$, which is a bundle map. It satisfies that $f \circ \pi_M = \pi_N \circ (Tf)$, that is, the following diagram commutes:
	\begin{center}
		\begin{tikzcd}
			TM \arrow[d, swap, "\pi_M"] \arrow[r, "Tf"] & TN \arrow[d, "\pi_N"] \\
			M \arrow[r, "f"] & N
		\end{tikzcd}
	\end{center}
	The \textbf{functoriality encodes the chain rule of the derivative}.
	
	\item In Algebraic Topology, the $n$-th Singular Homology $H_n$ defines a functor from the category of topological spaces to the category of abelian groups. Collectively, we a have a functor $H_*$ from the category of topological spaces to the category of graded abelian groups. There are many other functors that appear in Algebraic Topology, and they were important for the development of Category Theory.
\end{itemize}

\section{Natural transformations}

We can repeat the process, now treating functors as objects and define arrows between them, which are the natural transformations.

\begin{definition}{Natural transformation}{natural transformation}
	Let $\cat{C}$ and $\cat{D}$ be categories. Let also $F, G \colon \cat{C} \to \cat{D}$ be functors. A \textbf{natural transformation} $\eta \colon F \to G$ is a family of arrows $(\eta_X \colon F(X) \to G(X))_{X \in \cat{C}_0}$ in $\cat{C}$ such that, for any arrow $f \colon X \to Y$ in $\cat{C}$, the following diagram commutes:
	\begin{center}
		\begin{tikzcd}
			F(X) \arrow[r, "F(f)"] \arrow[d, swap, "\eta_X"] & F(Y) \arrow[d, "\eta_Y"] \\
			G(X) \arrow[r, "G(f)"] & G(Y)
		\end{tikzcd}
	\end{center}
\end{definition}

Historically, formalizing the concept of natural transformation was one of the reasons to develop Category Theory. Natural transformations will be important in this work in the definition of monoidal category.

We can compose natural transformations by composing the arrows for each object. This lets us interpret natural transformations as arrows in a category whose objects are functors. Then a natural transformation is an isomorphism in such category iff it is a family of isomorphisms. To keep the text simpler, we won't work directly with the category of functors and won't compose natural transformations explicitly, but this discussion gives motivation for the definition of natural isomorphisms. Natural isomorphisms will be used in the definition of monoidal category.

\begin{definition}{Natural isomorphism}{natural isomorphism}
	Let $\cat{C}$ and $\cat{D}$ be categories. Let also $F, G \colon \cat{C} \to \cat{D}$ be functors and $\eta \colon F \to G$ be a natural transformation. Then $\eta$ is called a \textbf{natural isomorphism} if each arrow $\eta_X \colon F(X) \to G(X)$ is an isomorphism in $\cat{D}$, for each object $X \in \cat{C}_0$.
\end{definition}

\section{Monoidal or tensor categories}

The category of finite dimensional vector spaces have another operation other than composition, which is the tensor product $\otimes$. There is a specialization of the definition of category that includes a second operation, which are the \textbf{monoidal categories or tensor categories}, whose names depend on the author. For any field $F$, $\finvec_F$ is a monoidal category, as will be proved later. More generally, for any unital ring $R$, the category of left $R$-modules and homomorphisms is a monoidal category (example 1.1.3 of \cite{MonoidalCatsAndTFT}).

Any monoidal category admits a diagrammatic notation called \textbf{string diagrams}. They are particularly useful for monoidal categories with duals, such as rigid, pivotal or spherical categories. These duals are categorical abstractions of the dual of a vector space. We won't delve into these types of categories, rather we'll focus on the category $\finvec_{\complexNumbers}$, due to its relation with Quantum Computation and Quantum Information in the Quantum Circuit model. More details about monoidal categories with duals and their string diagrams can be found in \cite{Coecke_Kissinger_2017, MonoidalCatsAndTFT}.

\begin{definition}{Monoidal category}{monoidal category}
	A \textbf{monoidal or tensor category} is a category $\cat{C}$ with the following additional data:
	\begin{itemize}
		\item A functor $\otimes \colon \cat{C} \times \cat{C} \to \cat{C}$, called the monoidal product.
		
		\item A special object $\unitObj \in \cat{C}_0$, called the unit object.
		
		\item A natural isomorphism of associativity
		\begin{equation}
			a = (a_{X,Y,Z} \colon (X \otimes Y) \otimes Z \stackrel{\cong}{\to} X \otimes (Y \otimes Z))_{X,Y,Z \in \cat{C}_0}.
		\end{equation}
		
		\item A natural isomorphism of left unitality
		\begin{equation}
			l = (l_X \colon \unitObj \otimes X \stackrel{\cong}{\to} X)_{X \in \cat{C}_0}.
		\end{equation}
		
		\item A natural isomorphism of right unitality
		\begin{equation}
			r = (r_X \colon X \otimes \unitObj \stackrel{\cong}{\to} X)_{X \in \cat{C}_0}.
		\end{equation}
	\end{itemize}
	The functors which are the domains and codomains of these natural isomorphisms won't be of much importance. These functors are defined in \cite{MonoidalCatsAndTFT}, in section 1.2.1, where monoidal categories are defined. We'll just show the diagrams that must commute for them to be natural transformations. That $a$, $l$ and $r$ are natural transformations, it means the following:
	\begin{itemize}
		\item $a$ is natural: For any $f \in \cat{C}(X, Y)$, $f' \in \cat{C}(X', Y')$ and $f'' \in \cat{C}(X'', Y'')$, the next diagram commutes:
		\begin{center}
			\begin{tikzcd}
				(X \otimes X') \otimes X'' \arrow[rr, "(f \otimes f')\otimes f''"] \arrow[d, swap, "a_{X, X', X''}"] & {} &(Y \otimes Y') \otimes Y'' \arrow[d, "a_{Y, Y', Y''}"] \\
				X \otimes (X' \otimes X'') \arrow[rr, "f \otimes (f' \otimes f'')"] & {} & Y \otimes (Y' \otimes Y'')
			\end{tikzcd}
		\end{center}
		
		\item $l$ is natural: For any $f \in \cat{C}(X, Y)$, the next diagram commutes:
		\begin{center}
			\begin{tikzcd}
				\unitObj \otimes X \arrow[r, "\id_{\unitObj} \otimes f"] \arrow[d, swap, "l_X"] & \unitObj \otimes Y \arrow[d, "l_Y"] \\
				X \arrow[r, "f"] & Y
			\end{tikzcd}
		\end{center}
		
		\item $r$ is natural: For any $f \in \cat{C}(X, Y)$, the next diagram commutes:
		\begin{center}
			\begin{tikzcd}
				X \otimes \unitObj \arrow[r, "f \otimes \id_{\unitObj}"] \arrow[d, swap, "r_X"] & Y \otimes \unitObj \arrow[d, "r_Y"] \\
				X \arrow[r, "f"] & Y
			\end{tikzcd}
		\end{center}
	\end{itemize}
	The natural isomorphisms $a$, $l$ and $r$ are required to satisfy some properties, which are the hypothesis used to prove Mac Lane's coherence theorem. Informally, this theorem says that monoidal categories can be treated as if the unitality and associativity isomorphisms were identities. These properties are the following:
	\begin{itemize}
		\item Pentagon coherence: for all objects $X, Y, Z, W \in \cat{C}_0$, the following diagram commutes:
		\begin{center}

			\tikzset{every picture/.style={line width=0.75pt}} 
			
			\begin{tikzpicture}[x=0.75pt,y=0.75pt,yscale=-1,xscale=1]
				
				\draw    (100,90) -- (100,52) ;
				\draw [shift={(100,50)}, rotate = 90] [color={rgb, 255:red, 0; green, 0; blue, 0 }  ][line width=0.75]    (10.93,-4.9) .. controls (6.95,-2.3) and (3.31,-0.67) .. (0,0) .. controls (3.31,0.67) and (6.95,2.3) .. (10.93,4.9)   ;
				\draw    (100,120) -- (128.34,138.89) ;
				\draw [shift={(130,140)}, rotate = 213.69] [color={rgb, 255:red, 0; green, 0; blue, 0 }  ][line width=0.75]    (10.93,-4.9) .. controls (6.95,-2.3) and (3.31,-0.67) .. (0,0) .. controls (3.31,0.67) and (6.95,2.3) .. (10.93,4.9)   ;
				\draw    (240,140) -- (278.21,120.89) ;
				\draw [shift={(280,120)}, rotate = 153.43] [color={rgb, 255:red, 0; green, 0; blue, 0 }  ][line width=0.75]    (10.93,-4.9) .. controls (6.95,-2.3) and (3.31,-0.67) .. (0,0) .. controls (3.31,0.67) and (6.95,2.3) .. (10.93,4.9)   ;
				\draw    (280,50) -- (280,88) ;
				\draw [shift={(280,90)}, rotate = 270] [color={rgb, 255:red, 0; green, 0; blue, 0 }  ][line width=0.75]    (10.93,-4.9) .. controls (6.95,-2.3) and (3.31,-0.67) .. (0,0) .. controls (3.31,0.67) and (6.95,2.3) .. (10.93,4.9)   ;
				\draw    (140,30) -- (228,30) ;
				\draw [shift={(230,30)}, rotate = 180] [color={rgb, 255:red, 0; green, 0; blue, 0 }  ][line width=0.75]    (10.93,-4.9) .. controls (6.95,-2.3) and (3.31,-0.67) .. (0,0) .. controls (3.31,0.67) and (6.95,2.3) .. (10.93,4.9)   ;
				
				\draw (1,92.4) node [anchor=north west][inner sep=0.75pt]    {$(( X\otimes Y) \otimes Z) \otimes W$};
				\draw (1,22.4) node [anchor=north west][inner sep=0.75pt]    {$( X\otimes ( Y\otimes Z)) \otimes W$};
				\draw (239,22.4) node [anchor=north west][inner sep=0.75pt]    {$X\otimes (( Y\otimes Z) \otimes W)$};
				\draw (241,92.4) node [anchor=north west][inner sep=0.75pt]    {$X\otimes ( Y\otimes ( Z\otimes W))$};
				\draw (131,142.4) node [anchor=north west][inner sep=0.75pt]    {$( X\otimes Y) \otimes ( Z\otimes W)$};
				\draw (159,2.4) node [anchor=north west][inner sep=0.75pt]    {$a_{X,Y\otimes Z,W}$};
				\draw (4,60.4) node [anchor=north west][inner sep=0.75pt]    {$a_{X,Y,Z} \otimes id_{W}$};
				\draw (291,60.4) node [anchor=north west][inner sep=0.75pt]    {$id_{X} \otimes a_{Y,Z,W}$};
				\draw (49,130.4) node [anchor=north west][inner sep=0.75pt]    {$a_{X\otimes Y,Z,W}$};
				\draw (282,123.4) node [anchor=north west][inner sep=0.75pt]    {$a_{X,Y,Z\otimes W}$};

			\end{tikzpicture}
			
		\end{center}
		This says that the two ways of constructing an isomorphism
		\begin{equation*}
			((X \otimes Y) \otimes Z) \otimes W \stackrel{\cong}{\to} X \otimes (Y \otimes (Z \otimes W))
		\end{equation*}
		using the associativity isomorphisms give the same isomorphism.
		
		\item Triangle coherence: for all objects $X, Y \in \cat{C}_0$, the following diagram commutes:
		\begin{center}
			\begin{tikzcd}
				(X \otimes \unitObj) \otimes Y \arrow[rr, "a_{X, \unitObj, Y}"] \arrow[rd, swap, "r_X \otimes \id_Y"] & {} & X \otimes (\unitObj \otimes Y) \arrow[ld, "\id_X \otimes l_Y"] \\
				{} & X \otimes Y & {}
			\end{tikzcd}
		\end{center}
		This says that the two ways of constructing an isomorphism
		\begin{equation*}
			(X \otimes \unitObj) \otimes Y \stackrel{\cong}{\to} X \otimes Y
		\end{equation*}
		using the associativity and unitality isomorphisms, one with $r_X$ and the other with $l_Y$, give the same isomorphism.
	\end{itemize}
	There are many other isomorphisms that we could construct using the associativity and unitality isomorphisms. The triangle and pentagon are a minimal set of commutative diagrams to ensure that all those other diagrams also commute, which is the content of Mac Lane's coherence theorem. More details about it can be found in chapter VII.Monoids, in section 2.Coherence of \cite{catswork}.
\end{definition}

In a monoidal category, an object $((\dots(X_1 \otimes X_2) \otimes \dots )\otimes X_{n-1}) \otimes X_n$ is isomorphic to $X_1 \otimes (X_2 \otimes (\dots \otimes (X_{n-1} \otimes X_n) \dots ))$. Any other ordering of the parenthesis also gives an isomorphic object. This is a consequence of Mac Lane's coherence theorem. They are not only isomorphic, but we can construct an isomorphism uniquely, which is the coherence part. For this reason, we typically don't write the parenthesis, as if the category was strict monoidal.

Next, let's continue the study of the category of finite dimensional vector spaces. We already saw that it is a category and that $\otimes$ is a functor, at least for a given choice of tensor product of vector spaces, which can be done using the standard construction from proposition \ref{proposition:tensor product existence general}. Now we prove that $\otimes$ turns it into a monoidal category.

\begin{proposition}{}{finvec as a monoidal category}
	Let $F$ be a field. By proposition \ref{proposition:finvec is a category}, we know that $\finvec_F$ is a category. By \ref{proposition:tensor is a functor}, we know that, given a choice of tensor product for each pair of vectors spaces, the tensor product defines a functor
	\begin{equation*}
		\otimes \colon \finvec_F \times \finvec_F \to \finvec_F .
	\end{equation*}
	We saw that such choice can be made using the standard construction of the tensor product from proposition \ref{proposition:tensor product existence general}. The category $\finvec_F$ is furthermore a \textbf{monoidal category}. Its monoidal structure is given by the following data:
	\begin{itemize}
		\item The monoidal product is the functor $\otimes \colon \finvec_F \times \finvec_F \to \finvec_F$.
		
		\item The unit object is $\unitObj \coloneqq F$.
		
		\item The associativity natural isomorphism 
		\begin{equation*}
			a = (a_{X, Y, Z}\colon (X\otimes Y)\otimes Z \stackrel{\cong}{\to} X \otimes (Y \otimes Z))_{X, Y, Z \in (\finvec_F )_0}
		\end{equation*}
		is given by
		\begin{align*}
			a_{X, Y, Z} \colon (X \otimes Y) \otimes Z &\longrightarrow X \otimes (Y \otimes Z) \\
			(x \otimes y) \otimes z &\longmapsto x \otimes (y \otimes z).
		\end{align*}
		
		\item The left unitality natural isomorphism
		\begin{equation*}
			l = (l_X \colon F \otimes X \stackrel{\cong}{\to} X)_{X \in (\finvec_F)_0}
		\end{equation*} 
		is given by
		\begin{align*}
			l_X \colon F \otimes X &\longrightarrow X \\
			\lambda \otimes x &\longmapsto \lambda x.
		\end{align*}
		
		\item The right unitality natural isomorphism
		\begin{equation*}
			r = (r_X \colon X \otimes F \stackrel{\cong}{\to} X)_{X \in (\finvec_F)_0}
		\end{equation*} 
		is given by
		\begin{align*}
			r_X \colon X \otimes F &\longrightarrow X \\
			x \otimes \lambda &\longmapsto \lambda x.
		\end{align*}
	\end{itemize}
	The isomorphisms $a_{X, Y, Z}$, $l_X$ and $r_X$ were constructed in corollary \ref{corollary:associativity} and in propositions \ref{proposition:left unitality} and \ref{proposition:right unitality}.
\end{proposition}
\begin{proof}
	First let's prove the commutativity of the diagrams that say that $a$, $l$ and $r$ are natural transformations.
	\begin{itemize}
		\item $a$ is natural: let $f \in X \to Y$, $f' \in X' \to Y'$ and $f'' \in X'' \to Y''$ be linear transformations in $\finvec_F$. The naturality diagram is the following:
		\begin{center}
			\begin{tikzcd}
				(X \otimes X') \otimes X'' \arrow[rr, "(f \otimes f')\otimes f''"] \arrow[d, swap, "a_{X, X', X''}"] & {} &(Y \otimes Y') \otimes Y'' \arrow[d, "a_{Y, Y', Y''}"] \\
				X \otimes (X' \otimes X'') \arrow[rr, "f \otimes (f' \otimes f'')"] & {} & Y \otimes (Y' \otimes Y'')
			\end{tikzcd}
		\end{center}
		It commutes iff
		\begin{equation}
			a_{Y, Y', Y''} [(f \otimes f') \otimes f''] = [f \otimes (f' \otimes f'')] a_{X, X', X''}.
		\end{equation}
		They are equal iff they coincide over elementary tensors. Let $x \in X$, $x' \in X'$ and $x'' \in X''$. For the left side, we have that
		\begin{align*}
			a_{Y, Y', Y''} [(f \otimes f') \otimes f''] ((x \otimes x') \otimes x'') &\stackrel{(i)}{=}  a_{Y, Y', Y''} [(f(x) \otimes f'(x')) \otimes f''(x'')] \\
			&\stackrel{(ii)}{=} f(x) \otimes (f(x') \otimes f''(x'')).
		\end{align*}
		In $(i)$ the linear transformations $f$, $f'$ and $f''$ were applied. In $(ii)$ was applied the associativity isomorphism $a_{Y, Y', Y''}$. Similarly for the right side, we have that
		\begin{align*}
			[f \otimes (f' \otimes f'')] a_{X, X', X''}((x \otimes x') \otimes x'') &\stackrel{(i)}{=} [f \otimes (f' \otimes f'')] (x \otimes (x' \otimes x'')) \\
			&\stackrel{(ii)}{=} f(x) \otimes (f'(x') \otimes f''(x'')) .
		\end{align*}
		In $(i)$ was applied the associativity $a_{X, X', X''}$. In $(ii)$ were applied $f$, $f'$ and $f''$. Since
		\begin{equation*}
			a_{Y, Y', Y''} [(f \otimes f') \otimes f''] ((x \otimes x') \otimes x'') = [f \otimes (f' \otimes f'')] a_{X, X', X''}((x \otimes x') \otimes x'')
		\end{equation*}
		for any $x \in X$, $x' \in X'$ and $x'' \in X''$, it follows that
		\begin{equation}
			a_{Y, Y', Y''} [(f \otimes f') \otimes f''] = [f \otimes (f' \otimes f'')] a_{X, X', X''},
		\end{equation}
		so $a$ is a natural transformation.
		
		\item $l$ is natural: let $f \colon X \to Y$ be a linear transformation in $\finvec_F$. The naturality diagram is the following:
		\begin{center}
			\begin{tikzcd}
				F \otimes X \arrow[r, "\id_F \otimes f"] \arrow[d, swap, "l_X"] & F \otimes Y \arrow[d, "l_Y"] \\
				X \arrow[r, "f"] & Y
			\end{tikzcd}
		\end{center}
		It commutes iff
		\begin{equation}
			f l_X = l_Y (\id_F \otimes f).
		\end{equation}
		Let $\lambda \in F$ and $x \in X$. For the left side, we have:
		\begin{align*}
			f l_X(\lambda \otimes x) &\stackrel{(i)}{=} f(\lambda x)
			\\
			&\stackrel{(ii)}{=} \lambda f(x).
		\end{align*}
		In $(i)$ was applied the unitality isomorphism $l_X$. In $(ii)$ was used that $f$ is linear, to pass $\lambda$ to outside of the function. Similarly for the right side, we have:
		\begin{align*}
			l_Y (\id_F \otimes f)(\lambda \otimes x) &\stackrel{(i)}{=} l_Y (\lambda \otimes f(x)) \\
			&\stackrel{(ii)}{=} \lambda f(x) .
		\end{align*}
		In $(i)$ was applied $\id_F \otimes f$. In $(ii)$ was applied $l_Y$. We proved that
		\begin{equation}
			f l_X (\lambda \otimes x) = l_Y (\id_F \otimes f) (\lambda \otimes x)
		\end{equation}
		for any $\lambda \in F$ and $x \in X$. From this it follows that
		\begin{equation}
			f l_X = l_Y (\id_F \otimes f), 
		\end{equation}
		so $l$ is a natural transformation.
		
		\item $r$ is natural: let $f \colon X \to Y$ be a linear transformation in $\finvec_F$. The naturality diagram is the following:
		\begin{center}
			\begin{tikzcd}
				X \otimes F \arrow[r, "f \otimes \id_F"] \arrow[d, swap, "r_X"] & Y \otimes F \arrow[d, "r_Y"] \\
				X \arrow[r, "f"] & Y
			\end{tikzcd}
		\end{center}
		It commutes iff
		\begin{equation}
			f r_X = r_Y (f \otimes \id_F).
		\end{equation}
		Let $\lambda \in F$ and $x \in X$. For the left side, we have:
		\begin{align*}
			f r_X(x \otimes \lambda) &\stackrel{(i)}{=} f(\lambda x)
			\\
			&\stackrel{(ii)}{=} \lambda f(x).
		\end{align*}
		In $(i)$ was applied the unitality isomorphism $r_X$. In $(ii)$ was used that $f$ is linear, to pass $\lambda$ to outside of the function. Similarly for the right side, we have:
		\begin{align*}
			r_Y (f \otimes \id_F)(x \otimes \lambda) &\stackrel{(i)}{=} r_Y ( f(x) \otimes \lambda) \\
			&\stackrel{(ii)}{=} \lambda f(x) .
		\end{align*}
		In $(i)$ was applied $f \otimes \id_F$. In $(ii)$ was applied $r_Y$. We proved that
		\begin{equation}
			f r_X (x \otimes \lambda) = r_Y ( f \otimes \id_F) (x \otimes \lambda)
		\end{equation}
		for any $\lambda \in F$ and $x \in X$. From this it follows that
		\begin{equation}
			f r_X = r_Y ( f \otimes \id_F), 
		\end{equation}
		so $r$ is a natural transformation.
	\end{itemize}
	
	Next we have to prove the pentagon and triangle coherences. The pentagon coherence corresponds to the equation
	\begin{equation}
		(\id_X \otimes a_{Y, Z, W})a_{X, Y\otimes Z, W}(a_{X, Y, Z} \otimes \id_W) = a_{X, Y, Z \otimes W} a_{X \otimes Y, Z, W},
	\end{equation}
	where $X, Y, Z, W \in (\finvec_F)_0$ are arbitrary. Let $x \in X$, $y \in Y$, $z \in Z$ and $w \in W$. For the left side, we have
	\begin{align*}
		(\id_X \otimes a_{Y, Z, W}) a_{X, Y\otimes Z, W} & (a_{X, Y, Z} \otimes \id_W)[((x \otimes y)\otimes z) \otimes w]\\
		&\stackrel{(i)}{=} (\id_X \otimes a_{Y, Z, W})a_{X, Y\otimes Z, W}[(x \otimes (y \otimes z))\otimes w] \\
		&\stackrel{(ii)}{=} (\id_X \otimes a_{Y, Z, W}) [x \otimes ((y\otimes z)\otimes w)] \\
		&\stackrel{(iii)}{=} x \otimes (y \otimes (z \otimes w)) .
	\end{align*}
	In $(i)$ was applied $a_{X, Y, Z} \otimes \id_W$; in $(ii)$ was applied $a_{X, Y\otimes Z, W}$; in $(iii)$ was applied $\id_X \otimes a_{Y, Z, W}$. Similarly for the right side, we have:
	\begin{align*}
		a_{X, Y, Z \otimes W} a_{X \otimes Y, Z, W} [((x \otimes y)\otimes z) \otimes w] &\stackrel{(i)}{=} a_{X, Y, Z \otimes W} [(x \otimes y)\otimes (z \otimes w)]\\
		&\stackrel{(ii)}{=} x \otimes (y\otimes (z \otimes w)) .
	\end{align*}
	Since
	\begin{align*}
		(\id_X \otimes a_{Y, Z, W}) a_{X, Y\otimes Z, W}(a_{X, Y, Z}& \otimes \id_W) [((x \otimes y) \otimes z) \otimes w] = \\
		&a_{X, Y, Z \otimes W} a_{X \otimes Y, Z, W} [((x \otimes y) \otimes z) \otimes w],
	\end{align*}
	for any $x \in X$, $y \in Y$, $z \in Z$ and $w \in W$, it follows that
	\begin{equation}
		(\id_X \otimes a_{Y, Z, W})a_{X, Y\otimes Z, W}(a_{X, Y, Z} \otimes \id_W) = a_{X, Y, Z \otimes W} a_{X \otimes Y, Z, W}.
	\end{equation}
	This proves the pentagon coherence.
	
	Finally, let's prove the triangle coherence. Let $X, Y \in (\finvec_F )_0$. The triangle coherence corresponds to the equation
	\begin{equation}
		(\id_X \otimes l_Y) a_{X, F, Y} = r_X \otimes \id_Y .
	\end{equation}
	Let $x \in X$, $y \in Y$ and $\lambda \in F$. For the left side, we have:
	\begin{align*}
		(\id_X \otimes l_Y) a_{X, F, Y} ((x \otimes \lambda) \otimes y) &\stackrel{(i)}{=} (\id_X \otimes l_Y) (x \otimes (\lambda \otimes y)) \\
		&\stackrel{(ii)}{=} x \otimes (\lambda y) \\
		&\stackrel{(iii)}{=} \lambda (x \otimes y).
	\end{align*}
	In $(i)$ was applied $a_{X, F, Y}$; in $(ii)$ was applied $\id_X \otimes l_Y$; in $(ii)$ was used the bilinearity of $\otimes$ to pass $\lambda$ to outside. Similarly for the right side, we have:
	\begin{align*}
		(r_X \otimes \id_Y) ((x \otimes \lambda)\otimes y) &\stackrel{(i)}{=} (\lambda x) \otimes y \\
		&\stackrel{(ii)}{=} \lambda (x \otimes y).
	\end{align*}
	In $(i)$ was applied $r_X \otimes \id_Y$; in $(ii)$ was used the bilinearity of $\otimes$ to pass $\lambda$ to outside. Since
	\begin{equation}
		(\id_X \otimes l_Y) a_{X, F, Y} ((x \otimes \lambda) \otimes y) = (r_X \otimes \id_Y) ((x \otimes \lambda)\otimes y)
	\end{equation}
	for any $x \in X$, $y \in Y$ and $\lambda \in F$, it follows that
	\begin{equation}
		(\id_X \otimes l_Y) a_{X, F, Y} = r_X \otimes \id_Y .
	\end{equation}
	This proves the triangle coherence and concludes the proof that $\finvec_F$ is a monoidal category.
\end{proof}

\section{String diagrams}
\label{section: string diagrams}

We have a convenient diagrammatic notation for the arrows in a monoidal category, which are the string diagrams. String diagrams treat the monoidal category as if it was strict, that is, as if the associativity isomorphisms $a_{X,Y,Z}$ and unitality isomorphisms $l_X$ and $r_X$ were identity arrows. This is possible due to Mac Lane's coherence theorem.  More details about string diagrams can be found in \cite{MonoidalCatsAndTFT, Turaev+2016, Coecke_Kissinger_2017}. Details about Mac Lane's coherence theorem can be found in \cite{catswork}. Here will be given a brief explanation of what are string diagrams and how to use them. The details of why string diagrams are correct won't be explained here. For this reason, they also won't be used often in this work. The string diagrams will follow closely the ones from \cite{MonoidalCatsAndTFT}, but with the difference that the standard string orientation will be upwards instead of downwards.

\subsection{String diagrams for the category of finite dimensional vector spaces}

Many types of monoidal categories have some type of string diagram notation. Instead of introducing each type of monoidal category and its string diagrams, we'll rather show how string diagrams can be used specifically for the category $\finvec_F$, for some field $F$. This is because of the heavy use of $\finvec_F$ in Quantum Computation and Quantum Information, at least for the Quantum Circuit model. Many string diagrams can be made in the same way for vector spaces of any dimension, or even for modules. The finite dimensions become important when dealing with duals. As seen in proposition \ref{proposition:ev is non degenerate pairing}, every finite dimensional space $X$ has a non-degenerate pairing $\lev_X \colon X^* \otimes X \to F$. The pairing have a kind of inverse $\lcoev_X \colon F \to X \otimes X^*$, in the sense of proposition \ref{proposition:ev is non degenerate pairing}. It can be computed from a basis, as in definition \ref{definition:coevaluations}. This definition doesn't work in infinite dimensions, because it would require an infinite summation of orthogonal vectors whose norms don't converge to zero. Such summation necessarily diverges. This is one of the reasons for restricting to the finite dimensional case.

We typically write a mathematical expression as a sequence of symbols. In a monoidal category we have two operations, which are the composition ($\circ$) and the monoidal product ($\otimes$). The most important idea in string diagrams is to not write expressions as a sequence, but rather to use different orthogonal directions for different operations. Doing so could introduce ambiguity on the meaning of an expression, but the functoriality of $\otimes$ will guarantee that the result doesn't depend on how we interpret the string diagram.

Let $F$ be a field. We are mostly interested in the cases where $F = \realNumbers$ or $F = \complexNumbers$. A linear transformation $f \in \finvec_F (X, Y)$ has the following string diagram representation:
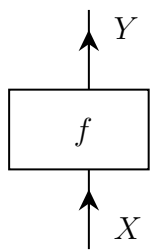
\begin{center}
	\tikzset{every picture/.style={line width=0.75pt}} 
	
	\begin{tikzpicture}[x=0.75pt,y=0.75pt,yscale=-1,xscale=1]
		
		\draw   (0,40) -- (70,40) -- (70,80) -- (0,80) -- cycle ;
		\draw    (40,0) -- (40,40) ;
		\draw    (40,20) -- (40,13) ;
		\draw [shift={(40,10)}, rotate = 90] [fill={rgb, 255:red, 0; green, 0; blue, 0 }  ][line width=0.08]  [draw opacity=0] (10.72,-5.15) -- (0,0) -- (10.72,5.15) -- (7.12,0) -- cycle    ;
		\draw    (40,80) -- (40,120) ;
		\draw    (40,100) -- (40,93) ;
		\draw [shift={(40,90)}, rotate = 90] [fill={rgb, 255:red, 0; green, 0; blue, 0 }  ][line width=0.08]  [draw opacity=0] (10.72,-5.15) -- (0,0) -- (10.72,5.15) -- (7.12,0) -- cycle    ;
		
		\draw (31,52.4) node [anchor=north west][inner sep=0.75pt]    {$f$};
		\draw (51,2.4) node [anchor=north west][inner sep=0.75pt]    {$Y$};
		\draw (51,102.4) node [anchor=north west][inner sep=0.75pt]    {$X$};

	\end{tikzpicture}
\end{center}
The linear transformation $f$ is represented by a box decorated with the symbol $f$. Some authors say colored instead of decorated. The domain and codomain of $f$ are represented diagrammatically by the strings attached to the box. The strings attached from below represent the domain, and the strings attached from above represent the codomain. The orientations of the strings are related to the dual. The standard orientation is upwards, the downwards orientation will be explained later.

Composition is represented by vertical concatenation. That is, the composition of linear transformations $f \in \finvec_F (X, Y)$ and $g \in \finvec_F (Y, Z)$ is represented as
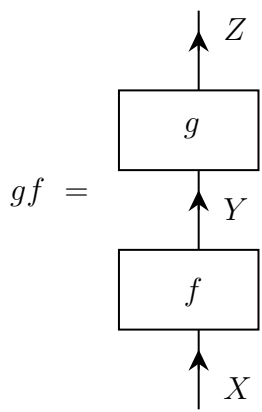
\begin{center}

	\tikzset{every picture/.style={line width=0.75pt}} 
	
	\begin{tikzpicture}[x=0.75pt,y=0.75pt,yscale=-1,xscale=1]
		
		\draw   (60,40) -- (130,40) -- (130,80) -- (60,80) -- cycle ;
		\draw    (100,0) -- (100,40) ;
		\draw    (100,20) -- (100,13) ;
		\draw [shift={(100,10)}, rotate = 90] [fill={rgb, 255:red, 0; green, 0; blue, 0 }  ][line width=0.08]  [draw opacity=0] (10.72,-5.15) -- (0,0) -- (10.72,5.15) -- (7.12,0) -- cycle    ;
		\draw    (100,80) -- (100,120) ;
		\draw    (100,100) -- (100,93) ;
		\draw [shift={(100,90)}, rotate = 90] [fill={rgb, 255:red, 0; green, 0; blue, 0 }  ][line width=0.08]  [draw opacity=0] (10.72,-5.15) -- (0,0) -- (10.72,5.15) -- (7.12,0) -- cycle    ;
		\draw   (60,120) -- (130,120) -- (130,160) -- (60,160) -- cycle ;
		\draw    (100,160) -- (100,200) ;
		\draw    (100,180) -- (100,173) ;
		\draw [shift={(100,170)}, rotate = 90] [fill={rgb, 255:red, 0; green, 0; blue, 0 }  ][line width=0.08]  [draw opacity=0] (10.72,-5.15) -- (0,0) -- (10.72,5.15) -- (7.12,0) -- cycle    ;
		
		\draw (91,52.4) node [anchor=north west][inner sep=0.75pt]    {$g$};
		\draw (111,2.4) node [anchor=north west][inner sep=0.75pt]    {$Z$};
		\draw (111,92.4) node [anchor=north west][inner sep=0.75pt]    {$Y$};
		\draw (91,132.4) node [anchor=north west][inner sep=0.75pt]    {$f$};
		\draw (111,182.4) node [anchor=north west][inner sep=0.75pt]    {$X$};
		\draw (4,82.4) node [anchor=north west][inner sep=0.75pt]    {$gf\ =$};

	\end{tikzpicture}
\end{center}
As such, identity functions are conveniently represented as a string without a box:
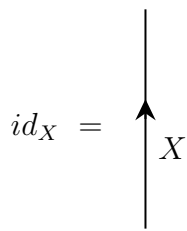
\begin{center}

	\tikzset{every picture/.style={line width=0.75pt}} 
	
	\begin{tikzpicture}[x=0.75pt,y=0.75pt,yscale=-1,xscale=1]
		
		\draw    (70,0) -- (70,110) ;
		\draw    (70,55) -- (70,48) ;
		\draw [shift={(70,45)}, rotate = 90] [fill={rgb, 255:red, 0; green, 0; blue, 0 }  ][line width=0.08]  [draw opacity=0] (10.72,-5.15) -- (0,0) -- (10.72,5.15) -- (7.12,0) -- cycle    ;
		
		\draw (75,62.4) node [anchor=north west][inner sep=0.75pt]    {$X$};
		\draw (1,50.4) node [anchor=north west][inner sep=0.75pt]    {$id_{X} \ =$};

	\end{tikzpicture}
\end{center}

The tensor product is represented by horizontal concatenation. If $f \in \finvec_F (X, Y)$ and $g \in \finvec_F (Z, W)$, then $f\otimes g \colon X \otimes Z \to Y \otimes W$ is represented as
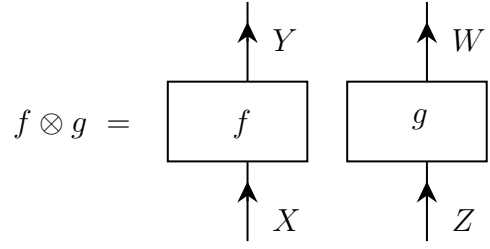
\begin{center}

	\tikzset{every picture/.style={line width=0.75pt}} 
	
	\begin{tikzpicture}[x=0.75pt,y=0.75pt,yscale=-1,xscale=1]
		
		\draw    (120,0) -- (120,40) ;
		\draw    (120,20) -- (120,13) ;
		\draw [shift={(120,10)}, rotate = 90] [fill={rgb, 255:red, 0; green, 0; blue, 0 }  ][line width=0.08]  [draw opacity=0] (10.72,-5.15) -- (0,0) -- (10.72,5.15) -- (7.12,0) -- cycle    ;
		\draw   (80,40) -- (150,40) -- (150,80) -- (80,80) -- cycle ;
		\draw    (120,80) -- (120,120) ;
		\draw    (120,100) -- (120,93) ;
		\draw [shift={(120,90)}, rotate = 90] [fill={rgb, 255:red, 0; green, 0; blue, 0 }  ][line width=0.08]  [draw opacity=0] (10.72,-5.15) -- (0,0) -- (10.72,5.15) -- (7.12,0) -- cycle    ;
		\draw   (170,40) -- (240,40) -- (240,80) -- (170,80) -- cycle ;
		\draw    (210,0) -- (210,40) ;
		\draw    (210,20) -- (210,13) ;
		\draw [shift={(210,10)}, rotate = 90] [fill={rgb, 255:red, 0; green, 0; blue, 0 }  ][line width=0.08]  [draw opacity=0] (10.72,-5.15) -- (0,0) -- (10.72,5.15) -- (7.12,0) -- cycle    ;
		\draw    (210,80) -- (210,120) ;
		\draw    (210,100) -- (210,93) ;
		\draw [shift={(210,90)}, rotate = 90] [fill={rgb, 255:red, 0; green, 0; blue, 0 }  ][line width=0.08]  [draw opacity=0] (10.72,-5.15) -- (0,0) -- (10.72,5.15) -- (7.12,0) -- cycle    ;
		
		\draw (131,12.4) node [anchor=north west][inner sep=0.75pt]    {$Y$};
		\draw (111,52.4) node [anchor=north west][inner sep=0.75pt]    {$f$};
		\draw (131,102.4) node [anchor=north west][inner sep=0.75pt]    {$X$};
		\draw (201,52.4) node [anchor=north west][inner sep=0.75pt]    {$g$};
		\draw (221,12.4) node [anchor=north west][inner sep=0.75pt]    {$W$};
		\draw (221,102.4) node [anchor=north west][inner sep=0.75pt]    {$Z$};
		\draw (1,52.4) node [anchor=north west][inner sep=0.75pt]    {$f\otimes g\ =$};

	\end{tikzpicture}
\end{center}
Mac Lane's coherence theorem guarantees that taking the tensor product in different orders give isomorphic results, and furthermore we can construct a unique isomorphism relating two results. For this reason, the tensor product is treated in string diagrams as if it was strict. As such, we won't use parenthesis to indicate the order in which the tensor product should be taken, and also typically won't write the associativity isomorphism $a_{X,Y,Z}$ in the string diagrams. It doesn't mean that the isomorphism can be completely discarded, but that it can be put back in a consistent manner, no matter how a string diagram is interpreted.

In proposition \ref{proposition:tensor is a functor} was shown that $\otimes$ is a functor. One of the conditions for being a functor is equation \ref{equation: tensor respects composition}, which says the following. For any linear transformations $f \colon X \to Y$, $g \colon Y \to Z$, $f'\colon X' \to Y'$ and $g' \colon Y' \to Z'$, we have that
\begin{equation*}
	(g f) \otimes (g' f') = (g \otimes g')(f \otimes f').
\end{equation*}
For string diagrams, this means that both ways of interpreting the following diagram are equivalent:
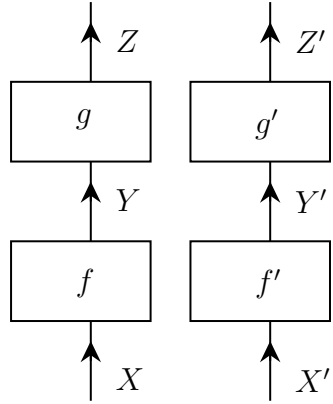
\begin{center}

	\tikzset{every picture/.style={line width=0.75pt}} 
	
	\begin{tikzpicture}[x=0.75pt,y=0.75pt,yscale=-1,xscale=1]
		
		\draw    (40,0) -- (40,40) ;
		\draw    (40,20) -- (40,13) ;
		\draw [shift={(40,10)}, rotate = 90] [fill={rgb, 255:red, 0; green, 0; blue, 0 }  ][line width=0.08]  [draw opacity=0] (10.72,-5.15) -- (0,0) -- (10.72,5.15) -- (7.12,0) -- cycle    ;
		\draw   (0,40) -- (70,40) -- (70,80) -- (0,80) -- cycle ;
		\draw    (40,80) -- (40,120) ;
		\draw    (40,100) -- (40,93) ;
		\draw [shift={(40,90)}, rotate = 90] [fill={rgb, 255:red, 0; green, 0; blue, 0 }  ][line width=0.08]  [draw opacity=0] (10.72,-5.15) -- (0,0) -- (10.72,5.15) -- (7.12,0) -- cycle    ;
		\draw   (90,40) -- (160,40) -- (160,80) -- (90,80) -- cycle ;
		\draw    (130,0) -- (130,40) ;
		\draw    (130,20) -- (130,13) ;
		\draw [shift={(130,10)}, rotate = 90] [fill={rgb, 255:red, 0; green, 0; blue, 0 }  ][line width=0.08]  [draw opacity=0] (10.72,-5.15) -- (0,0) -- (10.72,5.15) -- (7.12,0) -- cycle    ;
		\draw    (130,80) -- (130,120) ;
		\draw    (130,100) -- (130,93) ;
		\draw [shift={(130,90)}, rotate = 90] [fill={rgb, 255:red, 0; green, 0; blue, 0 }  ][line width=0.08]  [draw opacity=0] (10.72,-5.15) -- (0,0) -- (10.72,5.15) -- (7.12,0) -- cycle    ;
		\draw   (0,120) -- (70,120) -- (70,160) -- (0,160) -- cycle ;
		\draw    (40,160) -- (40,200) ;
		\draw    (40,180) -- (40,173) ;
		\draw [shift={(40,170)}, rotate = 90] [fill={rgb, 255:red, 0; green, 0; blue, 0 }  ][line width=0.08]  [draw opacity=0] (10.72,-5.15) -- (0,0) -- (10.72,5.15) -- (7.12,0) -- cycle    ;
		\draw   (90,120) -- (160,120) -- (160,160) -- (90,160) -- cycle ;
		\draw    (130,160) -- (130,200) ;
		\draw    (130,180) -- (130,173) ;
		\draw [shift={(130,170)}, rotate = 90] [fill={rgb, 255:red, 0; green, 0; blue, 0 }  ][line width=0.08]  [draw opacity=0] (10.72,-5.15) -- (0,0) -- (10.72,5.15) -- (7.12,0) -- cycle    ;
		
		\draw (51,12.4) node [anchor=north west][inner sep=0.75pt]    {$Z$};
		\draw (31,52.4) node [anchor=north west][inner sep=0.75pt]    {$g$};
		\draw (51,92.4) node [anchor=north west][inner sep=0.75pt]    {$Y$};
		\draw (121,52.4) node [anchor=north west][inner sep=0.75pt]    {$g'$};
		\draw (141,12.4) node [anchor=north west][inner sep=0.75pt]    {$Z'$};
		\draw (141,92.4) node [anchor=north west][inner sep=0.75pt]    {$Y'$};
		\draw (31,132.4) node [anchor=north west][inner sep=0.75pt]    {$f$};
		\draw (51,182.4) node [anchor=north west][inner sep=0.75pt]    {$X$};
		\draw (121,132.4) node [anchor=north west][inner sep=0.75pt]    {$f'$};
		\draw (141,182.4) node [anchor=north west][inner sep=0.75pt]    {$X'$};

	\end{tikzpicture}
\end{center}
That is, we can interpret this diagram as first concatenating vertically to obtain $gf$ and $g'f'$, and then concatenating horizontally to obtain $(gf)\otimes (g'f')$; or we can interpret it as first concatenating horizontally to obtain $g\otimes g'$ and $f \otimes f'$, and then concatenating vertically to obtain $(g\otimes g')(f \otimes f')$.

Another condition for $\otimes$ to be a functor is equation \ref{equation: tensor preserves identities}, which says that
\begin{equation*}
	\id_{X \otimes Y} = \id_X \otimes \id_Y.
\end{equation*}  
This gives the following diagrammatic equation:
\begin{center}
	\tikzset{every picture/.style={line width=0.75pt}} 
	
	\begin{tikzpicture}[x=0.75pt,y=0.75pt,yscale=-1,xscale=1]
		
		\draw    (10,0) -- (10,110) ;
		\draw    (10,55) -- (10,48) ;
		\draw [shift={(10,45)}, rotate = 90] [fill={rgb, 255:red, 0; green, 0; blue, 0 }  ][line width=0.08]  [draw opacity=0] (10.72,-5.15) -- (0,0) -- (10.72,5.15) -- (7.12,0) -- cycle    ;
		\draw    (120,0) -- (120,110) ;
		\draw    (120,55) -- (120,48) ;
		\draw [shift={(120,45)}, rotate = 90] [fill={rgb, 255:red, 0; green, 0; blue, 0 }  ][line width=0.08]  [draw opacity=0] (10.72,-5.15) -- (0,0) -- (10.72,5.15) -- (7.12,0) -- cycle    ;
		\draw    (170,0) -- (170,110) ;
		\draw    (170,55) -- (170,48) ;
		\draw [shift={(170,45)}, rotate = 90] [fill={rgb, 255:red, 0; green, 0; blue, 0 }  ][line width=0.08]  [draw opacity=0] (10.72,-5.15) -- (0,0) -- (10.72,5.15) -- (7.12,0) -- cycle    ;
		
		\draw (16,62.4) node [anchor=north west][inner sep=0.75pt]    {$X\otimes Y$};
		\draw (76,52.4) node [anchor=north west][inner sep=0.75pt]    {$=$};
		\draw (125,62.4) node [anchor=north west][inner sep=0.75pt]    {$X$};
		\draw (175,62.4) node [anchor=north west][inner sep=0.75pt]    {$Y$};

	\end{tikzpicture}
\end{center}
This diagram leads naturally to the following convention. If a string is oriented upwards and is decorated by a tensor product, then we can represent such tensor product as horizontal concatenation of strings. For any linear transformation $f \in  \finvec_F (X_1 \otimes \dots \otimes X_m, Y_1 \otimes \dots \otimes Y_n)$, we can represent it diagrammatically as
\begin{center}

	\tikzset{every picture/.style={line width=0.75pt}} 
	
	\begin{tikzpicture}[x=0.75pt,y=0.75pt,yscale=-1,xscale=1]
		
		\draw   (218,40) -- (288,40) -- (288,80) -- (218,80) -- cycle ;
		\draw    (228,0) -- (228,40) ;
		\draw    (228,20) -- (228,13) ;
		\draw [shift={(228,10)}, rotate = 90] [fill={rgb, 255:red, 0; green, 0; blue, 0 }  ][line width=0.08]  [draw opacity=0] (10.72,-5.15) -- (0,0) -- (10.72,5.15) -- (7.12,0) -- cycle    ;
		\draw    (278,0) -- (278,40) ;
		\draw    (278,20) -- (278,13) ;
		\draw [shift={(278,10)}, rotate = 90] [fill={rgb, 255:red, 0; green, 0; blue, 0 }  ][line width=0.08]  [draw opacity=0] (10.72,-5.15) -- (0,0) -- (10.72,5.15) -- (7.12,0) -- cycle    ;
		\draw    (228,80) -- (228,120) ;
		\draw    (228,100) -- (228,93) ;
		\draw [shift={(228,90)}, rotate = 90] [fill={rgb, 255:red, 0; green, 0; blue, 0 }  ][line width=0.08]  [draw opacity=0] (10.72,-5.15) -- (0,0) -- (10.72,5.15) -- (7.12,0) -- cycle    ;
		\draw    (278,80) -- (278,120) ;
		\draw    (278,100) -- (278,93) ;
		\draw [shift={(278,90)}, rotate = 90] [fill={rgb, 255:red, 0; green, 0; blue, 0 }  ][line width=0.08]  [draw opacity=0] (10.72,-5.15) -- (0,0) -- (10.72,5.15) -- (7.12,0) -- cycle    ;
		\draw   (0,40) -- (70,40) -- (70,80) -- (0,80) -- cycle ;
		\draw    (40,0) -- (40,40) ;
		\draw    (40,20) -- (40,13) ;
		\draw [shift={(40,10)}, rotate = 90] [fill={rgb, 255:red, 0; green, 0; blue, 0 }  ][line width=0.08]  [draw opacity=0] (10.72,-5.15) -- (0,0) -- (10.72,5.15) -- (7.12,0) -- cycle    ;
		\draw    (40,80) -- (40,120) ;
		\draw    (40,100) -- (40,93) ;
		\draw [shift={(40,90)}, rotate = 90] [fill={rgb, 255:red, 0; green, 0; blue, 0 }  ][line width=0.08]  [draw opacity=0] (10.72,-5.15) -- (0,0) -- (10.72,5.15) -- (7.12,0) -- cycle    ;
		
		\draw (249,52.4) node [anchor=north west][inner sep=0.75pt]    {$f$};
		\draw (199,2.4) node [anchor=north west][inner sep=0.75pt]    {$Y_{1}$};
		\draw (287,2.4) node [anchor=north west][inner sep=0.75pt]    {$Y_{n}$};
		\draw (239,2.4) node [anchor=north west][inner sep=0.75pt]    {$\dotsc $};
		\draw (199,100.4) node [anchor=north west][inner sep=0.75pt]    {$X_{1}$};
		\draw (289,102.4) node [anchor=north west][inner sep=0.75pt]    {$X_{m}$};
		\draw (239,102.4) node [anchor=north west][inner sep=0.75pt]    {$\dotsc $};
		\draw (31,52.4) node [anchor=north west][inner sep=0.75pt]    {$f$};
		\draw (51,2.4) node [anchor=north west][inner sep=0.75pt]    {$Y_{1} \otimes \dotsc \otimes Y_{n}$};
		\draw (161,52.4) node [anchor=north west][inner sep=0.75pt]    {$=$};
		\draw (51,100.4) node [anchor=north west][inner sep=0.75pt]    {$X_{1} \otimes \dotsc \otimes X_{m}$};

	\end{tikzpicture}
\end{center}

Mac Lane's coherence theorem was used to justify treating the tensor product as strictly associative. We also have the unitality isomorphisms defined in propositions \ref{proposition:left unitality} and \ref{proposition:right unitality}. For any finite dimensional $F$-vector space $X$, we have a left unitality isomorphism
\begin{equation*}
	l_X \colon F \otimes X \to X,
\end{equation*}
and a right unitality isomorphism
\begin{equation*}
	r_X \colon X \otimes F \to X.
\end{equation*}
As done with the associativity isomorphism, we typically won't write the unitality isomorphism $l_X$ and $r_X$ in the string diagrams. Again, this doesn't mean that the isomorphisms can be discarded, but that we can restore the isomorphisms in a consistent manner, no matter how the string diagram is interpreted. 
Mac Lane's coherence theorem lets us treat these isomorphisms as if they were identities, that is, as if
\begin{equation*}
	F \otimes X = X = X \otimes F.
\end{equation*}
Diagrammatically, this corresponds to being allowed to attach new upwards oriented strings decorated with $F$, or removing any upwards oriented string decorated with $F$. For example, for any linear transformations $f \in \finvec_F (F,$ $X)$, $g \in \finvec_F (X, F)$ and $h \in \finvec(F, F)$, we can write the following:
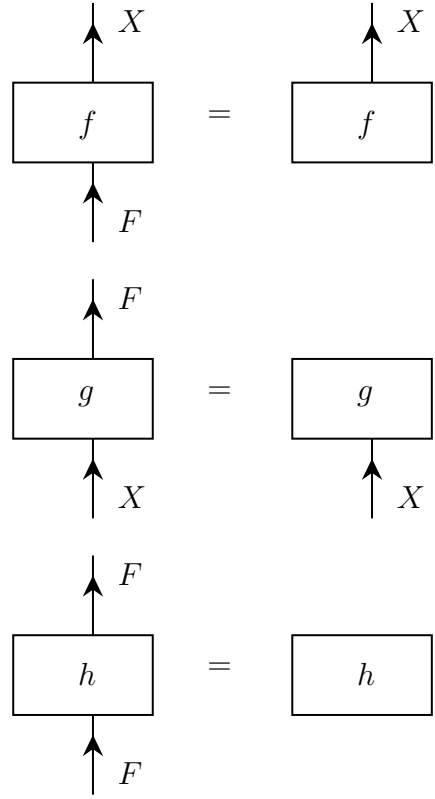
\begin{center}
	\tikzset{every picture/.style={line width=0.75pt}} 
	
	\begin{tikzpicture}[x=0.75pt,y=0.75pt,yscale=-1,xscale=1]
		
		\draw   (0,40) -- (70,40) -- (70,80) -- (0,80) -- cycle ;
		\draw    (40,0) -- (40,40) ;
		\draw    (40,20) -- (40,13) ;
		\draw [shift={(40,10)}, rotate = 90] [fill={rgb, 255:red, 0; green, 0; blue, 0 }  ][line width=0.08]  [draw opacity=0] (10.72,-5.15) -- (0,0) -- (10.72,5.15) -- (7.12,0) -- cycle    ;
		\draw    (40,80) -- (40,120) ;
		\draw    (40,100) -- (40,93) ;
		\draw [shift={(40,90)}, rotate = 90] [fill={rgb, 255:red, 0; green, 0; blue, 0 }  ][line width=0.08]  [draw opacity=0] (10.72,-5.15) -- (0,0) -- (10.72,5.15) -- (7.12,0) -- cycle    ;
		\draw   (140,40) -- (210,40) -- (210,80) -- (140,80) -- cycle ;
		\draw    (180,0) -- (180,40) ;
		\draw    (180,20) -- (180,13) ;
		\draw [shift={(180,10)}, rotate = 90] [fill={rgb, 255:red, 0; green, 0; blue, 0 }  ][line width=0.08]  [draw opacity=0] (10.72,-5.15) -- (0,0) -- (10.72,5.15) -- (7.12,0) -- cycle    ;
		
		\draw (31,52.4) node [anchor=north west][inner sep=0.75pt]    {$f$};
		\draw (51,2.4) node [anchor=north west][inner sep=0.75pt]    {$X$};
		\draw (51,102.4) node [anchor=north west][inner sep=0.75pt]    {$F$};
		\draw (96,52.4) node [anchor=north west][inner sep=0.75pt]    {$=$};
		\draw (171,52.4) node [anchor=north west][inner sep=0.75pt]    {$f$};
		\draw (191,2.4) node [anchor=north west][inner sep=0.75pt]    {$X$};

	\end{tikzpicture}
\end{center}

\begin{center}

	\tikzset{every picture/.style={line width=0.75pt}} 
	
	\begin{tikzpicture}[x=0.75pt,y=0.75pt,yscale=-1,xscale=1]
		
		\draw   (0,40) -- (70,40) -- (70,80) -- (0,80) -- cycle ;
		\draw    (40,0) -- (40,40) ;
		\draw    (40,20) -- (40,13) ;
		\draw [shift={(40,10)}, rotate = 90] [fill={rgb, 255:red, 0; green, 0; blue, 0 }  ][line width=0.08]  [draw opacity=0] (10.72,-5.15) -- (0,0) -- (10.72,5.15) -- (7.12,0) -- cycle    ;
		\draw    (40,80) -- (40,120) ;
		\draw    (40,100) -- (40,93) ;
		\draw [shift={(40,90)}, rotate = 90] [fill={rgb, 255:red, 0; green, 0; blue, 0 }  ][line width=0.08]  [draw opacity=0] (10.72,-5.15) -- (0,0) -- (10.72,5.15) -- (7.12,0) -- cycle    ;
		\draw   (140,40) -- (210,40) -- (210,80) -- (140,80) -- cycle ;
		\draw    (180,80) -- (180,120) ;
		\draw    (180,100) -- (180,93) ;
		\draw [shift={(180,90)}, rotate = 90] [fill={rgb, 255:red, 0; green, 0; blue, 0 }  ][line width=0.08]  [draw opacity=0] (10.72,-5.15) -- (0,0) -- (10.72,5.15) -- (7.12,0) -- cycle    ;
		
		\draw (31,52.4) node [anchor=north west][inner sep=0.75pt]    {$g$};
		\draw (51,2.4) node [anchor=north west][inner sep=0.75pt]    {$F$};
		\draw (51,102.4) node [anchor=north west][inner sep=0.75pt]    {$X$};
		\draw (96,52.4) node [anchor=north west][inner sep=0.75pt]    {$=$};
		\draw (171,52.4) node [anchor=north west][inner sep=0.75pt]    {$g$};
		\draw (191,102.4) node [anchor=north west][inner sep=0.75pt]    {$X$};

	\end{tikzpicture}
\end{center}

\begin{center}

	\tikzset{every picture/.style={line width=0.75pt}} 
	
	\begin{tikzpicture}[x=0.75pt,y=0.75pt,yscale=-1,xscale=1]
		
		\draw   (0,40) -- (70,40) -- (70,80) -- (0,80) -- cycle ;
		\draw    (40,0) -- (40,40) ;
		\draw    (40,20) -- (40,13) ;
		\draw [shift={(40,10)}, rotate = 90] [fill={rgb, 255:red, 0; green, 0; blue, 0 }  ][line width=0.08]  [draw opacity=0] (10.72,-5.15) -- (0,0) -- (10.72,5.15) -- (7.12,0) -- cycle    ;
		\draw    (40,80) -- (40,120) ;
		\draw    (40,100) -- (40,93) ;
		\draw [shift={(40,90)}, rotate = 90] [fill={rgb, 255:red, 0; green, 0; blue, 0 }  ][line width=0.08]  [draw opacity=0] (10.72,-5.15) -- (0,0) -- (10.72,5.15) -- (7.12,0) -- cycle    ;
		\draw   (140,40) -- (210,40) -- (210,80) -- (140,80) -- cycle ;
		
		\draw (31,52.4) node [anchor=north west][inner sep=0.75pt]    {$h$};
		\draw (51,2.4) node [anchor=north west][inner sep=0.75pt]    {$F$};
		\draw (51,102.4) node [anchor=north west][inner sep=0.75pt]    {$F$};
		\draw (96,52.4) node [anchor=north west][inner sep=0.75pt]    {$=$};
		\draw (171,52.4) node [anchor=north west][inner sep=0.75pt]    {$h$};

	\end{tikzpicture}
\end{center}
We can also attach new upwards oriented strings decorated with $F$. For any linear transformation $f \in \finvec_F (X,Y)$, we have that
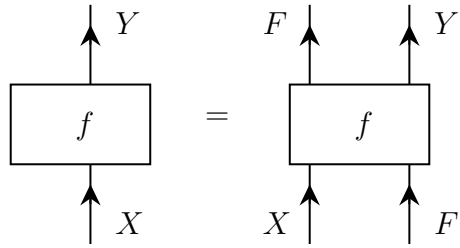
\begin{center}

	\tikzset{every picture/.style={line width=0.75pt}} 
	
	\begin{tikzpicture}[x=0.75pt,y=0.75pt,yscale=-1,xscale=1]
		
		\draw   (0,40) -- (70,40) -- (70,80) -- (0,80) -- cycle ;
		\draw    (40,0) -- (40,40) ;
		\draw    (40,20) -- (40,13) ;
		\draw [shift={(40,10)}, rotate = 90] [fill={rgb, 255:red, 0; green, 0; blue, 0 }  ][line width=0.08]  [draw opacity=0] (10.72,-5.15) -- (0,0) -- (10.72,5.15) -- (7.12,0) -- cycle    ;
		\draw    (40,80) -- (40,120) ;
		\draw    (40,100) -- (40,93) ;
		\draw [shift={(40,90)}, rotate = 90] [fill={rgb, 255:red, 0; green, 0; blue, 0 }  ][line width=0.08]  [draw opacity=0] (10.72,-5.15) -- (0,0) -- (10.72,5.15) -- (7.12,0) -- cycle    ;
		\draw   (140,40) -- (210,40) -- (210,80) -- (140,80) -- cycle ;
		\draw    (200,0) -- (200,40) ;
		\draw    (200,20) -- (200,13) ;
		\draw [shift={(200,10)}, rotate = 90] [fill={rgb, 255:red, 0; green, 0; blue, 0 }  ][line width=0.08]  [draw opacity=0] (10.72,-5.15) -- (0,0) -- (10.72,5.15) -- (7.12,0) -- cycle    ;
		\draw    (150,80) -- (150,120) ;
		\draw    (150,100) -- (150,93) ;
		\draw [shift={(150,90)}, rotate = 90] [fill={rgb, 255:red, 0; green, 0; blue, 0 }  ][line width=0.08]  [draw opacity=0] (10.72,-5.15) -- (0,0) -- (10.72,5.15) -- (7.12,0) -- cycle    ;
		\draw    (150,0) -- (150,40) ;
		\draw    (150,20) -- (150,13) ;
		\draw [shift={(150,10)}, rotate = 90] [fill={rgb, 255:red, 0; green, 0; blue, 0 }  ][line width=0.08]  [draw opacity=0] (10.72,-5.15) -- (0,0) -- (10.72,5.15) -- (7.12,0) -- cycle    ;
		\draw    (200,80) -- (200,120) ;
		\draw    (200,100) -- (200,93) ;
		\draw [shift={(200,90)}, rotate = 90] [fill={rgb, 255:red, 0; green, 0; blue, 0 }  ][line width=0.08]  [draw opacity=0] (10.72,-5.15) -- (0,0) -- (10.72,5.15) -- (7.12,0) -- cycle    ;
		
		\draw (31,52.4) node [anchor=north west][inner sep=0.75pt]    {$f$};
		\draw (51,2.4) node [anchor=north west][inner sep=0.75pt]    {$Y$};
		\draw (51,102.4) node [anchor=north west][inner sep=0.75pt]    {$X$};
		\draw (96,52.4) node [anchor=north west][inner sep=0.75pt]    {$=$};
		\draw (171,52.4) node [anchor=north west][inner sep=0.75pt]    {$f$};
		\draw (211,2.4) node [anchor=north west][inner sep=0.75pt]    {$Y$};
		\draw (125,102.4) node [anchor=north west][inner sep=0.75pt]    {$X$};
		\draw (125,2.4) node [anchor=north west][inner sep=0.75pt]    {$F$};
		\draw (211,102.4) node [anchor=north west][inner sep=0.75pt]    {$F$};

	\end{tikzpicture}
\end{center}
As said earlier, the associativity and unitality isomorphisms are typically implicit in these diagrams. The right side of this last diagrammatic equation can be interpreted as the linear transformation
\begin{equation*}
	l_Y^{-1} f r_X  \colon X \otimes F \to F \otimes Y .
\end{equation*}
Of course, $f$ may not be equal to $l_Y^{-1} f r_X$, but they differ only by the associativity and unitality isomorphisms. In other words, \textbf{the diagrammatic equatility isn't strict, rather it is equality up to associativity and unitality isomorphisms}. Because of Mac Lane's coherence theorem, strict equality can be obtained by putting associativity and unitality isomorphism such that both sides of the equation have same domain and codomain.

Until this point we only used upwards oriented strings. The orientation of the strings are related to the dual. Upwards is the standard orientation, for which nothing special is required. A downwards oriented string means that we have to take the dual of the space. For example, if $f \in \finvec(X^* \otimes Y, Z \otimes W^*)$, we have that
\begin{center}

	\tikzset{every picture/.style={line width=0.75pt}} 
	
	\begin{tikzpicture}[x=0.75pt,y=0.75pt,yscale=-1,xscale=1]
		
		\draw   (23,40) -- (93,40) -- (93,80) -- (23,80) -- cycle ;
		\draw    (33,0) -- (33,40) ;
		\draw    (33,20) -- (33,13) ;
		\draw [shift={(33,10)}, rotate = 90] [fill={rgb, 255:red, 0; green, 0; blue, 0 }  ][line width=0.08]  [draw opacity=0] (10.72,-5.15) -- (0,0) -- (10.72,5.15) -- (7.12,0) -- cycle    ;
		\draw    (83,0) -- (83,40) ;
		\draw    (83,20) -- (83,13) ;
		\draw [shift={(83,10)}, rotate = 90] [fill={rgb, 255:red, 0; green, 0; blue, 0 }  ][line width=0.08]  [draw opacity=0] (10.72,-5.15) -- (0,0) -- (10.72,5.15) -- (7.12,0) -- cycle    ;
		\draw    (33,80) -- (33,120) ;
		\draw    (33,100) -- (33,93) ;
		\draw [shift={(33,90)}, rotate = 90] [fill={rgb, 255:red, 0; green, 0; blue, 0 }  ][line width=0.08]  [draw opacity=0] (10.72,-5.15) -- (0,0) -- (10.72,5.15) -- (7.12,0) -- cycle    ;
		\draw    (83,80) -- (83,120) ;
		\draw    (83,100) -- (83,93) ;
		\draw [shift={(83,90)}, rotate = 90] [fill={rgb, 255:red, 0; green, 0; blue, 0 }  ][line width=0.08]  [draw opacity=0] (10.72,-5.15) -- (0,0) -- (10.72,5.15) -- (7.12,0) -- cycle    ;
		\draw   (190,40) -- (260,40) -- (260,80) -- (190,80) -- cycle ;
		\draw    (200,0) -- (200,40) ;
		\draw    (200,20) -- (200,13) ;
		\draw [shift={(200,10)}, rotate = 90] [fill={rgb, 255:red, 0; green, 0; blue, 0 }  ][line width=0.08]  [draw opacity=0] (10.72,-5.15) -- (0,0) -- (10.72,5.15) -- (7.12,0) -- cycle    ;
		\draw    (250,0) -- (250,40) ;
		\draw    (250,20) -- (250,27) ;
		\draw [shift={(250,30)}, rotate = 270] [fill={rgb, 255:red, 0; green, 0; blue, 0 }  ][line width=0.08]  [draw opacity=0] (10.72,-5.15) -- (0,0) -- (10.72,5.15) -- (7.12,0) -- cycle    ;
		\draw    (200,80) -- (200,120) ;
		\draw    (200,100) -- (200,107) ;
		\draw [shift={(200,110)}, rotate = 270] [fill={rgb, 255:red, 0; green, 0; blue, 0 }  ][line width=0.08]  [draw opacity=0] (10.72,-5.15) -- (0,0) -- (10.72,5.15) -- (7.12,0) -- cycle    ;
		\draw    (250,80) -- (250,120) ;
		\draw    (250,100) -- (250,93) ;
		\draw [shift={(250,90)}, rotate = 90] [fill={rgb, 255:red, 0; green, 0; blue, 0 }  ][line width=0.08]  [draw opacity=0] (10.72,-5.15) -- (0,0) -- (10.72,5.15) -- (7.12,0) -- cycle    ;
		
		\draw (126,52.4) node [anchor=north west][inner sep=0.75pt]    {$=$};
		\draw (54,52.4) node [anchor=north west][inner sep=0.75pt]    {$f$};
		\draw (6,2.4) node [anchor=north west][inner sep=0.75pt]    {$Z$};
		\draw (92,2.4) node [anchor=north west][inner sep=0.75pt]    {$W^{*}$};
		\draw (5,102.4) node [anchor=north west][inner sep=0.75pt]    {$X^{*}$};
		\draw (94,102.4) node [anchor=north west][inner sep=0.75pt]    {$Y$};
		\draw (221,52.4) node [anchor=north west][inner sep=0.75pt]    {$f$};
		\draw (169,2.4) node [anchor=north west][inner sep=0.75pt]    {$Z$};
		\draw (255,2.4) node [anchor=north west][inner sep=0.75pt]    {$W$};
		\draw (171,102.4) node [anchor=north west][inner sep=0.75pt]    {$X$};
		\draw (257,102.4) node [anchor=north west][inner sep=0.75pt]    {$Y$};

	\end{tikzpicture}
\end{center}
This notation by itself isn't very helpful. Its utility comes from combining it with the string diagrams for the evaluation and coevaluation maps. These maps were defined in proposition \ref{proposition:evaluations} and definition \ref{definition:coevaluations}. For any finite dimensional $F$-vector space $X$, we have evaluation maps
\begin{equation*}
	\lev_X \colon X^* \otimes X \to F ,
\end{equation*}
\begin{equation*}
	\rev_X \colon X \otimes X^* \to F,
\end{equation*}
and coevaluation maps
\begin{equation*}
	\lcoev_X \colon F \to X \otimes X^* ,
\end{equation*}
\begin{equation*}
	\rcoev_X \colon F \to X^* \otimes X .
\end{equation*}
The evaluations will be represented by caps and the coevaluations by cups. The arrows above these maps indicate the orientation of the cap or cup. We have the following string diagrams:
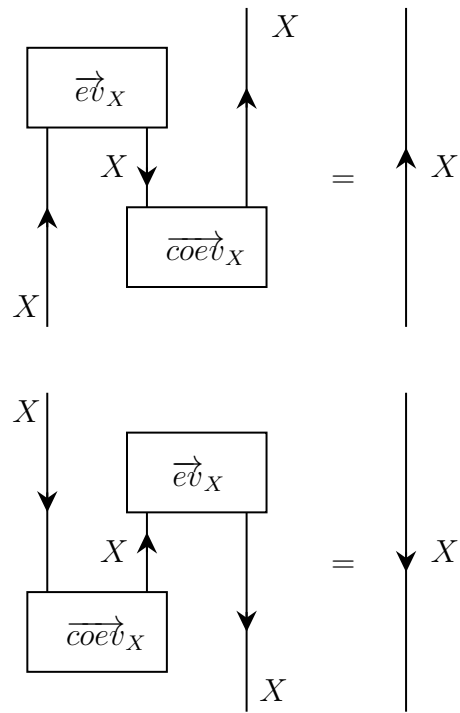
\begin{center}

	\tikzset{every picture/.style={line width=0.75pt}} 
	
	\begin{tikzpicture}[x=0.75pt,y=0.75pt,yscale=-1,xscale=1]
		
		\draw   (76,0) -- (146,0) -- (146,40) -- (76,40) -- cycle ;
		\draw    (86,40) -- (86,80) ;
		\draw    (86,60) -- (86,67) ;
		\draw [shift={(86,70)}, rotate = 270] [fill={rgb, 255:red, 0; green, 0; blue, 0 }  ][line width=0.08]  [draw opacity=0] (10.72,-5.15) -- (0,0) -- (10.72,5.15) -- (7.12,0) -- cycle    ;
		\draw    (136,40) -- (136,80) ;
		\draw    (136,60) -- (136,53) ;
		\draw [shift={(136,50)}, rotate = 90] [fill={rgb, 255:red, 0; green, 0; blue, 0 }  ][line width=0.08]  [draw opacity=0] (10.72,-5.15) -- (0,0) -- (10.72,5.15) -- (7.12,0) -- cycle    ;
		\draw  [draw opacity=0] (220,50) .. controls (220,33.43) and (233.43,20) .. (250,20) .. controls (266.57,20) and (280,33.43) .. (280,50) -- (250,50) -- cycle ; \draw   (220,50) .. controls (220,33.43) and (233.43,20) .. (250,20) .. controls (266.57,20) and (280,33.43) .. (280,50) ;  
		\draw    (250,20) -- (243,20) ;
		\draw [shift={(240,20)}, rotate = 360] [fill={rgb, 255:red, 0; green, 0; blue, 0 }  ][line width=0.08]  [draw opacity=0] (10.72,-5.15) -- (0,0) -- (10.72,5.15) -- (7.12,0) -- cycle    ;
		
		\draw (98,12.4) node [anchor=north west][inner sep=0.75pt]    {$\overleftarrow{ev}_{X}$};
		\draw (61,62.4) node [anchor=north west][inner sep=0.75pt]    {$X$};
		\draw (145,62.4) node [anchor=north west][inner sep=0.75pt]    {$X$};
		\draw (1,32.4) node [anchor=north west][inner sep=0.75pt]    {$\overleftarrow{ev}_{X} \ =$};
		\draw (176,32.4) node [anchor=north west][inner sep=0.75pt]    {$=$};
		\draw (285,32.4) node [anchor=north west][inner sep=0.75pt]    {$X$};

	\end{tikzpicture}
\end{center}
\begin{center}

	\tikzset{every picture/.style={line width=0.75pt}} 
	
	\begin{tikzpicture}[x=0.75pt,y=0.75pt,yscale=-1,xscale=1]
		
		\draw   (80,0) -- (150,0) -- (150,40) -- (80,40) -- cycle ;
		\draw    (90,40) -- (90,66.6) -- (90,80) ;
		\draw    (90,60) -- (90,56.3) ;
		\draw [shift={(90,53.3)}, rotate = 90] [fill={rgb, 255:red, 0; green, 0; blue, 0 }  ][line width=0.08]  [draw opacity=0] (10.72,-5.15) -- (0,0) -- (10.72,5.15) -- (7.12,0) -- cycle    ;
		\draw    (140,40) -- (140,80) ;
		\draw    (140,60) -- (140,67) ;
		\draw [shift={(140,70)}, rotate = 270] [fill={rgb, 255:red, 0; green, 0; blue, 0 }  ][line width=0.08]  [draw opacity=0] (10.72,-5.15) -- (0,0) -- (10.72,5.15) -- (7.12,0) -- cycle    ;
		\draw  [draw opacity=0] (220,50) .. controls (220,33.43) and (233.43,20) .. (250,20) .. controls (266.57,20) and (280,33.43) .. (280,50) -- (250,50) -- cycle ; \draw   (220,50) .. controls (220,33.43) and (233.43,20) .. (250,20) .. controls (266.57,20) and (280,33.43) .. (280,50) ;  
		\draw    (250,20) -- (257,20) ;
		\draw [shift={(260,20)}, rotate = 180] [fill={rgb, 255:red, 0; green, 0; blue, 0 }  ][line width=0.08]  [draw opacity=0] (10.72,-5.15) -- (0,0) -- (10.72,5.15) -- (7.12,0) -- cycle    ;
		
		\draw (101,12.4) node [anchor=north west][inner sep=0.75pt]    {$\overrightarrow{ev}_{X}$};
		\draw (65,62.4) node [anchor=north west][inner sep=0.75pt]    {$X$};
		\draw (151,62.4) node [anchor=north west][inner sep=0.75pt]    {$X$};
		\draw (1,32.4) node [anchor=north west][inner sep=0.75pt]    {$\overrightarrow{ev}_{X} \ =$};
		\draw (176,32.4) node [anchor=north west][inner sep=0.75pt]    {$=$};
		\draw (285,32.4) node [anchor=north west][inner sep=0.75pt]    {$X$};

	\end{tikzpicture}
\end{center}
\begin{center}

	\tikzset{every picture/.style={line width=0.75pt}} 
	
	\begin{tikzpicture}[x=0.75pt,y=0.75pt,yscale=-1,xscale=1]
		
		\draw   (100,40) -- (170,40) -- (170,80) -- (100,80) -- cycle ;
		\draw    (110,0) -- (110,40) ;
		\draw    (160,20) -- (160,27) ;
		\draw [shift={(160,30)}, rotate = 270] [fill={rgb, 255:red, 0; green, 0; blue, 0 }  ][line width=0.08]  [draw opacity=0] (10.72,-5.15) -- (0,0) -- (10.72,5.15) -- (7.12,0) -- cycle    ;
		\draw    (160,0) -- (160,40) ;
		\draw    (110,20) -- (110,13) ;
		\draw [shift={(110,10)}, rotate = 90] [fill={rgb, 255:red, 0; green, 0; blue, 0 }  ][line width=0.08]  [draw opacity=0] (10.72,-5.15) -- (0,0) -- (10.72,5.15) -- (7.12,0) -- cycle    ;
		\draw  [draw opacity=0] (310,30) .. controls (310,30) and (310,30) .. (310,30) .. controls (310,30) and (310,30) .. (310,30) .. controls (310,46.57) and (296.57,60) .. (280,60) .. controls (263.43,60) and (250,46.57) .. (250,30) -- (280,30) -- cycle ; \draw   (310,30) .. controls (310,30) and (310,30) .. (310,30) .. controls (310,30) and (310,30) .. (310,30) .. controls (310,46.57) and (296.57,60) .. (280,60) .. controls (263.43,60) and (250,46.57) .. (250,30) ;  
		\draw    (280,60) -- (273,60) ;
		\draw [shift={(270,60)}, rotate = 360] [fill={rgb, 255:red, 0; green, 0; blue, 0 }  ][line width=0.08]  [draw opacity=0] (10.72,-5.15) -- (0,0) -- (10.72,5.15) -- (7.12,0) -- cycle    ;
		
		\draw (118,52.4) node [anchor=north west][inner sep=0.75pt]    {$\overleftarrow{coev}_{X}$};
		\draw (81,2.4) node [anchor=north west][inner sep=0.75pt]    {$X$};
		\draw (171,2.4) node [anchor=north west][inner sep=0.75pt]    {$X$};
		\draw (1,32.4) node [anchor=north west][inner sep=0.75pt]    {$\overleftarrow{coev}_{X} \ =$};
		\draw (206,32.4) node [anchor=north west][inner sep=0.75pt]    {$=$};
		\draw (315,32.4) node [anchor=north west][inner sep=0.75pt]    {$X$};

	\end{tikzpicture}
\end{center}
\begin{center}

	\tikzset{every picture/.style={line width=0.75pt}} 
	
	\begin{tikzpicture}[x=0.75pt,y=0.75pt,yscale=-1,xscale=1]
		
		\draw   (100,40) -- (170,40) -- (170,80) -- (100,80) -- cycle ;
		\draw    (110,0) -- (110,40) ;
		\draw    (110,20) -- (110,27) ;
		\draw [shift={(110,30)}, rotate = 270] [fill={rgb, 255:red, 0; green, 0; blue, 0 }  ][line width=0.08]  [draw opacity=0] (10.72,-5.15) -- (0,0) -- (10.72,5.15) -- (7.12,0) -- cycle    ;
		\draw    (160,0) -- (160,40) ;
		\draw    (160,20) -- (160,13) ;
		\draw [shift={(160,10)}, rotate = 90] [fill={rgb, 255:red, 0; green, 0; blue, 0 }  ][line width=0.08]  [draw opacity=0] (10.72,-5.15) -- (0,0) -- (10.72,5.15) -- (7.12,0) -- cycle    ;
		\draw  [draw opacity=0] (310,30) .. controls (310,30) and (310,30) .. (310,30) .. controls (310,30) and (310,30) .. (310,30) .. controls (310,46.57) and (296.57,60) .. (280,60) .. controls (263.43,60) and (250,46.57) .. (250,30) -- (280,30) -- cycle ; \draw   (310,30) .. controls (310,30) and (310,30) .. (310,30) .. controls (310,30) and (310,30) .. (310,30) .. controls (310,46.57) and (296.57,60) .. (280,60) .. controls (263.43,60) and (250,46.57) .. (250,30) ;  
		\draw    (280,60) -- (287,60) ;
		\draw [shift={(290,60)}, rotate = 180] [fill={rgb, 255:red, 0; green, 0; blue, 0 }  ][line width=0.08]  [draw opacity=0] (10.72,-5.15) -- (0,0) -- (10.72,5.15) -- (7.12,0) -- cycle    ;
		
		\draw (118,52.4) node [anchor=north west][inner sep=0.75pt]    {$\overrightarrow{coev}_{X}$};
		\draw (81,2.4) node [anchor=north west][inner sep=0.75pt]    {$X$};
		\draw (171,2.4) node [anchor=north west][inner sep=0.75pt]    {$X$};
		\draw (1,32.4) node [anchor=north west][inner sep=0.75pt]    {$\overrightarrow{coev}_{X} \ =$};
		\draw (206,32.4) node [anchor=north west][inner sep=0.75pt]    {$=$};
		\draw (315,32.4) node [anchor=north west][inner sep=0.75pt]    {$X$};

	\end{tikzpicture}
\end{center}

Both $\lev_X$ and $\rev_X$ are non degenerate pairings with inverses $\lcoev_X$ and $\rcoev_X$, respectively. This amounts to the equations proved in proposition \ref{proposition:ev is non degenerate pairing}. Keeping the associativity and unitality isomorphisms implicit, the equations of this proposition become
\begin{gather}
	(\lev_X \otimes \id_{X^*})(\id_{X^*} \otimes \lcoev_X) = \id_{X^*}, \\
	(\id_X \otimes \lev_X) (\lcoev_X \otimes \id_X ) = \id_X, \\
	(\rev_X \otimes \id_X) (\id_X \otimes \rcoev_X) = \id_X , \\
	(\id_{X^*} \otimes \rev_X) (\rcoev_X \otimes \id_{X^*} ) = \id_{X^*} .
\end{gather}
As string diagrams, without yet using the caps and cups, these become the following diagrammatic equations:
\begin{center}

	\tikzset{every picture/.style={line width=0.75pt}} 
	
	\begin{tikzpicture}[x=0.75pt,y=0.75pt,yscale=-1,xscale=1]
		
		\draw   (10,20) -- (80,20) -- (80,60) -- (10,60) -- cycle ;
		\draw    (20,60) -- (20,160) ;
		\draw    (20,110) -- (20,117) ;
		\draw [shift={(20,120)}, rotate = 270] [fill={rgb, 255:red, 0; green, 0; blue, 0 }  ][line width=0.08]  [draw opacity=0] (10.72,-5.15) -- (0,0) -- (10.72,5.15) -- (7.12,0) -- cycle    ;
		\draw    (70,60) -- (70,100) ;
		\draw    (70,80) -- (70,73) ;
		\draw [shift={(70,70)}, rotate = 90] [fill={rgb, 255:red, 0; green, 0; blue, 0 }  ][line width=0.08]  [draw opacity=0] (10.72,-5.15) -- (0,0) -- (10.72,5.15) -- (7.12,0) -- cycle    ;
		\draw   (60,100) -- (130,100) -- (130,140) -- (60,140) -- cycle ;
		\draw    (120,0) -- (120,100) ;
		\draw    (120,50) -- (120,57) ;
		\draw [shift={(120,60)}, rotate = 270] [fill={rgb, 255:red, 0; green, 0; blue, 0 }  ][line width=0.08]  [draw opacity=0] (10.72,-5.15) -- (0,0) -- (10.72,5.15) -- (7.12,0) -- cycle    ;
		\draw    (200,0) -- (200,160) ;
		\draw    (200,70) -- (200,77) ;
		\draw [shift={(200,80)}, rotate = 270] [fill={rgb, 255:red, 0; green, 0; blue, 0 }  ][line width=0.08]  [draw opacity=0] (10.72,-5.15) -- (0,0) -- (10.72,5.15) -- (7.12,0) -- cycle    ;
		
		\draw (32,32.4) node [anchor=north west][inner sep=0.75pt]    {$\overleftarrow{ev}_{X}$};
		\draw (1,142.4) node [anchor=north west][inner sep=0.75pt]    {$X$};
		\draw (45,72.4) node [anchor=north west][inner sep=0.75pt]    {$X$};
		\draw (78,112.4) node [anchor=north west][inner sep=0.75pt]    {$\overleftarrow{coev}_{X}$};
		\draw (131,2.4) node [anchor=north west][inner sep=0.75pt]    {$X$};
		\draw (161,82.4) node [anchor=north west][inner sep=0.75pt]    {$=$};
		\draw (211,72.4) node [anchor=north west][inner sep=0.75pt]    {$X$};

	\end{tikzpicture}
\end{center}
\begin{center}

	\tikzset{every picture/.style={line width=0.75pt}} 
	
	\begin{tikzpicture}[x=0.75pt,y=0.75pt,yscale=-1,xscale=1]
		
		\draw   (60,20) -- (130,20) -- (130,60) -- (60,60) -- cycle ;
		\draw    (120,60) -- (120,160) ;
		\draw    (120,110) -- (120,103) ;
		\draw [shift={(120,100)}, rotate = 90] [fill={rgb, 255:red, 0; green, 0; blue, 0 }  ][line width=0.08]  [draw opacity=0] (10.72,-5.15) -- (0,0) -- (10.72,5.15) -- (7.12,0) -- cycle    ;
		\draw    (70,60) -- (70,100) ;
		\draw    (70,80) -- (70,87) ;
		\draw [shift={(70,90)}, rotate = 270] [fill={rgb, 255:red, 0; green, 0; blue, 0 }  ][line width=0.08]  [draw opacity=0] (10.72,-5.15) -- (0,0) -- (10.72,5.15) -- (7.12,0) -- cycle    ;
		\draw   (10,100) -- (80,100) -- (80,140) -- (10,140) -- cycle ;
		\draw    (20,0) -- (20,100) ;
		\draw    (20,60) -- (20,53) ;
		\draw [shift={(20,50)}, rotate = 90] [fill={rgb, 255:red, 0; green, 0; blue, 0 }  ][line width=0.08]  [draw opacity=0] (10.72,-5.15) -- (0,0) -- (10.72,5.15) -- (7.12,0) -- cycle    ;
		\draw    (200,0) -- (200,160) ;
		\draw    (200,80) -- (200,73) ;
		\draw [shift={(200,70)}, rotate = 90] [fill={rgb, 255:red, 0; green, 0; blue, 0 }  ][line width=0.08]  [draw opacity=0] (10.72,-5.15) -- (0,0) -- (10.72,5.15) -- (7.12,0) -- cycle    ;
		
		\draw (81,32.4) node [anchor=north west][inner sep=0.75pt]    {$\overleftarrow{ev}_{X}$};
		\draw (1,2.4) node [anchor=north west][inner sep=0.75pt]    {$X$};
		\draw (45,72.4) node [anchor=north west][inner sep=0.75pt]    {$X$};
		\draw (28,112.4) node [anchor=north west][inner sep=0.75pt]    {$\overleftarrow{coev}_{X}$};
		\draw (125,142.4) node [anchor=north west][inner sep=0.75pt]    {$X$};
		\draw (161,82.4) node [anchor=north west][inner sep=0.75pt]    {$=$};
		\draw (211,72.4) node [anchor=north west][inner sep=0.75pt]    {$X$};

	\end{tikzpicture}
\end{center}
\begin{center}

	\tikzset{every picture/.style={line width=0.75pt}} 
	
	\begin{tikzpicture}[x=0.75pt,y=0.75pt,yscale=-1,xscale=1]
		
		\draw   (10,20) -- (80,20) -- (80,60) -- (10,60) -- cycle ;
		\draw    (20,60) -- (20,160) ;
		\draw    (20,110) -- (20,103) ;
		\draw [shift={(20,100)}, rotate = 90] [fill={rgb, 255:red, 0; green, 0; blue, 0 }  ][line width=0.08]  [draw opacity=0] (10.72,-5.15) -- (0,0) -- (10.72,5.15) -- (7.12,0) -- cycle    ;
		\draw    (70,60) -- (70,100) ;
		\draw    (70,80) -- (70,87) ;
		\draw [shift={(70,90)}, rotate = 270] [fill={rgb, 255:red, 0; green, 0; blue, 0 }  ][line width=0.08]  [draw opacity=0] (10.72,-5.15) -- (0,0) -- (10.72,5.15) -- (7.12,0) -- cycle    ;
		\draw   (60,100) -- (130,100) -- (130,140) -- (60,140) -- cycle ;
		\draw    (120,0) -- (120,100) ;
		\draw    (120,50) -- (120,43) ;
		\draw [shift={(120,40)}, rotate = 90] [fill={rgb, 255:red, 0; green, 0; blue, 0 }  ][line width=0.08]  [draw opacity=0] (10.72,-5.15) -- (0,0) -- (10.72,5.15) -- (7.12,0) -- cycle    ;
		\draw    (200,0) -- (200,160) ;
		\draw    (200,80) -- (200,73) ;
		\draw [shift={(200,70)}, rotate = 90] [fill={rgb, 255:red, 0; green, 0; blue, 0 }  ][line width=0.08]  [draw opacity=0] (10.72,-5.15) -- (0,0) -- (10.72,5.15) -- (7.12,0) -- cycle    ;
		
		\draw (32,32.4) node [anchor=north west][inner sep=0.75pt]    {$\overrightarrow{ev}_{X}$};
		\draw (1,142.4) node [anchor=north west][inner sep=0.75pt]    {$X$};
		\draw (45,72.4) node [anchor=north west][inner sep=0.75pt]    {$X$};
		\draw (78,112.4) node [anchor=north west][inner sep=0.75pt]    {$\overrightarrow{coev}_{X}$};
		\draw (131,2.4) node [anchor=north west][inner sep=0.75pt]    {$X$};
		\draw (161,82.4) node [anchor=north west][inner sep=0.75pt]    {$=$};
		\draw (211,72.4) node [anchor=north west][inner sep=0.75pt]    {$X$};

	\end{tikzpicture}
\end{center}
\begin{center}

	\tikzset{every picture/.style={line width=0.75pt}} 
	
	\begin{tikzpicture}[x=0.75pt,y=0.75pt,yscale=-1,xscale=1]
		
		\draw   (60,20) -- (130,20) -- (130,60) -- (60,60) -- cycle ;
		\draw    (120,60) -- (120,160) ;
		\draw    (120,110) -- (120,117) ;
		\draw [shift={(120,120)}, rotate = 270] [fill={rgb, 255:red, 0; green, 0; blue, 0 }  ][line width=0.08]  [draw opacity=0] (10.72,-5.15) -- (0,0) -- (10.72,5.15) -- (7.12,0) -- cycle    ;
		\draw    (70,60) -- (70,100) ;
		\draw    (70,80) -- (70,73) ;
		\draw [shift={(70,70)}, rotate = 90] [fill={rgb, 255:red, 0; green, 0; blue, 0 }  ][line width=0.08]  [draw opacity=0] (10.72,-5.15) -- (0,0) -- (10.72,5.15) -- (7.12,0) -- cycle    ;
		\draw   (10,100) -- (80,100) -- (80,140) -- (10,140) -- cycle ;
		\draw    (20,0) -- (20,100) ;
		\draw    (20,50) -- (20,57) ;
		\draw [shift={(20,60)}, rotate = 270] [fill={rgb, 255:red, 0; green, 0; blue, 0 }  ][line width=0.08]  [draw opacity=0] (10.72,-5.15) -- (0,0) -- (10.72,5.15) -- (7.12,0) -- cycle    ;
		\draw    (200,0) -- (200,160) ;
		\draw    (200,80) -- (200,87) ;
		\draw [shift={(200,90)}, rotate = 270] [fill={rgb, 255:red, 0; green, 0; blue, 0 }  ][line width=0.08]  [draw opacity=0] (10.72,-5.15) -- (0,0) -- (10.72,5.15) -- (7.12,0) -- cycle    ;
		
		\draw (81,32.4) node [anchor=north west][inner sep=0.75pt]    {$\overrightarrow{ev}_{X}$};
		\draw (1,2.4) node [anchor=north west][inner sep=0.75pt]    {$X$};
		\draw (45,72.4) node [anchor=north west][inner sep=0.75pt]    {$X$};
		\draw (28,112.4) node [anchor=north west][inner sep=0.75pt]    {$\overrightarrow{coev}_{X}$};
		\draw (125,142.4) node [anchor=north west][inner sep=0.75pt]    {$X$};
		\draw (161,82.4) node [anchor=north west][inner sep=0.75pt]    {$=$};
		\draw (211,72.4) node [anchor=north west][inner sep=0.75pt]    {$X$};

	\end{tikzpicture}
\end{center}
If we use the caps and cups, then we obtain the following diagrams:
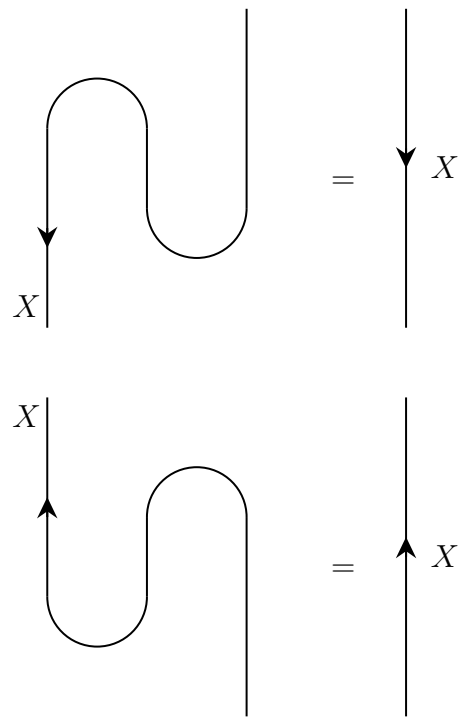
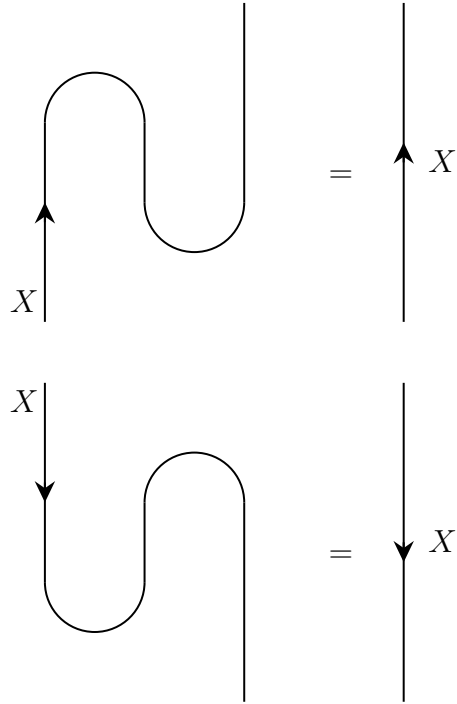
\begin{center}

	\tikzset{every picture/.style={line width=0.75pt}} 
	
	\begin{tikzpicture}[x=0.75pt,y=0.75pt,yscale=-1,xscale=1]
		
		\draw    (20,60) -- (20,160) ;
		\draw    (20,110) -- (20,117) ;
		\draw [shift={(20,120)}, rotate = 270] [fill={rgb, 255:red, 0; green, 0; blue, 0 }  ][line width=0.08]  [draw opacity=0] (10.72,-5.15) -- (0,0) -- (10.72,5.15) -- (7.12,0) -- cycle    ;
		\draw    (70,60) -- (70,100) ;
		\draw    (120,0) -- (120,100) ;
		\draw    (200,0) -- (200,160) ;
		\draw    (200,70) -- (200,77) ;
		\draw [shift={(200,80)}, rotate = 270] [fill={rgb, 255:red, 0; green, 0; blue, 0 }  ][line width=0.08]  [draw opacity=0] (10.72,-5.15) -- (0,0) -- (10.72,5.15) -- (7.12,0) -- cycle    ;
		\draw  [draw opacity=0] (20,60) .. controls (20,60) and (20,60) .. (20,60) .. controls (20,46.19) and (31.19,35) .. (45,35) .. controls (58.81,35) and (70,46.19) .. (70,60) -- (45,60) -- cycle ; \draw   (20,60) .. controls (20,60) and (20,60) .. (20,60) .. controls (20,46.19) and (31.19,35) .. (45,35) .. controls (58.81,35) and (70,46.19) .. (70,60) ;  
		\draw  [draw opacity=0] (120,100) .. controls (120,100) and (120,100) .. (120,100) .. controls (120,113.81) and (108.81,125) .. (95,125) .. controls (81.19,125) and (70,113.81) .. (70,100) -- (95,100) -- cycle ; \draw   (120,100) .. controls (120,100) and (120,100) .. (120,100) .. controls (120,113.81) and (108.81,125) .. (95,125) .. controls (81.19,125) and (70,113.81) .. (70,100) ;  
		
		\draw (1,142.4) node [anchor=north west][inner sep=0.75pt]    {$X$};
		\draw (161,82.4) node [anchor=north west][inner sep=0.75pt]    {$=$};
		\draw (211,72.4) node [anchor=north west][inner sep=0.75pt]    {$X$};

	\end{tikzpicture}
\end{center}
\begin{center}

	\tikzset{every picture/.style={line width=0.75pt}} 
	
	\begin{tikzpicture}[x=0.75pt,y=0.75pt,yscale=-1,xscale=1]
		
		\draw    (120,60) -- (120,160) ;
		\draw    (70,60) -- (70,100) ;
		\draw    (20,0) -- (20,100) ;
		\draw    (20,60) -- (20,53) ;
		\draw [shift={(20,50)}, rotate = 90] [fill={rgb, 255:red, 0; green, 0; blue, 0 }  ][line width=0.08]  [draw opacity=0] (10.72,-5.15) -- (0,0) -- (10.72,5.15) -- (7.12,0) -- cycle    ;
		\draw    (200,0) -- (200,160) ;
		\draw    (200,80) -- (200,73) ;
		\draw [shift={(200,70)}, rotate = 90] [fill={rgb, 255:red, 0; green, 0; blue, 0 }  ][line width=0.08]  [draw opacity=0] (10.72,-5.15) -- (0,0) -- (10.72,5.15) -- (7.12,0) -- cycle    ;
		\draw  [draw opacity=0] (70,60) .. controls (70,60) and (70,60) .. (70,60) .. controls (70,46.19) and (81.19,35) .. (95,35) .. controls (108.81,35) and (120,46.19) .. (120,60) -- (95,60) -- cycle ; \draw   (70,60) .. controls (70,60) and (70,60) .. (70,60) .. controls (70,46.19) and (81.19,35) .. (95,35) .. controls (108.81,35) and (120,46.19) .. (120,60) ;  
		\draw  [draw opacity=0] (70,100) .. controls (70,100) and (70,100) .. (70,100) .. controls (70,113.81) and (58.81,125) .. (45,125) .. controls (31.19,125) and (20,113.81) .. (20,100) -- (45,100) -- cycle ; \draw   (70,100) .. controls (70,100) and (70,100) .. (70,100) .. controls (70,113.81) and (58.81,125) .. (45,125) .. controls (31.19,125) and (20,113.81) .. (20,100) ;  
		
		\draw (1,2.4) node [anchor=north west][inner sep=0.75pt]    {$X$};
		\draw (161,82.4) node [anchor=north west][inner sep=0.75pt]    {$=$};
		\draw (211,72.4) node [anchor=north west][inner sep=0.75pt]    {$X$};

	\end{tikzpicture}
\end{center}
\begin{center}

	\tikzset{every picture/.style={line width=0.75pt}} 
	
	\begin{tikzpicture}[x=0.75pt,y=0.75pt,yscale=-1,xscale=1]
		
		\draw    (20,60) -- (20,160) ;
		\draw    (20,110) -- (20,103) ;
		\draw [shift={(20,100)}, rotate = 90] [fill={rgb, 255:red, 0; green, 0; blue, 0 }  ][line width=0.08]  [draw opacity=0] (10.72,-5.15) -- (0,0) -- (10.72,5.15) -- (7.12,0) -- cycle    ;
		\draw    (70,60) -- (70,100) ;
		\draw    (120,0) -- (120,100) ;
		\draw    (200,0) -- (200,160) ;
		\draw    (200,80) -- (200,73) ;
		\draw [shift={(200,70)}, rotate = 90] [fill={rgb, 255:red, 0; green, 0; blue, 0 }  ][line width=0.08]  [draw opacity=0] (10.72,-5.15) -- (0,0) -- (10.72,5.15) -- (7.12,0) -- cycle    ;
		\draw  [draw opacity=0] (20,60) .. controls (20,60) and (20,60) .. (20,60) .. controls (20,46.19) and (31.19,35) .. (45,35) .. controls (58.81,35) and (70,46.19) .. (70,60) -- (45,60) -- cycle ; \draw   (20,60) .. controls (20,60) and (20,60) .. (20,60) .. controls (20,46.19) and (31.19,35) .. (45,35) .. controls (58.81,35) and (70,46.19) .. (70,60) ;  
		\draw  [draw opacity=0] (120,100) .. controls (120,100) and (120,100) .. (120,100) .. controls (120,113.81) and (108.81,125) .. (95,125) .. controls (81.19,125) and (70,113.81) .. (70,100) -- (95,100) -- cycle ; \draw   (120,100) .. controls (120,100) and (120,100) .. (120,100) .. controls (120,113.81) and (108.81,125) .. (95,125) .. controls (81.19,125) and (70,113.81) .. (70,100) ;  
		
		\draw (1,142.4) node [anchor=north west][inner sep=0.75pt]    {$X$};
		\draw (161,82.4) node [anchor=north west][inner sep=0.75pt]    {$=$};
		\draw (211,72.4) node [anchor=north west][inner sep=0.75pt]    {$X$};

	\end{tikzpicture}
\end{center}
\begin{center}

	\tikzset{every picture/.style={line width=0.75pt}} 
	
	\begin{tikzpicture}[x=0.75pt,y=0.75pt,yscale=-1,xscale=1]
		
		\draw    (120,60) -- (120,160) ;
		\draw    (70,60) -- (70,100) ;
		\draw    (20,0) -- (20,100) ;
		\draw    (20,50) -- (20,57) ;
		\draw [shift={(20,60)}, rotate = 270] [fill={rgb, 255:red, 0; green, 0; blue, 0 }  ][line width=0.08]  [draw opacity=0] (10.72,-5.15) -- (0,0) -- (10.72,5.15) -- (7.12,0) -- cycle    ;
		\draw    (200,0) -- (200,160) ;
		\draw    (200,80) -- (200,87) ;
		\draw [shift={(200,90)}, rotate = 270] [fill={rgb, 255:red, 0; green, 0; blue, 0 }  ][line width=0.08]  [draw opacity=0] (10.72,-5.15) -- (0,0) -- (10.72,5.15) -- (7.12,0) -- cycle    ;
		\draw  [draw opacity=0] (70,60) .. controls (70,60) and (70,60) .. (70,60) .. controls (70,46.19) and (81.19,35) .. (95,35) .. controls (108.81,35) and (120,46.19) .. (120,60) -- (95,60) -- cycle ; \draw   (70,60) .. controls (70,60) and (70,60) .. (70,60) .. controls (70,46.19) and (81.19,35) .. (95,35) .. controls (108.81,35) and (120,46.19) .. (120,60) ;  
		\draw  [draw opacity=0] (70,100) .. controls (70,100) and (70,100) .. (70,100) .. controls (70,113.81) and (58.81,125) .. (45,125) .. controls (31.19,125) and (20,113.81) .. (20,100) -- (45,100) -- cycle ; \draw   (70,100) .. controls (70,100) and (70,100) .. (70,100) .. controls (70,113.81) and (58.81,125) .. (45,125) .. controls (31.19,125) and (20,113.81) .. (20,100) ;  
		
		\draw (1,2.4) node [anchor=north west][inner sep=0.75pt]    {$X$};
		\draw (161,82.4) node [anchor=north west][inner sep=0.75pt]    {$=$};
		\draw (211,72.4) node [anchor=north west][inner sep=0.75pt]    {$X$};

	\end{tikzpicture}
\end{center}
These diagrammatic equations can be interpreted as saying that we can straighten any curved strings in a string diagram. For this reason, these are also called \textbf{yanking equations}.

Since the category is symmetric, we have the symmetry isomorphism
\begin{align*}
	\tau_{X, Y} \colon X \otimes Y &\longrightarrow Y \otimes X \\
	x \otimes y &\longmapsto y \otimes x ,
\end{align*}
as proved in proposition \ref{proposition:tensor symmetry}. The symmetry $\tau_{X,Y}$ is represented diagrammatically as two strings crossing each other:
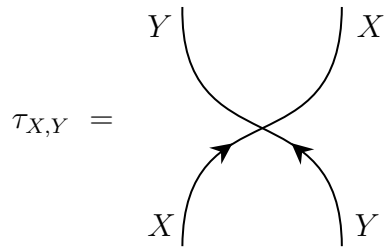
\begin{center}
	\tikzset{every picture/.style={line width=0.75pt}} 
	
	\begin{tikzpicture}[x=0.75pt,y=0.75pt,yscale=-1,xscale=1]
		
		\draw    (89.03,0) .. controls (87.73,80.6) and (168.73,40.6) .. (169.03,120) ;
		\draw    (89.03,120) .. controls (89.73,39.6) and (168.73,82.6) .. (169.03,0) ;
		\draw    (148.95,72.84) -- (145.98,70.27) ;
		\draw [shift={(143.72,68.3)}, rotate = 40.93] [fill={rgb, 255:red, 0; green, 0; blue, 0 }  ][line width=0.08]  [draw opacity=0] (10.72,-5.15) -- (0,0) -- (10.72,5.15) -- (7.12,0) -- cycle    ;
		\draw    (108.15,72.84) -- (111.73,70.21) ;
		\draw [shift={(114.15,68.44)}, rotate = 143.75] [fill={rgb, 255:red, 0; green, 0; blue, 0 }  ][line width=0.08]  [draw opacity=0] (10.72,-5.15) -- (0,0) -- (10.72,5.15) -- (7.12,0) -- cycle    ;
		
		\draw (70.02,102.4) node [anchor=north west][inner sep=0.75pt]    {$X$};
		\draw (174,102.4) node [anchor=north west][inner sep=0.75pt]    {$Y$};
		\draw (70.02,2.4) node [anchor=north west][inner sep=0.75pt]    {$Y$};
		\draw (175,2.4) node [anchor=north west][inner sep=0.75pt]    {$X$};
		\draw (2,52.4) node [anchor=north west][inner sep=0.75pt]    {$\tau _{X,Y} \ =$};
	\end{tikzpicture}
\end{center}
If some string is decorated with a dual space, this can be represented by a downwards oriented string, as done for other diagrams:
\begin{center}
	\tikzset{every picture/.style={line width=0.75pt}} 
	
	\begin{tikzpicture}[x=0.75pt,y=0.75pt,yscale=-1,xscale=1]
		
		\draw    (20.02,0) .. controls (18.72,80.6) and (99.72,40.6) .. (100.02,120) ;
		\draw    (20.02,120) .. controls (20.72,39.6) and (99.72,82.6) .. (100.02,0) ;
		\draw    (79.94,72.84) -- (76.97,70.27) ;
		\draw [shift={(74.7,68.3)}, rotate = 40.93] [fill={rgb, 255:red, 0; green, 0; blue, 0 }  ][line width=0.08]  [draw opacity=0] (10.72,-5.15) -- (0,0) -- (10.72,5.15) -- (7.12,0) -- cycle    ;
		\draw    (39.14,72.84) -- (42.72,70.21) ;
		\draw [shift={(45.14,68.44)}, rotate = 143.75] [fill={rgb, 255:red, 0; green, 0; blue, 0 }  ][line width=0.08]  [draw opacity=0] (10.72,-5.15) -- (0,0) -- (10.72,5.15) -- (7.12,0) -- cycle    ;
		\draw    (202.02,0) .. controls (200.72,80.6) and (281.72,40.6) .. (282.02,120) ;
		\draw    (202.02,120) .. controls (202.72,39.6) and (281.72,82.6) .. (282.02,0) ;
		\draw    (261.94,72.84) -- (258.97,70.27) ;
		\draw [shift={(256.7,68.3)}, rotate = 40.93] [fill={rgb, 255:red, 0; green, 0; blue, 0 }  ][line width=0.08]  [draw opacity=0] (10.72,-5.15) -- (0,0) -- (10.72,5.15) -- (7.12,0) -- cycle    ;
		\draw    (221.14,72.84) -- (218.99,74.66) ;
		\draw [shift={(216.7,76.6)}, rotate = 319.71] [fill={rgb, 255:red, 0; green, 0; blue, 0 }  ][line width=0.08]  [draw opacity=0] (10.72,-5.15) -- (0,0) -- (10.72,5.15) -- (7.12,0) -- cycle    ;
		
		\draw (-1,102.4) node [anchor=north west][inner sep=0.75pt]    {$X^{*}$};
		\draw (104.98,102.4) node [anchor=north west][inner sep=0.75pt]    {$Y$};
		\draw (1,2.4) node [anchor=north west][inner sep=0.75pt]    {$Y$};
		\draw (105.98,2.4) node [anchor=north west][inner sep=0.75pt]    {$X^{*}$};
		\draw (146,52.4) node [anchor=north west][inner sep=0.75pt]    {$=$};
		\draw (181,102.4) node [anchor=north west][inner sep=0.75pt]    {$X$};
		\draw (286.98,102.4) node [anchor=north west][inner sep=0.75pt]    {$Y$};
		\draw (183,2.4) node [anchor=north west][inner sep=0.75pt]    {$Y$};
		\draw (287.98,2.4) node [anchor=north west][inner sep=0.75pt]    {$X$};
	\end{tikzpicture}
\end{center}
We don't need to draw an upwards oriented string decorated with $\complexNumbers$, the unit object. Diagrammatically, such symmetry is treated as an identity function:
\begin{center}
	\tikzset{every picture/.style={line width=0.75pt}} 
	
	\begin{tikzpicture}[x=0.75pt,y=0.75pt,yscale=-1,xscale=1]
		
		\draw    (20.02,0) .. controls (18.72,80.6) and (99.72,40.6) .. (100.02,120) ;
		\draw    (20.02,120) .. controls (20.72,39.6) and (99.72,82.6) .. (100.02,0) ;
		\draw    (79.94,72.84) -- (76.97,70.27) ;
		\draw [shift={(74.7,68.3)}, rotate = 40.93] [fill={rgb, 255:red, 0; green, 0; blue, 0 }  ][line width=0.08]  [draw opacity=0] (10.72,-5.15) -- (0,0) -- (10.72,5.15) -- (7.12,0) -- cycle    ;
		\draw    (39.14,72.84) -- (42.72,70.21) ;
		\draw [shift={(45.14,68.44)}, rotate = 143.75] [fill={rgb, 255:red, 0; green, 0; blue, 0 }  ][line width=0.08]  [draw opacity=0] (10.72,-5.15) -- (0,0) -- (10.72,5.15) -- (7.12,0) -- cycle    ;
		\draw    (190,0) -- (190,120) ;
		\draw    (190,60) -- (190,53) ;
		\draw [shift={(190,50)}, rotate = 90] [fill={rgb, 255:red, 0; green, 0; blue, 0 }  ][line width=0.08]  [draw opacity=0] (10.72,-5.15) -- (0,0) -- (10.72,5.15) -- (7.12,0) -- cycle    ;
		
		\draw (1,102.4) node [anchor=north west][inner sep=0.75pt]    {$X$};
		\draw (104.98,102.4) node [anchor=north west][inner sep=0.75pt]    {$\complexNumbers$};
		\draw (1,2.4) node [anchor=north west][inner sep=0.75pt]    {$\complexNumbers$};
		\draw (105.98,2.4) node [anchor=north west][inner sep=0.75pt]    {$X$};
		\draw (141,52.4) node [anchor=north west][inner sep=0.75pt]    {$=$};
		\draw (201,72.4) node [anchor=north west][inner sep=0.75pt]    {$X$};
	\end{tikzpicture}
\end{center}
An important property of the symmetries is that they form a natural isomorphism. More precisely, for any $f \colon X \to X'$ and $g \colon Y \to Y'$, we have that
\begin{equation}
	\tau_{X', Y'} (f\otimes g) = (g \otimes f) \tau_{X, Y},
\end{equation}
that is, the following diagram commutes:
\begin{center}
	\begin{tikzcd}
		X \otimes Y \arrow[r, "f\otimes g"] \arrow[d, "\tau_{X, Y}"] & X' \otimes Y' \arrow[d, "\tau_{X', Y'}"] \\
		Y \otimes X \arrow[r, "g \otimes f"] & Y' \otimes X'
	\end{tikzcd}
\end{center}
Diagrammatically, this means that boxes can pass through strings:
\begin{center}
	\tikzset{every picture/.style={line width=0.75pt}} 
	
	\begin{tikzpicture}[x=0.75pt,y=0.75pt,yscale=-1,xscale=1]
		
		\draw    (219.02,80) .. controls (217.72,160.6) and (298.72,120.6) .. (299.02,200) ;
		\draw    (219.02,200) .. controls (219.72,119.6) and (298.72,162.6) .. (299.02,80) ;
		\draw    (278.94,152.84) -- (275.97,150.27) ;
		\draw [shift={(273.7,148.3)}, rotate = 40.93] [fill={rgb, 255:red, 0; green, 0; blue, 0 }  ][line width=0.08]  [draw opacity=0] (10.72,-5.15) -- (0,0) -- (10.72,5.15) -- (7.12,0) -- cycle    ;
		\draw    (238.14,152.84) -- (241.72,150.21) ;
		\draw [shift={(244.14,148.44)}, rotate = 143.75] [fill={rgb, 255:red, 0; green, 0; blue, 0 }  ][line width=0.08]  [draw opacity=0] (10.72,-5.15) -- (0,0) -- (10.72,5.15) -- (7.12,0) -- cycle    ;
		\draw   (199,40) -- (239,40) -- (239,80) -- (199,80) -- cycle ;
		\draw   (279,40) -- (319,40) -- (319,80) -- (279,80) -- cycle ;
		\draw    (29.03,0) .. controls (27.73,80.6) and (108.73,40.6) .. (109.03,120) ;
		\draw    (29.03,120) .. controls (29.73,39.6) and (108.73,82.6) .. (109.03,0) ;
		\draw    (88.95,72.84) -- (85.98,70.27) ;
		\draw [shift={(83.72,68.3)}, rotate = 40.93] [fill={rgb, 255:red, 0; green, 0; blue, 0 }  ][line width=0.08]  [draw opacity=0] (10.72,-5.15) -- (0,0) -- (10.72,5.15) -- (7.12,0) -- cycle    ;
		\draw    (48.15,72.84) -- (51.73,70.21) ;
		\draw [shift={(54.15,68.44)}, rotate = 143.75] [fill={rgb, 255:red, 0; green, 0; blue, 0 }  ][line width=0.08]  [draw opacity=0] (10.72,-5.15) -- (0,0) -- (10.72,5.15) -- (7.12,0) -- cycle    ;
		\draw    (299,40) -- (299,0) ;
		\draw    (219,40) -- (219,0) ;
		\draw    (299,20) -- (299,13) ;
		\draw [shift={(299,10)}, rotate = 90] [fill={rgb, 255:red, 0; green, 0; blue, 0 }  ][line width=0.08]  [draw opacity=0] (10.72,-5.15) -- (0,0) -- (10.72,5.15) -- (7.12,0) -- cycle    ;
		\draw    (219,20) -- (219,13) ;
		\draw [shift={(219,10)}, rotate = 90] [fill={rgb, 255:red, 0; green, 0; blue, 0 }  ][line width=0.08]  [draw opacity=0] (10.72,-5.15) -- (0,0) -- (10.72,5.15) -- (7.12,0) -- cycle    ;
		\draw   (10,120) -- (50,120) -- (50,160) -- (10,160) -- cycle ;
		\draw   (90,120) -- (130,120) -- (130,160) -- (90,160) -- cycle ;
		\draw    (109,200) -- (109,160) ;
		\draw    (29,200) -- (29,160) ;
		\draw    (109,180) -- (109,173) ;
		\draw [shift={(109,170)}, rotate = 90] [fill={rgb, 255:red, 0; green, 0; blue, 0 }  ][line width=0.08]  [draw opacity=0] (10.72,-5.15) -- (0,0) -- (10.72,5.15) -- (7.12,0) -- cycle    ;
		\draw    (29,180) -- (29,173) ;
		\draw [shift={(29,170)}, rotate = 90] [fill={rgb, 255:red, 0; green, 0; blue, 0 }  ][line width=0.08]  [draw opacity=0] (10.72,-5.15) -- (0,0) -- (10.72,5.15) -- (7.12,0) -- cycle    ;
		
		\draw (191,182.4) node [anchor=north west][inner sep=0.75pt]    {$X$};
		\draw (303.98,182.4) node [anchor=north west][inner sep=0.75pt]    {$Y$};
		\draw (216,52.4) node [anchor=north west][inner sep=0.75pt]    {$g$};
		\draw (296,52.4) node [anchor=north west][inner sep=0.75pt]    {$f$};
		\draw (121,182.4) node [anchor=north west][inner sep=0.75pt]    {$Y$};
		\draw (5,182.4) node [anchor=north west][inner sep=0.75pt]    {$X$};
		\draw (310,2.4) node [anchor=north west][inner sep=0.75pt]    {$X'$};
		\draw (191,2.4) node [anchor=north west][inner sep=0.75pt]    {$Y'$};
		\draw (107,132.4) node [anchor=north west][inner sep=0.75pt]    {$g$};
		\draw (27,132.4) node [anchor=north west][inner sep=0.75pt]    {$f$};
		\draw (120,2.4) node [anchor=north west][inner sep=0.75pt]    {$X'$};
		\draw (1,2.4) node [anchor=north west][inner sep=0.75pt]    {$Y'$};
		\draw (151,82.4) node [anchor=north west][inner sep=0.75pt]    {$=$};
	\end{tikzpicture}
\end{center}

String diagrams can be quite useful when dealing with monoidal categories, for more details see \cite{MonoidalCatsAndTFT, Turaev+2016, Coecke_Kissinger_2017}.

\subsection{Some useful properties and tricks}

Now that string diagrams were introduced, how can we compute something with them more easily than using regular equations? Here we list some properties that make them useful in computations.

An important property is that we can slide a box up or down. This corresponds to the following diagrammatic equation:
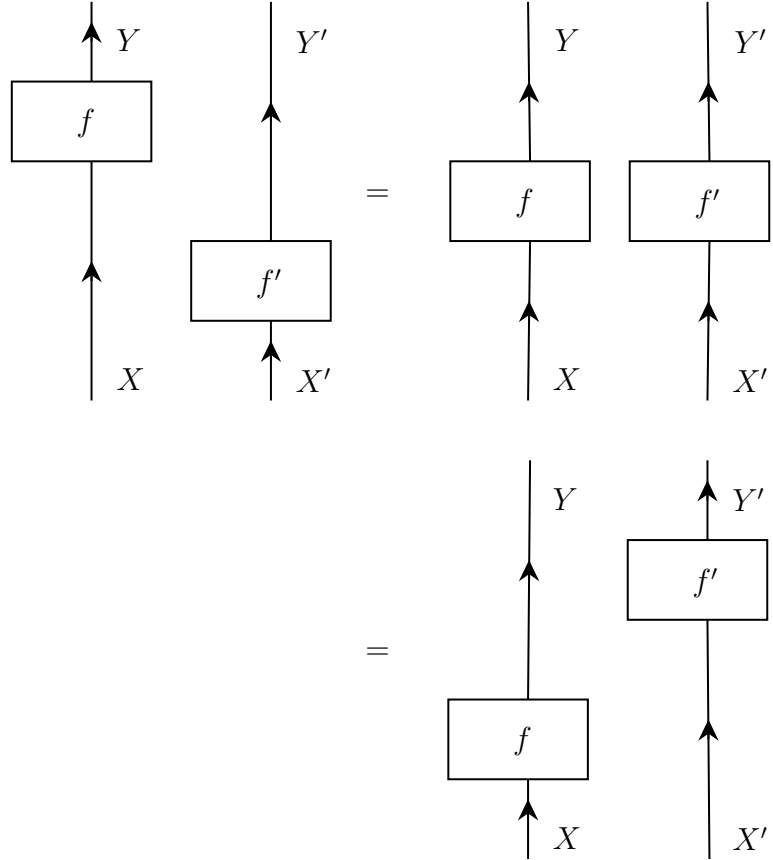
\begin{center}

	\tikzset{every picture/.style={line width=0.75pt}} 
	
	\begin{tikzpicture}[x=0.75pt,y=0.75pt,yscale=-1,xscale=1]
		
		\draw    (40,0) -- (40,40) ;
		\draw    (40,20) -- (40,13) ;
		\draw [shift={(40,10)}, rotate = 90] [fill={rgb, 255:red, 0; green, 0; blue, 0 }  ][line width=0.08]  [draw opacity=0] (10.72,-5.15) -- (0,0) -- (10.72,5.15) -- (7.12,0) -- cycle    ;
		\draw   (0,40) -- (70,40) -- (70,80) -- (0,80) -- cycle ;
		\draw    (130,0) -- (130,120) ;
		\draw    (130,60) -- (130,53) ;
		\draw [shift={(130,50)}, rotate = 90] [fill={rgb, 255:red, 0; green, 0; blue, 0 }  ][line width=0.08]  [draw opacity=0] (10.72,-5.15) -- (0,0) -- (10.72,5.15) -- (7.12,0) -- cycle    ;
		\draw    (40,80) -- (40,200) ;
		\draw    (40,140) -- (40,133) ;
		\draw [shift={(40,130)}, rotate = 90] [fill={rgb, 255:red, 0; green, 0; blue, 0 }  ][line width=0.08]  [draw opacity=0] (10.72,-5.15) -- (0,0) -- (10.72,5.15) -- (7.12,0) -- cycle    ;
		\draw   (90,120) -- (160,120) -- (160,160) -- (90,160) -- cycle ;
		\draw    (130,160) -- (130,200) ;
		\draw    (130,180) -- (130,173) ;
		\draw [shift={(130,170)}, rotate = 90] [fill={rgb, 255:red, 0; green, 0; blue, 0 }  ][line width=0.08]  [draw opacity=0] (10.72,-5.15) -- (0,0) -- (10.72,5.15) -- (7.12,0) -- cycle    ;
		\draw    (259,0) -- (260,80) ;
		\draw    (259.5,50) -- (259.5,43) ;
		\draw [shift={(259.5,40)}, rotate = 90] [fill={rgb, 255:red, 0; green, 0; blue, 0 }  ][line width=0.08]  [draw opacity=0] (10.72,-5.15) -- (0,0) -- (10.72,5.15) -- (7.12,0) -- cycle    ;
		\draw    (349,0) -- (350,80) ;
		\draw    (349.5,50) -- (349.5,43) ;
		\draw [shift={(349.5,40)}, rotate = 90] [fill={rgb, 255:red, 0; green, 0; blue, 0 }  ][line width=0.08]  [draw opacity=0] (10.72,-5.15) -- (0,0) -- (10.72,5.15) -- (7.12,0) -- cycle    ;
		\draw   (220,80) -- (290,80) -- (290,120) -- (220,120) -- cycle ;
		\draw    (260,120) -- (259,200) ;
		\draw    (259.5,160) -- (259.5,153) ;
		\draw [shift={(259.5,150)}, rotate = 90] [fill={rgb, 255:red, 0; green, 0; blue, 0 }  ][line width=0.08]  [draw opacity=0] (10.72,-5.15) -- (0,0) -- (10.72,5.15) -- (7.12,0) -- cycle    ;
		\draw   (310,80) -- (380,80) -- (380,120) -- (310,120) -- cycle ;
		\draw    (350,120) -- (349,200) ;
		\draw    (349.5,160) -- (349.5,153) ;
		\draw [shift={(349.5,150)}, rotate = 90] [fill={rgb, 255:red, 0; green, 0; blue, 0 }  ][line width=0.08]  [draw opacity=0] (10.72,-5.15) -- (0,0) -- (10.72,5.15) -- (7.12,0) -- cycle    ;
		\draw    (260,230) -- (259,350) ;
		\draw    (259.5,290) -- (259.5,283) ;
		\draw [shift={(259.5,280)}, rotate = 90] [fill={rgb, 255:red, 0; green, 0; blue, 0 }  ][line width=0.08]  [draw opacity=0] (10.72,-5.15) -- (0,0) -- (10.72,5.15) -- (7.12,0) -- cycle    ;
		\draw   (309,270) -- (379,270) -- (379,310) -- (309,310) -- cycle ;
		\draw    (349,230) -- (349,270) ;
		\draw    (349,250) -- (349,243) ;
		\draw [shift={(349,240)}, rotate = 90] [fill={rgb, 255:red, 0; green, 0; blue, 0 }  ][line width=0.08]  [draw opacity=0] (10.72,-5.15) -- (0,0) -- (10.72,5.15) -- (7.12,0) -- cycle    ;
		\draw    (349,310) -- (350,430) ;
		\draw    (349.5,370) -- (349.5,363) ;
		\draw [shift={(349.5,360)}, rotate = 90] [fill={rgb, 255:red, 0; green, 0; blue, 0 }  ][line width=0.08]  [draw opacity=0] (10.72,-5.15) -- (0,0) -- (10.72,5.15) -- (7.12,0) -- cycle    ;
		\draw   (219,350) -- (289,350) -- (289,390) -- (219,390) -- cycle ;
		\draw    (259,390) -- (259,430) ;
		\draw    (259,410) -- (259,403) ;
		\draw [shift={(259,400)}, rotate = 90] [fill={rgb, 255:red, 0; green, 0; blue, 0 }  ][line width=0.08]  [draw opacity=0] (10.72,-5.15) -- (0,0) -- (10.72,5.15) -- (7.12,0) -- cycle    ;
		
		\draw (51,12.4) node [anchor=north west][inner sep=0.75pt]    {$Y$};
		\draw (31,52.4) node [anchor=north west][inner sep=0.75pt]    {$f$};
		\draw (141,12.4) node [anchor=north west][inner sep=0.75pt]    {$Y'$};
		\draw (51,182.4) node [anchor=north west][inner sep=0.75pt]    {$X$};
		\draw (121,132.4) node [anchor=north west][inner sep=0.75pt]    {$f'$};
		\draw (141,182.4) node [anchor=north west][inner sep=0.75pt]    {$X'$};
		\draw (176,92.4) node [anchor=north west][inner sep=0.75pt]    {$=$};
		\draw (271,12.4) node [anchor=north west][inner sep=0.75pt]    {$Y$};
		\draw (361,12.4) node [anchor=north west][inner sep=0.75pt]    {$Y'$};
		\draw (251,92.4) node [anchor=north west][inner sep=0.75pt]    {$f$};
		\draw (270,182.4) node [anchor=north west][inner sep=0.75pt]    {$X$};
		\draw (341,92.4) node [anchor=north west][inner sep=0.75pt]    {$f'$};
		\draw (360,182.4) node [anchor=north west][inner sep=0.75pt]    {$X'$};
		\draw (176,322.4) node [anchor=north west][inner sep=0.75pt]    {$=$};
		\draw (270,242.4) node [anchor=north west][inner sep=0.75pt]    {$Y$};
		\draw (340,282.4) node [anchor=north west][inner sep=0.75pt]    {$f'$};
		\draw (360,242.4) node [anchor=north west][inner sep=0.75pt]    {$Y'$};
		\draw (250,362.4) node [anchor=north west][inner sep=0.75pt]    {$f$};
		\draw (270,412.4) node [anchor=north west][inner sep=0.75pt]    {$X$};
		\draw (360,412.4) node [anchor=north west][inner sep=0.75pt]    {$X'$};

	\end{tikzpicture}
\end{center}
It can be interpreted as the equation
\begin{equation}
	(f \otimes \id_{Y'})(\id_X \otimes f') = f \otimes f' = (\id_Y \otimes f')(f \otimes \id_{X'}).
\end{equation}
This is a direct consequence of equation \ref{equation: tensor respects composition}, which is part of the functoriality of $\otimes$. This result together with the yanking equations give us much liberty to deform the string diagrams.

An important trick for this work is using the caps and cups, which are the string diagrams for the evaluations and coevaluations, to move an object from the domain to the codomain, and vice-versa, taking duals as necessary. This defines isomorphisms of hom-sets in $\finvec_F$. Let $f \in \finvec_F (X \otimes Y, Z)$, whose string diagram is the following is the following:
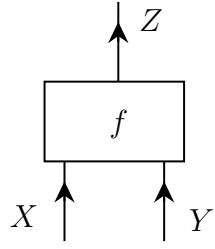
\begin{center}

	\tikzset{every picture/.style={line width=0.75pt}} 
	
	\begin{tikzpicture}[x=0.75pt,y=0.75pt,yscale=-1,xscale=1]
		
		\draw   (23,40) -- (93,40) -- (93,80) -- (23,80) -- cycle ;
		\draw    (60,0) -- (60,40) ;
		\draw    (60,20) -- (60,13) ;
		\draw [shift={(60,10)}, rotate = 90] [fill={rgb, 255:red, 0; green, 0; blue, 0 }  ][line width=0.08]  [draw opacity=0] (10.72,-5.15) -- (0,0) -- (10.72,5.15) -- (7.12,0) -- cycle    ;
		\draw    (33,80) -- (33,120) ;
		\draw    (33,100) -- (33,93) ;
		\draw [shift={(33,90)}, rotate = 90] [fill={rgb, 255:red, 0; green, 0; blue, 0 }  ][line width=0.08]  [draw opacity=0] (10.72,-5.15) -- (0,0) -- (10.72,5.15) -- (7.12,0) -- cycle    ;
		\draw    (83,80) -- (83,120) ;
		\draw    (83,100) -- (83,93) ;
		\draw [shift={(83,90)}, rotate = 90] [fill={rgb, 255:red, 0; green, 0; blue, 0 }  ][line width=0.08]  [draw opacity=0] (10.72,-5.15) -- (0,0) -- (10.72,5.15) -- (7.12,0) -- cycle    ;
		
		\draw (54,52.4) node [anchor=north west][inner sep=0.75pt]    {$f$};
		\draw (69,2.4) node [anchor=north west][inner sep=0.75pt]    {$Z$};
		\draw (4,100.4) node [anchor=north west][inner sep=0.75pt]    {$X$};
		\draw (94,102.4) node [anchor=north west][inner sep=0.75pt]    {$Y$};

	\end{tikzpicture}
\end{center}
Combining this diagram with a cup and identities, we can pass $Y$ to the codomain as $Y^*$, obtaining a linear transformation $\varphi(f) \in \finvec_F (X, Z \otimes Y^*)$:
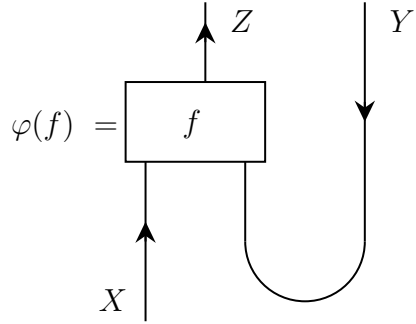
\begin{center}
	\tikzset{every picture/.style={line width=0.75pt}} 
	
	\begin{tikzpicture}[x=0.75pt,y=0.75pt,yscale=-1,xscale=1]
		
		\draw   (70,40) -- (140,40) -- (140,80) -- (70,80) -- cycle ;
		\draw    (110,0) -- (110,40) ;
		\draw    (110,20) -- (110,13) ;
		\draw [shift={(110,10)}, rotate = 90] [fill={rgb, 255:red, 0; green, 0; blue, 0 }  ][line width=0.08]  [draw opacity=0] (10.72,-5.15) -- (0,0) -- (10.72,5.15) -- (7.12,0) -- cycle    ;
		\draw    (80,80) -- (80,160) ;
		\draw    (80,120) -- (80,113) ;
		\draw [shift={(80,110)}, rotate = 90] [fill={rgb, 255:red, 0; green, 0; blue, 0 }  ][line width=0.08]  [draw opacity=0] (10.72,-5.15) -- (0,0) -- (10.72,5.15) -- (7.12,0) -- cycle    ;
		\draw    (130,80) -- (130,120) ;
		\draw  [draw opacity=0] (190,120) .. controls (190,120) and (190,120) .. (190,120) .. controls (190,120) and (190,120) .. (190,120) .. controls (190,136.57) and (176.57,150) .. (160,150) .. controls (143.43,150) and (130,136.57) .. (130,120) -- (160,120) -- cycle ; \draw   (190,120) .. controls (190,120) and (190,120) .. (190,120) .. controls (190,120) and (190,120) .. (190,120) .. controls (190,136.57) and (176.57,150) .. (160,150) .. controls (143.43,150) and (130,136.57) .. (130,120) ;  
		\draw    (190,0) -- (190,120) ;
		\draw    (190,50) -- (190,57) ;
		\draw [shift={(190,60)}, rotate = 270] [fill={rgb, 255:red, 0; green, 0; blue, 0 }  ][line width=0.08]  [draw opacity=0] (10.72,-5.15) -- (0,0) -- (10.72,5.15) -- (7.12,0) -- cycle    ;
		
		\draw (97,52.4) node [anchor=north west][inner sep=0.75pt]    {$f$};
		\draw (121,2.4) node [anchor=north west][inner sep=0.75pt]    {$Z$};
		\draw (55,142.4) node [anchor=north west][inner sep=0.75pt]    {$X$};
		\draw (201,2.4) node [anchor=north west][inner sep=0.75pt]    {$Y$};
		\draw (11,52.4) node [anchor=north west][inner sep=0.75pt]    {$\varphi(f)\ =$};

	\end{tikzpicture}
\end{center}
Keeping associativity and unitality isomorphisms implicit, $\varphi(f)$ can be expressed as
\begin{equation}
	\varphi(f) = (f \otimes \id_{Y^*})(\id_X \otimes \lcoev_Y ).
\end{equation}
Mac Lane's coherence theorem guarantees that we can recover the implicit isomorphisms in a consistent manner. To do so, we just need to put isomorphisms such that the domains of both sides of the equation are the same, and such that all compositions make sense. The codomain of the right side is already $Z \otimes Y^*$, which is also the codomain of $\varphi(f)$. The domain of the right side is $X \otimes F$, while the domain of $\varphi(f)$ is $X$. This is corrected by the unitality isomorphism
\begin{equation*}
	r_X^{-1} \colon X \to X \otimes F .
\end{equation*}
Also, the domain of $f \otimes \id_{Y^*}$ is $(X \otimes Y) \otimes Y^*$, while the codomain of $\id_X \otimes \lcoev_Y$ is $X \otimes (Y \otimes Y^*)$. They differ by the order in which the $\otimes$ is taken. This is corrected by the associativity isomorphism
\begin{equation*}
	a_{X,Y,Y^*}^{-1} \colon X \otimes (Y \otimes Y^*) \to (X \otimes Y) \otimes Y^* .
\end{equation*}
Making everything explicit, $\varphi(f)$ is expressed as
\begin{equation}
	\varphi(f) = (f \otimes \id_{Y^*}) a_{X,Y,Y^*}^{-1} (\id_X \otimes \lcoev_Y ) r_X^{-1} .
\end{equation}
This defines the linear transformation
\begin{align*}
	\varphi \colon \finvec_F (X \otimes Y, Z) &\longrightarrow \finvec_F (X, Z \otimes Y^*)\\
	f &\longmapsto (f \otimes \id_{Y^*}) a_{X,Y,Y^*}^{-1} (\id_X \otimes \lcoev_Y ) r_X^{-1} .
\end{align*}
We can see that $\varphi$ has a complicated expression, but with a very simple diagrammatic representation. It just pulls up the string decorated with $Y$.

Similarly, we can define a function that pulls down the string decorated with $Y$. Let $g \in \finvec_F (X, Z \otimes Y^*)$, with the following string diagram:
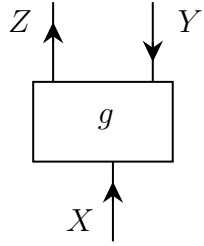
\begin{center}

	\tikzset{every picture/.style={line width=0.75pt}} 
	
	\begin{tikzpicture}[x=0.75pt,y=0.75pt,yscale=-1,xscale=1]
		
		\draw   (20,40) -- (90,40) -- (90,80) -- (20,80) -- cycle ;
		\draw    (30,0) -- (30,40) ;
		\draw    (30,20) -- (30,13) ;
		\draw [shift={(30,10)}, rotate = 90] [fill={rgb, 255:red, 0; green, 0; blue, 0 }  ][line width=0.08]  [draw opacity=0] (10.72,-5.15) -- (0,0) -- (10.72,5.15) -- (7.12,0) -- cycle    ;
		\draw    (60,80) -- (60,120) ;
		\draw    (60,100) -- (60,93) ;
		\draw [shift={(60,90)}, rotate = 90] [fill={rgb, 255:red, 0; green, 0; blue, 0 }  ][line width=0.08]  [draw opacity=0] (10.72,-5.15) -- (0,0) -- (10.72,5.15) -- (7.12,0) -- cycle    ;
		\draw    (80,0) -- (80,40) ;
		\draw    (80,20) -- (80,27) ;
		\draw [shift={(80,30)}, rotate = 270] [fill={rgb, 255:red, 0; green, 0; blue, 0 }  ][line width=0.08]  [draw opacity=0] (10.72,-5.15) -- (0,0) -- (10.72,5.15) -- (7.12,0) -- cycle    ;
		
		\draw (51,52.4) node [anchor=north west][inner sep=0.75pt]    {$g$};
		\draw (6,2.4) node [anchor=north west][inner sep=0.75pt]    {$Z$};
		\draw (35,102.4) node [anchor=north west][inner sep=0.75pt]    {$X$};
		\draw (91,2.4) node [anchor=north west][inner sep=0.75pt]    {$Y$};

	\end{tikzpicture}
\end{center}
Pulling down the string decorated with $Y$ gives a linear transformation $\varphi^{-1}(g) \in \finvec_F (X \otimes Y, Z)$:
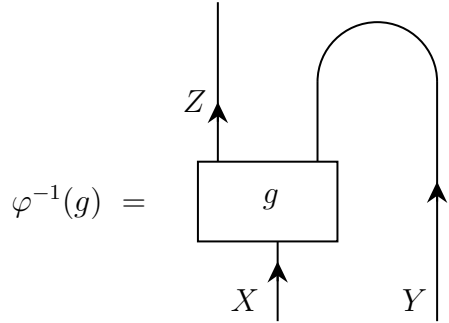
\begin{center}
	\tikzset{every picture/.style={line width=0.75pt}} 
	
	\begin{tikzpicture}[x=0.75pt,y=0.75pt,yscale=-1,xscale=1]
		
		\draw   (100,80) -- (170,80) -- (170,120) -- (100,120) -- cycle ;
		\draw    (110,0) -- (110,80) ;
		\draw    (110,60) -- (110,53) ;
		\draw [shift={(110,50)}, rotate = 90] [fill={rgb, 255:red, 0; green, 0; blue, 0 }  ][line width=0.08]  [draw opacity=0] (10.72,-5.15) -- (0,0) -- (10.72,5.15) -- (7.12,0) -- cycle    ;
		\draw    (140,120) -- (140,160) ;
		\draw    (140,140) -- (140,133) ;
		\draw [shift={(140,130)}, rotate = 90] [fill={rgb, 255:red, 0; green, 0; blue, 0 }  ][line width=0.08]  [draw opacity=0] (10.72,-5.15) -- (0,0) -- (10.72,5.15) -- (7.12,0) -- cycle    ;
		\draw    (160,40) -- (160,80) ;
		\draw  [draw opacity=0] (160,40) .. controls (160,23.43) and (173.43,10) .. (190,10) .. controls (206.57,10) and (220,23.43) .. (220,40) -- (190,40) -- cycle ; \draw   (160,40) .. controls (160,23.43) and (173.43,10) .. (190,10) .. controls (206.57,10) and (220,23.43) .. (220,40) ;  
		\draw    (220,40) -- (220,160) ;
		\draw    (220,100) -- (220,93) ;
		\draw [shift={(220,90)}, rotate = 90] [fill={rgb, 255:red, 0; green, 0; blue, 0 }  ][line width=0.08]  [draw opacity=0] (10.72,-5.15) -- (0,0) -- (10.72,5.15) -- (7.12,0) -- cycle    ;
		
		\draw (131,92.4) node [anchor=north west][inner sep=0.75pt]    {$g$};
		\draw (91,42.4) node [anchor=north west][inner sep=0.75pt]    {$Z$};
		\draw (115,142.4) node [anchor=north west][inner sep=0.75pt]    {$X$};
		\draw (201,142.4) node [anchor=north west][inner sep=0.75pt]    {$Y$};
		\draw (5,90.4) node [anchor=north west][inner sep=0.75pt]    {$\varphi ^{-1}( g) \ =$};

	\end{tikzpicture}
\end{center}
It is already being called $\varphi^{-1} (g)$, because it is the inverse of $\varphi$, but this have yet to be proved. As done for $\varphi$, we can express $\varphi^{-1}$ as follows:
\begin{align*}
	\varphi^{-1} \colon \finvec_F (X, Z \otimes Y^*) &\longrightarrow \finvec_F (X \otimes Y, Z) \\
	g &\longmapsto r_Z (\id_Z \otimes \lev_Y) a_{Z, Y^*, Y} (g \otimes \id_Y) .
\end{align*}

Lastly, last show that $\varphi^{-1}$ is the inverse of $\varphi$. The proof will employ string diagrams, but it can also be done using the explicit non diagrammatic definitions of $\varphi$ and $\varphi^{-1}$ given previously. For any $f \in \finvec_F (X \otimes Y, Z)$, we have that
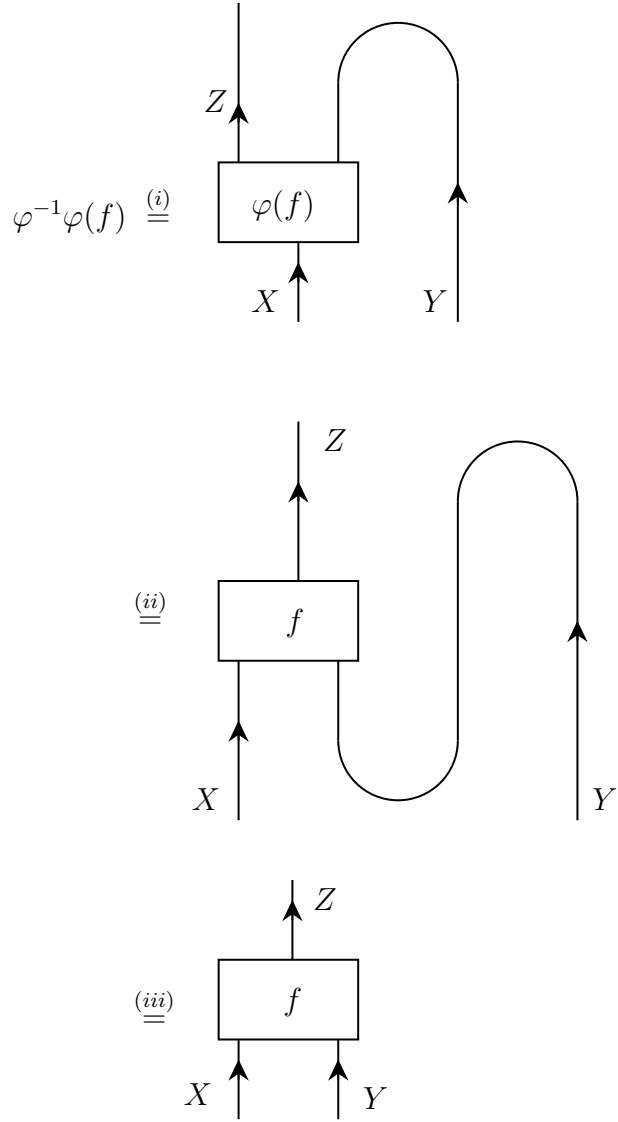
\begin{center}

	\tikzset{every picture/.style={line width=0.75pt}} 
	
	\begin{tikzpicture}[x=0.75pt,y=0.75pt,yscale=-1,xscale=1]
		
		\draw   (110,80) -- (180,80) -- (180,120) -- (110,120) -- cycle ;
		\draw    (120,0) -- (120,80) ;
		\draw    (120,60) -- (120,53) ;
		\draw [shift={(120,50)}, rotate = 90] [fill={rgb, 255:red, 0; green, 0; blue, 0 }  ][line width=0.08]  [draw opacity=0] (10.72,-5.15) -- (0,0) -- (10.72,5.15) -- (7.12,0) -- cycle    ;
		\draw    (150,120) -- (150,160) ;
		\draw    (150,140) -- (150,133) ;
		\draw [shift={(150,130)}, rotate = 90] [fill={rgb, 255:red, 0; green, 0; blue, 0 }  ][line width=0.08]  [draw opacity=0] (10.72,-5.15) -- (0,0) -- (10.72,5.15) -- (7.12,0) -- cycle    ;
		\draw    (170,40) -- (170,80) ;
		\draw  [draw opacity=0] (170,40) .. controls (170,23.43) and (183.43,10) .. (200,10) .. controls (216.57,10) and (230,23.43) .. (230,40) -- (200,40) -- cycle ; \draw   (170,40) .. controls (170,23.43) and (183.43,10) .. (200,10) .. controls (216.57,10) and (230,23.43) .. (230,40) ;  
		\draw    (230,40) -- (230,160) ;
		\draw    (230,100) -- (230,93) ;
		\draw [shift={(230,90)}, rotate = 90] [fill={rgb, 255:red, 0; green, 0; blue, 0 }  ][line width=0.08]  [draw opacity=0] (10.72,-5.15) -- (0,0) -- (10.72,5.15) -- (7.12,0) -- cycle    ;
		\draw   (110,290) -- (180,290) -- (180,330) -- (110,330) -- cycle ;
		\draw    (150,210) -- (150,290) ;
		\draw    (150,250) -- (150,243) ;
		\draw [shift={(150,240)}, rotate = 90] [fill={rgb, 255:red, 0; green, 0; blue, 0 }  ][line width=0.08]  [draw opacity=0] (10.72,-5.15) -- (0,0) -- (10.72,5.15) -- (7.12,0) -- cycle    ;
		\draw    (120,330) -- (120,410) ;
		\draw    (120,370) -- (120,363) ;
		\draw [shift={(120,360)}, rotate = 90] [fill={rgb, 255:red, 0; green, 0; blue, 0 }  ][line width=0.08]  [draw opacity=0] (10.72,-5.15) -- (0,0) -- (10.72,5.15) -- (7.12,0) -- cycle    ;
		\draw    (170,330) -- (170,370) ;
		\draw  [draw opacity=0] (230,370) .. controls (230,370) and (230,370) .. (230,370) .. controls (230,386.57) and (216.57,400) .. (200,400) .. controls (183.43,400) and (170,386.57) .. (170,370) -- (200,370) -- cycle ; \draw   (230,370) .. controls (230,370) and (230,370) .. (230,370) .. controls (230,386.57) and (216.57,400) .. (200,400) .. controls (183.43,400) and (170,386.57) .. (170,370) ;  
		\draw    (230,250) -- (230,370) ;
		\draw    (290,320) -- (290,313) ;
		\draw [shift={(290,310)}, rotate = 90] [fill={rgb, 255:red, 0; green, 0; blue, 0 }  ][line width=0.08]  [draw opacity=0] (10.72,-5.15) -- (0,0) -- (10.72,5.15) -- (7.12,0) -- cycle    ;
		\draw  [draw opacity=0] (230,250) .. controls (230,250) and (230,250) .. (230,250) .. controls (230,233.43) and (243.43,220) .. (260,220) .. controls (276.57,220) and (290,233.43) .. (290,250) -- (260,250) -- cycle ; \draw   (230,250) .. controls (230,250) and (230,250) .. (230,250) .. controls (230,233.43) and (243.43,220) .. (260,220) .. controls (276.57,220) and (290,233.43) .. (290,250) ;  
		\draw    (290,250) -- (290,410) ;
		\draw   (110,480) -- (180,480) -- (180,520) -- (110,520) -- cycle ;
		\draw    (147,440) -- (147,480) ;
		\draw    (147,460) -- (147,453) ;
		\draw [shift={(147,450)}, rotate = 90] [fill={rgb, 255:red, 0; green, 0; blue, 0 }  ][line width=0.08]  [draw opacity=0] (10.72,-5.15) -- (0,0) -- (10.72,5.15) -- (7.12,0) -- cycle    ;
		\draw    (120,520) -- (120,560) ;
		\draw    (120,540) -- (120,533) ;
		\draw [shift={(120,530)}, rotate = 90] [fill={rgb, 255:red, 0; green, 0; blue, 0 }  ][line width=0.08]  [draw opacity=0] (10.72,-5.15) -- (0,0) -- (10.72,5.15) -- (7.12,0) -- cycle    ;
		\draw    (170,520) -- (170,560) ;
		\draw    (170,540) -- (170,533) ;
		\draw [shift={(170,530)}, rotate = 90] [fill={rgb, 255:red, 0; green, 0; blue, 0 }  ][line width=0.08]  [draw opacity=0] (10.72,-5.15) -- (0,0) -- (10.72,5.15) -- (7.12,0) -- cycle    ;
		
		\draw (125,92.4) node [anchor=north west][inner sep=0.75pt]    {$\varphi ( f)$};
		\draw (101,42.4) node [anchor=north west][inner sep=0.75pt]    {$Z$};
		\draw (125,142.4) node [anchor=north west][inner sep=0.75pt]    {$X$};
		\draw (211,142.4) node [anchor=north west][inner sep=0.75pt]    {$Y$};
		\draw (5,90.4) node [anchor=north west][inner sep=0.75pt]    {$\varphi ^{-1} \varphi (f) \ \stackrel{(i)}{=}$};
		\draw (66,292.4) node [anchor=north west][inner sep=0.75pt]    {$\stackrel{(ii)}{=}$};
		\draw (142,302.4) node [anchor=north west][inner sep=0.75pt]    {$f$};
		\draw (161,212.4) node [anchor=north west][inner sep=0.75pt]    {$Z$};
		\draw (95,392.4) node [anchor=north west][inner sep=0.75pt]    {$X$};
		\draw (296,392.4) node [anchor=north west][inner sep=0.75pt]    {$Y$};
		\draw (141,492.4) node [anchor=north west][inner sep=0.75pt]    {$f$};
		\draw (156,442.4) node [anchor=north west][inner sep=0.75pt]    {$Z$};
		\draw (91,540.4) node [anchor=north west][inner sep=0.75pt]    {$X$};
		\draw (181,542.4) node [anchor=north west][inner sep=0.75pt]    {$Y$};
		\draw (66,492.4) node [anchor=north west][inner sep=0.75pt]    {$\stackrel{(iii)}{=}$};
	\end{tikzpicture}
\end{center}
In $(i)$ was used the diagrammatic expression for $\varphi^{-1}$, which consists of pulling down the string decorated with $Y$. In $(ii)$ was used the diagrammatic expression for $\varphi$, which consists of pulling up the string decorated with $Y$. In $(iii)$ was essentially used the yanking equations, which enable us to straighten the string decorated with $Y$. This proves that
\begin{equation}
	\varphi^{-1} \varphi = \id_{\finvec_F (X \otimes Y, Z)}.
\end{equation}
We also have that
\begin{equation}
	\varphi \varphi^{-1} = \id_{\finvec_F (X, Z \otimes Y^*)},
\end{equation}
which has a similar proof using the yanking equations. We proved the following proposition.

\begin{proposition}{Pulling strings}{pulling up or down}
	Let $F$ be a field. Let also $X$, $Y$ and $Z$ be finite dimensional $F$-vector spaces. We have a linear isomorphism
	\begin{align*}
		\varphi \colon \finvec_F (X \otimes Y, Z) &\longrightarrow \finvec_F (X, Z \otimes Y^*)\\
		f &\longmapsto (f \otimes \id_{Y^*}) a_{X,Y,Y^*}^{-1} (\id_X \otimes \lcoev_Y ) r_X^{-1},
	\end{align*}
	with inverse
	\begin{align*}
		\varphi^{-1} \colon \finvec_F (X, Z \otimes Y^*) &\longrightarrow \finvec_F (X \otimes Y, Z) \\
		g &\longmapsto r_Z (\id_Z \otimes \lev_Y) a_{Z, Y^*, Y} (g \otimes \id_Y) .
	\end{align*}
	They can be expressed with string diagrams as follows:
	\begin{center}
		\tikzset{every picture/.style={line width=0.75pt}} 
		
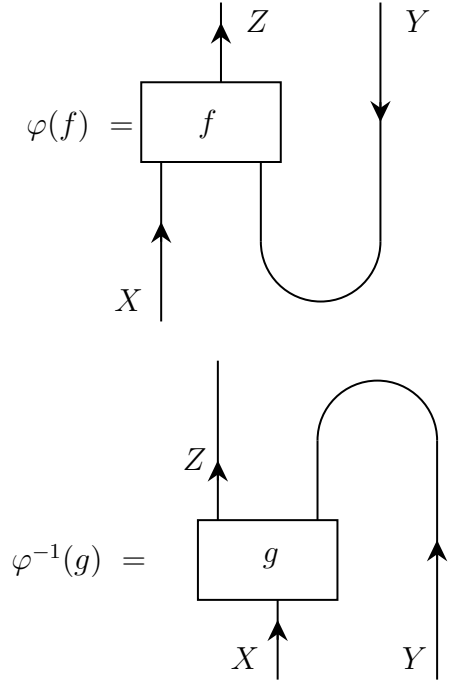
		\begin{tikzpicture}[x=0.75pt,y=0.75pt,yscale=-1,xscale=1]
			
			\draw   (70,40) -- (140,40) -- (140,80) -- (70,80) -- cycle ;
			\draw    (110,0) -- (110,40) ;
			\draw    (110,20) -- (110,13) ;
			\draw [shift={(110,10)}, rotate = 90] [fill={rgb, 255:red, 0; green, 0; blue, 0 }  ][line width=0.08]  [draw opacity=0] (10.72,-5.15) -- (0,0) -- (10.72,5.15) -- (7.12,0) -- cycle    ;
			\draw    (80,80) -- (80,160) ;
			\draw    (80,120) -- (80,113) ;
			\draw [shift={(80,110)}, rotate = 90] [fill={rgb, 255:red, 0; green, 0; blue, 0 }  ][line width=0.08]  [draw opacity=0] (10.72,-5.15) -- (0,0) -- (10.72,5.15) -- (7.12,0) -- cycle    ;
			\draw    (130,80) -- (130,120) ;
			\draw  [draw opacity=0] (190,120) .. controls (190,120) and (190,120) .. (190,120) .. controls (190,120) and (190,120) .. (190,120) .. controls (190,136.57) and (176.57,150) .. (160,150) .. controls (143.43,150) and (130,136.57) .. (130,120) -- (160,120) -- cycle ; \draw   (190,120) .. controls (190,120) and (190,120) .. (190,120) .. controls (190,120) and (190,120) .. (190,120) .. controls (190,136.57) and (176.57,150) .. (160,150) .. controls (143.43,150) and (130,136.57) .. (130,120) ;  
			\draw    (190,0) -- (190,120) ;
			\draw    (190,50) -- (190,57) ;
			\draw [shift={(190,60)}, rotate = 270] [fill={rgb, 255:red, 0; green, 0; blue, 0 }  ][line width=0.08]  [draw opacity=0] (10.72,-5.15) -- (0,0) -- (10.72,5.15) -- (7.12,0) -- cycle    ;
			
			\draw (97,52.4) node [anchor=north west][inner sep=0.75pt]    {$f$};
			\draw (121,2.4) node [anchor=north west][inner sep=0.75pt]    {$Z$};
			\draw (55,142.4) node [anchor=north west][inner sep=0.75pt]    {$X$};
			\draw (201,2.4) node [anchor=north west][inner sep=0.75pt]    {$Y$};
			\draw (11,52.4) node [anchor=north west][inner sep=0.75pt]    {$\varphi(f)\ =$};

		\end{tikzpicture}
	\end{center}
	\begin{center}
		\tikzset{every picture/.style={line width=0.75pt}} 
		
		\begin{tikzpicture}[x=0.75pt,y=0.75pt,yscale=-1,xscale=1]
			
			\draw   (100,80) -- (170,80) -- (170,120) -- (100,120) -- cycle ;
			\draw    (110,0) -- (110,80) ;
			\draw    (110,60) -- (110,53) ;
			\draw [shift={(110,50)}, rotate = 90] [fill={rgb, 255:red, 0; green, 0; blue, 0 }  ][line width=0.08]  [draw opacity=0] (10.72,-5.15) -- (0,0) -- (10.72,5.15) -- (7.12,0) -- cycle    ;
			\draw    (140,120) -- (140,160) ;
			\draw    (140,140) -- (140,133) ;
			\draw [shift={(140,130)}, rotate = 90] [fill={rgb, 255:red, 0; green, 0; blue, 0 }  ][line width=0.08]  [draw opacity=0] (10.72,-5.15) -- (0,0) -- (10.72,5.15) -- (7.12,0) -- cycle    ;
			\draw    (160,40) -- (160,80) ;
			\draw  [draw opacity=0] (160,40) .. controls (160,23.43) and (173.43,10) .. (190,10) .. controls (206.57,10) and (220,23.43) .. (220,40) -- (190,40) -- cycle ; \draw   (160,40) .. controls (160,23.43) and (173.43,10) .. (190,10) .. controls (206.57,10) and (220,23.43) .. (220,40) ;  
			\draw    (220,40) -- (220,160) ;
			\draw    (220,100) -- (220,93) ;
			\draw [shift={(220,90)}, rotate = 90] [fill={rgb, 255:red, 0; green, 0; blue, 0 }  ][line width=0.08]  [draw opacity=0] (10.72,-5.15) -- (0,0) -- (10.72,5.15) -- (7.12,0) -- cycle    ;
			
			\draw (131,92.4) node [anchor=north west][inner sep=0.75pt]    {$g$};
			\draw (91,42.4) node [anchor=north west][inner sep=0.75pt]    {$Z$};
			\draw (115,142.4) node [anchor=north west][inner sep=0.75pt]    {$X$};
			\draw (201,142.4) node [anchor=north west][inner sep=0.75pt]    {$Y$};
			\draw (5,90.4) node [anchor=north west][inner sep=0.75pt]    {$\varphi ^{-1}( g) \ =$};

		\end{tikzpicture}
	\end{center}
\end{proposition}
\begin{proof}
	Done previously in the main text.
\end{proof}

This result is almost the property that $\finvec_F$ is a closed monoidal category. A closed monoidal category $\cat{C}$ has the property that
\begin{equation}
	\cat{C}(X \otimes Y, Z) \cong \cat{C} (X, [Y,Z]),
\end{equation}
where $[Y,Z]$ is an internal hom. In the case of $\finvec_F$, we take as internal hom the vector space
\begin{equation}
	[Y,Z] \coloneqq Y^* \otimes Z.
\end{equation}
Therefore, the isomorphism becomes
\begin{equation}
	\label{equation: isomorphism of closed monoidal for finvec}
	\finvec_F (X \otimes Y, Z) \cong \finvec_F (X, Y^* \otimes Z).
\end{equation}
But $\finvec_F$ is a symmetric category, so we have a symmetry isomorphism
\begin{equation*}
	\tau_{Z, Y^*} \colon Z \otimes Y^* \to Y^* \otimes Z .
\end{equation*}
Combining $\varphi$ from proposition \ref{proposition:pulling up or down} with $\tau_{Z, Y^*}$, we get the isomorphism of equation \ref{equation: isomorphism of closed monoidal for finvec}.

There is much more that can be done with string diagrams, but this will be enough for this work. For more, see \cite{MonoidalCatsAndTFT, Turaev+2016, Coecke_Kissinger_2017}.

\cleardoublepage
\chapter{Convex Sets in Finite Dimensions}
\label{appendix: convex sets}

Due to the probabilistic nature of Quantum Mechanics, many convex sets arise naturally. Here we define briefly what are convex sets and extreme points. Then we show some results about compact convex sets.

Given a (real or complex) vector space $X$, a convex set $C \subseteq X$ is a subset $C$ that is closed by convex combinations. A convex combination of vectors $v_1,\dots,v_n$ is a linear combination $\sum_{i=1}^n p_i v_i$, where $p_i \geq 0$ and $\sum_{i=1}^n p_i = 1$. For any subset $S\subseteq X$, its convex hull is the set of convex combinations of points of $S$. Therefore, a subset $C\subseteq X$ is convex if it equals its convex hull. An extreme point of a convex set $C$ is one that can't be written as a convex combination of different points of $C$. For example, $[0,1]$ is a convex set with 0 and 1 as extreme points. By a theorem of Krein and Milman, every compact convex set in a finite dimensional vector space is the convex hull of its extreme points (theorem 3.3 of \cite{barvinokcourse}). A similar result is also true for infinite dimensional spaces, but this work will only use the finite dimensional case.

\begin{theorem}{Krein-Milman}{convex hull}
	Let $X$ be a finite dimensional real or complex vector space and let $C \subseteq X$ be a compact convex subset. Then $C$ is the convex hull of its extreme points.
\end{theorem}

Of course, this isn't true for general convex sets. For example, $[0,+\infty)$ is a convex set, but it isn't compact. Also, it has 0 as its unique extreme point, whose convex hull is $\{0\}$, which isn't $[0,+\infty)$.

Theorem \ref{theorem:convex hull} doesn't say anything about the topology of the compact convex set. In the finite dimensional case, which is the one we are interested in, we can say more about it. The next proposition is an exercise in Topology (footnote of page 2 of \cite{rotman_AT}). From this proposition we can prove a corollary that says that any compact convex set in $\realNumbers^n$ is homeomorphic to some closed ball of dimension at most $n$.

\begin{proposition}{}{compact convex set with non empty interior}
	Let $C \subseteq \realNumbers^n$ be a compact convex subset with $\interior{C} \neq \varnothing$. Then $C$ is homeomorphic to the closed unit ball $\overline{B(0,1)} \subseteq \realNumbers^n$.
\end{proposition}
\begin{proof}
	Let $z \in \interior C$ and define $C' \coloneqq C-z = \{x-z \mid x \in C\}$. Then $C'$ is a compact convex set homeomorphic to $C$, with $0 \in \interior C'$. The idea is to pick a line segment of $C'$ which starts at 0 and ends at the furthest point along some direction which is still inside $C'$, and then rescale such line segment so that its length becomes 1. First we define a function $R$ that gives the length $R(x)$ of the line segment determined by the direction of a unit vector $x \in S^{n-1}$, where
	\begin{equation}
		S^{n-1} = \{x \in \realNumbers^n \mid ||x|| = 1 \}
	\end{equation}
	is the unit sphere. Since $0 \in \interior C'$, there exists some $r_{min} > 0$ such that
	\begin{equation}
		\label{equation: smaller ball}
		\overline{B(0,r_{min})} \subseteq C'.
	\end{equation}
	$C'$ is also compact, so it is bounded, which implies that there is some $r_{max} \geq r_{min}$ such that
	\begin{equation}
		\label{equation: bigger ball}
		C' \subseteq \overline{B(0,r_{max})}.
	\end{equation}
	For this reason, we must have that
	\begin{equation}
		r_{min} \leq R(x) \leq r_{max}.
	\end{equation}
	The function $R$ is then defined as follows:
	\begin{align*}
		R \colon S^{n-1} &\longrightarrow [r_{min}, r_{max}] \\
		x &\longmapsto \max \left\{\lambda > 0 \mid \lambda x \in C' \right\}.
	\end{align*}
	We must prove that this definition is correct, that is, that the maximum exists and that it is in the interval $[r_{min}, r_{max}]$. For the existence, first notice that
	\begin{equation}
		r_{min} \in \{\lambda > 0 \mid \lambda x \in C' \},
	\end{equation}
	because of equation \ref{equation: smaller ball}. Equation \ref{equation: bigger ball} implies that
	\begin{equation}
		\{\lambda > 0 \mid \lambda x \in C' \} \subseteq (0,r_{max}].
	\end{equation}
	This shows that
	\begin{equation*}
		\{\lambda > 0 \mid \lambda x \in C' \}
	\end{equation*}
	is a non empty and bounded set, therefore it has a supremum. The supremum is a maximum iff it is in the set
	\begin{equation*}
		\{\lambda > 0 \mid \lambda x \in C' \}.
	\end{equation*}
	Let
	\begin{equation}
		s \coloneqq \sup \{\lambda > 0 \mid \lambda x \in C' \}.
	\end{equation}
	From the definition of supremum, there is an increasing sequence in the set that converges to it. That is, there is an increasing sequence $(\lambda_m)_{m \geq 1}$, with
	\begin{equation}
		\lambda_m > 0,
	\end{equation}
	\begin{equation}
		\lambda_m x \in C',
	\end{equation}
	such that
	\begin{equation}
		\lim_{m \to \infty} \lambda_m = s.
	\end{equation}
	Since $C'$ is compact, it is also closed. This implies that
	\begin{equation}
		\lim_{m \to \infty} \lambda_m x \in C',
	\end{equation}
	so
	\begin{equation}
		s x \in C'.
	\end{equation}
	Also, since the sequence $(\lambda_m)_{m \geq 1}$ is increasing and $\lambda_m > 0$, it follows that
	\begin{equation}
		s \geq \lambda_1 > 0.
	\end{equation}
	This proves that the supremum is in the set:
	\begin{equation}
		s \in \{\lambda > 0 \mid \lambda x \in C' \}.
	\end{equation}
	Therefore, the maximum from the definition of $R$ exists. Since the set
	\begin{equation*}
		\{\lambda > 0 \mid \lambda x \in C' \}
	\end{equation*}
	contains $r_{min}$ and is bounded from above by $r_{max}$, it follows that
	\begin{equation}
		\label{equation: R bounds}
		R(x) \in [r_{min}, r_{max}].
	\end{equation}
	This concludes that the definition of $R$ is correct.
	
	Having defined the function $R$, we can use it to rescale the line segments. The function that does the rescaling is defined as follows:
	\begin{align*}
		\varphi \colon C' &\longrightarrow \overline{B(0,1)}\\
		x &\longmapsto \begin{cases}
			\frac{x}{R(x/||x||)} &\text{ if } x \in C' \setminus \{0\}\\
			0 &\text{ if } x=0.
		\end{cases}
	\end{align*}
	To finish the proof, we need to prove that $\varphi$ is a homeomorphism. First we'll prove that $R$ is continuous.
	
	Let $x, x' \in S^{n-1}$. Taking them to be close enough, they must satisfy that
	\begin{equation}
		\label{equation: x x' > 0}
		x \cdot x' > 0.
	\end{equation}
	For instance, this is the case if we assume that
	\begin{equation}
		\label{equation: x-x' close enough}
		||x-x'|| \leq ||x|| = 1.
	\end{equation}
	With such assumption, we have that
	\begin{align*}
		x \cdot x' &\stackrel{(i)}{=} \frac{||x||^2 +||x'||^2 -||x-x'||^2}{2} \\
		&\stackrel{(ii)}{\geq} \frac{||x'||^2}{2} \\
		&\stackrel{(iii)}{>} 0.
	\end{align*}
	In $(i)$ was used that $||y||^2 = y \cdot y$, to find and expression of the scalar product in terms of the norm; in $(ii)$ was used equation \ref{equation: x-x' close enough}; in $(iii)$ was used that $x' \neq 0$. With this result, the vectors $x$ and $x'$ are in the same half of the unit sphere $S^{n-1}$.
	
	Next we find a lower bound for $R(x')$ from the information about $x$ and $C'$. By symmetry, this will also give a lower bound for $R(x)$. Combining both bounds we'll get that $R$ is continuous. First rescale $x$ and $x'$ to the maximum norm possible, but remaining inside $C'$. We get the following vectors:
	\begin{equation}
		\widetilde{x} \coloneqq R(x) x,
	\end{equation}
	\begin{equation}
		\widetilde{x'} \coloneqq R(x') x'.
	\end{equation}
	We have a special case when $R(x) = r_{min}$. To avoid separating into multiple cases, we'll work with a strictly smaller radius. Let $r_s \in \realNumbers$ be any such that
	\begin{equation}
		0 < r_s < r_{min}.
	\end{equation}
	In this way,
	\begin{equation}
		R(x) > r_s
	\end{equation}
	for any $x \in S^{n-1}$. Since $x, x' \in S^{n-1}$, these vectors are linearly dependent iff $x' = \pm x$. For computing $\lim_{x' \to x} R(x')$ we can assume that $x' \neq x$. We are assuming that $x \cdot x' > 0$, as in equation \ref{equation: x x' > 0}, so $x' \neq -x$. This implies that $(x,x')$ is linearly independent. Consider the following picture:
	\begin{center}
		\tikzset{every picture/.style={line width=0.75pt}} 
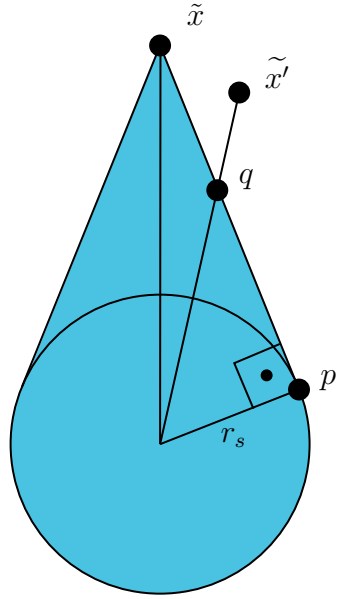
		\begin{tikzpicture}[x=0.75pt,y=0.75pt,yscale=-1,xscale=1]
			
			\draw  [draw opacity=0][fill={rgb, 255:red, 74; green, 194; blue, 226 }  ,fill opacity=1 ] (175.2,25) -- (244.7,197.6) -- (105.7,197.6) -- cycle ;
			\draw  [fill={rgb, 255:red, 74; green, 194; blue, 226 }  ,fill opacity=1 ] (100,225) .. controls (100,183.58) and (133.58,150) .. (175,150) .. controls (216.42,150) and (250,183.58) .. (250,225) .. controls (250,266.42) and (216.42,300) .. (175,300) .. controls (133.58,300) and (100,266.42) .. (100,225) -- cycle ;
			\draw    (175,225) -- (244.7,197.6) ;
			\draw  [fill={rgb, 255:red, 0; green, 0; blue, 0 }  ,fill opacity=1 ] (170,25) .. controls (170,22.24) and (172.24,20) .. (175,20) .. controls (177.76,20) and (180,22.24) .. (180,25) .. controls (180,27.76) and (177.76,30) .. (175,30) .. controls (172.24,30) and (170,27.76) .. (170,25) -- cycle ;
			\draw    (105.7,196.2) -- (175,25) ;
			\draw    (244.7,197.6) -- (175,25) ;
			\draw    (175,225) -- (214.7,48.6) ;
			\draw  [fill={rgb, 255:red, 0; green, 0; blue, 0 }  ,fill opacity=1 ] (209.7,48.6) .. controls (209.7,45.84) and (211.94,43.6) .. (214.7,43.6) .. controls (217.46,43.6) and (219.7,45.84) .. (219.7,48.6) .. controls (219.7,51.36) and (217.46,53.6) .. (214.7,53.6) .. controls (211.94,53.6) and (209.7,51.36) .. (209.7,48.6) -- cycle ;
			\draw  [fill={rgb, 255:red, 0; green, 0; blue, 0 }  ,fill opacity=1 ] (198.7,97.6) .. controls (198.7,94.84) and (200.94,92.6) .. (203.7,92.6) .. controls (206.46,92.6) and (208.7,94.84) .. (208.7,97.6) .. controls (208.7,100.36) and (206.46,102.6) .. (203.7,102.6) .. controls (200.94,102.6) and (198.7,100.36) .. (198.7,97.6) -- cycle ;
			\draw    (175,225) -- (175.2,25) ;
			\draw   (221.84,207.23) -- (212.21,184.37) -- (235.07,174.74) ;
			\draw  [fill={rgb, 255:red, 0; green, 0; blue, 0 }  ,fill opacity=1 ] (225.95,190.48) .. controls (225.95,189.1) and (227.07,187.98) .. (228.45,187.98) .. controls (229.83,187.98) and (230.95,189.1) .. (230.95,190.48) .. controls (230.95,191.86) and (229.83,192.98) .. (228.45,192.98) .. controls (227.07,192.98) and (225.95,191.86) .. (225.95,190.48) -- cycle ;
			\draw  [fill={rgb, 255:red, 0; green, 0; blue, 0 }  ,fill opacity=1 ] (239.7,197.6) .. controls (239.7,194.84) and (241.94,192.6) .. (244.7,192.6) .. controls (247.46,192.6) and (249.7,194.84) .. (249.7,197.6) .. controls (249.7,200.36) and (247.46,202.6) .. (244.7,202.6) .. controls (241.94,202.6) and (239.7,200.36) .. (239.7,197.6) -- cycle ;
			
			\draw (187,2.4) node [anchor=north west][inner sep=0.75pt]    {$\tilde{x}$};
			\draw (226,29.4) node [anchor=north west][inner sep=0.75pt]    {$\widetilde{x'}$};
			\draw (213,85.4) node [anchor=north west][inner sep=0.75pt]    {$q$};
			\draw (204,215.4) node [anchor=north west][inner sep=0.75pt]    {$r_s$};
			\draw (254,186.4) node [anchor=north west][inner sep=0.75pt]    {$p$};
		\end{tikzpicture}
	\end{center}
	Since $C'$ contains both $\widetilde{x}$ and $\overline{B(0, r_s)}$, and since $C'$ is convex, it contains the convex closure of $\{x\} \cup \overline{B(0, r_s)}$. This convex closure is represented by the blue region in the image. There is a vector $q$ with the biggest norm in the intersection of the line segment $\overline{0 \ \widetilde{x'}}$ with the convex closure. Since $\widetilde{x'}$ has the biggest norm possible, it is either in the convex closure and coincides with $q$, or it is outside the convex closure and $q$ has a norm lower than $R(x')$. Estimating the norm of $q$ then gives a lower bound for $R(x')$. From the picture it is clear that the lower bound will approach $R(x)$ from below as $x'$ approaches $x$. The line that contains $0$ and $\widetilde{x}$ divides the plane $\langle \widetilde{x}, \widetilde{x'} \rangle_\realNumbers$ in two halves. There is a unique vector $p$ in the same half plane of $\widetilde{x'}$ for which the line containing both $\widetilde{x}$ and $p$ is tangent to the sphere. The vector $q$ is a convex combination of $\widetilde{x}$ and $p$. We'll determine the weights of this convex combination using similarity of triangles.
	
	We need some convenient way to express the vectors. All of them are in the plane generated by $x$ and $x'$. We can complete $x$ to a basis of $\langle x, x' \rangle_\realNumbers$ using the Gram-Schmidt process, obtaining another vector
	\begin{equation}
		x^\perp \coloneqq \frac{x' -x \cdot x'}{||x' -x \cdot x'||}.
	\end{equation}
	The vectors $x$ and $x^\perp$ are an orthonormal basis for $\langle x, x' \rangle_\realNumbers$. Then we can write $\widetilde{x}$, $x'$ and $\widetilde{x'}$ as
	\begin{equation}
		\label{equation: tilde x}
		\widetilde{x} = R(x) x,
	\end{equation}
	\begin{equation}
		x' = x'_1 x^\perp + x'_2 x,
	\end{equation}
	\begin{equation}
		\label{equation: x' in base}
		\widetilde{x'} = \widetilde{x'}_1 x^\perp + \widetilde{x'}_2 x,
	\end{equation}
	for some $x'_1, x'_2, \widetilde{x'}_1, \widetilde{x'}_2 \geq 0$. Also, $p$ can be written as
	\begin{equation}
		p = p_1 x^\perp + p_2 x ,
	\end{equation}
	for some $p_1, p_2 \geq 0$. We can compute $p_i$ using trigonometry. Consider the following picture:
	\begin{center}
		\tikzset{every picture/.style={line width=0.75pt}} 
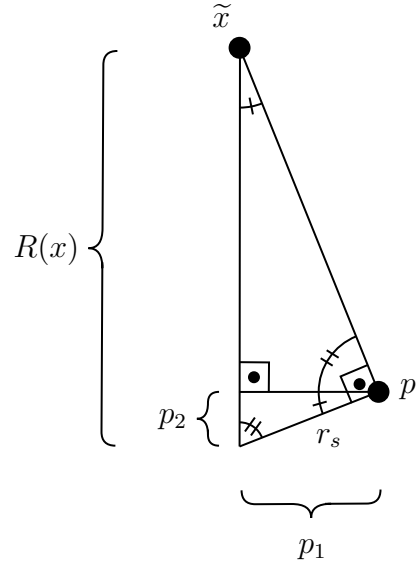
		\begin{tikzpicture}[x=0.75pt,y=0.75pt,yscale=-1,xscale=1]
			
			\draw    (175,225) -- (244.7,197.6) ;
			\draw  [fill={rgb, 255:red, 0; green, 0; blue, 0 }  ,fill opacity=1 ] (170,25) .. controls (170,22.24) and (172.24,20) .. (175,20) .. controls (177.76,20) and (180,22.24) .. (180,25) .. controls (180,27.76) and (177.76,30) .. (175,30) .. controls (172.24,30) and (170,27.76) .. (170,25) -- cycle ;
			\draw    (244.7,197.6) -- (175,25) ;
			\draw    (175,225) -- (175.2,25) ;
			\draw   (231.32,203.23) -- (225.69,189.86) -- (239.07,184.22) ;
			\draw  [fill={rgb, 255:red, 0; green, 0; blue, 0 }  ,fill opacity=1 ] (232.7,193.73) .. controls (232.7,192.35) and (233.81,191.23) .. (235.2,191.23) .. controls (236.58,191.23) and (237.7,192.35) .. (237.7,193.73) .. controls (237.7,195.11) and (236.58,196.23) .. (235.2,196.23) .. controls (233.81,196.23) and (232.7,195.11) .. (232.7,193.73) -- cycle ;
			\draw  [fill={rgb, 255:red, 0; green, 0; blue, 0 }  ,fill opacity=1 ] (239.7,197.6) .. controls (239.7,194.84) and (241.94,192.6) .. (244.7,192.6) .. controls (247.46,192.6) and (249.7,194.84) .. (249.7,197.6) .. controls (249.7,200.36) and (247.46,202.6) .. (244.7,202.6) .. controls (241.94,202.6) and (239.7,200.36) .. (239.7,197.6) -- cycle ;
			\draw    (174.7,197.65) -- (244.7,197.6) ;
			\draw   (174.83,182.7) -- (189.77,182.83) -- (189.65,197.78) ;
			\draw  [fill={rgb, 255:red, 0; green, 0; blue, 0 }  ,fill opacity=1 ] (179.72,190.26) .. controls (179.72,188.88) and (180.83,187.76) .. (182.22,187.76) .. controls (183.6,187.76) and (184.72,188.88) .. (184.72,190.26) .. controls (184.72,191.64) and (183.6,192.76) .. (182.22,192.76) .. controls (180.83,192.76) and (179.72,191.64) .. (179.72,190.26) -- cycle ;
			\draw  [draw opacity=0] (186.39,52.84) .. controls (182.93,54.23) and (179.16,55) .. (175.2,55) .. controls (175.04,55) and (174.87,55) .. (174.71,55) -- (175.2,25) -- cycle ; \draw   (186.39,52.84) .. controls (182.93,54.23) and (179.16,55) .. (175.2,55) .. controls (175.04,55) and (174.87,55) .. (174.71,55) ;  
			\draw  [draw opacity=0] (216.78,208.59) .. controls (215.44,205.19) and (214.7,201.48) .. (214.7,197.6) .. controls (214.7,185) and (222.47,174.21) .. (233.48,169.77) -- (244.7,197.6) -- cycle ; \draw   (216.78,208.59) .. controls (215.44,205.19) and (214.7,201.48) .. (214.7,197.6) .. controls (214.7,185) and (222.47,174.21) .. (233.48,169.77) ;  
			\draw  [draw opacity=0] (175.21,212.89) .. controls (180.31,212.98) and (184.64,216.22) .. (186.34,220.75) -- (175,225) -- cycle ; \draw   (175.21,212.89) .. controls (180.31,212.98) and (184.64,216.22) .. (186.34,220.75) ;  
			\draw    (180,50) -- (181.7,58.4) ;
			\draw    (182.7,210.48) -- (177.7,217.48) ;
			\draw    (185.7,212.48) -- (180.7,219.48) ;
			\draw    (218.7,202.48) -- (211.7,204.48) ;
			\draw    (224.7,180.48) -- (218.7,176.73) ;
			\draw    (221.7,184.48) -- (215.7,180.73) ;
			\draw   (176,247) .. controls (176.03,251.67) and (178.37,253.99) .. (183.04,253.96) -- (200.89,253.86) .. controls (207.56,253.82) and (210.9,256.13) .. (210.93,260.8) .. controls (210.9,256.13) and (214.22,253.78) .. (220.89,253.74)(217.89,253.76) -- (238.74,253.64) .. controls (243.41,253.61) and (245.73,251.27) .. (245.7,246.6) ;
			\draw   (164.7,197.6) .. controls (160.99,197.6) and (159.14,199.45) .. (159.14,203.16) -- (159.14,203.16) .. controls (159.14,208.45) and (157.29,211.1) .. (153.58,211.1) .. controls (157.29,211.1) and (159.14,213.75) .. (159.14,219.04)(159.14,216.66) -- (159.14,219.04) .. controls (159.14,222.75) and (160.99,224.6) .. (164.7,224.6) ;
			\draw   (113.7,26.6) .. controls (109.03,26.57) and (106.69,28.89) .. (106.66,33.56) -- (106.25,115.56) .. controls (106.22,122.23) and (103.87,125.55) .. (99.2,125.53) .. controls (103.87,125.55) and (106.18,128.89) .. (106.15,135.56)(106.16,132.56) -- (105.73,217.56) .. controls (105.71,222.23) and (108.03,224.57) .. (112.7,224.6) ;
			
			\draw (160,1.4) node [anchor=north west][inner sep=0.75pt]    {$\widetilde{x}$};
			\draw (211.85,214.7) node [anchor=north west][inner sep=0.75pt]    {$r_s$};
			\draw (254,190) node [anchor=north west][inner sep=0.75pt]    {$p$};
			\draw (133,205) node [anchor=north west][inner sep=0.75pt]    {$p_2$};
			\draw (203,270) node [anchor=north west][inner sep=0.75pt]    {$p_1$};
			\draw (60,116.4) node [anchor=north west][inner sep=0.75pt]    {$R(x)$};
		\end{tikzpicture}
	\end{center}
	We'll use similarity of triangles for the biggest and smallest triangles in the picture. The quotient of the lengths of the sides opposite to the angle with a single stroke and the one opposite to the right angle is the same for each triangle. This implies that
	\begin{equation}
		\frac{p_2}{r_s} = \frac{r_s}{R(x)},
	\end{equation}
	so
	\begin{equation}
		p_2 = \frac{r_s^2}{R(x)}.
	\end{equation}
	The other coordinate $p_1$ is computed using Pythagoras's theorem:
	\begin{equation}
		p_1 = \sqrt{r_s^2-p_2^2} = r_s \ \sqrt{1 - \frac{r_s^2}{R(x)^2}}.
	\end{equation}
	Then
	\begin{equation}
		\label{equation: p}
		p = r_s \left( \sqrt{1 - \frac{r_s^2}{R(x)^2}} \ x^\perp +  \frac{r_s}{R(x)} x \right).
	\end{equation}
	
	Next we find an expression for $q$. It is a convex combination of $\widetilde{x}$ and $p$, so it can be written as
	\begin{equation}
		q = t_1 \widetilde{x} + t_2 p,
	\end{equation}
	where $t_1, t_2 \geq 0$ and $t_1 + t_2 = 1$. The weights $t_1$ and $t_2$ are determined using trigonometry. Writing $q$ with respect to the vectors $x$ and $x^\perp$, we get that
	\begin{align*}
		q &\stackrel{(i)}{=} t_1 R(x) x  + t_2 r_s \left( \sqrt{1 - \frac{r_s^2}{R(x)^2}} \ x^\perp +  \frac{r_s}{R(x)} x \right) \\
		&\stackrel{(ii)}{=} t_2 r_s \sqrt{1 - \frac{r_s^2}{R(x)^2}} \ x^\perp + \left( t_1 R(x) + t_2 \frac{r_s^2}{R(x)} \right) x.
	\end{align*}
	In $(i)$ were used equations \ref{equation: tilde x} and \ref{equation: p}; in $(ii)$ the coefficients of $x^\perp$ and $x$ were grouped together. Since $q$ is a multiple of $\widetilde{x'}$, the quotient of its coordinates must equal the quotient of the coordinates of $\widetilde{x'}$. This is enough to determine the weights $t_i$. We have that
	\begin{equation}
		\frac{t_1 R(x) + t_2 \frac{r_s^2}{R(x)} }{ t_2 r_s \sqrt{1 - \frac{r_s^2}{R(x)^2}} } = \frac{|\widetilde{x'}_2|}{|\widetilde{x'}_1|}.
	\end{equation}
	We can pass the $t_2$ in the denominator to the nominator, dividing it:
	\begin{equation}
		\frac{\frac{t_1}{t_2} R(x) + \frac{r_s^2}{R(x)} }{ r_s \sqrt{1 - \frac{r_s^2}{R(x)^2}} } = \frac{|\widetilde{x'}_2|}{|\widetilde{x'}_1|}.
	\end{equation}
	Isolating $t_1 / t_2$, we get that
	\begin{equation}
		\label{equation: weights quotient 1}
		\frac{t_1}{t_2} = \frac{|\widetilde{x'}_2|}{|\widetilde{x'}_1|} \frac{r_s}{R(x)} \sqrt{1 - \frac{r_s^2}{R(x)^2}} -\frac{r_s^2}{R(x)^2}.
	\end{equation}
	Let's give a name to the right side. Define the function
	\begin{equation}
		\label{equation: f}
		f(x, x') \coloneqq \frac{|\widetilde{x'}_2|}{|\widetilde{x'}_1|} \frac{r_s}{R(x)} \sqrt{1 - \frac{r_s^2}{R(x)^2}} -\frac{r_s^2}{R(x)^2} .
	\end{equation}
	Equation \ref{equation: weights quotient 1} then becomes
	\begin{equation}
		\frac{t_1}{t_2} = f(x, x').
	\end{equation}
	Isolating $t_1$, we get that
	\begin{equation}
		\label{equation: weights quotient 2}
		t_1 = f(x, x') t_2.
	\end{equation}
	The weights also sum to 1, so
	\begin{equation}
		\label{equation: t2 = 1-t1}
		t_2 = 1-t_1.
	\end{equation}
	Substituting this expression in equation \ref{equation: weights quotient 2}, we get that
	\begin{equation}
		t_1 = f(x, x') (1-t_1) = f(x, x') - f(x, x') t_1.
	\end{equation}
	Isolating $t_1$, we get that
	\begin{equation}
		\label{equation: t1}
		t_1 = \frac{f(x, x')}{1 + f(x, x')}.
	\end{equation}
	We obtain $t_2$ from equation \ref{equation: t2 = 1-t1}:
	\begin{align*}
		t_2 &\stackrel{(i)}{=} 1-t_1 \\
		&\stackrel{(ii)}{=} 1 -\frac{f(x, x')}{1 + f(x, x')} \\
		&\stackrel{(iii)}{=} \frac{1+f(x, x') - f(x, x')}{1 + f(x, x')} \\
		&\stackrel{(iv)}{=} \frac{1}{1+f(x, x')}.
	\end{align*}
	In $(i)$ was used equation \ref{equation: t2 = 1-t1}; in $(ii)$ was used equation \ref{equation: t1}; in $(iii)$ the subtraction of numbers was turned into a single fraction; in $(iv)$ the $f(x')$ in the nominator was canceled. This proves that
	\begin{equation}
		\label{equation: t2}
		t_2 = \frac{1}{1+f(x, x')}.
	\end{equation}
	Previously we proved that
	\begin{equation}
		\label{equation: q}
		q = t_2 r_s \sqrt{1 - \frac{r_s^2}{R(x)^2}} \ x^\perp + \left( t_1 R(x) + t_2 \frac{r_s^2}{R(x)} \right) x.
	\end{equation}
	Having determined the weights $t_i$, we finished determining $q$.
	
	We'll use $q$ to estimate $R(x')$, since $R(x') \geq ||q||$. We'll see later that $\lim_{x' \to x} t_1 = 1$ and $\lim_{x' \to x} t_2 = 0$, so the relevant term of $q$ is $t_1 R(x)$. We can estimate $||q||$ as follows. From Pythagoras's theorem, we have that
	\begin{equation}
		||q||^2 = \left( t_2 r_s \sqrt{1 - \frac{r_s^2}{R(x)^2}} \ \right)^2 + \left( t_1 R(x) + t_2 \frac{r_s^2}{R(x)} \right)^2.
	\end{equation}
	Dropping the first square in the sum of squares and using that every term is non negative, we get that
	\begin{equation}
		||q|| \geq t_1 R(x) + t_2 \frac{r_s^2}{R(x)}.
	\end{equation}
	The term that contains $t_2$ is non negative, dropping it we get that
	\begin{equation}
		||q|| \geq t_1 R(x).
	\end{equation}
	Since $R(x') \geq ||q||$, this implies that
	\begin{equation}
		R(x') \geq t_1 R(x).
	\end{equation}
	Let's make the dependence on $x$ and $x'$ explicit. Using equation \ref{equation: t1}, we get that
	\begin{equation}
		R(x') \geq \frac{f(x,x')}{1+f(x,x')} R(x).
	\end{equation}
	Swapping $x$ and $x'$, we also have that
	\begin{equation}
		R(x) \geq \frac{f(x',x)}{1+f(x',x)} R(x').
	\end{equation}
	Isolating $R(x')$, we get that
	\begin{equation}
		\frac{1+f(x',x)}{f(x',x)} R(x) \geq R(x').
	\end{equation}
	The inequalities together imply that
	\begin{equation}
		\label{equation: R(x') bounds}
		\frac{f(x,x')}{1+f(x,x')} R(x) \leq R(x') \leq \frac{1+f(x',x)}{f(x',x)} R(x).
	\end{equation}
	To prove that $\lim_{x' \to x} R(x') = R(x)$, we'll show that:
	\begin{equation}
		\lim_{x' \to x} \frac{f(x,x')}{1+f(x,x')} = 1,
	\end{equation}
	\begin{equation}
		\lim_{x' \to x} \frac{1+f(x',x)}{f(x',x)} = 1.
	\end{equation}
	
	Before computing these limits, let's find a coordinate free expression for $f(x,x')$. The function was defined as
	\begin{equation}
		f(x, x') = \frac{|\widetilde{x'}_2|}{|\widetilde{x'}_1|} \frac{r_s}{R(x)} \sqrt{1 - \frac{r_s^2}{R(x)^2}} -\frac{r_s^2}{R(x)^2} .
	\end{equation}
	The relevant part is the fraction $|\widetilde{x'}_2| / |\widetilde{x'}_1|$. The components are related to $||\widetilde{x'}||$ by Pythagoras's theorem:
	\begin{equation}
		\label{equation: ||x' tilde|| pythagoras}
		||\widetilde{x'}||^2 = |\widetilde{x'}_1|^2 + |\widetilde{x'}_2|^2 .
	\end{equation}
	Using that $||\widetilde{x'}|| = R(x')$, we have that
	\begin{equation}
		\label{equation: x' tilde 1}
		|\widetilde{x'}_1| = \sqrt{R(x')^2 - |\widetilde{x'}_2|^2 }.
	\end{equation}
	From equation \ref{equation: x' in base}, the coordinate $\widetilde{x'}_2$ is given by the inner product
	\begin{equation}
		\widetilde{x'}_2 = \widetilde{x'} \cdot x.
	\end{equation}
	Taking the absolute value and using that $\widetilde{x'} = R(x') x'$, we get that:
	\begin{equation}
		|\widetilde{x'}_2 | = R(x') \ x' \cdot x.
	\end{equation}
	Again, by equation \ref{equation: x x' > 0}, we are assuming that $x \cdot x' > 0$. Substituting in equation \ref{equation: x' tilde 1}, we get that
	\begin{align*}
		|\widetilde{x'}_1| &= \sqrt{R(x')^2 - R(x')^2 (x' \cdot x )^2 } \\
		&= R(x') \sqrt{1 - (x' \cdot x )^2 }.
	\end{align*}
	In the last equality, the term $R(x')$ was factored. Next we use this result to find an expression for $|\widetilde{x'}_2|/|\widetilde{x'}_1|$. Using again equation \ref{equation: ||x' tilde|| pythagoras} and that $||\widetilde{x'}|| = R(x')$, we have that
	\begin{equation}
		\label{equation: x'tilde 2}
		|\widetilde{x'}_2| = \sqrt{R(x')^2 - |\widetilde{x'}_1|^2 }.
	\end{equation}
	Dividing by $|\widetilde{x'}_1 |$, we get that
	\begin{align*}
		\frac{|\widetilde{x'}_2|}{|\widetilde{x'}_1|} &\stackrel{(i)}{=} \sqrt{\frac{R(x')^2}{|\widetilde{x'}_1|^2} - 1 } \\
		&\stackrel{(ii)}{=} \sqrt{ \frac{1}{1 - (x' \cdot x )^2} -1}.
	\end{align*}
	In $(i)$ was used equation \ref{equation: x'tilde 2}; in $(ii)$ was used the expression for $|\widetilde{x'}_1|$. Substituting the expression for $|\widetilde{x'}_2| / |\widetilde{x'}_1|$ into the definition of $f$, as in equation \ref{equation: f}, we get that
	\begin{equation}
		\label{equation: f coordinate free}
		f(x,x') = \sqrt{ \frac{1}{1 - (x' \cdot x)^2} -1} \ \frac{r_s}{R(x)} \sqrt{1 - \frac{r_s^2}{R(x)^2}} -\frac{r_s^2}{R(x)^2}
	\end{equation}
	
	Next, let's compute $\lim_{x' \to x} f(x,x')$. As $x'$ approaches $x$, $x' \cdot x$ approaches $1$ from below. Therefore, $1 - (x' \cdot x)^2$ approaches $0$ from above as $x'$ approaches $x$. Then
	\begin{equation}
		\label{equation: limit square root}
		\lim_{x' \to x} \sqrt{ \frac{1}{1 - (x' \cdot x )^2} -1} = +\infty .
	\end{equation}
	Since
	\begin{equation}
		\frac{r_s}{R(x)} \sqrt{1 - \frac{r_s^2}{R(x)^2}} > 0,
	\end{equation}
	it follows that
	\begin{equation}
		\lim_{x' \to x} \sqrt{ \frac{1}{1 - (x' \cdot x )^2} -1} \ \frac{r_s}{R(x)} \sqrt{1 - \frac{r_s^2}{R(x)^2}} = +\infty.
	\end{equation}
	The remaining term of $f(x, x')$ doesn't depend on $x'$, so
	\begin{equation}
		\lim_{x' \to x} f(x, x') = +\infty.
	\end{equation}
	Then
	\begin{align*}
		\lim_{x' \to x} \frac{f(x,x')}{1+f(x,x')} &\stackrel{(i)}{=} \lim_{x' \to x} \frac{1+f(x,x')-1}{1+f(x,x')} \\
		&\stackrel{(ii)}{=} 1-\lim_{x' \to x} \frac{1}{1+f(x,x')} \\
		&\stackrel{(iii)}{=} 1 - 0 \\
		&= 1.
	\end{align*}
	In $(i)$ was added and subtracted 1 from the numerator; in $(ii)$ the term $1+f(x,x')$ from the numerator was divided by the denominator, obtaining 1; in $(iii)$ was used that $\lim_{x' \to x} f(x, x') = +\infty$. This proves that
	\begin{equation}
		\label{equation: limit fraction left}
		\lim_{x' \to x} \frac{f(x,x')}{1+f(x,x')} = 1.
	\end{equation}
	As said earlier, we would compute $\lim_{x' \to x} t_1$, which is the limit just computed. Since $t_2 = 1 - t_1$, of course $\lim_{x' \to x} t_2 = 0$.
	
	Now let's compute $\lim_{x' \to x} f(x', x)$. Swapping $x$ and $x'$ in equation \ref{equation: f coordinate free}, we get that
	\begin{equation}
		f(x', x) = \sqrt{ \frac{1}{1 - (x \cdot x' )^2} -1} \ \frac{r_s}{R(x')} \sqrt{1 - \frac{r_s^2}{R(x')^2}} -\frac{r_s^2}{R(x')^2}.
	\end{equation}
	The first square root is invariant by the swap, so we already know its limit, which is $+\infty$, as shown in equation \ref{equation: limit square root}. This time the other terms depend on $x'$, which makes it harder to compute the limit. By equation \ref{equation: R bounds}, $R(x')$ is bounded by $r_{min} \leq R(x') \leq r_{max}$. It follows that
	\begin{equation}
		\label{equation: bound 1}
		\frac{r_s}{R(x')} \sqrt{1 - \frac{r_s^2}{R(x')^2}} \geq \frac{r_s}{r_{max}} \sqrt{1 - \frac{r_s^2}{r_{min}^2}},
	\end{equation}
	\begin{equation}
		\label{equation: bound 2}
		-\frac{r_s^2}{R(x')^2} \geq -\frac{r_s^2}{r_{min}^2}.
	\end{equation}
	Notice that if we used $r_{min}$ instead of $r_s$, we would get the bound
	\begin{equation}
		\frac{r_{min}}{R(x')} \sqrt{1 - \frac{r_{min}^2}{R(x')^2}} \geq \frac{r_{min}}{r_{max}} \sqrt{1 - \frac{r_{min}^2}{r_{min}^2}} = 0,
	\end{equation}
	which is useless for computing the limit. Using the limit from equation \ref{equation: limit square root} and the bounds from equations \ref{equation: bound 1} and \ref{equation: bound 2}, it follows that
	\begin{equation}
		\lim_{x' \to x} f(x', x) = +\infty.
	\end{equation}
	Then
	\begin{align*}
		\lim_{x' \to x} \frac{1+f(x',x)}{f(x',x)} &\stackrel{(i)}{=} \frac{1}{f(x', x)} +1\\
		&\stackrel{(ii)}{=} 0+1 \\
		&= 1.
	\end{align*}
	In $(i)$ the $f(x', x)$ from the numerator was divided by the denominator, obtaining 1; in $(ii)$ was used that $\lim_{x' \to x} f(x', x) = +\infty$. This proves that
	\begin{equation}
		\label{equation: limit fraction right}
		\lim_{x' \to x} \frac{1+f(x',x)}{f(x',x)} = 1.
	\end{equation}
	
	The inequalities from equation \ref{equation: R(x') bounds} together with the limits from equations \ref{equation: limit fraction left} and \ref{equation: limit fraction right} imply that
	\begin{equation}
		\lim_{x' \to x} R(x') = R(x).
	\end{equation}
	This concludes that $R$ is a continuous function.
	
	It is immediate that $\varphi$ is continuous over $C' \setminus \{0\}$, because it is constructed from continuous functions. To prove that $\varphi$ is continuous, it remains to show the continuity at 0, that is,
	\begin{equation}
		\lim_{x \to 0} \varphi(x) = 0.
	\end{equation}
	We can assume when computing the limit that $x \neq 0$, so
	\begin{equation}
		\varphi(x) = \frac{x}{R(x/||x||)}.
	\end{equation}
	Taking the norm, we have that
	\begin{equation}
		||\varphi(x)|| = \frac{||x||}{R(x/||x||)}.
	\end{equation}
	By equation \ref{equation: R bounds}, we have that $R(x/||x||) \geq r_{min}$, so
	\begin{equation}
		||\varphi(x)|| \leq \frac{||x||}{r_{min}}.
	\end{equation}
	It follows from this inequality that
	\begin{equation}
		\lim_{x \to 0} \varphi(x) = 0.
	\end{equation}
	This concludes that $\varphi$ is a continuous function. The inverse is given by
	\begin{equation}
		\varphi^{-1}(x) = \begin{cases}
			R\left(\frac{x}{||x||}\right) x, &\text{ if } x \in \overline{B(0,1)} \setminus \{0\}, \\
			0, &\text{ if } x = 0.
		\end{cases}
	\end{equation}
	It continuous over $\overline{B(0,1)} \setminus \{0\}$, because it is constructed from continuous functions. From equation \ref{equation: R bounds}, we know that $R(x/||x||) \leq r_{max}$, so
	\begin{equation}
		||\varphi^{-1}(x)|| \leq r_{max} ||x||,
	\end{equation}
	for $x \in \overline{B(0,1)} \setminus \{0\}$. It follows that
	\begin{equation}
		\lim_{x \to 0} \varphi^{-1} (x) = 0.
	\end{equation}
	Therefore, $\varphi^{-1}$ is a continuous function, so $\varphi$ is a homeomorphism.
\end{proof}

\begin{corollary}{}{compact convex set is homeomorphic to a closed ball}
	Let $C \subseteq \realNumbers^n$ be a non empty compact convex subset. Then it is homeomorphic to a closed ball of dimension at most $n$. If $x \in C$, then the dimension of $C$ is
	\begin{equation}
		\dim_\realNumbers C = \dim_\realNumbers \langle C - x \rangle_\realNumbers,
	\end{equation}
	where $C-x \coloneqq \{c-x \mid c \in C\}$ and $\langle C - x \rangle_\realNumbers$ is the $\realNumbers$-vector space generated by $C - x$.
\end{corollary}
\begin{proof}
	Let $C' \coloneqq C-x$, then $0 \in C'$. Let also $V \coloneqq \langle C' \rangle_\realNumbers$. If $C' = \{0\}$, then $V = \{0\}$. In this case, $V \cong \realNumbers^0 = \{0\}$ and the closed ball $\overline{B(0,1)}$ in $\realNumbers^0$ is $\{0\}$, so $C'$ is already a closed ball in $V$. Of course, its dimension is $\dim_\realNumbers C' = 0 = \dim_\realNumbers V$.
	
	Now suppose that $\{0\} \subsetneq C'$, so that $\dim_\realNumbers V \geq 1$. Since $C'$ generates $V$, there is a basis $(c_1,\dots,c_d)$ for $V$, where $c_i \in C'$ and $d = \dim_\realNumbers V$. Since $0,c_1,\dots, c_d \in C'$ and $C'$ is convex, the convex hull of $\{0,c_1, \dots, c_d\}$ is contained in $C'$. This convex hull is the set
	\begin{equation}
		\left\{\sum_{i=1}^d t_i c_i \, \bigg| \, 0 \leq t_i \leq 1, \ \sum_{i=1}^d t_i \leq 1 \right\}.
	\end{equation}
	Notice that a convex combination of $d+1$ points would be parametrized by real numbers $t_0, t_1,\dots, t_d$, but, in this case, $t_0$ would be multiplied by the vector 0, so we only need the numbers $t_1, \dots, t_d$. Since $(c_1,\dots,c_d)$ is linearly independent, the convex hull of $\{0,c_1, \dots, c_d\}$ is a non degenerate $d$-simplex. As such, the subset $\{\sum_{i=1}^d t_i c_i \mid 0 <t_i < 1, \ \sum_{i=1}^d t_i < 1 \}$ is an open subset of $V$, so $C'$ has a non empty interior as a subspace of $V$. Since $V$ is a finite dimensional $\realNumbers$-vector space of dimension $d$, it is isomorphic to $\realNumbers^d$. Then, by proposition \ref{proposition:compact convex set with non empty interior}, $C'$ is homeomorphic to the closed unit ball $\overline{B(0,1)}$ in $\realNumbers^d$. The dimension of $C$ is then
	\begin{equation}
		\dim_\realNumbers C = \dim_\realNumbers C' = \dim_\realNumbers \overline{B(0,1)} = d = \dim_\realNumbers V = \dim_\realNumbers \langle C-x\rangle_\realNumbers.
	\end{equation}
	This concludes the proof.
\end{proof}

We'll also review some simple results. We prove that a product of two convex sets is also convex. Then we show that the extreme points of the product are the pairs of extreme points. Next we prove that a linear map that restricts to a bijection between two convex sets is also a bijection between their extreme points.

\begin{proposition}{}{product of convex sets is convex}
	Let $C \subseteq \realNumbers^n$ and $C' \subseteq \realNumbers^m$. If $C$ and $C'$ are convex sets, then $C \times C'$ is also a convex subset of $\realNumbers^n \times \realNumbers^m$.
\end{proposition}
\begin{proof}
	Let $t_1,t_2 \in [0,1]$ such that $t_1+t_2 = 1$. Let also $(x_1,x'_1), (x_2,x'_2) \in C \times C'$. Then
	\begin{equation}
		\sum_i t_i (x_i, x'_i) = \left(\sum_i t_i x_i, \sum_i t_i x'_i\right).
	\end{equation}
	Since $C$ and $C'$ are convex sets, then
	\begin{equation}
		\sum_i t_i x_i \in C,
	\end{equation}
	\begin{equation}
		\sum_i t_i x'_i \in C'.
	\end{equation}
	This proves that
	\begin{equation}
		\sum_i t_i (x_i, x'_i) \in C \times C'.
	\end{equation}
	This shows that $C \times C'$ is convex.
\end{proof}

\begin{proposition}{}{extreme points of product of convex sets}
	Let $C \subseteq \realNumbers^n$ and $C' \subseteq \realNumbers^m$, with $C$ and $C'$ convex sets. Then we have that
	\begin{equation*}
		\text{extreme points of } C\times C' = (\text{extreme points of } C) \times (\text{extreme points of } C').
	\end{equation*}
\end{proposition}
\begin{proof}
	Let $(x,x') \in C \times C'$. Suppose $x$ is not an extreme point of $C$, then we can write it as a non trivial convex combination
	\begin{equation}
		x = t_1 x_1 + t_2 x_2,
	\end{equation}
	where $t_i \in (0,1)$, $t_1 + t_2 = 1$, $x_i \in C$ and $x_1 \neq x_2$. This implies that
	\begin{equation}
		(x,x') = \left(\sum_i t_i x_i, \sum_i t_i x'\right) = \sum_i t_i (x_i,x').
	\end{equation}
	This is a non trivial convex combination, so $(x,x')$ isn't an extreme point of $C \times C'$. Similarly, if $x'$ isn't an extreme point of $C'$, then $(x,x')$ isn't an extreme point of $C \times C'$. This shows that
	\begin{equation*}
		\text{extreme points of } C\times C' \subseteq (\text{extreme points of } C) \times (\text{extreme points of } C').
	\end{equation*}
	
	Now suppose that $(x,x')$ isn't an extreme point of $C \times C'$. Then we can write it as a non trivial convex combination
	\begin{equation}
		(x,x') = t_1 (x_1, x'_1) + t_2 (x_2, x'_2),
	\end{equation}
	where $t_i \in (0,1)$, $t_1 + t_2 = 1$, $x_i \in C$, $x'_i \in C'$ and $(x_1,x'_1) \neq (x_2,x'_2)$. The convex combination above implies that
	\begin{equation}
		x = t_1 x_1 + t_2 x_2
	\end{equation}
	and
	\begin{equation}
		x' = t_1 x'_1 + t_2 x'_2.
	\end{equation}
	Since $(x_1,x'_1) \neq (x_2,x'_2)$, we must have $x_1 \neq x_2$ or $x'_1 \neq x'_2$. If $x_1 \neq x_2$, then $x$ is a non trivial convex combination, so $x$ isn't extreme. In the same way, if $x'_1 \neq x'_2$ then $x'$ isn't extreme. This proves that
	\begin{equation*}
		\text{extreme points of } C\times C' \supseteq (\text{extreme points of } C) \times (\text{extreme points of } C').
	\end{equation*}
\end{proof}

\begin{proposition}{}{linear bijection between extreme points}
	Let $T \colon \realNumbers^m \to \realNumbers^n$ be a $\realNumbers$-linear map. Let also $C \subseteq \realNumbers^m$ and $C' \subseteq \realNumbers^n$ be convex sets. If $T$ restricts to a bijection $T\colon C \to C'$, then it also restricts to a bijection between their extreme points.
\end{proposition}
\begin{proof}
	Let $x \in C$ and suppose $x$ isn't an extreme point of $C$. Then it can be written as a non trivial convex combination
	\begin{equation}
		x = t_1 x_1 + t_2 x_2,
	\end{equation}
	where $t_i \in (0,1)$, $t_1+t_2 = 1$, $x_i \in C$ and $x_1 \neq x_2$. Applying $T$ we get that
	\begin{equation}
		T(x) = t_1 T(x_1) + t_2 T(x_2).
	\end{equation}
	Since $T$ is injective over $C$, and since $x_1 \neq x_2$, we have that $T(x_1) \neq T(x_2)$. Therefore, $T(x)$ is a non trivial convex combination of points in $C'$. This proves that
	\begin{equation}
		T(x) \text{ is extreme } \implies x \text{ is extreme}.
	\end{equation}
	
	Now suppose that $T(x)$ isn't an extreme point of $C'$. Then we can write it as a non trivial convex combination
	\begin{equation}
		T(x) = t_1 y_1 + t_2 y_2,
	\end{equation}
	where  $t_i \in (0,1)$, $t_1+t_2 = 1$, $y_i \in C'$ and $y_1 \neq y_2$. Since $T(C) = C'$, there exists $x_i \in C$ such that $y_i = T(x_i)$. By linearity of $T$, we have
	\begin{equation}
		T(x) = t_1 T(x_1) + t_2 T(x_2) = T(t_1 x_1 + t_2 x_2).
	\end{equation}
	Since $C$ is convex, we have that $t_1 x_1 + t_2 x_2 \in C$. Since $T$ is injective over $C$, then
	\begin{equation}
		x = t_1 x_1 + t_2 x_2.
	\end{equation}
	Also, since $y_1 \neq y_2$, we must have $x_1 \neq x_2$. Therefore, $x$ is a non trivial convex combination of points in $C$, so it isn't an extreme point of $C$. This shows that
	\begin{equation}
		x \text{ is extreme } \implies T(x) \text{ is extreme}.
	\end{equation}
\end{proof}

\cleardoublepage

\chapter{Examples of Extreme UCPTP Maps}

\label{appendix: examples extreme ucptp}

In section \ref{section: method 2 choice of parameters} was presented a method that generated a large number of examples of extreme UCPTP maps. The examples were computed using a C program. The results are exact, which was possible due to the Fast Library for Number Theory (FLINT) C library. Examples were computed for dimensions $d$ with $3 \leq d \leq 22$. The method was applied for each Choi-rank $r$ with $3 \leq r \leq \sqrt{2}d$. The upper bound $\sqrt{2}d$ is the maximum Choi-rank allowed by corollary \ref{corollary:bound on rank of extreme ucptp}. The method sometimes failed to produce examples for a given $r$, we only show in this appendix the cases for which it succeeded in producing examples. As in section \ref{section: method 2 choice of parameters}, for a given  $(d,r)$, each example is parameterized by $N$, $A$ and $B$. As said in section \ref{section: method 2 choice of parameters}, $B$ is taken to be
\begin{equation}
	B = AN.
\end{equation}
Also, $A$ is taken to be the biggest number such that
\begin{equation}
	d+1\leq A \leq dr
\end{equation}
and
\begin{equation}
	\gcd(A, dr) = 1,
\end{equation}
if it exists. In this way, the only relevant numbers are $d$, $r$ and $N$, since $A$ and $B$ can be deduced from them. For completeness, the values of $A$ and $B$ are also presented. The computed examples are split into many tables, starting from table \ref{table: method 2 parameters part 1}. There is a total of 915 examples.

\begin{table}
	\caption{Parameters for the examples of UCPTP maps obtained using the method of section \ref{section: method 2 choice of parameters}, part 1.}
	\label{table: method 2 parameters part 1}
	\begin{center}
\begin{tabular}{ |c|c|c|c|c|c| }
			\hline
			number&d&r&N&A&B\\
			\hline
			1&3&3&2&4&8\\
			\hline
			2&4&4&3&5&15\\
			\hline
			3&5&3&2&7&14\\
			\hline
			4&5&5&4&6&24\\
			\hline
			5&5&6&3&7&21\\
			\hline
			6&5&6&4&7&28\\
			\hline
			7&6&3&2&7&14\\
			\hline
			8&6&5&4&7&28\\
			\hline
			9&6&6&5&7&35\\
			\hline
			10&7&3&2&10&20\\
			\hline
			11&7&4&3&9&27\\
			\hline
			12&7&5&3&11&33\\
			\hline
			13&7&5&4&8&32\\
			\hline
			14&7&7&5&9&45\\
			\hline
			15&7&7&6&8&48\\
			\hline
			16&7&8&6&9&54\\
			\hline
			17&7&9&6&10&60\\
			\hline
			18&8&3&2&11&22\\
			\hline
			19&8&4&3&9&27\\
			\hline
			20&8&6&3&13&39\\
			\hline
			21&8&6&4&11&44\\
			\hline
			22&8&7&6&9&54\\
			\hline
			23&8&8&5&11&55\\
			\hline
			24&8&8&7&9&63\\
			\hline
			25&8&9&5&13&65\\
			\hline
			26&8&9&6&11&66\\
			\hline
			27&8&10&7&11&77\\
			\hline
			28&9&3&2&13&26\\
			\hline
			29&9&4&3&11&33\\
			\hline
			30&9&5&3&14&42\\
			\hline
			31&9&5&4&11&44\\
			\hline
			32&9&6&4&13&52\\
			\hline
			33&9&7&4&13&52\\
			\hline
			34&9&7&5&11&55\\
			\hline
			35&9&7&6&10&60\\
			\hline
		\end{tabular}
	\end{center}
\end{table}

\begin{table}
	\label{table: method 2 parameters part 2}
	\caption{Parameters for the examples of UCPTP maps obtained using the method of section \ref{section: method 2 choice of parameters}, part 2.}
	\begin{center}
		\begin{tabular}{ |c|c|c|c|c|c| }
			\hline
			number&d&r&N&A&B\\
			\hline
			36&9&8&5&13&65\\
			\hline
			37&9&8&6&11&66\\
			\hline
			38&9&9&6&13&78\\
			\hline
			39&9&9&7&11&77\\
			\hline
			40&9&9&8&10&80\\
			\hline
			41&9&10&7&11&77\\
			\hline
			42&9&10&8&11&88\\
			\hline
			43&9&11&8&10&80\\
			\hline
			44&9&12&8&13&104\\
			\hline
			45&10&3&2&13&26\\
			\hline
			46&10&4&3&13&39\\
			\hline
			47&10&5&3&13&39\\
			\hline
			48&10&5&4&11&44\\
			\hline
			49&10&6&4&13&52\\
			\hline
			50&10&7&5&13&65\\
			\hline
			51&10&7&6&11&66\\
			\hline
			52&10&8&7&11&77\\
			\hline
			53&10&9&6&13&78\\
			\hline
			54&10&9&7&11&77\\
			\hline
			55&10&9&8&11&88\\
			\hline
			56&10&10&7&13&91\\
			\hline
			57&10&10&9&11&99\\
			\hline
			58&10&11&6&17&102\\
			\hline
			59&10&11&7&13&91\\
			\hline
			60&10&11&8&13&104\\
			\hline
			61&10&12&8&13&104\\
			\hline
			62&10&12&9&13&117\\
			\hline
			63&11&3&2&16&32\\
			\hline
			64&11&4&3&13&39\\
			\hline
			65&11&5&4&13&52\\
			\hline
			66&11&6&3&19&57\\
			\hline
			67&11&6&4&13&52\\
			\hline
			68&11&6&5&13&65\\
			\hline
			69&11&7&5&15&75\\
			\hline
			70&11&7&6&12&72\\
			\hline
		\end{tabular}
	\end{center}
\end{table}

\begin{table}
	\label{table: method 2 parameters part 3}
	\caption{Parameters for the examples of UCPTP maps obtained using the method of section \ref{section: method 2 choice of parameters}, part 3.}
	\begin{center}
		\begin{tabular}{ |c|c|c|c|c|c| }
			\hline
			number&d&r&N&A&B\\
			\hline
			71&11&8&5&17&85\\
			\hline
			72&11&8&6&13&78\\
			\hline
			73&11&9&4&23&92\\
			\hline
			74&11&9&5&19&95\\
			\hline
			75&11&9&6&16&96\\
			\hline
			76&11&10&6&17&102\\
			\hline
			77&11&10&7&13&91\\
			\hline
			78&11&10&8&13&104\\
			\hline
			79&11&11&9&13&117\\
			\hline
			80&11&11&10&12&120\\
			\hline
			81&11&12&7&17&119\\
			\hline
			82&11&12&8&13&104\\
			\hline
			83&11&12&9&13&117\\
			\hline
			84&11&12&10&13&130\\
			\hline
			85&11&13&8&17&136\\
			\hline
			86&11&13&9&15&135\\
			\hline
			87&11&13&10&14&140\\
			\hline
			88&11&14&10&15&150\\
			\hline
			89&11&15&10&16&160\\
			\hline
			90&11&15&12&13&156\\
			\hline
			91&12&3&2&17&34\\
			\hline
			92&12&4&3&13&39\\
			\hline
			93&12&5&3&19&57\\
			\hline
			94&12&5&4&13&52\\
			\hline
			95&12&6&4&17&68\\
			\hline
			96&12&6&5&13&65\\
			\hline
			97&12&7&5&13&65\\
			\hline
			98&12&8&7&13&91\\
			\hline
			99&12&9&5&19&95\\
			\hline
			100&12&9&6&17&102\\
			\hline
			101&12&9&7&13&91\\
			\hline
			102&12&9&8&13&104\\
			\hline
			103&12&10&9&13&117\\
			\hline
			104&12&11&7&17&119\\
			\hline
			105&12&11&8&13&104\\
			\hline
		\end{tabular}
	\end{center}
\end{table}

\begin{table}
	\label{table: method 2 parameters part 4}
	\caption{Parameters for the examples of UCPTP maps obtained using the method of section \ref{section: method 2 choice of parameters}, part 4.}
	\begin{center}
		\begin{tabular}{ |c|c|c|c|c|c| }
			\hline
			number&d&r&N&A&B\\
			\hline
			106&12&11&9&13&117\\
			\hline
			107&12&11&10&13&130\\
			\hline
			108&12&12&7&19&133\\
			\hline
			109&12&12&11&13&143\\
			\hline
			110&12&13&7&19&133\\
			\hline
			111&12&13&9&17&153\\
			\hline
			112&12&14&9&17&153\\
			\hline
			113&12&14&11&13&143\\
			\hline
			114&12&16&11&17&187\\
			\hline
			115&12&16&13&13&169\\
			\hline
			116&13&3&2&19&38\\
			\hline
			117&13&4&3&17&51\\
			\hline
			118&13&5&3&21&63\\
			\hline
			119&13&5&4&16&64\\
			\hline
			120&13&6&4&19&76\\
			\hline
			121&13&7&4&22&88\\
			\hline
			122&13&7&6&15&90\\
			\hline
			123&13&8&5&19&95\\
			\hline
			124&13&8&6&17&102\\
			\hline
			125&13&9&6&19&114\\
			\hline
			126&13&9&7&16&112\\
			\hline
			127&13&9&8&14&112\\
			\hline
			128&13&10&5&23&115\\
			\hline
			129&13&10&6&21&126\\
			\hline
			130&13&10&7&17&119\\
			\hline
			131&13&11&6&23&138\\
			\hline
			132&13&11&8&17&136\\
			\hline
			133&13&11&9&15&135\\
			\hline
			134&13&11&10&14&140\\
			\hline
			135&13&12&7&19&133\\
			\hline
			136&13&12&8&19&152\\
			\hline
			137&13&12&9&17&153\\
			\hline
			138&13&13&9&18&162\\
			\hline
			139&13&13&10&16&160\\
			\hline
			140&13&13&11&15&165\\
			\hline
		\end{tabular}
	\end{center}
\end{table}

\begin{table}
	\label{table: method 2 parameters part 5}
	\caption{Parameters for the examples of UCPTP maps obtained using the method of section \ref{section: method 2 choice of parameters}, part 5.}
	\begin{center}
		\begin{tabular}{ |c|c|c|c|c|c| }
			\hline
			number&d&r&N&A&B\\
			\hline
			141&13&13&12&14&168\\
			\hline
			142&13&14&8&19&152\\
			\hline
			143&13&14&9&19&171\\
			\hline
			144&13&14&10&17&170\\
			\hline
			145&13&14&11&15&165\\
			\hline
			146&13&14&12&15&180\\
			\hline
			147&13&15&8&23&184\\
			\hline
			148&13&15&11&17&187\\
			\hline
			149&13&15&12&16&192\\
			\hline
			150&13&16&11&17&187\\
			\hline
			151&13&16&12&17&204\\
			\hline
			152&13&17&12&18&216\\
			\hline
			153&13&17&14&15&210\\
			\hline
			154&14&3&2&19&38\\
			\hline
			155&14&4&3&17&51\\
			\hline
			156&14&5&4&17&68\\
			\hline
			157&14&6&3&25&75\\
			\hline
			158&14&6&4&19&76\\
			\hline
			159&14&7&4&23&92\\
			\hline
			160&14&7&5&19&95\\
			\hline
			161&14&7&6&15&90\\
			\hline
			162&14&8&5&19&95\\
			\hline
			163&14&8&6&17&102\\
			\hline
			164&14&9&6&19&114\\
			\hline
			165&14&9&7&17&119\\
			\hline
			166&14&10&8&17&136\\
			\hline
			167&14&11&10&15&150\\
			\hline
			168&14&12&8&19&152\\
			\hline
			169&14&12&9&17&153\\
			\hline
			170&14&13&8&19&152\\
			\hline
			171&14&13&9&19&171\\
			\hline
			172&14&13&10&17&170\\
			\hline
			173&14&13&11&15&165\\
			\hline
			174&14&13&12&15&180\\
			\hline
			175&14&14&8&23&184\\
			\hline
		\end{tabular}
	\end{center}
\end{table}

\begin{table}
	\label{table: method 2 parameters part 6}
	\caption{Parameters for the examples of UCPTP maps obtained using the method of section \ref{section: method 2 choice of parameters}, part 6.}
	\begin{center}
		\begin{tabular}{ |c|c|c|c|c|c| }
			\hline
			number&d&r&N&A&B\\
			\hline
			176&14&14&9&19&171\\
			\hline
			177&14&14&10&19&190\\
			\hline
			178&14&14&11&17&187\\
			\hline
			179&14&14&13&15&195\\
			\hline
			180&14&15&12&17&204\\
			\hline
			181&14&16&11&19&209\\
			\hline
			182&14&16&12&17&204\\
			\hline
			183&14&16&13&17&221\\
			\hline
			184&14&17&11&19&209\\
			\hline
			185&14&17&12&19&228\\
			\hline
			186&14&17&13&15&195\\
			\hline
			187&14&18&11&19&209\\
			\hline
			188&14&18&13&19&247\\
			\hline
			189&14&19&13&17&221\\
			\hline
			190&14&19&15&17&255\\
			\hline
			191&15&3&2&22&44\\
			\hline
			192&15&4&3&19&57\\
			\hline
			193&15&5&4&17&68\\
			\hline
			194&15&6&4&19&76\\
			\hline
			195&15&6&5&17&85\\
			\hline
			196&15&7&5&19&95\\
			\hline
			197&15&7&6&17&102\\
			\hline
			198&15&8&6&19&114\\
			\hline
			199&15&8&7&17&119\\
			\hline
			200&15&9&4&32&128\\
			\hline
			201&15&9&6&22&132\\
			\hline
			202&15&9&7&19&133\\
			\hline
			203&15&9&8&16&128\\
			\hline
			204&15&10&6&23&138\\
			\hline
			205&15&10&7&19&133\\
			\hline
			206&15&10&8&17&136\\
			\hline
			207&15&11&6&26&156\\
			\hline
			208&15&11&7&23&161\\
			\hline
			209&15&11&8&19&152\\
			\hline
			210&15&11&9&17&153\\
			\hline
		\end{tabular}
	\end{center}
\end{table}

\begin{table}
	\label{table: method 2 parameters part 7}
	\caption{Parameters for the examples of UCPTP maps obtained using the method of section \ref{section: method 2 choice of parameters}, part 7.}
	\begin{center}
		\begin{tabular}{ |c|c|c|c|c|c| }
			\hline
			number&d&r&N&A&B\\
			\hline
			211&15&11&10&16&160\\
			\hline
			212&15&12&7&23&161\\
			\hline
			213&15&12&8&19&152\\
			\hline
			214&15&12&10&17&170\\
			\hline
			215&15&13&8&23&184\\
			\hline
			216&15&13&9&19&171\\
			\hline
			217&15&13&10&19&190\\
			\hline
			218&15&13&11&17&187\\
			\hline
			219&15&13&12&16&192\\
			\hline
			220&15&14&9&23&207\\
			\hline
			221&15&14&12&17&204\\
			\hline
			222&15&15&9&23&207\\
			\hline
			223&15&15&10&22&220\\
			\hline
			224&15&15&11&19&209\\
			\hline
			225&15&15&12&17&204\\
			\hline
			226&15&15&13&17&221\\
			\hline
			227&15&15&14&16&224\\
			\hline
			228&15&16&11&19&209\\
			\hline
			229&15&16&12&19&228\\
			\hline
			230&15&16&13&17&221\\
			\hline
			231&15&16&14&17&238\\
			\hline
			232&15&17&12&19&228\\
			\hline
			233&15&17&13&19&247\\
			\hline
			234&15&17&14&16&224\\
			\hline
			235&15&18&11&23&253\\
			\hline
			236&15&18&12&19&228\\
			\hline
			237&15&18&13&19&247\\
			\hline
			238&15&18&14&19&266\\
			\hline
			239&15&19&13&17&221\\
			\hline
			240&15&19&14&17&238\\
			\hline
			241&15&19&15&17&255\\
			\hline
			242&15&19&16&17&272\\
			\hline
			243&15&20&14&19&266\\
			\hline
			244&15&20&16&17&272\\
			\hline
			245&15&20&17&17&289\\
			\hline
		\end{tabular}
	\end{center}
\end{table}

\begin{table}
	\label{table: method 2 parameters part 8}
	\caption{Parameters for the examples of UCPTP maps obtained using the method of section \ref{section: method 2 choice of parameters}, part 8.}
	\begin{center}
		\begin{tabular}{ |c|c|c|c|c|c| }
			\hline
			number&d&r&N&A&B\\
			\hline
			246&15&21&16&19&304\\
			\hline
			247&15&21&17&17&289\\
			\hline
			248&16&3&2&23&46\\
			\hline
			249&16&4&3&21&63\\
			\hline
			250&16&5&3&23&69\\
			\hline
			251&16&5&4&19&76\\
			\hline
			252&16&6&4&23&92\\
			\hline
			253&16&6&5&19&95\\
			\hline
			254&16&7&4&27&108\\
			\hline
			255&16&7&5&19&95\\
			\hline
			256&16&7&6&17&102\\
			\hline
			257&16&8&5&25&125\\
			\hline
			258&16&8&7&17&119\\
			\hline
			259&16&9&5&25&125\\
			\hline
			260&16&9&6&23&138\\
			\hline
			261&16&9&7&19&133\\
			\hline
			262&16&10&7&21&147\\
			\hline
			263&16&10&8&19&152\\
			\hline
			264&16&10&9&17&153\\
			\hline
			265&16&11&8&21&168\\
			\hline
			266&16&11&9&19&171\\
			\hline
			267&16&11&10&17&170\\
			\hline
			268&16&12&7&25&175\\
			\hline
			269&16&12&8&23&184\\
			\hline
			270&16&12&9&19&171\\
			\hline
			271&16&12&11&17&187\\
			\hline
			272&16&13&10&19&190\\
			\hline
			273&16&13&11&17&187\\
			\hline
			274&16&13&12&17&204\\
			\hline
			275&16&14&9&23&207\\
			\hline
			276&16&14&10&19&190\\
			\hline
			277&16&14&11&19&209\\
			\hline
			278&16&14&13&17&221\\
			\hline
			279&16&15&9&23&207\\
			\hline
			280&16&15&10&23&230\\
			\hline
		\end{tabular}
	\end{center}
\end{table}

\begin{table}
	\label{table: method 2 parameters part 9}
	\caption{Parameters for the examples of UCPTP maps obtained using the method of section \ref{section: method 2 choice of parameters}, part 9.}
	\begin{center}
		\begin{tabular}{ |c|c|c|c|c|c| }
			\hline
			number&d&r&N&A&B\\
			\hline
			281&16&15&11&19&209\\
			\hline
			282&16&15&12&19&228\\
			\hline
			283&16&15&13&17&221\\
			\hline
			284&16&15&14&17&238\\
			\hline
			285&16&16&9&27&243\\
			\hline
			286&16&16&12&21&252\\
			\hline
			287&16&16&13&19&247\\
			\hline
			288&16&16&15&17&255\\
			\hline
			289&16&17&11&23&253\\
			\hline
			290&16&17&13&19&247\\
			\hline
			291&16&17&14&19&266\\
			\hline
			292&16&18&11&25&275\\
			\hline
			293&16&18&12&23&276\\
			\hline
			294&16&18&13&19&247\\
			\hline
			295&16&18&14&19&266\\
			\hline
			296&16&18&15&19&285\\
			\hline
			297&16&19&13&23&299\\
			\hline
			298&16&19&14&21&294\\
			\hline
			299&16&19&15&17&255\\
			\hline
			300&16&20&13&23&299\\
			\hline
			301&16&20&14&21&294\\
			\hline
			302&16&20&15&21&315\\
			\hline
			303&16&20&17&17&289\\
			\hline
			304&16&21&13&25&325\\
			\hline
			305&16&21&15&19&285\\
			\hline
			306&16&21&17&19&323\\
			\hline
			307&16&22&17&19&323\\
			\hline
			308&16&22&18&19&342\\
			\hline
			309&16&22&19&17&323\\
			\hline
			310&17&3&2&25&50\\
			\hline
			311&17&4&3&21&63\\
			\hline
			312&17&5&4&21&84\\
			\hline
			313&17&6&3&31&93\\
			\hline
			314&17&6&4&25&100\\
			\hline
			315&17&6&5&19&95\\
			\hline
		\end{tabular}
	\end{center}
\end{table}

\begin{table}
	\label{table: method 2 parameters part 10}
	\caption{Parameters for the examples of UCPTP maps obtained using the method of section \ref{section: method 2 choice of parameters}, part 10.}
	\begin{center}
		\begin{tabular}{ |c|c|c|c|c|c| }
			\hline
			number&d&r&N&A&B\\
			\hline
			316&17&7&4&29&116\\
			\hline
			317&17&7&5&23&115\\
			\hline
			318&17&7&6&19&114\\
			\hline
			319&17&8&6&21&126\\
			\hline
			320&17&8&7&19&133\\
			\hline
			321&17&9&5&29&145\\
			\hline
			322&17&9&6&25&150\\
			\hline
			323&17&9&7&20&140\\
			\hline
			324&17&9&8&19&152\\
			\hline
			325&17&10&6&27&162\\
			\hline
			326&17&10&7&23&161\\
			\hline
			327&17&10&8&21&168\\
			\hline
			328&17&11&7&26&182\\
			\hline
			329&17&11&8&23&184\\
			\hline
			330&17&11&9&20&180\\
			\hline
			331&17&11&10&18&180\\
			\hline
			332&17&12&5&37&185\\
			\hline
			333&17&12&8&25&200\\
			\hline
			334&17&12&9&19&171\\
			\hline
			335&17&12&10&19&190\\
			\hline
			336&17&13&8&27&216\\
			\hline
			337&17&13&9&24&216\\
			\hline
			338&17&13&12&18&216\\
			\hline
			339&17&14&8&29&232\\
			\hline
			340&17&14&9&25&225\\
			\hline
			341&17&14&10&23&230\\
			\hline
			342&17&14&11&19&209\\
			\hline
			343&17&14&12&19&228\\
			\hline
			344&17&15&8&31&248\\
			\hline
			345&17&15&10&23&230\\
			\hline
			346&17&15&11&23&253\\
			\hline
			347&17&15&12&19&228\\
			\hline
			348&17&15&13&19&247\\
			\hline
			349&17&16&9&29&261\\
			\hline
			350&17&16&11&23&253\\
			\hline
		\end{tabular}
	\end{center}
\end{table}

\begin{table}
	\label{table: method 2 parameters part 11}
	\caption{Parameters for the examples of UCPTP maps obtained using the method of section \ref{section: method 2 choice of parameters}, part 11.}
	\begin{center}
		\begin{tabular}{ |c|c|c|c|c|c| }
			\hline
			number&d&r&N&A&B\\
			\hline
			351&17&16&12&21&252\\
			\hline
			352&17&16&13&19&247\\
			\hline
			353&17&16&14&19&266\\
			\hline
			354&17&17&10&28&280\\
			\hline
			355&17&17&13&22&286\\
			\hline
			356&17&17&14&20&280\\
			\hline
			357&17&17&15&19&285\\
			\hline
			358&17&17&16&18&288\\
			\hline
			359&17&18&10&29&290\\
			\hline
			360&17&18&11&25&275\\
			\hline
			361&17&18&13&23&299\\
			\hline
			362&17&18&14&19&266\\
			\hline
			363&17&18&15&19&285\\
			\hline
			364&17&18&16&19&304\\
			\hline
			365&17&19&12&26&312\\
			\hline
			366&17&19&13&24&312\\
			\hline
			367&17&19&15&21&315\\
			\hline
			368&17&19&16&20&320\\
			\hline
			369&17&20&13&23&299\\
			\hline
			370&17&20&14&23&322\\
			\hline
			371&17&20&15&21&315\\
			\hline
			372&17&20&16&21&336\\
			\hline
			373&17&21&13&26&338\\
			\hline
			374&17&21&15&23&345\\
			\hline
			375&17&21&16&22&352\\
			\hline
			376&17&21&17&20&340\\
			\hline
			377&17&21&18&19&342\\
			\hline
			378&17&22&15&23&345\\
			\hline
			379&17&22&16&23&368\\
			\hline
			380&17&22&17&21&357\\
			\hline
			381&17&22&18&19&342\\
			\hline
			382&17&22&19&19&361\\
			\hline
			383&17&23&16&24&384\\
			\hline
			384&17&23&18&21&378\\
			\hline
			385&17&23&19&20&380\\
			\hline
		\end{tabular}
	\end{center}
\end{table}

\begin{table}
	\label{table: method 2 parameters part 12}
	\caption{Parameters for the examples of UCPTP maps obtained using the method of section \ref{section: method 2 choice of parameters}, part 12.}
	\begin{center}
		\begin{tabular}{ |c|c|c|c|c|c| }
			\hline
			number&d&r&N&A&B\\
			\hline
			386&17&23&20&19&380\\
			\hline
			387&18&3&2&25&50\\
			\hline
			388&18&4&3&23&69\\
			\hline
			389&18&5&3&29&87\\
			\hline
			390&18&5&4&19&76\\
			\hline
			391&18&6&4&25&100\\
			\hline
			392&18&6&5&19&95\\
			\hline
			393&18&7&6&19&114\\
			\hline
			394&18&8&5&25&125\\
			\hline
			395&18&8&6&23&138\\
			\hline
			396&18&8&7&19&133\\
			\hline
			397&18&9&4&37&148\\
			\hline
			398&18&9&5&31&155\\
			\hline
			399&18&9&6&25&150\\
			\hline
			400&18&9&8&19&152\\
			\hline
			401&18&10&5&31&155\\
			\hline
			402&18&10&7&23&161\\
			\hline
			403&18&11&7&25&175\\
			\hline
			404&18&11&8&23&184\\
			\hline
			405&18&11&9&19&171\\
			\hline
			406&18&11&10&19&190\\
			\hline
			407&18&12&7&29&203\\
			\hline
			408&18&12&8&25&200\\
			\hline
			409&18&12&9&23&207\\
			\hline
			410&18&12&11&19&209\\
			\hline
			411&18&13&9&25&225\\
			\hline
			412&18&13&10&23&230\\
			\hline
			413&18&13&11&19&209\\
			\hline
			414&18&13&12&19&228\\
			\hline
			415&18&14&7&31&217\\
			\hline
			416&18&14&11&19&209\\
			\hline
			417&18&14&13&19&247\\
			\hline
			418&18&15&8&31&248\\
			\hline
			419&18&15&10&23&230\\
			\hline
			420&18&15&11&23&253\\
			\hline
		\end{tabular}
	\end{center}
\end{table}

\begin{table}
	\label{table: method 2 parameters part 13}
	\caption{Parameters for the examples of UCPTP maps obtained using the method of section \ref{section: method 2 choice of parameters}, part 13.}
	\begin{center}
		\begin{tabular}{ |c|c|c|c|c|c| }
			\hline
			number&d&r&N&A&B\\
			\hline
			421&18&15&13&19&247\\
			\hline
			422&18&15&14&19&266\\
			\hline
			423&18&16&10&25&250\\
			\hline
			424&18&16&11&25&275\\
			\hline
			425&18&16&12&23&276\\
			\hline
			426&18&16&13&19&247\\
			\hline
			427&18&16&15&19&285\\
			\hline
			428&18&17&10&29&290\\
			\hline
			429&18&17&11&25&275\\
			\hline
			430&18&17&12&25&300\\
			\hline
			431&18&17&13&23&299\\
			\hline
			432&18&17&14&19&266\\
			\hline
			433&18&17&15&19&285\\
			\hline
			434&18&17&16&19&304\\
			\hline
			435&18&18&11&29&319\\
			\hline
			436&18&18&13&23&299\\
			\hline
			437&18&18&17&19&323\\
			\hline
			438&18&19&12&25&300\\
			\hline
			439&18&19&13&25&325\\
			\hline
			440&18&19&14&23&322\\
			\hline
			441&18&20&13&23&299\\
			\hline
			442&18&20&14&23&322\\
			\hline
			443&18&20&15&23&345\\
			\hline
			444&18&20&17&19&323\\
			\hline
			445&18&21&14&25&350\\
			\hline
			446&18&21&15&25&375\\
			\hline
			447&18&21&16&23&368\\
			\hline
			448&18&21&17&19&323\\
			\hline
			449&18&22&13&29&377\\
			\hline
			450&18&22&15&25&375\\
			\hline
			451&18&22&16&23&368\\
			\hline
			452&18&22&17&23&391\\
			\hline
			453&18&22&19&19&361\\
			\hline
			454&18&23&15&25&375\\
			\hline
			455&18&23&16&25&400\\
			\hline
		\end{tabular}
	\end{center}
\end{table}

\begin{table}
	\label{table: method 2 parameters part 14}
	\caption{Parameters for the examples of UCPTP maps obtained using the method of section \ref{section: method 2 choice of parameters}, part 14.}
	\begin{center}
		\begin{tabular}{ |c|c|c|c|c|c| }
			\hline
			number&d&r&N&A&B\\
			\hline
			456&18&23&17&19&323\\
			\hline
			457&18&23&19&19&361\\
			\hline
			458&18&24&15&25&375\\
			\hline
			459&18&24&17&25&425\\
			\hline
			460&18&25&17&23&391\\
			\hline
			461&18&25&19&23&437\\
			\hline
			462&18&25&21&19&399\\
			\hline
			463&19&3&2&28&56\\
			\hline
			464&19&4&3&25&75\\
			\hline
			465&19&5&3&31&93\\
			\hline
			466&19&5&4&23&92\\
			\hline
			467&19&6&4&25&100\\
			\hline
			468&19&7&5&26&130\\
			\hline
			469&19&7&6&22&132\\
			\hline
			470&19&8&5&29&145\\
			\hline
			471&19&8&6&25&150\\
			\hline
			472&19&8&7&21&147\\
			\hline
			473&19&9&4&41&164\\
			\hline
			474&19&9&6&28&168\\
			\hline
			475&19&9&7&23&161\\
			\hline
			476&19&9&8&20&160\\
			\hline
			477&19&10&6&31&186\\
			\hline
			478&19&10&8&23&184\\
			\hline
			479&19&10&9&21&189\\
			\hline
			480&19&11&6&34&204\\
			\hline
			481&19&11&7&29&203\\
			\hline
			482&19&11&9&23&207\\
			\hline
			483&19&11&10&20&200\\
			\hline
			484&19&12&7&31&217\\
			\hline
			485&19&12&8&25&200\\
			\hline
			486&19&12&9&25&225\\
			\hline
			487&19&13&8&30&240\\
			\hline
			488&19&13&9&27&243\\
			\hline
			489&19&13&10&24&240\\
			\hline
			490&19&13&11&22&242\\
			\hline
		\end{tabular}
	\end{center}
\end{table}

\begin{table}
	\label{table: method 2 parameters part 15}
	\caption{Parameters for the examples of UCPTP maps obtained using the method of section \ref{section: method 2 choice of parameters}, part 15.}
	\begin{center}
		\begin{tabular}{ |c|c|c|c|c|c| }
			\hline
			number&d&r&N&A&B\\
			\hline
			491&19&13&12&20&240\\
			\hline
			492&19&14&9&29&261\\
			\hline
			493&19&14&10&25&250\\
			\hline
			494&19&14&11&23&253\\
			\hline
			495&19&15&7&37&259\\
			\hline
			496&19&15&8&34&272\\
			\hline
			497&19&15&9&31&279\\
			\hline
			498&19&15&11&23&253\\
			\hline
			499&19&15&12&23&276\\
			\hline
			500&19&16&10&29&290\\
			\hline
			501&19&16&11&27&297\\
			\hline
			502&19&16&12&25&300\\
			\hline
			503&19&16&13&23&299\\
			\hline
			504&19&16&14&21&294\\
			\hline
			505&19&17&9&35&315\\
			\hline
			506&19&17&12&26&312\\
			\hline
			507&19&17&13&24&312\\
			\hline
			508&19&17&15&21&315\\
			\hline
			509&19&17&16&20&320\\
			\hline
			510&19&18&12&25&300\\
			\hline
			511&19&18&13&25&325\\
			\hline
			512&19&18&14&23&322\\
			\hline
			513&19&19&11&32&352\\
			\hline
			514&19&19&13&27&351\\
			\hline
			515&19&19&14&25&350\\
			\hline
			516&19&19&16&22&352\\
			\hline
			517&19&19&17&21&357\\
			\hline
			518&19&19&18&20&360\\
			\hline
			519&19&20&11&33&363\\
			\hline
			520&19&20&15&23&345\\
			\hline
			521&19&20&16&23&368\\
			\hline
			522&19&20&17&21&357\\
			\hline
			523&19&20&18&21&378\\
			\hline
			524&19&21&12&32&384\\
			\hline
			525&19&21&13&29&377\\
			\hline
		\end{tabular}
	\end{center}
\end{table}

\begin{table}
	\label{table: method 2 parameters part 16}
	\caption{Parameters for the examples of UCPTP maps obtained using the method of section \ref{section: method 2 choice of parameters}, part 16.}
	\begin{center}
		\begin{tabular}{ |c|c|c|c|c|c| }
			\hline
			number&d&r&N&A&B\\
			\hline
			526&19&21&14&26&364\\
			\hline
			527&19&21&15&26&390\\
			\hline
			528&19&21&16&23&368\\
			\hline
			529&19&21&17&23&391\\
			\hline
			530&19&21&18&22&396\\
			\hline
			531&19&22&13&31&403\\
			\hline
			532&19&22&15&27&405\\
			\hline
			533&19&22&16&25&400\\
			\hline
			534&19&22&17&23&391\\
			\hline
			535&19&22&18&23&414\\
			\hline
			536&19&23&16&27&432\\
			\hline
			537&19&23&17&25&425\\
			\hline
			538&19&23&18&24&432\\
			\hline
			539&19&23&19&22&418\\
			\hline
			540&19&23&20&21&420\\
			\hline
			541&19&24&14&31&434\\
			\hline
			542&19&24&15&29&435\\
			\hline
			543&19&24&16&25&400\\
			\hline
			544&19&24&17&25&425\\
			\hline
			545&19&24&18&25&450\\
			\hline
			546&19&24&19&23&437\\
			\hline
			547&19&25&17&27&459\\
			\hline
			548&19&25&18&26&468\\
			\hline
			549&19&25&20&23&460\\
			\hline
			550&19&25&21&22&462\\
			\hline
			551&19&25&22&21&462\\
			\hline
			552&19&26&18&27&486\\
			\hline
			553&19&26&20&23&460\\
			\hline
			554&19&26&22&21&462\\
			\hline
			555&20&3&2&29&58\\
			\hline
			556&20&4&3&23&69\\
			\hline
			557&20&5&4&23&92\\
			\hline
			558&20&6&3&37&111\\
			\hline
			559&20&6&4&29&116\\
			\hline
			560&20&6&5&23&115\\
			\hline
		\end{tabular}
	\end{center}
\end{table}

\begin{table}
	\label{table: method 2 parameters part 17}
	\caption{Parameters for the examples of UCPTP maps obtained using the method of section \ref{section: method 2 choice of parameters}, part 17.}
	\begin{center}
		\begin{tabular}{ |c|c|c|c|c|c| }
			\hline
			number&d&r&N&A&B\\
			\hline
			561&20&7&3&43&129\\
			\hline
			562&20&7&5&27&135\\
			\hline
			563&20&7&6&23&138\\
			\hline
			564&20&8&6&23&138\\
			\hline
			565&20&8&7&21&147\\
			\hline
			566&20&9&6&29&174\\
			\hline
			567&20&9&7&23&161\\
			\hline
			568&20&10&7&27&189\\
			\hline
			569&20&10&8&23&184\\
			\hline
			570&20&10&9&21&189\\
			\hline
			571&20&11&6&31&186\\
			\hline
			572&20&11&7&31&217\\
			\hline
			573&20&11&8&27&216\\
			\hline
			574&20&11&9&23&207\\
			\hline
			575&20&12&7&31&217\\
			\hline
			576&20&12&8&29&232\\
			\hline
			577&20&12&9&23&207\\
			\hline
			578&20&12&10&23&230\\
			\hline
			579&20&13&8&31&248\\
			\hline
			580&20&13&9&27&243\\
			\hline
			581&20&13&10&23&230\\
			\hline
			582&20&13&11&23&253\\
			\hline
			583&20&13&12&21&252\\
			\hline
			584&20&14&8&33&264\\
			\hline
			585&20&14&10&27&270\\
			\hline
			586&20&14&11&23&253\\
			\hline
			587&20&14&12&23&276\\
			\hline
			588&20&15&9&31&279\\
			\hline
			589&20&15&10&29&290\\
			\hline
			590&20&15&11&23&253\\
			\hline
			591&20&15&12&23&276\\
			\hline
			592&20&16&9&33&297\\
			\hline
			593&20&16&12&23&276\\
			\hline
			594&20&16&13&23&299\\
			\hline
			595&20&16&15&21&315\\
			\hline
		\end{tabular}
	\end{center}
\end{table}

\begin{table}
	\label{table: method 2 parameters part 18}
	\caption{Parameters for the examples of UCPTP maps obtained using the method of section \ref{section: method 2 choice of parameters}, part 18.}
	\begin{center}
		\begin{tabular}{ |c|c|c|c|c|c| }
			\hline
			number&d&r&N&A&B\\
			\hline
			596&20&17&11&29&319\\
			\hline
			597&20&17&12&27&324\\
			\hline
			598&20&17&13&23&299\\
			\hline
			599&20&17&14&23&322\\
			\hline
			600&20&17&15&21&315\\
			\hline
			601&20&17&16&21&336\\
			\hline
			602&20&18&11&31&341\\
			\hline
			603&20&18&12&29&348\\
			\hline
			604&20&18&13&23&299\\
			\hline
			605&20&18&14&23&322\\
			\hline
			606&20&18&15&23&345\\
			\hline
			607&20&19&11&33&363\\
			\hline
			608&20&19&12&31&372\\
			\hline
			609&20&19&13&29&377\\
			\hline
			610&20&19&15&23&345\\
			\hline
			611&20&19&16&23&368\\
			\hline
			612&20&19&17&21&357\\
			\hline
			613&20&19&18&21&378\\
			\hline
			614&20&20&13&29&377\\
			\hline
			615&20&20&14&27&378\\
			\hline
			616&20&20&15&23&345\\
			\hline
			617&20&20&16&23&368\\
			\hline
			618&20&20&17&23&391\\
			\hline
			619&20&20&19&21&399\\
			\hline
			620&20&21&12&31&372\\
			\hline
			621&20&21&13&31&403\\
			\hline
			622&20&21&15&23&345\\
			\hline
			623&20&21&16&23&368\\
			\hline
			624&20&21&17&23&391\\
			\hline
			625&20&21&18&23&414\\
			\hline
			626&20&22&13&31&403\\
			\hline
			627&20&22&14&31&434\\
			\hline
			628&20&22&17&23&391\\
			\hline
			629&20&22&18&23&414\\
			\hline
			630&20&22&19&23&437\\
			\hline
		\end{tabular}
	\end{center}
\end{table}

\begin{table}
	\label{table: method 2 parameters part 19}
	\caption{Parameters for the examples of UCPTP maps obtained using the method of section \ref{section: method 2 choice of parameters}, part 19.}
	\begin{center}
		\begin{tabular}{ |c|c|c|c|c|c| }
			\hline
			number&d&r&N&A&B\\
			\hline
			631&20&23&13&33&429\\
			\hline
			632&20&23&14&31&434\\
			\hline
			633&20&23&15&29&435\\
			\hline
			634&20&23&16&27&432\\
			\hline
			635&20&23&18&21&378\\
			\hline
			636&20&23&19&21&399\\
			\hline
			637&20&24&17&23&391\\
			\hline
			638&20&24&18&23&414\\
			\hline
			639&20&24&19&23&437\\
			\hline
			640&20&25&18&27&486\\
			\hline
			641&20&25&19&23&437\\
			\hline
			642&20&25&20&23&460\\
			\hline
			643&20&25&21&23&483\\
			\hline
			644&20&26&17&29&493\\
			\hline
			645&20&26&18&27&486\\
			\hline
			646&20&26&19&27&513\\
			\hline
			647&20&26&20&23&460\\
			\hline
			648&20&26&21&23&483\\
			\hline
			649&20&26&22&23&506\\
			\hline
			650&20&26&23&21&483\\
			\hline
			651&20&27&19&23&437\\
			\hline
			652&20&27&21&23&483\\
			\hline
			653&20&27&22&23&506\\
			\hline
			654&20&28&21&23&483\\
			\hline
			655&20&28&22&23&506\\
			\hline
			656&20&28&23&23&529\\
			\hline
			657&21&3&2&31&62\\
			\hline
			658&21&4&3&25&75\\
			\hline
			659&21&5&3&34&102\\
			\hline
			660&21&5&4&26&104\\
			\hline
			661&21&6&4&31&124\\
			\hline
			662&21&6&5&25&125\\
			\hline
			663&21&7&4&34&136\\
			\hline
			664&21&7&5&29&145\\
			\hline
			665&21&7&6&23&138\\
			\hline
		\end{tabular}
	\end{center}
\end{table}

\begin{table}
	\label{table: method 2 parameters part 20}
	\caption{Parameters for the examples of UCPTP maps obtained using the method of section \ref{section: method 2 choice of parameters}, part 20.}
	\begin{center}
		\begin{tabular}{ |c|c|c|c|c|c| }
			\hline
			number&d&r&N&A&B\\
			\hline
			666&21&8&5&31&155\\
			\hline
			667&21&8&6&25&150\\
			\hline
			668&21&8&7&23&161\\
			\hline
			669&21&9&5&37&185\\
			\hline
			670&21&9&6&31&186\\
			\hline
			671&21&9&7&26&182\\
			\hline
			672&21&9&8&23&184\\
			\hline
			673&21&10&6&31&186\\
			\hline
			674&21&10&8&23&184\\
			\hline
			675&21&10&9&23&207\\
			\hline
			676&21&11&6&38&228\\
			\hline
			677&21&11&7&32&224\\
			\hline
			678&21&11&8&26&208\\
			\hline
			679&21&11&9&25&225\\
			\hline
			680&21&11&10&23&230\\
			\hline
			681&21&12&8&31&248\\
			\hline
			682&21&12&9&25&225\\
			\hline
			683&21&12&10&25&250\\
			\hline
			684&21&13&9&29&261\\
			\hline
			685&21&13&10&25&250\\
			\hline
			686&21&13&11&23&253\\
			\hline
			687&21&13&12&22&264\\
			\hline
			688&21&14&8&31&248\\
			\hline
			689&21&14&9&31&279\\
			\hline
			690&21&14&10&29&290\\
			\hline
			691&21&14&11&25&275\\
			\hline
			692&21&14&12&23&276\\
			\hline
			693&21&15&8&38&304\\
			\hline
			694&21&15&9&34&306\\
			\hline
			695&21&15&10&31&310\\
			\hline
			696&21&15&11&26&286\\
			\hline
			697&21&15&12&26&312\\
			\hline
			698&21&15&13&23&299\\
			\hline
			699&21&15&14&22&308\\
			\hline
			700&21&16&10&31&310\\
			\hline
		\end{tabular}
	\end{center}
\end{table}

\begin{table}
	\label{table: method 2 parameters part 21}
	\caption{Parameters for the examples of UCPTP maps obtained using the method of section \ref{section: method 2 choice of parameters}, part 21.}
	\begin{center}
		\begin{tabular}{ |c|c|c|c|c|c| }
			\hline
			number&d&r&N&A&B\\
			\hline
			701&21&16&11&29&319\\
			\hline
			702&21&16&12&25&300\\
			\hline
			703&21&16&13&25&325\\
			\hline
			704&21&16&14&23&322\\
			\hline
			705&21&17&9&38&342\\
			\hline
			706&21&17&10&32&320\\
			\hline
			707&21&17&11&32&352\\
			\hline
			708&21&17&12&29&348\\
			\hline
			709&21&17&13&26&338\\
			\hline
			710&21&17&14&25&350\\
			\hline
			711&21&17&15&23&345\\
			\hline
			712&21&17&16&22&352\\
			\hline
			713&21&18&10&37&370\\
			\hline
			714&21&18&11&31&341\\
			\hline
			715&21&18&14&25&350\\
			\hline
			716&21&18&15&25&375\\
			\hline
			717&21&18&16&23&368\\
			\hline
			718&21&19&10&37&370\\
			\hline
			719&21&19&11&34&374\\
			\hline
			720&21&19&12&32&384\\
			\hline
			721&21&19&13&29&377\\
			\hline
			722&21&19&14&26&364\\
			\hline
			723&21&19&15&26&390\\
			\hline
			724&21&19&16&23&368\\
			\hline
			725&21&19&17&23&391\\
			\hline
			726&21&19&18&22&396\\
			\hline
			727&21&20&12&31&372\\
			\hline
			728&21&20&13&31&403\\
			\hline
			729&21&20&15&23&345\\
			\hline
			730&21&20&16&23&368\\
			\hline
			731&21&20&17&23&391\\
			\hline
			732&21&20&18&23&414\\
			\hline
			733&21&21&13&32&416\\
			\hline
			734&21&21&15&29&435\\
			\hline
			735&21&21&16&26&416\\
			\hline
		\end{tabular}
	\end{center}
\end{table}

\begin{table}
	\label{table: method 2 parameters part 22}
	\caption{Parameters for the examples of UCPTP maps obtained using the method of section \ref{section: method 2 choice of parameters}, part 22.}
	\begin{center}
		\begin{tabular}{ |c|c|c|c|c|c| }
			\hline
			number&d&r&N&A&B\\
			\hline
			736&21&21&17&25&425\\
			\hline
			737&21&21&18&23&414\\
			\hline
			738&21&21&19&23&437\\
			\hline
			739&21&21&20&22&440\\
			\hline
			740&21&22&13&31&403\\
			\hline
			741&21&22&15&29&435\\
			\hline
			742&21&22&16&25&400\\
			\hline
			743&21&22&17&25&425\\
			\hline
			744&21&22&18&25&450\\
			\hline
			745&21&22&19&23&437\\
			\hline
			746&21&22&20&23&460\\
			\hline
			747&21&23&16&29&464\\
			\hline
			748&21&23&17&26&442\\
			\hline
			749&21&23&18&26&468\\
			\hline
			750&21&23&19&25&475\\
			\hline
			751&21&23&20&22&440\\
			\hline
			752&21&24&17&29&493\\
			\hline
			753&21&24&18&25&450\\
			\hline
			754&21&24&19&25&475\\
			\hline
			755&21&24&20&25&500\\
			\hline
			756&21&25&17&29&493\\
			\hline
			757&21&25&18&29&522\\
			\hline
			758&21&25&19&26&494\\
			\hline
			759&21&25&20&26&520\\
			\hline
			760&21&25&21&23&483\\
			\hline
			761&21&25&22&23&506\\
			\hline
			762&21&26&16&31&496\\
			\hline
			763&21&26&17&31&527\\
			\hline
			764&21&26&18&29&522\\
			\hline
			765&21&26&19&25&475\\
			\hline
			766&21&26&20&25&500\\
			\hline
			767&21&26&21&25&525\\
			\hline
			768&21&26&22&23&506\\
			\hline
			769&21&26&23&23&529\\
			\hline
			770&21&27&17&32&544\\
			\hline
		\end{tabular}
	\end{center}
\end{table}

\begin{table}
	\label{table: method 2 parameters part 23}
	\caption{Parameters for the examples of UCPTP maps obtained using the method of section \ref{section: method 2 choice of parameters}, part 23.}
	\begin{center}
		\begin{tabular}{ |c|c|c|c|c|c| }
			\hline
			number&d&r&N&A&B\\
			\hline
			771&21&27&19&29&551\\
			\hline
			772&21&27&20&26&520\\
			\hline
			773&21&27&22&25&550\\
			\hline
			774&21&27&23&23&529\\
			\hline
			775&21&27&24&23&552\\
			\hline
			776&21&28&19&29&551\\
			\hline
			777&21&28&20&29&580\\
			\hline
			778&21&28&22&25&550\\
			\hline
			779&21&28&23&25&575\\
			\hline
			780&21&28&24&23&552\\
			\hline
			781&21&28&25&23&575\\
			\hline
			782&21&29&22&26&572\\
			\hline
			783&21&29&24&25&600\\
			\hline
			784&21&29&26&23&598\\
			\hline
			785&22&3&2&31&62\\
			\hline
			786&22&4&3&29&87\\
			\hline
			787&22&5&3&31&93\\
			\hline
			788&22&5&4&27&108\\
			\hline
			789&22&6&4&31&124\\
			\hline
			790&22&6&5&25&125\\
			\hline
			791&22&7&4&37&148\\
			\hline
			792&22&7&5&29&145\\
			\hline
			793&22&7&6&25&150\\
			\hline
			794&22&8&7&25&175\\
			\hline
			795&22&9&5&37&185\\
			\hline
			796&22&9&6&31&186\\
			\hline
			797&22&9&7&25&175\\
			\hline
			798&22&9&8&23&184\\
			\hline
			799&22&10&6&31&186\\
			\hline
			800&22&10&7&31&217\\
			\hline
			801&22&10&8&27&216\\
			\hline
			802&22&10&9&23&207\\
			\hline
			803&22&11&6&39&234\\
			\hline
			804&22&11&7&31&217\\
			\hline
			805&22&11&8&29&232\\
			\hline
		\end{tabular}
	\end{center}
\end{table}

\begin{table}
	\label{table: method 2 parameters part 24}
	\caption{Parameters for the examples of UCPTP maps obtained using the method of section \ref{section: method 2 choice of parameters}, part 24.}
	\begin{center}
		\begin{tabular}{ |c|c|c|c|c|c| }
			\hline
			number&d&r&N&A&B\\
			\hline
			806&22&11&9&25&225\\
			\hline
			807&22&11&10&23&230\\
			\hline
			808&22&12&7&37&259\\
			\hline
			809&22&12&8&31&248\\
			\hline
			810&22&12&9&29&261\\
			\hline
			811&22&12&10&25&250\\
			\hline
			812&22&13&7&37&259\\
			\hline
			813&22&13&8&35&280\\
			\hline
			814&22&13&9&31&279\\
			\hline
			815&22&13&10&27&270\\
			\hline
			816&22&13&11&25&275\\
			\hline
			817&22&13&12&23&276\\
			\hline
			818&22&14&8&37&296\\
			\hline
			819&22&14&9&31&279\\
			\hline
			820&22&14&10&29&290\\
			\hline
			821&22&14&11&27&297\\
			\hline
			822&22&14&12&25&300\\
			\hline
			823&22&14&13&23&299\\
			\hline
			824&22&15&9&31&279\\
			\hline
			825&22&15&10&31&310\\
			\hline
			826&22&15&11&29&319\\
			\hline
			827&22&15&13&23&299\\
			\hline
			828&22&15&14&23&322\\
			\hline
			829&22&16&11&31&341\\
			\hline
			830&22&16&12&29&348\\
			\hline
			831&22&16&15&23&345\\
			\hline
			832&22&17&13&27&351\\
			\hline
			833&22&17&14&25&350\\
			\hline
			834&22&17&15&23&345\\
			\hline
			835&22&17&16&23&368\\
			\hline
			836&22&18&10&37&370\\
			\hline
			837&22&18&12&31&372\\
			\hline
			838&22&18&13&29&377\\
			\hline
			839&22&18&14&25&350\\
			\hline
			840&22&18&15&25&375\\
			\hline
		\end{tabular}
	\end{center}
\end{table}

\begin{table}
	\label{table: method 2 parameters part 25}
	\caption{Parameters for the examples of UCPTP maps obtained using the method of section \ref{section: method 2 choice of parameters}, part 25.}
	\begin{center}
		\begin{tabular}{ |c|c|c|c|c|c| }
			\hline
			number&d&r&N&A&B\\
			\hline
			841&22&18&17&23&391\\
			\hline
			842&22&19&12&31&372\\
			\hline
			843&22&19&13&31&403\\
			\hline
			844&22&19&14&29&406\\
			\hline
			845&22&19&15&27&405\\
			\hline
			846&22&19&16&25&400\\
			\hline
			847&22&19&17&23&391\\
			\hline
			848&22&19&18&23&414\\
			\hline
			849&22&20&12&31&372\\
			\hline
			850&22&20&13&31&403\\
			\hline
			851&22&20&14&31&434\\
			\hline
			852&22&20&15&29&435\\
			\hline
			853&22&20&16&27&432\\
			\hline
			854&22&20&17&23&391\\
			\hline
			855&22&20&19&23&437\\
			\hline
			856&22&21&12&37&444\\
			\hline
			857&22&21&13&31&403\\
			\hline
			858&22&21&14&31&434\\
			\hline
			859&22&21&15&29&435\\
			\hline
			860&22&21&16&25&400\\
			\hline
			861&22&21&17&25&425\\
			\hline
			862&22&21&18&25&450\\
			\hline
			863&22&21&19&23&437\\
			\hline
			864&22&21&20&23&460\\
			\hline
			865&22&22&14&31&434\\
			\hline
			866&22&22&15&31&465\\
			\hline
			867&22&22&16&29&464\\
			\hline
			868&22&22&17&27&459\\
			\hline
			869&22&22&18&25&450\\
			\hline
			870&22&22&19&25&475\\
			\hline
			871&22&22&21&23&483\\
			\hline
			872&22&23&13&37&481\\
			\hline
			873&22&23&14&35&490\\
			\hline
			874&22&23&15&31&465\\
			\hline
			875&22&23&16&31&496\\
			\hline
		\end{tabular}
	\end{center}
\end{table}

\begin{table}
	\label{table: method 2 parameters part 26}
	\caption{Parameters for the examples of UCPTP maps obtained using the method of section \ref{section: method 2 choice of parameters}, part 26.}
	\begin{center}
		\begin{tabular}{ |c|c|c|c|c|c| }
			\hline
			number&d&r&N&A&B\\
			\hline
			876&22&23&17&29&493\\
			\hline
			877&22&23&18&27&486\\
			\hline
			878&22&23&19&25&475\\
			\hline
			879&22&23&20&25&500\\
			\hline
			880&22&24&18&29&522\\
			\hline
			881&22&24&19&25&475\\
			\hline
			882&22&24&20&25&500\\
			\hline
			883&22&24&21&25&525\\
			\hline
			884&22&25&15&31&465\\
			\hline
			885&22&25&17&31&527\\
			\hline
			886&22&25&18&29&522\\
			\hline
			887&22&25&19&27&513\\
			\hline
			888&22&25&20&27&540\\
			\hline
			889&22&25&21&23&483\\
			\hline
			890&22&26&16&35&560\\
			\hline
			891&22&26&17&31&527\\
			\hline
			892&22&26&19&29&551\\
			\hline
			893&22&26&20&27&540\\
			\hline
			894&22&26&21&27&567\\
			\hline
			895&22&26&22&25&550\\
			\hline
			896&22&26&23&23&529\\
			\hline
			897&22&27&17&31&527\\
			\hline
			898&22&27&20&29&580\\
			\hline
			899&22&27&21&25&525\\
			\hline
			900&22&27&22&25&550\\
			\hline
			901&22&27&23&25&575\\
			\hline
			902&22&28&17&31&527\\
			\hline
			903&22&28&19&31&589\\
			\hline
			904&22&28&20&29&580\\
			\hline
			905&22&28&21&29&609\\
			\hline
			906&22&28&23&25&575\\
			\hline
			907&22&28&24&25&600\\
			\hline
			908&22&28&25&23&575\\
			\hline
			909&22&29&20&31&620\\
			\hline
			910&22&29&21&27&567\\
			\hline
		\end{tabular}
	\end{center}
\end{table}

\begin{table}[t]
	\label{table: method 2 parameters part 27}
	\caption{Parameters for the examples of UCPTP maps obtained using the method of section \ref{section: method 2 choice of parameters}, part 27.}
	\begin{center}
		\begin{tabular}{ |c|c|c|c|c|c| }
			\hline
			number&d&r&N&A&B\\
			\hline
			911&22&29&22&27&594\\
			\hline
			912&22&29&23&27&621\\
			\hline
			913&22&29&24&25&600\\
			\hline
			914&22&29&25&25&625\\
			\hline
			915&22&30&21&31&651\\
			\hline
		\end{tabular}
	\end{center}
\end{table}
\clearpage

\newpage
\bibliographystyle{alpha}
\bibliography{bibliography}

\end{document}